\documentclass[]{iopart}

\expandafter\let\csname equation*\endcsname\relax
\expandafter\let\csname endequation*\endcsname\relax

\usepackage{graphicx}
\usepackage{epstopdf}
\usepackage{cite}
\usepackage[perpage,marginal]{footmisc}
\usepackage{etoolbox}

\graphicspath{{sec-alt-th/}{sec-astro/}{sec-binaries/}{sec-compobj/}{sec-gr-tests/plots/}{sec-nr/}}

\usepackage{pdflscape}
\usepackage[unicode,colorlinks=true,linkcolor=blue,urlcolor=blue,citecolor=gray]{hyperref}
\usepackage[all]{hypcap}
\usepackage{amsmath,amsfonts,amssymb}
\usepackage{color}
\usepackage[usenames,dvipsnames]{xcolor}
\usepackage{bm}
\usepackage{longtable}
\usepackage[T1]{fontenc}
\usepackage[utf8]{inputenc}  
\usepackage{verbatim}
\usepackage{pifont}
\usepackage{yfonts}
\usepackage{url}
\usepackage{multirow}
\usepackage{microtype}

\makeatletter
\def\url@leostyle{%
  \@ifundefined{selectfont}{\def\UrlFont{\sf}}{\def\UrlFont{\small\ttfamily}}}
\makeatother
\newcommand{\be}{\begin{equation}}
\newcommand{\ee}{\end{equation}}
\newcommand{\bel}[1]{\begin{equation}\label{#1}}
\newcommand{\ba}{\begin{eqnarray}}
\newcommand{\bea}{\begin{eqnarray}}
\newcommand{\ea}{\end{eqnarray}}
\newcommand{\eea}{\end{eqnarray}}
\newcommand{\bal}[1]{\begin{eqnarray}\label{#1}}

\makeatletter
\def\@mkboth#1#2{}
\newlength\appendixwidth
\preto\appendix{\addtocontents{toc}{\protect\patchl@section}}
\newcommand{\patchl@section}{%
  \settowidth{\appendixwidth}{\textbf{Appendix }}%
  \addtolength{\appendixwidth}{1.5em}%
  \patchcmd{\l@section}{1.5em}{\appendixwidth}{}{\ddt}%
}
\makeatother

\def\a{\alpha}
\def\b{\beta}
\def\r{\rho}
\def\s{\sigma}

\def\m{\mu}
\def\n{\nu}

\def\Om{\Omega}

\def\d{\delta}

\def\pd{\partial}
\def\pa{\partial}

\def\f{\frac}
\def\nn{\nonumber}

\def\f{\frac}

\def\mpl{M_{\rm Pl}}

\usepackage{xspace}

\definecolor{darkgreen}{rgb}{0.2,0.7,0.2}

\usepackage[normalem]{ulem}
\usepackage{fancyhdr}

\makeatletter
\renewcommand\footnoterule{%
  \kern-3\p@
  \hrule\@width2.5cm
  \kern2.6\p@}
\makeatother

\numberwithin{equation}{section}
\numberwithin{figure}{section}

\newcommand{\sAE}{\ifmmode\text{\AE}\else\AE\fi}

\begin{document}

\pagestyle{fancy}
\lhead{Testing General Relativity}
\chead{}
\rhead{\thepage}
\lfoot{}
\cfoot{}
\rfoot{}

\begin{center}
\topical{Testing General Relativity with Present and Future Astrophysical Observations}
\end{center}

\author{%
Emanuele~Berti$^{1,2}$,
Enrico~Barausse$^{3,4}$,
Vitor~Cardoso$^{2,5}$,
Leonardo~Gualtieri$^{6}$,
Paolo~Pani$^{6,2}$,
Ulrich~Sperhake$^{7,8,1}$,
Leo~C.~Stein$^{8,9}$,
Norbert~Wex$^{10}$,
Kent~Yagi$^{11,12}$,
Tessa~Baker$^{13}$,
C.~P.~Burgess$^{14,5}$,
Fl\'avio~S.~Coelho$^{15}$,
Daniela~Doneva$^{16}$,
Antonio~De~Felice$^{17,18}$,
Pedro~G.~Ferreira$^{13}$,
Paulo~C.~C.~Freire$^{10}$,
James~Healy$^{19}$,
Carlos~Herdeiro$^{15}$,
Michael~Horbatsch$^{1}$,
Burkhard~Kleihaus$^{20}$,
Antoine~Klein$^{1}$,
Kostas~Kokkotas$^{16}$,
Jutta~Kunz$^{20}$,
Pablo~Laguna$^{21}$,
Ryan~N.~Lang$^{22,23,24}$,
Tjonnie~G.~F.~Li$^{25,26}$,
Tyson~Littenberg$^{27}$,
Andrew~Matas$^{28}$,
Saeed~Mirshekari$^{29}$,
Hirotada~Okawa$^{2}$,
Eugen~Radu$^{15}$,
Richard~O'Shaughnessy$^{19,22}$,
Bangalore~S.~Sathyaprakash$^{30}$,
Chris~Van~Den~Broeck$^{31}$,
Hans~A.~Winther$^{13}$,
Helvi~Witek$^{7}$,
Mir~Emad~Aghili$^{1}$,
Justin~Alsing$^{32}$,
Brett~Bolen$^{33}$,
Luca~Bombelli$^{1}$,
Sarah~Caudill$^{22}$,
Liang~Chen$^{1}$,
Juan~Carlos~Degollado$^{15}$,
Ryuichi~Fujita$^{2}$,
Caixia~Gao$^{1}$,
Davide~Gerosa$^{7}$,
Saeed~Kamali$^{1}$,
Hector~O.~Silva$^{1}$,
Jo\~ao~G.~Rosa$^{15}$,
Laleh~Sadeghian$^{22}$,
Marco~Sampaio$^{15}$,
Hajime~Sotani$^{34}$
and
Miguel~Zilhao$^{19}$
}

\address{$^{1}$~Department of Physics and Astronomy, The University of Mississippi, University, MS 38677-1848, USA}
\address{$^{2}$~CENTRA, Departamento de F\'isica, Instituto Superior T\'ecnico, Universidade de Lisboa, Avenida Rovisco Pais 1, 1049 Lisboa, Portugal}
\address{$^{3}$~CNRS, UMR 7095, Institut d'Astrophysique de Paris, 98bis Bd Arago, 75014 Paris, France}
\address{$^{4}$~Sorbonne Universit{\'e}s, UPMC Univ Paris 06, UMR 7095, 98bis Bd Arago, 75014 Paris, France}
\address{$^{5}$~Perimeter Institute for Theoretical Physics, Waterloo, Ontario N2L 2Y5, Canada}
\address{$^{6}$~Dipartimento di Fisica, ``Sapienza'' Universit\`a di Roma \& Sezione INFN Roma 1, P.le A. Moro 2, 00185 Roma, Italy}
\address{$^{7}$~Department of Applied Mathematics and Theoretical Physics, Centre for Mathematical Sciences, University of Cambridge, Wilberforce Road, Cambridge CB3 0WA, UK}
\address{$^{8}$~Theoretical Astrophysics 350-17, California Institute of Technology, Pasadena, CA 91125, USA}
\address{$^{9}$~Cornell Center for Astrophysics and Planetary Science, Cornell University, Ithaca, NY 14853, USA}
\address{$^{10}$~Max-Planck-Institut f\"ur Radioastronomie, Auf dem H\"ugel 69, D-53121 Bonn, Germany}
\address{$^{11}$~Department of Physics, Princeton University, Princeton, NJ 08544, USA}
\address{$^{12}$~Department of Physics, Montana State University, Bozeman, Montana 59717, USA}
\address{$^{13}$~Astrophysics, University of Oxford, DWB, Keble Road, Oxford, OX1 3RH, UK}
\address{$^{14}$~Department of Physics \& Astronomy, McMaster University, Hamilton ON, Canada}
\address{$^{15}$~Departamento de F\'isica da Universidade de Aveiro and CIDMA Campus de Santiago, 3810-183 Aveiro, Portugal}
\address{$^{16}$~Theoretical Astrophysics, Eberhard Karls University of T\"ubingen, T\"ubingen 72076, Germany}
\address{$^{17}$~ThEP's CRL, NEP, The Institute for Fundamental Study, Naresuan University, Phitsanulok 65000, Thailand}
\address{$^{18}$~Thailand Center of Excellence in Physics, Ministry of Education, Bangkok 10400, Thailand}
\address{$^{19}$~Center for Computational Relativity and Gravitation, School of Mathematical Sciences, Rochester Institute of Technology, 85 Lomb Memorial Drive, Rochester, NY 14623, USA}
\address{$^{20}$~Institut f\"ur Physik, Universit\"at Oldenburg, D-26111 Oldenburg, Germany}
\address{$^{21}$~Center for Relativistic Astrophysics and School of Physics, Georgia Institute of Technology, Atlanta, GA 30332, USA}
\address{$^{22}$~Leonard E Parker Center for Gravitation, Cosmology, and Astrophysics, University of Wisconsin-Milwaukee, Milwaukee, WI 53211, USA}
\address{$^{23}$~Department of Physics, University of Florida, Gainesville, FL 32611, USA}
\address{$^{24}$~Department of Physics, University of Illinois at Urbana-Champaign, Urbana, IL 61801, USA}
\address{$^{25}$~Department of Physics, The Chinese University of Hong Kong, Shatin, N.T., Hong Kong, China}
\address{$^{26}$~LIGO --- California Institute of Technology, Pasadena, CA 91125, USA}
\address{$^{27}$~Center for Interdisciplinary Exploration and Research in Astrophysics (CIERA) \& Dept. of Physics and Astronomy, Northwestern University, 2145 Sheridan Rd., Evanston, IL 60208, USA}
\address{$^{28}$~CERCA/Department of Physics, Case Western Reserve University, 10900 Euclid Ave, Cleveland, OH 44106, USA}
\address{$^{29}$~ICTP South American Institute for Fundamental Research \& Instituto de F\'{\i}sica Te\'orica, UNESP - Universidade Estadual Paulista, Rua Dr.~Bento T.~Ferraz 271 - 01140-070  S\~ao Paulo, SP, Brazil}
\address{$^{30}$~School of Physics and Astronomy, Cardiff University, 5, The Parade, Cardiff CF24 3AA, UK}
\address{$^{31}$~Nikhef, Science Park, 1098 XG Amsterdam, The Netherlands}
\address{$^{32}$~Imperial Centre for Inference and Cosmology, Imperial College London, Prince Consort Road, London SW7 2AZ, UK}
\address{$^{33}$~Department of Physics, Grand Valley State University, Allendale, MI 49401-9403, USA}
\address{$^{34}$~Yukawa Institute for Theoretical Physics, Kyoto University, Kyoto 606-8502, Japan}

\vspace{6em}

\begin{abstract}
One century after its formulation, Einstein's general relativity has
made remarkable predictions and turned out to be compatible with all
experimental tests. Most of these tests probe the theory
in the weak-field regime, and there are theoretical and experimental
reasons to believe that general relativity should be modified when
gravitational fields are strong and spacetime curvature is large. The
best astrophysical laboratories to probe strong-field gravity are
black holes and neutron stars, whether isolated or in binary
systems. We review the motivations to consider extensions of general
relativity. We present a (necessarily incomplete) catalog of modified
theories of gravity for which strong-field predictions have been
computed and contrasted to Einstein's theory, and we summarize our
current understanding of the structure and dynamics of compact objects
in these theories. We discuss current bounds on modified gravity from
binary pulsar and cosmological observations, and we highlight the
potential of future gravitational wave measurements to inform us on
the behavior of gravity in the strong-field regime.
\end{abstract}

\pacs{04.20.-q, 04.30.Tv, 04.40.Dg, 04.70.-s, 04.80.Cc, 04.80.Nn}

\clearpage
{\tableofcontents}

\clearpage

\section{Introduction}
\label{sec:intro}

Einstein's theory of general relativity (GR), together with quantum
mechanics, is one of the pillars of modern physics. The theory has
passed all precision tests to date with flying colors. Most of these
tests -- with the possible exception of
binary pulsar observations -- are probes of {\em weak-field} gravity;
more precisely, they probe gravity at intermediate length
($1\,\mu{\rm m}\lesssim \ell \lesssim 1$~AU $\sim 10^{11}$m)
and therefore intermediate energy scales. Laboratory experiments and
astrophysical observations verify the so-called ``Einstein equivalence
principle'' (i.e.~the weak equivalence principle supplemented by local
Lorentz invariance and local position invariance) and they set
constraints on hypothetical weak-field deviations from GR, as encoded
in the parametrized post-Newtonian (PPN) formalism (see~\cite{PW:2014}
for an introduction, and~\cite{Will:2014xja} for a review of the state
of the art on experimental tests of GR).

The conceptual foundations of GR are so elegant and solid that when
asked what he would do if Eddington's expedition to the island of
Principe failed to match his theory, Einstein famously replied: ``I
would feel sorry for the good Lord.  The theory is correct.''
Chandrasekhar made a similar private remark to Clifford Will when Will
was a postdoc in Chicago: ``Why do you spend so much time and energy
testing GR? We {\em know} that the theory is right.''
Giving up the fundamental, well tested principles underlying
Einstein's theory has dramatic consequences, often spoiling the beauty
and relative simplicity of Einstein's theory. However, there is
growing theoretical and experimental evidence that modifications of GR
at small and large energies are somehow inevitable.

From a theoretical point of view, GR is a purely classical
theory. Power-counting arguments indicate that GR is not
renormalizable in the standard quantum field theory
sense. Strong-field modifications may provide a solution to this
problem: it has long been known that the theory becomes renormalizable
if we add quadratic curvature terms -- i.e.,
high-energy/high-curvature corrections -- to the Einstein-Hilbert
action~\cite{Stelle:1976gc}. Furthermore, high-energy corrections can
avoid the formation of singularities that are inevitable in classical
GR, as shown by the Hawking-Penrose singularity
theorems~\cite{Hawking:1969sw}. Candidate theories of quantum gravity
(such as string theory and loop quantum gravity) make specific and
potentially testable predictions of {\em how} GR must be modified at
high energies.

From an observational point of view, cosmological measurements are
usually interpreted as providing evidence for dark matter and a
nonzero cosmological constant (``dark energy''). This interpretation
poses serious conceptual issues, including the cosmological constant
problem (``why is the observed value of the cosmological constant so
small in Planck units?'') and the coincidence problem (``why is the
energy density of the cosmological constant so close to the present
matter density?''). No dynamical solution of the cosmological constant
problem is possible within GR~\cite{Weinberg:1988cp}. It seems
reasonable that ultraviolet corrections to GR would inevitably
``leak'' down to cosmological scales, showing up as low-energy
(infrared) corrections.

The arguments summarized above suggest that GR should be modified at
both low and high energies. This is a serious challenge for theorists.
Einstein's theory is the unique interacting theory of a
Lorentz-invariant massless helicity-2 particle~\cite{Deser:1969wk,Wald:1986bj},
and therefore new physics in the gravitational sector must introduce
additional degrees of freedom.
Any additional degrees of freedom must modify the theory at low and/or
high energies {\em while being consistent with GR in the
  intermediate-energy regime}, i.e.~at length scales $1\,\mu{\rm
  m}\lesssim \ell \lesssim 10^{11}$m, where the theory is extremely
well tested.  Laboratory, Solar System and binary-pulsar experiments
verify the Einstein equivalence principle to remarkable accuracy; they
force PPN parameters to be extremely close to their GR values; and (as
we will see below) they place stringent bounds on popular extensions
of GR, such as scalar-tensor theories and Lorentz-violating theories
(see~\cite{Will:2014xja,Wex:2014nva} for reviews).

Some confusion exists about how to link tests of gravity that take
place in the very different regimes described above. For example,
though it is agreed that strong-field constraints on GR do not rule
out cosmological modifications (or vice-versa), it is not immediately
obvious how to express this statement quantitatively, except perhaps in
specific models. One method of resolving this problem was recently put
forward in~\cite{Baker:2014zba}. There the authors place a wide range
of laboratory, astrophysical and cosmological systems on a
two-dimensional parameter space, where the axis quantities are the
approximate gravitational potential and Kretschmann scalar of a
system. The Kretschmann scalar is used because it gives a rough
measure of how relativistic the system is and does not vanish in
vacuum (the diagnostic power of the Ricci scalar is limited for this
reason).
Many orders of magnitude of Kretschmann curvature separate the classic
PPN tests of gravity from both the strong-field regime and the
cosmological regime, so we cannot simply take existing Solar System
constraints as comprehensive.

The main focus of this review is on present and future tests of {\em
  strong-field gravity}. It is useful to classify ``tests of
strong-field gravity'' as belonging to two qualitatively different
categories.  ``External tests'' are laboratory experiments,
astrophysical and cosmological observations that can be used to
determine whether GR (as opposed to any of the numerous proposed
extensions) is the correct theory of gravity. ``Internal tests'' are
observations that tell us whether some key predictions of GR (e.g.~the
Kerr solution of the Einstein equations in vacuum, or the radiative
dynamics of compact objects) are ``internally'' consistent with
astrophysical observations.

Compact objects such as black holes (BHs) and neutron stars (NSs) are
our best natural laboratories to constrain strong gravity. In these
celestial bodies gravity prevails over all other interactions, and
collapse leads to large-curvature, strong-gravity environments (see
e.g.~\cite{Psaltis:2008bb}). The Kerr metric is
a solution of the vacuum field equations in a large class of
modified gravity theories, but theories that differ from GR
generically predict different {\em dynamics} and different
gravitational-wave (GW) signatures when compact objects are displaced
from equilibrium and/or when they merge. This is the reason why a
large part of this review will be devoted to the structure and
dynamics of BHs and NSs, whether isolated or in binary systems.

\subsection{Taxonomy of proposed extensions of general relativity}

To frame external tests in terms of hypothesis testing, one would like
to have one or more valid alternatives to GR. What constitutes a
``valid alternative'' is, of course, a matter of taste. From our
perspective (i.e., in terms of tests of strong-field gravity) the
alternative should be a cosmologically viable fundamental theory
passing intermediate energy tests, with a well-posed initial value
formulation, and field equations that follow from an action
principle. Furthermore, the theory should be simple enough to make
definite, calculable prediction in the strong-field regime: ideally,
it should allow us to predict the structure and dynamics of compact
objects and the gravitational radiation that they emit, whether
isolated or in binary systems.

This is a very stringent set of requirements. There are countless
attempts to modify
GR~\cite{Sotiriou:2008rp,DeFelice:2010aj,Nojiri:2010wj,Capozziello:2011et,Clifton:2011jh,Hinterbichler:2011tt,deRham:2014zqa},
but (for the reasons listed above) in several cases the modifications
introduce some screening mechanism in order to be viable at
intermediate energies. Screening mechanisms include chameleons,
symmetrons, dilatons,
MOND-like dynamics,
the Vainshtein mechanism, etcetera,
depending on whether the screening is set by the local value of the
field or by its derivatives~\cite{Joyce:2014kja}.

Chapter~\ref{sec:alt-th} reviews various theories that have been
explored in some detail as phenomenological alternatives to GR in the
strong-field regime.
The chapter begins with a discussion of Lovelock's theorem, a
``uniqueness theorem'' for the field equations of GR. Uniqueness is
based on a small set of definite assumptions. The interest of
Lovelock's theorem from a pragmatic point of view is that it can be
``turned around,'' and used to classify extensions of GR based on
which of the underlying assumptions of Lovelock's theorem they
violate. Within this classification framework, we list and discuss
several theories that have been seriously considered as plausible
alternatives to GR in the context of strong-field tests. This
selection is necessarily incomplete, and the authors of this review
have different opinions on the intrinsic merits, viability and
aesthetic appeal of these theories. The main criterion we used to
choose these particular theories is that they are simple enough to
make definite (and sometimes ``orthogonal'') predictions for the
strong-field dynamics of compact objects. The theories we discuss
include:

\begin{itemize}
\item[1)] scalar-tensor theories and their generalizations (including
  tensor-multiscalar and Horndeski theories);
\item[2)] $f(R)$ theories;
\item[3)] theories whose action contains terms quadratic in the
  curvature, including in particular Einstein-dilaton-Gauss-Bonnet
  (EdGB) and dynamical Chern-Simons (dCS) theories;
\item[4)] Lorentz-violating theories, including Einstein-\AE ther,
  Ho\v rava and $n$-Dirac-Born-Infeld ($n$-DBI) gravity;
\item[5)] massive gravity theories;
\item[6)] theories involving nondynamical fields, including the
  Palatini formulation of $f(R)$ gravity and Eddington-inspired
  Born-Infeld (EiBI) gravity.
\end{itemize}
    
This broad classification will be a leitmotif of the
review. Table~\ref{tab:theories} lists some key references to the
literature on the
various theories listed above, plus others that are not considered in
depth here. The table is an incomplete (but hopefully useful) ``bird's
eye'' reference guide for further study. Similar tables following the
same classification scheme will support our discussion of the
structure and stability of compact objects.

Since we do not have a full theory of quantum gravity, an effective
field theory (EFT) approach is often invoked when constructing
phenomenological alternatives to
GR~\cite{Burgess:2003jk,Burgess:2007pt}. For example, not all theories
of gravity with action quadratic in the curvature (item 3 in the list)
are acceptable: the equations are of second order in the
strong-coupling limit (a very desirable feature, given that
higher-order derivatives are vulnerable to the so-called Ostrogradskii
instability~\cite{Woodard:2006nt}) only if the quadratic invariants
appear in the special ``Gauss-Bonnet'' combination. To avoid
higher-order derivatives in the equations of motion one must generally
assume that couplings are small, and work in an EFT framework. A more
detailed discussion of EFTs and further references can be found in
Section~\ref{sec:EFT}.

\subsection{Compact objects in modified theories of gravity}

Investigations of compact objects, binary pulsars, cosmology and
gravitational radiation vary in depth and scope for the various
classes of theories listed above. The best studied examples include
scalar-tensor theories and some forms of quadratic gravity.
Chapters~\ref{sec:BHs},~\ref{sec:NSs} and~\ref{sec:CB} are devoted to
a discussion of isolated BHs, isolated NSs and compact binary systems
in various theories.

\paragraph{Isolated black holes.}
In Chapter~\ref{sec:BHs} we discuss BHs, one of the most striking
predictions of GR. There is a consensus in the astronomy community
that the massive compact objects in galactic centers, as well as the
compact objects with mass larger than about $3M_\odot$ found in some
low-mass X-ray binaries, are well described by the Kerr solution in
GR. However, this ``BH paradigm'' rests on somewhat shaky foundations.

From a theorist's point of view, one of the most convincing arguments
in favor of the BH paradigm is that the alternatives are either
unstable (as in the case of dense star clusters, fermion stars or
naked singularities), unnatural (e.g.~``exotic'' matter violating some
of the energy conditions), contrived (such as gravastars), more
implausible than BHs as the end-point of gravitational collapse (boson
stars) or nearly indistinguishable from Kerr.

The experimental evidence that astronomical BH candidates possess
event horizons (more correctly, apparent horizons) rather than solid
surfaces usually rests on plausibility arguments based on accretion
physics~\cite{Narayan:2005ie,Narayan:2013gca}. All of these arguments
are model-dependent, and they leave room for some skepticism~(see
e.g.~\cite{Abramowicz:2002vt}).

Strictly speaking, any tests that probe the Kerr {\em metric} alone
(such as tests based on matter accretion or ray tracing of photon
trajectories) are of little value as internal tests of GR. The reason
is that a large number of extensions of GR admit the Kerr metric as a
solution, and the theories that do not (e.g.~EdGB, dCS and some
Lorentz-violating gravity theories) predict BH solutions that differ
from GR by amounts that may not be astrophysically measurable.
Despite this somewhat pessimistic caveat, many ``quasi-Kerr metrics''
have been proposed to perform GR tests, and we will review these
proposals in Chapter~\ref{sec:BHs}. Most deformations of the Kerr
metric should be viewed as unnatural strawmen: they often have serious
pathologies, and they are therefore unacceptable even for the limited
scope of parametrizing deviations from the Kerr metric~\cite{Johannsen:2013rqa}.

The prospects for testing GR with BHs look brighter when we recall
that all extensions of GR predict different {\em dynamics} and
different GW signatures when compact objects are perturbed away from
equilibrium and/or when they merge. These arguments suggest that the
most promising way to verify whether the compact objects in galactic
centers or low-mass X-ray binaries are actually Kerr BHs is via direct
observation of gravitational radiation, especially in the strong-field
merger/ringdown phase.

Last but not least, astrophysical BHs can be used to constrain
modifications of GR in a different way. Many proposed modifications of
Einstein's theory and extension of the Standard Model of particle
physics predict the existence of light bosonic degrees of
freedom. Light bosons can trigger a superradiant instability, that
extracts angular momentum from rotating BHs. By setting the
superradiant instability timescale equal to the typical timescale for
accretion to spin up the hole (say, the Salpeter time) one can get
very stringent constraints on the allowed masses of light bosons
(e.g.~axions, Proca fields or massive gravitons).

Table~\ref{tab:BHsummary} is a quick reference guide to BH solutions
and stability in various modified theories of gravity, organized in
the same way as Table~\ref{tab:theories}.

\paragraph{Isolated neutron stars.}
In Chapter~\ref{sec:NSs} we discuss NS solutions and their stability
in various extensions of GR. Among other topics, we review the
possibility that NSs in scalar-tensor theory may significantly deviate
from their GR counterparts in the presence of ``spontaneous
scalarization'' (a phase transition akin to spontaneous
magnetization~\cite{Damour:1993hw}), we discuss controversial claims on the existence of
NSs in $f(R)$ theories, and we review the somewhat surprising
``no-hair'' properties of NSs in quadratic gravity.

A major problem in carrying out strong-gravity tests with NSs is the
degeneracy between our ignorance of the equation of state (EOS) of
high-density matter and strong-gravity effects. A possibility to lift
the degeneracy consists of using universal relations between the
moment of inertia, Love number (a measure of tidal deformability) and
quadrupole moment of a NS -- the so-called ``I-Love-Q'' relations~\cite{Yagi:2013bca} -- as well as EOS-independent relations between
the lowest three multipole moments and those of higher order~\cite{Pappas:2013naa,Yagi:2014bxa}. Section~\ref{subsec:ILQ} overviews
the promises and challenges of this approach.

A property of isolated NSs that plays an important role in many
extensions of GR is their ``sensitivity.'' The sensitivity is a
measure of how the gravitational mass of the NS (or any
self-gravitating object) varies as it moves within the nonhomogeneous
extra field(s) mediating the gravitational interactions -- or in other
words, a measure of the violation of the strong equivalence principle
in the theory in question. Chapter~\ref{sec:sensitivities} is a review
of sensitivity calculations, that play an important role in binary
dynamics.

In Table~\ref{tab:NSsummary} we give a quick reference guide to NS
solutions and their stability in various modified theories of gravity.

\paragraph{Compact binaries.}
In preparation for binary pulsar tests (covered in
Chapter~\ref{sec:BP}) and GW tests (Chapter~\ref{sec:GW}), in
Chapter~\ref{sec:CB} we review calculations of compact binary dynamics
in some extensions of GR. The equations of motion and GW fluxes have
been derived using the post-Newtonian (PN) expansion -- an expansion
in powers of $v/c$, where $v$ is the orbital velocity of the binary --
in scalar-tensor theory, $f(R)$ gravity, specific forms of quadratic
gravity (including EdGB and dCS) and Lorentz-violating theories. In
comparison, numerical work is much less developed: at the moment of
writing this review, simulations of compact binary mergers were
carried out only for some of the simplest scalar-tensor theories.

\subsection{Present and future tests of strong gravity}

Chapters~\ref{sec:BP} and~\ref{sec:GW} capitalize on the material
covered in previous chapters. Chapter~\ref{sec:BP} reviews {\em
  present} astrophysical tests of GR, more specifically those coming
from binary pulsar and cosmological observations. Chapter~\ref{sec:GW}
focuses on the potential payoff of {\em future} GW observations, and
on how astrophysical modeling will affect our ability to perform tests
of strong-field gravity in this context.

The first part of Chapter~\ref{sec:BP} is an overview of the
spectacular progress of GR tests from binary pulsars. These
extraordinary natural laboratories can be utilized to probe with high
precision various nonradiative strong-field effects, as well as
radiative aspects of gravity~\cite{1992Natur.355..132T}. For instance,
pulsars are now able to test Einstein’s quadrupole formula for GW
emission to an accuracy of less than $0.1\%$. They provide stringent
bounds on dipolar radiation and on violations of the strong
equivalence principle by strongly self-gravitating bodies (the best
tests coming from pulsar-white dwarf systems), and they tightly
constrain hypothetical violations of local Lorentz invariance of
gravity. The near future in this field is particularly
bright. Facilities such as the Five-hundred meter Aperture Spherical
radio Telescope (FAST) and the Square Kilometer Array (SKA) are
expected to come online soon. They should provide drastic improvements
in the precision of current tests, qualitatively new tests with
already known systems, and the discovery of many new ``pulsar
laboratories'' (possibly including the first pulsar-BH system).

The second part of Chapter~\ref{sec:BP} reviews cosmological tests of
GR. In the last few decades, a remarkable wealth of astronomical data
has constrained the expansion rate of the Universe and provided
accurate maps of large-scale structure and the cosmic microwave
background, placing ever-tightening constraints on cosmological
parameters.
In particular, anisotropies in the cosmic microwave background encode
information on the geometry of the Universe, its material constituents
and the initial conditions for structure formation. If GR is assumed
to be correct, 96\% of the material content of the Universe must
consist of dark matter and dark energy. Since the evidence for these
dark constituents of the Universe is purely gravitational, there have
been countless attempts at finding theories in which dark matter and
dark energy arise from modifications of gravity. These modifications
affect the expansion rate of the Universe, but they should also affect
gravitational clustering in a way that might be distinguishable from
GR. The proliferation of alternative theories of gravity has led to
the development of model-independent cosmological tests of modified
gravity somewhat similar to the PPN framework, which are now one of
the primary drivers for future surveys of large scale structure. In
the linear regime, these model-independent tests can be grouped in
three classes, corresponding to three manifestations of a gravity
theory: the action, the field equations derived from that action, and
the combinations of those field equations which influence observable
quantities. Chapter~\ref{sec:BP} reviews these tests as well as recent
progress in the nonlinear regime, where screening effects are
important and numerical simulations are necessary.

Last but not least, in Chapter~\ref{sec:GW} we turn our attention to
the future of strong-gravity tests, focusing on the promise of GW
observations by Earth- and space-based detectors. The main target for
both classes of detectors is the inspiral and merger of compact
binaries. A technique called matched filtering, based on a careful
monitoring of the GW phase to extract the (generally weak) signal from
the detector's noise, is used to observe these systems and to measure
their parameters. GR makes very specific and testable predictions on
the GW phasing of compact binaries as they inspiral, and on the
oscillation frequencies of the compact objects that they produce as a
result of the merger. If observed, any deviations from these
predictions may identify problems in Einstein's theory, and even point
us to specific ways in which it could be modified.

There are several comprehensive reviews on GW-based tests of GR. In
particular, the recent {\em Living Reviews in Relativity} article by
Yunes and Siemens~\cite{Yunes:2013dva} provides an excellent
introduction to the literature on GR tests with Earth-based detectors
(such as Advanced LIGO/Virgo, LIGO-India and KAGRA) and Pulsar Timing
Arrays, and the review by Gair et al.~\cite{Gair:2012nm} expounds the
great potential of future space-based detectors such as eLISA.
We find it unnecessary to reproduce that material here, and therefore
we focus on aspects that are not covered in detail in those reviews,
namely: (1) the data analysis implementation of GR tests in advanced
Earth-based detectors (the TIGER framework), arguably our best hope to
constrain modified gravity using GW observations in the near future;
and (2) an analysis of how astrophysical effects can limit (or
sometimes enhance) our ability to test strong-field gravity with GW
observations.

As a rule, in this paper we use geometrical units where the
gravitational constant and the speed of light are set to unity:
$G_N=c=1$.  Factors of $G_N$ and $c$ are occasionally reinstated for
clarity, and in isolated cases (e.g. in Section~\ref{sec:EFT}) we
switch to units such that $\hbar=c=1$. We adopt the mostly positive
signature for the metric, and the same conventions as in Misner,
Thorne, and Wheeler~\cite{MTW} for the Riemann tensor.

\begin{landscape}
{
\newcommand{\yes}{Yes}
\newcommand{\no}{\bf{No}}
\renewcommand{\yes}{\checkmark}
\renewcommand{\no}{\ding{55}}

\newcommand{\mypbox}[2]{\multicolumn{1}{p{#1}}{\centering #2}}

\begin{table}[t]
\capstart
\begin{tabular}{@{}l|c|ccccc|cc@{}}
\hline \noalign{\smallskip}\hline \noalign{\smallskip}
  \multicolumn{1}{c|}{Theory}
  &\multicolumn{1}{p{1.5cm}|}{\centering Field \\ content}
  &\mypbox{1.3cm}{\centering Strong \\ EP}
  &\mypbox{1.3cm}{ Massless \\ graviton}
  &\mypbox{1.3cm}{\centering Lorentz \\ symmetry}
  &\mypbox{1.1cm}{\centering Linear \\ $T_{\mu\nu}$}
  &\multicolumn{1}{p{1.1cm}|}{\centering Weak \\ EP}
  &\mypbox{1.3cm}{\centering Well- \\ posed?}
  & \mypbox{2cm}{\centering Weak-field \\ constraints}\\ 
\noalign{\smallskip}
\hline 
Extra scalar field\\	
\hspace{0.5cm} Scalar-tensor    &S		&\no &\yes &\yes &\yes &\yes &\yes\cite{Salgado:2008xh}           &\cite{Bertotti:2003rm,Alsing:2011er,Freire:2012mg}\\
\hspace{0.5cm} Multiscalar      &S	&\no &\yes &\yes &\yes &\yes  	&\yes\cite{Choquet-Bruhat:2009xil}          &\cite{Damour:1992we}\\
\hspace{0.5cm} Metric $f(R)$    &S	&\no &\yes &\yes &\yes &\yes  	&\yes\cite{LanahanTremblay:2007sg,Paschalidis:2011ww}           &\cite{Berry:2011pb}\\
\hspace{0.5cm} Quadratic gravity   	\\
\hspace{1cm} Gauss-Bonnet  	&S	&\no &\yes &\yes &\yes &\yes  	&\yes?          &\cite{Yagi:2012gp}\\
\hspace{1cm} Chern-Simons 	&P	&\no &\yes &\yes &\yes &\yes  	&\no\yes?~\cite{Delsate:2014hba}              &\cite{AliHaimoud:2011fw}\\
\hspace{1cm} Generic   		&S/P	&\no &\yes &\yes &\yes &\yes  	&?          &\\
\hspace{0.5cm} Horndeski 	&S &\no &\yes &\yes &\yes &\yes  	&\yes?          &\\
Lorentz-violating\\	
\hspace{0.5cm} \AE-gravity   &SV		&\no &\yes &\no &\yes &\yes  	&\yes?           &\cite{Foster:2005dk,Jacobson:2008aj,Yagi:2013qpa,Yagi:2013ava}\\
\hspace{0.5cm} Khronometric/&&&&&&&&\\
\hspace{0.5cm} Ho\v{r}ava-Lifshitz  &S 	&\no &\yes &\no &\yes &\yes 	&\yes?           &\cite{Blas:2010hb,Blas:2011zd,Yagi:2013qpa,Yagi:2013ava}\\
\hspace{0.5cm} n-DBI   	&S	&\no &\yes &\no &\yes &\yes  	&?           &none~(\cite{Coelho:2013dya}) \\
Massive gravity\\	
\hspace{0.5cm} dRGT/Bimetric   	&SVT	&\no &\no &\yes &\yes &\yes  	&?           &\cite{deRham:2014zqa}\\
\hspace{0.5cm} Galileon 	&S &\no &\yes &\yes &\yes &\yes  	&\yes?          &\cite{deRham:2012fw,deRham:2014zqa}\\
Nondynamical fields\\	
\hspace{0.5cm} Palatini $f(R)$  &-- 	&\yes &\yes &\yes &\no &\yes  	&\yes           &none\\
\hspace{0.5cm} Eddington-Born-Infeld  &-- 	&\yes &\yes &\yes &\no &\yes  	&?           &none\\
\hline
Others, not covered here\\
\hspace{0.5cm} TeVeS   	&SVT		&\no &\yes &\yes &\yes &\yes  	&?           &\cite{Freire:2012mg}\\
\hspace{0.5cm} $f(R){\cal L}_m$  &? 	&\no &\yes &\yes &\yes &\no  	&?           &\\
\hspace{0.5cm} $f(T)$   	 &?	&\no &\yes &\no &\yes &\yes &? 	           &\cite{Iorio:2012cm}\\
\noalign{\smallskip}\hline \noalign{\smallskip} \hline
\end{tabular}
\vskip 8pt 
\centering \caption{Catalog of several theories of gravity and their
  relation with the assumptions of Lovelock's theorem. Each theory
  violates at least one assumption (see also
  Figure~\ref{fig:dissectingLovelock}), and can be seen as a proxy for
  testing a specific principle underlying GR. See text for details of
  the entries.  Key to abbreviations: S: scalar; P: pseudoscalar; V:
  vector; T: tensor; ?: unknown; \yes?: not explored in detail or not
  rigorously proven, but there exist arguments to expect \yes. The
  occurrence of \no\yes? means that there exist arguments in favor of
  well-posedness within the EFT formulation, and against
  well-posedness for the full theory.  Weak-field constraints (as
  opposed to strong-field constraints, which are the main topic of
  this review) refer to Solar System and binary pulsar tests.  Entries
  below the last horizontal line are not covered in this review.}
\label{tab:theories}
\end{table}

\begin{table}[t]
\capstart
\begin{tabular}{@{}l|cccc@{}}
\hline \noalign{\smallskip}\hline \noalign{\smallskip}
Theory                       
& Solutions                    
& Stability                                  
& Geodesics
& Quadrupole
\\ 
\noalign{\smallskip}
\hline 
Extra scalar field\\	
\hspace{0.5cm} Scalar-tensor    
&$\equiv$GR~\cite{Hawking:1972qk,Heusler:1995qj,Jacobson:1999vr,HeuslerBook,Sotiriou:2011dz,Graham:2014ina}     
&\cite{Anabalon:2014lea,Damour:1976kh,Detweiler:1980uk,Zouros:1979iw,Cardoso:2004nk,Shlapentokh-Rothman:2013ysa,Cardoso:2013krh}                   
&--
&--
\\
\hspace{0.5cm} Multiscalar/Complex scalar 
&$\supset$GR~\cite{Heusler:1995qj,Herdeiro:2014goa,Herdeiro:2015gia}
&?	
&?	
&\cite{Herdeiro:2014goa,Herdeiro:2015gia}
\\
\hspace{0.5cm} Metric $f(R)$   
&$\supset$GR~\cite{HeuslerBook,Sotiriou:2011dz}    
& \cite{Hersh:1985hz,Nzioki:2014oaa}       
&?	
&?
\\
\hspace{0.5cm} Quadratic gravity   	\\
\hspace{1cm} Gauss-Bonnet  	
& NR~\cite{Mignemi:1992nt,Kanti:1995vq,Yunes:2011we}; SR~\cite{Pani:2009wy,Ayzenberg:2014aka}; FR~\cite{Kleihaus:2011tg}
& \cite{Torii:1998gm,Ayzenberg:2013wua}   
& SR~\cite{Pani:2009wy,Pani:2011gy,Maselli:2014fca}; FR~\cite{Kleihaus:2011tg}
&\cite{Kleihaus:2014lba,Ayzenberg:2014aka}
\\
\hspace{1cm} Chern-Simons 	
& SR~\cite{Yunes:2009hc,Konno:2009kg,Yagi:2012ya}; FR~\cite{Stein:2014xba}
& NR~\cite{Cardoso:2009pk,Molina:2010fb,Garfinkle:2010zx,Konno:2014qua}; SR~\cite{Ayzenberg:2013wua}  
&\cite{Yunes:2011we,Vincent:2013uea}
&\cite{Yagi:2012ya} 	
\\ 
\hspace{1cm} Generic   		
& SR~\cite{Pani:2011gy}          
&?        
& \cite{Pani:2011gy}            
& Eq.~\eqref{quadrupole_quadratic}
\\
\hspace{0.5cm} Horndeski 	
& \cite{Sotiriou:2013qea,Sotiriou:2014pfa,Babichev:2013cya}                       
&?~\cite{Kobayashi:2012kh,Kobayashi:2014wsa} 
&?	
&?
\\
Lorentz-violating\\	
\hspace{0.5cm} \AE-gravity     
& NR~\cite{Eling:2006ec,Barausse:2011pu,Barausse:2013nwa}	
&?	
& \cite{Barausse:2011pu,Barausse:2013nwa}	
&?
\\
\hspace{0.5cm} Khronometric/
&&&&\\
\hspace{0.5cm} Ho\v{r}ava-Lifshitz  & NR, SR~\cite{Barausse:2011pu,Wang:2012nv,Barausse:2012qh,Barausse:2013nwa}   
& ?~\cite{Blas:2011ni} 
& \cite{Barausse:2011pu,Barausse:2013nwa}  
&?
\\
\hspace{0.5cm} n-DBI   	
&NR\cite{Herdeiro:2011im,Coelho:2013zq}	
&?	
&?	
&?
\\
Massive gravity\\	
\hspace{0.5cm} dRGT/Bimetric		
& $\supset$GR, NR~\cite{Brito:2013xaa,Babichev:2014fka,Babichev:2014tfa,Volkov:2014ooa}
&\cite{Babichev:2013una,Brito:2013wya,Brito:2013yxa,Babichev:2014oua}
&?
&?
\\
\hspace{0.5cm} Galileon 	
&\cite{Hui:2012qt}
&?   
&?	
&?
\\
Nondynamical fields\\	
\hspace{0.5cm} Palatini $f(R)$  &$\equiv$GR	&--	&--	&--\\
\hspace{0.5cm} Eddington-Born-Infeld  &$\equiv$GR	&--	&--	&--\\
\noalign{\smallskip}\hline \noalign{\smallskip} \hline
\end{tabular}
\vskip 8pt 
\centering \caption{Catalogue of BH properties in several theories of
  gravity. The column ``Solutions'' refers to asymptotically-flat,
  regular solutions. Legend: ST=``Scalar-Tensor,'' $\equiv$GR=``Same
  solutions as in GR,'' $\supset$GR=``GR solutions are also solutions
  of the theory,'' NR=``Non rotating,'' SR=``Slowly rotating,''
  FR=``Fast rotating/Generic rotation,'' ?=unknown or uncertain.}
\label{tab:BHsummary}
\end{table}

\begin{table}[t]
\capstart
\begin{tabular}{@{}l|ccccccc@{}}
\hline \noalign{\smallskip}\hline\noalign{\smallskip}
Theory
& \multicolumn{3}{c}{Structure}
&Collapse 
&Sensitivities 
&Stability 
&Geodesics\\
 & NR & SR & FR & & & &  \\
\hline 
Extra scalar field\\	
\hspace{0.5cm} Scalar-Tensor    &\cite{Salmona:1967zz,1974GReGr...5..663H,Damour:1993hw,Damour:1996ke,Tsuchida:1998jw,Salgado:1998sg} & \cite{Damour:1996ke,Sotani:2012eb,Pani:2014jra} & \cite{Doneva:2013qva,Doneva:2014uma,Doneva:2014faa}
&\cite{Scheel:1994yr,Scheel:1994yn,Shibata:1994qd,Harada:1996wt,Novak:1997hw,Novak:1998rk,Kerimo:1998qu,Novak:1999jg}
&\cite{Zaglauer:1992bp}
&\cite{1969ApJ...155..999N,Harada:1997mr,Harada:1998,Sotani:2004rq,Sotani:2005qx,Lima:2010na,Pani:2010vc,Mendes:2013ija,Landulfo:2014wra,Sotani:2014tua,Silva:2014ora}
&\cite{DeDeo:2004kk,Doneva:2014uma}\\
\hspace{0.5cm} Multiscalar      &? & ? & ?	&?	&?	&? &?\\
\hspace{0.5cm} Metric $f(R)$   
&\cite{Kobayashi:2008tq,Cooney:2009rr,Upadhye:2009kt,Babichev:2009td,Babichev:2009fi,Miranda:2009rs,Jaime:2010kn,Astashenok:2013vza,Astashenok:2014pua,Astashenok:2014gda,Yazadjiev:2014cza,Arapoglu:2010rz,Alavirad:2013paa} & \cite{Staykov:2014mwa} & \cite{Yazadjiev:2015zia}
&\cite{Cembranos:2012fd,Borisov:2011fu}
&?
&\cite{Seifert:2007fr,Kainulainen:2008pr}
&?
\\
\hspace{0.5cm} Quadratic gravity   	\\
\hspace{1cm} Gauss-Bonnet  	
&\cite{Pani:2011xm} & \cite{Pani:2011xm} & \cite{Kleihaus:2014lba}
&?
&?
&?
&?
\\
\hspace{1cm} Chern-Simons 	
&$\equiv$ GR & \cite{Yunes:2009ch,AliHaimoud:2011fw,Yagi:2013bca,Yagi:2013mbt,Yagi:2013awa} & ?
&?
&\cite{Yagi:2013mbt}
&?
&?
\\
\hspace{0.5cm} Horndeski 	&? & ? & ?	&?	&?	&? &?\\
Lorentz-violating\\	
\hspace{0.5cm} \AE-gravity      
&\cite{Eling:2006df,Eling:2007xh} & ? & ?
&\cite{Garfinkle:2007bk}
&\cite{Yagi:2013qpa,Yagi:2013ava}
&\cite{Seifert:2007fr}
&?
\\
\hspace{0.5cm} Khronometric/&&&&&&&\\
\hspace{0.5cm} Ho\v{r}ava-Lifshitz  
&\cite{Greenwald:2009kp} & ? & ?
&?
&\cite{Yagi:2013qpa,Yagi:2013ava}
&?
&?
\\
\hspace{0.5cm} n-DBI   	&? & ? & ?	&?	&?	&? &?\\
Massive gravity\\	
\hspace{0.5cm} dRGT/Bimetric   		&\cite{Babichev:2010jd,Gruzinov:2011mm} & ? & ?	&?	&?	&? &?\\
\hspace{0.5cm} Galileon 	&\cite{Chagoya:2014fza}	& \cite{Chagoya:2014fza} & ? &\cite{Bellini:2012qn,Barreira:2013xea} 	&?	&? &?\\
Nondynamical fields\\	
\hspace{0.5cm} Palatini $f(R)$  &\cite{Kainulainen:2006wz,Kainulainen:2007bt,Barausse:2007pn,Barausse:2007ys,Olmo:2008pv} & ? & ?	&?	&--	&? &?\\
\hspace{0.5cm} Eddington-Born-Infeld  
&\cite{Pani:2011mg,Pani:2012qb,Pani:2012qd,Sham:2013sya,Harko:2013wka,Sotani:2014goa,Casanellas:2011kf} & \cite{Pani:2011mg,Pani:2012qb} & ?
&\cite{Pani:2012qb}
&--
&\cite{Sham:2012qi,Sotani:2014xoa}
&?
\\
\noalign{\smallskip}\hline \noalign{\smallskip} \hline
\end{tabular}
\vskip 8pt 
\centering \caption{Catalog of NS properties in several theories of
  gravity. Symbols and abbreviations are the same as in
  Table~\ref{tab:BHsummary}.}
\label{tab:NSsummary}
\end{table}
}

\end{landscape}

\clearpage
\section{Extensions of general relativity: motivation and overview}
\label{sec:alt-th}

\subsection{A compass to navigate the modified-gravity atlas}
\label{subsec:review-modgrav}
\begin{figure}
\capstart
\vspace{-0.3em}
\centering
\includegraphics[width=\textwidth,clip=true]{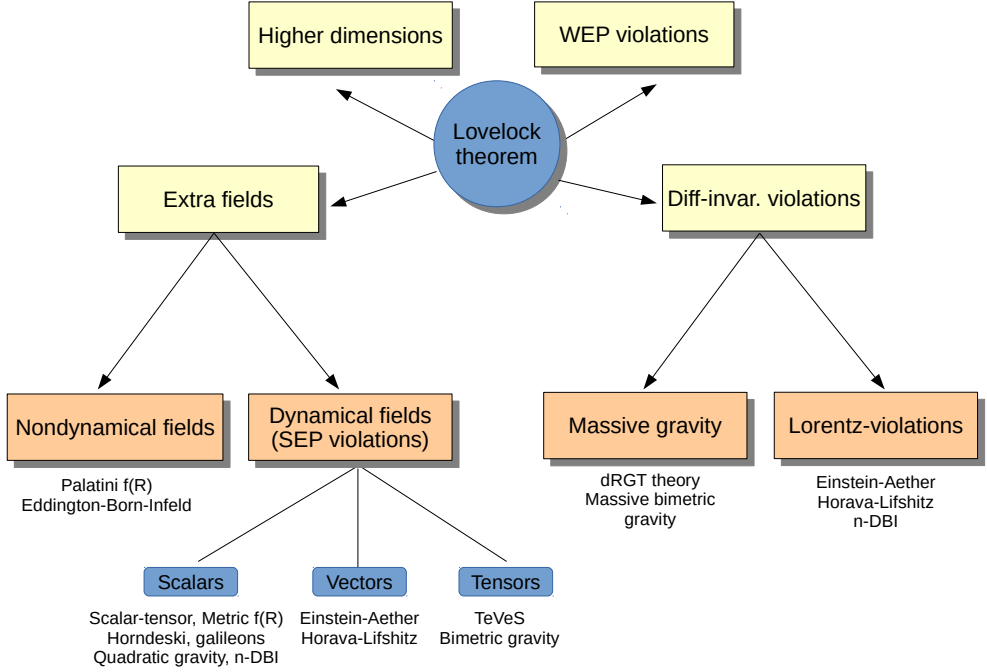}
\caption{This diagram illustrates how Lovelock's theorem serves as a
  guide to classify modified theories of gravity. Each of the yellow boxes 
connected to the circle
  represents a class of modified theories of gravity that arises from
  violating one of the assumptions underlying the theorem. A theory
  can, in general, belong to multiple classes. See
  Table~\ref{tab:theories} for a more precise classification.
\label{fig:dissectingLovelock}}
\end{figure}

There are countless inequivalent ways to modify GR, many of them
leading to theories that can be designed to agree with current
observations. Cosmological observations and fundamental physics
considerations suggest that GR must be modified at very low and/or
very high energies. Experimental searches for beyond-GR physics are a
particularly active and well motivated area of research, so it is
natural to look for a guiding principle: if we were to find
experimental hints of modifications of GR, which of the assumptions
underlying Einstein's theory should be abandoned?

Such a guiding principle can be found by examining the building blocks
of Einstein's theory. Lovelock's
theorem~\cite{Lovelock:1971yv,Lovelock:1972vz} (the generalization of
a theorem due to Cartan~\cite{Cartan:1922}) is particularly useful in
this context. In simple terms, the theorem states that GR emerges as
the unique theory of gravity under specific assumptions. More
precisely, it can be articulated as follows:
\begin{quote}
 \emph{In four spacetime dimensions the only divergence-free symmetric
   rank-2 tensor constructed solely from the metric $g_{\mu\nu}$ and
   its derivatives up to second differential order, and preserving
   diffeomorphism invariance, is the Einstein tensor plus a
   cosmological term.}
\end{quote}

Lovelock's theorem suggests a natural route to Einstein's equations
\begin{equation}
 G_{\mu\nu}+\Lambda g_{\mu\nu} = 8\pi T_{\mu\nu}\,, \label{Einstein}
\end{equation}
where $G_{\mu\nu}\equiv R_{\mu\nu}-\frac{1}{2}Rg_{\mu\nu}$ is the
Einstein tensor, and $T_{\mu\nu}$ is the matter stress-energy tensor.
Indeed, the divergence-free
nature of the Einstein tensor (that follows from the Bianchi
identities) implies that $T_{\mu\nu}$ is also divergence free,
$\nabla_\mu T^{\mu\nu}=0$. This property is necessary for geodesic
motion and it guarantees the validity of the weak equivalence
principle, i.e.~the universality of free fall
(cf.~\cite{Sotiriou:2007zu,DiCasola:2013yga} for further
discussion). If we assume that the equations of motion for the
gravitational field and the matter fields follow from a Lagrangian,
the arguments above single out the Einstein-Hilbert action
\begin{equation}
S=\frac{1}{16\pi}\int d^4 x \sqrt{-g} R +S_{M}[\Psi,\,g_{\mu\nu}]\,,\label{EHaction}
\end{equation}
where $\Psi$ collectively denotes the matter fields, which couple
minimally to $g_{\mu\nu}$, so that $S_M$ reduces to the Standard Model
action in a freely falling frame.

As it stands, Lovelock's theorem seems to leave little room for
modifying the gravitational theory~\eqref{EHaction}. However, when
analyzed in detail, the theorem contains a number of nontrivial
assumptions~\cite{Sotiriou:2015lxa}. Giving up each of these
assumptions provides a way to circumvent the theorem and gives rise to
different classes of modified theories of gravity, as illustrated in
Figure~\ref{fig:dissectingLovelock}.
Specifically, there are at least four inequivalent ways to circumvent
Lovelock's theorem:
\begin{enumerate}
\item[(1)] {\bf Additional fields.}
{
\addtolength{\parskip}{-1em}
  \paragraph{Dynamical fields.}~The simplest and most beaten path
  to circumvent Lovelock's theorem consists of adding extra degrees of
  freedom. This leaves more options to construct the left-hand side of
  Einstein's equations~\eqref{Einstein}, including more than just the
  metric and connection. Lifting this assumption paves
  the way for countless possibilities, where the metric tensor
  $g_{\mu\nu}$ is coupled to extra fundamental (scalar, vector,
  tensor) fields. Similar corrections arise from lifting the
  assumption of second differential order.\footnote{Indeed,
    higher-order equations can always be brought to second-order form
    by adding an arbitrary number of (effective) extra fields. A
    representative example is metric $f(R)$
    gravity~\cite{Sotiriou:2008rp}, see Section~\ref{subsec:f(R)}.} Because
  of the coupling with extra dynamical fields, these theories usually
  violate the strong equivalence principle~\cite{Will:2014xja}.  It is
  not straightforward to construct theories with extra fields
  nonminimally coupled to gravity that avoid instabilities associated
  to the new degrees of freedom, as generically predicted by
  Ostrogradski's theorem~\cite{Woodard:2006nt}. Because such degrees
  of freedom remain undetected to date, a major challenge for these
  theories has been to tame the behavior of the extra fields, so as to
  evade current experimental constraints related to their
  existence~\cite{Will:2014xja}.
  \paragraph{Nondynamical fields.}~Lovelock's theorem implicitly
  assumes that the matter stress-energy tensor $T_{\mu\nu}$ enters the
  field equations~\eqref{Einstein} \emph{linearly}. By dropping this
  assumption, it is possible to construct theories where the left-hand
  side of Eq.~\eqref{Einstein} is precisely the Einstein tensor,
  whereas the right-hand side is a nonlinear combination of
  $T_{\mu\nu}$ such that its covariant divergence vanishes, i.e., that
  $\nabla_\mu T^{\mu\nu}=0$ remains an
  identity~\cite{Pani:2013qfa}. These theories satisfy the weak
  equivalence principle and are equivalent to GR in vacuum, but differ
  from it in the coupling to matter. Due to such nonlinear couplings,
  they resolve some of the curvature singularities that afflict fluid
  collapse and early-time cosmology in GR~\cite{Banados:2010ix}. The
  only theories belonging to this class known to date are special
  classes of theories which modify GR by adding only auxiliary
  (i.e.~nondynamical) fields, the prototypical example being the
  Palatini formulation of $f({\cal R})$
  gravity~\cite{Sotiriou:2008rp,Olmo:2011uz}. Here ${\cal R}=g^{\mu\nu}{\cal R}_{\mu\nu}$, where ${\cal R}_{\mu\nu}$ denotes
  the Ricci tensor built from the connection, to distinguish it from
  the Ricci tensor $R_{\mu\nu}$ in the metric formalism: cf.~the discussion
  below Eq.~\eqref{actionPalatini}.}
\item[(2)] {\bf Violations of diffeomorphism invariance.}
{
\addtolength{\parskip}{-1em}
  \paragraph{Lorentz invariance.}~One particular form of
  diffeomorphism invariance, namely Lorentz invariance, has been
  tested with remarkable precision in the Standard Model sector, and
  it is widely believed to be a necessary ingredient of viable
  gravitational theories. However, if we assume that Lorentz
  invariance is just an emergent symmetry that is broken at high
  energies in the gravitational sector, a new class of gravity
  theories can be built.
  Some of these theories were found to possess a better ultraviolet
  behavior than GR~\cite{Horava:2009uw}. Violations of Lorentz
  invariance are typically encoded in some extra field(s), so that
  theories of this class usually also belong to category (1) above.
  \paragraph{Massive gravity.}~The assumption of diffeomorphism
  invariance is also crucial because it implies that gravity should be
  mediated by a \emph{massless} spin-2 field. Understanding how the
  graviton can acquire mass is a century-old problem, and strong
  constraints on the graviton mass are in
  place~\cite{Jacobson:2005bg}. Massive gravity theories are currently
  under intense scrutiny, mostly because of their applications in the
  context of the cosmological constant problem
  (see~\cite{deRham:2014zqa} for a review).  }
\item[(3)] {\bf Higher dimensions.} 
{
\\[0.5em]
Even retaining all other assumptions
  of Lovelock's theorem, the Einstein-Hilbert action~\eqref{EHaction}
  is not unique in higher dimensions. Gravitational theories built in
  dimensions other than four have a strong theoretical interest for
  several reasons, including the formulation of consistent string
  theories or understanding how the field equations depend on an extra
  parameter, i.e.~the spacetime dimension.\footnote{%
    The Cauchy problem in $D$-dimensional Gauss-Bonnet gravity was
    first investigated by Choquet-Bruhat
    in~\cite{ChoquetBruhat:1988dw} (see
    also~\cite{Choquet-Bruhat:2009xil}).  Reall et al.~showed that
    Lovelock theories in $D>4$ spacetime dimensions allow for acausal
    propagation of physical degrees of freedom in some backgrounds,
    including BH spacetimes~\cite{Reall:2014pwa}. They concluded that
    higher-dimensional Lovelock theories may or may not be hyperbolic
    depending on the background spacetime.
    Willison~\cite{Willison:2014era} showed that Lovelock gravity is
    locally well-posed in arbitrary backgrounds, but global
    hyperbolicity is still an open problem.}  These theories may even
  offer a resolution of the hierarchy problem, because they predict
  that the fundamental Planck mass can be several orders of magnitude
  smaller than the effective four-dimensional
  $\mpl\approx 10^{19}{\rm GeV}$.  Some extra-dimensional models are
  severely constrained from an experimental point of view (see
  e.g.~\cite{Kanti:2004nr}), and their relevance for beyond-GR effects
  in astrophysics is limited. However, some quantum-gravity
  corrections might be accessible through astrophysical observations,
  as we discuss in Section~\ref{sec:EFT}. GR in higher dimensions
  leads naturally to additional fields if the theory is reduced to
  $D=4$ dimensions: additional scalar and gauge fields emerge by
  performing a Kaluza-Klein or dimensional reduction from $D>4$ to
  $D=4$ dimensions. We discuss theories in higher dimensions only
  marginally and refer the interested reader to other reviews,
  e.g.~\cite{Clifton:2011jh}. }
\item[(4)] {\bf WEP violations.}
{
\\[0.5em]
The requirement that the left-hand side of Einstein's equation be
divergence-free is dictated by the desire of having a divergence-free
$T_{\mu\nu}$ and, in turn, by the weak equivalence principle. Various
classes of theories that circumvent Lovelock's theorem only by
postulating a nonminimal coupling to the matter sector (and thus
violating the weak equivalence principle) have been proposed
(see e.g.~\cite{Bertolami:2007gv}, and~\cite{Capozziello:2011et} for a
review). Nevertheless, because the equivalence principle has been
tested with the astonishing precision of one part in
$10^{13}$~\cite{Schlamminger:2007ht}, we will seldom discuss theories
where it is violated. }

\end{enumerate}
Although rather elementary, the classification proposed above has the
virtue of simplicity. In this review we are mainly interested in
understanding \emph{how to test GR}, and especially \emph{what} we can
test, rather than attempting a comprehensive classification of
alternative theories. Our point of view is therefore very practical:
any modified theory of gravity will necessarily fall into one of the
categories above, and therefore it will violate one or more of the
fundamental principles underlying GR; these violations will determine
the new effects predicted by the theory, and the payoff of a
hypothetical observation of these effects.

In Table~\ref{tab:theories} we give a schematic (and necessarily
incomplete) summary of proposed extensions of GR that are of interest
for astrophysics, i.e.~those that provide potential means to test the
fundamental principles of GR with current and near-future
astrophysical observations.  In the rest of this review we discuss
these theories (and their implications for experimental verifications
of GR in the weak-field and strong-field regimes) in more detail.

Regardless of the manner in which Lovelock's theorem is violated, all
these theories face a common challenge: how to modify the behaviour of
gravity at extreme energy scales, whilst leaving the (tightly
constrained) intermediate energy regime unchanged?
The hypothetical solution to this problem is termed
``screening.'' When used in a general sense, the word is simply a
label for unknown physics, much like the phrase ``dark energy.'' Three
concrete kinds of screening mechanisms are known, corresponding to
density-dependent modifications of the three kinds of terms appearing
in the action of a scalar field: (i) kinetic terms (including derivative
self-interactions), (ii) potential terms, and (iii) couplings to matter fields
\cite{Joyce:2014kja}. These modifications lead to, respectively, the
symmetron/dilaton screening mechanisms, the chameleon mechanism and
the Vainshtein mechanism; see Section~\ref{subsec:cosmology} for
further mathematical details.

It is likely that these exhaust the possibilities for a gravity theory
with one additional scalar field.
However, it is far from clear that one of these three mechanisms can
be embedded in every gravity theory in the current literature. One
could argue that any theory lacking an ``in-built'' screening
mechanism is disfavoured or, at best, incomplete. However, given the
rapidly-evolving nature of this research area, it would seem hasty to
discard all nonscreening theories at this stage.

If GR is not the fully correct theory of gravity, then
we are forced to accept one of the following propositions:
\begin{itemize}
\item[a)] The
  true theory must incorporate one of the three known mechanisms; or
\item[b)] There exist yet-unknown screening mechanisms, which require
  more than a single scalar field to operate; or
\item[c)] Deviations from GR do exist in the intermediate energy
  regime, but are below the current detection threshold of PPN and
  binary pulsar constraints.
\end{itemize}

It is worth noting that, in addition to the strong-field and
cosmological tests of gravity described in this review, screening
mechanisms have spawned a wave of new laboratory and astrophysical
tests of gravity. Laboratory examples include experiments to detect
the chameleon mechanism using cold atom interferometry
\cite{Burrage:2014oza,Hamilton:2015zga}, and the ``afterglow'' of a
chameleon field interacting with the electromagnetic field inside a
radio frequency cavity \cite{Steffen:2010ze}. New astrophysical tests
include searches for a potential mismatch between distance indicators
such as cepheid variables and tip-of-the-red-giant-branch (TRGB) stars
in unscreened dwarf galaxies \cite{Jain:2012tn}.

\subsection{Scalar-tensor gravity}
\label{subsec:ST}
One of the most natural extensions of GR is {\it scalar-tensor
  gravity}, in which one or more scalar degrees of freedom are
included in the gravitational sector of the theory, through a {\it
  nonminimal coupling} (i.e., the Ricci scalar in the
Einstein-Hilbert action is multiplied by a function of the scalar
field(s)). Several reviews provide extensive discussions on the
subject, see
e.g.~\cite{Damour:1992we,Chiba:1997ms,2003sttg.book.....F,Faraoni:2004pi,Sotiriou:2015lxa}.

Scalar fields with nonminimal couplings to gravity appear in several
contexts, such as in string theory~\cite{polchinski1998string}, in
Kaluza-Klein-like theories~\cite{Duff:1994tn} or in braneworld
scenarios~\cite{Randall:1999ee,Randall:1999vf}. They also have
important applications in cosmology~\cite{Clifton:2011jh}. Therefore,
scalar-tensor gravity is a good framework to study phenomenological
aspects of several possible fundamental theories.

\subsubsection{The Bergmann-Wagoner formulation.}
The most general action of scalar-tensor gravity with one scalar field
which is at most quadratic in derivatives of the fields was studied by
Bergmann and Wagoner~\cite{Bergmann:1968ve,Wagoner:1970vr}, and can be
written (after an appropriate field redefinition) as:
\begin{equation}
  S=\frac{1}{16\pi}\int d^4x\sqrt{-g}\left[\phi R-\frac{\omega(\phi)}{\phi}g^{\mu\nu}
    \left(\partial_\mu\phi\right)\left(\partial_\nu\phi\right)
    -U(\phi)\right]+S_M[\Psi,g_{\mu\nu}]\,,\label{STactionJ}
\end{equation}
where $\omega$ and $U$ are arbitrary functions of the scalar field
$\phi$, and $S_M$ is the action of the matter fields $\Psi$. When
$\omega(\phi)=\omega_{BD}$ is constant and $U(\phi)=0$, the theory
reduces to (Jordan-Fierz-)Brans-Dicke
gravity~\cite{Jordan:1959eg,Fierz:1956zz,Brans:1961sx}, an extension
of GR which was proposed in the mid-$20^{th}$ century
(see~\cite{Goenner:2012cq,Brans:2008zz,Brans:2005ra} for a historical
account).

The Bergmann-Wagoner theory \eqref{STactionJ} can be expressed in a
different form through a scalar field redefinition
$\varphi=\varphi(\phi)$ and a conformal transformation of the metric
$g_{\mu\nu}\rightarrow g^\star_{\mu\nu}=A^{-2}(\varphi)
g_{\mu\nu}$. In particular, fixing $A(\varphi)=\phi^{-1/2}$, the
action~(\ref{STactionJ}) -- generally referred to as the {\it Jordan-frame} 
action -- transforms into the {\it Einstein-frame} action
\begin{equation}
  S=\frac{1}{16\pi}\int d^4x\sqrt{-g^\star}\left[R^\star-2g^{\star\mu\nu}
    \left(\partial_\mu\varphi\right)\left(\partial_\nu\varphi\right)
    -V(\varphi)\right]+S_M[\Psi,A^2(\varphi)g^\star_{\mu\nu}]\,,\label{STactionE}
\end{equation}
where $g^\star$ and $R^\star$ are the determinant and Ricci scalar of
$g_{\mu\nu}^{\star}$, respectively, and the potential 
$V(\varphi) \equiv A^4(\varphi)U(\phi(\varphi))$.
The price paid for the minimal
coupling of the scalar field in the gravitational sector is the
nonminimal coupling in the matter sector of the action: particle
masses and fundamental constants depend on the scalar field.

We remark that the actions~\eqref{STactionJ} and \eqref{STactionE} are
just different representations of the same theory: the outcome of an
experiment will not depend on the chosen representation, as long as
one takes into account that the units of physical quantities do scale
with powers of the conformal factor
$A$~\cite{Flanagan:2004bz,Sotiriou:2007zu}.  It is then legitimate,
when
modeling a physical process, to choose the conformal frame in which
calculations are simpler: for instance, in vacuum the Einstein-frame
action \eqref{STactionE} formally reduces to the GR action minimally
coupled with a scalar field. It may then be necessary to change the
conformal frame when extracting physically meaningful statements
(since the scalar field is minimally coupled to matter in the Jordan
frame, test particles follow geodesics of the {\em Jordan-frame}
metric, not of the Einstein-frame metric).

The relation between Jordan-frame and Einstein-frame quantites is
simply $\phi=A^{-2}(\varphi)$, $3+2\omega(\phi)=\alpha(\varphi)^{-2}$,
where $\alpha(\varphi)\equiv d(\ln
A(\varphi))/d\varphi$~\cite{Will:2014xja}. Note that the theory is
fixed once the function $\omega(\phi)$ -- or, equivalently,
$\alpha(\varphi)$ -- is fixed, and the form of the scalar potential is
chosen. Moreover, many phenomenological studies neglect the scalar
potential. This approximation corresponds to neglecting the
cosmological term, the mass of the scalar field and any possible
scalar self-interaction. In an asymptotically flat spacetime the
scalar field tends to a constant $\phi_0$ at spatial infinity,
corresponding to a minimum of the potential. Taylor expanding
$U(\phi)$ around $\phi_0$ yields a cosmological constant and a mass
term for the scalar field to the lowest
orders~\cite{Wagoner:1970vr,Alsing:2011er}.

Scalar-tensor theory with a vanishing scalar potential is
characterized by a single function $\alpha(\varphi)$. The expansion of
this function around the asymptotic value $\varphi_0$ can be written
in the form
\begin{equation}
\alpha(\varphi)=\alpha_0+\beta_0(\varphi-\varphi_0)+\dots\label{DEalphabeta}
\end{equation}
As mentioned above, the choice $\alpha(\varphi)=\alpha_0=$constant
(i.e., $\omega(\phi)=$constant) corresponds to Brans-Dicke theory. A
more general formulation, proposed by Damour and Esposito-Far\`ese, is
parametrized by $\alpha_0$ and $\beta_0$~\cite{Damour:1993hw,Damour:1996ke}. Another simple variant is massive
Brans-Dicke theory, in which $\alpha(\varphi)$ is constant, but the
potential is nonvanishing and has the form
$U(\phi)=\f{1}{2}U''(\phi_0)(\phi-\phi_0)^2$, so that the scalar field
has a mass $m_s^2\sim U''(\phi_0)$. Note that since the scalar field
$\varphi$ in the action (\ref{STactionE}) is dimensionless, the
function $\alpha(\varphi)$ and the constants $\alpha_0$, $\beta_0$ are
dimensionless as well.

The field equations of scalar-tensor theory in the Jordan frame are
(see e.g.~\cite{Eardley:1975,will1993theory})
\begin{subequations}
\label{ST:Jordan}
\begin{align}
G_{\mu \nu} &= \frac{8\pi}{\phi}T_{\mu \nu}+\frac{\omega(\phi)}{\phi^2}\left(\partial_\mu\phi\partial_\nu\phi
  -\frac{1}{2}g_{\mu\nu}\partial_\lambda\phi\partial^\lambda\phi\right)+\frac{1}{\phi}(\nabla_\mu\nabla_\nu\phi-g_{\mu\nu}\Box_g \phi)
  -\frac{U(\phi)}{2\phi}g_{\mu\nu}\, ,
\label{eq:tensoreqnJ}\\
\Box_g \phi &= \frac{1}{3+2\omega(\phi)}\left(8\pi T - 16\pi\phi\frac{\partial T}{\partial \phi}
-\frac{d\omega}{d\phi}\partial_\lambda\phi\partial^\lambda\phi+\phi\frac{dU}{d\phi}-2U(\phi)\right) \, ,
\label{eq:scalareqnJ}
\end{align}
\end{subequations}
where $T^{\mu\nu}=-2(-g)^{-1/2}\delta S_M(\Psi,g_{\mu\nu})/\delta
g_{\mu\nu}$ is the Jordan-frame stress-energy tensor of matter
fields, and $T=g^{\mu\nu}T_{\mu\nu}$.

In the Einstein frame, the field equations are
\begin{subequations}
\label{ST:Einstein}  
\begin{align}
  G^\star_{\mu \nu} &=2\left(\partial_\mu\varphi\partial_\nu\varphi-
  \frac{1}{2}g^\star_{\mu\nu}\partial_\sigma\varphi\partial^\sigma\varphi\right)-
  \frac{1}{2}g^\star_{\mu\nu}V(\varphi)+8\pi T^\star_{\mu\nu}\,,\label{eq:tensoreqnE}  \\
\Box_{g^\star} \varphi &=-4\pi\alpha(\varphi)T^\star+\frac{1}{4}\frac{dV}{d\varphi}\,,\label{eq:scalareqnE}
\end{align}
\end{subequations}
where $T^{\star\,\mu\nu}=-2(-g)^{-1/2} \delta
S_M(\Psi,A^2g^\star_{\mu\nu})/\delta g^\star_{\mu\nu}$ is the Einstein-frame
stress-energy tensor of matter fields and
$T^\star=g^{\star\,\mu\nu}T^\star_{\mu\nu}$ (see
e.g.~\cite{Damour:1992we}). Eq.~\eqref{eq:scalareqnE} shows that
$\alpha(\varphi)$ couples the scalar fields to
matter~\cite{Damour:1995kt}, as does $(3+2\omega(\phi))^{-1}$ in the
Jordan frame: cf.~Eq.~\eqref{eq:scalareqnJ}].

Astrophysical observations set bounds on the parameter space of
scalar-tensor theories. In the case of Brans-Dicke theory, the best
observational bound ($\alpha_0<3.5\times 10^{-3}$) comes from the
Cassini measurement of the Shapiro time delay. In the more general
case with $\beta_0\neq0$, current constraints on $(\alpha_0,\beta_0)$
have been obtained by observations of NS-NS and NS-WD binary
systems~\cite{Freire:2012mg}, and will be discussed in
Section~\ref{sec:BP} (cf.~Figure~\ref{fig:stg}). Observations of compact
binary systems also constrain massive Brans-Dicke theory, leading to
exclusion regions in the $(\alpha_0,m_s)$ plane~\cite{Alsing:2011er}.

An interesting feature of scalar-tensor gravity is the prediction of
certain characteristic physical phenomena which do not occur at all in
GR. Even though we know from observations that $\alpha_0\ll1$ and that
GR deviations are generally small, these phenomena may lead to
observable consequences. There are at least three possible smoking
guns of scalar-tensor gravity. The first is the emission of dipolar
gravitational radiation from compact binary
systems~\cite{Eardley:1975,Will:1989sk}, which will be discussed in
Section~\ref{sec:binaries/scal-tens-theor}. Dipolar gravitational
radiation is ``pre-Newtonian,'' i.e.~it occurs at lower PN order than
quadrupole radiation, and it does not exist in GR. The second is the
existence of nonperturbative NS solutions in which the scalar field
amplitude is finite even for $\alpha_0\ll1$. This {\it spontaneous
  scalarization} phenomenon~\cite{Damour:1993hw,Damour:1996ke} will be
discussed in detail in Section~\ref{sec:NS_ST}. Here we only remark
that spontaneous scalarization would significantly affect the mass and
radius of a NS, and therefore the orbital motion of a compact binary
system, even far from coalescence.  The third example is also
nonperturbative, and it involves massive fields. The coupling of
massive scalar fields to matter in orbit around rotating BHs leads to
a surprising effect: because of superradiance, matter can hover into
``floating orbits'' for which the net gravitational energy loss at
infinity is entirely provided by the BH's rotational
energy~\cite{Cardoso:2011xi}.

The phenomenology of scalar-tensor theory in vacuum spacetimes, such
as BH spacetimes, is less interesting. When the matter action $S_M$
can be neglected, the Einstein-frame formulation of the theory is
equivalent to GR minimally coupled to a scalar field. BHs in
Bergmann-Wagoner theories satisfy the same {\it no-hair theorem} as in
GR, and thus the stationary BH solutions in the two theories
coincide~\cite{Heusler:1995qj,Sotiriou:2011dz}. Moreover, dynamical
(vacuum) BH spacetimes satisfy a similar {\it generalized no-hair
  theorem}: the dynamics of a BH binary system in Bergmann-Wagoner
theory with vanishing potential are the same as in
GR~\cite{Damour:1992we}, up to at least $2.5$ PN order for generic
mass ratios~\cite{Mirshekari:2013vb} and at any PN order in the
extreme mass-ratio limit~\cite{Yunes:2011aa} (see
Section~\ref{sec:pn_st}). These no-hair theorems will be discussed in
Section~\ref{sec:ST_BHs}.

\subsubsection{Scalar-tensor theories with multiple scalar fields.} \label{sec:multiscalars}
When gravity is coupled with more than one scalar field, the
action~\eqref{STactionJ} has the more general form~\cite{Damour:1992we}
\begin{align}
S\,=\,\,& \frac1{16\pi}\int d^4x\sqrt{-g}\left(
F(\phi)R-\gamma_{ab}(\phi)g^{\mu\nu}\partial_\mu\phi^a
\partial_\nu\phi^b-V(\phi)\right)+S_M[\Psi,\,g_{\mu\nu}]\,,
\label{STactionJ_multi}
\end{align}
where $F,V$ are functions of the $N$ scalar fields $\phi^a$
($a=1\ldots N$). The scalar fields live on a manifold (the {\it target
  space}) with metric $\gamma_{ab}(\phi)$. The
action~\eqref{STactionJ_multi} is invariant not only under space-time
diffeomorphisms, but also under target-space diffeomorphisms,
i.e.~scalar field redefinitions. These theories have a richer
structure
than those with a single scalar field, since the geometry of the
target space can affect the dynamics.  For instance, the theories with
a complex scalar field discussed in Section~\ref{sec:complscal}, in
which the no-hair theorems can be circumvented, can also be seen as
multiscalar-tensor theories with $N=2$.

\subsubsection{Horndeski gravity.}\label{subsec:horndeski}

The most general scalar-tensor theory with second-order field
equations (and one scalar field) is Horndeski gravity~\cite{Horndeski:1974wa}.  The action of Horndeski gravity can be
written in terms of Galileon interactions (see~\cite{Deffayet:2011gz}
and Section~\ref{subsec:massgrav}) as
\begin{equation}
\begin{aligned}
  S={}&\int d^4x\sqrt{-g}\Big\{K(\phi,X)-G_3(\phi,X)\Box\phi\\
    &{}+G_4(\phi,X)R+G_{4,X}(\phi,X)\left[(\Box\phi)^2
      -(\nabla_\mu\nabla_\nu\phi)(\nabla^\mu\nabla^\nu\phi)\right]\\
    &{}+G_5(\phi,X)G_{\mu\nu}\nabla^\mu\nabla^\nu\phi
    -\frac{G_{5,X}(\phi,X)}{6}\big[
      (\Box\phi)^3-3\Box\phi(\nabla_\mu\nabla_\nu\phi)(\nabla^\mu\nabla^\nu\phi)
    \\
    &{}+2(\nabla_\mu\nabla_\nu\phi)(\nabla^\mu\nabla_\sigma\phi)
    (\nabla^\nu\nabla^\sigma\phi)\big]\Big\}\,,
\end{aligned}
    \label{action_horndeski}
\end{equation}
where $K$ and the $G_i$'s ($i=1\dots 5$) are functions of the scalar
field $\phi$ and of its kinetic term
$X=-1/2\partial^\mu\phi\partial_\mu\phi$, and $G_{i,X}$ are
derivatives of $G_i$ with respect to the kinetic term $X$. For a
particular choice of these functions, this theory coincides with
Gauss-Bonnet gravity (see Section~\ref{subsec:quadratic}).

As we shall discuss in Section~\ref{sec:ST_BHs}, in Horndeski theory
the no-hair theorem can be circumvented, and thus stationary BH
solutions can be different from GR.

\subsection{Metric $f(R)$ theories}
\label{subsec:f(R)}
The standard paradigm to explain the acceleration of the cosmic
expansion is to postulate the existence of a diffuse form of dark
energy described by an exotic equation of state ($P\approx -\rho$) and
amounting to roughly $70\%$ of the critical energy density. The
cosmological constant is the most natural candidate for this dark
``fluid,'' although its tiny value (as inferred by cosmological
observations) clashes with the value of vacuum energy as inferred from
particle physics. As mentioned above, this is one of the main problems in 
theoretical physics: the cosmological constant
problem~\cite{Weinberg:1988cp,Carroll:2000fy,Martin:2012bt}.

As an alternative to the standard $\Lambda$CDM ($\Lambda$-Cold Dark Matter) model, it has been
proposed that infrared modifications of gravity could be the
explanation for the cosmic acceleration. In this context, so-called
$f(R)$ modified gravities have a long history~\cite{Schmidt:2006jt}
and have been widely explored as prototypical infrared corrections to
GR.
The action for $f(R)$ gravity reads
\begin{equation}
S=\frac{1}{16\pi}\int d^4 x\,\sqrt{-g}f(R)+S_M\left[\Psi,\,g_{\mu\nu}\right]\,,\label{action}
\end{equation}
where $\Psi$ collectively denotes all matter fields and $f(R)$ is a
function of the scalar curvature $R$. It is customary to use a
simplified notation where $f_R\equiv f'(R)$, $f_{RR}\equiv f''(R)$ and
so on. We shall focus for the moment on the theory obtained from the
action above through a metric variational principle. Palatini $f(R)$
gravity is a completely different theory, that will be discussed in
Section~\ref{subsec:auxiliary} below.

Primarily, $f(R)$ theories attracted attention for their potential to
describe the cosmological acceleration of the Universe without a
fine-tuned cosmological constant~\cite{Sotiriou:2008rp}. Viable $f(R)$
models are usually chosen by ensuring that the field equations admit a
de Sitter solution with curvature radius $R_{\rm dS}$.
We refer the reader to specialized reviews~\cite{Sotiriou:2008rp,DeFelice:2010aj,Capozziello:2011et} for a more detailed
discussion of the theoretical aspects and of current experimental constraints.

\paragraph{Viable $f(R)$ theories.}
If one wishes to modify GR at cosmological scales, while leaving the
large curvature behavior essentially unaffected, very stringent
constraints are in place. Solar System observations and local tests
strongly constrain viable $f(R)$ models and rule out many candidates
(cf.~\cite{DeFelice:2010aj} for a review).  In general, $f(R)$ models
must be described by monotonically growing and convex functions,
i.e.~$f_R>0$ and $f_{RR}>0$, in order to avoid ghosts (i.e., negative
kinetic energy states) and tachyons. Furthermore, $f(R)$ gravity
theories are introduced to modify the infrared behavior of GR when
$R\lesssim R_c$, $R_c$ being some cosmological curvature scale of the
order of $R_{\rm dS}$. In order to recover Einstein's theory at higher
curvature and to pass Solar System tests, viable models usually have,
at leading order,
\begin{equation}
f\to R\,,\quad f_R\to 1\,,\quad f_{RR}\to0\,,\label{limit_f} \qquad R\gg R_c\,.
\end{equation}
In the following, we shall focus on classes of $f(R)$ theories of gravity that satisfy the above requirements.

\paragraph{Different formulations of $f(R)$ theories.}
It is well known that $f(R)$ theories are dynamically equivalent to a
specific class of scalar-tensor
theories~\cite{Teyssandier:1983zz,Barrow:1988xi,Barrow:1988xh,Wands:1993uu}
(see~\cite{Sotiriou:2008rp} for a review), so they propagate an
additional scalar degree of freedom. These theories allow for
different formulations depending on which quantity is identified as
the scalar field.  At least three different approaches to the study of
$f(R)$ theories have been proposed. While equivalent in principle,
each approach has different features and practical drawbacks. A common
choice in the literature is to transform the $f(R)$
action~\eqref{action} into a Brans-Dicke theory with $\omega=0$ in the
Jordan frame. If $f_{RR}\neq0$, the action~\eqref{action} is
dynamically equivalent to
\begin{equation}
 S_J=\frac{1}{16\pi}\int d^4 x\,\sqrt{-g}\left[\phi R-V_{\rm KM}(\phi)\right]+S_M\left[\Psi,\,g_{\mu\nu}\right]\,,\label{action_J}
\end{equation}
where $\phi=f_R$ and $V_{\rm KM}(\phi)=R f_R-f$ (the reason
for the KM subscript will be apparent shortly).
It should be stressed that, if $f_{RR}=0$ at some point, the
equivalence is not guaranteed and must be checked on a case-by-case
basis.
Scalar-tensor theories with a vanishing kinetic term of the
form~\eqref{action_J} were also studied by
O'Hanlon~\cite{O'Hanlon:1972ya} and others (see
e.g.~\cite{Fiziev:1999zt}).

In the context of compact objects, Kobayashi and
Maeda~\cite{Kobayashi:2008tq,Upadhye:2009kt} integrated the field
equations arising from the action~\eqref{action_J}, which read
\begin{align}
 \phi R_{\mu\nu}-\frac{1}{2}f g_{\mu\nu}-\nabla_\mu\nabla_\nu \phi+g_{\mu\nu}\square\phi&=8\pi T_{\mu\nu}\,, \label{RmunufR}\\
 \square \phi=\frac{8\pi}{3} T+\frac{1}{3}\left[2f(R(\phi))-\phi\, R(\phi)\right]&\equiv\frac{8\pi}{3} T+\frac{dV_{\rm KM}}{d\phi}\,, \label{scalareqKM}
\end{align}
where the evolution equation for the scalar degree of freedom $\phi$
is obtained from the trace of Eq.~(\ref{RmunufR}) above, and $R$ is
now an implicit function of $\phi$.

It is possible to recast $f(R)$ theory as a scalar-tensor theory in
the Einstein frame for a new scalar field
$\varphi\propto\log\phi$~\cite{Teyssandier:1983zz,Sotiriou:2006hs}. By
defining
\begin{equation}
 \varphi\equiv\frac{\sqrt{3}}{2}\log f_R\,,\quad  g^\star_{\mu\nu} = A^{-2}  g_{\mu\nu}\,,\quad A^{-2}\equiv f_R=e^{2\varphi/\sqrt{3}}\,,
\end{equation}
the action~\eqref{action_J} becomes
\begin{equation}
 S_E=\frac{1}{16\pi}\int d^4 x\sqrt{-g^\star}\left[R^\star-2\partial_\alpha\varphi\partial^\alpha\varphi-V_{\rm BL}(\varphi)\right]+S_M\left[\Psi,\,A^2  g^\star_{\mu\nu}\right]\,,\label{action_E}
\end{equation}
where the new scalar potential reads
\begin{equation}
 V_{\rm BL}(\varphi)=\frac{R f_R-f}{f_R^2}\,,\label{VBL}
\end{equation}
Here $R=R(\varphi)$, and we introduced a subscript ``BL'' because this
formulation was used by Babichev and Langlois in their study of
compact stars~\cite{Babichev:2009td,Babichev:2009fi}.

Besides the standard approaches discussed above, another formulation
of the theory was proposed by Jaime, Pati\~no and
Salgado~\cite{Jaime:2010kn}. In this approach, the Ricci curvature $R$
is considered as an independent scalar degree of freedom.  By
introducing a new scalar field $\psi$, the action~\eqref{action} is
dynamically equivalent to
\begin{equation}
 S_R=\frac{1}{16\pi}\int d^4 x\,\sqrt{-g}\left[f'(\psi)R-V_{\rm JPS}(\psi)\right]+S_M\left[\Psi,\,g_{\mu\nu}\right]\,,\label{action_R}
\end{equation}
where $V_{\rm JPS}(\psi)\equiv \psi f'(\psi)-f$. Variation with
respect to $\psi$ leads to $\psi=R$, if $f_{RR}\neq0$. In fact, this
is usually considered as an intermediate step in reducing the
action~\eqref{action} to the scalar-tensor theory~\eqref{action_J}:
see e.g.~\cite{Sotiriou:2008rp}. As in the case of Brans-Dicke theory
in the Jordan frame, the field equations for the scalar field simply
impose $\psi=R$, but the scalar evolution arises from the trace of
Einstein's equation. The field equations read
\begin{align}
 G_{\mu\nu}={}&\frac{1}{f_R}\left[f_{RR}\nabla_\mu\nabla_\nu R+f_{RRR}(\nabla_\mu R)(\nabla_\nu R)-\frac{g_{\mu\nu}}{6}\left(R f_R+f+16\pi T\right)+8\pi T_{\mu\nu}\right] \,,\nonumber
 \\
 \square R={}&\frac{dV^{\rm eff}_{\rm JPS}}{dR}\equiv\frac{8\pi}{3f_{RR}} T-\frac{f_{RRR}}{f_{RR}}(\nabla R)^2+{\frac{dV_{\rm JPS}}{dR}}\,,\label{Jaime2}
\end{align}
with ${dV_{\rm JPS}}/{dR}\equiv(2f-Rf_R)/3f_{RR}$.
As pointed out in~\cite{Jaime:2010kn}, in this formulation the
potential is as well defined as the function $f(R)$.

The field equations of $f(R)$ gravity are of fourth differential
order, but the theory admits a well-posed initial value problem by
virtue of its equivalence with scalar-tensor
gravity~\cite{LanahanTremblay:2007sg,Paschalidis:2011ww}.

\subsection{Quadratic gravity}
\label{subsec:quadratic}
One of the most pressing problems in theoretical physics is to
accommodate GR in the framework of quantum field theories. It has long
been known that Einstein's theory is not renormalizable in the
standard quantum field theory sense, and this is a major obstacle on
the route to quantum gravity. The situation changes if the
Einstein-Hilbert action is assumed to be only the first term in an
expansion containing all possible curvature invariants, as also
suggested by low-energy effective string theories. Already in the
1970s, Stelle showed that including quadratic curvature terms in the
action makes the theory renormalizable~\cite{Stelle:1976gc}. This
comes at the cost of having higher-derivative terms in the field
equations, which generically introduce ghosts or other pathologies
(but see~\cite{Biswas:2011ar,Talaganis:2014ida} for recent progress in
constructing a class of ghost-free, higher-derivative extensions of
GR).

At second order in the curvature, the only independent algebraic
curvature invariants are
\begin{equation}
 R^2\,,\quad R_{\mu\nu}^2\,,\quad R_{\mu\nu\rho\sigma}^2\,,\quad {}^{*}\!RR\,, \label{curvatureinvariants}
\end{equation}
where $R_{\mu\nu}^2\equiv R_{\mu\nu}R^{\mu\nu}$,
$R_{\mu\nu\rho\sigma}^2\equiv
R_{\mu\nu\rho\sigma}R^{\mu\nu\rho\sigma}$,
${}^{*}\!RR\equiv\frac{1}{2}R_{\mu\nu\rho\sigma}\epsilon^{\nu\mu\lambda\kappa}R^{\rho\sigma}{}_{\lambda\kappa}$
is the Pontryagin scalar, and $\epsilon^{\mu\nu\rho\sigma}$ is the
Levi-Civita tensor.  Of particular interest are the Gauss-Bonnet
scalar $R_{GB}^2\equiv R^2-4R_{\mu\nu}^2+R_{\mu\nu\rho\sigma}^2$ and
the Pontryagin scalar (also referred to as the Chern-Simons scalar)
defined above, because these terms can be shown to emerge in
low-energy realizations of string
theory~\cite{Moura:2006pz,polchinski1998string}. The Pontryagin scalar
also appears in loop quantum gravity~\cite{ashtekar1989cp}. However,
these terms alone
do not yield modifications to Einstein's equations in four spacetime
dimensions, because their integrals are four-dimensional topological invariants
and only account for boundary terms in the action. To circumvent this
problem one is thus forced to add extra dynamical fields, i.e., extra
propagating degrees of freedom (but cf.~Section~\ref{subsec:auxiliary}
below for a different strategy using nondynamical fields). The
simplest way to introduce nontrivial higher-order curvature
corrections is via coupling with a scalar field.

The most generic class of four-dimensional theories obtained by
including all quadratic algebraic curvature invariants coupled to a
single scalar field reads~\cite{Yunes:2011we,Pani:2011gy}
\begin{equation}
\begin{aligned}
 S={}&\frac{1}{16\pi}\int\sqrt{-g} d^4x 
\Big[R-2\nabla_\mu\phi\nabla^\mu\phi-V(\phi)\\
&{}+f_1(\phi)R^2 + f_2(\phi) R_{\mu\nu}R^{\mu\nu}+f_3(\phi) 
R_{\mu\nu\rho\sigma}R^{\mu\nu\rho\sigma}+f_4(\phi){}^{*}\!RR\Big]\\
&{}+S_\text{mat}\left[\Psi,\,\gamma(\phi)g_{\mu\nu}\right]\,,\label{eq:action_quadratic}
\end{aligned}
\end{equation}
where $V(\phi)$ is the scalar self-potential, $f_i(\phi)$
($i=1,\dots,4$) are coupling functions, and in the matter action
$S_\text{mat}$ we have included a nonminimal but universal metric
coupling, which thus satisfies the weak (but in general not the
strong) equivalence principle. The action~\eqref{eq:action_quadratic}
generically yields higher-order field equations that are prone to the
Ostrogradski instability and to the appearance of ghosts, unless the
various terms appear in the special combination corresponding to the
four-dimensional Gauss-Bonnet invariant (discussed in
Section~\ref{subsec:EdGB} below).
To avoid this instability, the theory~\eqref{eq:action_quadratic} must
be considered as an \emph{effective} action, obtained as the
truncation of a more general theory, valid only up to second
order in curvature.\footnote{Alternatively, one can circumvent the Ostrogradski instability
by expanding the phase-space of the (dynamical) variables
if the resulting equations of motion constitute a closed system of PDEs 
that are at most second order~\cite{Chen:2012au,Chen:2013aha}.}
In the \emph{decoupling limit} (where the effective theory is valid, see
Section~\ref{sec:EFT}), a perturbative approach is applicable and the
field equations remain of second differential order for generic
combinations of the curvature invariants.
For example, it has been shown that dCS gravity (introduced in
Section~\ref{sec:ChernSimons} below) does not exhibit any ghost-like
instabilities when treated order-by-order in the perturbation scheme
and, in fact, can be cast into a well-posed Cauchy problem in the
decoupling limit~\cite{Delsate:2014hba}.
We expect a similar argument to hold for EdGB gravity
(see~Section~\ref{subsec:EdGB}), but a rigorous proof in this case is
still missing.

The EFT
approach is not only motivated by the desire to avoid higher-order
derivatives in the field equations, but it arises naturally in some
low-energy expansion in string theory, which indeed contains the
Gauss-Bonnet and Chern-Simons terms coupled respectively to the
dilaton and axion at second
order in the curvature. In this approach the Einstein-Hilbert term is
considered as the first-order term in a (possibly infinite) series
expansion containing all possible curvature corrections. In this
sense, GR may be only accurate up to second-order terms in the curvature.

In the geometrical units adopted here, the scalar field entering the
action~\eqref{eq:action_quadratic} is dimensionless, whereas the
coupling functions $f_i(\phi)$ have the dimensions of a length
squared, i.e.~of an inverse curvature.  Thus, at variance with the
scalar-tensor theories previously discussed, quadratic-gravity
corrections may require introducing a new fundamental length scale.
If the length scale is taken to be the Planck length,
quadratic-gravity corrections would be negligible at the scales of
compact objects: they would be suppressed by a factor of
\begin{equation}
 \frac{\ell_{\rm Planck}^2}{\ell_{\rm BH}^2}\sim 10^{-78}\,,
\end{equation}
where $\ell_{\rm Planck}\sim 10^{-35}\,{\rm m}$ is the Planck length,
and $\ell_{\rm BH}\sim 10\,{\rm km}$ is the typical scale of a compact
object.
However, as discussed in the introduction, the region close to compact
objects has been poorly probed, particularly when the spacetime is
highly dynamical.  Moreover, assuming that GR is correct all the way
down to the Planck scale, with no new gravitational physics along the
way, would be a tremendous extrapolation.\footnote{As an illustration,
  the gravitational potential at the Earth's surface, where Newtonian
  gravity proved to be extremely successful, is only 4 orders of
  magnitude smaller than the gravitational potential at the Sun's
  surface, where relativistic effects are relevant, as shown by the
  classical tests of GR. In a particle physics context, even a very
  successful theory such as quantum electrodynamics cannot be
  extrapolated from atomic to nuclear energy scales, where the strong
  interaction dominates over electromagnetism; and again, these two
  scales are separated by just 6 orders of magnitude.  } When
considering quadratic gravity, the standard approach is to assume the
existence of this new fundamental scale, unrelated to the Planck
scale, and proceed with calculations of observables from compact
objects.  For experimental observations to differ from GR predictions
the ``new'' length scale must be comparable to astrophysical scales.
Here we will adopt this agnostic phenomenological point of view.

In this approach, quadratic-curvature terms may be important when
dealing with nonlinear, relativistic solutions. Clearly, within this
perturbative context we can only consider corrections which are small
compared to the leading Einstein-Hilbert term.
In practice, the coupling functions $f_i$ are expanded as
\begin{equation}
 f_i(\phi)=\eta_i+\alpha_i \phi+{\cal O}(\phi^2)\,, \label{eq:quadratic_expansion}
\end{equation}
where $\eta_i$ and $\alpha_i$ ($ i=1,2,3,4$) are dimensionful coupling
constants. When the coupling functions are constant,
i.e.~$\alpha_i=0$, the theories above admit all vacuum GR
solutions~\cite{Yunes:2011we,Psaltis:2007cw}. However, even in this
case the background solutions generically have a different linear
response with respect to GR: for example, these theories predict a
different GW
emission~\cite{Yunes:2007ss,Barausse:2008xv,Molina:2010fb}. We will
mostly be interested in theories that modify the structure of BHs and
NSs, and we will consider the generic case $\alpha_i\neq0$. At any
rate, it is remarkable that in the weak-coupling limit (and provided
that the $f_i$'s are analytic functions) all viable quadratic
theories of gravity boil down to a small number of coupling
\emph{constants} that parametrize strong-curvature deviations from GR.

\subsubsection{Einstein-dilaton-Gauss-Bonnet gravity}\label{subsec:EdGB}
When $f_2=-4f_1=-4f_{3}$ and $f_4=0$, the
theory~\eqref{eq:action_quadratic} reduces to EdGB
gravity~\cite{Kanti:1995vq}, with action
\begin{equation}
  S=\frac{1}{16\pi}\int\sqrt{-g} d^4x \left[R-2\nabla_\mu\phi\nabla^\mu\phi-
    V(\phi)+f_1(\phi) R_{\rm GB}^2\right]\,, \label{EdGBaction}
\end{equation}
where $f_1(\phi)$ is a generic coupling function and the Gauss-Bonnet invariant $R_{\rm GB}^2$
has been defined below Eq.~(\ref{curvatureinvariants}). This is the only
quadratic theory of gravity whose field equations are of second
differential order for {\em any} coupling, and not just in the
weak-coupling limit. Indeed, when $f_1(\phi)=\alpha_{\rm GB}
e^{-2\phi}$, the theory reduces to the bosonic sector of heterotic
string theory~\cite{Gross:1986mw}.
Gauss-Bonnet gravity can also be seen as a particular case of
Horndeski gravity~\cite{Kobayashi:2011nu}, as mentioned in
Section~\ref{subsec:horndeski}. For instance, in the case
$f_1(\phi)=\alpha\phi$, the action~\eqref{EdGBaction} can be shown to
be equivalent to the action~\eqref{action_horndeski} with $K=X/2$,
$G_3=0$, $G_4=1/2$, $G_5=-2\alpha\ln|X|$~\cite{Sotiriou:2013qea}.

As in all of these theories, the coupling parameter is
\emph{dimensionful} and, specifically, it has dimensions of an inverse
curvature. 
It is thus natural to expect that the strongest constraints
on the theory should come from physical systems involving high
curvature: BHs, NSs and the early Universe.
We postpone a discussion
of BHs and NSs to Sections~\ref{sec:BHs} and~\ref{sec:NSs},
respectively. Here we anticipate the observational bounds that have
been derived.

Most bounds have been derived in the weak-coupling approximation, where one expects
\begin{equation}
 \sqrt{|\alpha_{\rm GB}|}\lesssim {\cal O}(L)\,,
\end{equation}
where $L$ is the typical curvature radius in the system under
consideration. Thus, Solar System constraints---such as those derived
by measuring the Shapiro time delay of the Cassini
probe~\cite{Bertotti:2003rm}---give a mild bound $\sqrt{|\alpha_{\rm
    GB}|}\lesssim10^{13}{\rm cm}$, which is in fact of the order of an
astronomical unit. On the other hand, as we shall discuss in
Section~\ref{sec:BHs}, BHs in this theory carry a scalar charge, and
observations of BH low-mass X-ray binaries give a constraint which is
six orders of magnitude stronger~\cite{Yagi:2012gp}:
\begin{equation}
 \sqrt{|\alpha_{\rm GB}|}\lesssim 5\times 10^6{\rm cm} \label{constraint_EdGB}
\end{equation}
(in the units of Eq.~\eqref{EdGBaction}).  As expected, this
constraint is comparable to the typical size of a stellar-mass BH.  On
the other hand, the only bound on EdGB gravity as an \emph{exact}
theory is of theoretical nature, because the existence of BH solutions
implies that $\sqrt{|\alpha_{\rm GB}|}$ be smaller than the BH horizon
size~\cite{Kanti:1995vq}; this bound implies
$\alpha_{\rm GB}/M^2\lesssim0.691$~\cite{Pani:2009wy}. Thus, the
observational constraint~\eqref{constraint_EdGB} is likely to be a
good estimate also for the exact EdGB gravity.

As previously mentioned, the bounds listed above are clearly satisfied
if one assumes that quadratic curvature corrections become relevant
only at the Planck scale. Nonetheless, they represent the best
constraints on quadratic gravity to date, and they were obtained
without any a priori assumptions on the regime in which deviations
from GR should be relevant.

\subsubsection{Chern-Simons gravity}
\label{sec:ChernSimons}

While the terms proportional to $f_1$, $f_2$ and $f_3$ in the
action~\eqref{eq:action_quadratic} are all associated with
qualitatively similar corrections to GR, the term proportional to
$f_4$ is peculiarly different, to the extent that the special case
$f_1=f_2=f_3=0$ describes a specific theory (Chern-Simons gravity)
which has been widely scrutinized in recent years
(see~\cite{Alexander:2009tp} for a review). At variance with EdGB
gravity, to avoid higher-order derivatives in the field equations
Chern-Simons theory must be considered as an EFT.  Almost all work so
far has focused on the special case $f_4=\alpha_{\rm CS}\phi$, working
perturbatively in the coupling constant $\alpha_{\rm CS}$. Then the
action reads
\begin{equation}
 S=\frac{1}{16\pi}\int\sqrt{-g} d^4x \left[R-2\nabla_\mu\phi\nabla^\mu\phi-V(\phi) +\alpha_{\rm CS}\, \phi\, {}^{*}\!RR\right]\,, \label{CSaction}
\end{equation}
and most of the literature considered the case of a vanishing scalar
potential: $V(\phi)=0$. Like the Gauss-Bonnet term, the
Chern-Simons term ${{}^{*}\!RR}$ is also a topological invariant, so that
if $f_4={\rm const}$ the theory is equivalent to GR.

For historical reasons, Chern-Simons gravity comes in two flavors: (i)
a nondynamical version in which the scalar kinetic term
in~\eqref{CSaction} is absent, and (ii) a theory where the scalar is a
true dynamical degree of freedom, that goes under the name of
dynamical Chern-Simons (dCS) gravity. These two theories are actually
very different from each other. Despite some confusion in the
literature, only the nondynamical theory is parity breaking, whereas
dCS gravity simply has different solutions than GR for spacetimes
which are not reflection-invariant, as in the case of spinning
objects. Furthermore, the nondynamical version introduces a
constraint, ${{}^{*}\!RR}=0$, arising from the variation of the CS
action with respect to the nondynamical scalar
field~\cite{Grumiller:2007rv}. This constraint limits the space of
solutions of the modified gravitational equations and introduces other
problems~\cite{Alexander:2009tp}. For these reasons, the dynamical
version of the theory has received much more attention in recent
years.

It can be shown that any spherically symmetric solution of GR is also
a solution of dCS gravity~\cite{Yunes:2007ss}, and this makes it
challenging to distinguish between the two theories. On the other
hand, dCS gravity is almost unique as an extension of GR, as it
predicts corrections only in the presence of a parity-odd source such
as rotation.  Among the most studied predictions of the theory are an
amplitude birefringence in GW propagation~\cite{Alexander:2009tp} and
modified spinning solutions, including corrections to Kerr BHs and
rotating NSs, to be discussed in detail in Sections~\ref{sec:BHs}
and~\ref{sec:NSs}.

To lowest order in the rotation rate, the CS modification to GR only
affects the gravitomagnetic sector of the metric. Tests of the theory
might therefore rely on frame-dragging effects.
Using the results of Gravity Probe B~\cite{Everitt:2011hp},
Ref.~\cite{AliHaimoud:2011fw} derived the bound
\begin{equation}
 \sqrt{|\alpha_{\rm CS}|}<{\cal O}(10^{13}){\rm cm}\,.\label{csobsb}
\end{equation}
As mentioned above, dCS gravity should be interpreted as an EFT, and
to have perturbative control requires $\alpha_{\rm CS}/M^2 \ll
1$. This requirement is stronger than the bound~\eqref{csobsb} for
BHs with masses $M\lesssim 10^8\,M_\odot$.

Similar bounds come from the Lense-Thirring effect as measured by the
LAGEOS satellites, which have also been used to constrain the
nondynamical version of the theory~\cite{Alexander:2009tp}. Note that
these bounds are of the order of an astronomical unit, as expected
from the previous dimensional analysis. The detection of GWs from an
extreme mass-ratio inspiral (EMRI) can potentially yield constraints
of the order $\sqrt{|\alpha_{\rm CS}|}<{\cal O}(10^{10}){\rm cm}$ or
even determine the Chern-Simons parameter with fractional errors below
5\%~\cite{Canizares:2012ji}.
Since large-curvature environments are expected to put
stronger bounds on the theory, the optimal systems to constrain
quadratic gravity are compact binaries. Indeed,
Ref.~\cite{Yagi:2012vf} derived projected bounds that are six orders
of magnitude more stringent than the one above by considering future
GW detection of the late inspiral of BH binaries. Similar bounds were
also recently estimated in Ref.~\cite{Stein:2014xba} by analyzing CS
corrections of rapidly-spinning Kerr BHs. Such corrections could be
constrained from electromagnetic observations of accreting stellar
mass BHs such as those in low-mass X-ray binaries, e.g.~GRO J1655$-$40
(cf.~e.g.~\cite{Shafee:2005ef}).

\subsection{Lorentz-violating theories}
\label{subsec:lorentz-viol}
\label{lv-theories}

While Lorentz invariance has been tested to high precision in the
matter
sector~\cite{Kostelecky:2003fs,Kostelecky:2008ts,Mattingly:2005re,Jacobson:2005bg,Liberati:2013xla},
constraints in the gravity sector are much weaker. Constraints on
Lorentz invariance in gravity beyond those obtainable in the Solar
System have received much interest after Ho\v
rava~\cite{Horava:2009uw} pointed out that a power-counting
renormalizable theory can be constructed by giving up Lorentz
invariance in gravity. 
We will focus here on Einstein-\AE ther and khronometric
gravity, which are the most generic theories violating boost symmetry in
gravity at low energies. A very clear review of these theories can be found 
in~\cite{Blas:2014aca}.\footnote{An alternative
 parametrized EFT approach to Lorentz violations in both the gravity
and matter sectors was developed by Kostelecky et
al.~\cite{Colladay:1998fq,Kostelecky:2003fs,Bailey:2006fd,Kostelecky:2010ze}.
For binary pulsar constraints in the EFT framework
of~\cite{Colladay:1998fq,Kostelecky:2003fs,Bailey:2006fd,Kostelecky:2010ze},
see~\cite{Shao:2014oha,Shao:2014bfa}.}

\subsubsection{Einstein-\AE ther}
To break boost invariance in the most generic way, one can describe
the gravitational degrees of freedom by means of a metric and a
timelike vector field, $\boldsymbol{u}$, usually referred to as the
``\ae ther.'' Up to total divergences, the most general action composed of
the metric and two or fewer derivatives of the \ae ther, and that
couples the \ae ther minimally to matter (so as to enforce the weak
equivalence principle and experimental evidence against the existence
of ``fifth forces'') is given by the Einstein-\AE ther
action~\cite{Jacobson:2000xp,Eling:2004dk,Jacobson:2008aj}
\begin{equation}
\label{eq:S-AE}
S_{\sAE} = \frac{1}{16\pi G_{\sAE}}\int \sqrt{-g}~ (R -M^{\a\b}{}_{\m\n} \nabla_\a u^\m \nabla_\b u^\n)~d^{4}x
+S_{{\rm mat}}[\Psi,\,g_{\mu\nu}]
\,,
\end{equation}
where
\begin{equation}
M^{\a\b}{}_{\m\n} = c_1 g^{\a\b}g_{\m\n}+c_2\d^{\a}_{\m}\d^{\b}_{\n}
+c_3 \d^{\a}_{\n}\d^{\b}_{\m}-c_4 u^\a u^\b g_{\m\n}\,,
\end{equation}
$c_i$ ($i=1,\dots,4$) are dimensionless couplings, and $\Psi$ denotes
the matter degrees of freedom. In this Section we do not assume
$G_N=1$; the ``bare'' $G_{\sAE}$ is related to the ``Newtonian''
gravitational constant $G_N$ measured locally by Cavendish-type
experiments via~\cite{Carroll:2004ai}
\begin{equation}
\label{eq:GN}
G_N=\frac{2 G_{\sAE}}{2-(c_1+c_4)}\,.
\end{equation}
To enforce the timelike character of the \ae ther, one has to impose
\begin{equation}
  \label{unit}
g_{\m\n}u^\m u^\n=-1\,,
\end{equation}
either implicitly or by adding a Lagrange multiplier $\ell(g_{\m\n}u^\m u^\n+1)$ 
in the variation of the action
above.

The field equations for Einstein-\AE ther theory can be derived by
varying the action \eqref{eq:S-AE} with respect to $g^{\a\b}$ and
$u^\mu$, while imposing the constraint \eqref{unit}. 
This results
in the following modified Einstein equations:
\begin{equation}\label{E_def}
E_{\a\b}\equiv G_{\a\b} - T^{\sAE}_{\a\b}-8\pi G_{\sAE} T^{\rm mat}_{\a\b}=0\,,
\end{equation}
where $ T^{\rm mat}_{\a\b}$ is the matter stress-energy tensor,
\begin{align}\label{Tae}
T^{\sAE}_{\a\b}&=-\nabla_\m\left(J^{\phantom{(\a}\m}_{(\a}u_{\b)}-J^\m_{\phantom{\m}(\a}u_{\b)}-J_{(\a\b)}u^\m\right)\nonumber
-c_1\,\left[
  (\nabla_\m u_\a)(\nabla^\m u_\b)-(\nabla_\a u_\m)(\nabla_\b u^\m)
  \right]\nonumber\\
&+\left[ u_\n(\nabla_\m J^{\m\n}){+}c_4 \dot{u}^2
  \right] u_\a u_\b +c_4 \dot{u}_\a \dot{u}_\b
-\frac{1}{2} M^{\s\r}{}_{\m\n} \nabla_\s u^\m \nabla_\r u^\n
g_{\a\b}\,,
\end{align}
$J^\alpha{}_{\mu}=M^{\alpha\beta}{}_{\mu\nu}\nabla_\beta u^\nu$,
and  $\dot{u}^{\alpha}\equiv u^{\beta}\nabla_{\beta}u^{\alpha}$.
These are completed by the \ae ther equations
\begin{equation}\label{AE_def}
  \sAE_\mu\equiv\left(\nabla_\a J^{\a\n}+c_4\dot{u}_\a\nabla^\n u^\a\right) \left(g_{\mu\nu}+u_\m u_\n\right)=0\,.
\end{equation}

Strong constraints on the coupling constants $c_i$ come from Solar
System tests. This is because Einstein-\AE ther theory predicts that
the (dimensionless) preferred-frame parameters $\alpha_1$ and
$\alpha_2$ of the PPN expansion will in general be nonzero functions
of the $c_i$'s~\cite{Foster:2005dk,Jacobson:2008aj}. Because Solar
System experiments constrain $|\alpha_1|\lesssim 10^{-4}$ and
$|\alpha_2|\lesssim 10^{-7}$~\cite{Will:2014xja}, one can expand the
theory in $\alpha_1$ and $\alpha_2$, and reduce the parameter space to
just two independent couplings $c_\pm\equiv c_1\pm c_3$.  The
remaining couplings are given by $c_2=(-2 c_1^2-c_1 c_3+c_3^2)/(3
c_1)+{\cal O}(\alpha_1,\alpha_2)$, $c_4=-c_3^2/c_1+{\cal
  O}(\alpha_1,\alpha_2)$~\cite{Foster:2005dk,Jacobson:2008aj}.
Further constraints on the two independent couplings $c_\pm$ come from
requiring that the theory should have positive energy (i.e.~no ghosts)
and that Minkowski space should be linearly stable (i.e.~no gradient
instabilities)~\cite{Will:2014xja}.

Einstein-\AE ther theory predicts the existence of not only spin-2
gravitational perturbations (like in GR), but also spin-1 and spin-0
gravitational perturbations. All these propagating modes have speeds
that are functions of the $c_i$, and which differ in general from the
speed of light~\cite{Jacobson:2004ts}. However, if these modes were
propagating at speeds lower than the speed of light, photons (or
relativistic particles) could Cherenkov radiate into the gravitational
field and lose energy to these modes, and this would lead to
(unobserved) experimental consequences~\cite{Elliott:2005va}.
Therefore, one has to impose that the speed of the spin-2, spin-1 and
spin-0 gravitons is larger than (or equal to) the speed of light.
Taking into account these constraints, one obtains the viable region
plotted in cyan in Figure~\ref{fig:LVconstraints} (left panel) for the
two independent couplings $c_\pm$.  As previewed in that figure and
discussed in Section~\ref{sec:pulsar_LV}, more stringent constraints on
$c_\pm$ come from binary pulsar data~\cite{Yagi:2013qpa,Yagi:2013ava}.

\subsubsection{Khronometric theory}
If we impose that the \ae ther is always hypersurface-orthogonal, one can express it as 
\begin{equation}
\label{ho}
u_\a=-\frac{\partial_\a T}{\sqrt{-g^{\m\n}\partial_\m T \partial_\n T}}\,,
\end{equation}
where $T$ is the hypersurface-defining scalar and the constraint
\eqref{unit} has already been enforced.
By assumption, surfaces of constant $T$ foliate the spacetime,
and one can re-express the action \eqref{eq:S-AE}
adapted to this ``preferred time'' $T$. This yields a different theory, described by the ``khronometric
theory'' action~\cite{Horava:2009uw,Blas:2009qj,Jacobson:2010mx}
\begin{multline}\label{action-K}
S_{K}=\frac{1-\beta}{16\pi G_{\sAE}}\!\int dT d^3x \, N\sqrt{h} \, \left(K_{ij}K^{ij} - \frac{1+\lambda}{1-\beta} K^2\right.
\\+\left.  \frac{1}{1-\beta}{}^{(3)}\!R +  \frac{\alpha}{1-\beta}\, a_ia^i\right)+S_{{\rm mat}}[\Psi,\,g_{\mu\nu}]\,,
\end{multline}
where $N=(-g^{TT})^{-1/2}$ is the lapse function, $K^{ij}$ is the extrinsic curvature of $T=$ constant hypersurfaces,
$h_{ij}$ is the induced spatial metric on those hypersurfaces,
${}^{(3)}\!R$ their three-dimensional Ricci curvature,
$a_i=\partial_i\ln{N}$, and the \ae ther is now related to the lapse via
$u_\a=-N\delta_{\a}^{T}  \,.$
We have also redefined the theory's parameters via
\begin{equation}\label{eq:EAtoKH}
\lambda\equiv c_2, \quad \beta\equiv c_3+c_1, \quad \alpha\equiv c_4+c_1.
\end{equation}
It should be stressed at this stage that the action \eqref{action-K} only depends on three couplings, as opposed to four
in the action~\eqref{eq:S-AE}. This is because the hypersurface-orthogonality constraint~\eqref{ho} makes it possible to
re-express one of those four couplings in terms of the remaining three without loss of generality.

The field equations of khronometric theory are obtained by replacing the hypersurface orthogonality constraint~\eref{ho}
in the action~\eqref{eq:S-AE}, and then varying the action with respect to $g_{\a\b}$ and $T$; they are
\begin{align}
\label{hl1}
E_{\alpha\beta}+2\sAE_{(\alpha}u_{\beta)}&=0\,,\\
 \label{hleq}
\nabla_\mu \left(\frac{\sAE^\mu}{\sqrt{-\nabla^\alpha T \nabla_\alpha T}}  \right)&=0\,.
\end{align}
Note that Eq.~\eref{hleq} actually follows from Eq.~\eref{hl1} and from the conservation of the matter stress-energy
tensor, i.e.~the only independent equations are the modified Einstein equations and the equations of motion of
matter~\cite{Jacobson:2010mx}. By comparing this set of equations with
the Einstein-\sAE ther equations~\eqref{E_def} and~\eqref{AE_def}, it
is easy to see that the hypersurface-orthogonal solutions of
Einstein-\AE ther theory will also be solutions of khronometric
theory. The converse is true in spherical
symmetry~\cite{Jacobson:2010mx,Blas:2011ni,Blas:2010hb,Barausse:2012ny},
but not in more general situations. For instance, slowly rotating BH
solutions are different in the two
theories~\cite{Barausse:2012ny,Barausse:2012qh,Barausse:2013nwa}.

As in the Einstein-\AE ther case, in Khronometric theory the PPN
preferred-frame parameters $\alpha_1$ and $\alpha_2$ are nonzero and
functions of the couplings. In light of the bounds $|\alpha_1|\lesssim
10^{-4}$ and $|\alpha_2|\lesssim 10^{-7}$~\cite{Will:2014xja}, one can
expand khronometric theory in $\alpha_1$ and $\alpha_2$. As a result,
one is left with two independent parameters (say $\beta$ and
$\lambda$), while the third parameter $\alpha$ is related to the first
two by $\alpha=2\beta+{\cal O}(\alpha_1,\alpha_2)$.\footnote{%
  Though it may seem that the conditions $\alpha_1=\alpha_2=0$ would
  reduce the dimensionality of the parameter space to a
  one-dimensional subspace, both $\alpha_1$ and $\alpha_2$ happen to
vanish for $\alpha=2\beta$ in
  Khronometric theory. Thus, the conditions $\alpha_1=\alpha_2=0$ still select 
a two-dimensional
  subspace.  
However, this only holds at the origin in $\alpha$-space,
  and so saturating the bounds to $|\alpha_1|\approx10^{-4}$ and
  $|\alpha_2|\approx10^{-7}$ reduces to a one-dimensional subspace. We
  refer the reader to~\cite{Yagi:2013ava} for a detailed discussion.}

From the hypersurface orthogonality constraint~\eqref{ho}, there are
no propagating spin-1 gravitational modes.  Requiring positive
energies, linear stability of Minkowski space, and the absence of
gravitational Cherenkov radiation for the remaining spin-0 and spin-2
degrees of freedom still selects a sizeable region of the parameter
space
$(\lambda,\beta)$~\cite{Blas:2010hb,Barausse:2011pu,Elliott:2005va},
shown in cyan in Figure~\ref{fig:LVconstraints} (right panel). Further
constraints come from requiring that the theoretically predicted Big
Bang nucleosyntesis elemental abundances agree with
observations~\cite{Carroll:2004ai,Yagi:2013qpa,Yagi:2013ava}; these
constraints are much stronger for Khronometric theory than for
Einstein-\AE{}ther~\cite{Carroll:2004ai,Jacobson:2008aj}, and are represented by 
the orange region in Figure~\ref{fig:LVconstraints} (right panel).  As
reviewed in that Figure and discussed
later in Section~\ref{sec:pulsar_LV}, even more stringent constraints
on $\lambda$ and $\beta$ come from binary pulsar
observations~\cite{Yagi:2013qpa,Yagi:2013ava}.

\subsubsection{Ho\v rava gravity}
The khronometric theory action \eqref{action-K} is particularly interesting because it is the low-energy (or infrared)
limit of Ho\v rava gravity~\cite{Horava:2009uw}, a renormalizable quantum field theory which has only spatial diffeomorphism invariance. The complete action of Ho\v rava gravity is~\cite{Blas:2009qj}
\begin{equation}
\label{SBPSHfull}
S_{\rm H}= \frac{1}{16\pi G_{\rm H}}\int dT d^3x \, N\sqrt{h}
\left(L_2+\frac{\hbar^2}{M_\star^2}L_4+\frac{\hbar^4}{M_\star^4}L_6\right)\,,
\end{equation}
where 
\begin{equation}\label{L2}
L_2=K_{ij}K^{ij} - \frac{1+\lambda}{1-\beta} K^2
+  \frac{1}{1-\beta}{}^{(3)}\!R +  \frac{\alpha}{1-\beta}\, a_ia^i
\end{equation}
is the Lagrangian density of Khronometric theory [c.f. Eq.~\eqref{action-K}],
$M_\star$ is a mass scale, and $L_4$ and $L_6$ are terms of fourth- and 
sixth-order in the spatial derivatives, but contain no
derivatives with respect to the preferred time $T$.

Complete constraints on $M_\star$ are somewhat elusive to obtain, and are probably one of the most important open
questions in Ho\v rava gravity~\cite{Liberati:2012jf}. The reason is that one would expect Lorentz violations to
percolate from gravity into the matter sector, where Lorentz symmetry has been verified to high precision by
particle physics and cosmic-ray
experiments~\cite{Kostelecky:2003fs,Kostelecky:2008ts,Mattingly:2005re,Jacobson:2005bg,Liberati:2013xla}. However,
several mechanisms have been put forward to suppress this percolation.  For instance, it has been suggested that the
operators that violate Lorentz symmetry in the matter sector might be finely tuned to much smaller values than those in
the gravity sector. Also, Lorentz invariance in matter might be an emergent property at low
energies~\cite{Froggatt:1991ft}, as an accidental symmetry~\cite{GrootNibbelink:2004za} or due to renormalization group
phenomena~\cite{Chadha:1982qq,Bednik:2013nxa} . Finally, it has been shown that two sectors with different Lorentz
violation degrees can easily coexist if their interaction is suppressed by a high mass scale~\cite{Pospelov:2010mp}, and
this could be the case for the gravity and matter sector. Therefore,
taking into account only the gravitational bounds
(i.e.~assuming that percolation of Lorentz violation into the matter sector is efficiently suppressed), one obtains
$M_{\star}\gtrsim 10^{-2}$ eV from sub-millimeter gravitational experiments. Also, perhaps surprisingly, $M_\star$
has an upper bound ($M_{\star}\lesssim 10^{16}$ GeV) from the requirement that the theory remains perturbative at all
scales~\cite{Papazoglou:2009fj,Kimpton:2010xi,Blas:2009ck}, so that the power-counting renormalizability arguments
proposed in~\cite{Horava:2009uw} apply.

Three things are worth stressing about the higher-order derivative terms $L_4$ and $L_6$ in the action. First, the
presence of sixth-order spatial derivatives is essential for power-counting renormalizability~\cite{Horava:2009uw}.
Second, the fourth and sixth order terms in the spatial derivatives generally lead to nonlinear dispersion relations
for the gravitational degrees of freedom of the theory, i.e.~the spin-2 and spin-0 gravitons (the latter present in
the theory because of the foliation-defining scalar $T$) satisfy
\begin{equation}
\label{mdsr}
\omega^2\propto k^2+\alpha_4 \left(\frac{\hbar}{M_\star}\right)^2 k^4+\alpha_6 \left(\frac{\hbar}{M_\star}\right)^4 k^6+\ldots\, ,
\end{equation}
where $\omega$ and $k$ are respectively the frequency and the wave-number, while $\alpha_4$ and $\alpha_6$
are dimensionless constants.  Because such a dispersion relation allows for infinite propagation speeds in the
ultraviolet limit, the notion of a BH may appear problematic in these theories. However, we will return to this problem
in Section~\ref{sec:BH_LV} and show that the presence of a dynamical foliation-defining scalar $T$ actually allows for BHs
to be defined in this theory as well~\cite{Barausse:2011pu,Blas:2011ni,Barausse:2013nwa}.  Third, aside from
instantaneous propagation at very high energies, the higher-order terms $L_4$ and $L_6$ are typically negligible in
astrophysical settings~\cite{Barausse:2013nwa}. These terms induce corrections on the spacetime geometry around
astrophysical objects that are of order ${\cal{O}}(G_{N}^{-2} M^{-2} M_\star^{-2}) \sim {\cal{O}}(\mpl^4/(M
M_{\star})^2)$ for an object of mass $M$, which translates into an error $\lesssim 10^{-16} (M_{\odot}/M)^2$.

\subsubsection{$n$-DBI gravity \label{subsec:n-dbi}}
Inspired by the approximate scale invariance of the Universe at early and late 
times, when it is believed to be approximately
de Sitter, Herdeiro et al.~\cite{Herdeiro:2011im,Herdeiro:2011km} proposed a modification of GR that automatically results in
inflation at early times, without the need for additional scalar fields. This model, dubbed $n$-DBI gravity, was
designed so that it yields the Dirac-Born-Infeld type conformal scalar theory when the Universe is conformally
flat and resembles Einstein's gravity
in weakly curved space-times.
Interestingly, not only does it result in inflation, but it can also accommodate a smooth transition to radiation- and
matter-dominated epochs, followed by late time acceleration. The two distinct accelerating periods, with two distinct
effective cosmological constants, are a manifestation that the cosmological constant can vary in this theory. Moreover,
a large hierarchy between these two cosmological constants can be naturally achieved if the naive cosmological constant
appearing in a weak-curvature expansion of the theory is associated to
the TeV scale, which also suggests a new mechanism to
address the cosmological constant problem. The action for $n$-DBI
gravity is~\cite{Herdeiro:2011im}
\begin{equation}
  S_{n{\rm DBI}}=-\frac{3\lambda}{4\pi G_N^2}\int d^4x\,\sqrt{-g}\,
  \left\{\sqrt{1+\frac{G_N}{6\lambda}(R+\mathcal{K})}-q\right\}\,,
  \qquad \mathcal{K}=-2\nabla_\mu(n^\mu\nabla_\nu n^\nu)\,.\label{action-nDBI}
\end{equation}
It contains two dimensionless parameters $\lambda$ and $q$ and an everywhere time-like vector field $\boldsymbol{n}$
coupled to the gravitational sector which breaks Lorentz invariance and makes the theory invariant under \emph{foliation
  preserving diffeomorphisms}, in a way similar to Ho\v{r}ava-Lifschitz gravity. Concretely, if we perform an
Arnowitt-Deser-Misner (ADM) decomposition~\cite{Arnowitt:1960es}, then
$\boldsymbol{n}$ determines the lapse function $N$
through $n_\mu=-N\,dt$. This gives rise to a scalar degree of freedom, in addition to the two tensor polarizations of
GR~\cite{Coelho:2012xi}. Remarkably, the term $\mathcal{K}$ in (\ref{action-nDBI}) allows for the equations of motion to remain at most
second order in time derivatives, despite an infinite power series in the Ricci curvature.

Any solution of Einstein's gravity with cosmological
constant plus matter, admitting a foliation with constant
$R+\mathcal{K}$, is also a solution of $n$-DBI gravity. Moreover,
any Einstein space admitting a foliation with constant
$^{(3)}R-N^{-1}\Delta N$ (where $^{(3)}R$ is
the Ricci scalar of the 3-dimensional hypersurfaces) is a solution of $n$-DBI gravity~\cite{Herdeiro:2011im}.  By
requiring spherical symmetry, one can explicitly obtain the Schwarzschild, Reissner-Nordstr\"om and (anti-)de Sitter BH
solutions, albeit in an unusual set of coordinates. Unlike GR, however, the cosmological constant is not determined at
the level of the action, but appears instead as an integration constant. The foliation condition of constant
$^{(3)}R-N^{-1}\Delta N$ can be interpreted as the {\it maximal slicing} gauge
condition common in numerical
relativity~\cite{Alcubierre:2008}, and it is then straightforward to show that 
the Kerr metric in Boyer-Lindquist
coordinates is also a solution of $n$-DBI gravity~\cite{Coelho:2013zq}.

Since $n$-DBI gravity has a preferred foliation, one might expect the
PPN preferred-frame parameters $\alpha_1$ and $\alpha_2$ to be
nonvanishing.
Then experimental bounds on the PPN parameters should provide a lower bound for $\lambda$ which, together with
the estimate coming from inflation, would in principle define a finite
interval of viability $\lambda_{\rm PPN}<\lambda<\lambda_{\rm inf}$.
However, in a perturbative expansion about Minkowski, solutions of
$n$-DBI gravity coincide with those of GR and exist for all values of
$\lambda,q$~\cite{Coelho:2013dya}. Thus, at least to first PN order,
we have $\alpha_{1}=\alpha_{2}=0$, and $n$-DBI is indistinguishable
from GR in the Solar System.

\subsection{Massive gravity and Galileons}
\label{subsec:massgrav}
de Rham-Gabadadze-Tolley (dRGT) massive gravity is an  infrared modification of GR  in which gravity is  described by a
local, Lorentz-invariant,  self-interacting, massive spin-2 field.  If the mass of  the graviton is of  order the Hubble
scale today, $m_g \sim 10^{-33}\rm{eV}$, massive gravity may explain the observed cosmic acceleration (see~\cite{deRham:2014zqa} for a
recent review). For clarity of presentation, in this Section we do not assume $G_N=1$, but the
gravitational constant is expressed in terms of the Planck mass $\mpl=(8\pi G_N)^{-1/2}$.

Historically, one of the main challenges in constructing a consistent theory of massive gravity has been preventing the
appearance of a scalar ghost mode in the spectrum. The existence of this spurious ghost mode follows from a simple
counting argument: a massive spin-2 particle should have five degrees of freedom, but there are six possible
polarizations for GWs carried by a symmetric tensor $h_{ij}$. Thus a theory of massive gravity needs to
contain a constraint so that this sixth allowed mode is not present. At the linear level, this problem was solved by
Fierz and Pauli by choosing a specific tuning in the mass term~\cite{Pauli:1939xp}. However, Boulware and Deser showed
that the sixth mode generically reappears as a ghost at the nonlinear level~\cite{Boulware:1973my}. When this mode
arises nonlinearly, it is referred to as a Boulware-Deser (BD) ghost.

dRGT massive gravity, originally proposed in~\cite{deRham:2010ik,deRham:2010kj}, was constructed to avoid the BD ghost
to all orders around any background. The action is
\begin{equation}
S_{\rm dRGT} = \int d^4 x \sqrt{-g} \left[ \frac{\mpl^2}{2}  R + \frac{\mpl^2 m_g^2}{4}
\sum_{n=0}^{4} \alpha_n \mathcal{L}_n(\mathcal{K}) \right]\,,
\label{eq:S-dRGT}
\end{equation}
where
$\mathcal{K}^\mu_{\ \ \nu}\equiv \delta^\mu_\nu -
\sqrt{g^{\mu\alpha}f_{\alpha \nu}} $,
$m_g$ is the graviton mass and $\alpha_i$ are constant
coefficients\footnote{The matrix square root
  $\sqrt{g^{\mu\alpha}f_{\alpha \nu}}\equiv{M^{\mu}}_\nu$ is defined
  in such a way that
  ${M^\mu}_\rho {M^\rho}_\nu = g^{\mu\alpha}f_{\alpha \nu}$.  For flat
  backgrounds, there is no problem defining the matrix square root in
  the action perturbatively around flat space using the infinite
  series expansion given in~\cite{deRham:2010kj}. It is possible to
  avoid dealing with matrix square roots by defining massive gravity
  in the vielbein
  language~\cite{Nibbelink:2006sz,Hinterbichler:2012cn}, in which case
  the mass term becomes a finite polynomial in the vielbeins, rather
  than a matrix square root. In this formalism there is no need to
  take any matrix square roots: cf. Eqs.~(6.1) and (6.2)
  of~\cite{deRham:2014zqa}.}.

The Lagrangians $\mathcal{L}_n$ are functions of symmetric tensors
$X_{\mu\nu}$:
\begin{equation}
\label{eq:Ln}
\mathcal{L}_n[X] = \epsilon^{\mu_1 \mu_2 \mu_3 \mu_4}\epsilon_{\nu_1 \nu_2 \nu_3 \nu_4}
\prod_{k=1}^{n} X^{\nu_k}_{\mu_k} \prod_{k'=n+1}^{4}\delta^{\nu_{k'}}_{\mu_{k'}},
\end{equation}
so that $\mathcal{L}_0[X] = 4!,\ \mathcal{L}_1[X] = 3!  X^\mu_\mu,\ \mathcal{L}_2[X]=2!\left(X^\mu_\mu X^\nu_\nu -
X^\mu_\nu X^\nu_\mu \right)$, etc. The existence of two second-class constraints (and thus the absence of the BD ghost)
has been confirmed by many authors, see for example~\cite{Hassan:2011tf,Hassan:2011hr}.

The metric $f_{\mu\nu}$ is a fixed, external metric called the ``reference metric.'' The reference metric is needed because
the only nontrivial scalar function that can be built out of a single metric is the determinant $\det(g)$ or functions
of the determinant, which simply give a cosmological constant or
a single new scalar mode~\cite{Boulware:1973my}. Recently
it has been shown that in certain frameworks the reference metric can be eliminated
altogether~\cite{Bernard:2014bfa}. We may also make the reference metric dynamical by adding a second Einstein-Hilbert
term $S_f=\frac{M_f^2}{2} \int d^4 x \sqrt{-f} R[f]$ to the 
action~\cite{Hassan:2011zd}. This is known as bigravity; in this case
the spectrum consists of one massless and one massive graviton.

In the following we will mostly focus on massive gravity with a fixed Minkowski reference metric.

\paragraph{Decoupling limit.}
Many classic tests of gravity take place at length scales much shorter than the Hubble scale, and in regions where the
gravitational field is weak: $|h/\mpl| \ll 1$. In this regime, we may study 
massive gravity in a simple approximation of the
fully nonlinear theory by considering the decoupling limit: $m_g\rightarrow 0$ and $\mpl \rightarrow \infty,$ with
$\Lambda_3 = (m_g^2 \mpl)^{1/3}$ fixed. To be relevant for cosmology
we need $m_g \sim 10^{-33} \ {\rm eV}$, so $\Lambda_3 \sim
(1000\ {\rm km})^{-1}$. In all cases where calculations have been done in the full theory and in the decoupling limit,
the decoupling limit turned out to be an excellent approximation: see e.g.~\cite{Babichev:2010jd}.  In this
limit, we may decompose the metric perturbation $H_{\mu\nu}=g_{\mu\nu}-\eta_{\mu\nu}$ into helicity eigenstates: two
helicity-2 modes, two helicity-1 modes and one helicity-0 mode (see~\cite{deRham:2010ik} for a derivation of the decoupling
limit). The helicity-2 modes have the same dynamics as in GR, while the helicity-1 modes are not sourced by matter in
this limit. Therefore, we will focus exclusively on the helicity-0 mode, whose dynamics is governed by
\begin{equation}
\label{eq:galileon}
S_{\rm gal}[\pi] = \int d^4 x \left\{-\frac{3}{4}(\pa \pi)^2 \!+\! \sum_{n=3}^5 
c_n
\mathcal{L}_n^{(g)}\left[\frac{1}{\Lambda_3^3}
\partial_\mu \partial_\nu \pi\right] \!+\! \frac{g_1}{\mpl} \pi T \!+\! \frac{g_2}{\mpl \Lambda_3^3} \partial_\mu \pi
\partial_\nu \pi T^{\mu\nu}\right\},
\end{equation}
where $T_{\mu\nu}$ is an external stress-energy tensor, $c_n$, $g_i$ are constant coefficients,
$\mathcal{L}_n^{(g)}= \pi {\cal L}_{n-1}$, and ${\cal L}_n$ are the same as those given in Eq.~\eqref{eq:Ln}. The
interactions for the $\pi$ field are called the Galileon interactions~\cite{Nicolis:2008in};
$\mathcal{L}_3^{(g)}, \mathcal{L}_4^{(g)}, \mathcal{L}_5^{(g)}$ are called the Cubic, Quartic, and Quintic Galileons, respectively. The
Galileon interactions ensure that $\pi$ has second-order equations of motion, which reflects the fact that the BD ghost
is not present.

As can be seen from Eq.~(\ref{eq:galileon}), the helicity-0 mode remains coupled to matter in the decoupling limit, in
which $m_g\rightarrow 0$. This surprising fact is known as the van Dam-Veltman-Zakharov (vDVZ)
discontinuity~\cite{vanDam:1970vg, Zakharov:1970cc}. This would appear to rule out massive gravity because the
helicity-0 mode would then source a fifth force of gravitational strength, so that, for example, the bending of light by
the Sun in massive gravity would differ from the GR prediction by $25\%$.

\paragraph{Vainshtein mechanism.}
The resolution to the vDVZ discontinuity, as originally proposed by
Vainshtein in~\cite{Vainshtein:1972sx}, is that we cannot ignore the
nonlinear self-interactions of the helicity-0 mode. These interactions
serve to suppress the coupling to matter, restoring continuity with GR
(see~\cite{Babichev:2013usa} for an introduction).  This is why it was
crucial to keep the scale $\Lambda_3$ fixed in the decoupling limit.

The Vainshtein mechanism was proved to work under specific assumptions
(e.g.  for spherically symmetric, static spacetimes). A general proof
of its validity is still lacking, but there have been some studies of
the Vainshtein mechanism in time-dependent situations, including
binary pulsars \cite{deRham:2012fw,deRham:2012fg} and cosmology
\cite{Chow:2009fm,deRham:2011by}. Furthermore, there is by now a fair
amount of numerical evidence that the Vainshtein mechanism operates
even beyond the spherically-symmetric static solutions for
Galileons. For example, Koyama and collaborators carried out numerical
simulations characterizing the strength of the Vainshtein mechanism
\cite{Li:2013tda,Li:2013nua}, considering in particular a two-body
system that breaks spherical symmetry~\cite{Hiramatsu:2012xj} and
using $N$-body simulations to study the growth of structures, such as
dark matter halos and cosmic webs~\cite{Falck:2014jwa,Falck:2015rsa}.

From a field-theoretic perspective, the Vainshtein mechanism may be
understood by considering fluctuations in the Galileon
($\pi = \pi_0 + \phi$) in a regime where the background is large, in
the sense that $\partial^2 \pi_0 \gg \Lambda_3^3$. Then, expanding to
quadratic order, the fluctuations have the quadratic action (using the
Cubic Galileon for definiteness)
\begin{equation}
S[\pi_0 + \phi] = \int d^4 x Z[\pi_0]^{\mu\nu} \partial_\mu \phi \partial_\nu \phi + \frac{1}{\mpl} \phi T + \cdots\,,
\end{equation}
where $Z\sim \partial^2 \pi_0 / \Lambda_3^3 \gg 1$. After canonically
normalizing, $\phi \rightarrow \phi / \sqrt{Z}$, the effective
coupling to matter is redressed:
$\mpl \rightarrow \mpl \sqrt{Z} \gg \mpl$. Thus the Galileon decouples
from matter once this effect is taken into account.

Solutions for Galileons around a static, spherically symmetric source
of mass $M$ that exhibit the Vainshtein screening mechanism have been
explicitly constructed in~\cite{Nicolis:2004qq, Nicolis:2008in,
  Brito:2014ifa}. The field profile has a characteristic length scale
called the Vainshtein radius,
$r_V \equiv \Lambda_3^{-1} (M/\mpl)^{1/3}$. For the Sun,
$r_{V,\odot}\sim\ 100\ {\rm pc}$. The Galileon generates a fifth force
sourced by the mass. At distances large compared to the Vainshtein
radius, the Galileon force is comparable to Newtonian gravity:
$F_\pi / F_N \rightarrow 1$. Yet this force is highly suppressed at
short distances: $F_\pi/F_N \sim (r/r_V)^\alpha$, where $\alpha=3/2$
for the Cubic Galileon and $\alpha=2$ for the Quartic Galileon (the
Quintic Galileon interactions vanish in the spherically symmetric
case). Perturbations to these solutions have been shown to be stable
for a wide variety of parameters~\cite{Nicolis:2008in}.

\paragraph{Hassan-Rosen bimetric theory and other nonlinear massive gravity theories.}
Recently, there has been revived interest in nonlinear theories of
massive gravity. Here we mostly focus on dRGT theory and on
the bimetric extension proposed by Hassan and
Rosen~\cite{Hassan:2011zd}. In bimetric massive gravity both metrics
are dynamical and the theory propagates seven degrees of freedom,
corresponding to a massive graviton and to a massless graviton.  A
different proposal is nonlocal massive gravity~\cite{Jaccard:2013gla},
in which the field equations are nonlocal but respect causality. This
theory propagates the five degrees of freedom of a massive graviton
plus a scalar ghost which, however, has the same mass as the massive
graviton. Therefore, the ghost is associated with a vacuum
instability, which is irrelevant even at cosmological
scales~\cite{Jaccard:2013gla,Maggiore:2013mea}. Because of its
nonlocal character, few phenomenological studies of this theory are
available to date.

\paragraph{Observational tests.}
A classic observational way to constrain massive gravity is to consider constraints on the Yukawa force law in the Solar System. In the weak-field
limit, the Yukawa gravitational potential of a point mass takes the form $V(r)\sim e^{-m_g r}/r$. Tests of deviations from
the inverse square law from Solar System experiments set the bound $m_g < 0.5\times 10^{-21}\ {\rm eV}$~\cite{Will:2014xja}.

The best bounds on Galileons come from Lunar Laser Ranging. In~\cite{Dvali:2002vf,Lue:2002sw} it was shown that the
fifth force discussed above in the context of spherically symmetric solutions leads to an anomalous precession of the
perihelion of the moon
\begin{equation}
\delta \phi \sim \frac{\delta\Phi}{\Phi} \sim \left(\frac{R}{r_V}\right)^{\alpha} \ {\rm radians / orbit},
\end{equation}
where $\Phi$ is the gravitational potential, $\delta\Phi$ is the extra 
contribution of the helicity-zero mode $\pi$ to the potential, $R$ is the 
semi-major radius, and $\alpha$ depends on the Galileon model 
(cf.~\cite{deRham:2014zqa} for a review). For the Cubic Galileon $\alpha = 
3/2$, leading to $\delta\phi \sim
10^{-12}\ {\rm rad/orbit}$ when using parameters relevant for cosmology. As the current observational precision is $\sim
0.5\times 10^{-11}\ {\rm rad/orbit}$~\cite{Will:2014xja}, next-generation 
experiments can potentially rule out the Cubic
Galileon. For the Quartic Galileon, $\alpha=2$. For $m_g \sim 10^{-33}$ this yields $\delta \phi \sim 10^{-16}\ {\rm
rad/orbit}$, which cannot be probed by current experiments. Turning this around and phrasing it as a bound on $\Lambda$, we find $\Lambda >
10^{-11}\ {\rm eV}$, or alternatively $m_g < 10^{-30}\ {\rm eV}$.

\subsection{Gravity with auxiliary fields}
\label{subsec:auxiliary}
All the theories previously discussed imply, in one way or another,
the existence of extra \emph{dynamical} fields.
This is the most common way to circumvent Lovelock's theorem.
There is, however, a more subtle way which does not require any additional degrees of freedom~\cite{Pani:2013qfa}. Specifically, it is
possible to construct a theory which modifies only the right-hand side of Einstein's equations, adding another
symmetric rank-2 tensor constructed solely from the metric
and the matter fields (i.e.~without introducing any new degrees of freedom). The
additional term must be identically divergence-free so as to not
compromise the weak equivalence principle.

These requirements may seem hard to satisfy simultaneously, but there is in fact a generic prescription to construct
such theories. In some special theories that include \emph{auxiliary} (i.e., nondynamical) fields, eliminating these fields leads
precisely (and generically) to the type of modification of Einstein's
equation just described, without modifying
the field equations of matter.  Two representative theories belonging to this
class are Palatini $f({\cal R})$
gravity~\cite{Sotiriou:2008rp,Olmo:2011uz} and the so-called
Eddington-inspired Born-Infeld (EiBI) gravity~\cite{Banados:2010ix},
as discussed below.

\paragraph{Palatini $f({\cal R})$ gravity.}
The action of $f({\cal R})$ gravity in the Palatini formalism reads
\begin{equation}
  S=\frac{1}{16\pi}\int d^4x \sqrt{-g} f({\cal R})+S_M[\Psi,\,g_{\mu\nu}]\,,
  \label{actionPalatini}
\end{equation}
where henceforth ${\cal R}=g^{\mu\nu}{\cal R}_{\mu\nu}$ and ${\cal
  R}_{\mu\nu}$ denotes the Ricci tensor built from the connection
$\Gamma_{\mu\nu}^\sigma$ (which is assumed to be symmetric), to
distinguish it from the Ricci tensor $R$ built from the Levi-Civita
connection of the metric $g_{\mu\nu}$, as in the metric formalism
discussed in Section~\ref{subsec:ST}.
Recall that in the Palatini (or affine)
approach the connection $\Gamma_{\mu\nu}^\sigma$ is considered as an
independent field, which enters the action~\eqref{actionPalatini} on
the same footing as the metric $g_{\mu\nu}$. This choice has dramatic
consequences for the theory, which is in fact completely different
from \emph{metric} $f(R)$ gravity. By varying the
action~\eqref{actionPalatini} with respect to the metric and the
independent connection, the field equations can be cast in the
form~\cite{Sotiriou:2008rp}
\begin{align}
 f'({\cal R}){\cal R}_{(\mu\nu)}-\frac{g_{\mu\nu}}{2}f({\cal R})&=8\pi T_{\mu\nu}\,, \label{palatini1}\\
 \tilde{\nabla}_\mu[\sqrt{-g}f'({\cal R})g^{\mu\nu}]&=0   \label{palatini2} \,,
\end{align}
where $\tilde{\nabla}_\mu$ is the covariant derivative associated with $\Gamma_{\mu\nu}^\sigma$, $T^{\mu\nu}\equiv
-2(-g)^{-1/2}\delta S_M/\delta g_{\mu\nu}$ is the standard stress-energy tensor,
whose indices are raised and lowered with
$g_{\mu\nu}$, and a prime denotes derivative with respect to ${\cal R}$. GR is recovered when $f({\cal R})={\cal R}$, because Eq.~\eqref{palatini2} becomes the definition of the
Levi-Civita connection; this implies ${\cal R}_{\mu\nu}=R_{\mu\nu}$
and, in turn, Eq.~\eqref{palatini1} yields the standard
Einstein equations. Remarkably, in this framework the fact that $\Gamma_{\mu\nu}^\sigma$ is the Levi-Civita connection
of $g_{\mu\nu}$ emerges \emph{dynamically}, and it is not imposed a priori as in the standard Einstein-Hilbert action.

Nevertheless, for generic functions $f({\cal R})$ the field equations~\eqref{palatini2} imply that
$\Gamma_{\mu\nu}^\sigma$ is the Levi-Civita connection of the conformal metric $h_{\mu\nu}=f'({\cal R})g_{\mu\nu}$, and
the dynamical content of the theory is very different from metric $f(R)$. The connection can be expressed
\emph{algebraically} in terms of $g_{\mu\nu}$ and the matter fields, and the field equations in
Palatini $f(\cal{R})$ gravity read~\cite{Sotiriou:2008rp}
\begin{equation}
\begin{aligned}
  G_{\mu\nu}={}&\frac{8\pi}{f'}T_{\mu\nu}-\frac{g_{\mu\nu}}{2}\left({\cal R}-\frac{f}{f'}\right)+
  \frac{1}{f'}(\nabla_\mu\nabla_\nu-g_{\mu\nu}\square)f'\\
  &{}-\frac{3}{2{f'}^2}\left[(\nabla_\mu f')(\nabla_\nu f')-
    \frac{g_{\mu\nu}}{2}(\nabla f')^2\right]\,,
\end{aligned} \label{palatiniF}
\end{equation}
where now $\nabla_\mu$ is the covariant derivative associated with the 
Levi-Civita connection of the metric.  Furthermore, by taking the trace of 
Eq.~\eqref{palatini1} one obtains
the following algebraic equation for ${\cal R}$:
\begin{equation}
 f'({\cal R}){\cal R}-2f({\cal R})=8\pi T\,,
\end{equation}
which reduces to the standard trace relation $R=-8\pi T$ in the GR limit. For a given $f$ the equation above can be solved for ${\cal
  R}$, and by plugging the solution back into Eq.~\eqref{palatiniF} one obtains 
a set of field equations that depend
only on the metric $g_{\mu\nu}$ and on the stress-energy tensor $T_{\mu\nu}$, with no extra degrees of freedom.

Thus, the theory has effectively the same degrees of freedom as GR, i.e.~it propagates only a massless spin-2
graviton. This is a striking difference with respect to metric $f(R)$ gravity, which propagates an extra scalar field
and is in fact equivalent to scalar-tensor theory. It can be shown that Palatini $f({\cal R})$ is also equivalent
to a scalar-tensor theory in the form~\eqref{STactionJ} with a potential that depends on the functional form of $f(R)$
and with $\omega(\phi)=0$, i.e.~the kinetic term is vanishing and the scalar field is nondynamical or
\emph{auxiliary}~\cite{Sotiriou:2008rp}.  This theory is \emph{equivalent} to GR in vacuum ($T_{\mu\nu}=0$),
but the gravitational field equations contain a nonlinear source through terms like $\square f'$.
Nonetheless,
$\nabla_\mu T^{\mu\nu}=0$ identically, as can be seen by the fact that matter fields are minimally coupled to the metric
in the action~\eqref{actionPalatini}.

\paragraph{EiBI gravity.} 
The idea behind EiBI gravity is to incorporate the Palatini approach into a gravitational analog of Born-Infeld
nonlinear electrodynamics~\cite{Tseytlin:1999dj} which removes the divergence of the electron self-energy by introducing
an upper bound of the electric field at the origin. Studies of similar proposals for a gravitational Born-Infeld-like
action are thus motivated by the prospect of resolving the curvature singularities that afflict GR in a similar fashion.

Inspired by Born-Infeld theory, EiBI gravity is described by the action~\cite{Banados:2010ix}
\begin{equation}
  S = \frac{1}{8\pi \kappa}\int d^4x\Big(\sqrt{|\det\left(g_{\mu\nu} + \kappa {\cal R}_{(\mu\nu)}\right)|}-
  (1+\kappa \Lambda) \sqrt{g}\Big)+S_M\left[g_{\mu\nu},\Psi\right]\,,\label{actionEiBI}
\end{equation}
where $g=|\det(g_{\mu\nu})|$, $\Lambda$ is the cosmological constant and $\kappa$ is a new EiBI
parameter with dimensions of length squared.  EiBI gravity is naturally based on the Palatini formulation because
in the metric approach the field equations contain ghosts, which must be eliminated by adding extra terms to the
action~\cite{Deser:1998rj,Vollick:2003qp}. Here we
will focus on the original EiBI proposal~\cite{Banados:2010ix} (cf.~\cite{Pani:2012qd} for a discussion).

When expanded at second order in $\kappa{\cal R}_{\mu\nu}$, the action~\eqref{actionEiBI} takes the form
\begin{equation}
  S=\frac{1}{16\pi}\int d^4x\sqrt{g}\left[{\cal R}-2\Lambda+\frac{\kappa}{4}\left({\cal R}^2-
    2 {\cal R}_{(\mu\nu)}{\cal R}^{(\mu\nu)}\right)\right]+S_M\left[g_{\mu\nu},\Psi\right]+{\cal O}(\kappa^2)\,, \label{action2}
\end{equation}
and to lowest order EiBI gravity reduces to the Palatini formulation of GR with a cosmological constant. To
next-to-leading order, quadratic corrections in the curvature tensor built from the independent connection appear in the
action~\eqref{action2}. The Palatini formulation guarantees that, despite these extra terms, no higher derivatives of
the metric would appear in the field equations. When expanded order by order in $\kappa$, the action~\eqref{actionEiBI}
takes the form of a specific Palatini $f({\cal R},{\cal R}_{\mu\nu})$ theory.

Beyond the perturbative level, independent variation of the action~\eqref{actionEiBI} with respect to the metric and the
connection yields
\begin{align}
\sqrt{q}q^{\mu\nu}&=\sqrt{g}\left[(1+\kappa\Lambda) g^{\mu\nu}-8\pi\kappa T^{\mu\nu}\right]\label{eqALG}\,,\\
0&=\tilde{\nabla}_\sigma[\sqrt{q} q^{(\mu\nu)}]-\tilde{\nabla}_\gamma[\sqrt{q} q^{\gamma(\mu}]\delta^{\nu)}_\sigma\,,\label{varGamma}
\end{align}
where we have defined $q_{\mu\nu}\equiv g_{\mu\nu}+\kappa {\cal R}_{(\mu\nu)}$, $\tilde{\nabla}_a$ is the
covariant derivative associated with $\Gamma_{\mu\nu}^\sigma$, and $q^{\mu\nu}$ 
is the inverse of $q_{\mu\nu}$. After
some manipulation, Eq.~\eqref{varGamma} implies that $\Gamma_{\mu\nu}^\sigma$ is 
the Levi-Civita connection of
$q_{\mu\nu}$ and, by using Eq.~\eqref{eqALG}, one obtains an {\em algebraic} equation that determines $q_{\mu\nu}$ in
terms of $g_{\mu\nu}$ and $T_{\mu\nu}$.
Similarly to the case of Palatini $f({\cal R})$ gravity, one can eliminate $\Gamma_{\mu\nu}^\sigma$ from the field
equations. The final set of equations is of second differential order in the metric $g_{\mu\nu}$ and contains
second derivatives of $T_{\mu\nu}$.  This is true in the full theory, but it becomes more explicit by working
perturbatively in the $\kappa {\cal R}_{\mu\nu}\ll1$ limit. To first order, the field equations read~\cite{Pani:2012qd}
\begin{equation}
  \begin{split}
 R_{\mu\nu}={}&\Lambda g_{\mu\nu}+8\pi\left(T_{\mu\nu}-\frac{1}{2}Tg_{\mu\nu}\right)\\
 &{}+\kappa\left[S_{\mu\nu}-\frac{1}{4}Sg_{\mu\nu}\right]+\frac{\kappa}{2}\left[\nabla_\mu\nabla_\nu \tau-2\nabla_\sigma\nabla_{(\mu}\tau_{\sigma\nu)}+\square\tau_{\mu\nu}\right]+{\cal O}(\kappa^2)\,, \label{eqg2}
\end{split}
\end{equation}
where $S_{\mu\nu} \equiv 64\pi^2[{T^\sigma}_\mu T_{\sigma \nu}-\frac{1}{2}T 
T_{\mu\nu}]$ and $\tau_{\mu\nu}\equiv 8\pi(
T_{\mu\nu}-\frac{1}{2}g_{\mu\nu}T)+\Lambda g_{\mu\nu}$.
Einstein's theory is recovered when $\kappa\to 0$. For $\kappa\neq 0$, Eq.~\eqref{eqg2} contains second derivatives of
$T_{\mu\nu}$.  This is in contrast to Einstein's theory, where the stress-energy tensor appears on the right-hand side
of Einstein's equations at zero differential order.  This different structure is also evident in the Newtonian limit of
the theory. The modified Poisson equation sourced by a matter density $\rho$ is~\cite{Banados:2010ix,Pani:2012qb}
\begin{equation}
 \nabla^2\Phi=4\pi{\rho}+2\pi{\kappa}\nabla^2\rho\,, \label{PoissonEiBI}
\end{equation}
whose solution reads $\Phi=\Phi_{\rm N}+2\pi {\kappa}\rho$, where $\Phi_{\rm N}$ is the standard Newtonian potential.

\paragraph{Generalized auxiliary field gravity.}
It has recently been pointed out that Palatini $f({\cal R})$ and EiBI gravity are only two examples of a generic class
of theories with auxiliary fields that can be constructed in a derivative expansion. Up to 4th order in derivatives, the
field equations of this theory read~\cite{Pani:2013qfa}
\begin{align}
\label{eqAuxiliary}
G_{\mu\nu} ={}& 8\pi T_{\mu\nu} - \Lambda g_{\mu\nu}   - 8\pi \beta_1 \Lambda\, g_{\mu\nu}\,T +
16\pi^2\left(1-2\beta_1  \Lambda\right)(\beta_1-\beta_4)\, g_{\mu\nu}\, T^2    \nn \\
&{}+  64\pi^2\left[\beta_4\left(1-2\beta_1\, \Lambda\right) - \beta_1\right] \, T\,
T_{\mu\nu} + 64\pi^2\left[\frac{1}{2}\beta_4 \, g_{\mu\nu}\, T_{\sigma\gamma}\, T^{\sigma\gamma}-
  2\beta_4\, T^\sigma\,_\mu\,T_{\sigma\nu}\right] \nn \\
&{}+ 8\pi\left[\beta_1\, \nabla_\mu\nabla_\nu\,T - \beta_1\, g_{\mu\nu}\, \Box\,T  -  \beta_4
  \, \Box\,T_{\mu\nu} + 2\beta_4\, \nabla^\sigma\nabla_{(\mu}\,T_{\nu)\sigma}\right]+\ldots
\end{align}
Remarkably, to this order the theory contains only two extra coupling constants, $\beta_1$ and $\beta_4$, which
completely parametrize \emph{any} theory belonging to this class. In this parametrization, EiBI gravity and Palatini
$f({\cal R})$ theories belong to ``orthogonal'' classes: the small-coupling limit of EiBI gravity corresponds
to $\beta_1=0$ and $\beta_4\propto\kappa$, whereas generic Palatini $f({\cal R})$ theories correspond to $\beta_4=0$,
with $\Lambda$ and $\beta_1$ depending on the specific $f({\cal R})$ model.

\paragraph{Main Results.}
Most applications of these theories were worked out for Palatini $f({\cal R})$ 
gravity. EiBI gravity 
has been investigated to a lesser extent, despite a recent surge of activity. We list some of the main findings below:
\begin{itemize}
 \item As mentioned, this class of theories is \emph{equivalent} to GR in vacuum. Hence, no BH-based tests can
   distinguish these theories from GR. However, nonperturbative
   effects can replace the singular interior of a \emph{charged} BH by a
   regular wormhole geometry~\cite{Olmo:2013gqa}, similarly to the resolution of the point-charge singularity in
   Born-Infeld electrodynamics.
 \item When applied to early cosmology, the Big Bang singularity that appears generically in GR cosmological models is
   replaced by a regular behavior~\cite{Barragan:2009sq,Banados:2010ix}.
 \item The critical mass of a NS can be much larger in these theories than in GR~\cite{Pani:2011mg} and
   gravitational collapse is suppressed~\cite{Pani:2012qb}, as discussed in Section~\ref{sec:NS_auxiliary}.
 \item When matter is described by a perfect fluid with a barotropic equation of state $P=P(\rho)$, the modified field
   equations are equivalent to GR sourced by an effective perfect fluid with a different equation of
   state, where the dependence $P(\rho)$ is highly nonlinear~\cite{Delsate:2012ky}. This allows for interesting
   configurations: for example a fluid may satisfy all energy conditions in flat spacetime but, when coupled to
   gravity, the effective stress-energy tensor [the right-hand side of Eq.~\eqref{eqAuxiliary}] can violate some energy
   conditions. Furthermore the degeneracy between different EOSs and beyond-GR corrections is maximal in
   these theories~\cite{Delsate:2012ky}.
 \item As we discuss in more detail in Section~\ref{sec:NSs}, these
   theories lead to curvature singularities when there are
   discontinuities in energy density, e.g.~at the interface between a
   solid body and vacuum~\cite{Barausse:2007pn,Barausse:2007ys,
     Barausse:2008nm,Sotiriou:2008dh,Pani:2012qd,Pani:2013qfa}. Whether
   or not this is a consequence of the often employed polytropic
   approximation, or whether such singularities can be avoided in other
   ways, is currently unclear~\cite{Kim:2013nna}.
 \item An analysis of the Newtonian limit of this class of theories
   was performed in~\cite{Pani:2013qfa}, with the result that the
   lowest-order PN solution does not fit into the standard PPN
   framework~\cite{Will:2014xja}. The PPN framework should therefore be
   extended to accommodate these theories.
 \item Flanagan~\cite{Flanagan:2003rb} pointed out that Palatini
   $f({\cal R})$ gravity can produce unacceptable deviations in the
   matter sector, and therefore it would be in conflict with the
   Standard Model of particle physics. This result is debated
   (cf.~\cite{Vollick:2003ic,Flanagan:2004bz} and~\cite{Olmo:2011uz}
   for a review) but, if correct, it should also apply to EiBI gravity
   and other theories belonging to this
   category~\cite{Pani:2012qd,Pani:2013qfa}. Another potential
   shortcoming of these theories is related to the averaging problem
   in cosmology~\cite{Flanagan:2003rb,Li:2008fa}. More detailed
   studies in these directions are necessary to assess the viability
   of theories with auxiliary fields.
\end{itemize}

\subsection{General relativity and quantum mechanics: an effective field theory approach}
\label{sec:EFT}
Since gravity is nonrenormalizable, a useful alternative point of view
on modifications of GR is provided by effective field theory (EFT),
since this is the framework widely used for understanding
nonrenormalizable theories elsewhere in physics. This approach
recognizes that such theories arise when one focuses on observables
involving only the lower of two well-separated scales, and so make
sense (even at the quantum level) only within the context of a
low-energy approximation. Indeed this is arguably the only way known
yet to make sense of such theories at the quantum level, and this
provides their main motivation. Typically quantum effects are
suppressed by the small ratio of scales, and so the classical
approximation itself fails if applied at too high energies. For
gravity the higher scale might be the mass of some hitherto
undiscovered particle, but -- for the reasons given below -- cannot be
higher than the Planck scale: energies comparable to
$M_{\rm Pl} \sim 10^{19}$~GeV. However, for some theories of gravity
(e.g. higher-dimensional theories) this scale can be much smaller.
Consequently quantum gravity corrections may be important at lower
energies, and so be accessible to astrophysical observations. In most
of the strong-field modifications of GR discussed in this review one
implicitly assumes the existence of a new fundamental scale, smaller
than $M_{\rm Pl}$, at which the modifications set in. As we now
discuss, EFT is extremely powerful in this context: simply by assuming
the existence of two different scales, the EFT framework provides a
prescription to obtain all viable corrections to the classical action,
even when the full quantum theory is unknown.

If a system involves two very different energy (or mass) scales, $M_1
\ll M_2$, a drastic simplification occurs when observables are
expanded in powers\footnote{It can sometimes happen that a Taylor
  expansion is inadequate (e.g.~when infrared divergences occur), and
  then more singular functional forms -- such as a logarithmic
  dependence $\sim \log(M_1/M_2)$ -- can also arise.} of the small
ratio $M_1/M_2$. The EFT formulation is designed to exploit this
simplification as early in a calculation as possible, focusing on
observables that directly involve energies only at the lower of the
two scales, $E \lesssim M_1$. In this case, because no ``heavy''
states at scale $M_2$ appear directly in the observables, they can
only influence the result as virtual states. As a result their effects
on longer wavelengths can always be incorporated as corrections to the
``effective'' Lagrangian (or Hamiltonian) used to describe the
evolution of the low-energy states. In particular, the same corrected
Lagrangian can be used to compute {\em all} low-energy observables, so
it is much more efficient to first compute the effective Lagrangian
once and for all, and later use this to compute implications for any
observables of interest.

In formulae, suppose a system is described by a theory having both
``heavy'' and ``light'' fields, $h$ and $l$, described by a classical
action $S(h,l)$. Suppose also that measurements are performed at low
energies, using quantities ${\cal O}_k(l)$ involving only the lighter
fields $l$. Observables can then be expressed in terms of functional
integrals of the form
\begin{align}
\left \langle {\cal O}_1 \cdots {\cal O}_n \right \rangle &=
\int {\cal D} h \, {\cal D} l \; \left[ {\cal O}_1(l) \cdots {\cal O}_n(l) \right] \; \exp
\Bigl[ i S(h, l) \Bigr] \nonumber\\
&= \int {\cal D} l \; \left[ {\cal O}_1(l)
\cdots {\cal O}_n(l) \right] \; \exp \Bigl[ i S_{\rm eff}(l) \Bigr] \,,
\end{align}
where the ${\cal O}_k$'s independence of $h$ allows the ${\cal D}h$
integration to be performed once and for all, ensuring all its effects
arise through the combination
\begin{equation}
\label{integratingout}
\exp \Bigl[ i S_{\rm eff}(l)
\Bigr] \equiv \int {\cal D} h \; \exp \Bigl[ i S(h, l) \Bigr] \,.
\end{equation}
Expanding in powers of $1/M_2$ in $S_{\rm eff}(l)$ then amounts to
writing it as a local expansion in derivatives of $l$, with more
complicated interactions being suppressed by higher powers of $1/M_2$.

For instance, if the light field is the metric, this leads to an
expansion of the form
\begin{align}
\label{gravaction}
S_{\rm eff} =& {} - \int {\rm d}^4 x
\; \sqrt{-g} \; \Bigl[ c_4 + c_{(2,1)} \, R + c_{(0,1)} R^2 +
c_{(0,2)} R_{\mu\nu} R^{\mu\nu} + c_{(0,3)} R_{\mu\nu\lambda\rho}
R^{\mu\nu\lambda\rho} \nonumber\\
&\qquad {}+ c_{(-2,1)} \, R^3 +
c_{(-2,2)} \, R R_{\mu\nu} R^{\mu\nu} + \cdots \Bigr] \,,
\end{align}
where all possible terms consistent with symmetries (such as general
covariance) are included.\footnote{Not all such terms need be
  independent of one another, making it useful in practice to identify
  a minimal basis of interactions of each dimension. For some reviews
  of gravity formulated as an EFT,
  see~\cite{Burgess:2003jk,Goldberger:2007hy,Donoghue:2012zc}.}  The
subscript $d$ on the constants $c_{(d,k)}$ means that they have
dimension (mass)${}^d$ in ``fundamental'' units (for which $\hbar = c
= 1$); $k$ simply labels the possible terms for each value of $d$.

By integrating out a heavy field with mass $M_2$, as in
Eq.~(\ref{integratingout}), one typically gets a contribution to
$c_{(d,k)}$ of order $c_{(d,k)} = \tilde c_{(d,k)} M^d_2$ (where the
dimensionless coefficients, $\tilde c_{(d,k)}$, might depend
logarithmically on $M_2$). It is because $M_2$ is assumed large and we
are interested in expanding in powers of $1/M_2$ that only positive
powers of curvature appear in this expression. If more than one such
fields are integrated out one might find a sum of contributions of
this form,
\begin{equation}
c_{(d,k)} \sim \sum_f \tilde c^{\,f}_{(d,k)} \, M_f^d \,,
\end{equation}
where $M_f$ is the mass of the corresponding particle. Clearly the
contribution coming from the field with the largest mass dominates in
$c_4$ and $c_{(2,1)}$, while the {\em smallest} mass wins in
$c_{(d,k)}$ for any $d < 0$. We therefore expect $c_{(d,k)}$ to be of
order $m^{-d}$, where $m$ is the mass of the lightest particle that
has been integrated out, perhaps the electron in applications to the
Solar System.\footnote{Notice that these arguments indicate that for
  practical applications $c_{(d,k)}$ is almost certainly {\em not} of
  order $\mpl^d$ when $d<0$, unlike the choice often made.} By
contrast, $c_4$ and $c_{(2,k)}$ should be potentially enormous, since
they are most sensitive to the most massive particles that are
present. This expectation is borne out for $c_{(2,1)}$, which can be
identified with $\mpl^2/2$, since $\mpl \approx 10^{19}$ GeV is the
highest fundamental energy scale we know in Nature. This argument
seems to fail for $c_4$, which cosmological observations indicate
cannot be larger than on the order of $(10^{-3} \; \hbox{eV})^4$. Why
$c_4$ should be so small is a long-standing unsolved problem: the
cosmological constant problem.\footnote{A classic review of the
  cosmological constant problem is given in~\cite{Weinberg:1988cp}. A
  review with a sturdy defense of anthropic approaches and issues
  raised by the ``landscape'' of solutions to quantum gravity theories
  is in~\cite{Polchinski:2006gy}; see also~\cite{Burgess:2013ara}.  }

\subsubsection{Power-counting and the semiclassical approximation.}

What does any of this have to do with the classical
approximation? The connection to EFTs arises for two reasons. First,
$S_{\rm eff}(l)$ enters into expressions in precisely the same way as
would a classical action; the influence of the heavy fields makes the
system behave at low energies {\em as if} its classical action were
$S_{\rm eff}(l)$. Second, much of the nitty gritty of EFT techniques
aims to identify how successive heavy-field corrections to $S_{\rm
eff}$ propagate through to contribute to observables, to make their
calculation as efficient as possible. The point is that these same
techniques can be used to track which combinations of parameters arise
order-by-order in the loop expansion, and so whose small size
ultimately justifies this expansion. Since classical physics is just
the leading (nonloop) contribution, such arguments also justify when
it suffices to stop with a classical result.

It is worth illustrating this with a specific example. A particularly
simple class of observables for a low-energy gravity theory consist of
the scattering of weakly coupled gravitons moving through a weakly
curved classical geometry. If $c_4 = 0$ we can take the background
geometry to be flat space,\footnote{The assumption of flatness here is
  purely for convenience, and the conclusions below apply equally well
  to curved spaces, since they rely essentially on dimensional
  arguments.} allowing us to expand the metric around the Minkowski
background: $g_{\mu\nu} = \eta_{\mu\nu} + h_{\mu\nu}$. We then ask how
each of the terms in the gravitational action, Eq.~\eqref{gravaction},
contribute. In particular, with a view to asking how large quantum
corrections can be, we can ask about the relative size of different
contributions to the amplitude, ${\cal A}(E)$, for 2 gravitons to
scatter into another 2 with energy $E$.

As shown in~\cite{Burgess:2003jk,Goldberger:2007hy,Donoghue:2012zc} in
some detail, the contribution to this amplitude of an $L$-loop Feynman
graph built using $V_{i,r}$ vertices built from a term in $S_{\rm eff}$
involving $r$ powers of the curvature tensor, involving the emission
or absorption of $i$ gravitons, is of order
\begin{equation}
\label{GRcount}
{\cal A}(E) \sim \left( {E \over \mpl} \right)^2 \left(
{E \over 4 \pi \mpl} \right)^{2 L} {\prod_{i} \prod_{r\ge 2}}
\left[{E^2 \over \mpl^2} \left( {E \over M} \right)^{(2r-4)}
\right]^{V_{i,r}} \,.
\end{equation}
Here $M$ is the mass that sets the dimensions of the coefficients
$c_{(d,k)} \propto M^{d}$ for $d < 0$, which is assumed for simplicity
to be the same order of magnitude for all negative
$d$. Eq.~(\ref{GRcount}) has several useful consequences.

First, because $r \ge 1$ for all terms in Eq.~(\ref{gravaction}) the
contribution to ${\cal A}$ contains no negative powers of $E$. This
illustrates how $S_{\rm eff}$ encapsulates how observables
simplify in the hierarchical low-energy limit, where $E \ll M$,
$E\ll \mpl$. In particular, this expression quantifies why the weakness of the
graviton's coupling follows purely from the low-energy approximation,
$E \ll \mpl$ and $E \ll M$.

Second, these expressions identify precisely which kinds of
interactions dominate scattering amplitudes at low energies. The
minimum suppression by powers of $E$ comes when $L = 0$ and we choose
$V_{i,r} = 0$ unless $r = 1$, and so is given by arbitrary tree graphs
constructed purely from the Einstein-Hilbert action. This tells us
what we would be inclined to believe in any case: it is $L = 0$
(no-loop) graphs built only from the Einstein-Hilbert action --
i.e.~classical GR -- which govern the low-energy dynamics of
GWs, giving a result of order $(E/\mpl)^2$.

But we may also identify the next-to-leading contributions. These
are proportional to $(E/\mpl)^4$ and can appear in one of two ways:
\begin{enumerate}
\item either: $L = 1$ and $V_{i,r} = 0$ for any $r \ne 1$,
\item or: $L = 0$ with $\sum_i V_{i,2} = 1$ and $V_{i,1}$ arbitrary, and
$V_{i,r} = 0$ for all $r \ge 3$.
\end{enumerate}
That is, the next-to-leading contribution is obtained by computing the
one-loop corrections using only Einstein gravity, or by working to
tree level and including precisely one curvature-squared interaction
in addition to any number of interactions from the Einstein-Hilbert
term. Both are suppressed compared to the leading term by a factor of
$(E/\mpl)^2$. At this order the ultraviolet divergences that famously
plague gravitational loops in option (i) above are absorbed into
renormalizations of the coefficients of the curvature-squared
contributions that appear in option (ii), and so on further down the
$E/\mpl$ expansion.

These conclusions are borne out by explicit calculations. At tree
level the only nonzero amplitudes are related by crossing symmetry to
the amplitude for which all graviton helicities have the same sign,
and this is given by~\cite{DeWitt:1967yk,DeWitt:1967ub,DeWitt:1967uc}:
\begin{equation}
-i {\cal A}_{(++,++)}^{\rm tree} = 8 \pi G \,\left(\frac{s^3}{tu} \right)\,,
\end{equation}
where $s$, $t$ and $u$ are the usual Mandelstam invariants built from
inner products of the graviton four-momenta, all of which are
proportional to the square of the center-of-mass energy, $E_{\rm
cm}$. This shows that it is the frame-independent center-of-mass
energy that appears in the $E/\mpl$ expansion of ${\cal A}$. The
one-loop corrections are also computed~\cite{Dunbar:1994bn}, and are
infrared divergent. These infrared divergences cancel in the usual way
with tree-level Bremsstrahlung diagrams~\cite{Weinberg:1965nx}, leading to a
finite result~\cite{Donoghue:1999qh}, which is suppressed as expected
relative to the tree contribution by terms of order $(E/\mpl)^2$, up to
logarithmic corrections.

It is expressions like the amplitude scaling~\eqref{GRcount} that make
the explicit connection between EFTs and the domain of validity of the
semi-classical (or loop) approximation. This expression reveals that
the loop expansion for gravity is secretly a low-energy
approximation. This turns out to be generic for any nonrenormalizable
field theory~\cite{Weinberg:1978kz}. For such theories the only known
way to extract sensible quantum corrections is within a low-energy
approximation, for which the classical action should be regarded as a
general derivative expansion along the lines of
Eq.~\eqref{gravaction}. All terms in this action consistent with
symmetries and field content are compulsory, since their presence is
required to renormalize the ultraviolet divergences that are generated
by loops involving terms arising at lower orders in the derivative
expansion.

\subsubsection{Modified gravity seen through EFT glasses.}

We can now return to our road map of modified gravity theories to see
what it leads us to expect. Following~\cite{Burgess:2009ri}, we argue
that EFT can provide useful guidelines.

\paragraph{New particles and/or dimensions.}
The most conservative modifications simply involve
the addition of new light particles or the addition of more dimensions
(or both), with the new additions resembling those about which we
already know. It is certainly true that such modifications can be
sensible in principle, and explicit examples exist (such as string
theory) for higher-energy physics that can produce such
modifications. It makes sense to constrain such possibilities
observationally.

There are also issues that can be expected to arise in such theories
if the new particles are light enough to be relevant over
astrophysical or cosmological distances. This is true in particular
for proposals meant to describe present-epoch dark
energy~\cite{Weinberg:1988cp,Polchinski:2006gy,Burgess:2013ara}. Such
particles are so close to massless that many of the constraints on
massless particles in practice are likely to apply. In particular, it
can be expected that in the Lorentz-invariant framework of Special
Relativity the new particles must be spin zero, half or must be gauge
particles with spin one or $\frac32$ or smaller.

Another problem potentially can also arise, associated with the size
of quantum corrections to the mass, particularly for spinless
particles represented by a scalar field, $\phi$. Then the low-energy
EFT contains a mass term of the form ${\cal L}_{\rm eff} = -
c_{(2,2)} \, \sqrt{-g} \; \phi^2$ whose coefficient $c_{(2,2)}
\propto M^2$ should be large, for the same reasons (given above) that
lead one to expect $c_{(2,1)}$ and $c_4$ are large. This is a
``hierarchy'' problem, similar to the cosmological constant problem;
very light spinless particles very rarely arise as the low-energy
limit of something more fundamental because their masses are sensitive
to quantum contributions from every heavy state at higher energies
that is integrated out.

A similar problem does not occur for spin-half particles, because
for these the particle mass can be forbidden by a chiral symmetry,
under which the fermion's left- and right-handed components rotate
differently: $\psi \to i \gamma_5 \psi$. Because of this it can only
receive quantum corrections from particles that also break this
symmetry. Only very few symmetries (supersymmetry and scale
invariance) are known that can forbid a scalar mass, making this
mechanism more challenging to use at low energies for spinless
particles.

\paragraph{Modifications to the Einstein equations.}
Short-distance modifications to the left-hand side
of Einstein's equations are also very plausible, since these can
easily be generated by integrating out various kinds of heavy
fields. In well understood situations these usually lead to modified
actions along the lines of Eq.~(\ref{gravaction}) that are
local polynomials of the metric and its derivatives, and involve all
possible kinds of interactions allowed by the assumed symmetries. In
particular, it should be noted that generic higher-derivative
interactions are allowed, and explicit
calculations~\cite{Burgess:2014lwa} show these need not take the
specific Horndeski
or Lovelock forms that are sometimes advocated as being required to
avoid the presence of ghosts.  What is hard to achieve in this way are
modifications like $f(R)$ gravity where $f(R)$ is an unusual function,
such as $1/R$. Proposals such as this one step away from the
underlying EFT understanding of the validity of the semi-classical
approximation, and so the onus is on proponents to justify the regime
of validity of any classical approximation. This is particularly so in
situations like dark energy proposals, where one of the basic problems
(the cosmological constant problem) cannot be seen until quantum
effects are examined.

\paragraph{Breaking diffeomorphism invariance.}
As described above, it is ultimately the consistency
of Lorentz invariance and quantum mechanics that drives many of the
consistency conditions for massless (and very light) particles,
including the requirements for the gauge invariance of their
interactions. However the constraints are no longer quite as exacting
once the particles are not exactly massless. In this case the general
consistency issues can be expected to persist if the particle is light
enough compared with the higher scales of the theory, $m \ll M$, but
can be evaded if this hierarchy is not too large.

This observation has prompted some to put aside until later
understanding the embedding into higher-energy physics, and instead to
explore the implications of relaxing the assumptions of gauge
invariance (and so usually also Lorentz invariance) at low
energies. The hope is to find a consistent low-energy effective
description that applies only up to relatively low energy scales, and
hope that once this is done a consistent ultraviolet completion can be
found. In most cases of this type, no candidate ultraviolet extension
is yet known.

Modifications to gravity provide a rich theoretical laboratory as to
how quantum field theories work, that display their tight consistency
issues in new and instructive situations. Sensible modifications --
i.e.~those that can be embedded into well-understood ultraviolet
completions -- are the goal, but are also not that easy to come
by. Together with their success in describing astrophysical and
cosmological observations, theoretical soundness should be regarded as
one part of the evidence to be used when assessing the likelihood of
such theories describing Nature.

\subsection{Open problems}
\label{op:theories}
Here we give a (necessarily biased) list of open problems regarding the modified theories of gravity discussed in this
chapter:
\paragraph{Scalar-tensor gravity and metric $f(R)$ theories.}
\begin{itemize}
 \item Multiscalar-tensor theories with a nontrivial target space
   geometry offer a barely explored, uncharted territory
   (cf.~\cite{Damour:1992we} for pioneering work). The presence of
   different nonperturbative phenomena and richer phenomenology
   awaits exploration.
 
  \item Building $f(R)$ theories which are observationally viable in
    the weak-field limit and differ from GR at cosmological scales (as
    needed to alleviate the difficulties associated with a
    cosmological constant: see Section~\ref{subsec:cosmology}) is a
    challenging
    task~\cite{DeFelice:2010aj,Clifton:2012ry,Clifton:2015ira}. Indeed,
    it seems that in $f(R)$ gravity one cannot produce any drastically
    different behavior at cosmological scales without simultaneously
    compromising the Newtonian limit at small
    scales~\cite{Clifton:2012ry}. If confirmed, this quite general
    result would cast serious doubts on the attractiveness of $f(R)$
    gravity as an alternative to GR.
  
  \item Even cosmologically viable $f(R)$ models seem to be disfavored
    against the $\Lambda$CDM paradigm. On the other hand, the simplest
    inflationary models can be framed as $f(R)$ theories, where the
    inflaton is the extra scalar degree of freedom.
\end{itemize} 

\paragraph{Quadratic gravity.}
\begin{itemize}
\item
  At variance with GR and scalar-tensor
   theories, a well-posed formulation of quadratic gravity is not
   available yet. This is important for the theoretical viability of
   the theory and for numerical simulations, as discussed in the
   remainder of this review.  In fact, while a well-posed formulation
   is expected to exist in the small-coupling limit, it is unclear
   whether these theories are well posed in their exact form
   (see~\cite{Delsate:2014hba} for an analysis of this problem in dCS
   gravity).

 \item
   Because very few studies have
  analyzed quadratic gravity beyond the perturbative regime, it is
  unknown whether such theories predict strong-field effects (akin to
  spontaneous scalarization in scalar-tensor theories) that are not
  captured by a perturbative analysis in the small-coupling
  regime. The results of~\cite{Pani:2011xm} seem to suggest that such
  effects may not occur, at least for isolated stars.

\item
  To the best of our knowledge, the effects of
   a scalar self-potential (and in particular of a mass term) have not been explored yet.

 \item
   Higher-order terms are suppressed by powers of the (dimensionful)
   coupling constants. The latter depend on the system under
   consideration, and if $\alpha\sim L$ (recall that $L$ is the
   typical size of the system) the perturbative analysis would break
   down, and higher-order curvature invariants would become
   increasingly more important. Furthermore, also other combinations
   of higher-order terms can give rise to second-order field
   equations, similarly to the Gauss-Bonnet term in quadratic gravity.

 \item
   Quadratic curvature terms might also be coupled to an extra
   fundamental vector (or higher-spin) field, which would allow for
   new scalar quantities in the action and presumably for a completely
   different phenomenology.

 \item A theory that received some interest is conformal
   gravity~\cite{Mannheim:1988dj}, where the Lagrangian ${\cal
     L}=R_{\mu\nu}R^{\mu\nu}-(1/3) R^2$ is constructed out of
   contractions of the Weyl tensor. This theory admits all
   \emph{vacuum} solutions of Einstein gravity with a cosmological
   costant (see e.g.~\cite{Varieschi:2014ata} and references therein
   for a study of the geodesic motion around spinning BHs in this
   theory). Having higher-order derivatives, the theory may be plagued
   by ghosts, but this issue is still debated. Another problem is that
   the field equations of conformal gravity imply $T=0$, so that only
   conformal matter can be consistently coupled to gravity. It is
   presently unclear whether realistic models of stars -- which are
   characterized by $T\neq0$ -- can be constructed in this theory.
\end{itemize} 
\paragraph{Lorentz-violating theories.}
\begin{itemize}   
 \item 
There are three main open issues in Lorentz-violating gravity. The
first is the relation between violations of Lorentz symmetry and
ultraviolet renormalizability. This relation has been shown only at
the power-counting level and for scalar-field toy
models~\cite{Horava:2009uw,Visser:2009fg} (and not yet for spin-2
gravitons). The second open issue has to do with the percolation of
Lorentz violations from the gravity sector into the matter sector,
where Lorentz symmetry has been tested to high
accuracy~\cite{Kostelecky:2003fs,Kostelecky:2008ts}. Observations of
the synchrotron radiation from the Crab nebula show that some
mechanism is needed to suppress this
percolation~\cite{Liberati:2012jf}. While several mechanisms have been
proposed (see e.g.~\cite{Liberati:2013xla} for a review), detailed
studies are necessary to assess their viability. The third open
problem has to do with the causal structure of BHs, and namely with
the existence of a universal horizon. Universal horizons have only
been found in spherical and slowly rotating BH
solutions~\cite{Barausse:2011pu,Blas:2011ni,Barausse:2013nwa}, but it
is unclear whether they exist in generic situations and whether they
are stable at the nonlinear level~\cite{Blas:2011ni}.

\item At the moment, the most important challenge faced by $n$-DBI
  gravity, from a theoretical point of view, is the prediction of a
  clear and distinct observable signature that sets it apart from GR
  and might enable constraints to be imposed on its validity. How
  generic is the property that GR solutions are also solutions of
  $n$-DBI gravity (with the same matter content)? This question can be
  recast very objectively as the existence of a foliation with a
  specific property. How generically can such a foliation be found?
  This subclass of solutions has a self-contained perturbation theory
  that can be made to coincide with that of GR in the PN regime. The
  analysis performed in~\cite{Coelho:2013dya}, however, was not
  exhaustive, in the sense that the solution provided, which matches that of GR, 
was not shown to be unique. This is an important
  open question.

\end{itemize} 
\paragraph{Massive gravity and Galileon theories.}
In order to address tests of gravity beyond the decoupling limit, one
needs exact or numerical solutions of the full nonlinear theory. Here
we will briefly discuss some of the progress searching for BH and
cosmological solutions in massive gravity.
\begin{itemize}   
 \item Any viable theory of modified gravity should have BH
   solutions. If we restrict ourselves to a flat reference metric and
   a dynamical metric $g_{\mu\nu}$ that is static and spherically
   symmetric, the most general solution is given by
   Schwarzschild-(anti-)de Sitter (see~\cite{Volkov:2014ooa} for a
   review of BH solutions in massive gravity and in bigravity). The
   cosmological constant of the asymptotic solutions is set by the
   graviton mass.  It is not possible to find asymptotically flat
   solutions.  Additionally, as discussed in~\cite{deRham:2014zqa},
   there may be physically interesting BH solutions that do not have
   exact spherical symmetry.

\item dRGT with a Minkowski reference metric has no nontrivial spatially flat FRW solutions~\cite{D'Amico:2011jj}. There
are several ways to address this problem. One approach is to look for backgrounds with spatial curvature, or where the
reference metric is FRW. It has turned out that these backgrounds exhibit instabilities or are infinitely strongly
coupled and cannot be trusted~\cite{deRham:2014zqa}.

\item Alternatively, one may accept that homogenous and isotropic
  solutions do not exist, and try to find inhomogeneous solutions. An
  exact solution of this kind is
  known~\cite{D'Amico:2011jj,Gratia:2012wt}. While the solution is
  infinitely strongly coupled, this solution is a proof of principle
  that there may be viable inhomogeneous cosmological solutions in
  massive gravity. As with BH solutions, the full space of
  possibilities is still being actively explored.

\item Another approach is to look for FRW solutions in theories that
  add new degrees of freedom to massive gravity. For example, in
  bigravity (where the reference metric is a dynamical field) one can
  find stable FRW solutions, as shown in~\cite{Fasiello:2013woa,DeFelice:2014nja}. Another idea along these
  lines is to consider adding a new scalar degree of freedom to
  massive gravity. Common examples are to allow the mass to be a
  dynamical scalar field~\cite{D'Amico:2011jj}, or to introduce a new
  scalar mode called the quasidilaton through the reference metric:
  see e.g.~\cite{D'Amico:2012zv,DeFelice:2013tsa}.
\end{itemize} 

\paragraph{Gravity with auxiliary fields.}
There are various open questions related to modified theories of
gravity with auxiliary fields, since this is a relatively new research
field.
\begin{itemize}
 \item The issue of curvature singularities appearing at the surface
   of compact stars~\cite{Barausse:2007pn,Pani:2012qd,Pani:2013qfa} is
   crucial to assess the theoretical viability of these
   theories. These singularities can be alleviated in some
   situations~\cite{Kim:2013nna}, but they seem to be
   ubiquitous. Solving this issue requires also an understanding of
   the ``average problem''~\cite{Li:2008fa}: given an ensemble of
   fundamental particles, is the standard stress-energy tensor for a
   perfect fluid a valid approximation in these theories?

\item Similar comments apply to the Cauchy problem in these theories, whose well-posedness is still under
 scrutiny~\cite{LanahanTremblay:2007sg,Sotiriou:2008rp,Olmo:2011uz}.

\item The fact that EiBI corrections due to a barotropic perfect fluid are completely degenerate with the EOS~\cite{Delsate:2012ky} makes it very difficult to test these theories or to rule them out with observations. To date it is unknown whether such degeneracy extends to other forms of matter.

\item As discussed, Palatini $f({\cal R})$ gravity and EiBI gravity are usually investigated under various assumptions,
 e.g.~assuming a symmetric Ricci tensor and a symmetric connection. A metric-affine~\cite{Sotiriou:2006qn} version of
 EiBI gravity is still lacking.
\end{itemize}

\clearpage
\section{Black holes}
\label{sec:BHs}

In this chapter we review BH solutions in the modified theories of
gravity described earlier. The next chapter is devoted to a similar
review of compact star solutions. Table~\ref{tab:BHsummary}
is meant to provide a practical guide to the literature on BH
solutions and their stability in various theories at the time of
writing.

\subsection{Black holes in general relativity}

One of the most remarkable predictions of GR is that regular,
stationary BHs in Einstein-Maxwell theory are extremely simple
objects, being defined by at most three parameters: mass, angular
momentum and electric charge. This was established by a series of
uniqueness theorems due to Hawking, Carter and Robinson
(see~\cite{Bekenstein:1996pn,Carter:1997im,Chrusciel:2012jk,Robinson}
for reviews), which imply that all isolated BHs in Einstein-Maxwell
theory are described by the Kerr-Newman family.
Astrophysical BHs are thought to be neutral to a very good
approximation because of quantum discharge
effects~\cite{Gibbons:1975kk}, electron-positron pair
production~\cite{1969ApJ...157..869G,1975ApJ...196...51R,Blandford:1977ds}
and charge neutralization by astrophysical plasmas.
Therefore the geometry of astrophysical BHs in GR is simply described
by the two-parameter Kerr metric~\cite{Kerr:1963ud}, which in standard
Boyer-Lindquist coordinates reads
\begin{equation}
\begin{aligned}
ds^2={}&-(1-2Mr/\rho^2)dt^2-4aMr\sin^2\theta/\rho^2 dt d\varphi
+\frac{\rho^2}{\Delta}\,dr^2+\rho^2\,d\theta^2\\
&{}+\left(r^2+a^2+2Ma^2r\sin^2\theta/\rho^2\right )\sin^2\theta
d\varphi^2 \,,
\end{aligned}
\label{metricKerrLambda}
\end{equation}
where
$\Delta\equiv r^2+a^2-2Mr$ and $\rho^2\equiv r^2+a^2 \cos^2\theta$.
This metric describes the gravitational field of a spinning BH of mass
$M$ and angular momentum $J=a M$.
The roots of $\Delta$ correspond to the event horizon
($r_+=M+\sqrt{M^2-a^2}$) and the Cauchy horizon
($r_-=M-\sqrt{M^2-a^2}$). The static surface $g_{tt}=0$ defines the
boundary of the ergosphere: $r_{\rm
  ergo}=M+\sqrt{M^2-a^2\cos^2\theta}$. The ``angular velocity of
the event horizon'' is
\begin{equation}
\Omega_{\rm H}\equiv a/(r_+^2+a^2)\,.
\label{eq:Omega-H}
\end{equation}

Because of the uniqueness theorem, and because NSs in GR cannot be
more massive than $\sim 3M_\odot$~\cite{Rhoades:1974fn}, any
observation of a compact object with mass larger than $\sim 3M_\odot$
with metric different from the Kerr geometry would inevitably signal a
departure from standard physics (either in the gravitational or in the matter sector).
Therefore tests of strong-field gravity targeting
BH systems aim at verifying the ``Kerr hyphothesis'' in various ways.
Teukolsky~\cite{Teukolsky:2014vca} recently compiled an excellent
review on the discovery of the Kerr metric and the impact of this
discovery in astrophysics. We refer interested readers to Teukolsky's
review and standard textbooks~\cite{Shapiro:1983du,MTB,Frolov:1998wf}
for surveys of our current understanding of BHs in GR; here we
summarize some considerations on the stability and no-hair properties
of GR BHs that should be kept in mind when we
discuss BH solutions in modified theories of gravity.

The key theoretical developments after Kerr's discovery were the
derivation of a separable equation -- the ``Teukolsky master
equation'' -- describing perturbations of scalar, neutrino,
electromagnetic and gravitational fields~\cite{Teukolsky:1972my}; the
use of this master equation to assess the mode stability of the
metric~\cite{Teukolsky:1973ha,Press:1973zz,Teukolsky:1974yv}; and
Whiting's work, that conclusively excluded the possibility of
exponentially growing modes~\cite{Whiting:1988vc}. The free
oscillation modes of Kerr BHs under the boundary conditions of ingoing
waves at the horizon and outgoing waves at infinity (whose frequencies
form the so-called ``quasinormal mode'' (QNM) spectrum~\cite{pressringdown})
were investigated extensively by Leaver~\cite{Leaver:1985ax} and
several other authors. The spectrum consists of an infinite discrete
set of complex-frequency modes (hence ``quasinormal''); the nonzero
imaginary part is due to gravitational radiation damping. QNMs
find important applications in various areas of physics, ranging
from quantum gravity to the gauge-gravity duality
(see~\cite{Kokkotas:1999bd,Nollert:1999ji,Berti:2009kk,Konoplya:2011qq}
for reviews). The Teukolsky equation is not self-adjoint, and
QNMs do not form a complete set. The absence of unstable
modes may be a good enough stability proof for a physicist, but not
for a mathematician: mode stability does not imply linear stability.
A rigorous proof of linear stability was carried out by Kay and Wald
for Schwarzschild BHs~\cite{Kay:1987ax}, but the extension of this
analysis to Kerr is still work in
progress~\cite{Dafermos:2008en,Dafermos:2009uq,Dafermos:2010hb,Dafermos:2014cua,Dafermos:2014jwa},
and there is now evidence for instability in extremal Kerr
BHs~\cite{Aretakis:2012ei} (see
also~\cite{Lucietti:2012sf,Yang:2012pj,Yang:2013uba,Cook:2014cta}). For our
purposes, and with the previous caveats, we will consider the absence
of unstable modes as a physically satisfactory stability criterion.

In the rest of this chapter we will first review the properties of BHs
in various extensions of GR, and then turn to a discussion of possible
ways to verify the Kerr hypothesis.

\subsection{Scalar-tensor theories}
\label{sec:ST_BHs}

\subsubsection{Real scalars and no-hair theorems}

Theoretical studies impose remarkable constraints and limitations on
BH solutions in scalar-tensor theories. No-scalar-hair theorems for
the simplest scalar-tensor theories were proved by various authors
Refs.~\cite{Hawking:1972qk,1971ApJ...166L..35T,Chase:1970,Bekenstein:1995un},
and state that stationary BH solutions in Brans-Dicke theory are the
same as those in GR. In other words, the scalar must be trivial and
the geometry must be described by the Kerr metric.

One way to understand this property is to recall our discussion in
Section~\ref{subsec:ST}. By means of field redefinitions, it is always
possible to reformulate the action of a scalar-tensor theory as the
action~\eqref{STactionE} of a minimally coupled sigma-model,
where matter fields have a nontrivial coupling with the scalar
field. However, in vacuum (and in particular in BH spacetimes) the
matter action can be discarded, and Eq.~\eqref{STactionE}
reduces to the Einstein-Hilbert action with a minimally coupled scalar
field.

Extensions of these uniqueness theorems to multiple scalars and to
more generic scalar-tensor theories have been established more
recently~\cite{Heusler:1995qj,Sotiriou:2011dz}. These results assume
the scalar field to be time-independent, a requirement recently shown
to be unnecessary for any scalar-tensor theory with a {\it real}
scalar~\cite{Graham:2014ina}.  The no-scalar-hair theorems have also
been confirmed by numerical studies of gravitational collapse to
nonrotating
BHs~\cite{Scheel:1994yr,Scheel:1994yn,Shibata:1994qd,Harada:1996wt,Novak:1997hw,Kerimo:1998qu}.

In summary, the Kerr family of vacuum BH solutions in GR is also the
most general vacuum solution in a rather general class of
scalar-tensor theories, although some exceptions exist, as we discuss in the next sections.

Furthermore, alternative theories with the same equilibrium solutions as GR have,
in general, different dynamics. The theorems summarized above imply
that Kerr BHs in GR are linearly stable, but they are unstable because
of superradiance in massive scalar-tensor theories (including
minimally coupled massive
scalars)~\cite{Damour:1976kh,Detweiler:1980uk,Zouros:1979iw,Cardoso:2004nk,Shlapentokh-Rothman:2013ysa,Cardoso:2013krh}.
Superradiance extracts energy from rotating BHs, and transfers this
energy to the perturbing field. For a monochromatic wave of frequency
$\omega$, the condition for superradiance
is~\cite{zeldovich1,zeldovich2,Cardoso:2013krh}
\be
0<\omega<m\Omega_H\,,\label{super_cond}
\ee
where $m>0$ is the azimuthal harmonic index and the angular velocity
of the BH horizon $\Omega_H$ was defined in Eq.~\eqref{eq:Omega-H}. If
the scalar is massive, superradiance triggers an
instability~\cite{Press:1972zz,Damour:1976kh,Detweiler:1980uk,Zouros:1979iw,Cardoso:2004nk,Shlapentokh-Rothman:2013ysa}:
the ergoregion amplifies the field, and the mass term ``traps it.''
The linear stages of the instability lead to the growth of a
non-spherically symmetric scalar ``cloud'' outside the horizon
[because the mechanism requires nontrivial azimuthal dependence, as
  seen from (\ref{super_cond})]. For a single {\it real} scalar field,
this leads to a nonzero quadrupole moment of the cloud resulting in
periodic GW emission. Thus, the end-state is thought to be a Kerr BH
with lower
spin~\cite{Witek:2012tr,Okawa:2014nda,Cardoso:2013krh,Brito:2014wla}.
Note, however, that the instability time scale depends on the scalar
field's mass, and may be of the order of the Hubble time, leading to
what in practice amounts to hairy BH configurations.

\subsubsection{Complex scalars: new hairy rotating black holes}
\label{sec:complscal}
When the scalar is time-dependent, the assumptions behind the no-hair
theorems do not apply. Of course, the backreaction of a generic time-dependent 
scalar field will lead to a time-dependent geometry and not
an equilibrium BH state. But for a \textit{complex scalar field}
(which is equivalent to two scalar fields) there is a special type of
time dependence that yields a time-independent stress-energy tensor
and hence is compatible with a stationary metric. This time dependence
is simply a phase evolution, analogous to that of stationary states in
quantum mechanics: $\Psi(t,{\bf x})=e^{-i \omega t}\phi({\bf x})$. As
shown in~\cite{Pena:1997cy}, however, no spherically symmetric BHs
exist even with this time dependence. With the wisdom of hindsight
this is easy to understand. The null generator of the horizon
$\chi=\partial_t$ does not preserve $\Psi$. As such there is scalar
flux through the horizon and hence there can be no static geometry.

The latter argument can be circumvented by introducing
rotation for the BH spacetime and thus making the geometry
axisymmetric. Then, the null generator of the horizon gains an additional term:
$\chi=\partial_t+\Omega_H\partial_\varphi$, where $\Omega_H$ is the
angular velocity of the horizon (given in Eq.~\eqref{eq:Omega-H}), and $\partial_\varphi$ the Killing
vector field which generates the axial symmetry.
We must also introduce an azimuthally dependent phase for
the scalar field, $\Psi(t,\varphi,{\bf x})=e^{-i \omega t}e^{i m
  \varphi} \phi({\bf x})$, where $m\in \mathbb{Z}^{\pm}$ since
$\varphi$ is periodic with $\varphi\sim \varphi +2\pi$. Observe that,
again, the azimuthal dependence vanishes in the stress-energy
tensor. Then, $\mathcal{L}_\chi \Psi=0$, as long as
\begin{equation}
\omega=m\Omega_H \ . 
\label{cloudcond}
\end{equation}
Thus there is no scalar field flux through the horizon as long
as~\eqref{cloudcond} is obeyed, regardless of the value of $\phi({\bf
  x})$ on the horizon. This argument suggests the existence of
asymptotically flat rotating BHs with complex scalar hair. Such
solutions were indeed found in~\cite{Herdeiro:2014goa}.
The ultimate physical reason for the existence of these equilibrium
states -- in the sense that the geometry has an asymptotically
timelike Killing vector field, just like Kerr -- is that GW emission
is halted due to cancellations in the stress-energy tensor, which
becomes independent of the time and azimuthal variables, thus avoiding
GW emission and consequent angular momentum losses.

The condition~\eqref{cloudcond} for the existence of hairy BHs lies
precisely at the threshold of the superradiant
condition~\eqref{super_cond}. This is no accident. A test-field
analysis of the type that leads to condition~\eqref{super_cond}, for a
complex scalar field on the Kerr background, reveals that \textit{real
  frequency} bound states are possible precisely in between amplified
modes, which obey the superradiant condition~\eqref{super_cond}, and
decaying modes, which obey $\omega>m\Omega_H$. These are stationary
scalar
clouds~\cite{Hod:2012px,Hod:2013zza,Herdeiro:2014goa,Benone:2014ssa,Herdeiro:2014pka}. The
hairy BHs found in~\cite{Herdeiro:2014goa} can be thought of as
nonlinear realizations of these clouds, when the scalar field becomes
``heavy'' and backreacts (see also~\cite{Herdeiro:2014ima}).

The solutions found in~\cite{Herdeiro:2014goa} correspond to a
five-parameter family of the Einstein-(massive)-Klein-Gordon
theory. Three of the parameters are continuous: the ADM mass $M$, the
ADM angular momentum $J$, and a Noether charge $Q$. The latter is
obtained by integrating the time component of the scalar field
4-current on a spacelike slice and may be regarded as measuring the
amount of scalar hair outside the horizon. In fact, it proves
convenient to introduce a normalized Noether charge $q\equiv
Q/2J$. Then, $q$ is a compact parameter in the full space of
solutions: $0\le q\le 1$. The value $q=0$ corresponds to Kerr BHs,
showing that these hairy BHs are continuously connected to the
standard Kerr family. This is why the solutions
in~\cite{Herdeiro:2014goa} were dubbed ``Kerr BHs with scalar hair.''
The value $q=1$ corresponds to asymptotically flat, rotating boson
stars~\cite{Yoshida:1997qf,Kleihaus:2005me}. These are (horizonless)
gravitating solitons, which are kept in equilibrium by a balance
between the scalar field self-gravity and its wave-like dispersive
nature. Rotating boson stars sustained by a complex, massive field
have $Q=2J$, which justifies the normalization taken.  The two
remaining parameters of the solutions found in~\cite{Herdeiro:2014goa}
are discrete: the aforementioned azimuthal harmonic index $m\in
\mathbb{Z}^\pm$ and the node number $n\in \mathbb{N}_0$. The latter
counts the number of zeros of the scalar field radial profile. One may
regard $n=0$ as the fundamental configuration and $n\ge 1$ as excited
states.

The line element and scalar distribution describing Kerr BHs with scalar hair reads:
\begin{align}
\label{ansatz_hairy}
ds^2&=e^{2F_1}\left(\frac{dR^2}{N }+R^2 d\theta^2\right)+e^{2F_2}R^2 \sin^2\theta (d\varphi-W dt)^2-e^{2F_0} N dt^2, 
~~~~N\equiv 1-\frac{R_H}{R}, \nonumber \\
\Psi&=\phi(r,\theta)e^{i(m\varphi-\omega t)}.
\end{align} 
In this ansatz there are five functions of $(R,\theta)$:
$F_0,F_1,F_2,N,\phi$. To obtain them one numerically solves five
nonlinear, coupled PDEs, with appropriate boundary conditions that
ensure both asymptotic flatness and regularity at the horizon. The
latter requirement actually implies condition~\eqref{cloudcond}. The
parameter $R_H$ is the location of the event horizon in this
coordinate system. We remark that these are \textit{not}
Boyer-Lindquist coordinates in the Kerr limit. In order to write Kerr
in the form~\eqref{ansatz_hairy} one must change the radial coordinate
$r$ in~\eqref{metricKerrLambda} by the transformation
$R=r-{a^2}/{r_H}$, where $r_H=M+\sqrt{M^2-a^2}$ is the event horizon
location in Boyer-Lindquist coordinates.

\begin{figure}[h!]
\centering
\includegraphics[height=2.48in]{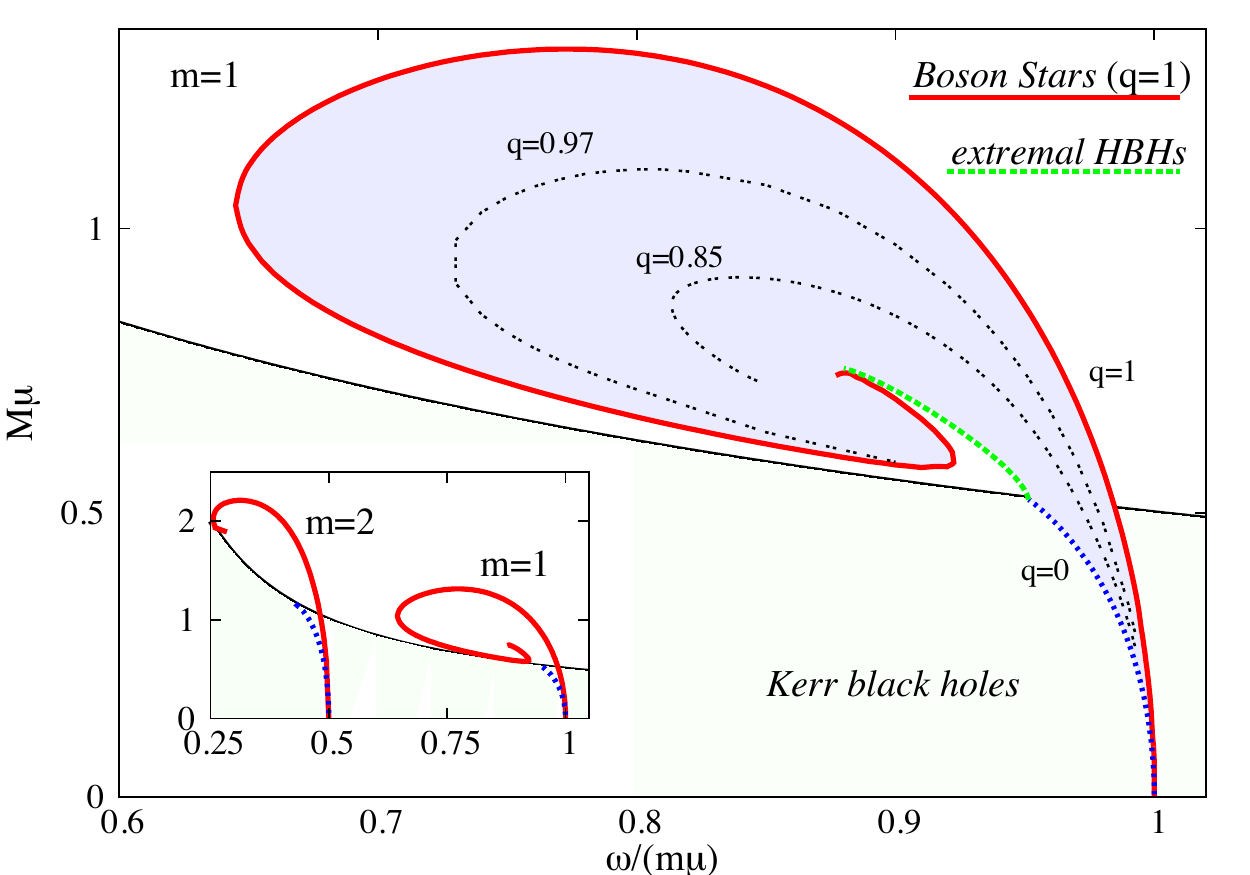}\\
\caption{%
  Domain of existence of hairy BHs for $n=0$,
  $m=1$ in $M$-$\omega$ space (shaded blue region). The black solid
  curve corresponds to extremal Kerr BHs, which obey
  $M={1}/(2\Omega_H)$; Kerr BHs exist below it. For $q=0$, the domain
  of existence connects to Kerr solutions (dotted blue line). For
  $q=1$, $R_H$ vanishes and hairy BHs reduce to boson stars (red solid
  line). The final line that delimits the domain of existence of the
  hairy BHs (dashed green line) corresponds to extremal BHs, i.e.~with
  zero temperature. (Inset) Boson star curves for $m=1,2$. The units
  in the axes are normalized to the scalar field mass $\mu$. [Adapted
  from~\cite{Herdeiro:2014goa}.]}
\label{hairy-pm}
\end{figure}

The parameter and phase space for the solutions with $n=0$, $m=1$ were
discussed in detail in~\cite{Herdeiro:2014goa} and are summarized in
Figure~\ref{hairy-pm}. There is a region of overlap of hairy and Kerr
BHs with the same $(M,\,J)$. In that sense there is nonuniqueness. The
degeneracy seems to be raised, however, by the introduction of $q$: no
two solutions were found with the same $(M,\,J,\,q)$. In the region of
nonuniqueness, hairy BHs have larger entropy than the corresponding
Kerr BHs. As such the former cannot decay into the latter
adiabatically. Also, hairy BHs can violate the Kerr bound: $J\le
M^2$. This violation is not surprising since it is known to occur for
rotating boson stars~\cite{Ryan:1996nk}, and hairy BHs are
continuously connected to boson stars. It is indeed a generic feature
that hairy BHs are more star-like, i.e.~less tightly constrained in
their physical properties than Kerr BHs. This observation also has
implications for possible astrophysical phenomenology of hairy BHs, an
aspect of special relevance for this review. It was observed
in~\cite{Herdeiro:2014goa} that both the quadrupole moment and the
angular frequency at the ISCO can differ significantly for hairy BHs,
as compared to the standard Kerr values. Finally, hairy BHs have a
richer structure of ergo-regions than Kerr, with the occurence of
\textit{ergo-Saturns}, besides ergo-spheres, in a region of parameter
space~\cite{Herdeiro:2014jaa}.

\begin{figure}[h!]
\centering
\includegraphics[height=1.78in]{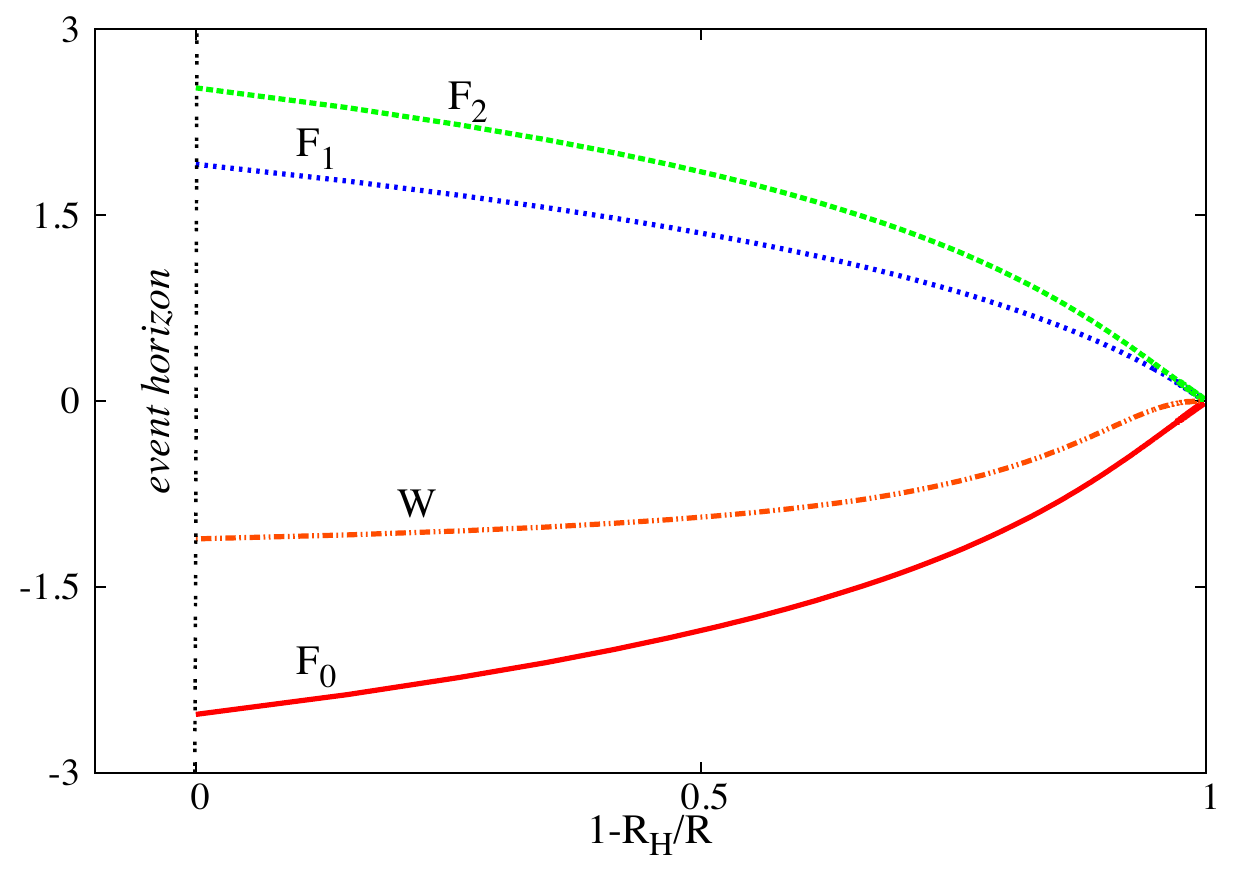}
\includegraphics[height=1.78in]{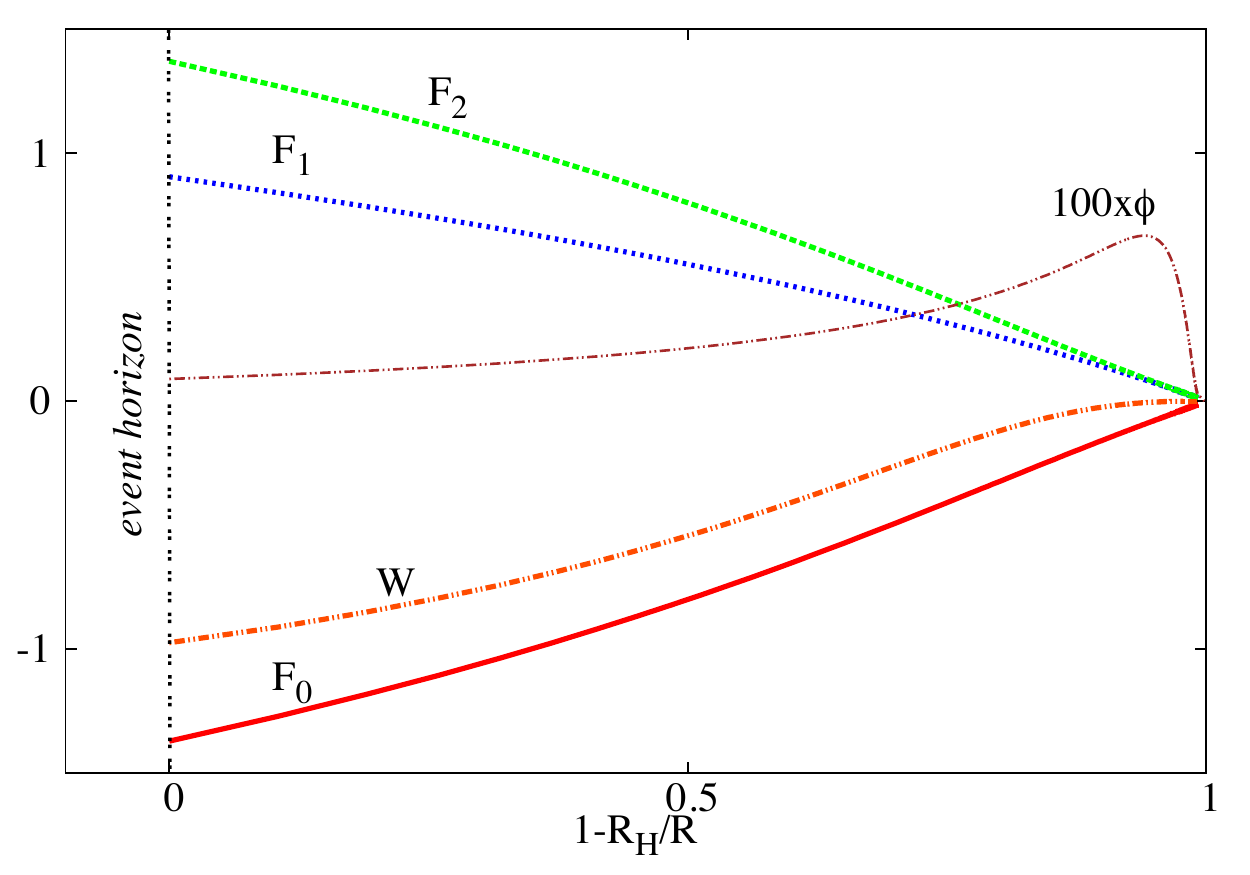}\\
\caption{Left panel: The metric functions in \eqref{ansatz_hairy} for
  a Kerr BH in the region of nonuniqueness. We have chosen its mass
  and angular momentum to be $M=0.415$, $J=0.172$; this corresponds to
  $r_H= 0.066$. Right panel: The metric and
  scalar field functions for a hairy BH in the region of
  nonuniqueness with the same $M,J$ as the Kerr BH. This hairy BH is
  Kerr-like and has $ \omega=0.975$ and $r_H=0.2$. The value of the
  scalar field profile function has been multiplied by a factor of
  100.}
\label{functions_plot}
\end{figure}

In Figure~\ref{functions_plot} we plot the five functions
in~\eqref{ansatz_hairy} for an example of a Kerr BH (left panel, for
which case $\phi=0$) and also for a hairy Kerr BH solution (right
panel), on the equatorial plane $\theta=\pi/2$ and in terms of a
compactified radial coordinate $1-R_H/R$. The behavior observed here
is quite generic. All metric functions are monotonic functions of
$R$. The scalar field profile function is nonzero on the horizon, has
one maximum and tends to zero asymptotically, also a generic behavior
for $n=0$ solutions. A set of ten example solutions (including the two
just mentioned) are available online as a supplement to this
review~\cite{ref:webpage}. The files provide all five metric functions
$(F_0,\,F_1,\,F_2,\,N,\,\phi)$ on a fine grid (see
also~\cite{Herdeiro:2015gia} for a detailed computation of these
solutions).

The stability of these solutions and the formation mechanism of hairy
BHs that deviate significantly from Kerr remain urgent open issues. A
recent analysis suggests that, should these solutions arise from a
superradiant instability of the Kerr metric, the energy-density of the
scalar field would be negligible and the geometry would be well
described by the Kerr solution~\cite{Brito:2014wla}.

\subsubsection{Evading no-hair theorems in Horndeski/Gauss-Bonnet gravity}
\label{sec:BH_Horndeski}
Hawking's no-hair theorem for stationary BHs in Brans-Dicke
theory~\cite{Hawking:1972qk} was recently extended by Sotiriou and
Faraoni to more general scalar-tensor theories~\cite{Sotiriou:2011dz}.
Hui and Nicolis~\cite{Hui:2012qt}
further extended these proofs to the most general scalar-tensor
theory leading to second-order field equations, i.e.~Horndeski's
theory (introduced in Section~\ref{subsec:horndeski}).  Hui and Nicolis
claimed that static, spherically symmetric, asymptotically flat BHs in
vacuum have no hair in Horndeski's theory, provided that the scalar
exhibits shift symmetry -- i.e., symmetry under $\phi\to\phi+{\rm
  constant}$. The conclusion follows from the fact that the scalar
field equation can be written as a conservation equation for the
Noether current $J^\mu$ associated with the shift symmetry, namely
\begin{equation}
 \nabla_\mu J^\mu=0\,. \label{eq:Noether}
\end{equation}
In a nutshell, the argument is the following.  If the scalar field
respects the symmetries of the metric,
\begin{equation}
 ds^2=-f(r)dt^2+f(r)^{-1}dr^2+R(r)^2(d\theta^2+\sin^2\theta d\varphi^2)\,,
\end{equation}
then the only nonvanishing component of $J^\mu$ is the radial one,
which gives the value of the invariant $J^\mu J_\mu=(J^r)^2/{f}$. Because
$f=0$ at the horizon, $J^r$ must also vanish there. Then
the conservation equation~\eqref{eq:Noether} implies that $J^r$ must be zero
everywhere. Finally, the last step of the proof is to argue that
$J^r=0$ everywhere implies that the scalar field must be constant, and
therefore the metric must satisfy Einstein's equations in vacuum.

This last step was criticized in~\cite{Sotiriou:2013qea}, where it was
shown that there exists a counterexample where the scalar field has a
nontrivial profile even though $J^r=0$. This happens precisely when
the scalar field is \emph{linearly coupled to the Gauss-Bonnet
  invariant} $R_{\rm GB}$ (see Section~\ref{subsec:quadratic}),
i.e.~for a theory with action
\begin{equation}
 S=\frac{1}{16\pi}\int\sqrt{-g} d^4x \left[R-2\nabla_a\phi\nabla^a\phi+\alpha \phi R_{\rm GB}^2\right]\,. \label{EdGBaction2}
\end{equation}
This special case of Horndeski's theory is also a special case of the
quadratic gravity theories discussed in Section~\ref{subsec:EdGB}.
BHs in this theory are indeed endowed with a nontrivial scalar profile
(cf.~Section~\ref{sec:BHquadratic}). The scalar charge is not an
independent quantity, but it depends on the BH mass, so these BHs are
said to have ``hair of the second kind,'' but they are nevertheless
different from their Schwarzschild counterparts.  These solutions were
discussed in~\cite{Sotiriou:2014pfa}. They are the only vacuum,
spherically symmetric hairy BHs in Horndeski's theory with shift
symmetry for which the scalar field respects the symmetries of the
metric.

If the scalar field is time-dependent, it is possible for BHs to
develop hair both in scalar-tensor theories~\cite{Jacobson:1999vr} and
in Horndeski theories~\cite{Babichev:2013cya}. Furthermore, as
discussed below, BHs in scalar-tensor theories can grow hair in the
presence of matter.

\subsubsection{BHs surrounded by matter}
Isolated BHs in scalar-tensor theories are the same as in GR, but the
situation changes completely in the presence of
matter. Refs.~\cite{Cardoso:2013fwa,Cardoso:2013opa} have investigated
the effects of simple models of accretion disks and halos around BHs
in generic scalar-tensor theories. In these theories the Klein-Gordon
equation on a Kerr BH surrounded by matter takes the form
\begin{equation}
 [\square-\mu_{\rm eff}^2]\Psi=0\,,
\end{equation}
where the effective mass $\mu_{\rm eff}$ depends on the specific
scalar-tensor theory, and it is proportional to the trace of the
stress-energy tensor.

Depending on the sign of the scalar coupling, $\mu_{\rm eff}^2$ can
either be positive or negative. In the latter case the system is prone
to a \emph{tachyonic} instability and spontaneously develops a scalar
hair supported by matter. This phenomenon is akin to the spontaneous
scalarization phenomenon in NSs (cf.~Section~\ref{sec:NS_ST}
below). On the other hand, when $\mu_{\rm eff}^2>0$ the scalar field
acquires a real effective mass and can trigger a ``spontaneous
superradiant instability'' similar to the one discussed
previously. The instability is much stronger than in the vacuum case,
because the presence of matter drastically affects the amplification
of scalar waves. In fact, superradiant amplification from spinning BHs
is strongly enhanced when Breit-Wigner resonances
occur~\cite{Cardoso:2013fwa,Cardoso:2013opa}. Possible astrophysical
implications of this amplification have not been investigated yet, but
they may yield phenomenological constraints on the parameter space of
scalar-tensor theories.

The effect of a matter distribution around BHs in scalar-tensor theory
was studied in~\cite{Davis:2014tea} in the context of theories with a
screening mechanism. In scalar-tensor theory, screening occurs when
the conformal factor $A(\varphi)$ and/or the potential $V(\varphi)$
suppress the effect of modified gravity in dense enviroments (such as
those of the Solar System or of our Galaxy), allowing for
modifications on a cosmological scale which are not in conflict with the
bounds from Solar System and binary pulsar tests~\cite{Brax:2012gr}
(see Section~\ref{subsec-cosm-nonlinear}). Using a simple spherically
symmetric model of an accretion disk or a galactic halo, and
neglecting the effect of matter and of the scalar field on the
spacetime metric, Ref.~\cite{Davis:2014tea} shows that matter induces
a nontrivial, spherically symmetric scalar field profile. Their
estimates suggest that the effect of the ``scalar fifth force'' on
test particles should be much smaller than effects due to the
quadrupole emission of GWs, and therefore that it is unlikely to
reveal the presence of a fifth force in this context.

\subsubsection{Stability}
\label{sec:bh-st-stability}
The stability of BHs in scalar-tensor theories is a nontrivial
issue. No-hair theorems do not apply to BH dynamics, which is
different from GR even in the simplest scalar-tensor theories.

To our knowledge, there are very few works on BH stability in
scalar-tensor theories. Kobayashi, Motohashi and Suyama studied the
linear perturbations of static, spherically symmetric BHs in Horndeski
gravity~\cite{Kobayashi:2012kh,Kobayashi:2014wsa} finding a set of
necessary conditions for BH stability. Quite interestingly, these
conditions impose restrictions on the general Horndeski action.
The stability of static, spherically symmetric BHs with respect to
linear odd-parity perturbations was demonstrated in the case of GR
minimally coupled with a scalar field (with an arbitrary
potential)~\cite{Anabalon:2014lea}; their result also applies to
Bergmann-Wagoner theory (see Section~\ref{subsec:ST}). On the other
hand, it has long been known that rotating BHs in GR minimally coupled
with a {\em massive} scalar field are unstable. We will return to this
point in Section~\ref{sec:superradiance}.

\subsection{f(R) theories}
Scalar-tensor theories include $f(R)$ theories as a special case.
The vacuum Kerr spacetime is a solution in $f(R)$ gravity by virtue of
the theorems that apply to generic scalar-tensor
theories~\cite{Sotiriou:2011dz}. Typically $f(R)$ theories propagate
massive degrees of freedom~\cite{Nzioki:2014oaa}. As a consequence,
rotating BHs may be prone to superradiant instabilities: this
possibility was discussed as early as 1985 by Hersh and
Ove~\cite{Hersh:1985hz}. Interestingly, in this case the effective
scalar field is related to the scalar curvature of the metric, which
grows exponentially through superradiance. This suggests that, at
variance with the case of real massive fields previously discussed,
the end-state of superradiant instabilities in $f(R)$ gravity might be
different from a Kerr BH~\cite{Hersh:1985hz}.

\subsection{Quadratic gravity}
\label{sec:BHquadratic}
New BH solutions can be found in theories with quadratic curvature
terms in the action.  We first discuss the perturbative approach in
generic quadratic theories before focusing on results specific to the
EdGB and dCS theories.

\subsubsection{Perturbative solutions in the slow-rotation limit}
\label{subsec:BHquadraticPert}

Consider the action~\eqref{eq:action_quadratic}, that includes EdGB
and dCS as special cases.  BH solutions in this theory are not known
in full generality (with the exception of EdGB gravity, see
Section~\ref{sec:BH_EdGB} below), but perturbative solutions were
obtained analytically when the coupling functions $f_i$ admit the
expansion~\eqref{eq:quadratic_expansion} and the BH is slowly rotating
(numerical solutions for rapid rotation will be discussed below). The
solution describing a static, spherically symmetric BH is a limiting
case (vanishing rotation) of this family.
Consider the following metric ansatz for the stationary, slow-rotation
limit:
\begin{equation}
\begin{aligned}
 ds^2={}&{}-f(r,\theta)dt^2+g(r,\theta)^{-1}dr^2-2\omega(r)\sin^2\theta
 dtd\varphi \\
 &{}+r^2\Theta(r,\theta)d\theta^2+r^2\sin^2\theta\Phi(r,\theta) d\varphi^2
 \,,
\end{aligned}
\label{metric_slowrot_quadratic}
\end{equation}
and let the scalar field have the dependence $\phi=\phi(r,\theta)$.
By solving the field equations one finds the following metric
functions that describe a slowly rotating BH
solution~\cite{Pani:2011gy}:
 \begin{align}
   f(r,\theta)={}& f^{(0)}+\frac{\alpha_3^2}{4}\left[-\frac{49}{40 M_0^3 r}+
     \frac{1}{3 M_0 r^3}+\frac{26}{3 r^4}+\frac{22 M_0}{5 r^5}+\frac{32 M_0^2}{5 r^6}-\frac{80 M_0^3}{3 r^7}\right]\,,\nonumber\\ 
   g(r,\theta)={}& g^{(0)}+\frac{\alpha_3^2}{4}\left[-\frac{49}{40 M^3_0 r}+
     \frac{r+M_0}{M_0^2 r^3}+\frac{52}{3 r^4}+\frac{2 M_0}{r^5}+\frac{16 M_0^2}{5 r^6}-\frac{368 M_0^3}{3 r^7}\right]\,,\nonumber\\ 
   \omega(r)={}&\frac{2 a M_0}{r}-\frac{a\alpha_3^2}{4}\left[\frac{3}{5 M_0 r^3}+
     \frac{28}{3 r^4}+\frac{6 M_0}{r^5}+\frac{48  M_0^2}{5 r^6}-\frac{80 M_0^3}{3 r^7}\right]\nonumber\\
&{}-a\alpha_4^2\frac{5}{2}\left[\frac{1}{r^4}+\frac{12 M_0}{7r^5}+\frac{27 M_0^2}{10r^6}\right] \,, \nonumber\\
   \Theta(r,\theta)={}& 1+\frac{\cos^2\theta}{r^2}a^2\,,\qquad \Phi(r,\theta)=
   1+\frac{r+2M_0\sin^2\theta}{r^3}a^2 \,.\nonumber
\end{align}
The scalar field solution is given by
\begin{align}
  \phi(r,\theta)={}&{\alpha_3}\left[\frac{1}{2M_0 r}+\frac{1}{2r^2}+\frac{2 M_0}{3 r^3}\right]+
  a\alpha_4 \frac{5\cos\theta}{8M_0}\left[\frac{1}{r^2}+\frac{2 M_0}{r^3}+\frac{18 M_0^2}{5r^4}\right]\nn\\
  &{}-\frac{\alpha_3 a^2}{2}\left[\frac{1}{10r^4}+\frac{1}{5M_0r^3}+\frac{M_0+r}{4M_0^3r^2}+
    \cos^2\theta \left(\frac{48M_0^2+21M_0r+7r^2}{5M_0r^5}\right)\right] \,.\nonumber
\end{align}
Here $f^{(0)}\equiv 1-2 M_0/r+2a^2M_0\cos^2\theta/r^3$ and
$g^{(0)}\equiv 1-2M_0/r+a^2\left(r-(r-2M_0)\cos^2\theta\right)/r^3$ are
the expansions of the Kerr metric coefficients of
Eq.~\eqref{metricKerrLambda} up to terms of order ${\cal O}(a^2)$.
Note that, as discussed in Section~\ref{subsec:quadratic}, $\alpha_3$
is the EdGB coupling constant, and $\alpha_4$ is the dCS coupling
constant.

The curvature invariants are regular in the exterior spacetime.  The
angular momentum is $J=aM_0$, whereas the physical (ADM) mass of the BH
is~\cite{Mignemi:1992nt,Yunes:2011we}
\begin{equation}
M=M_0 \left[1+\f{{49}\alpha_3^2}{320M_0^4}\right] \,. \label{EdGB:mass}
\end{equation}
The above solution is accurate up to order
${\cal O}(a^2/M^2,\alpha_i^2/M^4,a\alpha_i^2/M^5)$ in the metric,
and up to order
${\cal O}(a^2/M^2,\alpha_i^2/M^4,a\alpha_i^2/M^5,a^2\alpha_i/M^3)$ in
the scalar field.

Using the second-order in spin corrections obtained
in~\cite{Yagi:2012ya,Ayzenberg:2014aka}, the corrections to the Kerr
quadrupole moment $Q_{\rm Kerr}$ arising in quadratic gravity read
\begin{equation}
  Q=Q_{\rm Kerr}\left(1+\frac{4463}{2625}\frac{\alpha_3^2}{M^4}-\frac{201}{448}
  \frac{\alpha_4^2}{M^4} \right)\,, \label{quadrupole_quadratic}
\end{equation}
where the quadrupole moment is defined through a large-distance
expansion of the metric as
\begin{equation}
 g_{tt}\to -1+\frac{2M}{r}+\frac{\sqrt{3}}{2}\frac{Q}{r^3} Y_{20}(\theta)\,,
\end{equation}
and $Y_{20}$ is a spherical harmonic with $l=2$ and $m=0$.

Constraints on BH solutions are mostly derived by understanding how
matter moves in the vicinity of the BH.
Geodesic motion can be derived from the matter action for a point
particle:
\be
S_\text{mat}=-m\int dt\,\sqrt{-\gamma(\phi)g_{\mu\nu}\dot x^\mu\dot x^\nu}\,,
\label{eq:mataction}
\ee
where $m$ is the mass of the particle, and $\gamma(\phi)$ is the
coupling function between the matter and the scalar field. For the
low-energy limit of heterotic string theory, $\gamma=e^{\phi}$.
In the small-coupling limit we have
\begin{equation}
 \gamma(\phi)=1+2b\phi+{\cal O}(\phi^2)\,,
\end{equation}
where $b=0$ for minimal coupling and $b=1/2$ in heterotic string
theory. We focus on equatorial motion ($\theta=\pi/2$,
$\dot\theta\equiv0$).
Expanding geodesic quantities to the same order as the metric itself,
we find the following expressions for the ISCO location and the
frequency at the ISCO (both normalized by the physical mass
$M$):
\begin{align}
  \frac{R_\text{ISCO}}{M}={}&6-4 \sqrt{\frac{2}{3}} \frac{a}{M_0}-\frac{7 a^2}{18 M_0^2}+
  \frac{8}{9}\frac{b\alpha_3}{M_0^2}-\frac{17}{54}\sqrt{\frac{2}{3}}\frac{ba\alpha_3}{M_0^3}\nn\\
  &{}-\left(\frac{16297}{38880}-\frac{22267 a}{17496 \sqrt{6} M_0}\right) \frac{\alpha_3^2}{M_0^4}+
  \frac{77 a}{216 \sqrt{6} M_0^5}\alpha_4^2,\nn\\
       M\Omega_\text{ISCO}={}&\frac{1}{6 \sqrt{6}}+\frac{11 a}{216 M_0}+\frac{59 a^2}{648 \sqrt{6} M_0^2}
       -\frac{12113 a}{5225472 M_0^5}\alpha_4^2\nn\\
       &{}-\frac{29}{432 \sqrt{6}}\frac{b\alpha_3}{M_0^2}-\frac{169}{7776}\frac{ba\alpha_3}{M_0^3}+
       \left(\frac{32159}{2099520 \sqrt{6}}-\frac{49981 a}{75582720 M_0}\right) \frac{\alpha_3^2}{M_0^4}\,.\label{eq_romega_isco}
\end{align}
We have kept only dominant terms in $b$, and for simplicity we focused
on corotating orbits (the result for counterrotating orbits is
trivially obtained by inverting the sign of $a$).  Note that $a/M_0$ is
not the physical dimensionless angular momentum, $J/M^2$, but
it can be easily related to the latter to second order in
$\alpha_3/M^2$ using
Eq.~\eqref{EdGB:mass}~\cite{Maselli:2013mva}.
 
The behavior of the ISCO frequency depends on several coupling
parameters. For $b=0$, the dominant correction is of order ${\cal
  O}(\alpha_3^2)$ and tends to increase the ISCO frequency. The first
corrections proportional to the BH spin are ${\cal O}(a\alpha_3^2)$
and ${\cal O}(a\alpha_4^2)$, and they contribute to lower the
frequency. However, when a nonminimal coupling is turned on, its
effect is dominant~\cite{Pani:2009wy}. When $b\neq0$, the ISCO frequency gets 
corrections of order ${\cal O}(b\alpha_3).$\footnote{%
  Note that a nonminimal coupling to the
  matter sector violates the weak equivalence principle. Since the
  latter is tested within the astonishing precision of 1 part in
  $10^{13}$~\cite{Will:2014xja}, a very stringent bound on $b$ can be
  derived: $b\alpha_i/M^2<10^{-13}$.}

For null geodesics, the light-ring frequency $\Omega_{\rm LR}=L_{\rm
  LR}/E_{\rm LR}$ (related to the real part of the ringdown frequency
of the BH in the eikonal limit~\cite{Cardoso:2008bp}) and the
light-ring radius $R_{\rm LR}$ do not depend on the coupling $\gamma$:
\begin{align}
  \frac{R_{\rm LR}}{M}={}&3-\frac{2 a}{\sqrt{3} M_0}-\frac{2 a^2}{9 M^2_0}+
  \frac{31}{81 \sqrt{3}}\frac{a\alpha_4^2}{M_0^5}-\left(\frac{961}{3240}-
  \frac{33667 a}{174960 \sqrt{3} M_0}\right) \frac{\alpha_3^2}{M_0^4}\,,\\
       M\Omega_{\rm LR}={}&\frac{1}{3 \sqrt{3}}+\frac{2 a}{27 M_0}+\frac{11 a^2}{162 \sqrt{3} M_0^2}
       -\frac{131}{20412} \frac{a\alpha_4^2}{M^5_0}+\left(\frac{4397}{262440 \sqrt{3}}+
       \frac{24779 a}{4723920 M_0}\right) \frac{\alpha_3^2}{M^4_0} \,.
\end{align}
The dominant correction is ${\cal O}(\alpha_3^2)$ and increases the
frequency, whereas the ${\cal O}(a\alpha_3^2)$ and ${\cal
  O}(a\alpha_4^2)$ corrections have opposite relative signs.

\subsubsection{EdGB theory}\label{sec:BH_EdGB}
\paragraph{Static and slowly rotating solutions.}
Perturbative BH solutions in EdGB gravity can be obtained as special
cases of the results discussed above. Besides these perturbative results, an exact
static solution in EdGB gravity (i.e., a solution going beyond the perturbative level in $\alpha_3$) is also
known~\cite{Kanti:1995vq}. It has the form
\begin{equation}
  ds^2=-f(r)dt^2+g(r)^{-1}dr^2+r^2d\theta^2+r^2\sin^2\theta d\varphi^2
  \,,
\end{equation}
where the metric functions $f(r)$, $g(r)$ and the scalar field $\phi(r)$ can be found by solving a system of ordinary
differential equations. The solution (regular at the horizon and at infinity) only exists
when~\cite{Kanti:1995vq,Pani:2009wy}
\begin{equation}
 0<\alpha_3/M^2\lesssim0.691\,.
\end{equation}

Ref.~\cite{Pani:2009wy} extended these results to the slow-rotation
case, analyzing geodesics and QNMs in the eikonal
limit. Ref.~\cite{Maselli:2014fca} studied the epicyclic frequencies
of this solution, with the aim of constraining EdGB gravity through
observations of BH quasi-periodic oscillations.
Ref.~\cite{Ayzenberg:2014aka} analyzed corrections of second order in
the spin working in the small-coupling limit. The character of the
solution changes quite dramatically: the solution at first order in
spin is algebraically special (i.e.~of Petrov type D, just like the
Kerr metric), while the second-order solution is of Petrov type I. The
Petrov type of the metric is very important, because it is related to
the separability of the field equations. This example illustrates the
importance of obtaining {\em exact} BH solutions (rather than
perturbative expansions) when analyzing their features in modified
theories of gravity.

\paragraph{Rapidly rotating solutions.}
Rapidly rotating BHs in EdGB theory have been obtained
numerically~\cite{Kleihaus:2011tg} using the Lewis-Papapetrou ansatz
for a
stationary, axially symmetric spacetime.  The line element can be
parametrized as
\begin{equation}
ds^2
= - e^{2 \nu_0}dt^2
      +e^{2(\nu_1-\nu_0)}\left[e^{2 \nu_2}\left(dr^2+r^2 d\theta^2\right)
       +r^2 \sin^2\theta
          \left(d\varphi-\omega dt\right)^2\right] ,
\label{metric2} 
\end{equation}
where $\nu_0$, $\nu_1$, $\nu_2$ and $\omega$ are functions of $r$ and
$\theta$ only. BH solutions are asymptotically flat and possess the
expansion
\begin{align}
\nu_0 & =  -\frac{M}{r} + \frac{D_1 M}{3r^3} - \frac{M_2}{r^3} P_2(\cos\theta) +{\cal O}(r^{-4}) ,
\label{exnu0} \\
\nu_1 & =  \frac{D_1}{r^2} +{\cal O}(r^{-3}) ,
\label{exnu1} \\
\nu_2 & =  -\frac{4 M^2+16 D_1+ q^2}{8 r^2}\sin^2\theta +{\cal O}(r^{-3}) ,
\label{exnu2} \\
\omega & =  \frac{2 J}{r^3}  +{\cal O}(r^{-4}) ,
\label{exom} \\
\phi & =  \frac{q}{r}  +{\cal O}(r^{-2}) ,
\label{exdil}
\end{align}
where $P_2(\cos\theta)$ is a Legendre polynomial, and $\phi$ denotes the dilaton field.  The expansion constants $M$,
$J$, and $q$ denote the (ADM) mass, the angular momentum and the dilaton charge, respectively. The expansion
(\ref{exnu0})-(\ref{exdil}) also depends on the constants $D_1$, $M_2$.

The quadrupole moment $Q$ of EdGB BHs reads~\cite{Kleihaus:2014lba} 
\begin{equation}
Q
=  -M_2 +\frac{4}{3}\left[\frac{1}{4}+\frac{D_1}{M^2}
+\frac{q^2}{16M^2}\right] M^3 .
\label{Q}
\end{equation}
This has been obtained by extending the formalism of Geroch and
Hansen~\cite{Geroch:1970cd,Hansen:1974zz}.\footnote{The scalar field
  of the static BH in EdGB gravity decays as $1/r$ at large distances,
  similarly to the electric field of a Reissner-Nordstr\"om BH. The
  Geroch-Hansen formalism to compute multipole moments was extended to
  stationary electrovacuum spacetimes in
  Refs.~\cite{0264-9381-7-10-012,Sotiriou:2004ud}. Using the fact that
  the Gauss-Bonnet curvature term $R_{\rm GB}$ decays very quickly at large distances, the
  structure of the first multipoles can be shown to be equivalent to
  that of a Reissner-Nordstr\"om BH with a suitable indentification of
  the scalar charge.}

\begin{figure}
\begin{center}
\begin{tabular}{cc}
\includegraphics[width=.5\textwidth, angle=0, clip=true]{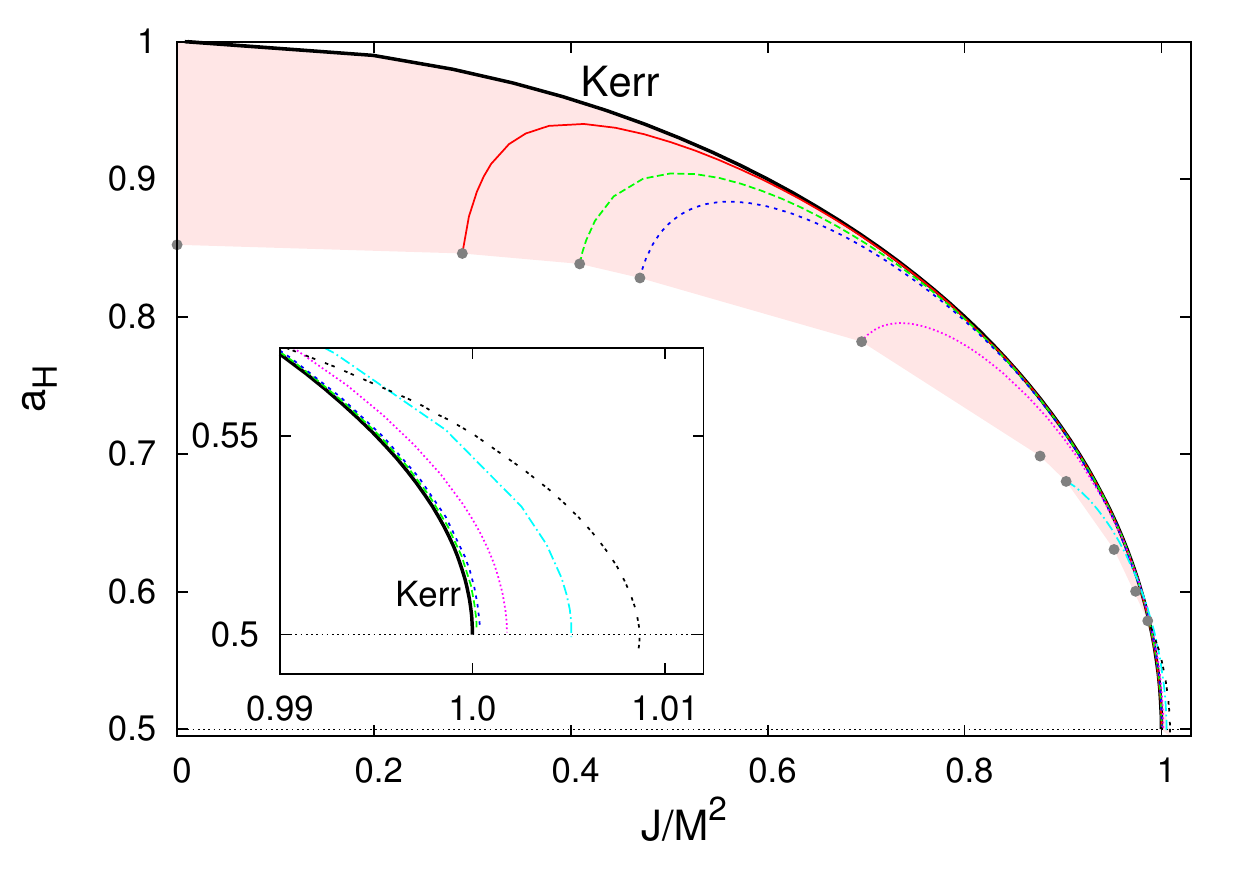}&
\includegraphics[width=.5\textwidth, angle=0, clip=true]{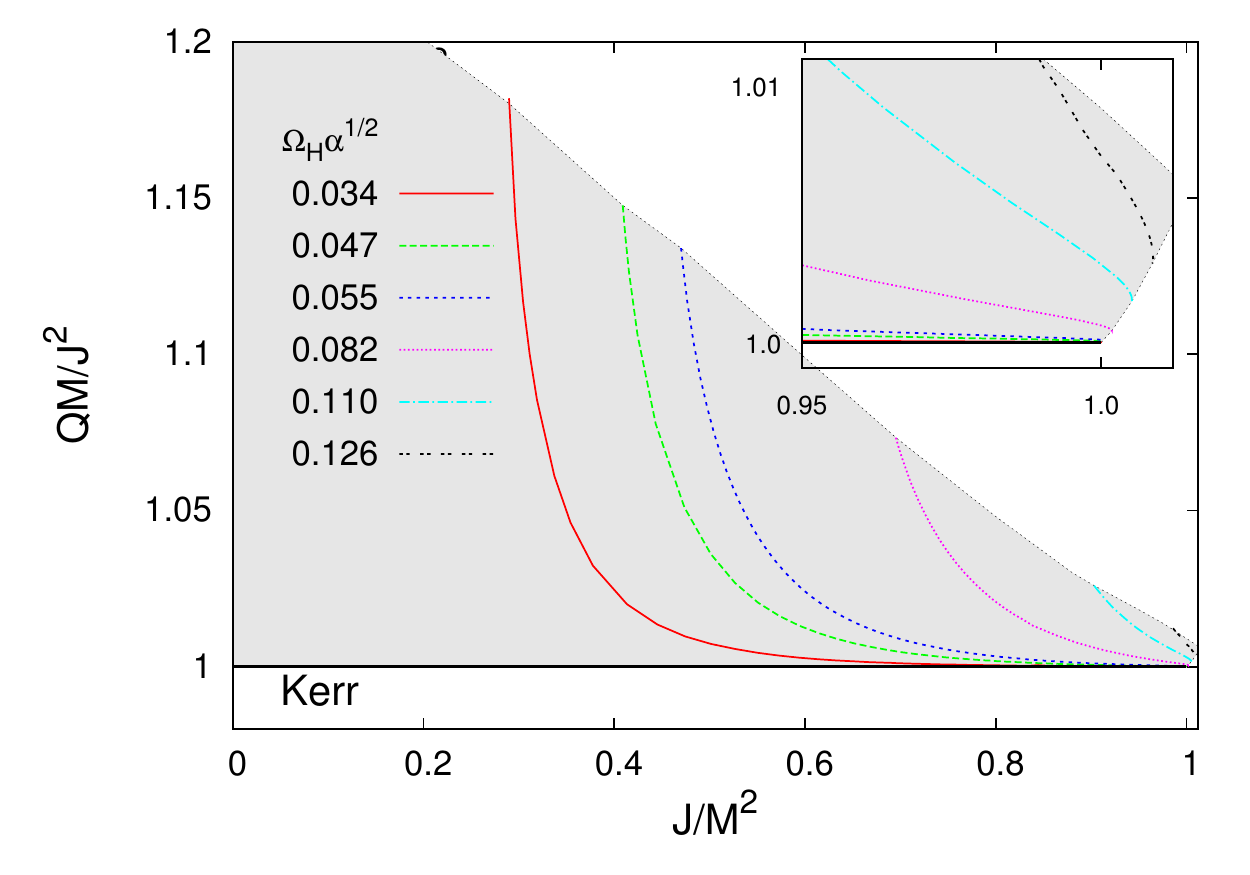}\\
\end{tabular}
\end{center}
\caption{The domain of existence of EDBG BHs (shaded area). We plot
  the scaled horizon area $a_{\rm H}=A_{\rm H}/M^2$ (left panel) and
  the scaled quadrupole moment $\hat{Q}=Q M/J^2$ (right panel) as
  functions of the scaled angular momentum $j=J/M^2$.  Different
  curves correspond to families of EDBG BHs with fixed scaled horizon
  angular velocity $\Omega_{\rm H}
  \alpha_3^{1/2}$. [From~\cite{Kleihaus:2011tg,Kleihaus:2014lba}.]}
\label{KKRfig1}
\end{figure}

The domain of existence of EdGB BHs is illustrated by the shaded area
in the left panel of Figure~\ref{KKRfig1}. The figure shows the scaled
horizon area $a_{\rm H}=A_{\rm H}/M^2$ (where $A_{\rm H}$ is the BH area)
as a function of the scaled
angular momentum $j=J/M^2$.  The upper-left edge corresponds to
Schwarzschild BHs, which are all mapped to the point $a_{\rm H}=1$,
$j=0$.  Likewise, Kerr BHs lie on a curve in this plot, i.e.~the upper
boundary of the domain of existence (except for $j\approx 1$: see
inset).
The lower boundary of the domain of existence corresponds to critical
BH solutions.  These arise when the argument of a square root in the
expansion of the dilaton function at the horizon
vanishes~\cite{Kanti:1995vq,Torii:1996yi,Kleihaus:2011tg}.  For a
given value of the coupling constant $\alpha_3$ and of the mass, EdGB BHs possess lower horizon area than
Kerr BHs.  A remarkable feature is that EdGB BHs can slightly exceed
the Kerr bound ($j \le 1$) for the dimensionless angular momentum.
Only the metric functions for EdGB solutions with $j\ge 1$ are well
defined, while the dilaton field diverges at the poles at the horizon.

The right panel of Figure~\ref{KKRfig1} shows the rescaled quadrupole
moment $\hat{Q}=Q M/J^2$~\cite{Kleihaus:2014lba}.  $\hat{Q}$ is
largest for slow rotation, and decreases with increasing $j$.  The
deviations of $\hat{Q}$ from the corresponding Kerr values can be up
to 20\% and more. Superspinning EdGB BHs with $j>1$ always have $\hat
Q>1$.

\begin{figure}
\begin{center}
\includegraphics[width=9.2cm, angle=0, clip=true]{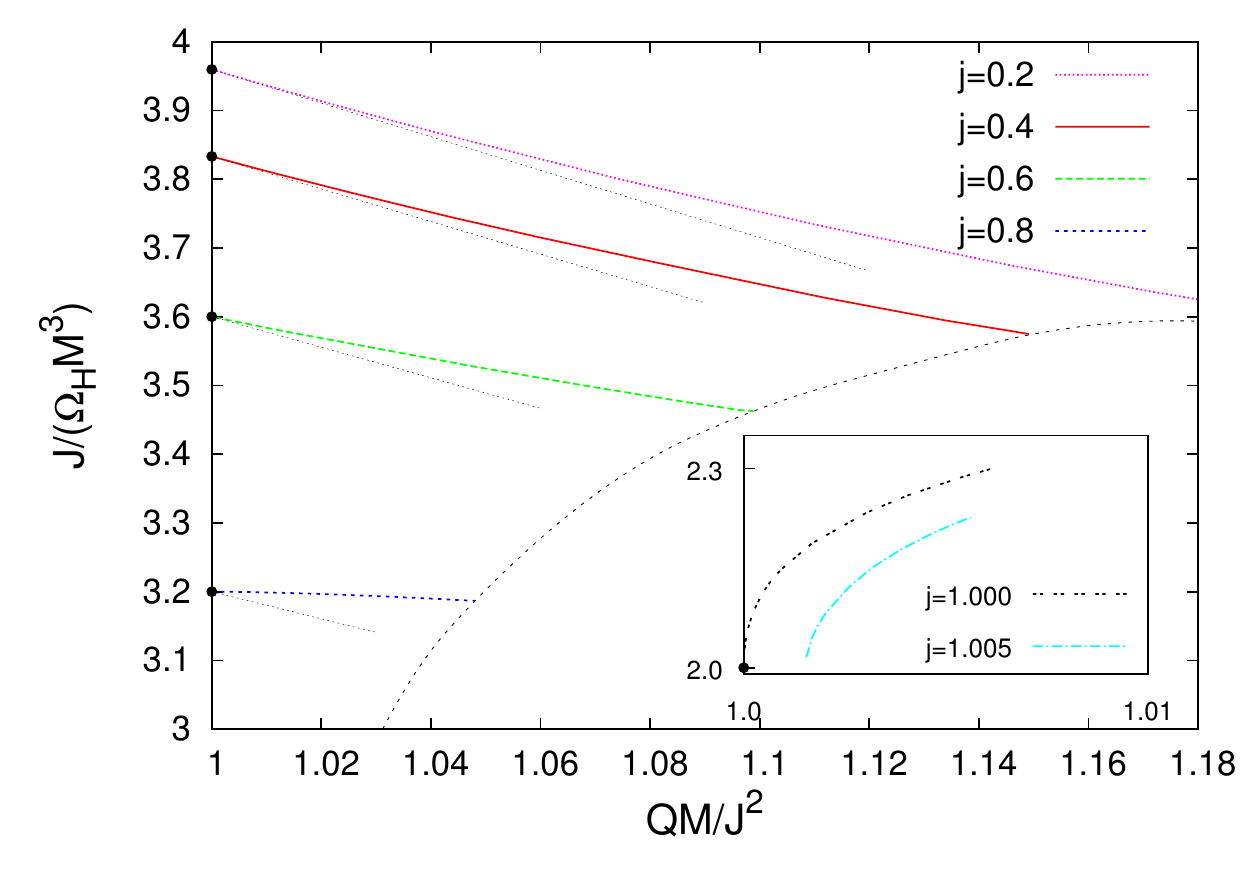}
\end{center}
\caption{The scaled moment of inertia $\hat{I} = J/(\Omega_{\rm H}
  M^3)$ is shown versus the scaled quadrupole moment $\hat{Q}$ for
  fixed values of $j$.  The Kerr BHs are
  indicated by the fat dots on the $\hat{I}$-axis.  The straight
  dotted lines represent the perturbative results
  of~\cite{Ayzenberg:2014aka}.  The critical BHs are represented by
  the dotted curve. [From~\cite{Kleihaus:2014lba}.]
}
\label{KKMfig1}
\end{figure}

Figure~\ref{KKMfig1} exhibits the scaled moment of inertia $\hat{I} =
J/(\Omega_{\rm H} M^3)$ versus the scaled quadrupole moment $\hat{Q}$
for fixed values of $j$~\cite{Kleihaus:2014lba}.  The Kerr values
$\hat{I}_{\rm Kerr}=2(1+\sqrt{1-j^2})$ and $\hat{Q}_{\rm Kerr}=1$ are
indicated by dots on the vertical axis of the plot, and represent the minimum
possible values for $\hat{Q}$. Families of EdGB solutions terminate at
the critical solutions represented by the dotted curve.  For
comparison, straight, dotted lines show the perturbative results
of~\cite{Ayzenberg:2014aka}, derived for small $\alpha_3$ and small
$j$.
The inset shows the region $j>1$, not present in GR.  The extraction
of higher multipole moments from the numerical solutions is still an
open problem.

The study of geodesic motion around rapidly rotating EdGB BHs unveiled
some interesting features~\cite{Pani:2009wy,Kleihaus:2011tg}. Timelike
geodesics for circular motion are obtained from the Lagrangian
$2 {\cal L} = e^{2 b \phi} g_{\mu\nu}\dot x^\mu  \dot x^\nu =-1,$
where again the constant $b$ fixes the coupling between the matter and
the dilaton field ($b=1/2$ corresponds to the low-energy limit of
heterotic string theory).

\begin{figure}
\begin{center}
\begin{tabular}{cc}
\includegraphics[width=.47\textwidth, angle=0, clip=true]{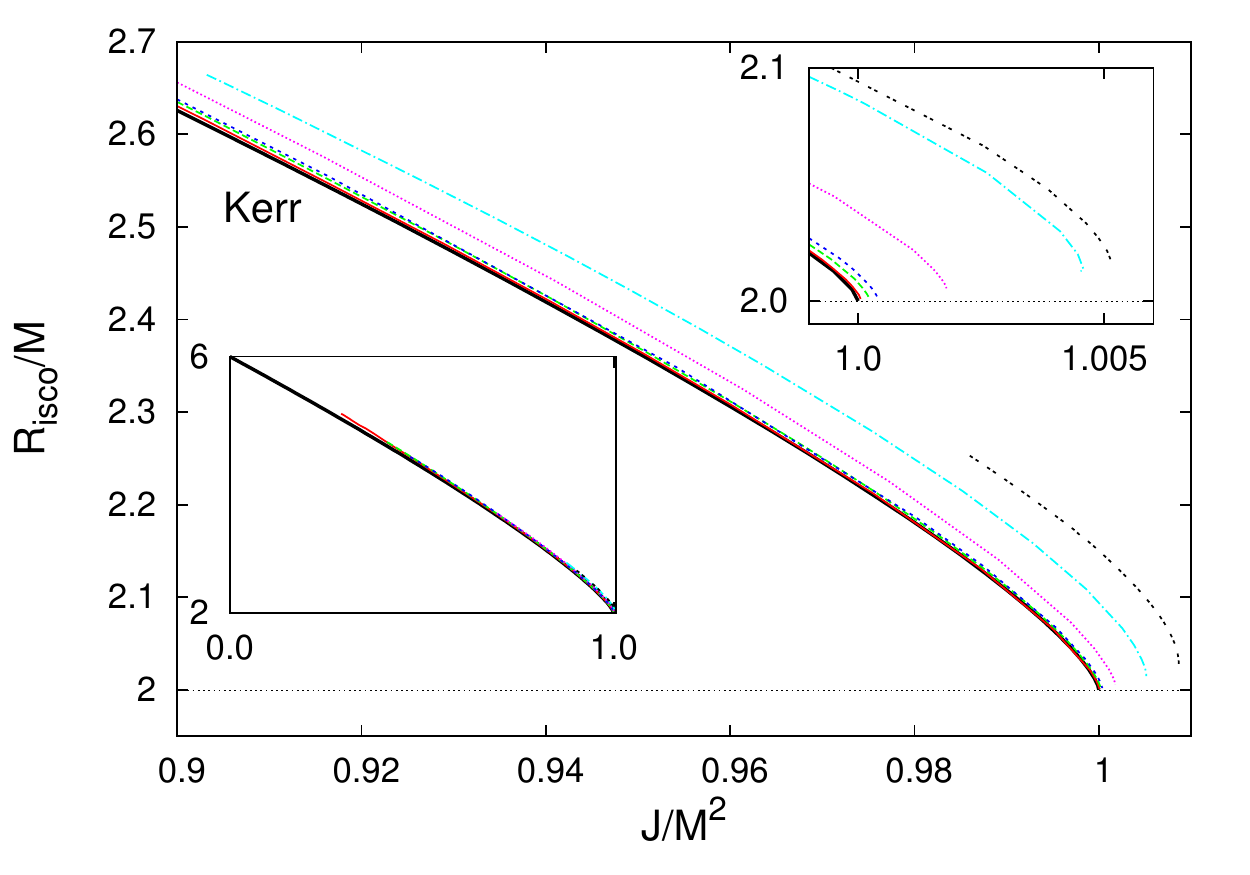}&
\includegraphics[width=.47\textwidth, angle=0, clip=true]{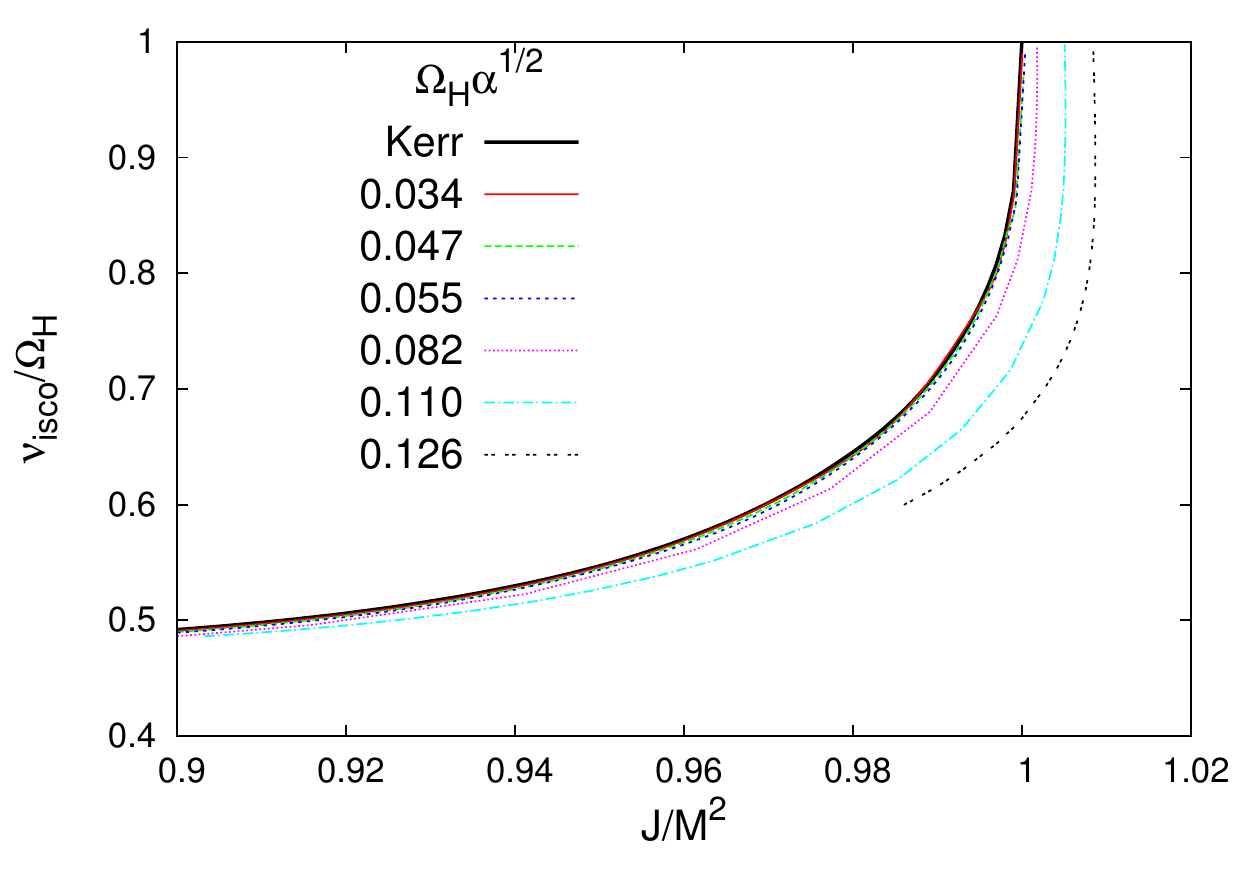}\\
\end{tabular}
\end{center}
\caption{%
  (Left) The scaled circumferential ISCO radius $R_{\rm ISCO}/M$
  and (right) the scaled ISCO frequency $\nu_{\rm  ISCO}/\Omega_{\rm H}$
  are shown versus the scaled angular
  momentum $j=J/M^2$ for dilaton matter coupling constant $b=1/2$
  for families of EDBG BHs with fixed scaled horizon angular
  velocity $\Omega_{\rm H} \alpha_3^{1/2}$. [From~\cite{Kleihaus:2014lba}.] }
\label{KKRfig2}
\end{figure}

Figure~\ref{KKRfig2} shows the scaled circumferential ISCO radius
$R_{\rm ISCO}/M$ versus the scaled angular momentum $j=J/M^2$ for a
coupling constant $b=1/2$.  The Kerr solutions and the extremal
EdGB solutions possess the smallest values for the scaled ISCO radius
$R_{\rm ISCO}/M$, whereas the maximal values of $R_{\rm ISCO}/M$ are
found for the critical EdGB solutions. Note that the ISCO radius is
not given in Boyer-Lindquist coordinates.
When the rescaled angular momentum is large, the deviation of $R_{\rm
  ISCO}/M$ from the Kerr value can be as large as 10\% for
$b=1/2$. Similarly, for large angular momentum the orbital
frequencies at the ISCO exhibit deviations from
the Kerr frequencies as large as 60\% [see~\eqref{eq_romega_isco} for a small-spin expansion].
Note, however, that the weak equivalence principle imposes $b=0$, so that
the corrections to the geodesic quantities are expected to be smaller.

Future space-based observations of the gravitational signal emitted by
extreme mass-ratio inspirals~\cite{Ryan:1995wh} and observations of
the electromagnetic signal associated to quasi-periodic oscillations
in low mass X-ray binaries~\cite{Feroci:2012qh,Pappas:2012nt,Vincent:2013uea} can be used to map the
spacetime around a BH. The key information is encoded in the orbital
frequency $\Omega_{\rm ISCO}$ and, more generally, in the epicyclic
frequencies ($\Omega_r$, $\Omega_\theta$, $\Omega_\phi$). These
quantities have been computed in~\cite{Vincent:2013uea} in the case of
dCS gravity, and in~\cite{Maselli:2014fca} for the case of EdGB gravity.

\paragraph{Stability.}
In the special cases investigated so far, BHs in EdGB gravity were
found to be linearly stable. Ref.~\cite{Kanti:1997br} studied radial
perturbations of static BHs, and Ref.~\cite{Pani:2009wy} considered
axial gravitational perturbations. The stability against polar
gravitational perturbations is an open problem.

\subsubsection{dCS theory}\label{sec:BH_dCS}

\begin{figure}[htb]
  \centering
  \includegraphics[width=0.7\columnwidth]{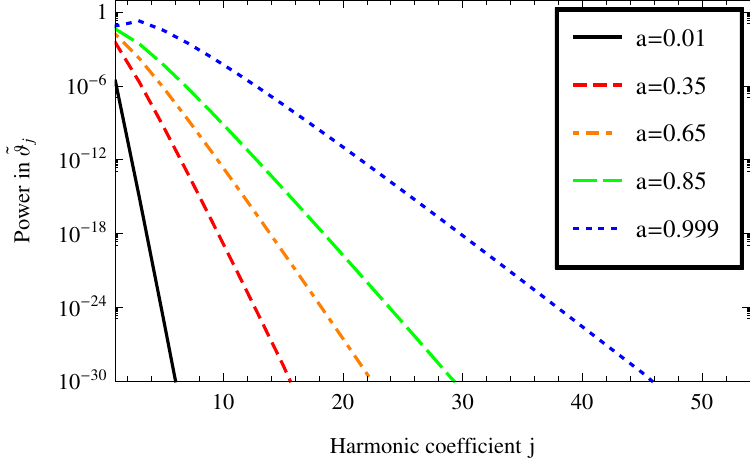}
  \caption{``Power'' in different multipole moments of the scalar
    field $\tilde{\theta}$ (rescaled with $\alpha_4/M^2$)
    around a rotating BH in dCS. The horizontal axis is the multipole
    moment number $j$, i.e.~the coefficient of $P_{j}(\cos\theta)$
    (only the odd coefficients are plotted).  The vertical axis is the
    $L^{2}$ norm of $\tilde{\theta}_{j}$ on a log scale.  As spin
    increases, the exponential convergence slows down, and there is
    more power in higher multipole moments. [From~\cite{Stein:2014xba}.]}
  \label{fig:BHs/dCS/power-theta}
\end{figure}

The field equations of dCS gravity in spherical symmetry reduce to
GR~\cite{Grumiller:2007rv,Molina:2010fb}, so static BH solutions are
given by the Schwarzschild metric.

Spinning BHs in dCS gravity are more interesting, because they are
endowed with a nontrivial scalar field sourced by a nonvanishing
Pontryagin density (${{}^{*}\!RR}\neq0$). Spinning solutions at first
order in a slow-rotation expansion were computed
in~\cite{Yunes:2009hc,Konno:2009kg}.  These solutions can be obtained from the
general slowly rotating BH solution for quadratic gravity discussed in
Section~\ref{subsec:BHquadraticPert} by setting $\alpha_3=0$, and they
have been extended to second order in the BH spin~\cite{Yagi:2012ya}.

In the slow-rotation limit, the scalar field is dominated by a dipole
moment which is proportional to the spin of the BH.  The correction to
the metric quadrupole moment was computed in~\cite{Yagi:2012ya}. The
geometry is of Petrov Type D at first order in rotation, and of Petrov
Type I at higher order.  For arbitrary rotation, the scalar field
profile and trace of the metric deformation were computed
in~\cite{Stein:2014xba}.  As rotation increases, the higher multipole
moments of the scalar field are sourced more strongly, as seen in
Figure~\ref{fig:BHs/dCS/power-theta}.

\paragraph{Stability.}
The linear stability of Schwarzschild BHs in dCS gravity was studied
in Refs.~\cite{Cardoso:2009pk,Molina:2010fb}, where it was found that
these solutions are mode-stable against all gravitational and scalar
perturbations.  The stability of slowly rotating solutions has not
been studied yet, but see~\cite{Garfinkle:2010zx,Ayzenberg:2013wua}
for a high-frequency analysis.

\subsection{Lorentz-violating theories\label{sec:BH_LV}}

The notion of a BH in Lorentz-violating gravity is at first glance less clear-cut than in GR. As mentioned in Section~\ref{lv-theories}, 
in the infrared limit the most generic Lorentz-violating gravity theories are Einstein-\AE ther and khronometric theory, which allow for spin-2 gravitons (like in GR), 
but also for spin-0 and (in Einstein-\AE ther theory but not in khronometric gravity) also spin-1 gravitons. These propagating gravitational modes have speeds that are functions of the coupling parameters $c_i$ of the theories, and are 
generally different from the speed of light appearing in the Maxwell equations and regulating the propagation of the electromagnetic field.
As a result, BHs in these theories, provided they exist, will present multiple horizons, namely: a ``matter horizon''
for the electromagnetic field and the other matter fields, which do not couple directly to the Lorentz violating \ae ther or khronon  field (so as to
enforce the weak equivalence principle, cf.~Section~\ref{lv-theories});
a spin-2 horizon for the spin-2 gravitons; a spin-0 horizon for the spin-0 gravitons; and (for Einstein-\AE ther theory only) a spin-1 horizon for the spin-1 gravitons.
These horizons will generally lie at different locations, depending on the propagation velocity of the corresponding field. However, because of the
requirement that there be no gravitational Cherenkov radiation in these theories (cf.~discussion in  Section~\ref{lv-theories}), for viable
values of the coupling constants the propagation speeds of the spin-2, spin-0 and (when present) spin-1 modes will be larger than (or equal to) the speed of
light, and therefore the spin-2, spin-1 and (when present) spin-0 horizons will be enclosed by the matter horizon.

The situation gets even more complicated if one interprets Einstein-\AE ther and khronometric theory as low-energy
limits of a more generic Lorentz-violating gravity theory containing higher-order spatial derivative terms in the
action. This is the case for instance in Ho\v rava gravity, whose action \eqref{SBPSHfull} reduces to that of
khronometric gravity in the infrared limit, but which contains fourth- and sixth-order spatial derivative terms
that are crucial for the power-counting renormalizability of the theory.  The presence of those terms, as mentioned in
Section~\ref{lv-theories}, causes gravitons to obey nonlinear dispersion 
relations [see Eq.~\eqref{mdsr}].
The matter degrees of freedom (and photons in particular) will also satisfy similar
nonlinear dispersion relations, although the coefficients
of the nonlinear terms
may be smaller than for gravitons,
and in particular sufficiently small to satisfy particle physics tests of Lorentz invariance~\cite{Kostelecky:2003fs,Kostelecky:2008ts,Mattingly:2005re,Jacobson:2005bg,Liberati:2013xla}, if the theory
efficiently suppresses percolation of the Lorentz violations from gravity to the matter sector (cf.~discussion in
Section~\ref{lv-theories}).  From a conceptual point of view, however, Eq.~\eqref{mdsr} makes the very concept of an
event horizon meaningless in the ultraviolet limit, because it implies diverging propagation speeds $d\omega/dk$ in the limit
$k\to\infty$. Therefore, the question arises of whether the multiple event horizons discussed above are simply
low-energy artifacts.

To answer this question, Ref.~\cite{Barausse:2011pu} (building on Ref.~\cite{Eling:2006ec}) looked first at BH solutions in the infrared limit of Lorentz-violating gravity, i.e.~in 
Einstein-\AE ther and khronometric theory. As mentioned in  Section~\ref{lv-theories}, the two theories have exactly the same solutions for
static, spherically symmetric, asymptotically flat BHs. More specifically, using ingoing Eddington-Finkelstein coordinates, the most generic
static and spherically symmetric ansatz for the metric and the \ae ther is given by
\begin{align}
\label{efmetric}
ds^2&=-f(r)dv^2+2 B(r)dv dr+r^2d\Omega^2\,,\\
\label{efaether}\boldsymbol{u}&=A(r) \partial_v-\frac{1-f(r) A^2(r)}{2 B(r) A(r)}\,\partial_r\,.
\end{align}
Solving the field equations perturbatively near spatial infinity, and imposing asymptotic flatness, one obtains the series-expanded solution~\cite{Barausse:2011pu,Eling:2006ec}
\begin{align}
f(r) &= 1-\frac{r_g}{r}-\frac{c_{1}+c_{4} }{48} \frac{r_g^3}{r^3}+ \cdots\,,   \label{asyF}\\
B(r) &= 1+\frac{c_{1}+c_{4}}{16}  \frac{r_g^2}{r^2}+\frac{c_{1}+c_{4}}{12} \frac{r_g^3}{r^3}+\cdots\,,  \label{asyB} \\
A(r) &= 1+\frac12 \frac{r_g}{r}+\frac{A_2 r_g^2}{r^2}-
\left(\frac{1}{16} -\frac{c_1+c_4}{96}-A_2\right) \frac{r_g^3}{r^3}+\cdots\,, \label{asyA}
\end{align}
where $r_g=2 G_N M/c^2$ [the locally measured gravitational constant $G_N$
being related to the ``bare'' one appearing in the action by Eq.~\eqref{eq:GN}], $M$
is the mass of the BH as measured by an observer far from the system, and $A_2$ is a dimensionless ``\ae{}ther charge.''
The latter can in principle take arbitrary values, but if one attempts to construct the BH solution corresponding
to a given $A_2$ value by integrating the field equations inwards starting from the asymptotic solution \eqref{asyF}--\eqref{asyA}, 
one obtains a solution that presents  a finite-area singularity on the spin-0 horizon~\cite{Eling:2006ec,Barausse:2011pu}. Only for a specific value $A_2=A_2^{\rm reg}$ (a function 
of the theory's coupling constants) is the BH regular everywhere except for the central $r=0$ singularity.
We stress that fully  nonlinear numerical simulations~\cite{Garfinkle:2007bk} have shown that spherically symmetric gravitational collapse does indeed select the regular $A_2=A_2^{\rm reg}$
BH solution. 
In practice, this regular BH solution is found
by solving the field equations perturbatively near the spin-0 horizon, imposing that the solution be regular there, and then selecting the solution that
matches the asymptotically flat perturbative solution~\eqref{asyF}--\eqref{asyA} by a shooting procedure (see~\cite{Barausse:2011pu,Barausse:2013nwa} for more details).
After the shooting procedure has selected the asymptotically flat solution, the behavior in the interior can be obtained by integrating inwards from the spin-0 horizon.

\begin{figure}[tb]
\capstart
\begin{center}
\begin{tabular}{lr}
\includegraphics[width=6.2cm,clip=true]{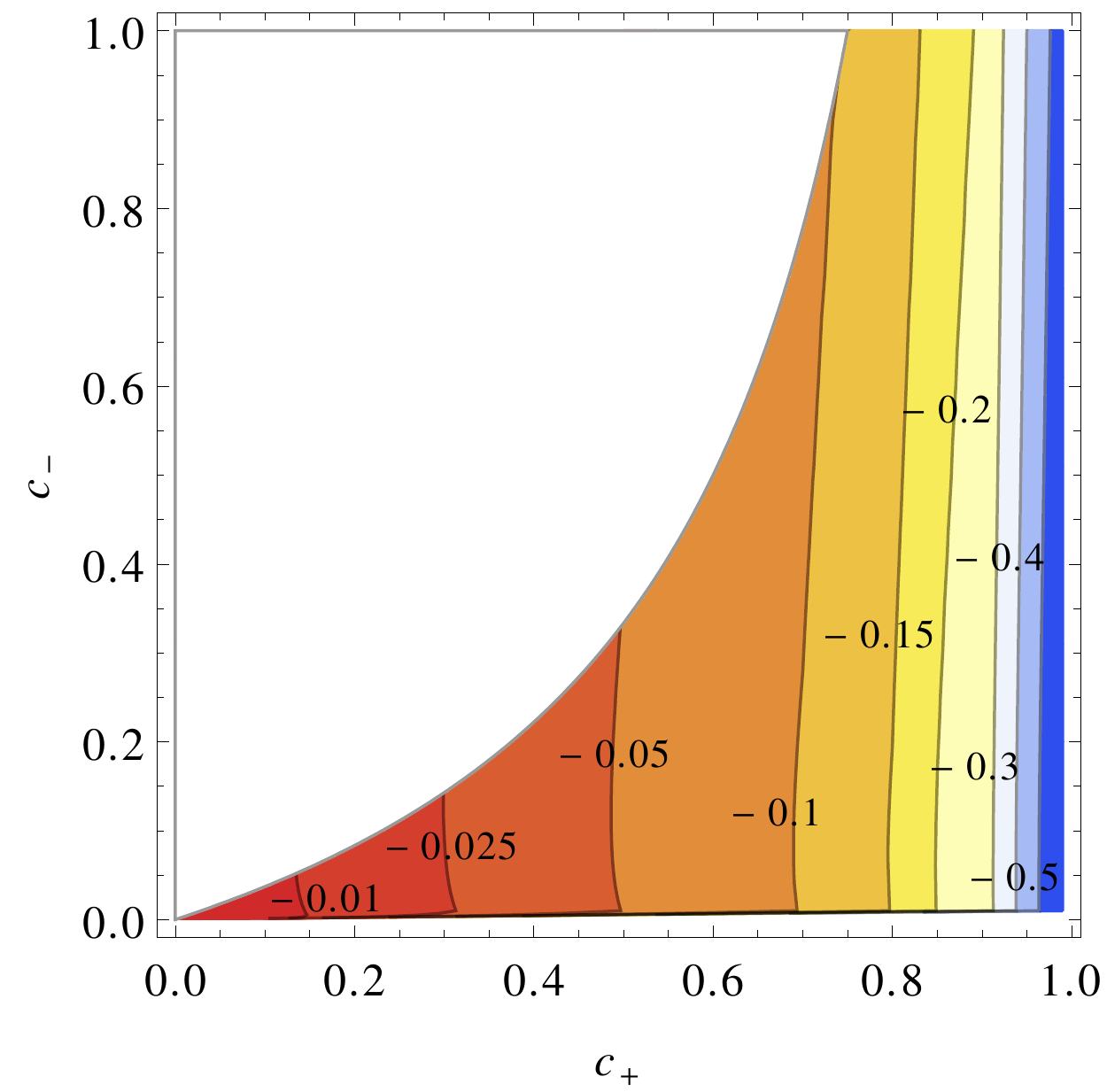}
& \includegraphics[width=6.2cm,clip=true]{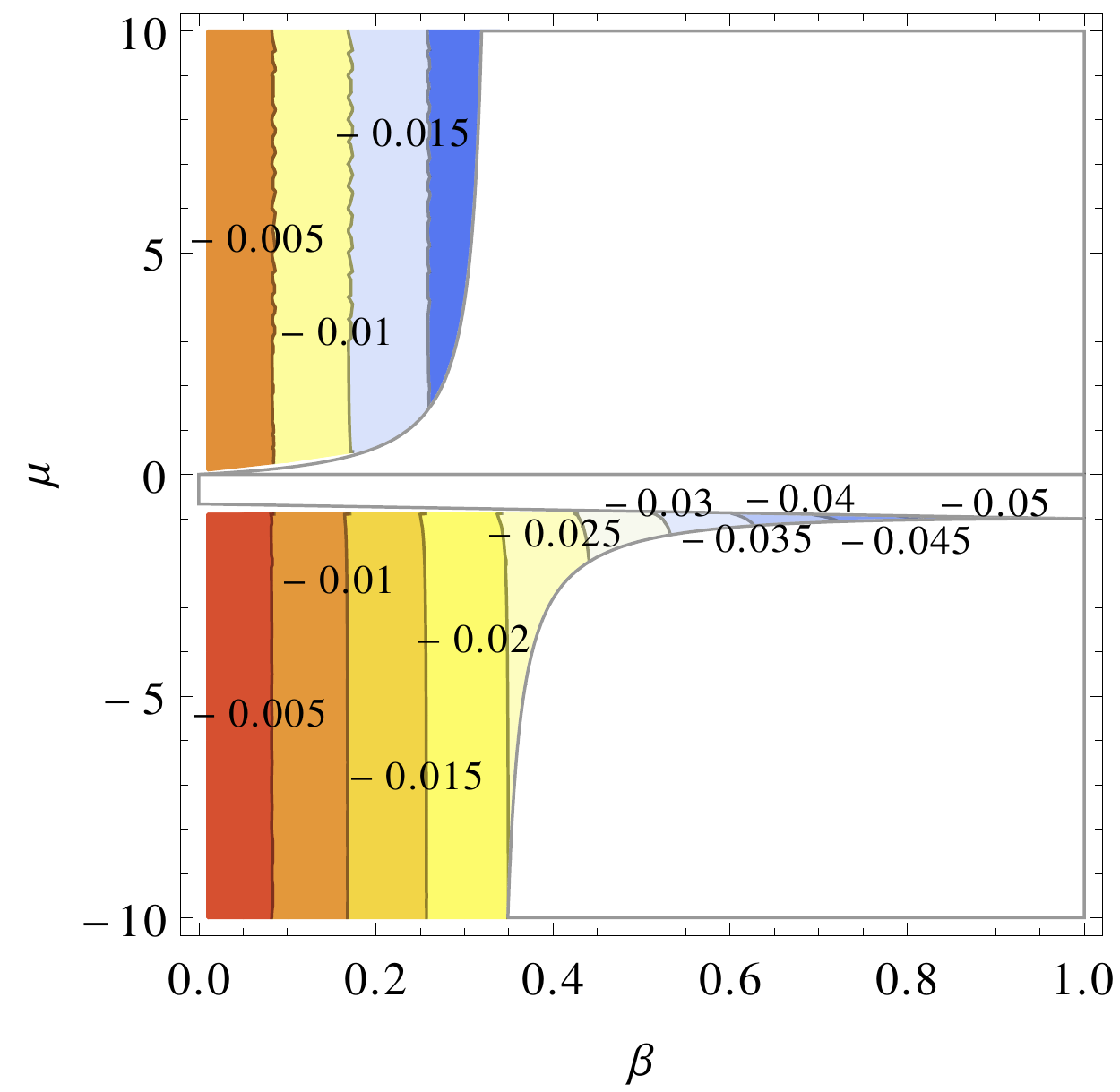}
\end{tabular}
\caption{Fractional deviation of the dimensionless combination $\omega_{_{\rm ISCO}} r_g$ from its GR value, in Einstein-\AE ther (left) and khronometric theory (right).
Negative values denote smaller values in Lorentz-violating gravity than in GR. The quantities $c_\pm$, $\mu$, $\beta$ are defined in Section~\ref{subsec:lorentz-viol}. 
 [From~\cite{Barausse:2013nwa,Barausse:2011pu}.]\label{isco}}
\end{center}
\end{figure}

\begin{figure}[htb]
\capstart
\begin{center}
\begin{tabular}{lr}
\includegraphics[width=6.2cm,clip=true]{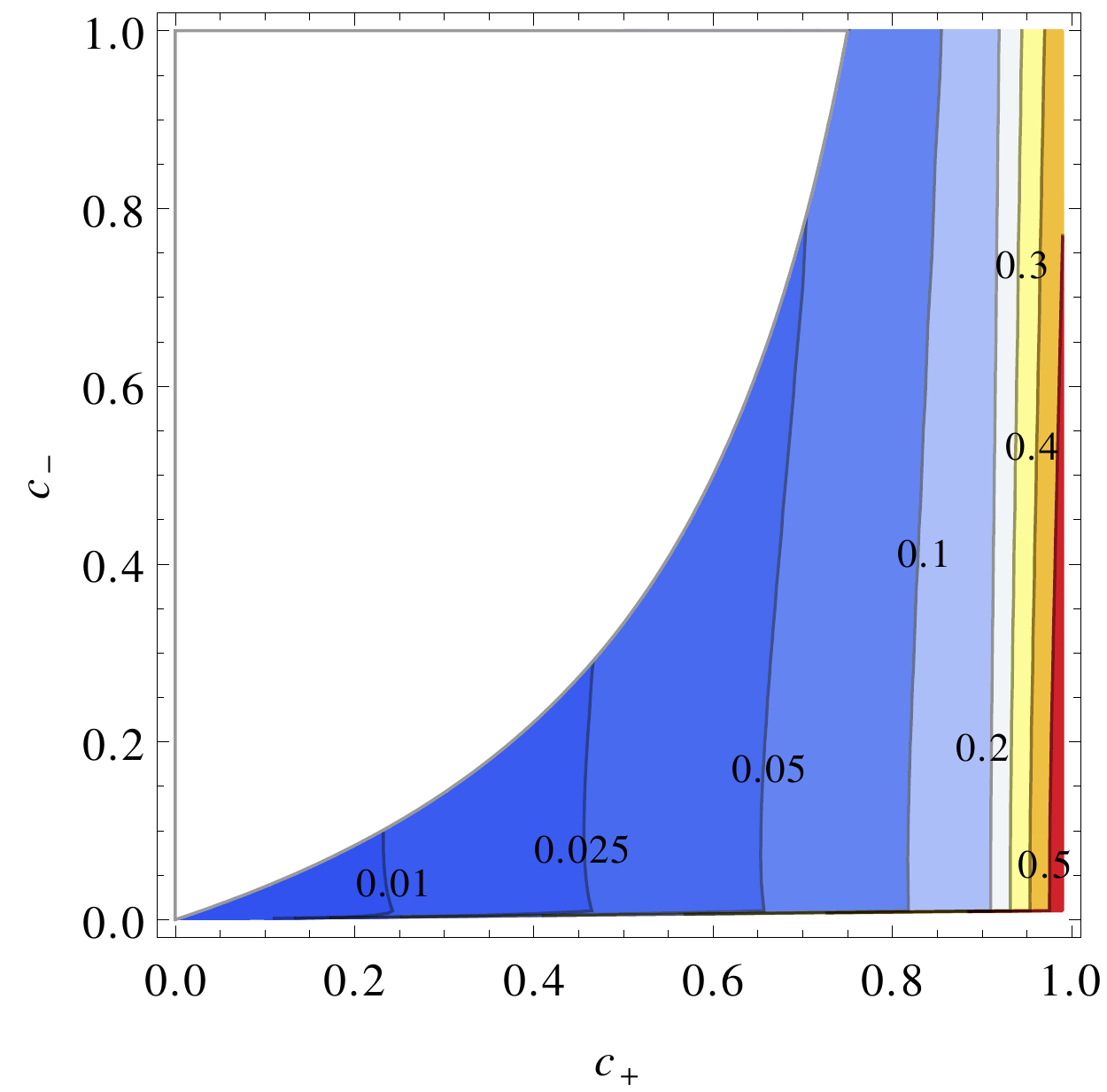}
& \includegraphics[width=6.2cm,clip=true]{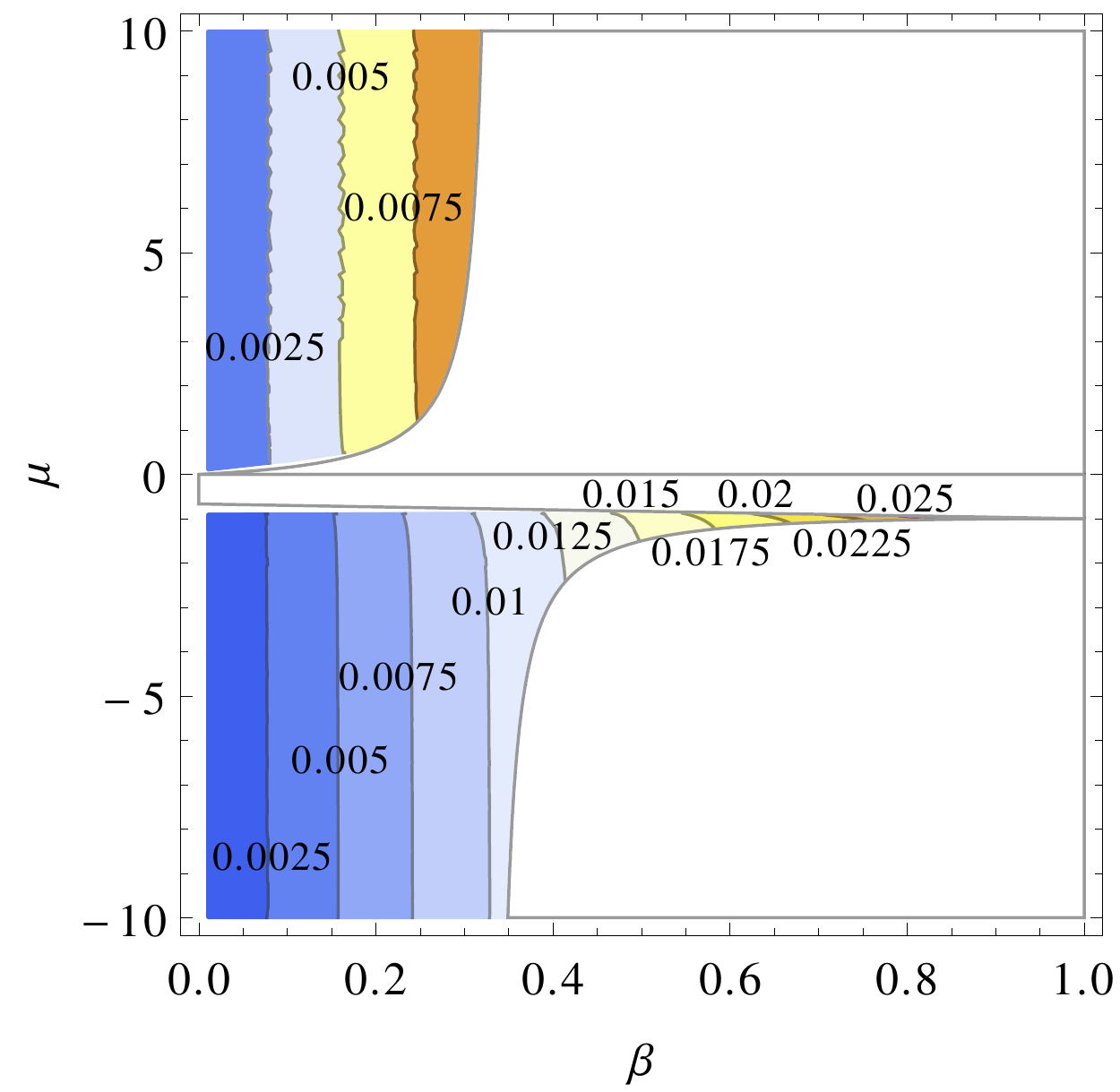}
\end{tabular}
\caption{Fractional deviation of the dimensionless combination $b_{\rm ph}/r_g$ from its GR value, in Einstein-\AE ther (left) and khronometric theory (right). 
Postive values denote larger values in in Lorentz-violating gravity than in GR.  The quantities $c_\pm$, $\mu$, $\beta$ are defined in Section~\ref{subsec:lorentz-viol}. 
[From~\cite{Barausse:2013nwa,Barausse:2011pu}.]\label{bph}}
\end{center}
\end{figure}

The resulting solutions will therefore describe the BHs of infrared Lorentz-violating gravity and present multiple horizons, as discussed above. However, in spite of
the causal structure differing from GR, the BH geometry outside the outermost horizon (i.e.~the matter one) is very similar to GR as far as
astrophysical tests are concerned. As two representative examples, Figures~\ref{isco} and~\ref{bph} show the fractional
deviation from GR of the dimensionless combinations $\omega_{_{\rm ISCO}} r_g$ and $b_{\rm ph}/r_g$, 
where $\omega_{_{\rm ISCO}}$ is the orbital frequency of the innermost stable circular orbit (ISCO) and 
$b_{\rm ph}$ is the impact parameter of the circular photon orbit. The former is measurable, at least in
principle, with GW observations of the inspiral of binary BH systems or with observations of iron-K$\alpha$ emission lines from
accretion disks, while the latter regulates the size of the BH ``shadow'' observable with future electromagnetic telescopes,
as well as the BH ringdown frequencies, in principle measurable with GW detectors.
For values of
the couplings allowed by binary pulsar observations (cf.~the purple region in Figure~\ref{fig:LVconstraints}), deviations from GR are below the percent level,
and thus outside the reach of electromagnetic observations (cf.~e.g.~\cite{Bambi:2011jq}), although probably within the reach of space-based detectors such as eLISA~\cite{Seoane:2013qna}.

\begin{figure}[tb]
\capstart
\begin{center}
\includegraphics[width=9.2cm,clip=true]{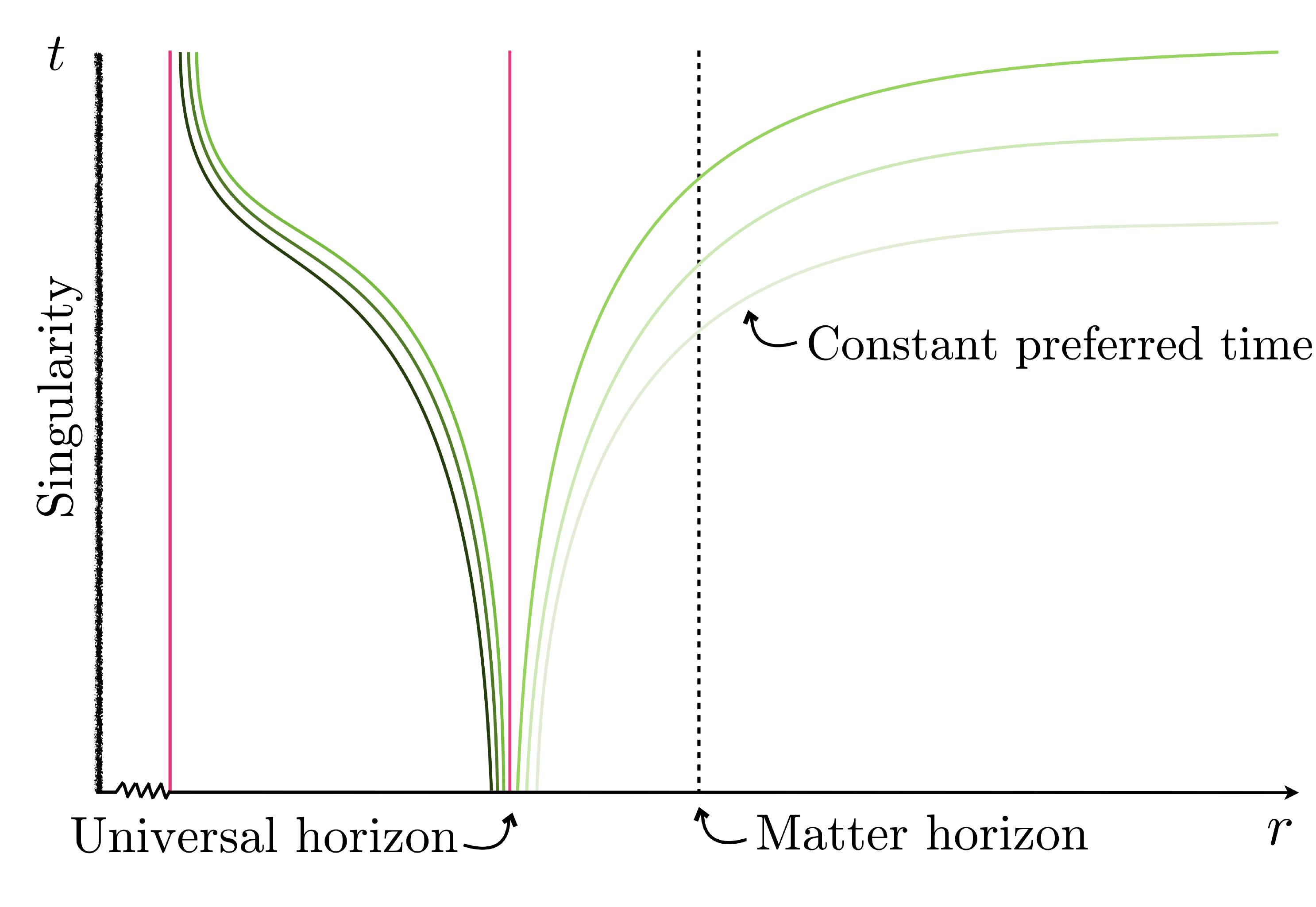}
\caption{Sketch of the causal structure of a BH possessing a universal horizon, marked with a red vertical line. 
The green curves are hypersurfaces of constant preferred time $T$ (the darker the color, the larger $T$).
The universal horizon itself is a hypersurface of constant $T$. As can be seen, a signal emitted at the universal horizon has to travel inwards, 
simply because it has to propagate in the future direction as defined by the 
preferred time $T$. Note that multiple universal horizons are generally present,
but here we are truncating the region between the first two and the central singularity.
[From~\cite{Barausse:2013nwa}.]\label{causal}}
\end{center}
\end{figure}

The behavior of these BH solutions is however very different from GR inside the matter horizon. Figure~\ref{causal} shows a
spacetime diagram that captures schematically the causal stucture of the BH solutions studied in~\cite{Barausse:2011pu}. Hypersurfaces of constant preferred time $T$ are represented by green lines that get darker and darker as the value of $T$ increases (i.e.~a darker green means that the curve
is farther in the future). Also shown (in red) are two very special hypersurfaces of constant $T$, namely ones that are also hypersurfaces of constant radius. Those 
hypersurfaces lie within the matter, spin-2, spin-0 and (when present) spin-1 horizons, and act as \textit{universal} horizons for signals of \textit{arbitrary} speed~\cite{Barausse:2011pu,Blas:2011ni,Barausse:2013nwa}. This can be
understood because a signal emitted at the universal horizon must propagate into the future, as defined by the preferred time $T$. As a result, as can be seen from Figure~\ref{causal},
such a signal must propagate inwards (i.e.~towards smaller radii). 
Note that the solutions of~\cite{Barausse:2011pu} present multiple universal horizons, but we are truncating Figure~\ref{causal}
to show just the outermost two. Also, it is clear that the concept of a universal horizon only makes sense in the presence of the \textit{ultraviolet} 
higher-order spatial derivative terms in the action,
which produce the nonlinear dispersion relation \eqref{mdsr}, thus allowing for the infinite-speed signals for which the universal horizon is relevant. 
As mentioned earlier, the BH
solutions of~\cite{Barausse:2011pu} were instead derived by solving the field equations for the \textit{infrared} limit of Lorentz-violating gravity theories. Nevertheless, the universal horizon of those
solutions lies very close to the matter horizon and far from the central singularity, and thus in a region of small curvature for the
BH masses that are relevant in astrophysics. Indeed, simple  dimensional arguments show that for astrophysical BHs the effect of the higher-order derivative terms is tiny at the location of the
universal horizon, leading to corrections $\lesssim 10^{-16} (M_{\odot}/M)^2$ ($M$ being the BH mass) away from the results obtained with the infrared limit of the theory. Universal horizons have been shown to be compatible with the
first~\cite{Berglund:2012bu} -- and possibly the
second~\cite{Cropp:2013sea} -- law of BH thermodynamics.

Finally, we stress that while the presence of such a universal horizon is  a remarkable feature of the theory, clues of its instability to nonlinear perturbations
have been reported in the decoupling limit (i.e.~neglecting the backreaction of the \ae ther/khronon on the metric) and for the low-energy limit of
Ho\v rava gravity (i.e.~khronometric theory)~\cite{Blas:2011ni}. A fully nonlinear analysis accounting for the \ae ther's/khronon's backreaction
has not been performed yet, and it is needed to draw definitive conclusions about stability. Similarly, the effect of the higher-order spatial derivative terms on the stability of the universal horizon is unknown.

\subsection{Massive gravity}
BH solutions in massive-gravity theories are still largely
unexplored. In massive bigravity theories there are asymptotically flat
solutions for which the reference metric $f_{\mu\nu}$ equals the
spacetime metric ($f_{\mu\nu}=g_{\mu\nu}$). One such family of
solutions includes the Kerr metric.

The superradiant instability responsible for hairy BH solutions in
theories of minimally coupled massive gravity
(cf.~Section~\ref{subsec:massgrav}) also destabilizes Kerr BHs in
massive gravity~\cite{Brito:2013wya}. Because of this instability,
astrophysical BH spin measurements imply that the graviton mass $\mu$
in any theory of massive gravity must be smaller than
$5\times 10^{-23}$ eV~\cite{Brito:2013wya}
(cf. Section~\ref{sec:superradiance} for a discussion of similar
constraints on the masses of ultralight scalar and vector fields).

Graviton masses $\mu$ {\it smaller} than a threshold value $\mu
M\leq0.438$ (in $G=c=\hbar=1$ units, with $M$ the BH mass) trigger
yet another instability against monopole fluctuations, that plagues
even nonrotating
BHs~\cite{Babichev:2013una,Brito:2013wya,Brito:2013yxa}. Quite
remarkably, the mass coupling $\mu M$ is well within the instability
region for values of $M$ and $\mu$ that are phenomenologically
relevant. In a cosmological context it is natural to consider the
graviton mass to be of the order of the Hubble constant, i.e.~%
$\mu\sim H\sim 10^{-33}{\rm eV}$~\cite{Hinterbichler:2011tt}. Such a
tiny graviton mass would destabilize any Schwarzschild BH with mass
smaller than $10^{22} M_\odot$!

Instabilities often signal the existence of a new family of
equilibrium solutions. Because there are no complex fields in massive
gravity, the superradiant instability presumably drives rotating BHs
to slower rotation rates. However, the monopole instability affecting
nonrotating BHs hints at the existence of a truly new family of BH
solutions~\cite{Brito:2013yxa,Brito:2013xaa}. These ``hairy''
solutions have metrics of the form
\begin{align}
g_{\mu\nu}dx^{\mu}dx^{\nu}&=-F(r)^2\, dt^2 + B(r)^{-2}\, dr^2 + r^2 d\Omega^2\,,\nonumber\\
f_{\mu\nu}dx^{\mu}dx^{\nu}&=-p(r)^2\, dt^2 + \left[U'(r)\right]^{2}/Y(r)^2\, dr^2 + \left[U(r)\right]^2 d\Omega^2\,,\nonumber
\end{align}
where $'\equiv d/dr$. Such asymptotically flat, hairy BH solutions were indeed 
found numerically in~\cite{Brito:2013xaa}; their properties depend on the 
particular theory under consideration, i.e.~on the values of the parameters 
$\alpha_3$ and $\alpha_4$ as defined in Eq.~\eqref{eq:S-dRGT}.
Notebooks to generate these hairy
solutions are available online~\cite{DyBHo:web}.

Additional BH solutions may exist when the fiducial metric is not
proportional to the spacetime metric. The only asymptotically flat
solutions found so far belong to the Kerr
family~\cite{Babichev:2014tfa,Volkov:2014ooa} and monopole
fluctuations are {\it stable} for these
configurations~\cite{Babichev:2014oua}. Stability against nonradial
modes and superradiant amplification has not been studied at the
time of writing.

\subsection{Gravity with auxiliary fields\label{BI_BF}}
As discussed in Section~\ref{subsec:auxiliary}, gravitational theories
that modify GR by adding solely nondynamical fields are
\emph{equivalent} to Einstein's theory in vacuum. In these theories,
vacuum BH solutions and their dynamics are the same as in GR. In
particular, any stationary, regular and asymptotically flat geometry
is described by the Kerr family. However, corrections to GR appear in
the coupling with matter. An interesting aspect of these theories is
that singularities may not form during
gravitational collapse~\cite{Pani:2011mg} and in early cosmology~\cite{Banados:2010ix}. 
Nonvacuum solutions -- such as charged BHs -- are generally different from
GR~\cite{Banados:2010ix}. Furthermore, similarly to the regularization
of the Coulomb field generated by a point charge in Born-Infeld
electromagnetism, the curvature singularity hosted in the BH interior
in EiBI gravity may be replaced by a regular, wormhole-like geometry
due to nonperturbative effects~\cite{Olmo:2013gqa}.

\subsection{Parametrized phenomenological deviations from the Kerr metric}

BH solutions and their properties are obviously dependent on the
theory they are derived from.  Although many theories -- some of which
were described previously -- share the Schwarzschild and Kerr geometry
as stationary solutions, even in these cases their dynamical
properties (stability, GW emission, etcetera) depend on the field content
of the theory. Unfortunately, in the context of alternatives to
Einstein's theory, the possibilities are endless. Each theory has its
own family (or families) of BH solutions, and in the absence of
observational data in the strong-field regime, choosing one's favorite
theory is largely a matter of taste.

Thus, some efforts focus on {\it parametrizing} generic spacetimes,
rather than exploring specific theories.  These efforts are in many
ways parallel to the PPN expansion, designed to parametrize
asymptotically flat spacetimes in the weak-field
regime~\cite{EddingtonBook,Will:1972zz,Nordtvedt:1972zz}.  The PPN
approach facilitates tests of the weak-field regime of GR and is
particularly well suited to perform tests in the Solar System, which
translate into constraints on alternative theories of gravity. For
example, one can show that any metric theory of gravity yielding an
asymptotically flat spacetime admits the
expansion~\cite{Will:2014xja}:
\begin{align}
 -g_{tt}&\to 1-\frac{2{M}}{r}+2(\beta-\gamma)\frac{{M^2}}{r^2}+{\cal O}(1/r^3)\,, \label{gttPPN}\\
  g_{ij}&\to \delta_{ij}\left[1+2\gamma\frac{M}{r}+{\cal O}(1/r^2)\right]\,,  \label{grrPPN}
\end{align}
where $M$ is the ADM mass, and the indices $(i,j)$ run over
asymptotically Cartesian coordinates.
The PPN parameters are very well constrained by observations,
$|\gamma-1|\lesssim10^{-5}$ and $|\beta-1|\lesssim
2.3\times10^{-4}$~\cite{Will:2014xja}. The success of the PPN approach
is rooted in the existence of an extensively studied, unique reference
metric, the Minkowski geometry. Because the metric is
post-Minkowskian, the meaning of the coordinates is clear, and so are
the physical predictions one can draw from the metric.

A comparable ``reference metric'' is lacking in the strong-field
regime. For this reason, developing a parametrized approach to
quantify deviations from GR is a nontrivial problem.  Existing
attempts 
deal with the construction of a generic parametrization of spinning
geometries which can be matched continuously onto the Kerr metric in
the strong- {\it and} in the weak-field regime. This is a formidable
task with no unique solution.
Several approaches have been proposed, each of them with their own
limitations, but all very similar in spirit (see
e.g.~\cite{Johannsen:2013rqa,Cardoso:2014rha,Rico_thesis}).  The
original ``bumpy BH'' formalism assumes Einstein's equations, and
perturbs the Kerr metric to find BHs {\it in GR} distorted by small
amounts of unspecified
matter~\cite{Collins:2004ex,Vigeland:2009pr}. The metric computed
within this approach is supposed to be valid only in vacuum. This
formalism cannot be extended in a straightforward manner to test
alternative theories of gravity (but see~\cite{Gair:2011ym} for some
improvements over the analysis of~\cite{Vigeland:2011ji}).
A similar approach was used to build ``quasi-Kerr'' spacetimes, by
expanding generic slowly rotating spacetimes up to the lowest
nontrivial quadrupole moment~\cite{Glampedakis:2005cf}. These
solutions are not regular at the horizon.  Stationary, axisymmetric
and asymptotically flat solutions of the vacuum Einstein equations
which do {\it not} describe BHs, most notably the so-called
Manko-Novikov spacetimes~\cite{1992CQGra...9.2477M}, have also been
used by several authors to model spacetimes in alternative theories
and to parametrize deviations from the Kerr
geometry~\cite{Gair:2007kr,Bambi:2011jq}.

To overcome some of the limitations of the parametrizations
above -- while introducing others -- it was recently proposed to build
on the
``Newman-Janis algorithm''\footnote{The Newman-Janis algorithm allows
  one to generate the Kerr family of BH starting from the
  Schwarzschild family~\cite{Newman:1965tw}. This approach works in GR, but is
  bound to fail in general for modified theories of gravity.} to
generate spinning BH solutions in arbitrary theories of
gravity~\cite{Johannsen:2011dh}. Using suitable choices of parameters,
these solutions consist of small deformations of the Kerr geometry. At
variance with previous studies, this approach does not assume the
validity of Einstein's equations, nor the existence of an approximate
Carter constant~\cite{Vigeland:2011ji}. Even though the procedure
makes use of the -- unjustified, because the field equations are
unknown -- Newman-Janis transformation (see e.g.~\cite{Hansen:2013owa}
for some criticism), the final transformed metric could as well be the
ad-hoc starting point for the investigation of deviations from
GR~\cite{Rico_thesis}.  Such parametrized metrics can {\it in
  principle} be suitable for tests involving observations of the
images of inner accretion flows, X-ray observations of
relativistically broadened iron lines or of the continuum spectra of
accretion disks, for which a regular behavior very close to the event
horizon is crucial~\cite{Bambi:2014sfa}.

The generalized deformed Kerr metric in this approach is~\cite{Cardoso:2014rha}
\begin{align}
g_{tt}&=-F(1+h^t), \label{eq:JP2_1}\\
g_{rr}&=\frac{(r^2+a^2\cos^2\theta) (1+h^r)}{\Delta+a^2 \sin^2\theta h^r},\label{eq:JP2_2}\\
g_{\theta\theta}&=r^2+a^2\cos^2\theta,\label{eq:JP2_3}\\
g_{\phi\phi}&=\sin^2\theta \left\{r^2+a^2\cos^2\theta + a^2 \sin^2\theta \left[2 H - F (1+h^t) \right] \right\},\label{eq:JP2_4}\\
g_{t\phi}&=-a\sin^2\theta \left[H-  F (1+h^t)\right], \label{eq:JP2_5}
\end{align}
where $F\equiv 1-2M_0r/\Sigma$, we have introduced $H\equiv\sqrt{(1+h^r)(1+h^t)}$, 
\begin{equation}
h^i(r,\theta)\equiv\sum^\infty_{k=0}\left(\epsilon_{2k}^i+\epsilon_{2k+1}^i\frac{M_0r}{\Sigma}\right)\left(\frac{M_0^2}{\Sigma}\right)^k \label{hJP}
\end{equation}
are the small deformation quantities parametrizing deviations
from the Kerr geometry in terms of dimensionless numbers $\epsilon_k^i$, and $\Sigma=r^2+a^2\cos^2\theta$, $\Delta=r^2+a^2-2M_0r$.
\begin{figure}
\begin{center}
\begin{tabular}{c}
\includegraphics[width=.7\textwidth, angle=0, clip=true]{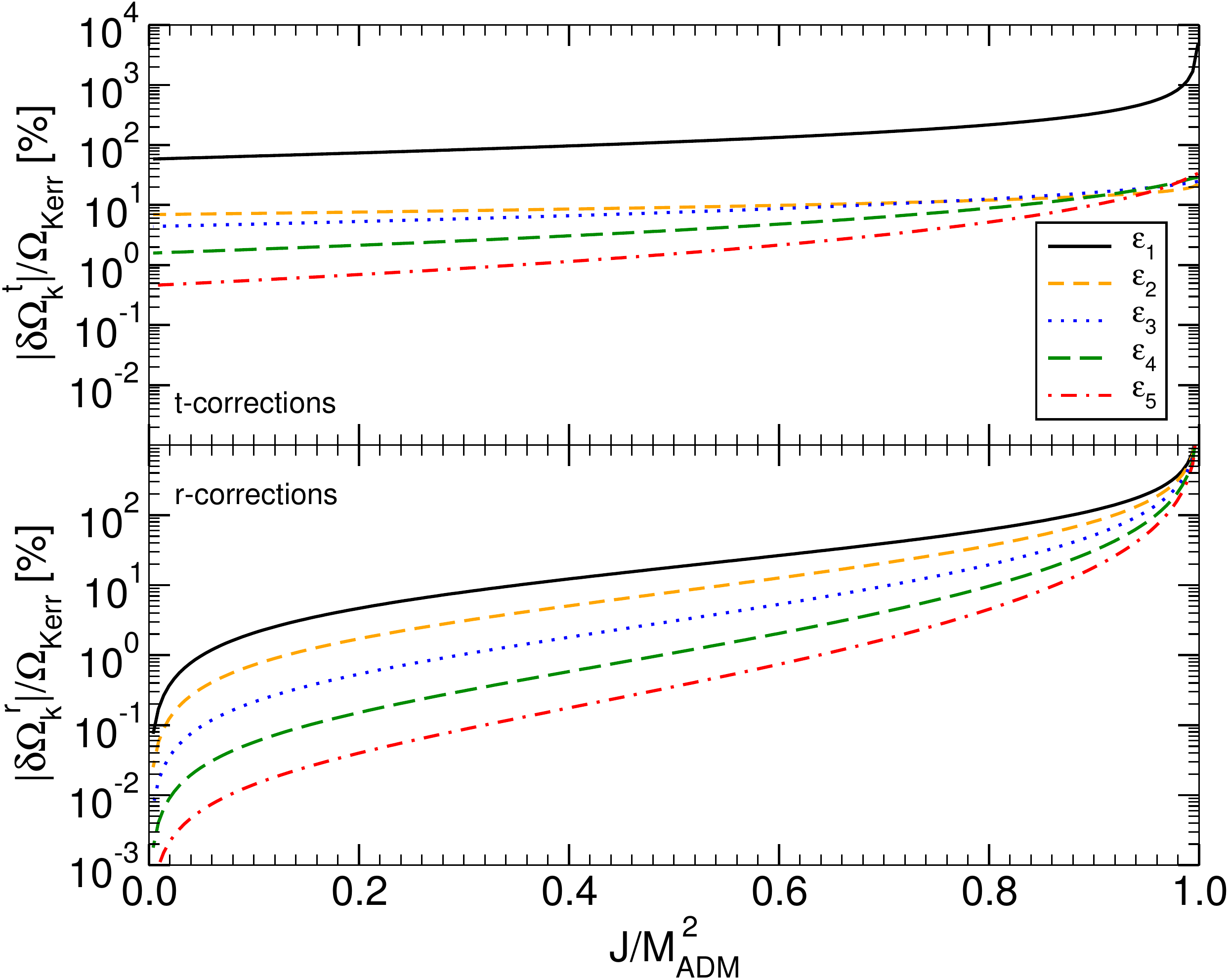}
\end{tabular}
\end{center}
\caption{
Relative corrections $\delta\Omega_k/\Omega_0$ to the ISCO frequency as a function of $J/{M^2}$ for the metric~\eqref{eq:JP2_1}--\eqref{eq:JP2_5} to linear order in $\epsilon_k\ll1$ up to $k=9$. The ISCO frequency reads $\Omega=\Omega_0+\sum_k\delta \Omega_k\epsilon_k$, where $\Omega_0$ is the ISCO frequency of a Kerr geometry. The small-coupling approximation requires $(\delta \Omega_k/\Omega_0)\epsilon_k\ll1$ for consistency.
Each $\epsilon_k-$line is built by setting to zero all other $\epsilon_i,\,i\neq k$.
The two panels refer to the corrections associated to $\epsilon_k^t$ (upper panel) and $\epsilon_k^r$ (lower panel), respectively. For ease of comparison, the range of the vertical axis is the same for both panels. In this case the total ISCO frequency reads $\Omega=\Omega_0+\sum_k\delta \Omega_k^t\epsilon_k^t+\sum_k\delta \Omega_k^r\epsilon_k^r$. The small-coupling approximation requires $(\delta \Omega_k^i/\Omega_0)\epsilon_k^i\ll1$ for consistency. [From~\cite{Cardoso:2014rha}.]
}
\label{fig:deltaOmegaISCO2}
\end{figure}

Imposing asymptotic flatness requires only
$\epsilon_0^t=\epsilon_0^r=0$, but does not imply any constraint on
$\epsilon_1^t$ and $\epsilon_1^r$. Expanding the metric
elements~\eqref{eq:JP2_1} and \eqref{eq:JP2_2} at infinity and
comparing with the PPN expansions~\eqref{gttPPN} and \eqref{grrPPN},
we can identify
\begin{align}
  {M}&=M_0\left(1-{\epsilon_1^t}/{2}\right)\,,\label{mass2}\\
  \epsilon_1^r&=-2-\gamma(\epsilon_1^t-2)\,,\\
  2\epsilon_2^t&=(\beta-\gamma)(\epsilon_1^t-2)^2+4\epsilon_1^t\,.
\end{align}
Therefore, even imposing the GR values $\beta=\gamma=1$ supported by
observations, the parameters $\epsilon_1^t$, $\epsilon_2^r$ and all
the $\epsilon_k^i$'s with $k>2$ $(i=t,r)$ are left unconstrained.
Figure~\ref{fig:deltaOmegaISCO2} shows the shifts of the ISCO frequency for the generalized metric~\eqref{eq:JP2_1}--\eqref{eq:JP2_5} in the small-$\epsilon_k^i$ limit. 
For low rotation rates the corrections associated to $\epsilon_k^t$
are larger than those associated to $\epsilon_k^r$, while the converse
is true for rapid rotation, i.e.~when $a/M\gtrsim 0.85$. An exception
to this behavior are the $\epsilon_1^i$ parameters, for which the
$t$--correction is larger than the $r$--correction for any spin.
The dominant corrections are the ones associated with $\epsilon_1^t$,
although in the fast-spinning case the corrections $\delta\Omega_k^r$
for different values of $k$ are all comparable to each other, and they
are also comparable to $\delta\Omega_1^t$. However, at least for
moderately large spin, the corrections $\delta\Omega_1^t$ and
$\delta\Omega_2^r$ are dominant. Note that both $\epsilon_1^t$ and
$\epsilon_2^r$ are currently unconstrained by observations, so that
their contribution would likely dominate the near-horizon geometry of
the deformed Kerr metric~\eqref{eq:JP2_1}--\eqref{eq:JP2_5}. A more
detailed analysis of this parametrization has recently appeared
in~\cite{Bambi:2014sfa,Bambi:2014mla}.

The approach summarized above relies on a Taylor expansion of the
unknown functions $h^t(r)$ and $h^r(r)$ in powers of
$M/r$. Continued-fraction resummations based on a compactified radial
coordinate have been recently proposed and explored for nonrotating
BHs~\cite{Rezzolla:2014mua}.

\subsection{BH mimickers}

Despite growing experimental evidence (see
e.g.~\cite{Broderick:2005xa}), at the moment an incontrovertible proof
that dark compact objects are indeed BHs (i.e.~that they possess an
event horizon or, at least, an apparent horizon) is lacking. In fact,
concerns have been raised on whether such a proof is possible at all
with electromagnetic observations~\cite{Abramowicz:2002vt}.

Our current understanding of stellar evolution strongly suggests that
even extreme forms of matter cannot support the enormous self-gravity
of massive and ultracompact objects, so that the latter are naturally
expected to be BHs. The above picture has been challenged by the
construction of exotic objects -- so-called ``BH mimickers'' -- relying
on different support mechanisms. These objects are all (almost) as
compact as BHs, but do not possess horizons. Among others, they
include \emph{boson stars}, consisting of self-gravitating massive
scalar fields~\cite{Liebling:2012fv,Macedo:2013qea}; gravitational
condensate stars or \emph{gravastars}~\cite{Mazur:2001fv}, supported
by an exotic EOS of the form $P(\rho)\approx -\rho$; and
\emph{superspinars}~\cite{Gimon:2007ur}, objects with angular momentum
exceeding the Kerr bound and with some form of matter replacing the
singular BH interior.

The key observational distinction between genuine BHs and
``mimickers'' is the presence of a surface. Experimental tests of this
property are challenging in the electromagnetic spectrum, but they should become
simpler in the context of future GW observations: the oscillation
modes of BHs have a very precise and well-known structure, which can
be tested against
observations~\cite{Berti:2005ys,Berti:2009kk,Berti:2006qt}, while the
presence of a surface will leave an imprint on the GWs generated
during the merger of two
objects~\cite{Kesden:2004qx,Macedo:2013qea,Pani:2009ss} (but see the
discussion about QNMs and ringdown modes in
Section~\ref{sec:environment} and Ref.~\cite{Barausse:2014tra}).

Some BH mimickers can be ruled out by purely theoretical arguments,
that generally rely on instabilities related to the absence
of the event horizon. Mimickers can be ruled out when these
instabilities grow on time scales much shorter than the age of the
Universe.

The theoretical foundation for the presence of these instabilities is
the work of Friedman, that showed how \emph{any spacetime with an
  ergoregion but without a horizon is linearly
  unstable}~\cite{1978CMaPh..63..243F}. The instability is due to
long-lived modes that exist for ultracompact objects whose radius is
$R\lesssim 3M$, and that might turn unstable because of the effects of
rotation~\cite{SchutzComins1978, Cardoso:2014sna}. Ultracompact
objects such as gravastars and boson stars become linearly unstable
when they possess an
ergoregion~\cite{SchutzComins1978,Cardoso:2007az}, with an instability
time scale that depends strongly on the compactness and
spin~\cite{Chirenti:2008pf}.  The same instability affects also
superspinars~\cite{Cardoso:2008kj,Pani:2010jz}.

In addition to the ergoregion instability, a new mechanism could
exclude {\it any} ultracompact ``star'' on the grounds that such an
object would be nonlinearly unstable~\cite{Keir:2014oka}. In this case
the instability is due to the existence of long-lived modes in the
linearized spectrum. These modes are trapped between the center of the
object and the light ring, and they are localized near a second, {\em
  stable} null geodesic~\cite{Cardoso:2014sna}. The long-lived modes
may become unstable under fragmentation via a
Dyson-Chandrasekhar-Fermi mechanism at the nonlinear
level. Alternatively, nonlinear interactions over their long life time
may lead to the formation of small BHs close to the stable light
ring~\cite{Cardoso:2014sna}.

If confirmed, the nonlinear instability results could soon give further support to the
BH hypothesis: the mere observation of a light ring -- a much simpler
task than the observation of the event horizon, and something that is
within the reach of upcoming
facilities~\cite{Johannsen:2015qca,Lu:2014zja,GRAVITY} -- would be
conclusive evidence for the existence of BHs.

\subsection{BHs as strong-gravity laboratories for exotic 
fields}\label{subsec-bh-exotic}
Besides being the optimal testbed for tests of GR in the strong-curvature 
regime, BHs can also be used to study exotic fields, as those appearing in 
extensions of the Standard Model of particle physics and as dark-matter 
candidates. This possibility stems from a surprising connection between 
strong-field gravity and particle physics. Although not immediately related with 
tests of GR, we conclude this chapter by discussing two examples in which the 
interplay between BHs and exotic fields is particularly dramatic.
We consider the dynamics of scalar fields in the framework of Einstein's GR but 
-- as it will be clear below -- the qualitative aspects of this analysis are 
mostly independent of the underlying theory of gravity.
\subsubsection{Collapse of self-interacting scalar fields}
\label{sec:scalarCollapse}
One of the most important phenomena in GR where the nonlinearity of
the theory plays a crucial role is that of gravitational collapse (for
a review, see~\cite{Gundlach:2007gc} and references therein). A
particularly intriguing result in this context has recently been
discovered numerically by Bizo{\'n} and
Rostworowski~\cite{Bizon:2011gg}, namely the collapse to a BH of
arbitrarily small spherically symmetric, massless scalar field
configurations in asymptotically anti-de Sitter (AdS) spacetimes.  The
AdS boundary plays a key role for the dynamics because, in contrast to
asymptotically flat spacetimes, the scalar field pulses reach spatial
infinity in finite time and get reflected back onto the coordinate
origin.  This effective {\em confinement} of the spacetime combined
with the nonlinear interaction of the wave modes results in a resonant
transfer of energy to higher frequencies, i.e.~shorter
wavelengths~\cite{Bizon:2011gg,Dias:2011ss,Buchel:2012uh} (see
also~\cite{Bizon:2013xha}). On the other hand, there exist
asymptotically AdS scalar-field solutions which do not collapse into a
BH, such as time-periodic solutions or boson
stars~\cite{Buchel:2013uba,Dias:2012tq,Maliborski:2013jca}.
Gravitational collapse in these spacetimes could also be prevented by
the formation of nonlinear bound states of {\em massive} fields. Such
bound states have been studied extensively in asymptotically flat
spacetimes~\cite{Seidel:1991zh,Seidel:1993zk,Page:2003rd,Cardoso:2005vk,Dolan:2007mj,Pani:2012vp,Pani:2012bp,Witek:2012tr,Okawa:2014nda}.

To study the possibility of
asymptotically flat spacetimes being unstable in the context of
confinement mechanisms~\cite{Choptuik:1992jv,Brady:1997fj}, consider the action
\begin{equation}
  S = \int d^4x \sqrt{-g}
      \left( \frac{R}{16\pi}
            -\frac{1}{2}\partial^{\mu}\varphi\, \partial_{\mu}\varphi
            -\frac{1}{2}\mu^2\varphi^2 \right)\,.
  \label{eq:actionOkawa}
\end{equation}
This choice corresponds to the special case $V(\varphi)/16\pi =
\frac{1}{2}\mu^2 \varphi^2$ and to a one-dimensional (hence flat)
target space in Eq.~\eqref{STactionE}, i.e., to a minimally coupled
massive scalar field of mass $\mu$.  Applying the
Arnowitt-Deser-Misner (ADM) decomposition to the field equations
resulting from (\ref{eq:actionOkawa}) for the special case of
spherical symmetry, one obtains two constraint equations and a set of
evolution equations which are given explicitly in Eqs.~(3), (4)
of~\cite{Okawa:2013jba}.  Initial data are constructed by analytically
solving the constraint equations for a Gaussian scalar pulse in a
Minkowski background.
\begin{figure}
\capstart{}
  \centering
  \includegraphics[width=5.cm]{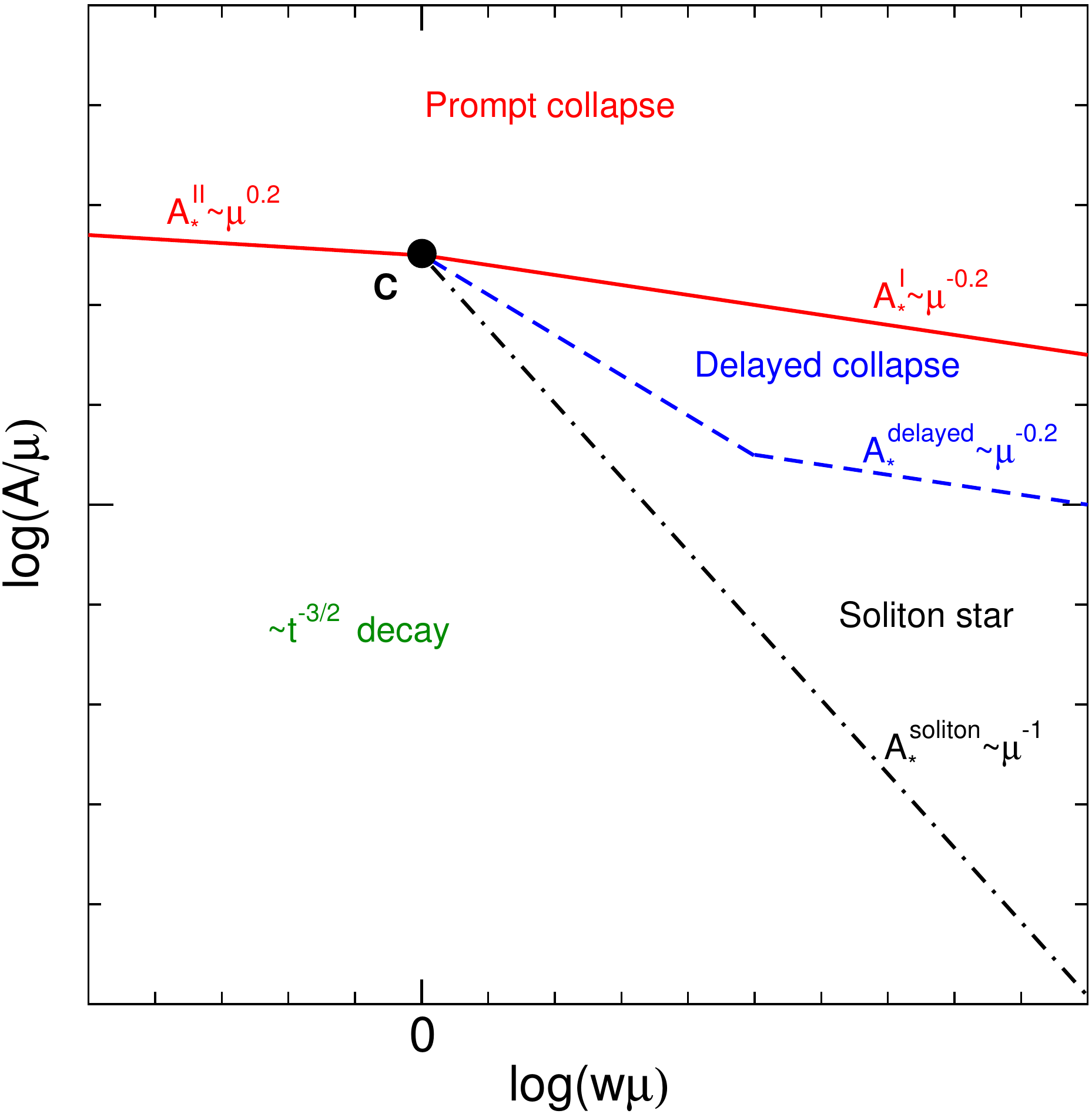}
  \includegraphics[width=7.5cm]{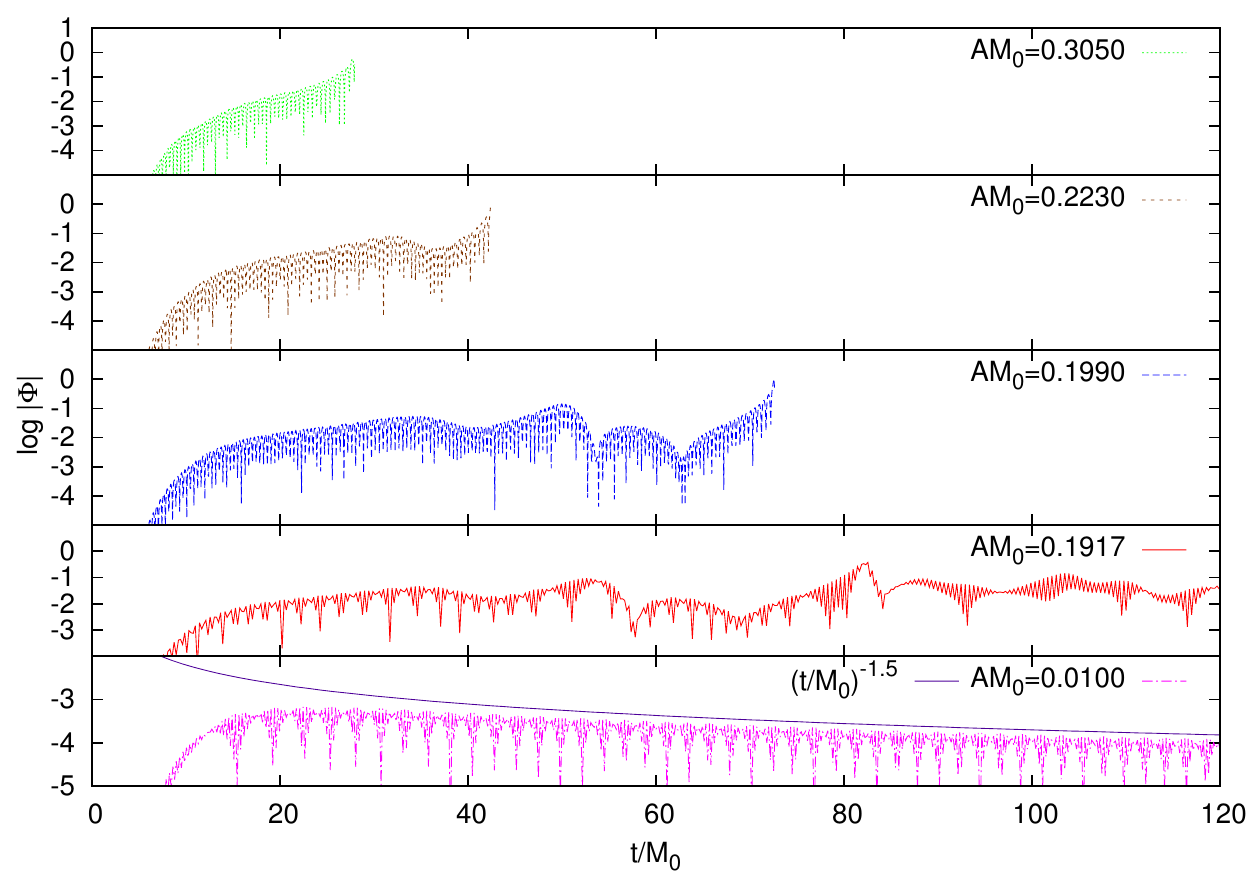}
  \caption{Left panel: Qualitative phase diagram for the spherically symmetric
           collapse of a massive scalar field in the amplitude $(A)$
           vs. width $(w)$ plane. Right panel: The scalar field amplitude
           at the coordinate origin as a function of time is shown for
           selected values of the amplitude $A$. $M_0$ denotes the
           ADM mass of the spacetime. [Adapted from~\cite{Okawa:2013jba}.]
          }
  \label{fig:okawa}
\end{figure}
The results of the numerical time evolutions are summarized in
Figure~\ref{fig:okawa}. The left panel of the figure shows the phase
diagram for the collapse of a massive scalar field in the
$w\mu$-$A/\mu$ plane, where $w\mu$ and $A/\mu$ represent the initial
pulse width to Compton wavelength ratio and the pulse amplitude,
respectively.  For small values of the width, the phase diagram
reveals a behavior similar to the massless case: collapse above a
threshold initial amplitude, and dispersion below that threshold. For
large $w\mu$, however, the phase diagram exhibits a much richer
phenomenology. This is also demonstrated in the right panel of the
figure, where the scalar field amplitude at the coordinate origin is
shown as a function of time for several configurations with varying
initial amplitude. For large initial amplitude the scalar field
collapses promptly to a BH, irrespective of whether a mass term
$U(\phi) = \frac{1}{2}\mu^2 \phi^2$ is included in the action or
not. For smaller amplitudes, however, this mass term leads to a
delayed collapse, similar to that observed in the AdS case. The mass
term introduces an effective confinement of the scalar field, which
gets reflected off the potential barrier instead of escaping to
infinity, and thus collapses after some number of reflections: cf.~the
top three plots in the right panel.  Additionally, there exist
meta-stable, long-lived oscillations (second lowest plot in the right
panel) which occur for smaller values of the initial
amplitudes. Finally, for very small $A$, the scalar field decays
(bottom plot).

In summary, the numerical evolutions demonstrate that for a massive
scalar field coupled to gravity there exist nonlinear bound states,
i.e.~meta-stable oscillations~\cite{Seidel:1991zh,Page:2003rd,Grandclement:2011wz,Okawa:2013jba}. This
implies that the mass term of the scalar field can lead to a
confinement-induced gravitational collapse similar to the AdS case,
but that in the asymptotically flat case, energy can escape to some
extent from the potential, which results (for some values of the
initial parameters) in a bound state, rather than BH formation through
gravitational collapse. Indeed, a rather generic class of arbitrarily
small initial data evolving in a totally confined geometry seems to be
generically unstable to BH formation~\cite{Okawa:2014nea}. If
confirmed, this result might have interesting implications for the
nonlinear stability of compact stars, whose fluid perturbations are
effectively confined within the stellar surface.

\subsubsection{Superradiant instabilities: black holes as observatories for beyond-standard-model physics}
\label{sec:superradiance}
One of the main reasons why BHs represent interesting
laboratories for the exploration of the properties of light bosonic fields is 
the superradiant instability of spinning 
BHs~\cite{zeldovich1,Press:1972zz,Cardoso:2004nk} (for a recent exhaustive 
overview on the subject, see Ref.~\cite{Brito:2015oca}). As previously 
discussed, superradiance
occurs in the interaction between BHs and fundamental fields
with frequencies $\omega \le m \Omega_{\rm H}$, where $\Omega_{H}$ is the
angular velocity of the BH horizon, and $m$ denotes the
azimuthal mode number. The interaction provides a {\em classical}
mechanism to reduce the mass and spin of the rotating hole as the
field taps into the rotational energy and gets amplified. A gedanken
experiment first proposed by Press and Teukolsky~\cite{Press:1972zz}
suggests that a superradiant system can be rendered unstable by a
run-away amplification of the field if it is immersed inside a
reflective cavity: this is a superradiant instability, or ``BH
bomb.''

Such a configuration naturally arises for the case of massive fields,
as in the action~\eqref{eq:actionOkawa}: as discussed above, the mass
term $\mu$ leads to a potential barrier, and thus acts as a
``mirror''~\cite{Damour:1976kh,Detweiler:1980uk,Zouros:1979iw}.  In
that case, modes with $\omega \lesssim \mu$ are trapped inside the
potential well, and the rotating BH can become (superradiantly)
unstable against these modes. The growth rate of the BH-bomb
instability is regulated by the coupling between the field's mass
$\mu$ and the BH mass $M$, and it is strongest when these parameters
satisfy the condition $M \mu \sim \mathcal{O}(1)$. In other words, the
interaction is maximized when the Compton wavelength of the field is
comparable to the size of the BH.  More specifically, perturbative
calculations predict that the strongest growth rates of scalar fields
surrounding a BH with dimensionless spin parameter $a/M \sim 0.99$ are
realized for the dipole mode with a coupling $M\mu\sim0.42$. The
$e$-folding times in this case are $\tau \sim 50
M/M_{\odot}$~\cite{Dolan:2007mj,Cardoso:2005vk}. That time
scale can decrease by several orders of magnitude if we consider
massive vector fields~\cite{Rosa:2011my,Pani:2012vp,Witek:2012tr} or
gravitons~\cite{Brito:2013wya}.

While the BH bomb mechanism is negligible for known composite or
fundamental scalar particles interacting with astrophysical BHs (the
mass coupling is $M \mu \ge 10^{18}$, yielding time scales longer than
the age of the Universe), it can play a significant role if the
field's mass is $10^{-22}~{\rm eV} \le \mu \le 10^{-8}~{\rm eV}$, as
might be the case for dark-matter candidates, ultra-light
axions~\cite{Arvanitaki:2009fg,Peccei:1977hh} or fundamental fields in
modified gravity
theories~\cite{Sotiriou:2008rp,Clifton:2011jh,Yunes:2013dva}. Given
that the superradiant amplification provides a mechanism to reduce the
energy and spin of a BH, one can argue that BHs with certain
parameters $(M,\,a/M)$ should not exist if they interact with fields
of mass $\mu$. Conversely, the observation of
BHs~\cite{McClintock:2009as,Reynolds:2013qqa,AmaroSeoane:2012km}
within these exclusion regions allows us to constrain the allowed
field masses. This effect has indeed been used to constrain the mass
of Proca fields by comparing the superradiant instability time scale
with the Salpeter time scale, which gives (roughly) the time it takes
to spin up and feed a BH through accretion. The bound obtained from
this study constrains the mass of a hypothetical light vector field,
$\mu_{\gamma} \le 10^{-20}~{\rm eV}$~\cite{Pani:2012vp}.  Strictly
speaking, superradiant instabilities only exclude mass intervals
(superradiance is ineffective at large boson
masses~\cite{Brito:2015oca}), but the quoted upper limit on the vector
field mass takes into account previous constraints obtained by other
means.  In principle this bound also applies to a hypothetical massive
photon, but in this case it may be necessary to model the interaction
of the photons with the surrounding accretion disk and plasma.

More solid bounds are in place for massive
gravitons~\cite{Brito:2013wya}, which are only weakly coupled to
matter. The superradiant instability under massive spin-2
perturbations is the strongest instability of the Kerr metric known to
date and, together with observations of rapidly spinning supermassive
BHs, imposes the constraint $\mu_g\lesssim5\times10^{-23}{\rm eV}$ on
the mass $\mu_g$ of the graviton~\cite{PDG}.

\begin{center}
\begin{table}[t]
 \begin{tabular}{ccccc}
  Field	& & Bounds &	& Reference \\
  \hline  
  \hline  
  Scalar & $\mu \lesssim 5\times 10^{-20} {\rm eV} $ & $\cup$ & $ \mu \gtrsim 10^{-11} {\rm eV}$ & \cite{Arvanitaki:2010sy,Arvanitaki:2014wva} \\
  Vector & $\mu_\gamma \lesssim 5\times 10^{-21} {\rm eV}  $ & $\cup$ & $ \mu_\gamma \gtrsim 10^{-11} {\rm eV}$ & \cite{Pani:2012vp} \\
  Tensor & $\mu_g \lesssim 5\times 10^{-23} {\rm eV}  $ & $\cup$ & $ \mu_g \gtrsim 10^{-11} {\rm eV}$ & \cite{Brito:2013wya} \\
  \hline  
 \end{tabular}
 \caption{Current bounds on the mass of ultralight bosonic degrees of  
   freedom arising from BH superradiant instabilities within a  
   linearized approximation  (cf.~\cite{Brito:2015oca} for details).}\label{tab:SR}
\end{table}
\end{center}
 
Table~\ref{tab:SR} summarizes the current bounds on the mass of
ultralight bosonic degrees of freedom arising from BH superradiant
instabilities (cf.~\cite{Brito:2015oca} for details).

These results follow from perturbative calculations, and leave various
questions unanswered. (i) What is the fate of the system if we include
back reaction? Does the instability persist or is the system driven
towards a stable regime? (ii) Is it possible to form in this manner a
``gravitational atom'' or ``pulsar,'' as has been suggested
in~\cite{Arvanitaki:2009fg,Arvanitaki:2010sy}? (iii) What would be the
observational signatures, including modifications of GW signals and
the radiation emitted by the field itself?

Reference~\cite{Brito:2014wla} has recently addressed these questions
by performing a quasi-adiabatic, fully relativistic evolution of the
superradiant instability of a Kerr BH including GW emission and gas
accretion. It turns out that GW emission does not have a significant
effect on the evolution of the BH, although it contributes to
dissipate the dipolar bosonic cloud that forms as a result of the
instability. The mass of the cloud can be a sizeable fraction of the
total BH mass, but its energy density is very low, because the cloud
typically extends over very large distances. This implies that
backreaction effects are always negligible: even in the presence of
effective bosonic ``hair'' (both for real and for complex fields), the
geometry remains close to Kerr. Thus, the prospects of imagining
deviations from Kerr due to superradiantly produced bosonic clouds in
the electromagnetic band are low, but such systems are a primary
source for observations aiming at testing the Kerr hypothesis through
GW detection.
Finally, the role of gas accretion is very important. On the one hand,
accretion competes against superradiant extraction of mass and angular
momentum. On the other hand, accretion might produce the optimal
conditions for superradiance, for example by increasing the BH spin
before the instability becomes effective or by increasing the
superradiant coupling $M \mu$.

In order to verify the theoretical bounds on the existence of light bosons~\cite{Arvanitaki:2010sy,Kodama:2011zc,Pani:2012vp,Brito:2013wya}, a relevant problem concerns the \emph{final} BH state at the time of observation in realistic situations. In other words, given the observation of an old BH and the measurement of its mass and spin, would these measurements be compatible with the evolution driven by superradiance, accretion and GW emission?
\begin{figure}[t]
\begin{center}
\begin{tabular}{c}
\includegraphics[width=0.7\textwidth]{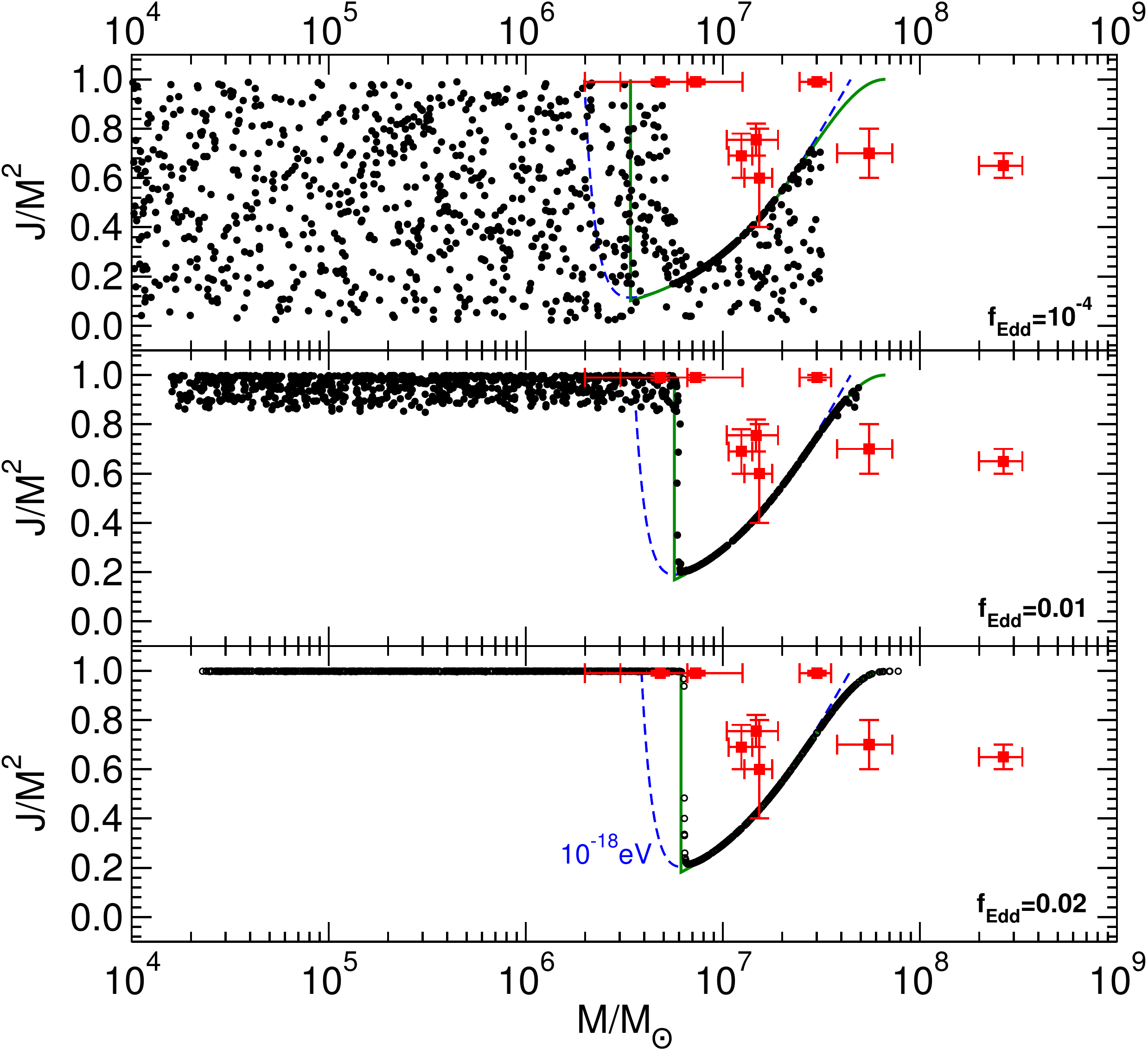}
\end{tabular}
\end{center}
\caption{\label{fig:ReggeMC} The final BH mass and spin in the Regge
  plane~\cite{Arvanitaki:2010sy} for initial data consisting of
  $N=10^3$ BHs with initial mass and spin randomly distributed between
  $\log_{10}M_0\in[4,7.5]$ and $J_0/M_0^2\in[0.001,0.99]$. The BH
  parameters are then extracted at $t=t_F$, where $t_F$ is distributed
  on a Gaussian centered at $\bar t_{F}\sim 2\times 10^9{\rm yr}$ with
  width $\sigma=0.1\bar t_{F}$. As an example we considered
  $\mu=10^{-18}{\rm eV}$, but similar results hold for other
  masses. The dashed blue line is the prediction of the linearized
  analysis obtained by comparing the superradiant instability time
  scale with the accretion time
  scale~\cite{Arvanitaki:2010sy,Pani:2012vp,Brito:2013wya}, whereas
  the solid green line is a new prediction computed
  in~\cite{Brito:2014wla}. Old BHs do not populate the region above
  the green threshold curve, especially for high accretion rates. The
  experimental points with error bars refer to the supermassive BHs
  listed in~\cite{Brenneman:2011wz}. [From~\cite{Brito:2014wla}.]  }
\end{figure}
This problem is addressed in Figure~\ref{fig:ReggeMC}, which shows the
final BH mass and spin in the Regge plane~\cite{Arvanitaki:2010sy}
(i.e.~a BH mass-spin diagram) for $N=10^3$ Monte Carlo evolutions for
a scalar field mass $\mu=10^{-18}{\rm eV}$. We consider three
different accretion rates $f_{\rm Edd}$ (defined as the fraction of
mass accretion rate relative to the Eddington limit) and, in each
panel, we superimpose the bounds derived from the linearized analysis,
i.e.~the threshold line when the instability time scale equals the
accretion time scale. As a comparison, in the same plot we include the
experimental points for the measured mass and spin of some
supermassive BHs listed in~\cite{Brenneman:2011wz}. These results
confirm that a very solid prediction of the existence of ultralight
bosons is the appearance of ``holes'' in the Regge
plane~\cite{Arvanitaki:2010sy}, i.e.~regions of the BH mass-spin
diagram which should not be populated by old BHs. We refer
to~\cite{Brito:2014wla} for a detailed discussion.

A quasi-adiabatic evolution is well suited to studying superradiant
instabilities because of the existence of two very different
scales~\cite{Brito:2014wla}. One is dictated by the oscillation time
$\tau_S\sim1/\mu$, the other by the instability growth time scale,
$\tau\gg\tau_S$. In the most favorable case for the instability (that
of a massive scalar field), $\tau\sim 10^6\tau_S\sim 10^6 M$ is the
minimum evolution time scale required for the superradiant effects to
become noticeable.  Thus, fully numerical simulations that capture the
effects of the instability are extremely challenging to perform.

\begin{figure}[t]
\capstart{}
  \centering
  \includegraphics[width=0.49\textwidth]{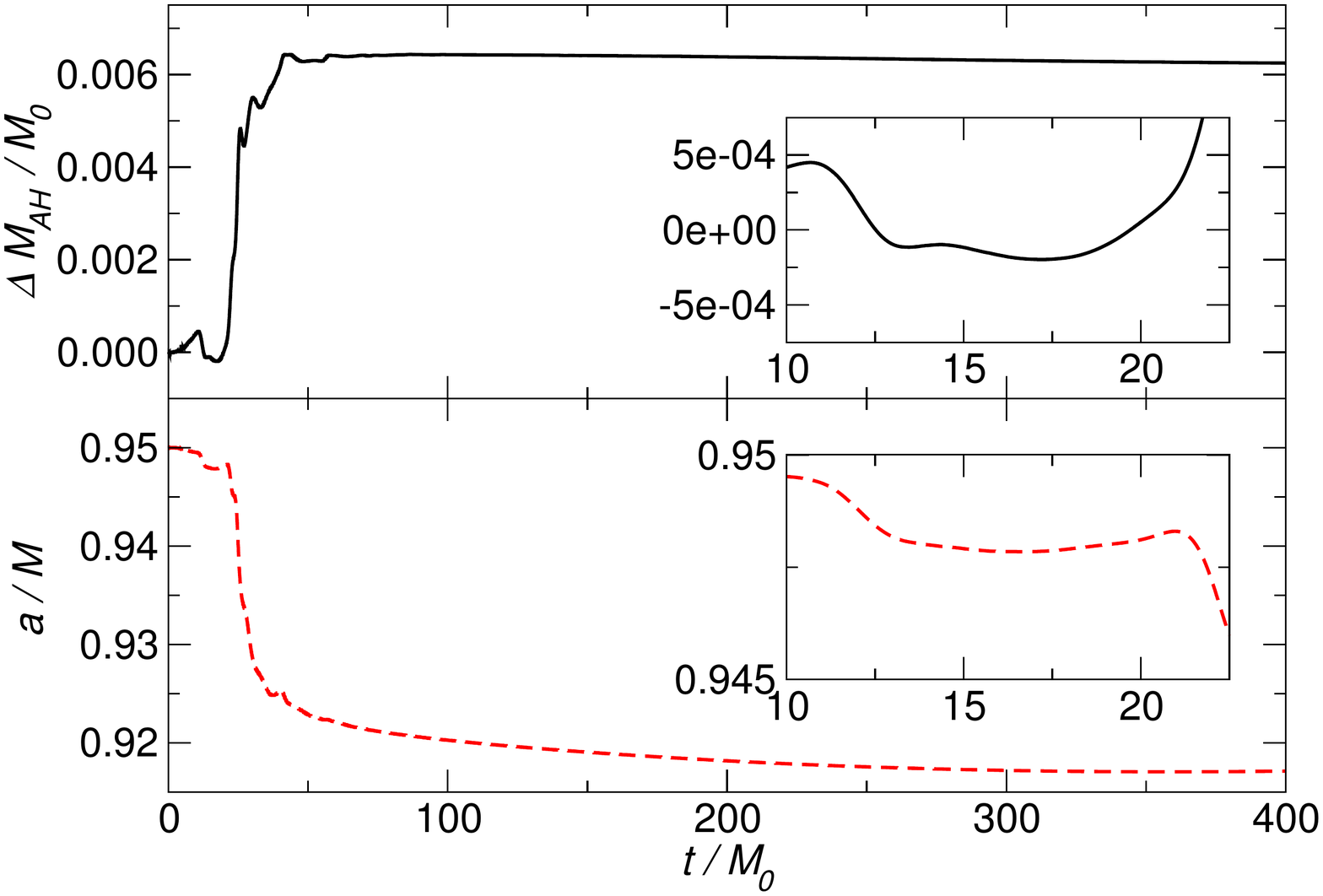}
  \includegraphics[width=0.49\textwidth]{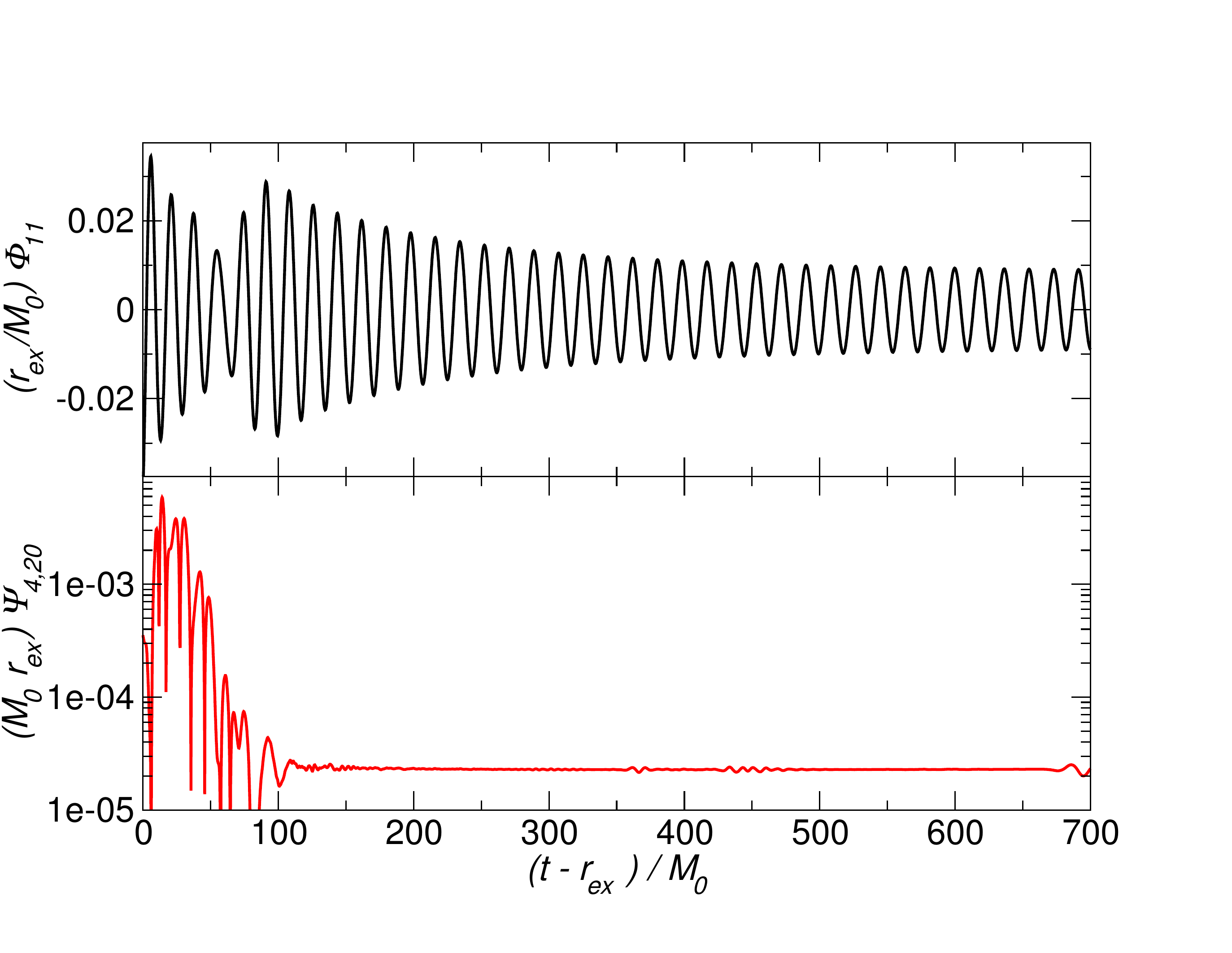}
  \caption{\label{fig:NonLin} Nonlinear evolutions of a scalar cloud
    with $M\mu=0.3$ around a BH with initial spin $a/M=0.95$.  Left:
    fractional variation in the BH mass (top) and dimensionless spin
    of the hole (bottom). At early times (shown in the inset), both
    quantities {\em decrease} hinting at superradiance scattering. At
    later times, the superradiance condition is no longer satisfied,
    resulting in an increase of the BH mass due to accretion of the
    scalar cloud.  Right: dominant scalar (top) and induced
    gravitational waveforms (bottom).  [From~\cite{Okawa:2014nda}.]}
\end{figure}

The first fully nonlinear evolutions of massive scalars coupled to
Kerr BHs~\cite{Okawa:2014nda} have found evidence for superradiant
scattering in the early stages of the interaction, which has also been
observed in simulations modeling the infall of GWs into a rotating
BH~\cite{East:2013mfa}. After this scattering, however, the spin of
the BH decreases so much that the system is driven out of the
superradiant regime, and slow accretion of the scalar field dominates
the ensuing evolution.\footnote{Note that these evolutions focused
  only on an isolated BH-scalar field system, neglecting the effects
  due to the presence of ordinary matter or accretion disks.}  These
features are illustrated in the left panel of
Figure~\ref{fig:NonLin}. Because of the mass term and the resulting
potential barrier the scalar field is trapped inside a region near the
BH, and therefore forms a scalar cloud which continuously leaks into
the rotating hole. This induces long-lived scalar and gravitational
radiation: cf.~right panel in Figure~\ref{fig:NonLin} (as well as the
animations available online at~\cite{DyBHo:web}). Although the
gravitational radiation itself is not sensitive to the mass potential
barrier, the continued interaction between the BH and the scalar cloud
excites a GW signal with about twice the scalar-field frequency, which
follows the beating pattern exhibited in the scalar
modes~\cite{Witek:2012tr,Okawa:2014nda}. Moreover, the frequencies in
both the scalar and GWs are of the order $f\sim
\mathcal{O}(10)~{\rm kHz} (M/M_{\odot})^{-1}$, implying that the
signals generated by stellar-mass or supermassive BHs are potentially
observable with the Advanced LIGO/VIRGO network~\cite{Aasi:2013wya} or
future space-based detectors such as eLISA~\cite{AmaroSeoane:2012km},
respectively (see~\cite{Arvanitaki:2014wva} for recent work on GW
signatures of bosonic clouds around BHs).

In addition to interesting constraints on particle masses,
superradiant mechanisms have the potential to test the existence and
geometry of extra dimensions~\cite{Rosa:2012uz}.  Finally, the
possibility of indirect observation of superradiance in a BH-pulsar
system has recently been proposed. The idea is that the pulsar's GW
and electromagnetic luminosities may exhibit a characteristic
modulation due to superradiant scattering that depends on the pulsar
position relative to the BH~\cite{Rosa:2015hoa}. If observed, this would
be the first -- albeit indirect -- observation of rotational
superradiance, which forms the basis of all the superradiant
instabilities discussed here~\cite{Brito:2015oca}.

\subsection{Open problems}\label{op:BHs}
Here we give a (necessarily biased) list of open problems regarding BH
physics in the context of tests of gravity:
\begin{itemize}
 \item The stability of the hairy BH solutions found
   in~\cite{Herdeiro:2014goa}, as well as
   their formation mechanism in astrophysical scenarios, remain an urgent open 
issue.
 \item In the context of stationary solutions, BHs in dCS gravity are
   known only for low spin and were obtained analytically within a
   perturbative scheme. Highly spinning dCS BHs have not been
   constructed numerically yet.
 \item The stability of BHs in quadratic gravity has been only
   partially investigated. Notable missing investigations include:
   polar gravitational perturbations of static EdGB BHs, any
   perturbations of slowly rotating EdGB BHs, and the gravito-scalar
   perturbations of slowly rotating dCS BHs. The latter might be
   relevant in the context of Ostrogradski instabilities and the
   well-posedness of the dCS gravity.
 \item Similarly, the stability of BHs in Lorentz-violating theories
   under gravitational perturbations has not been studied, even in
   the static case. Such an analysis might be interesting in the
   context of stability of the universal horizons.
 \item The phenomenology of BHs in Horndeski theory, Galileon theory and
   massive gravity has not been studied in detail yet. In theories
   admitting BH solutions other than Kerr, it would be interesting to
   understand whether such solutions are formed as the result of
   gravitational collapse.
 \item Despite various attempts, a solid strong-field parametrization
   of spinning BHs in generic modified gravity is not yet available.
 \item The endpoint of the superradiant instability of spinning BHs
   triggered by massive bosons in full GR is unknown, due to the
   long time scales of the problem. Nonlinear effects, such as
   bosenova collapse~\cite{Yoshino:2012kn}, should be taken into
   account in numerical simulations.
\end{itemize}

\clearpage
\section{Neutron stars}
\label{sec:NSs}

Our discussion of BH solutions in modified theories of gravity shows
that there is a significant problem in testing GR with present and
future astrophysical observations: the Kerr metric is a solution of
the field equations in many proposed alternatives to GR, and even when
it is not, deviations from the structure and dynamics of the Kerr
metric are expected to be suppressed by some (typically small)
coupling constant. Unlike BHs, the structure of compact stars depends
on the coupling of gravity with matter in strong-field
regions. Therefore NSs are a valuable alternative to BHs in tests of
strong-field gravity, because they can probe (and possibly rule out)
those theories that are close to GR in vacuum, but differ in the
description of the coupling between matter and gravity. In fact, the
enormous gravitational field of NSs, the high density of matter at
their cores and the existence of pulsars with fast spin and large
magnetic fields make them ideal laboratories to study all fundamental
interactions~\cite{Lattimer:2004pg,Lattimer:2006xb,Ozel:2010fw,Steiner:2010fz,Steiner:2012xt,Hebeler:2013nza,Psaltis:2013fha}.

The study of compact stars in alternative theories of gravity has a
long history. In this chapter we give a concise overview of what we
consider to be the state of the art, and for each specific theory we
present a catalog of all\footnote{One exception are the so-called NS
  ``sensitivities'': extra charges that the NS can acquire in modified
  gravity due to the violation of the strong equivalence
  principle. These are discussed in Section~\ref{sec:sensitivities},
  in preparation for our review of binary dynamics in modified gravity
  (Chapter~\ref{sec:CB}).} the NS observables that have been computed
at the moment of writing. Table~\ref{tab:NSsummary} is meant to
provide a practical guide to the literature. A glance at the table
shows that while NS equilibrium configurations have been explored in
many classes of modified theories of gravity, their stability and
their dynamical properties are still largely unknown.

\subsection{General-relativistic stellar models} \label{sec:NS_GR}

Before discussing equilibrium stellar solutions in modified gravity,
it is convenient to present a brief summary of the basic properties of
relativistic stars in Einstein's theory (we refer the reader
to~\cite{Stergioulas:2003yp,Shapiro:1983du,FriedmanStergioulas} for
excellent treatments of the subject).

In GR, the equilibrium of spherically symmetric, static (nonrotating)
stars is governed by the Tolman-Oppenheimer-Volkoff (TOV) equations,
that follow from
Einstein's equations for a perfect-fluid stress energy tensor.  When
supplemented with an equation of state (EOS) relating the fluid's
density and pressure, Einstein's equations form a closed system of
ordinary differential equations. The NS EOS encodes the
thermodynamical behavior of matter in the extreme conditions
prevailing in the NS interior. Despite recent progress (see
e.g.~\cite{Lattimer:2004pg,Lattimer:2006xb,Ozel:2010fw,Steiner:2010fz,Steiner:2012xt,Hebeler:2013nza,Psaltis:2013fha}),
the EOS is still largely unknown at the supranuclear densities
characterizing the NS core.

The solutions to the TOV equations are obtained, in general, by
numerical integration. They form a single-parameter family, where the
parameter labeling different solutions can be chosen to be (say) the
central density. Relativistic equilibrium configurations are
characterized by a maximum mass and a maximum compactness that depend
on the EOS. Uncertainties in the EOS translate into uncertainties in
the NS mass-radius relation. For example, for a typical NS mass $M\sim
1.4 M_\odot$, EOSs compatible with our current knowledge of nuclear
physics predict radii ranging from $\sim$6 to $\sim$16
kilometers~\cite{Steiner:2012xt}.

Generic rotating stellar models are more difficult to
construct. However, ``old'' NSs are expected to rotate rather slowly,
unless they are spun up by accretion from a companion; therefore,
perturbative calculations using a slow-rotation expansion are reliable
in many situations of astrophysical interest. The formalism to
construct slowly rotating NS models, developed in the seminal work by
Hartle and Thorne~\cite{Hartle:1967he,Hartle:1968ht}, has been pushed
up to fourth order in rotation~\cite{Yagi:2014bxa}. Various works~\cite{Berti:2004ny,Benhar:2005gi,Yagi:2014bxa} have shown that the
equilibrium properties of slowly rotating solutions compare favorably
with numerical codes that solve Einstein's equations in full
generality to construct models of relativistic stars with arbitrary
rotation rates (cf.~\cite{Stergioulas:2003yp,FriedmanStergioulas} for
reviews).

Linear perturbations of stellar configurations are complex and interesting,
even for static objects. Because of GW emission, the modes of
relativistic NSs (just like the modes of BHs) have a dissipative
component, i.e.~they are QNMs. In addition to the
standard fluid modes, that have well studied counterparts in the
Newtonian limit~\cite{1989nos..book.....U}, compact stars also possess
characteristic modes of oscillation (the so-called $w$-modes)
associated to pure spacetime perturbations, rather than fluid
displacements
(see~\cite{Kokkotas:1999bd,Ferrari:2007dd,Andersson:2009yt,FriedmanStergioulas}
for reviews). These $w$-modes are similar in nature to BH QNMs.
Therefore NS QNMs carry information about the
stellar geometry (as in the BH case), but in addition they can also be
used to infer the properties of the NS EOS. In fact, one of the main
scientific goals of third-generation Earth-based GW detectors is their
potential to fulfill the promise of ``GW
asteroseismology''~\cite{Andersson:1997rn,Benhar:2004xg,Andersson:2009yt}:
accurate GW measurements of the oscillation frequencies would allow us
to reconstruct the properties of the NS (something that is routinely
done in helioseismology) and therefore to constrain nuclear physics in
regimes that are out of reach for laboratory experiments. Rotating
compact stars are characterized by various instabilities (most
notably, the Chandrasekhar-Friedman-Schutz
instability~\cite{Chandrasekhar:1992pr,Friedman:1978hf} and the
related r-mode instability~\cite{Andersson:2000mf}) that have
important implications for their spin rate and evolution. This topic
is largely unexplored in modified gravity, but there is a very large
body of work on these instabilities and their implications in GR~\cite{Kokkotas:1999bd,Andersson:2006nr,Andersson:2009yt,FriedmanStergioulas}.

\subsection{Scalar-tensor theories} \label{sec:NS_ST}
It should come as no surprise that most of the work on NSs in modified
theories of gravity has focused on the simplest and arguably most natural extensions of GR,
namely scalar-tensor theories. In these theories, the properties of
static and spinning NSs and of their oscillation modes are well
understood.

The modified TOV equations of hydrostatic equilibrium in Brans-Dicke
theory were first studied by Salmona~\cite{Salmona:1967zz}. Soon
after, Nutku~\cite{1969ApJ...155..999N} explored the radial stability
of stellar models using a PN treatment.  Hillebrandt and
Heintzmann~\cite{1974GReGr...5..663H} analyzed incompressible
(constant density) configurations.
These studies found that corrections to NS structure are typically
suppressed by a factor $1/\omega_{\rm BD}$, where $\omega_{\rm BD}$ is
the Brans-Dicke coupling constant. The current best bound $\omega_{\rm
  BD}>40,000$~\cite{Will:2014xja} implies that the bulk properties of
NSs in the original Brans-Dicke theory deviate from GR by unmeasurable
amounts.

However, as pointed out by Damour and
Esposito-Far\`ese~\cite{Damour:1993hw,Damour:1996ke}, for a particular
class of scalar-tensor theories that is indistinguishable from GR in
the weak field regime [more precisely, when $\alpha_0=0$ and
  $\beta_0<0$ in the expansion of the Einstein-frame coupling function
  \eqref{DEalphabeta}], a nonlinear phenomenon called ``spontaneous
scalarization'' can occur, introducing macroscopically (and
observationally) significant modifications to the structure of the
star\footnote{For a comprehensive study of analytic solutions and an
  extensive bibliography, see~\cite{Horbatsch:2010hj}. Note in
  particular that Tsuchida et al.~\cite{Tsuchida:1998jw} extended the
  Buchdahl inequality ($M/R\leq 4/9$ for incompressible stars) to
  generalized scalar-tensor theories.}.  In addition the solutions
become nonunique: for certain ranges of the parameter space, NS
solutions in GR coexist with scalarized NSs.  One of the most
interesting observations in~\cite{Damour:1993hw,Damour:1996ke} is that
scalarized configurations are energetically favored over their GR
counterparts. Scalarization occurs also for BHs in the presence of
matter
fields~\cite{Stefanov:2008,Doneva:2010,Cardoso:2013fwa,Cardoso:2013opa}.

A simple way to illustrate the principle behind spontaneous
scalarization is by taking the limit in which the scalar field
$\varphi$ is a small perturbation around a GR solution.  Expanding
around the constant value $\varphi_{0}$ to first order in
$\hat\varphi\equiv\varphi-\varphi_{0}\ll1$, the field equations in the
Einstein frame~(\ref{eq:tensoreqnE}), (\ref{eq:scalareqnE}) read (see
e.g.~\cite{Yunes:2011aa})
\begin{align}
  &G_{\mu\nu}^\star=8\pi T_{\mu\nu}^\star
  \,, \label{Einsteinlin} \\
&\square^\star\hat\varphi=-4\pi \alpha_0 T^\star-4\pi \beta_0\hat\varphi T^\star\,. \label{KGlin}
\end{align}
Here we have assumed analyticity around $\varphi\sim \varphi_{0}$ and
used Eq.~\eqref{DEalphabeta}, where $A(\varphi)$ is the nonminimal
coupling to the matter fields in the Einstein frame, as defined by the
action~\eqref{STactionE}.

It is clear from Eq.~\eqref{KGlin} that $\alpha_0$ controls the
effective coupling between the scalar and matter. Various
observations, such as weak-gravity constraints and tests of violations
of the strong equivalence principle, require $\alpha_0$ to be
negligibly small when the scalar tends to its asymptotic value~\cite{Damour:1998jk,Damour:1996ke,Freire:2012mg}. This implies that a
configuration in which the scalar $\varphi\approx\varphi_0$ and $\alpha_0\approx 0$
should be at least an approximate solution in most viable
scalar-tensor theories. A detailed study
of the connection between the perturbative and nonperturbative scalarized
solutions can be found in~\cite{Stefanov:2008,Doneva:2010}. 

With $\alpha_0=0$, any background GR solution solves the field
equations above at first order in the scalar field. At this order, the
Klein-Gordon equation reads
\be
\left[\square^\star-\mu_s^2(x^\nu  )\right]\hat\varphi=0\,,\qquad \mu_s^2(x^\nu)\equiv -4\pi\beta_0 T^\star\,. \label{effectivemass}
\ee
Thus, the coupling of the scalar field to matter is equivalent to an
effective position-dependent mass. Depending on the sign of $\beta_0
T^\star$, the effective mass squared can be negative. Because $-T^\star\approx
\rho^\star>0$, this happens when $\beta_0<0$.  When $\mu_s^2<0$ in a
sufficiently large region inside the NS, scalar perturbations of a GR
equilibrium solution develop a tachyonic instability (i.e., the
perturbations propagate superluminally, as particles with imaginary
mass). This instability is associated with an exponentially growing
mode, which causes the growth of scalar hair in a process akin to
ferromagnetism~\cite{Damour:1993hw,Damour:1996ke}.

Spherically symmetric NSs develop spontaneous scalarization for
$\beta_0\lesssim-4.35$~\cite{Harada:1998}. Detailed investigations of
stellar structure~\cite{Damour:1996ke,Salgado:1998sg}, numerical
simulations of collapse~\cite{Shibata:1994qd,Harada:1996wt,Novak:1997hw} and stability studies~\cite{Harada:1997mr,Harada:1998} confirmed that spontaneously
scalarized configurations would indeed be the end-state of stellar
collapse in these theories.  In fact, spontaneously scalarized
configurations may also be the result of semiclassical vacuum
instabilities~\cite{Lima:2010na,Pani:2010vc,Mendes:2013ija,Landulfo:2014wra}.

The nonradial oscillation modes of spontaneously scalarized,
nonrotating stars were studied
in~\cite{Sotani:2004rq,Sotani:2005qx,Sotani:2014tua,Silva:2014ora}. The
bottom line of these studies is that the oscillation frequencies can
differ by a large amount from their GR counterparts if spontaneous
scalarization modifies the equilibrium properties of the star (e.g.,
the mass-radius relation) by appreciable amounts. However, current
binary pulsar observations yield very tight constraints on spontaneous
scalarization, and the oscillation modes of scalarized stars for
viable theory parameters are unlikely to differ from the corresponding
GR modes by any measurable amount.

\begin{figure*}[t]
\begin{center}
\begin{tabular}{ll}
\includegraphics[width=0.5\textwidth]{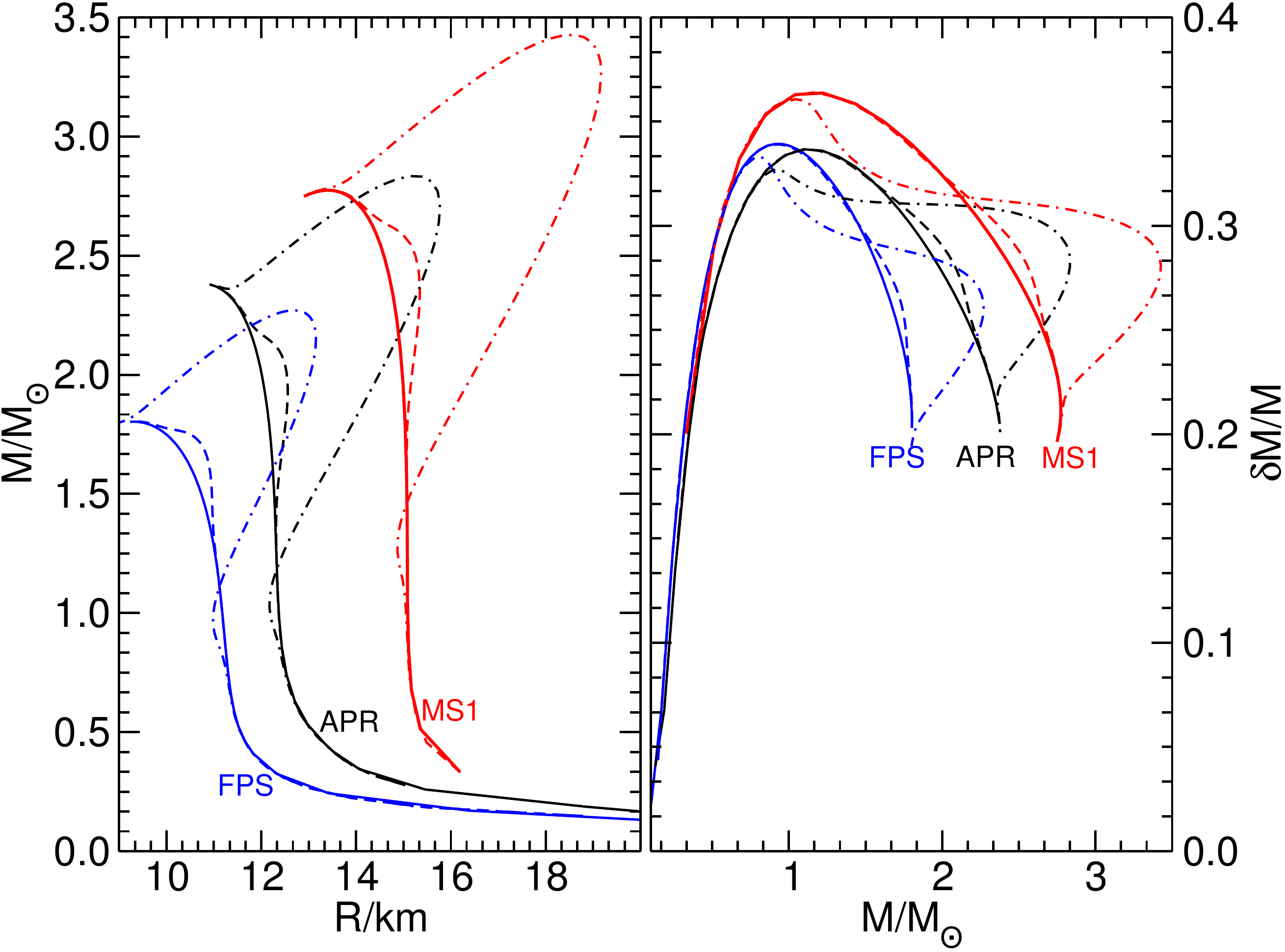}&
\includegraphics[width=0.5\textwidth]{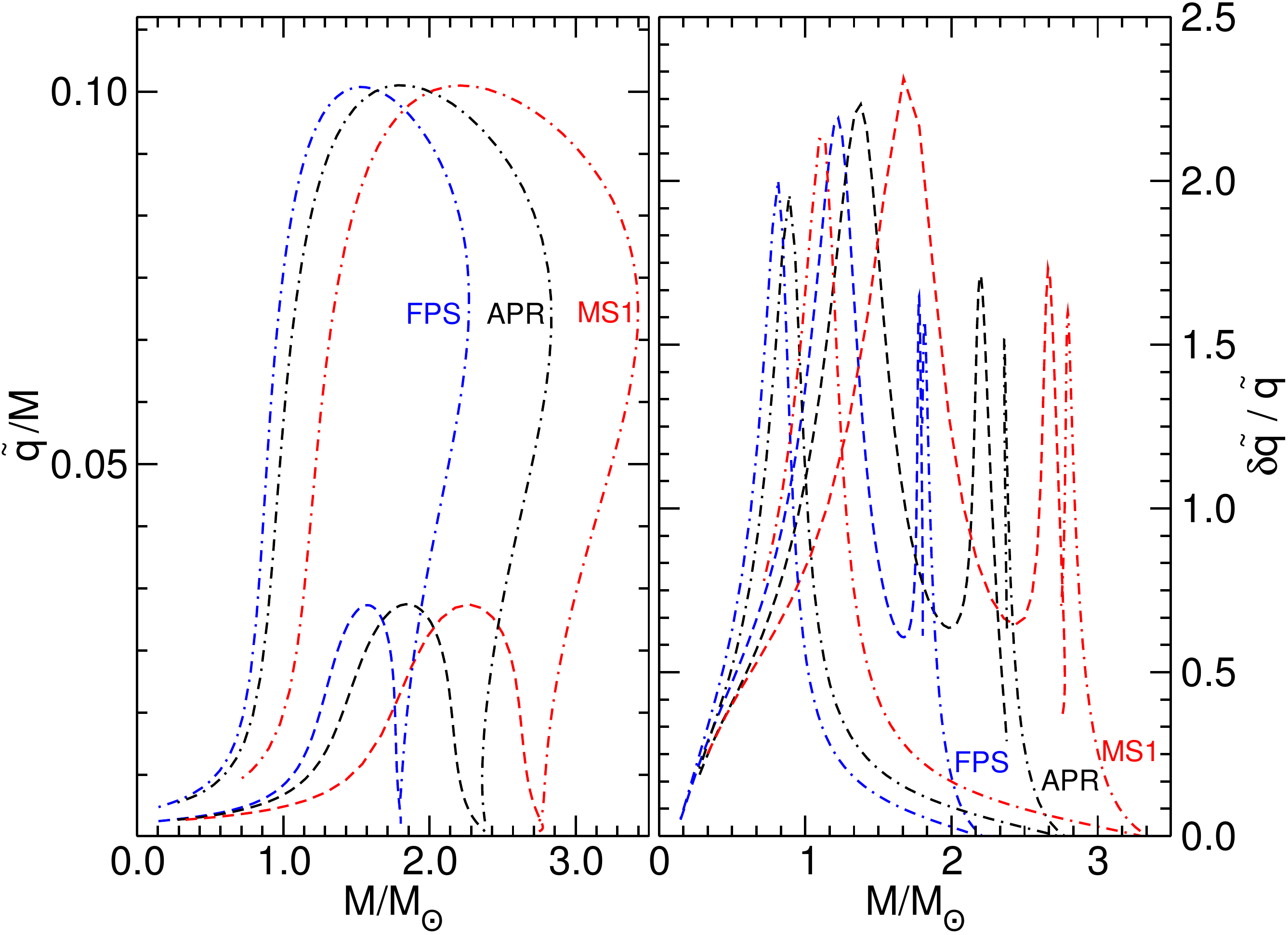}\\
\includegraphics[width=0.5\textwidth]{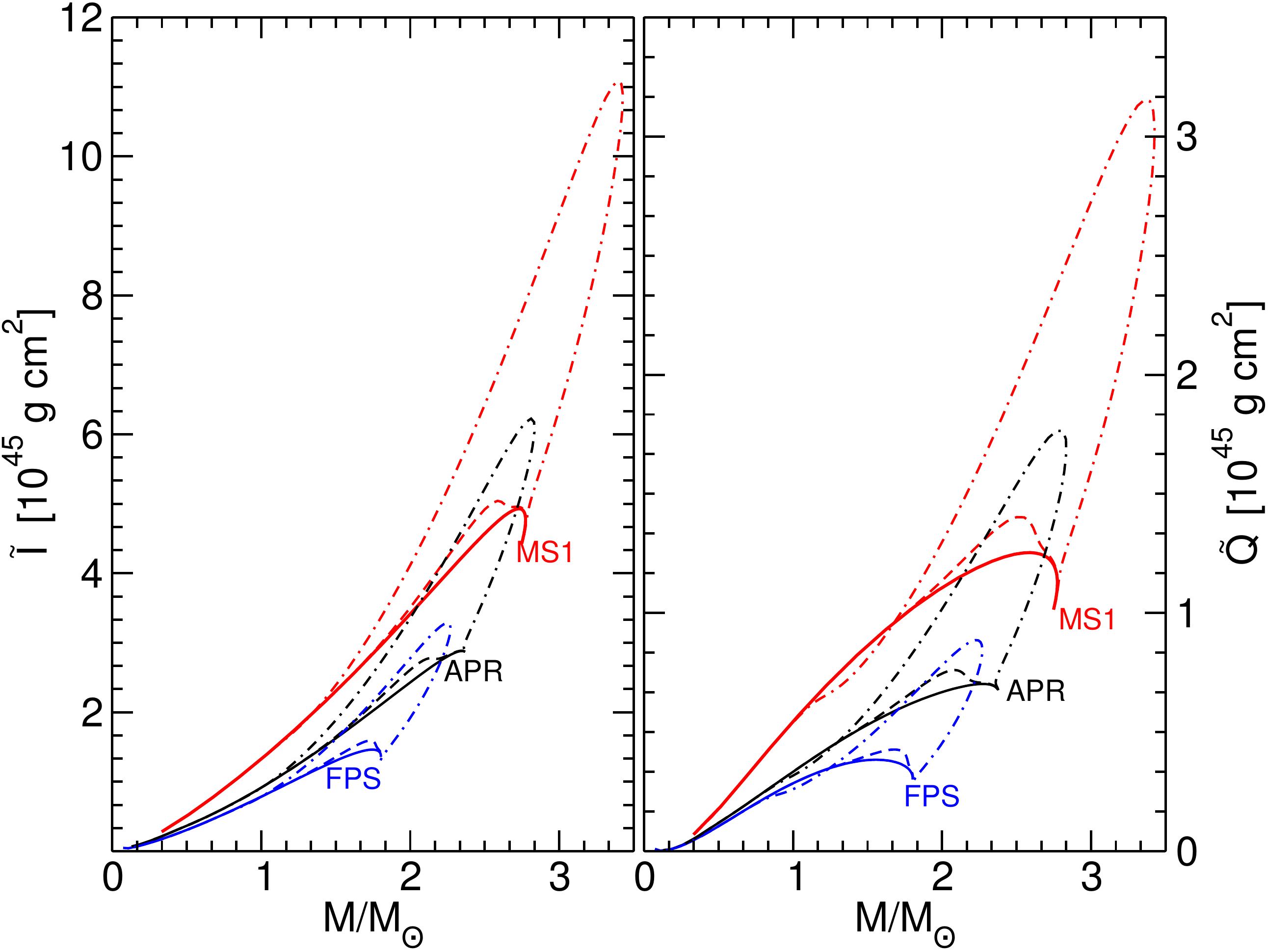}&
\includegraphics[width=0.5\textwidth]{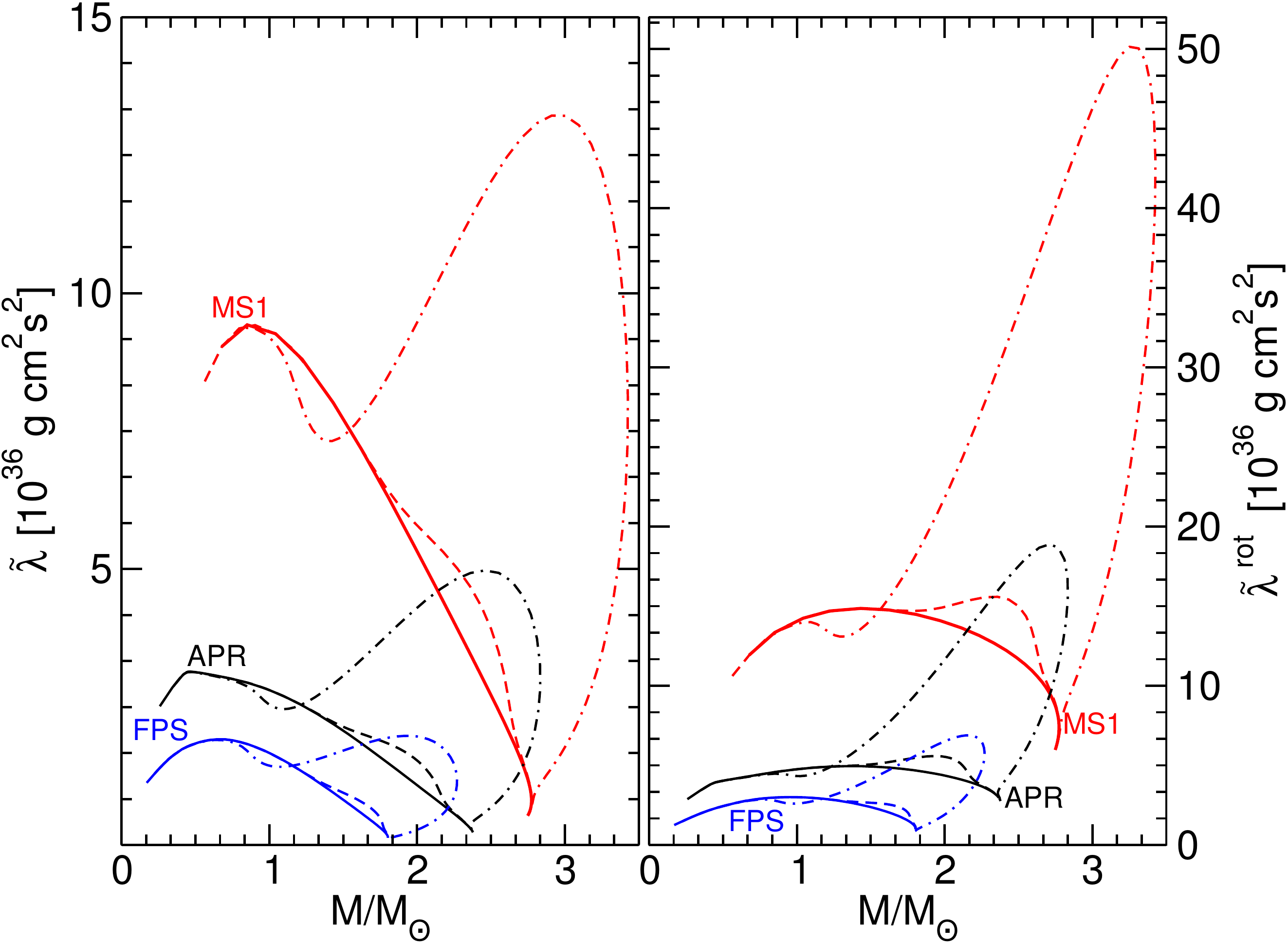}\\
\end{tabular}
\caption{NS configurations in GR (solid lines) and in two
  scalar-tensor theories defined by Eq.~\eqref{STactionE} with
  $A(\varphi)\equiv e^{\frac{1}{2}\beta_0 \varphi^2}$ and $V(\varphi)\equiv0$. Dashed
  lines refer to $\beta_0=-4.5$, $\varphi_0^\infty/\sqrt{4\pi}=10^{-3}$; dash-dotted
  lines refer to $\beta_0=-6$, $\varphi_0^\infty/\sqrt{4\pi}=10^{-3}$. Each panel
  shows results for three different EOS models (\texttt{FPS},
  \texttt{APR} and \texttt{MS1}).
  Top-left panel, left inset: relation between the nonrotating mass
  $M$ and the radius $R$ in the Einstein frame. Top-left panel, right inset:
  relative mass correction $\delta M/M$ induced by rotation as a
  function of the mass $M$ of a nonspinning star with the same central
  energy density.
  Top-right panel, left inset: scalar charge $\tilde{q}/M$ as a function
  of $M$. Top-right panel, right inset: relative correction to the
  scalar charge $\delta \tilde{q}/\tilde{q}$ induced by rotation as a
  function of $M$.
  Bottom-left panel: Jordan-frame moment of inertia $\tilde{I}$ (left
  inset) and Jordan-frame quadrupole moment $\tilde{Q}$ (right inset)
  as functions of $M$.
  Bottom-right panel: Jordan-frame tidal ($\tilde{\lambda}$) and
  rotational ($\tilde{\lambda}^{\rm rot}$) Love numbers as functions
  of $M$. [From~\cite{Pani:2014jra}.]
\label{fig:NS_ST2}}
\end{center}
\end{figure*}
\paragraph{Spontaneous scalarization and quantum instabilities in scalar-tensor theories with a conformal coupling.}
An interesting class of scalar-tensor theories that has been
recently investigated in the context of NS physics is the following:
\begin{equation}
  S=\frac{1}{16\pi}\int d^4x\sqrt{-g}\left[R
    -2g^{\mu\nu}\varphi_{,\mu}\varphi_{,\nu}-\xi\,R\,\varphi^2\right]+S_{\rm perfect\,fluid}\,, 
\end{equation}
where $\xi$ is the conformal coupling parameter. For $\xi =1/12$ the
scalar field equations are invariant under conformal transformations
($g_{\mu\nu}\rightarrow
\gamma^2g_{\mu\nu}\,,\varphi\rightarrow\gamma^{-1}\phi$), whereas for
$\xi=0$ one recovers the usual minimally coupled massless scalar.
The theory above can be obtained as a particular case of the
action~\eqref{STactionJ} after a field redefinition.

Lima, Matsas and Vanzella showed that the vacuum expectation value of
nonminimally coupled scalar fields can grow exponentially in
relativistic stars~\cite{Lima:2010na}. At the classical level, this
quantum instability can be interpreted in terms of the spontaneous
scalarization discussed above~\cite{Pani:2010vc}. The instability can
occur for both positive and negative values of $\xi$. When $\xi<0$ and
$|\xi|$ is large enough, the instability can occur even for Newtonian
stars. For a detailed analysis of the approach to the classical limit
and of the relation between the quantum and classical nature of the
final state, see~\cite{Mendes:2013ija,Landulfo:2014wra}.

\paragraph{Slowly rotating solutions.}

Spinning NSs at first order in the Hartle-Thorne slow-rotation
approximation were studied by Damour and
Esposito-Far\`ese~\cite{Damour:1996ke} and later by
Sotani~\cite{Sotani:2012eb}. At first order in rotation, the scalar
field only affects the moment of inertia, mass and radius of the
NS. Second-order calculations~\cite{Pani:2014jra} are necessary to
compute corrections to the spin-induced quadrupole moment, tidal and
rotational Love numbers, as well as higher-order corrections to the NS
mass and to the scalar charge. Figure~\ref{fig:NS_ST2} shows
representative examples of the properties of NSs in a scalar-tensor
theory with spontaneous scalarization at second order in the rotation
parameter.

\paragraph{Rapidly rotating solutions.}

\begin{figure}[ht!]
\centering
\begin{tabular}{cc}
\includegraphics[width=0.5\textwidth]{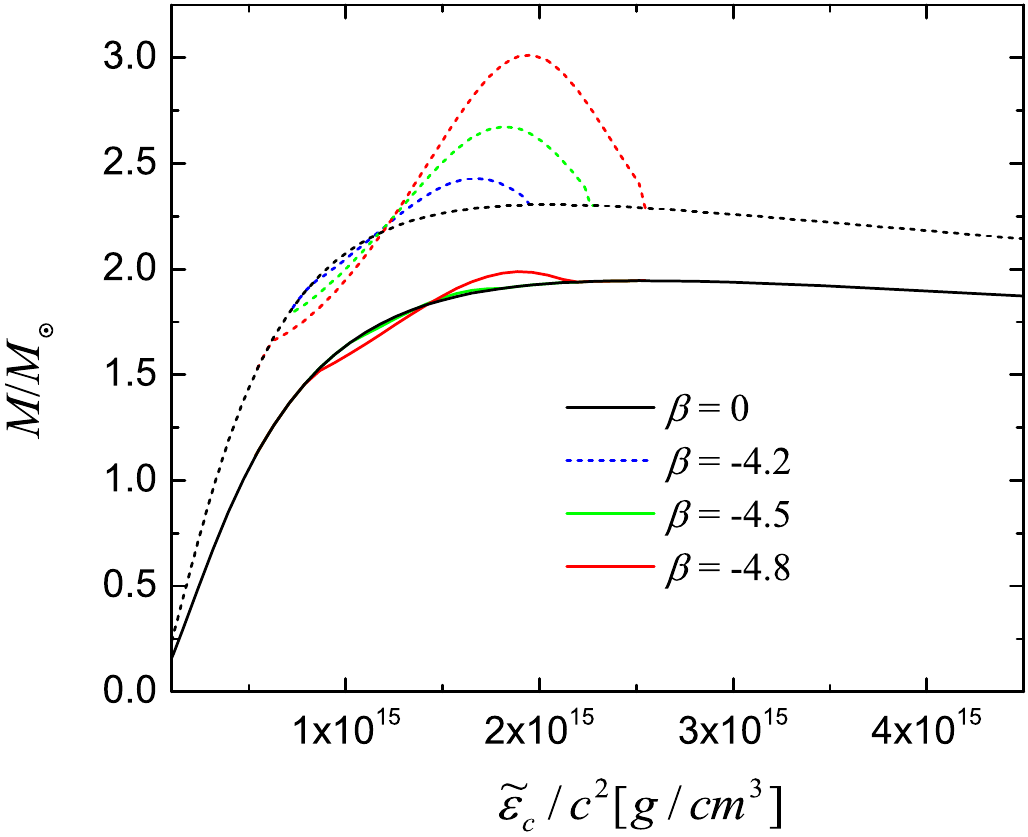}&
\includegraphics[width=0.5\textwidth]{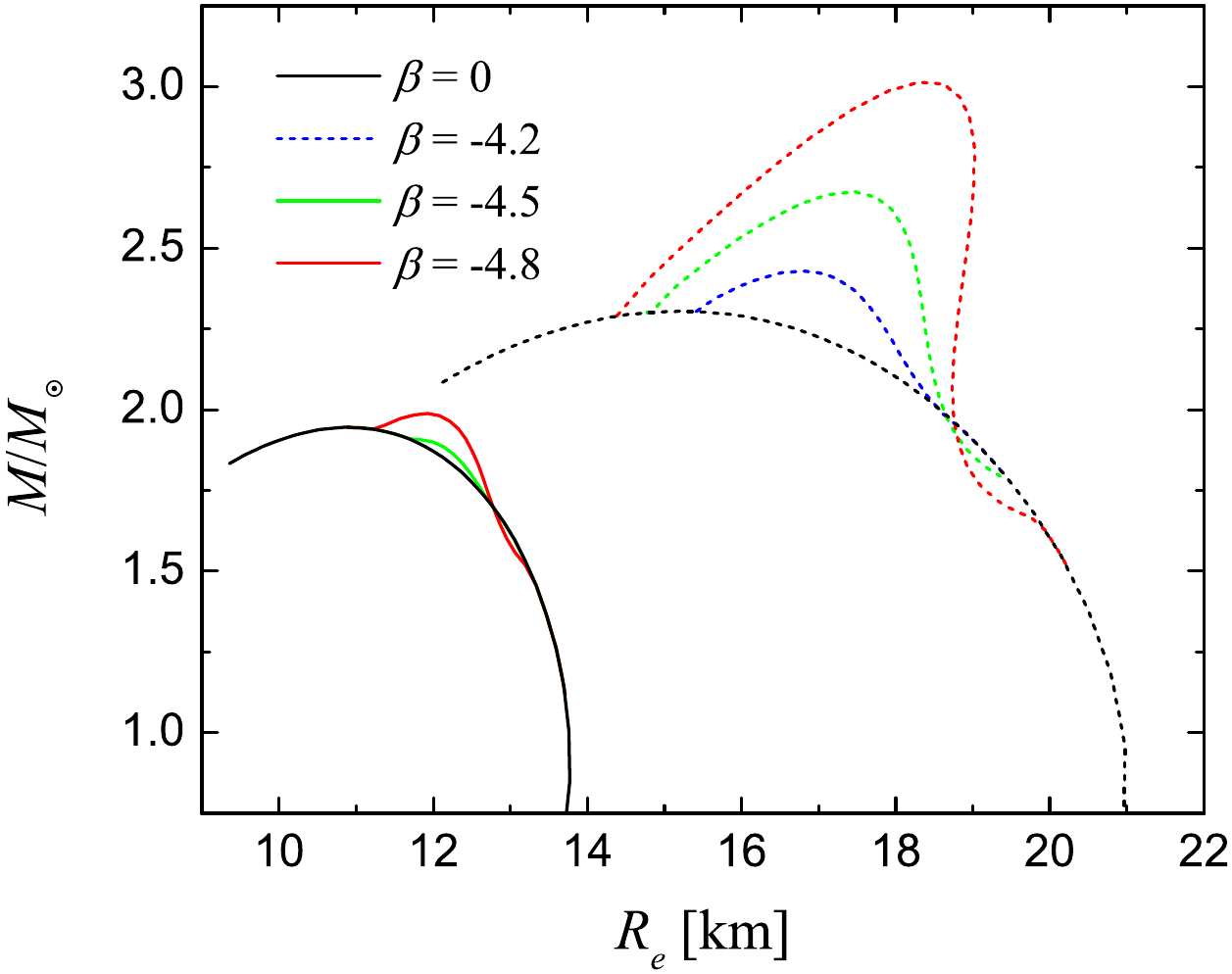}
\end{tabular}
\caption{The NS mass as a function of the central energy density (left
  panel) and of the radius (right panel) for static sequences of NSs
  (solid lines) and sequences of stars rotating at the mass-shedding
  limit (dotted lines).  The trivial solutions coincide with the GR
  limit ($\beta=\beta_0=0$). For $\beta=\beta_0=-4.2$ nontrivial
  solutions do not exist in the nonrotating
  case. [From~\cite{Doneva:2013qva}.]}
\label{fig:NS_ST_fast}
\end{figure}

Rapidly rotating NSs in scalar-tensor theories were recently
constructed in~\cite{Doneva:2013qva} by extending the {\tt RNS}
code~\cite{Stergioulas:2003yp}. The results shown in
Figure~\ref{fig:NS_ST_fast} illustrate that scalarization effects are
stronger for rapidly rotating stars, and deviations from GR are
sensibly larger for fast-spinning NSs. One of the reasons is that the
stress-energy tensor (which acts as a source for the scalar field)
gets a contribution from the rotational energy of the star. The
nontrivial scalar field has a strong effect also on the NS angular
momentum and moment of inertia, which can differ by as much as a
factor of two from their GR values at the Kepler limit~\cite{Doneva:2013qva}. In addition,
there exists a larger range of parameters for which scalarization
occurs, and the critical value of the coupling constant $\beta_0$
where a nontrivial scalar field can develop increases substantially.
For example, for the polytropic EOS considered in~\cite{Doneva:2013qva} the critical value of $\beta_0$ increases from
$\beta_0\gtrsim-4.35$ in the nonrotating case to $\beta_0\gtrsim-3.9$
for rapid rotation.  For realistic EOSs scalarization can occur for
even larger $\beta_0$~\cite{Doneva:2014uma}. Binary pulsar
observations imply $\beta_0\gtrsim-4.5$ (see Section~\ref{sec:BP}).
For a marginally allowed $\beta_0$, nonrotating scalarized NSs would
not differ considerably from the GR solutions, whereas rapid rotation
can produce significant deviations that can potentially set even
stronger astrophysical constraints on scalar-tensor theories~\cite{Doneva:2014uma,Doneva:2014faa}. Other proposed mechanisms that
can amplify the effects of scalarization include anisotropy~\cite{Silva:2014fca} and ``dynamical scalarization'' for merging NSs
in scalar-tensor theories, that will be discussed in
Section~\ref{sec:CB}~\cite{Barausse:2012da,Palenzuela:2013hsa,Shibata:2013pra,Taniguchi:2014fqa}. In
the last stages before merger the rotational frequencies of each NS
may approach the Kepler limit.

\subsection{f(R) theories} \label{NS_fofR}

In principle $f(R)$ theories can be mapped to a specific form of the
action in scalar-tensor theory~\cite{Sotiriou:2008rp,DeFelice:2010aj},
but this mapping involves subtleties and technicalities that justify a
separate discussion of NS solutions in metric $f(R)$ gravity. In fact
the literature on NS solutions in metric $f(R)$ gravity is quite
extensive, and it contains several apparently controversial claims
~\cite{Frolov:2008uf,Kobayashi:2008tq,Upadhye:2009kt,Babichev:2009td,Babichev:2009fi,Jaime:2010kn}.

The recent interest in $f(R)$ theories is due to their potential to
explain cosmological observations without introducing dark matter or
dark energy. In terms of compact objects, this means that one is
usually interested in matching the stellar interior to a de Sitter
metric with an effective cosmological constant
\begin{equation}
  \Lambda_\text{eff}=R_{\rm dS}/4\,,\label{Lambdaeff}
\end{equation}
where $R_{\rm dS}$ is the curvature at the de Sitter point, and $R\to
R_{\rm dS}$ far from the star. The problem involves two completely
different density (or curvature) scales, because the central density
of a NS ($\rho_0\sim10^{14} \text{g cm}^{-3}$) is enormously larger
than the density associated to the cosmological constant
($\rho_\Lambda=\Lambda/(8\pi G)\sim 10^{-29}\text{g cm}^{-3}$):
$\rho_\Lambda/\rho_0 \sim10^{-43}\label{ratio_f(R)}$.
In practice, only much larger values
($\rho_\Lambda/\rho_0\sim10^{-10}-10^{-6}$) can be used in
numerical codes. This issue is not specific to $f(R)$ theories: it
would also arise in GR with a positive cosmological constant if one
tries to match a NS interior with a de Sitter exterior. In fact, the
large disparity in density (or curvature) scales is not a problem if
one assumes that the cosmological scale has no sensible influence on
local physics. In other words, one would expect local observables such
as the NS mass and radius to be insensitive to $\rho_\Lambda/\rho_0$,
as long as this ratio is small enough:
$\rho_\Lambda/\rho_0\sim10^{-10}$ (say) would be practically
indistinguishable from $\rho_\Lambda/\rho_0\sim10^{-43}$, except for
giving an unrealistically large cosmological constant.

Calculations of NS structure in $f(R)$ theory used different
approaches, reaching different conclusions on the very existence of
relativistic compact stars. Here we try to clarify some critical
issues in the literature, pointing the reader to the original
references for more details.

\paragraph{Singular potential.}
When $f(R)$ is reformulated as a scalar-tensor theory, the potential
for the scalar degree of freedom can, and in general will, be
singular~\cite{Frolov:2008uf}.  The scalar-field
equation~\eqref{scalareqKM} can be recast in the form
\begin{equation}
\square\phi=V_{\rm KM}'(\phi)-{\cal F}\,, 
\end{equation}
where a prime denotes a derivative with respect to $\phi$ and
${\cal F}=8\pi/3 (\rho-3P)$ plays the role of a matter-driven force
term.  For various solutions describing late-time cosmology or compact
objects, the ``force'' ${\cal F}$ pushes the scalar field towards the
unprotected curvature singularity, which is at finite distance (in
field and energy space) from the equilibrium configuration. As
discussed below, this happens precisely for those $f(R)$ models which
are otherwise theoretically and observationally
viable~\cite{Frolov:2008uf}.
This singular character of the potential can cast doubts on the
viability of $f(R)$ gravity and on the existence of compact objects in
these theories.  However, this premature conclusion depends on the
choice of a specific scalar-tensor formulation of $f(R)$ gravity.

Kobayashi and Maeda~\cite{Kobayashi:2008tq} reported that the field
equations inside relativistic stars are plagued by singularities, but
subsequent work~\cite{Upadhye:2009kt,Babichev:2009td} claimed that
such singularities were unphysical and due to numerical
instabilities. Indeed, the scalar field in the interior of compact
objects can be very close to the value that corresponds to the
singular potential, but does not necessarily need to end up in the
singularity.  This makes the integration challenging from a numerical
point of view, but it does not necessarily imply a pathology in the
underlying theory, as we show below.

First of all, note that any $f(R)$ model that meets the minimum
requirements to satisfy Solar System constraints -- i.e., it satisfies
the requirements of Eq.~\eqref{limit_f} -- is such that the first and
second derivatives of the scalar potential defined
in~\cite{Kobayashi:2008tq,Upadhye:2009kt,Babichev:2009td} are {\em
  divergent} in the $R\to\infty$ limit. This limit corresponds to a
\emph{finite} value of the scalar field and to a \emph{finite} value of the
potential at the singular point. In the following, we shall
generically denote the scalar degree of freedom (in the various formulations discussed in Section~\ref{subsec:f(R)}) by
$\Phi$ and the scalar potential by $V_\Phi$, whereas $\Phi_s$ denotes
the value of the scalar at the singular point,
i.e.~$\Phi_s=\Phi(R\to\infty)$. Let us parametrize the large-curvature
expansion by using the rather generic expression (see Eq.~\eqref{limit_f})
\begin{equation}
  f(R)\sim R+R_c\left[a+b\left(\frac{R_c}{R}\right)^c +
    d\log\left(\frac{R_c}{R}\right)\right]\,,\quad R\gg R_c\,, \label{expansion}
\end{equation}
where $(a,b,c,d)$ are dimensionless real constants, $c\geq0$,
$d\geq0$, $R_c$ is some curvature scale of the order of $R_{\rm dS}$,
and we have kept only the dominant terms in a large-curvature
expansion.
Eq.~\eqref{expansion} is a good approximation in the interior of a NS,
where the curvature is much larger than the cosmological curvature
$R_{\rm dS}$, and indeed most of the models considered in the
compact object literature belong to this class. Using
Eq.~\eqref{expansion}, it is straightforward to prove that
\begin{equation}
 V_\Phi(R\gg R_c)\to R_c\left[{\rm const}+d\log\left(\frac{R_c}{R}\right)\right]
\end{equation}
up to a constant (that can be adjusted to eliminate the constant term
above). In conclusion, in the limit $\Phi\to\Phi_s$ the potential
$V_\Phi$ is finite if $d=0$ and diverges logarithmically if $d\neq0$,
but its derivative $V_\Phi'\to\infty$ in any case.
Therefore, the energy density needed to make the singularity energetically
accessible is roughly
\begin{equation}
 V_\Phi'(\Phi\to\Phi_s)\sim R\,.
\end{equation}
In models in which corrections to GR are relatively small for
$R\gg R_c$, this quantity is parametrically of the same order as the
matter energy density $\rho_c$ in the interior of a NS. This can be
seen by taking the trace of the modified Einstein equations [see
discussion around Eq.~\eqref{Rmin} below]. This simple argument seems
to suggest that the singularity should be accessible, as discussed
in~\cite{Frolov:2008uf}. However, as we will see below, in $f(R)$
theories a subtle mechanism prevents such singularities to be
accessible, at least in various situations. The price to pay is that
numerical integrations for realistic values of the theory parameters
are extremely challenging.

To this end, it is important to remark that the singular behavior of the scalar
potential is not an \emph{intrinsic} ingredient of the theory, but
just a prerogative of specific formulations of $f(R)$ gravity.
Indeed, this severe problem does not arise in the approach developed
by Jaime et al.~\cite{Jaime:2010kn}, in which the potential defined in
Eq.~\eqref{Jaime2} is regular for finite curvature.
When $R\gg R_c$, using Eq.~\eqref{expansion} we get
\begin{equation}\label{VJPS}
 V_{\rm JPS}(R)\sim \left\{\begin{array}{ll}
  R_c^{3}\left({R}/{R_c}\right)^{c+4}\,, &d=0\\
  R_c^{3}\left({R}/{R_c}\right)^4\,, &d\neq0
\end{array} \right.\,.
\end{equation}
Remarkably, in this formulation the singularity at $R\to\infty$ is protected by an \emph{infinite}
potential well for any viable model satisfying Eq.~\eqref{limit_f}.
Furthermore, dimensional arguments show that in this case the energy density needed
to reach the singularity is of order
\begin{equation}
V_{\rm JPS}'(R)R^{-1}\sim \left\{\begin{array}{ll}
  R_c\left({\rho_c}/{R_c}\right)^{c+2}\,, &d=0\\
  R_c\left({\rho_c}/{R_c}\right)^2\,, &d\neq0
\end{array} \right.\,,
\end{equation}
which is always much larger than the internal energy density of a
NS. In other words, in this formulation the singularity at infinity is protected by an infinite potential
barrier~\cite{Jaime:2010kn}, as in any well-behaved mechanical model (e.g.~the harmonic
oscillator).

\paragraph{Chameleon mechanism.}
The existence of singular scalar configurations accessible at finite
energies is potentially dangerous. As we just discussed, in the case
of $f(R)$ gravity it is not the theory itself to be potentially
problematic, but only some particular formulations of the theory.
Clearly, the viability of the theory cannot depend on the particular
formulation chosen in~\cite{Jaime:2010kn}, so there must exist a
mechanism that prevents singular behavior also in other formulations.
Indeed, even in these potentially ill-defined formulations, a
subtle~\emph{chameleon mechanism}~\cite{Khoury:2003aq,Gubser:2004uf}
keeps the scalar field away from the singularity.  The chameleon
mechanism is related to the generation of an infinitely large mass
term of the scalar field, due to self-interactions and to interactions
with other matter fields (see e.g.~\cite{Upadhye:2009kt}).

At high curvature $dV_\Phi/d\Phi \sim R$ and $dV_\Phi^{\rm
  eff}/d\Phi\to R+8\pi T$,
so that the effective potential has a minimum at
\begin{equation}
 R_{\rm min}\sim -8\pi T\,,\label{Rmin}
\end{equation}
which corresponds to some $\Phi_{\rm min}=\Phi(R_{\rm min})$.  For any
$f(R)$ gravity theory satisfying the viability
conditions~\eqref{limit_f}, assuming $T\sim{\rm const}$, the scalar
mass in the large-curvature limit reads
\begin{equation}\label{chameleon_R}
 m^2_\Phi\equiv\left.\frac{d^2V_\Phi^{\rm eff}(\Phi)}{d\Phi^2}\right|_{\Phi=\Phi_{\rm min}} \sim \left\{\begin{array}{ll}
  R_c\left({-8\pi T}/{R_c}\right)^{c+2}\,, & d=0\\
  R_c\left({-8\pi T}/{R_c}\right)^2\,, & d\neq0
\end{array} \right.\,.
\end{equation}
It is crucial to realize that the dimensionless quantity $m_\Phi R^{-1/2}$ 
is an estimate of the ratio between the curvature lengthscale and the
Compton wavelength of the graviton; therefore, it
is proportional to the number of integration steps needed to resolve the
dynamics of the chameleon field in a region of approximately constant
curvature $R$~\cite{Upadhye:2009kt}. Evaluating this quantity inside a
NS, where $R\sim -8\pi T\sim\rho_c$, one obtains
\begin{equation}\label{chameleon_R2}
 m_\Phi^2 \rho_c^{-1}\sim \left\{\begin{array}{ll}
  \left(x_{\rm dS}{\rho_c}/{\Lambda_{\rm eff}}\right)^{c+1}\,, & d=0\\
  \left(x_{\rm dS}{\rho_c}/{\Lambda_{\rm eff}}\right)\,, & d\neq0
 \end{array} \right.
\,,
\end{equation}
where $x_{\rm dS}=R_{\rm dS}/R_c$.  The larger the ratio
${\rho_c}/{\Lambda_{\rm eff}}$, the larger the effective chameleon
mass and the number of steps needed for the integration in the stellar
interior. For a realistic NS embedded in a de Sitter universe the
effective mass is extremely large in cosmological units, and so is the
number of integration steps.
Such a heavy field is challenging to treat numerically. Indeed, as
discussed by~\cite{Upadhye:2009kt}, Yukawa-like error modes grow as
$e^{m_\Phi r}/r$, and dominate if the Compton scale $m_\Phi^{-1}$
becomes much smaller than the computational domain. Hence, in the ``wrong'' formulations the
integration of the field equations becomes practically impossible in
realistic situations, but this does not imply that the solutions are singular.

Note also that the chameleon mass depends on powers of $x_{\rm dS}$,
which may be a large quantity, making the integration even more
challenging. For example, the bound $x_{\rm dS}\gtrsim10^3$ must be
imposed for a popular model (the simplest version of the Starobinsky model~\cite{Starobinsky:2007hu}, where $f(R)=R[1-\mu R_c R/(R_c^2+R^2)]$) to
satisfy local tests~\cite{Hu:2007nk,Upadhye:2009kt}. Curiously, this
fact is often overlooked and unrealistic values $x_{\rm dS}={\cal
  O}(1)$ are commonly used (but see~\cite{Upadhye:2009kt} for an
exception).

In summary, the challenge of constructing compact models in $f(R)$
gravity depends to a large extent on the formulation adopted and, in many cases, is usually overcome by restricting the discussion to
unrealistic stars with $\rho_c\sim 10^2 \Lambda_{\rm eff}$, which is
much smaller than the realistic density one should use ($\rho_c\sim
10^{43} \Lambda_{\rm eff}$), and even in this case numerical
integrations can be very difficult.
The chameleon mechanism provides an elegant way to keep the scalar
field away from the singular point. In fact, it is remarkable that the
same nonlinearities that make the scalar potential singular also keep
the field away from the singularity. As Babichev and Langlois put it
in their paper, the chameleon mechanism forces the scalar field to
stay ``attached to a track very near a precipice, without falling into
it''~\cite{Babichev:2009fi}. The distance of the track from the
precipice can be tiny (as small as $10^{-43}$ in cosmological units),
so that it is practically impossible to follow the evolution of the
scalar field numerically without being contaminated by the nearby
singularity.

\paragraph{Multivalued potential.}

Another known and controversial issue with the approaches
by~\cite{Upadhye:2009kt,Babichev:2009td} is multivaluedness. When we
recast $f(R)$ theories as scalar-tensor theories, the scalar-field
potential can be
multi-valued~\cite{Frolov:2008uf,Kobayashi:2008tq,Jaime:2010kn}. This
conclusion is model-dependent, but it applies e.g.~to the Starobinsky
model (cf.~Figure~1 of~\cite{Frolov:2008uf}) and to any model for
which $f_R$ is not a monotonic function of $R$. In particular, the
cosmological branch of the Starobinsky potential is discontinuous at
the singular point, i.e.~as $\Phi\to\Phi_s$.

This problem is usually ignored on the grounds that stellar structure
calculations only refer to the structure of the potential around a
local minimum, and that possible multiple branches are harmless if the
entire cosmological evolution of the scalar field is confined within a
single-valued branch of the potential.  While this is true,
multivaluedness might seem to make $f(R)$ theories less natural and attractive,
especially if the scalar field is extremely close to the singular and
discontinuous point of the potential (as in the interior of compact
objects).

However --~similarly to the issue with the singular scalar potential~-- also multivaluedness is
formulation-dependent, as pointed out in~\cite{Jaime:2010kn}. Indeed, it turns 
out that the potential~\eqref{VJPS} is \emph{not}
multivalued. This is simply due to the fact that the ``true''
dynamical degree of freedom is the curvature $R$, not $f_R$, and no
inversion $R=R(\Phi)$ is needed in the approach
of~\cite{Jaime:2010kn}.

\paragraph{Main results.}
The mapping of $f(R)$ gravity to scalar-tensor theories -- both in the
Jordan and in the Einstein frame -- is plagued by several potentially
dangerous issues, including singularities and multivaluedness in the
scalar potential and a diverging effective mass for the field. These
issues are intertwined, and in fact they can be seen as ``features,''
as they are needed for the theoretical safety of the theory (an
example being the chameleon mechanism that keeps the scalar field away
from the singularity). The numerical challenges they introduce may
also serve as a motivation to develop more efficient integration
methods. These same issues make the study of compact objects in $f(R)$
gravity particularly difficult, especially for realistic
configurations
~\cite{Cooney:2009rr,Arapoglu:2010rz,Alavirad:2013paa,Astashenok:2013vza,Astashenok:2014pua,Yazadjiev:2014cza,Staykov:2014mwa,Yazadjiev:2015zia}.

Most studies of NS configurations in $f(R)$ gravity consider
\emph{perturbative} corrections to the Einstein-Hilbert action of the
form $f(R)=R+\epsilon f_1(R)$, with $\epsilon\ll1$~\cite{Cooney:2009rr,Astashenok:2013vza,Arapoglu:2010rz,Alavirad:2013paa,Astashenok:2014pua}. This
expansion is similar in spirit to the EFT approach discussed elsewhere
in this review, and it bypasses some of the difficulties listed
above. Some of these models~\cite{Astashenok:2013vza} predict NSs with
large compactness (NS radii as small as $\sim9$ km), which are
difficult to obtain in GR, even taking into account current
uncertainties in the EOS.

Yazadjiev et al.~\cite{Yazadjiev:2014cza} recently went beyond the
perturbative level constructing static equilibrium models of NSs in a
theory of the form $f(R)=R+\lambda R^2$, where the coupling $\lambda$
is not necessarily small. They found that deviations from GR are
comparable with the variations due to uncertainties in the EOS, even
for large values of $\lambda$. Subsequent work by Staykov et
al.~\cite{Staykov:2014mwa} extended the analysis to first order in the
slow-rotation approximation, finding that the NS moment of inertia can
be up to $30\%$ larger than its GR counterpart. This correction is
larger than that introduced by uncertainties in the EOS, and (in
principle) it can be used to break the EOS degeneracy
(cf.~Section~\ref{subsec:ILQ} below). Yazadjiev et
al.~\cite{Yazadjiev:2015zia} constructed rapidly rotating NSs in
nonperturbative $f(R)=R+\lambda R^2$ gravity. For fast rotation, the
maximum NS mass and moment of inertia can be up to $\sim 16\%$ and
$60\%$ larger than in GR, respectively. These corrections to the NS
properties are large enough that, if observed, they may be used to
constrain the parameter $\lambda$.

\subsection{Quadratic gravity}
\label{sec:NSs/quadratic-gravity}

A remarkable consequence of recent work is that, from the point of
view of the properties of compact objects, scalar-tensor theories and
quadratic gravity are ``orthogonal'' extensions of GR. Isolated BHs in
scalar-tensor theories are described by the Kerr solution, just like
in GR, whereas NS configurations can acquire a scalar charge through
spontaneous scalarization. On the contrary, BH solutions in quadratic
gravity theories are endowed with a scalar field that is supported by
the higher-order curvature terms, but a ``no-scalar-monopole-hair''
theorem holds for NSs in EdGB and dCS gravity (at least in the
\emph{perturbative} regime).

There is a simple heuristic proof of the NS no-hair theorem in
quadratic gravity~\cite{Yagi:2011xp}. Integrating the scalar equation
of motion in EdGB gravity (within the perturbative EFT expansion)
yields
\begin{equation}
  \int d^4x \sqrt{-g}\, \square\phi \propto \int d^4x\sqrt{-g} R_{\rm GB}^2 \,,
  \label{proofNSs}
\end{equation}
where $R_{\rm GB}^2$ is the Gauss-Bonnet term (a similar conclusion applies to
dCS gravity, as long as we replace $R_{\rm GB}^2$ with the Pontryagin
density ${}^*RR$). Because $R_{\rm GB}^2$ (and ${}^*RR$) are
topological invariants, the right-hand side of Eq.~\eqref{proofNSs}
vanishes identically for any simply connected, asymptotically flat
domain. Furthermore isolated NSs are stationary, so the time
dependence of the left-hand side can be neglected, yielding an
integral over the volume. Finally, Stokes' theorem yields
\begin{equation}
 \int \sqrt{-g} (\partial_i\phi) n^i dS = \int \sqrt{-g}(\partial_r \phi) 
dS=0\,,
\label{StokesQuad}
\end{equation}
where the unit vector $n^i$ is normal to the surface $S$, taken to be
a 2-sphere at infinity. In order to have finite energy, the scalar
field must go to zero at large distances ($r\to \infty$). If we had
$\phi\to Q/r$, where $Q$ is a constant related to some hypothetical
scalar charge, Eq.~\eqref{StokesQuad} would imply $Q=0$. Therefore the
scalar field must decay faster than $1/r$ at infinity, and isolated
NSs in EdGB gravity and dCS gravity have no scalar monopole charge.

\subsubsection{EdGB theory}

The proof given above is valid only in the \emph{perturbative} regime,
i.e.~when the $R_{\rm GB}^2$ or ${}^*RR$ terms in the action are
coupled \emph{linearly} to the scalar field, so that the right-hand
side of Eq.~\eqref{proofNSs} vanishes identically. If we consider the
EdGB action of Eq.~\eqref{EdGBaction} with a generic coupling to the
Gauss-Bonnet term, ``baldness'' does not necessarily apply to NS
solutions. Furthermore, even in the perturbative regime the proof does
not exclude the possibility that NSs may be endowed with a nontrivial
scalar field profile that just happens to have vanishing scalar
monopole charge. Indeed, it is easy to check that the source term of
the scalar field equation does not vanish in general ($R_{\rm
  GB}^2\neq0$): a constant (or zero) scalar field is not a solution of
the field equations, and therefore we would expect the NS properties
to be affected by a nontrivial scalar field.
These modifications were studied in~\cite{Pani:2011xm} for the case of
standard EdGB gravity with an exponential coupling of the form
\begin{equation}
 f_1(\phi)=\alpha_{\rm GB} e^{\frac{\beta}{\sqrt{8\pi}}\phi}\,
\end{equation}
(in the notation of~\eqref{EdGBaction}), where $\alpha_{\rm GB}$ and
$\beta$ are coupling constants. For $\beta/\sqrt{8\pi}=2$ the model
reduces to the bosonic sector of heterotic string theory~\cite{Gross:1986mw}.

\begin{figure*}[htb]
\begin{center}
\begin{tabular}{cc}
\includegraphics[width=0.5\textwidth]{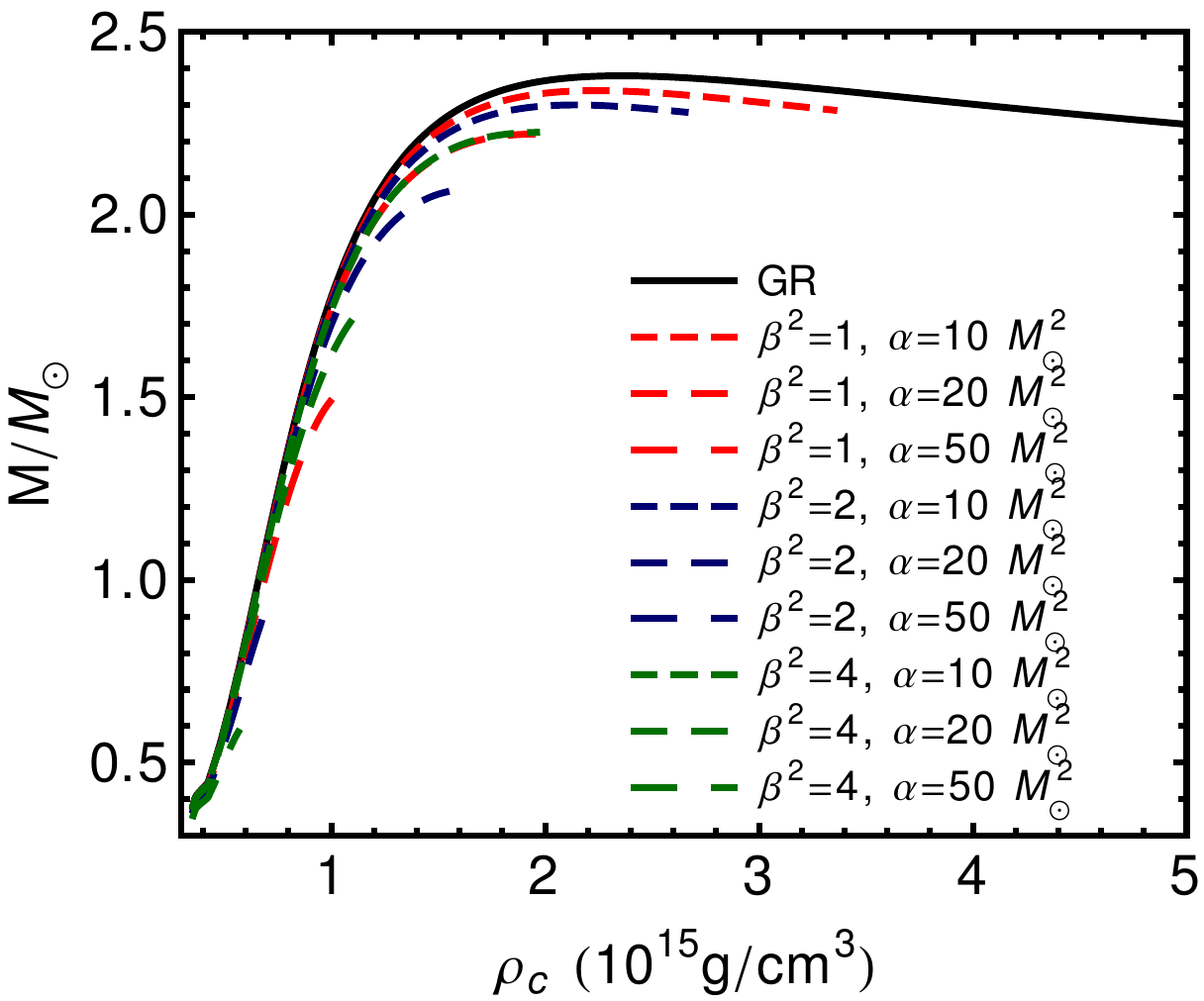}&
\includegraphics[width=0.5\textwidth]{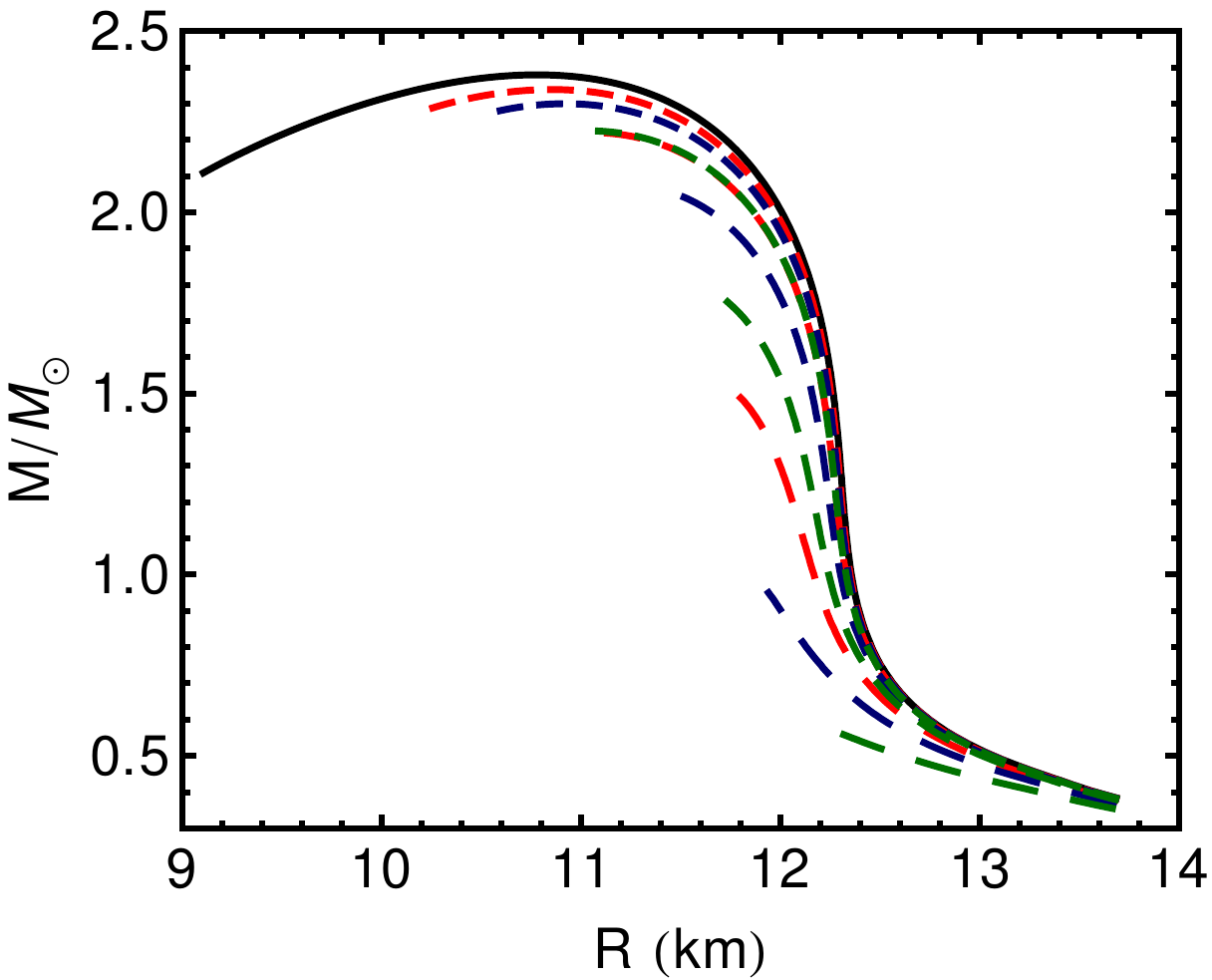}\\
\includegraphics[width=0.5\textwidth]{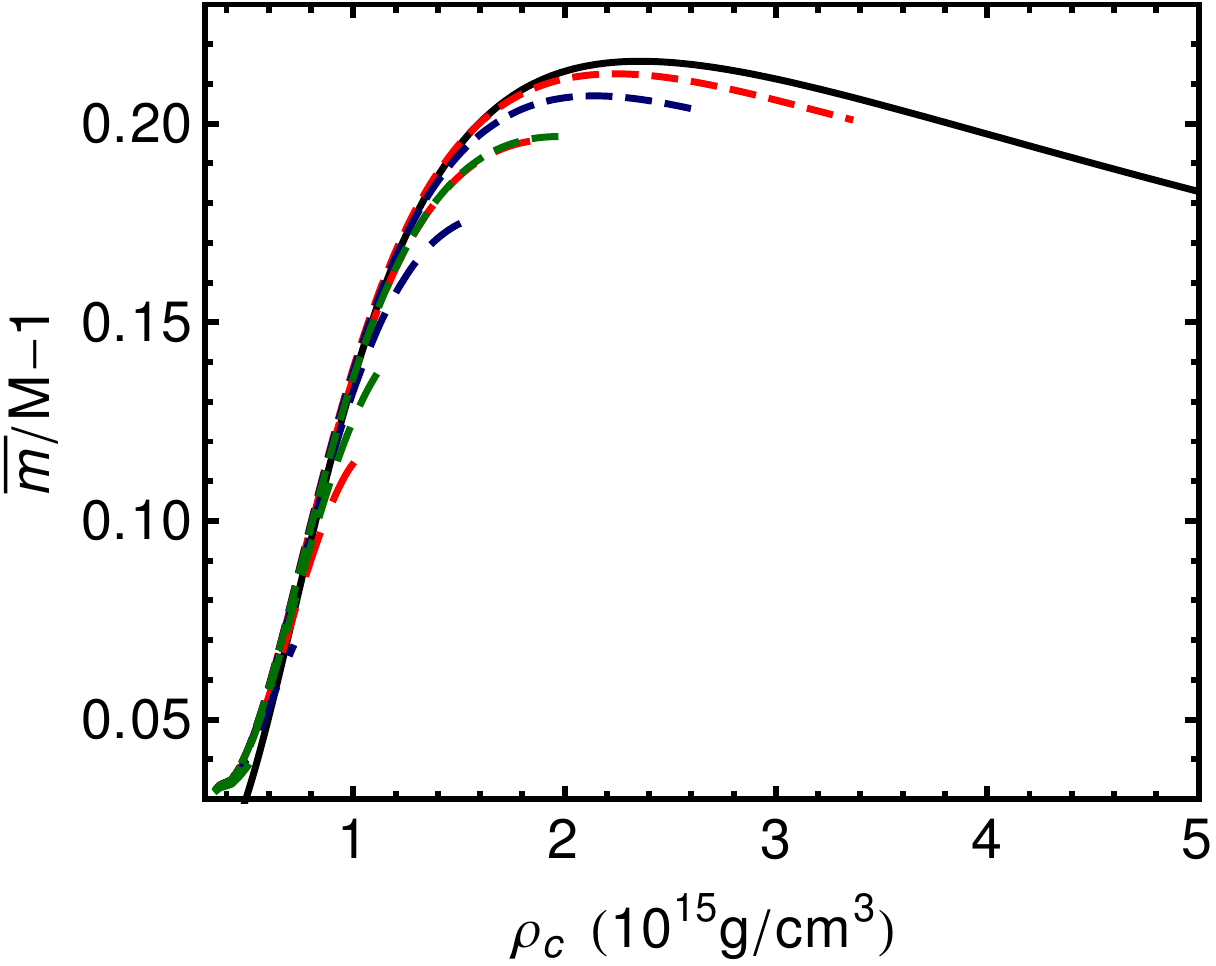}&
\includegraphics[width=0.5\textwidth]{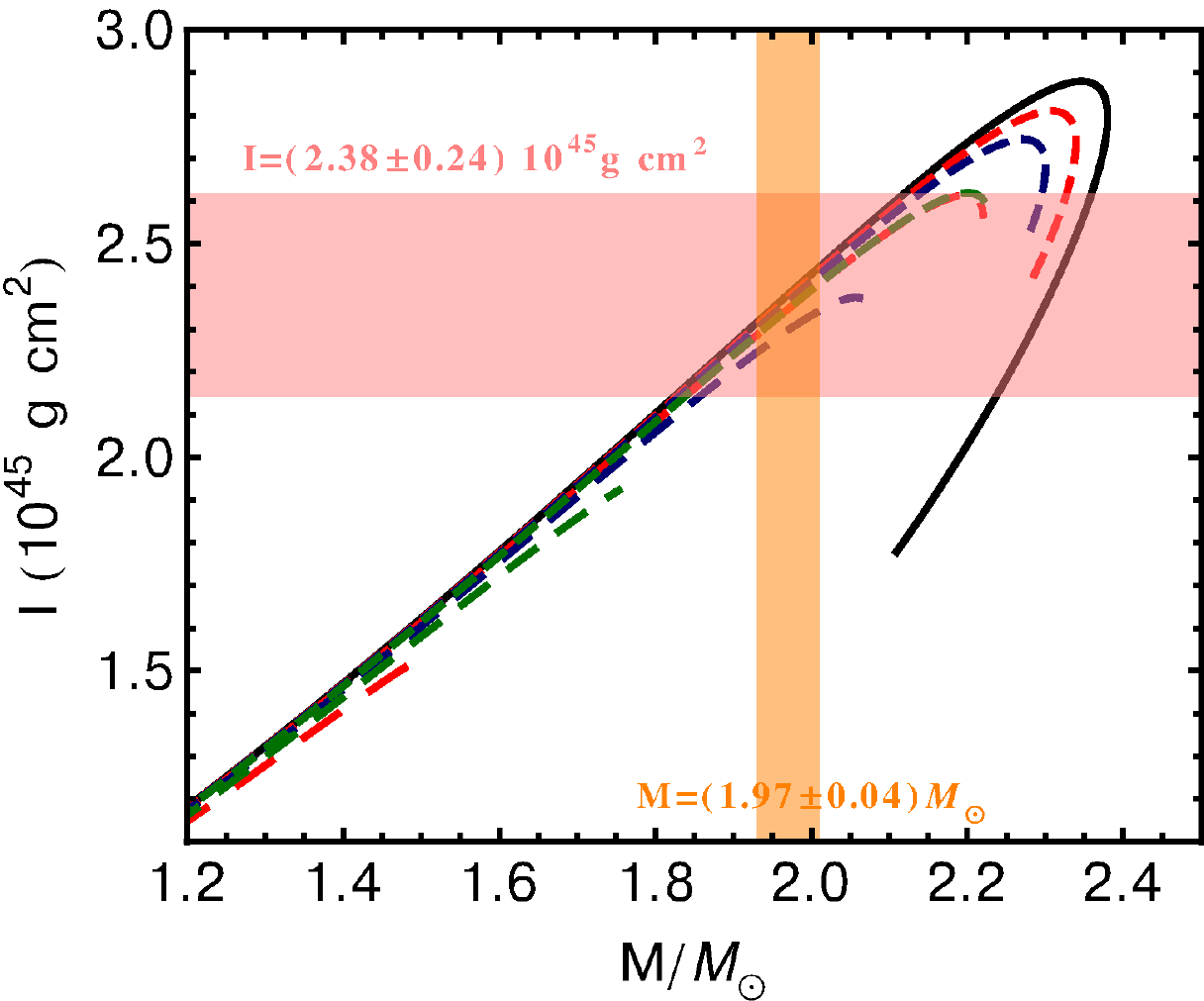}
\end{tabular}
\caption{Compact star models in EdGB gravity for different values of
  the parameters $\alpha$ (denoted by $\alpha_{\rm GB}$ in the text)
  and $\beta$, using the APR EOS. The bottom-left panel shows the NS
  binding energy as a function of its central energy density.  The
  bottom-right panel displays the moment of inertia as a function of
  the NS mass, together with the observation of a NS with $M\approx
  2M_\odot$~\cite{Demorest:2010bx} (see
  also~\cite{Antoniadis:2013pzd}) and a putative future observation of
  the moment of inertia in agreement with GR within
  $10\%$~\cite{Lattimer:2004nj}. Curves terminate when the
  condition~\eqref{alpha_max} is not
  fulfilled. [From~\cite{Pani:2011xm}.]
\label{fig:NSsEdGB}}
\end{center}
\end{figure*}

\paragraph{Static and slowly rotating solutions.}

\begin{figure}[htb]
\begin{center}
\begin{tabular}{cc}
\includegraphics[width=0.5\textwidth]{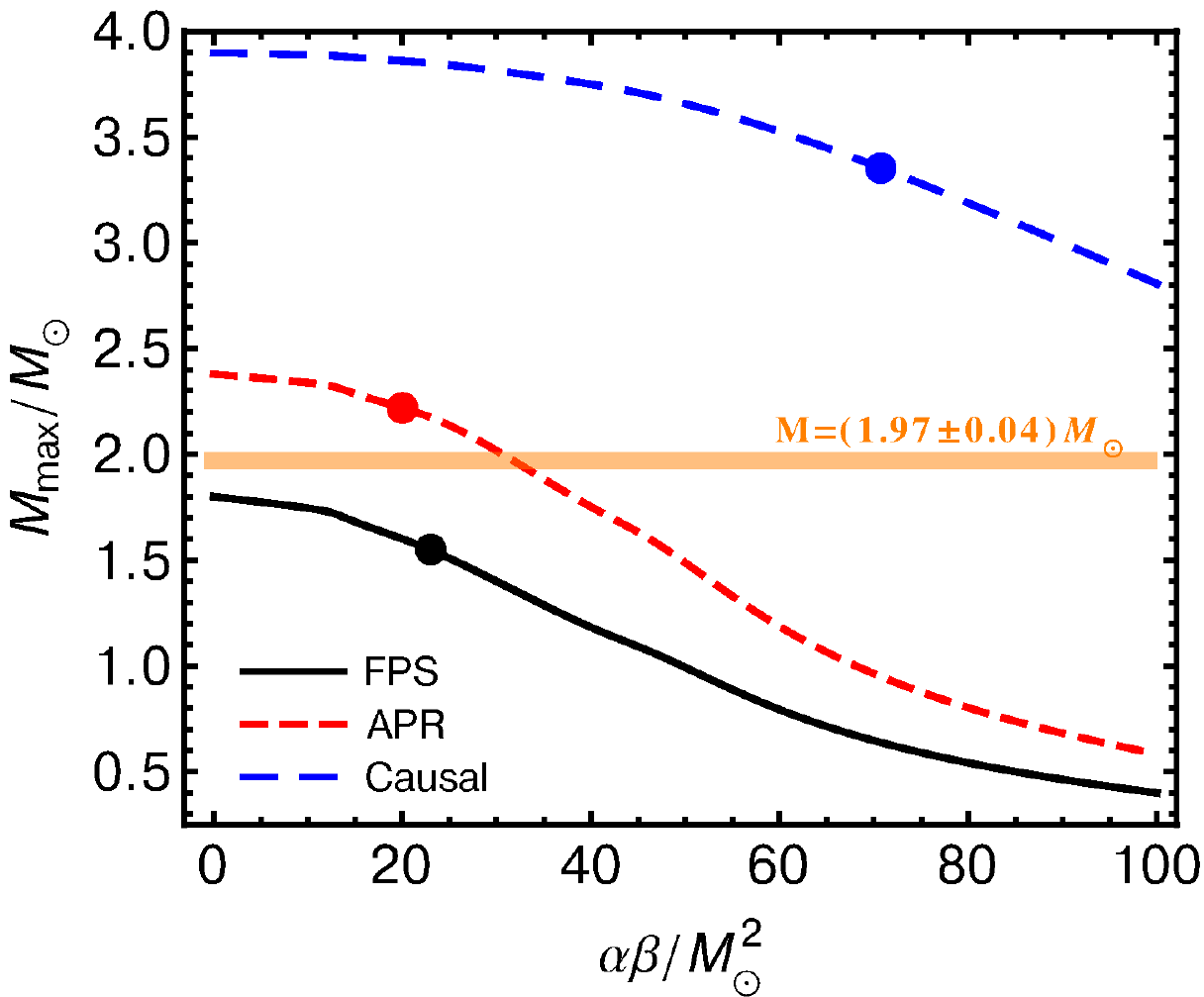}&
\includegraphics[width=0.5\textwidth]{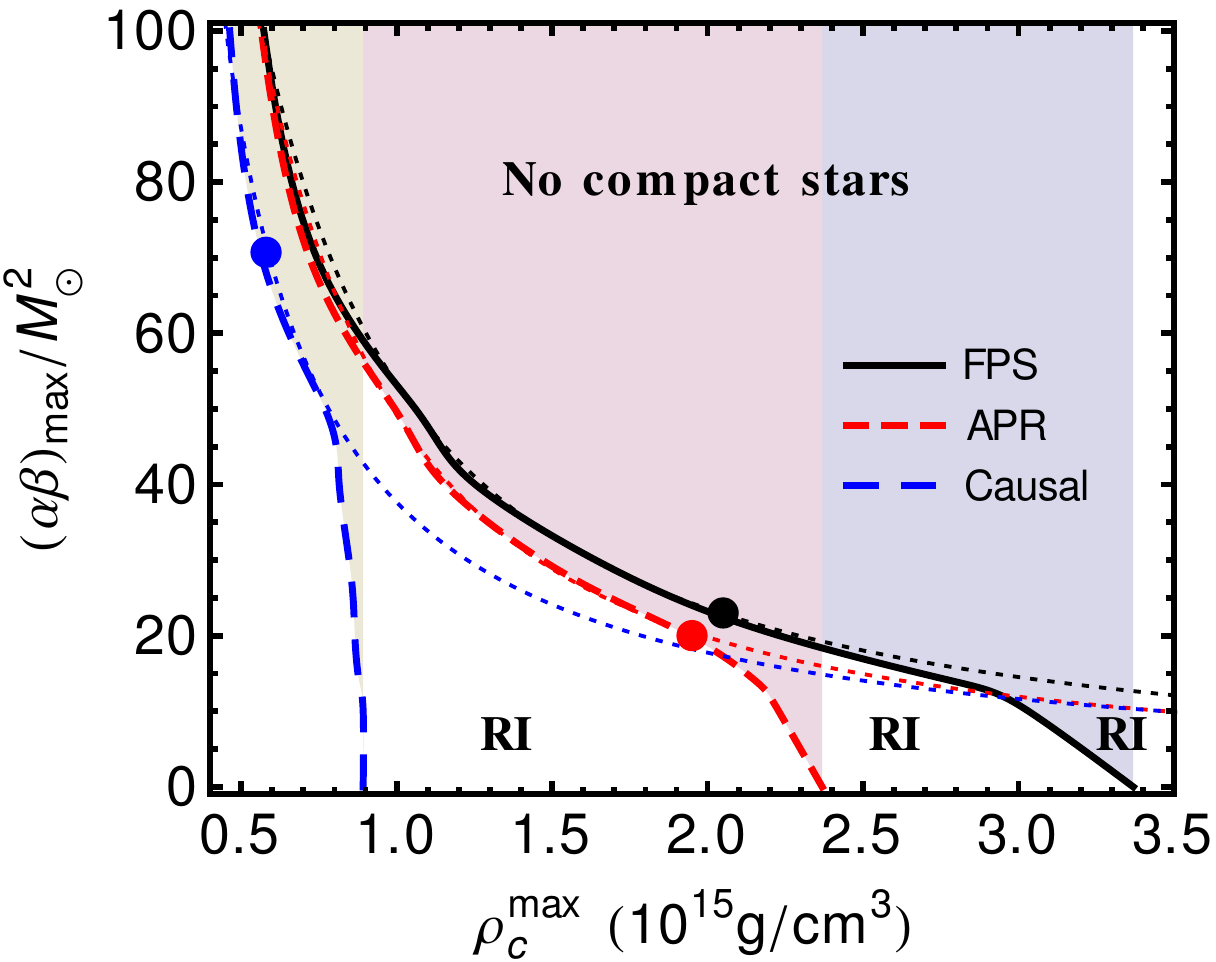}
\end{tabular}
\caption{Left panel: Maximum mass as a function of the product
  $\alpha_{\rm GB}\beta$ of the EdGB coupling parameters, for
  different EOS models and in the nonrotating case . To the left of
  the filled circle, this maximum mass corresponds to the radial
  stability criterion; to the right, it corresponds to the maximum
  central density for which we can construct static equilibrium
  models. The maximum observed NS mass $M\approx2 M_\odot$ is marked
  by a horizontal
  line~\cite{Demorest:2010bx,Antoniadis:2013pzd}. Right panel:
  Exclusion plot for the EdGB coupling in the small-field limit and
  for nonrotating models. In the region above the dotted lines no
  compact star solution can be constructed
  (cf.~Eq.~\eqref{alpha_max}). In the region above the thick lines
  (marked as ``RI,'' Radial Instability), static configurations are
  unstable against radial perturbations. Markers indicate the maximum
  central density of radially stable stars in
  GR. [From~\cite{Pani:2011xm}.]
\label{fig:NSsEdGB2}}
\end{center}
\end{figure}

Some properties of static and rotating solutions (computed at first
order in the slow-rotation approximation) are shown in
Figure~\ref{fig:NSsEdGB}.
Regardless of the EOS and for any value of $\alpha_{\rm GB}$, the
coupling to the dilaton tends to {\em reduce} the importance of
relativistic effects: this is again in contrast with the case of
scalar-tensor theory, where spontaneous scalarization increases the
relevance of relativistic effects (cf.~e.g.~the mass-radius curves in
Figure~\ref{fig:NS_ST2}). This trend is confirmed in the left panel of
Figure~\ref{fig:NSsEdGB2}, showing that the maximum gravitational mass
$M_{\rm max}$ decreases monotonically as a function of the product $\alpha_{\rm GB}\beta$ of the EdGB
coupling parameters.
Thus in EdGB gravity -- as in GR -- soft EOS models should be ruled
out by observations of high-mass NSs. As we will see, similar
conclusions apply to other theories.

An interesting feature of EdGB gravity is that, for fixed values of
$\alpha_{\rm GB}$ and $\beta$, there exists a constraint on the
central density $\rho_c$ (or central pressure $P_c$) that allows for
the existence of NSs. In the small-field limit, the constraint
reads~\cite{Pani:2011xm}
\begin{equation}
\label{alpha_max}
\begin{split}
 \alpha_{\rm GB}^2\beta^2<\frac{1}{7776 \pi  P_c^4 \rho_c}
\Bigg[&128 \rho_c^3-27 P_c^2 \rho_c+288 P_c \rho_c^2+54 P_c^3 \\
&{}-2\sqrt{ (3 P_c+\rho_c)
\left(3 P_c-8 \rho_c\right)^2 (3 P_c+4 \rho_c)^3}\Bigg].
\end{split}
\end{equation}
If we fix $\alpha_{\rm GB}\beta$, the condition above implies that no
NSs exist above some critical maximum central density.

Quite interestingly, the requirement that the theory should support a
maximum mass $M_{\rm max}$ larger than some fiducial observational
value can place rather stringent upper bounds on the EdGB
coupling. Under rather mild assumptions, Pani et
al.~\cite{Pani:2011xm} estimated that $\alpha_{\rm GB}\lesssim {\cal
  O}(10) M_\odot^2$, the precise number depending on the EOS and on
the value of $\beta$. This bound is slightly more stringent than the
purely theoretical bound that results from requiring the existence of
BHs in the theory~\cite{Kanti:1995vq,Pani:2009wy}.

Unfortunately, future observations of the moment of inertia are
unlikely to place even tighter bounds on the theory. This is because
deviations of the moment of inertia from its GR value are at most
$\sim 5\%$, at least in the slow-rotation limit~\cite{Pani:2011xm},
while the precision of future observations is expected to be $\sim
10\%$ in optimistic scenarios~\cite{Lattimer:2004nj}. At least for
EdGB gravity, we expect the most stringent constraints to come from
mass measurements, rather than from measurements of the moment of
inertia.

\begin{figure}[t!]
\begin{center}
\begin{tabular}{cc}
\includegraphics[width=0.7\textwidth, angle=0, clip=true]{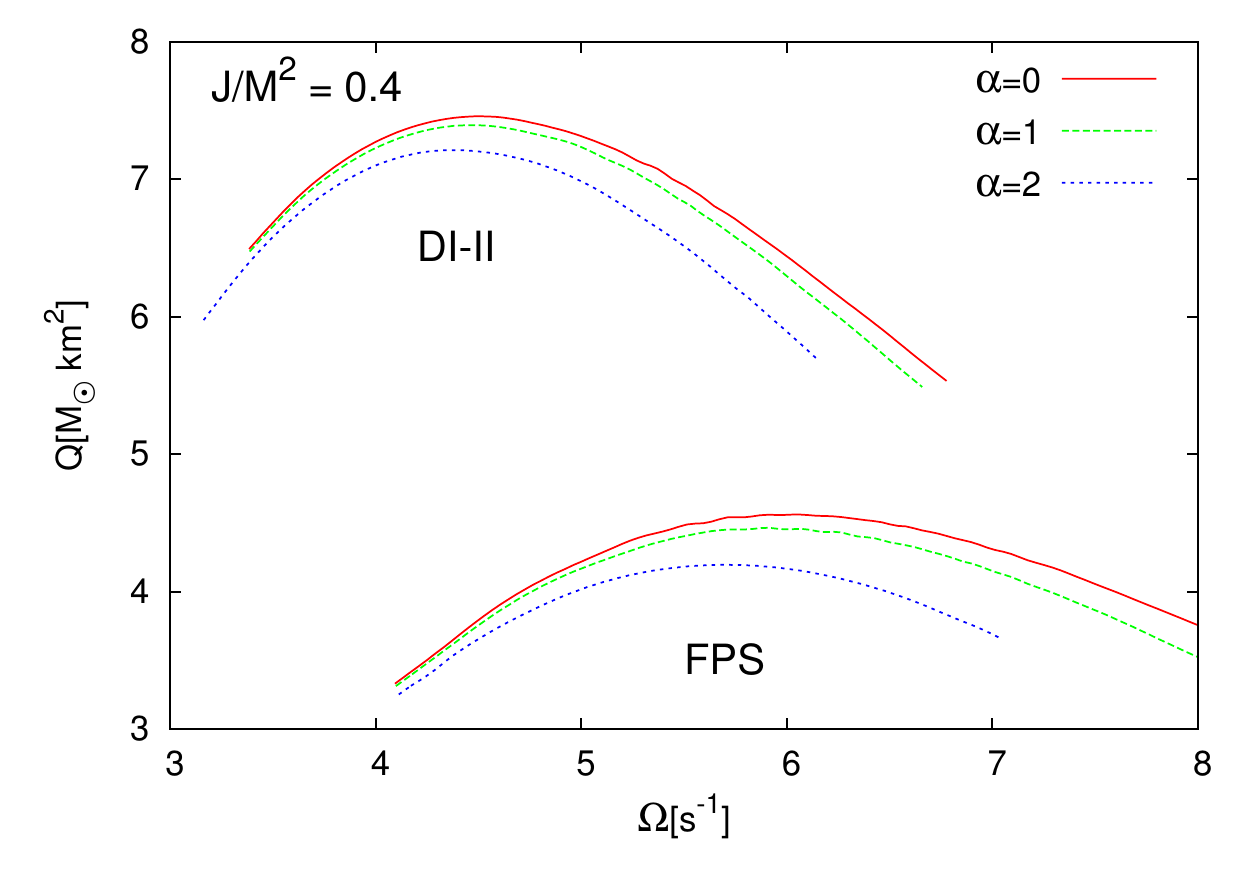}
\end{tabular}
\end{center}
\caption{The quadrupole moment $Q$ (in units of $M_\odot \cdot {\rm
    km}^2$) is shown versus the angular velocity $\Omega$ (in Hz) for
  the scaled angular momentum $j=0.4$ and EdGB couplings $\alpha\equiv
  \alpha_{\rm GB}/M_\odot^2=0$, 1 and 2~\cite{Kleihaus:2014lba}. The
  upper and lower sets of curves refer to the polytropic EOS
  DI-II~\cite{Diaz-Alonso:1985} and to an approximation of EOS
  FPS~\cite{Haensel:2004nu}, respectively.}
\label{KKMfig2}
\end{figure}

\paragraph{Rapidly rotating solutions.}
The previous analysis was limited to small (first-order) corrections
in the NS spin. Rapidly rotating NSs in EdGB gravity were considered
by Kleihaus et al.~\cite{Kleihaus:2014lba}.
Figure~\ref{KKMfig2} shows the quadrupole moment $Q$ in units of
$M_\odot \cdot {\rm km}^2$ for NSs at fixed angular momentum ($j\equiv
J/M^2=0.4$) versus the angular velocity $\Omega$.  The EdGB results
with coupling
$\alpha_{\rm GB}/M_\odot^2=1,2$ are compared to the case of GR ($\alpha_{\rm GB}=0$) for two
EOSs: the polytropic EOS from~\cite{Diaz-Alonso:1985} (DI-II) and the
FPS EOS~\cite{Haensel:2004nu} (fitted by a polytrope).
The quadrupole moment is more sensitive to the EOS than to the Gauss-Bonnet
coupling $\alpha_{\rm GB}$, so that even a putative measurement of the
quadrupole moment of a fast-spinning NS
can not be used to constrain the theory.

\begin{figure}[tb]
\begin{center}
\begin{tabular}{c}
  \includegraphics[width=0.7\textwidth,clip=true]{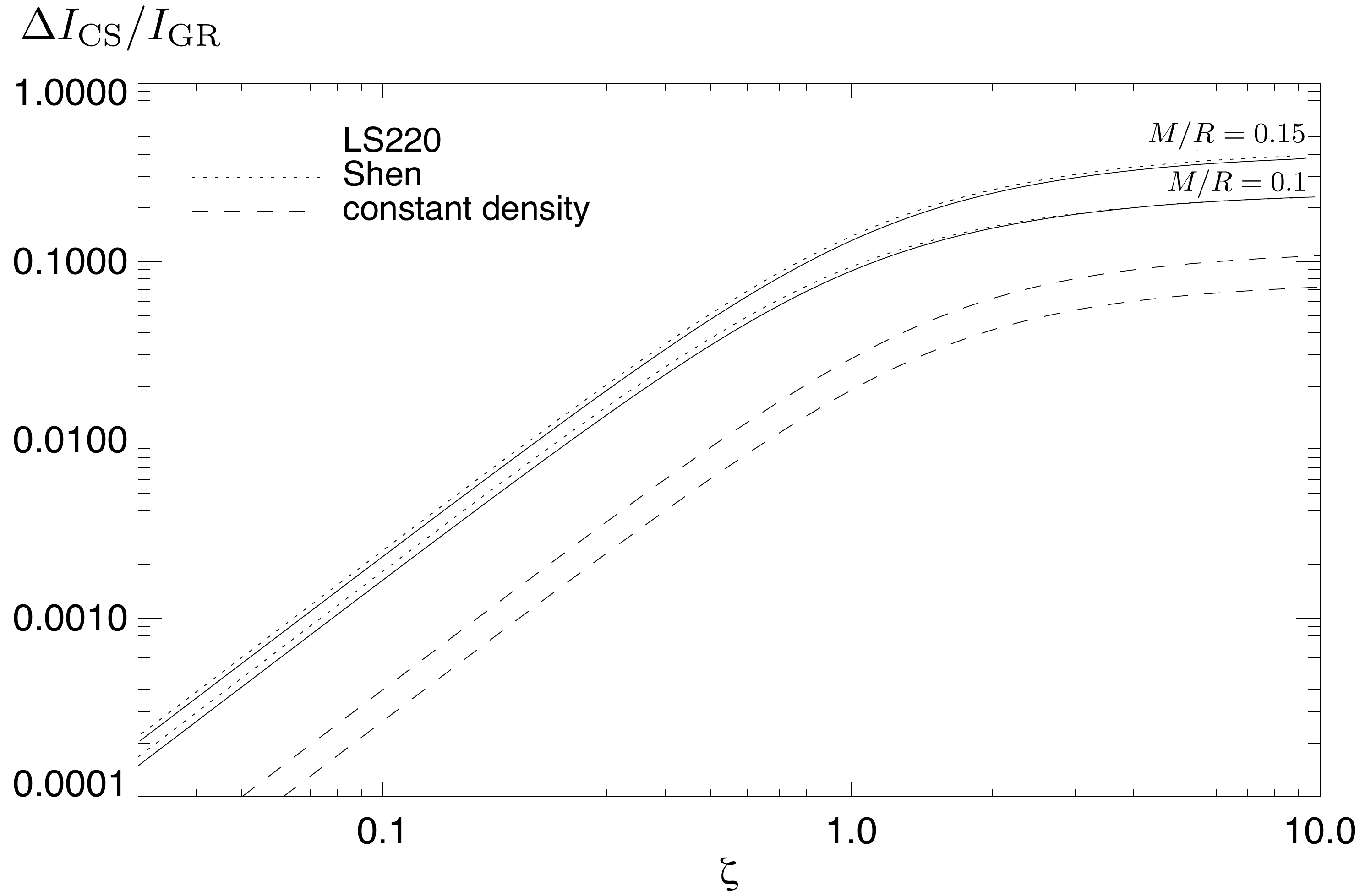}\\
 \includegraphics[width=0.7\textwidth,clip=true]{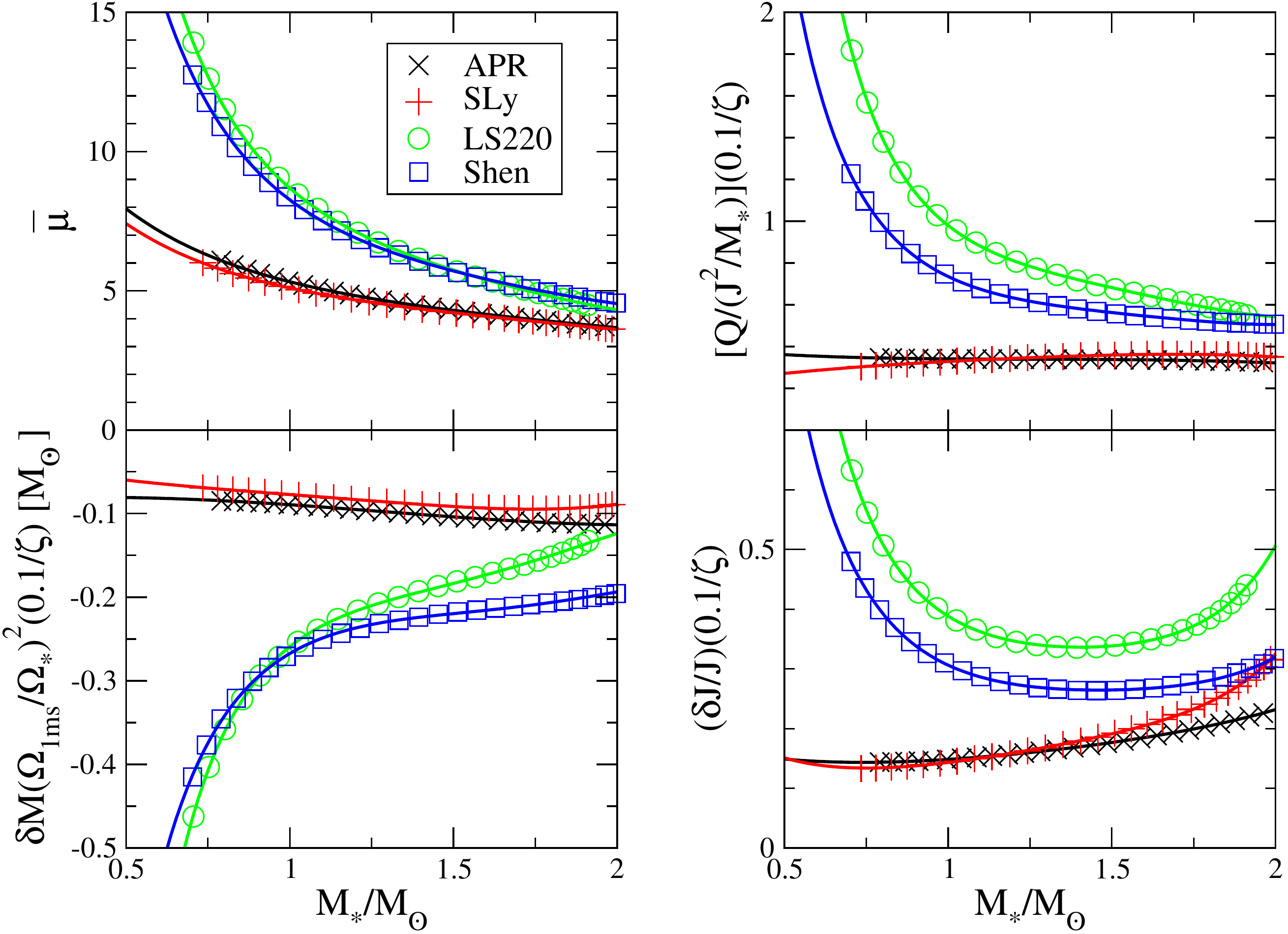}
\end{tabular}
\caption{\label{fig:NSsCS} Properties of slowly rotating NSs in dCS
  gravity. 
  The CS coupling $\zeta$ is related to $\alpha_{\rm CS}$, introduced in
  Eq.~\eqref{CSaction}, by $\zeta\equiv 4 \sqrt{2} \alpha_{\rm CS}
  M/R^3$ in the top panel, and $\zeta \equiv 4 \alpha_{CS}^2 M^2/R^6$
  in the bottom panels.
Top panel: change in the moment of inertia induced by the CS
modification as a function of the CS coupling for two different EOSs
and two different values of the NS compactness. The dashed lines show
the analytic result for a nonrelativistic constant-density
star~\cite{AliHaimoud:2011fw}.
Bottom panels: numerical (symbols) and fitted (curve) results for the
dimensionless scalar dipole charge $\bar{\mu}$ (top left), the
quadrupole correction $Q$ (top right), the mass shift $\delta M$
(bottom left), and the angular momentum shift $\delta J$ (bottom
right), as functions of the NS mass $M_\star$. The y-axes are rescaled as explained in~\cite{Yagi:2013mbt}.
[From~\cite{AliHaimoud:2011fw,Yagi:2013mbt}].
}
\end{center}
\end{figure}

\subsubsection{dCS theory}\label{sec:ns-dcs}

The CS term does not introduce any modification to GR for spherically
symmetric configuration: isolated NSs differ from their GR
counterparts only when they are rotating. So far, rotating solutions
have been computed only in the slow-rotation approximation. The
``baldness theorem'' discussed above implies that the scalar monopole
will vanish, but NSs can still support a nontrivial scalar field. In
fact, spinning NSs have a {\em scalar dipole} ``hair,'' which modifies
the geometry and the properties of the star. The CS correction to the
NS moment of inertia was calculated
in~\cite{AliHaimoud:2011fw,Yunes:2009ch} (at first order in slow
rotation). Calculations of the CS correction to the NS quadrupole
moment require going to second order in rotation, and they can be
found in~\cite{Yagi:2013mbt}.
A summary of the main results is presented in
Figure~\ref{fig:NSsCS}. The plot shows: (1) the CS corrections to the
moment of inertia, (2) the scalar dipole susceptibility (which is
related to the NS sensitivity, discussed in
Section~\ref{sec:sensitivities} below), (3) the CS corrections to the
quadrupole moment and (4) the mass and angular momentum shifts induced
by rotation at second order in the spin.  The mass shift is always
negative, while the quadrupole moment deformation is always positive.

\subsection{Lorentz-violating theories}

\label{sec:NS_LV}

\begin{figure}[htb]
\capstart
\begin{center}
\includegraphics[width=0.7\textwidth,clip=true]{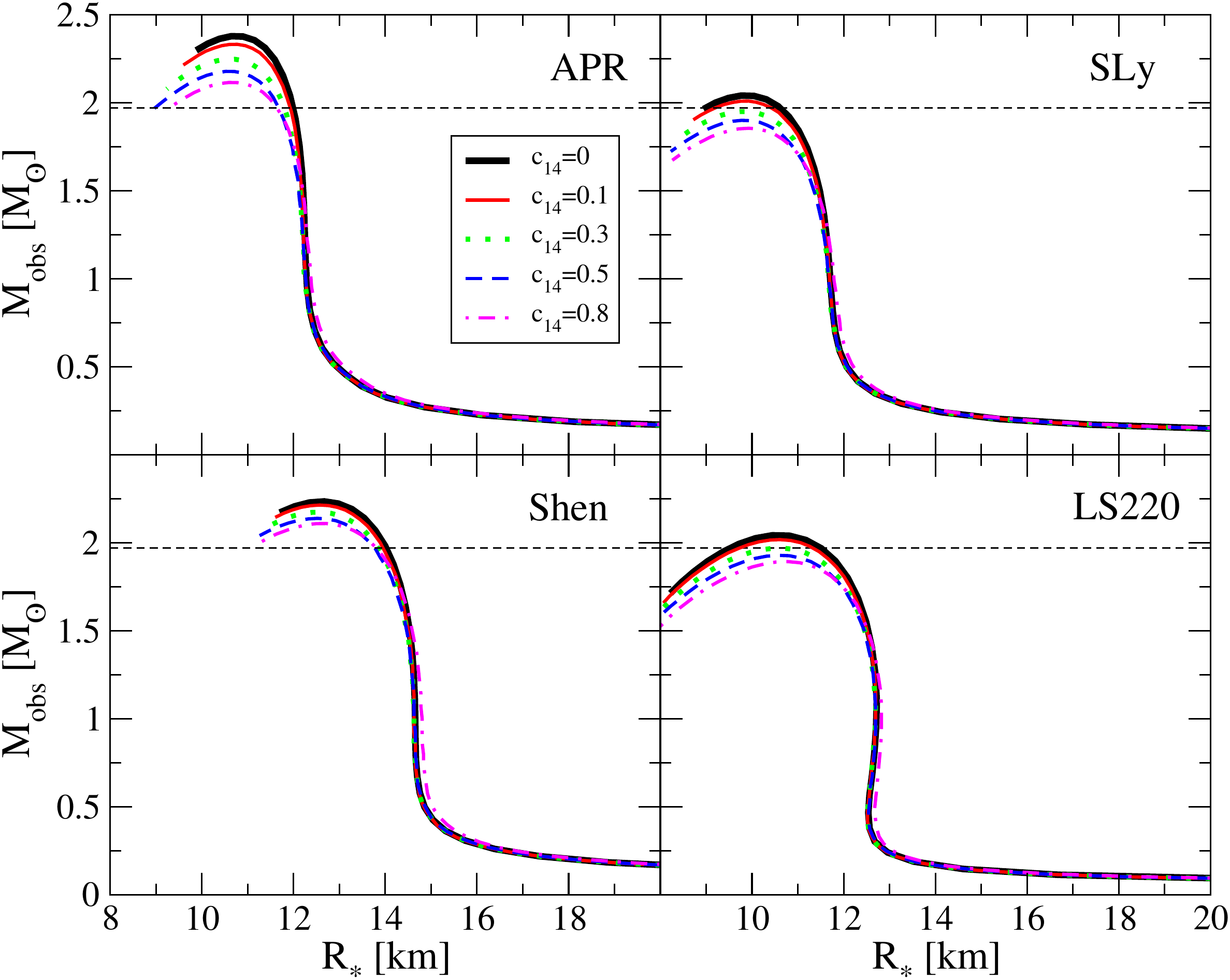}
\caption{\label{fig:MR-AE} Mass-radius relations in Einstein-\AE ther
  theory with various coupling constants and EOS models
  APR~\cite{Akmal:1998cf} (top left), SLy~\cite{2001A&A...380..151D}
  (top right), Shen~\cite{Shen:1998gq,Shen:1998by} (bottom left) and
  LS220~\cite{Lattimer:1991nc} (bottom right). The thick black curve
  is the GR result, and the horizontal dashed line corresponds to the
  lower mass bound from PSR J0348+0432~\cite{Antoniadis:2013pzd} (see
  also~\cite{Demorest:2010bx}). [From~\cite{Yagi:2013ava}.]}
\end{center}
\end{figure}

Static, spherically symmetric NS solutions in Einstein-\AE ther
theory were first constructed by Eling et al.~\cite{Eling:2007xh} (see
also~\cite{Eling:2006df,Greenwald:2009kp,Yagi:2013qpa,Yagi:2013ava}).
Figure~\ref{fig:MR-AE} presents the mass-radius relation in
Einstein-\AE ther theory for various coupling constants and EOSs. The
NS mass decreases (at fixed radius) as a function of the coupling
constant $c_{14}$. Spherically symmetric solutions in khronometric
theory are identical to those of Einstein-\AE ther
theory~\cite{Jacobson:2010mx,Blas:2011ni,Blas:2010hb,Barausse:2012ny},
therefore the conclusions drawn for Einstein-\AE ther theory also
apply to khronometric theory.

Stringent tests on Lorentz-violating theories come from binary
dynamics, which depends on the NS sensitivity
(cf.~Section~\ref{sec:sensitivities} below). In preparation for a
discussion of these tests, it is useful to examine stationary
configurations of nonspinning NS solutions in slow motion with
velocity $v^i$ with respect to the \AE ther field.

To be specific, let us consider Einstein-\AE ther theory.
Appropriately choosing a time coordinate\footnote{Following the
  analysis of Ref.~\cite{Yagi:2013ava}, here we consider a coordinate
  system that is comoving with respect to the NS fluid elements by
  aligning the time coordinate vector to the fluid 4-velocity
  $U^\mu$. We adopt asymptotically spherical coordinates in which the
  velocity direction of the \AE ther field agrees with the polar axis,
  and impose invariance under the simultaneous reflection of $v^i
  \mapsto -v^i$ and $t \mapsto -t$.}, the metric and \AE ther-field
ansatz read
\begin{align}
\label{metric-ansatz}
ds^2 ={}& -e^{\lambda (r)} dt^2 + e^{\mu (r)} d r^2 + r^2 \left( d\theta^2
+ \sin^2 \theta d \varphi^2 \right) 
\nn \\ 
&{}- 2 v V(r,\theta) dt dr -
2 v r S(r,\theta) dt d\theta + \mathcal{O}(v^2)\,,  
\\
\label{aether-ansatz} 
u_\mu ={}& -e^{\lambda(r)/2} \delta^t_\mu- v W(r,\theta ) \delta_\mu^r +
\mathcal{O}(v^2)\,, 
\end{align}
while the fluid four-velocity $U^\mu=e^{-\lambda (r)/2} \delta^\mu_t$ is the 
same as the static
solution, since we are working in the comoving frame; $\mu$ and
$\lambda$ are of zeroth order in velocity, while $V$, $S$ and $W$ are
linear in velocity. We decompose the latter set of functions using
Legendre polynomials as $V(r,\theta) = \sum_\ell k_\ell (r) P_\ell
(\cos \theta)$, $S(r,\theta) = \sum_\ell s_\ell (r) \frac{d P_\ell
  (\cos \theta)}{d\theta}$ and $W(r,\theta) = \sum_\ell w_\ell (r)
P_\ell (\cos \theta)$
and insert them into the field equations, that reduce to a set of
ordinary differential equations. One can show that all modes
decouple~\cite{Yagi:2013ava}.
Such equations are solved in both the interior and exterior, imposing
the matching condition that the solutions should be continuous and
differentiable at the NS surface.  As discussed in
Section~\ref{sec:sensitivities}, to extract the sensitivities, it is
useful to consider the asymptotic behavior of $k_1(r)$ at spatial
infinity given by
\be
\label{eq:k1-asympt}
k_1(r) = -1 + A \frac{M}{r} + \mathcal{O} \left( \frac{M^2}{r^2} \right)\,, 
\ee
where $A$ is a coefficient that is determined by the matching. In
Section~\ref{sec:sensitivities} we will show that this coefficient defines
the sensitivity in this theory: cf.~Eq.~\eqref{eq:sensitivity-AE}.

In Section~\ref{sec:sensitivities} we will also discuss NS
sensitivities in khronometric theory. In this case, thanks to the
additional condition of hypersurface-orthogonality of the vector
field, one can eliminate $W$ from the vector field ansatz and repeat
the same analysis as in the Einstein-\AE ther case
(cf.~\cite{Yagi:2013ava} for details).

\subsection{Massive gravity and Galileons}
There are very few phenomenological studies of NSs in massive
gravity. This is probably due to the technical difficulties related to
the Vainshtein mechanism: cf.~Section~\ref{subsec:massgrav}, and~\cite{Babichev:2013usa} for a review. In a nutshell, the helicity-0
graviton mode becomes strongly coupled at distances smaller than the
so-called ``Vainshtein radius,'' a characteristic length scale that
depends on both the theory parameters and the source. The Vainshtein
effect resolves the vDVZ discontinuity afflicting Fierz-Pauli theory,
but the presence of a new scale (which might be parametrically larger
than the stellar radius and is much smaller than the Compton
wavelength of the graviton) makes it difficult to obtain nonvacuum
solutions without resorting to approximations.

Before the recent developments in nonlinear, ghost-free massive
gravity~\cite{deRham:2010kj}, Damour et al.~\cite{Damour:2002gp}
reconsidered the vDVZ discontinuity problem and its possible
resolution through Vainshtein's nonlinear resummation of nonlinear
effects. As part of this study, the authors investigated the viability
of spherically symmetric stars in a nonlinear version of Fierz-Pauli
theory.  They found that some solutions show physical singularities,
but also that there exist regular solutions interpolating between a
modified GR interior and a de Sitter exterior, with curvature
proportional to the square of the graviton mass.
Another relevant study in this context was performed by Babichev et
al., who considered the problem of recovering GR from the decoupling
limit of the theory in the case of static, spherically symmetric
sources~\cite{Babichev:2009jt,Babichev:2010jd}.
To the best of our knowledge, the only attempt to construct static
stellar configurations in full dRGT massive gravity was carried out by
Gruzinov and Mirbabayi~\cite{Gruzinov:2011mm}, but phenomenological
studies of observational constraints (including stellar rotation and
realistic EOSs) are still lacking.
In the context of the cubic Galileon model
(cf. Section~\ref{subsec:massgrav}), in which the Vainshtein mechanism
suppresses the scalar field interactions with matter,
Ref.~\cite{Chagoya:2014fza} studied nonrelativistic, slowly rotating
stars and static relativistic stars, finding that deviations from GR
are suppressed at high densities.
Spherically-symmetric gravitational collapse in Galileon theories was studied 
in~\cite{Bellini:2012qn,Barreira:2013xea}.

\subsection{Gravity with auxiliary fields}\label{sec:NS_auxiliary}
Gravitational theories with auxiliary fields are equivalent to GR in
vacuum and only differ from GR in their coupling to matter. Therefore
these theories may look like the prototypical example of modified
theories of gravity whose phenomenology can be explored using NSs, but
not BHs. While this is true, it turns out that it is quite difficult
to put observational constraints on these theories, due to a severe
degeneracy between the nuclear matter EOS and beyond-GR
effects. Furthermore these theories do not violate the strong
equivalence principle, and NSs do not acquire extra charges that could
leave an imprint in binary dynamics. For this reason, the very concept
of ``sensitivity'' (discussed in Section~\ref{sec:sensitivities}
below) is meaningless in these theories. We will now briefly review
the literature on NSs in various theories with auxiliary fields.

\paragraph{Palatini $f({\cal R})$.}

The Palatini formulation of $f({\cal R})$ gravity has been
investigated in detail in other respects, but the literature on NS
solutions is not very extensive. Most studies dealt with the problem
of curvature singularities (a problem shared with EiBI gravity: see Refs.~\cite{Kainulainen:2006wz,Kainulainen:2007bt,Barausse:2007pn,Barausse:2007ys,Olmo:2008pv} and
below for a unified discussion), but there is no detailed analysis of
NS properties in Palatini gravity and of possible strong-field
tests. It is reasonable to expect that most of the phenomenology
should be at least qualitatively similar to EiBI theory, reviewed
below. This is confirmed by the findings of Pani et
al.~\cite{Pani:2013qfa}, who showed that Palatini $f({\cal R})$
gravity and EiBI gravity are only two representative examples of a
{\em single} class of theories that modify GR by adding nondynamical
fields.

\paragraph{Eddington-inspired Born-Infeld gravity.}

\begin{figure*}[tb]
 \begin{center}
 \begin{tabular}{cc}
 \includegraphics[width=0.5\textwidth]{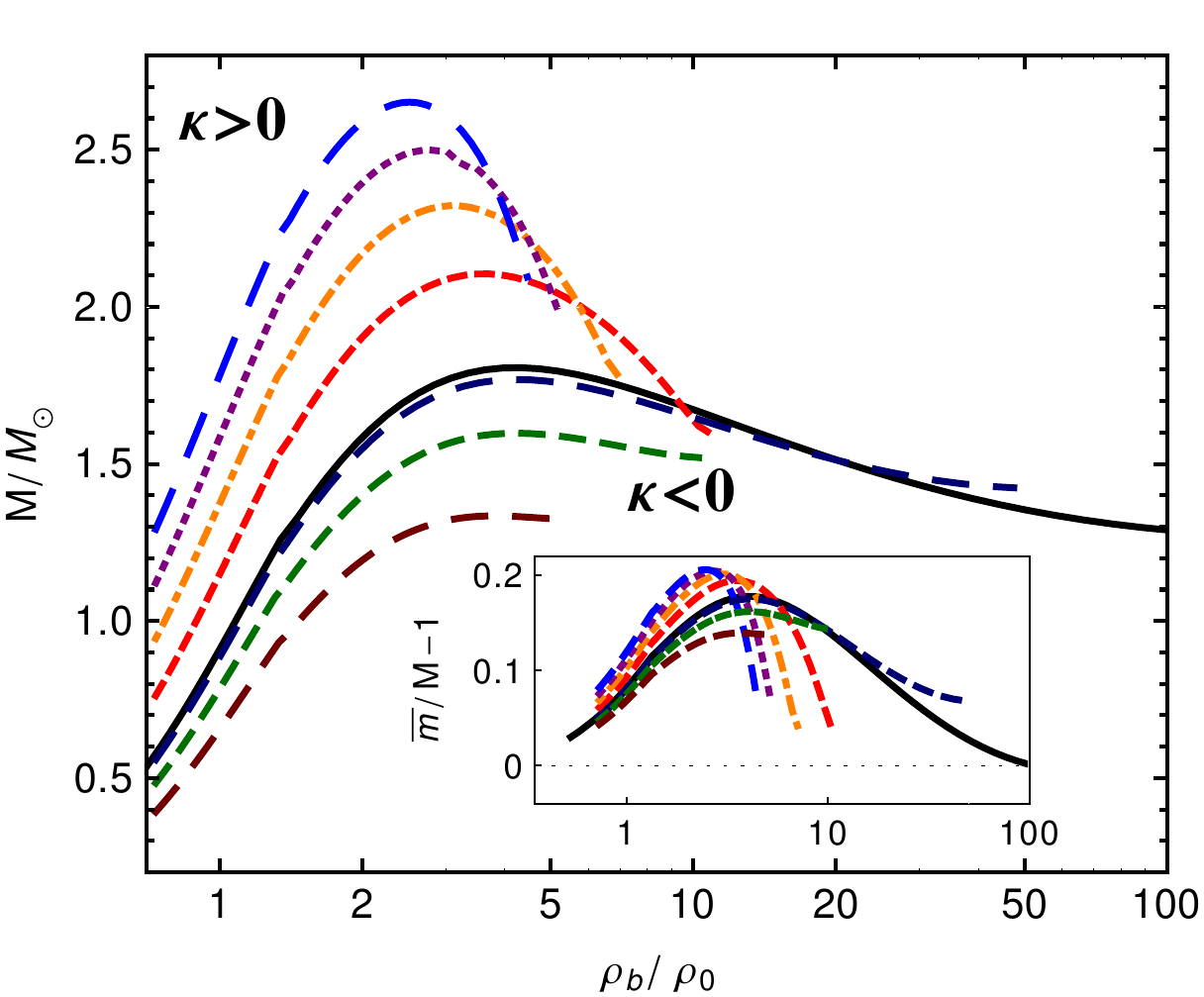}
 \includegraphics[width=0.5\textwidth]{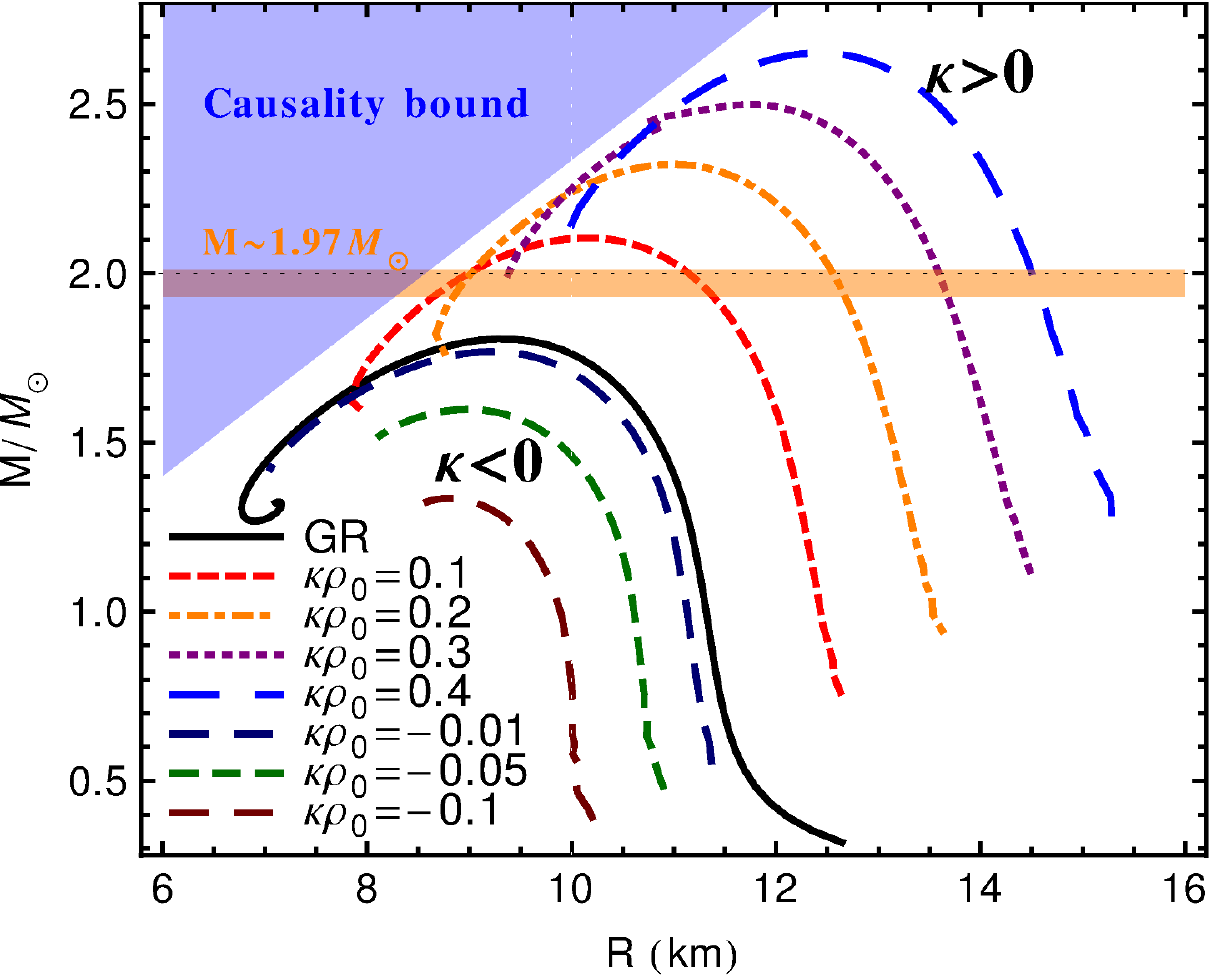}\\
 \end{tabular}
 \includegraphics[width=0.5\textwidth]{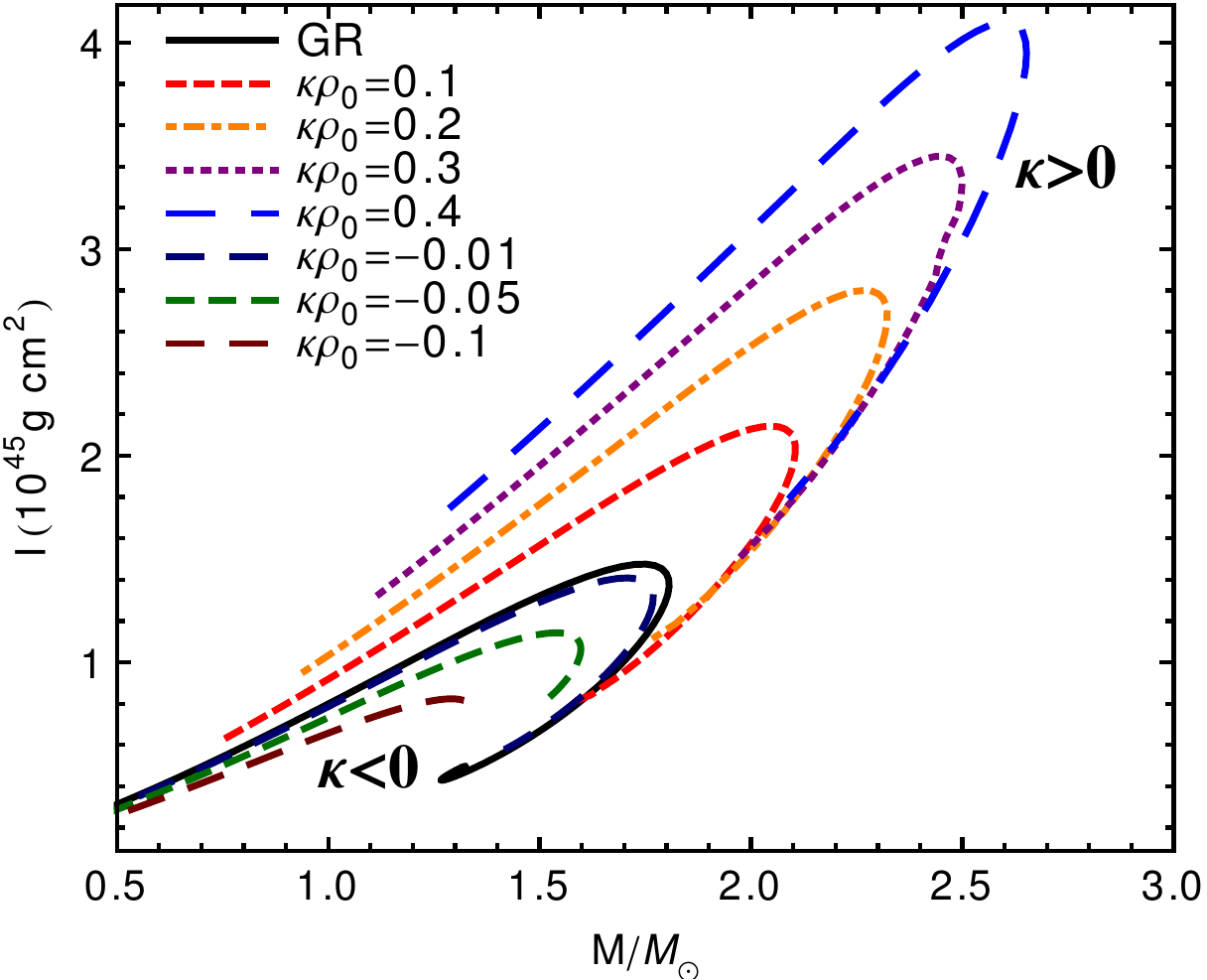}
 \caption{Compact stars in EiBI theory constructed using the FPS EOS
   and different values of the EiBI parameter $\kappa$
   [cf.~Eq.~\eqref{actionEiBI}]. Top-left panel: mass as a function of
   the central baryonic density $\rho_b$ (inset: binding energy as a
   function of $\rho_b$). Top-right panel: mass-radius
   relation. Bottom panel: moment of inertia as a function of the
   stellar mass. The central density and coupling constant are
   normalized by the typical density of nuclear matter,
   $\rho_0=8\times 10^{17}$~kg\ m$^{-3}$. Curves terminate when either
   condition~\eqref{lim1} or \eqref{lim2} are not fulfilled, and
   self-gravitating objects can not exist. In the top-right panel, a
   horizontal band corresponds to the maximum observed NS mass
   $M\simeq 2M_\odot$~\cite{Demorest:2010bx,Antoniadis:2013pzd},
   whereas the shaded region ($R\gtrsim 2.9 M$) is excluded by
   causality~\cite{Lattimer:2006xb}. [Adapted
     from~\cite{Pani:2012qb}.]
 \label{fig:NSsEiBI}}
 \end{center}
 \end{figure*}

Static and slowly rotating NSs in EiBI gravity were constructed
in~\cite{Pani:2011mg}, and their phenomenology was studied
in~\cite{Pani:2012qb,Harko:2013wka,Sham:2013sya}. For a given EOS, the
maximum mass of equilibrium NS configurations can be twice as large as
the corresponding GR model for experimentally viable values of the
EiBI parameter. An example of NS properties for a piecewise polytropic
EOS~\cite{Read:2008iy} that reproduces the FPS EOS is shown in Figure~\ref{fig:NSsEiBI} (cf.~\cite{Pani:2012qb} for
details).
Within GR, the FPS EOS is ruled out by observations of NSs with masses
$M=(1.97\pm0.04)M_\odot$~\cite{Demorest:2010bx} (horizontal band in
Figure~\ref{fig:NSsEiBI}) and
$M=(2.01\pm0.04)M_\odot$~\cite{Antoniadis:2013pzd}. In EiBI gravity
the maximum mass of a NS can be much larger than in GR, and
observations of high-mass NSs can be accommodated without invoking a
stiffer EOS.

An interesting property of EiBI gravity is that, for a given value of
$\kappa$, no self-gravitating objects can exist above some critical
central density $\rho_c$ (or pressure $P_c$). More specifically, one
must have
\begin{align}
P_c\kappa<1\,&\quad\text{for $\kappa>0$} \,,\label{lim1}\\
\rho_c|\kappa|<1\,&\quad\text{for $\kappa<0$}\,,\label{lim2}
\end{align}
where $P_c$ and $\rho_c$ are the central pressure and density,
respectively~\cite{Pani:2011mg}. Assuming that NSs can reach central
densities $\rho_c\sim10^{18}$~kg\ m$^{-3}$ and $P_c\sim
10^{34}$~N\ m$^{-2}$, these bounds would constrain the theory,
yielding $|\kappa|\lesssim 1 \mbox{ m}^5\mbox{kg}^{-1}\mbox{s}^{-2}$.
Avelino~\cite{Avelino:2012ge} derived even stronger constraints,
$|\kappa|<R^2$, from the existence of a self-gravitating astrophysical
objects of size $R$. For a typical NS this translates into the bound
\begin{equation}
 |\kappa|\lesssim 10^{-2} \mbox{ m}^5\mbox{kg}^{-1}\mbox{s}^{-2}\,.
\end{equation}
However these constraints are only indicative, because they are based
on the (untestable) assumption that matter in the NS core reaches
nuclear densities in EiBI gravity (as it does in GR). Furthermore,
there is no constraint from causality (cf.~the shaded region in the
top-right panel of Figure~\ref{fig:NSsEiBI}) because causality
constraints are always satisfied, even for large values of
$\kappa$. This is due to the existence of a maximum compactness
($M/R\lesssim0.3$) and it can be understood by looking at the theory's
effective stress-energy tensor~\cite{Delsate:2012ky}.

Sham et al.~\cite{Sham:2012qi} studied the radial stability of
relativistic stars in EiBI. Sotani~\cite{Sotani:2014xoa} investigated
nonradial oscillations in the Cowling approximation, finding that the
observation of the fundamental mode frequency may be used to
distinguish EiBI gravity from GR if the coupling is sufficiently
large. Because of the peculiar nonlinear coupling to matter, exotic
star-like solutions (such as pressureless
stars~\cite{Pani:2011mg,Pani:2012qb}, wormholes~\cite{Harko:2013aya}
and geons~\cite{Olmo:2013gqa}) exist in this theory. Furthermore (for
$\kappa>0$) the Chandrasekhar limit $M\lesssim 1.4 M_\odot$ on the
mass of white dwarfs is replaced by a minimum radius
condition~\cite{Pani:2012qb}: $R_{\rm WD}>\sqrt{{3\kappa}/({16\pi})}$.

Pani et al.~\cite{Pani:2012qb} studied nonrelativistic stellar
collapse in EiBI, finding that for any $\kappa>0$ the final state of
the collapse is a pressureless star (instead of a singularity, as in
GR). No relativistic simulations of collapse in the full theory are
available to date.

\paragraph{Degeneracy between NS EOS and nonlinear matter coupling.}

The degeneracy between beyond-GR effects and uncertainties in the NS
EOS is an intrinsic limitation in our ability to carry out precision
tests of gravity with NS observations. Uncertainties in the EOS often
translate into uncertainties in macroscopic observables (such as
masses, radii or oscillation frequencies) which are larger than
putative deviations from GR. This is usually the case for theories
(such as scalar-tensor theory) that are well constrained in the
weak-field limit.

Because gravitational theories with auxiliary fields are essentially
unconstrained in the weak-field limit, and they only modify GR in
their coupling to matter, NSs may look like ideal laboratories to
constrain them. Unfortunately, in this case GR modifications are
``maximally degenerate'' with EOS uncertainties. The reason is that
these theories do not contain extra dynamical fields, so the
right-hand side of the gravitational field
equations~\eqref{eqAuxiliary} can be interpreted as the stress-energy
tensor of an ``effective'' fluid with a contrived EOS. In particular,
Delsate et al.~\cite{Delsate:2012ky} proved analytically that EiBI
gravity coupled to a perfect fluid is \emph{equivalent} to GR with an
``effective'' perfect fluid stress-energy tensor.  If the original
$T_{\mu\nu}$ satisfies the energy conditions usually imposed in GR,
the same is not necessarily true for the effective stress-energy
tensor.  For this reason the singularity theorems of GR do not
apply~\cite{Delsate:2012ky}, and theories with auxiliary fields have
singularity-avoidance properties. The intrinsic ambiguity between
modifications in the gravitational coupling and variations in the EOS
makes it difficult (if not conceptually impossible) to constrain these
theories.

\paragraph{Curvature singularities.}

Palatini $f({\cal R})$ gravity and EiBI gravity are experimentally
viable theories passing all weak-field tests. 
However, it has been pointed out that their peculiar coupling with
matter might lead to the appearance of curvature singularities at the
surface of macroscopic objects, where large gradients are
present\footnote{As
  discussed in Section~\ref{subsec:auxiliary}, Palatini $f({\cal R})$
  gravity may be affected by other pathologies, including potential
  conflicts with the Standard Model~\cite{Flanagan:2003rb} and issues
  with averaging in cosmology~\cite{Flanagan:2003rb,Li:2008fa}. These
  problems are still debated (cf.~\cite{Olmo:2011uz} for a review),
  but they cast serious doubts on the viability of this class of
  theories.}~\cite{Barausse:2007pn,Pani:2012qd,Pani:2013qfa}. The root of the 
problem lies in the fact that
Eq.~\eqref{eqAuxiliary} contains third-order derivatives of the matter
fields. This is in contrast to GR, where at most first derivatives of
the matter fields appear on the right-hand side of Einstein's
equations.

This different structure is also evident in the Newtonian limit of the
theory. The solution of the modified Poisson
equation~\eqref{PoissonEiBI}, $\Phi=\Phi_{\rm N}+2\pi {\kappa}\rho$,
shows that the gravitational potential $\Phi$ is algebraically related
to~$\rho$.
Any matter configuration which is discontinuous or just not smooth
enough will produce discontinuities in the metric, as well as
singularities in the curvature invariants that depend on the second
derivatives of~$\Phi$, and ultimately lead to unacceptable
phenomenology~\cite{Barausse:2007pn,Barausse:2007ys,Barausse:2008nm,Pani:2012qd}. For
example, the Ricci curvature $R$ of a self-gravitating barotropic
perfect fluid would depend on the second derivatives of the pressure
field, whereas in GR it simply reads $R=8\pi(\rho-3P)$. If the
function $P(r)$ is continuous but not differentiable at the stellar
surface, then $P'(r)$ would be discontinuous at the radius and
$P''(r)$ would introduce an unacceptable Dirac-delta contribution to
the curvature.
This was shown explicitly in~\cite{Barausse:2007pn,Pani:2012qd} by
assuming a polytropic EOS near the stellar surface: for rather
standard values of the polytropic index, the scalar curvature is
actually \emph{divergent} near the NS radius due to the presence of
higher-order derivatives of the matter fields.
Strong deviations from GR are expected when the curvature becomes
unbound near the stellar surface. For example, surface singularities
would give rise to tidal forces which can be orders of magnitude
larger than in Einstein's theory. This and other consequences of
curvature singularities in theories with auxiliary fields are
discussed in~\cite{Barausse:2007pn,Barausse:2007ys,Barausse:2008nm,Pani:2012qd,Pani:2013qfa}.

The appearance of curvature singularities may look like a fatal blow
for these theories, but there is some controversy surrounding this
issue. Kainulainen et al.~\cite{Kainulainen:2007bt} criticized the
analysis of Palatini $f({\cal R})$ gravity carried out
in~\cite{Barausse:2007pn}, whereas
Refs.~\cite{Barausse:2007ys,Barausse:2008nm} argued that the original
analysis is essentially correct. Olmo~\cite{Olmo:2008pv} showed that
in a prototypical Palatini theory with $f({\cal R})={\cal R}+\lambda
{\cal R}^2$, where $\lambda$ is of the order of the Planck length
squared, the curvature invariant grows only at extremely small
densities, so that the theory is practically viable (but in this case
the macroscopic properties of NSs are indistinguishable from their GR
counterparts, and no interesting phenomenology can be probed with
astrophysical observations).
Finally, Kim~\cite{Kim:2013nna} pointed out that strong tidal forces
near the stellar surface can modify the effective EOS in such a way
that curvature singularities are removed. This result suggests that
the gravitational backreaction on the matter dynamics can modify the
effective description of the fluid. Further analysis is needed to test
the generality of this conclusion and the viability of these theories.

\subsection{Strong-field tests of gravity with universal relations in NSs and quark stars}\label{subsec:ILQ}

NSs and quark stars (QSs) are, in principle, excellent laboratories to
test strong-field gravity~\cite{Stairs:2003yja,Wex:2014nva}.  However,
the EOS of matter prevailing in the interior of NSs and QSs is poorly
known, and this poses an immediate difficulty if we want to probe the
nature of gravity with observations of compact stars. Measurable
macroscopic properties of these objects (such as the mass-radius
relation) are sensitive to both the EOS and the underlying
gravitational theory. Even if we could measure NS masses and radii
with a precision comparable to the deviations induced by gravitational
physics beyond GR, constraints on modifications of GR would demand a
knowledge of the EOS that is currently unavailable. Recently, Yagi et
al.~\cite{Yagi:2013bca,Yagi:2013awa} have shown that it is possible to
use certain nearly universal (i.e., almost EOS-independent) relations
between macroscopic observable properties of NSs to break this
degeneracy and carry out test of strong-field gravity {\em without
  prior detailed knowledge of the high-density EOS}.

\noindent
\paragraph{I-Love-Q and ``three-hair'' relations in GR.} 
Various nearly universal relations are known to hold in GR. For
example, relations among NS oscillation mode frequencies and certain
combinations of the NS mass and radius were found and discussed
in~\cite{Andersson:1997rn,Benhar:1998au,Kokkotas:1999bd,Tsui:2004qd},
and they were later extended to the case of rapid rotation in
\cite{Gaertig2010,Doneva2013}. Lau et al.~\cite{Lau:2009bu} found
universal relations between the NS moment of inertia and oscillation
frequencies. Ravenhall and Pethick~\cite{1994ApJ...424..846R} first
proposed a possible relation between the moment of inertia and the
stellar compactness $C=M/R$. Subsequently, more accurate relations
were found in~\cite{Lattimer:2000nx,Bejger:2002ty,Lattimer:2004nj}. A
relation between the quadrupole moment and the compactness for NSs and
QSs was reported in~\cite{2013MNRAS.433.1903U}. A functional relation
between the NS binding energy and the compactness was found
in~\cite{Lattimer:1989zz}, confirmed in~\cite{Prakash:1996xs} and
refined in~\cite{Lattimer:2000nx}. For rotating stars, the
mass-shedding (Keplerian) frequency was found to be related to the
compactness of nonrotating equilibrium models
in~\cite{Lattimer:2004pg}. This relation was confirmed
in~\cite{Haensel:2009wa}, where the authors pointed out an additional
relation between the equatorial radius for rotating configurations and
the radius for nonrotating configurations.

\begin{figure}[htb]
\capstart
\begin{center}
\begin{tabular}{l}
\includegraphics[width=7.5cm,clip=true]{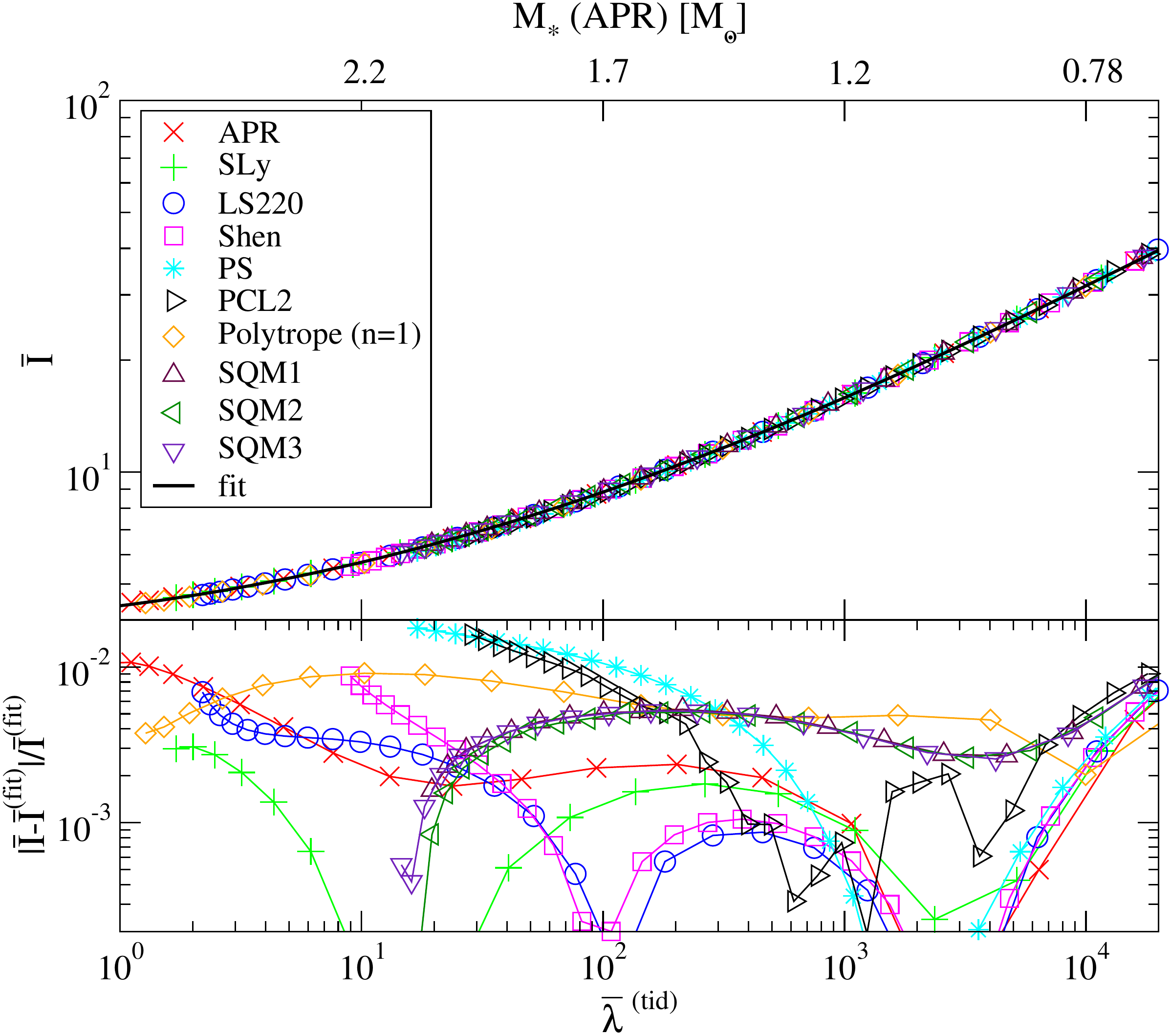}
\end{tabular}
\caption{(Top) Universal relation between the dimensionless moment of
  inertia $\bar{I}$ and the dimensionless tidal Love number
  $\bar{\lambda}^{\mathrm{(tid)}}$ for various EOSs
  (APR~\cite{Akmal:1998cf}, SLy~\cite{2001A&A...380..151D},
  LS220~\cite{Lattimer:1991nc}, Shen~\cite{Shen:1998gq,Shen:1998by},
  PS~\cite{1975NuPhA.237..507P}, PCL2~\cite{Prakash:1995uw} and the
  $n=1$ polytropic EOS for NSs; SQM1-3~\cite{Prakash:1995uw} for
  QSs), together with a fitting curve (solid).  The top $x$-axis shows
  the corresponding NS mass for the APR EOS. The parameter varied
  along each curve is the NS or QS central density, or equivalently
  the stellar compactness, with the latter increasing to the left of
  the plots. (Bottom) Fractional errors between the fitting curve and
  numerical results. The EOS-independence holds within a
  few percent accuracy. [From~\cite{Yagi:2013bca}.]}
\label{fig:I-Love}
\end{center}
\end{figure}

Yagi and Yunes~\cite{Yagi:2013bca,Yagi:2013awa} found new universal
relations among macroscopic
quantities that characterize slowly and uniformly rotating
unmagnetized NSs and QSs. If $M$ is the mass of a nonrotating model
and $\chi=J/M^2$ the dimensionless spin of the star, the universal
relations connect the following three quantities: the normalized
moment of inertia $\bar I=I/M^3$, the normalized tidal Love number (a
measure of stellar deformability) $\bar \lambda=\lambda/M^5$ and the
normalized quadrupole moment $\bar Q= - Q/(M^3\chi^2)$. For example,
Figure~\ref{fig:I-Love} shows the ``I-Love'' relation. These
``I-Love-Q'' relations are remarkably independent of the EOS -- more
so than the relations listed earlier\footnote{Several works relaxed
  some of the assumptions made in the original papers. The
  universality was first confirmed in~\cite{Lattimer:2012xj} using a
  wider range of EOSs. Maselli et al.~\cite{Maselli:2013mva} studied
  the I-Love relations for merging binary NSs, finding small
  deviations ($\sim 10\%$) from the relations that are valid for
  isolated NSs. Haskell et al.~\cite{2014MNRAS.438L..71H} found that
  universality holds also for magnetized NSs, as long as the magnetic
  fields are smaller than $\sim 10^{12} \mathrm{G}$ and the spin
  periods smaller than 0.1s. Doneva et al.~\cite{Doneva:2013rha}
  relaxed the slow-rotation approximation.
Using the RNS code~\cite{Stergioulas:1994ea}, they found that the I-Q
relation is spin-dependent, and that for a fixed spin frequency, the
universality only holds for a subclass of NS EOSs, concluding that the
universality is lost for rapidly rotating NSs and QSs. However,
subsequent work~\cite{Pappas:2013naa,Chakrabarti:2013tca,Yagi:2014bxa}
showed that the universality is still preserved for fixed {\em
  dimensionless} spin parameters. Finally, Martinon et
al.~\cite{Martinon:2014uua} investigated universality for the
nonbarotropic EOSs typical of proto-NSs. Deviations from universality
can be as large as $\sim 30\%$ in the early stages of NS formation,
but they decrease as soon as the entropy gradients smooth out.}.

Pappas and Apostolatos~\cite{Pappas:2013naa} studied the relation
between the NS current octupole and mass quadrupole moments and found
that it is insensitive to both, the EOS and the NS spin. Stein et
al.~\cite{Stein:2013ofa} confirmed and extended this finding by
investigating relations among multipole moments for
uniformly rotating, Newtonian polytropes, under the additional
assumption that the isodensity contours are self-similar ellipsoids~\cite{Lai:1993ve}.
They found the following ``Newtonian 3-hair
relation'' for the mass-multipole moments $M_\ell$ and the
current-multipole moments $S_\ell$:
\begin{equation}
\label{eq:3-hair}
M_\ell + i \frac{q}{a} S_\ell = \bar{B}_{n,\lfloor \frac{\ell -1}{2} \rfloor} M_0 (i q)^\ell\,, 
\end{equation}
where $a \equiv S_1/M_0$, $i q \equiv \sqrt{M_2/M_0}$, $\lfloor x
\rfloor$ stands for the largest integer that does not exceed $x$, and
the $\bar{B}_{n,\lfloor \frac{\ell -1}{2} \rfloor}$'s are constant
coefficients determined by solving (in general numerically) the
Lane-Emden equation.
Eq.~(\ref{eq:3-hair}) states that, given a polytropic EOS, all
multipole moments are functions of the first three moments -- namely
the mass, spin and quadrupole moment.
It is an approximate generalization of the BH no-hair
relation~\cite{Hansen:1974zz}, $M_\ell + i S_\ell = M_0 (i a)^\ell$ ,
which states that all multipole moments of a stationary and
axisymmetric (uncharged) BH in GR can be
expressed in terms of its mass and spin.

\begin{figure}[tb]
\capstart
\begin{center}
\begin{tabular}{l}
\includegraphics[width=7.5cm,clip=true]{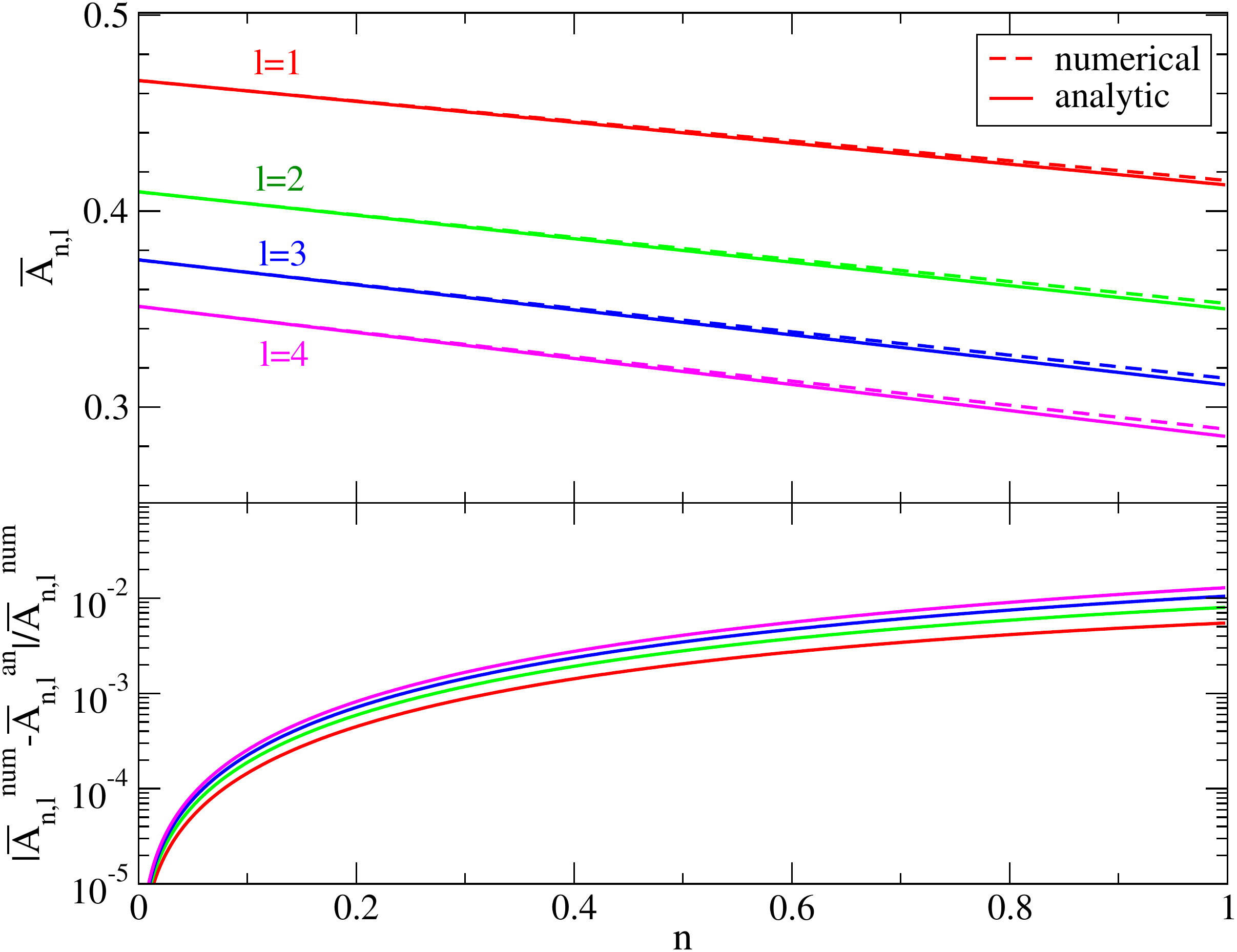}
\end{tabular}
\caption{Top inset: $\bar A_{n,\ell} \equiv (\bar B_{n,\ell})^{\ell}$,
  where $\bar{B}_{n,\ell}$ represents the coefficient of the 3-hair
  relation in Eq.~\eqref{eq:3-hair}, is plotted against the polytropic
  index $n$.  Dashed lines are numerical solutions of the Lane-Emden
  equation; solid lines are analytic, perturbed Lane-Emden solutions
  around $n=0$. Observe that $\bar{A}_{n,\ell}$ only changes by less
  than 10\% from $n=0$ to $n=1$. Bottom inset: fractional differences
  between the numerical and analytic relations. The analytic result is
  within $\mathcal{O}(1)\%$ of the numerical result even for
  $n=1$. [From~\cite{Chatziioannou:2014tha}.]}
\label{fig:Abar}
\end{center}
\end{figure}

The dashed curves in the top panel of Figure~\ref{fig:Abar} show that
this relation is EOS-independent to better than $10\%$ accuracy for
low-order multipoles ($\ell \leq 10$).
Ref.~\cite{Chatziioannou:2014tha} showed that the relation also
applies to Newtonian stars with piecewise polytropic EOSs (a good
approximation to realistic NS EOSs~\cite{Read:2008iy}) and
derived a purely analytic relation (solid curves in the top panel of
Figure~\ref{fig:Abar}) 
by perturbing $n=0$ polytropes and using the
perturbed Lane-Emden solution in~\cite{1978SvA....22..711S}.  
The analytic relation approximates numerical results within
$\mathcal{O}(1)\%$ accuracy, as shown in the bottom panel of
Figure~\ref{fig:Abar}. The validity of three-hair relations was
confirmed in the relativistic regime up to hexadecapole order for both
NSs and QSs~\cite{Yagi:2014bxa}, and its origin was investigated
in~\cite{Yagi:2014qua}.

The I-Love-Q relations and their generalizations to higher multipoles
have various interesting applications. Any astrophysical measurement
of either $\bar I$, $\bar \lambda$ or $\bar Q$ will automatically
yield the other two quantities, regardless of the uncertainty in the
EOS.
For instance, the tidal Love number $\lambda$ may be measured
with future GW observations (see Section~\ref{sec:sub:stellarmass}),
and such a measurement would give information on $I$ and $Q$.
On the nuclear physics front, the universal relations allow us to
break degeneracies among parameters in NS observations. For example,
Psaltis et al.~\cite{Psaltis:2013zja} showed that the X-ray pulse
profile emitted by hot spots on the NS surface depends not only on the
NS mass and radius, but also on its moment of inertia, quadrupole
moment, eccentricity, etcetera. By using the universal relations
reported in~\cite{Baubock:2013gna}, including the I-Q relation, one
can eliminate some of the model parameters. This breaks degeneracies
in parameter estimation and may allow future X-ray satellites such as
NICER~\cite{2012SPIE.8443E..13G} and
LOFT~\cite{2012AAS...21924906R,Feroci:2012qh} to measure the NS mass
and radius within $\sim 5\%$ accuracy~\cite{Psaltis:2013fha}, as long
as systematic errors are under control~\cite{Lo:2013ava}.

\noindent
\paragraph{Universal relations in other theories of gravity and tests of GR.} 

In the context of this review, the I-Love-Q and ``three-hair''
relations are interesting because they can break the degeneracy
between the uncertainties in nuclear and gravitational physics, and
allow us to perform strong-field tests of gravity with NSs.
Since in general the relations depend on the underlying gravitational
theory (but see below for caveats), if one can measure {\em any two}
of the I-Love-Q quantities independently, one can in principle perform
a model-independent consistency test of GR or test a specific
alternative theory~\cite{Yagi:2013bca}.
Model-independent tests can also be obtained by measuring the first
four multipole moments, i.e., mass, angular momentum, mass quadrupole
and spin octupole~\cite{Pappas:2013naa}.

\begin{figure*}[tb]
\capstart
\begin{center}
\begin{tabular}{l}
\includegraphics[width=6.2cm,clip=true]{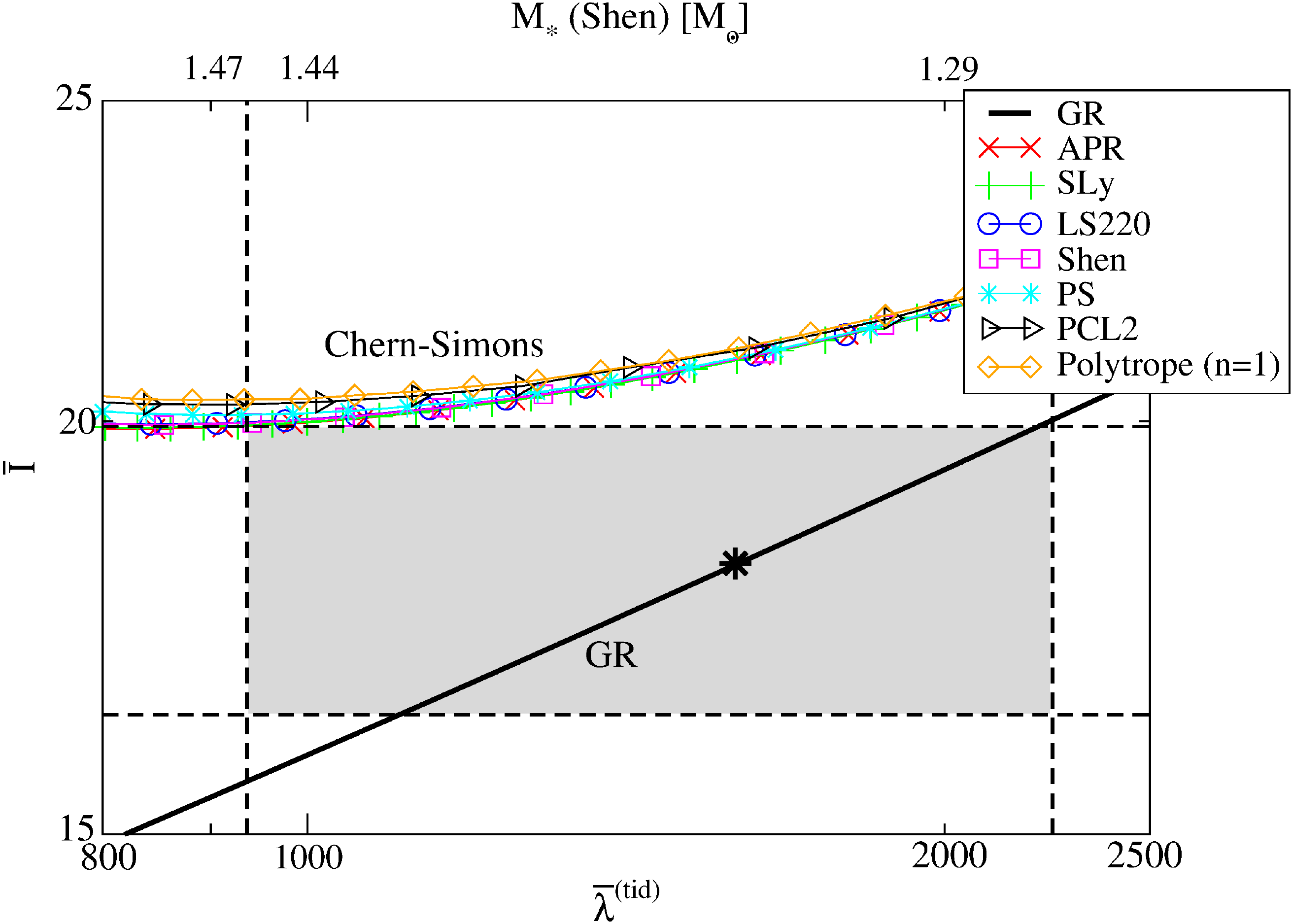}
\includegraphics[width=6.2cm,clip=true]{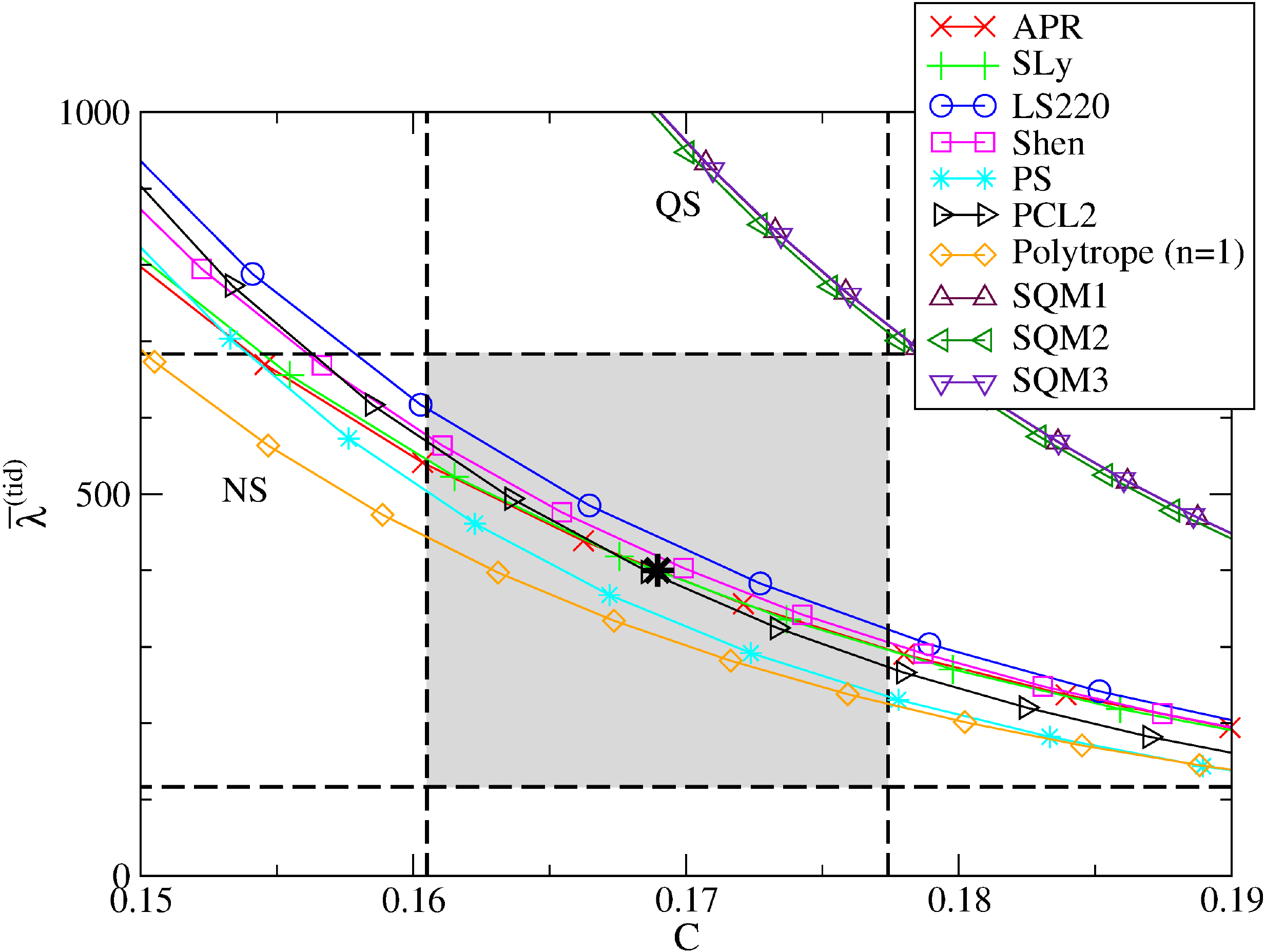}
\end{tabular}
\caption{Left panel: The I-Love relation in dCS gravity for various
  EOSs. The shaded region represents a hypothetical error box in the
  I-Love plane resulting from independent measurements of the moment
  of inertia and of the tidal Love number (the black asterisk marks
  the hypothetically measured values). The black solid line shows the
  I-Love relation in GR. The top axis shows the NS mass $M_*$ for the
  Shen EOS. An alternative theory is consistent with the measurement
  only if the modified I-Love relation passes through the error
  box. Right panel: Love-compactness relation in GR for various
  EOSs. The shaded region represents a hypothetical measurement error
  box in the Love-compactness plane. Different NS EOSs are consistent
  with the error box, and therefore it may be possible to carry out
  tests of GR (this conclusion does not apply to QSs). [From~\cite{Yagi:2013bca}.]}
\label{fig:I-Love-error}
\end{center}
\end{figure*}

As an example, Refs.~\cite{Yagi:2013bca,Yagi:2013awa} studied the
I-Love-Q relations in dCS
gravity~\cite{jackiw:2003:cmo,Yunes:2009hc,Alexander:2009tp}, whose
action is given by Eq.~\eqref{CSaction}.
Corrections to the NS moment of inertia and quadrupole moment were
calculated in~\cite{Yunes:2009ch,AliHaimoud:2011fw}
and~\cite{Yagi:2013mbt} by constructing slowly rotating NS solutions
in dCS gravity that are valid to linear and quadratic order in the
spin, respectively (the $\ell =2$ electric tidal Love number is the
same as in GR~\cite{Yagi:2011xp}).
Yagi et al.~\cite{Yagi:2013bca,Yagi:2013awa} found that an
EOS-independent I-Q relation holds in dCS gravity for a fixed
$\xi^{1/4}/M_*$ (where $\xi\equiv 4\alpha_{\rm CS}^2$ in the notation of Eq.~\eqref{CSaction}), as shown in the left panel of
Figure~\ref{fig:I-Love-error}, and that this relation generally differs
from GR. The dCS relation shown in Figure~\ref{fig:I-Love-error} is the
marginally allowed case, corresponding to $\xi^{1/4} = 1.85 \times 10^4
M_*$: this constraint is six orders of magnitude stronger than the
current best bound from Solar System and table-top
experiments~\cite{AliHaimoud:2011fw,Yagi:2012ya}.

This hypothetical test would not be so constraining for other theories
of gravity. For example, Sham et al.~\cite{Sham:2013cya} calculated
universal I-Love-Q relations (as well as similar relations involving
$f$-mode oscillation frequencies) in Eddington-inspired Born-Infeld
(EiBI) gravity~\cite{Banados:2010ix}, finding that they are almost the
same as in GR. The reason is that, in this theory, modified-gravity
effects are equivalent to modifying the EOS within
GR~\cite{Delsate:2012ky}, and we already know that the relations are
insensitive to the EOS.

\begin{figure*}[tb]
\capstart
\begin{center}
\begin{tabular}{cc}
\includegraphics[width=6cm]{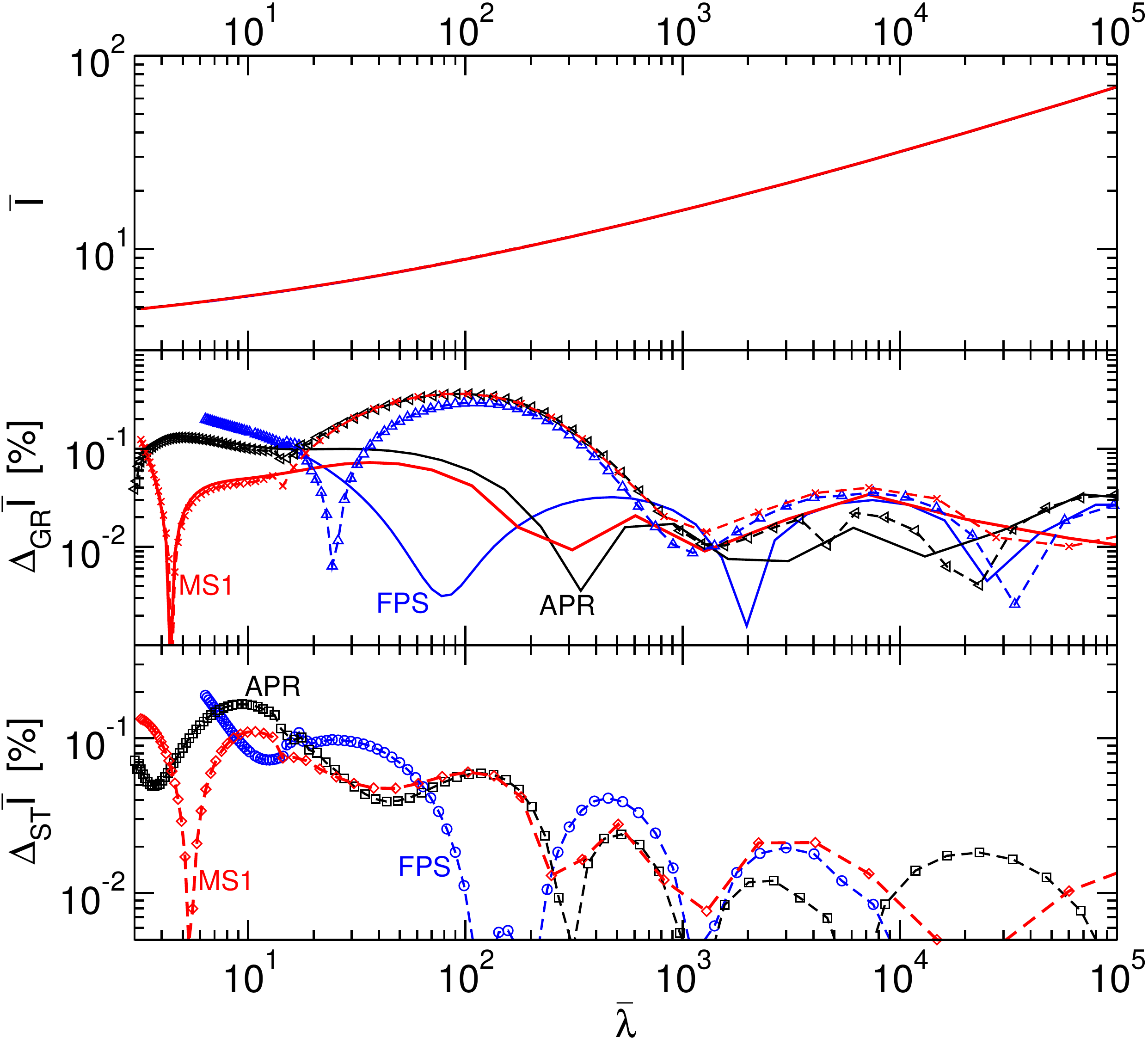}&
\includegraphics[width=6cm]{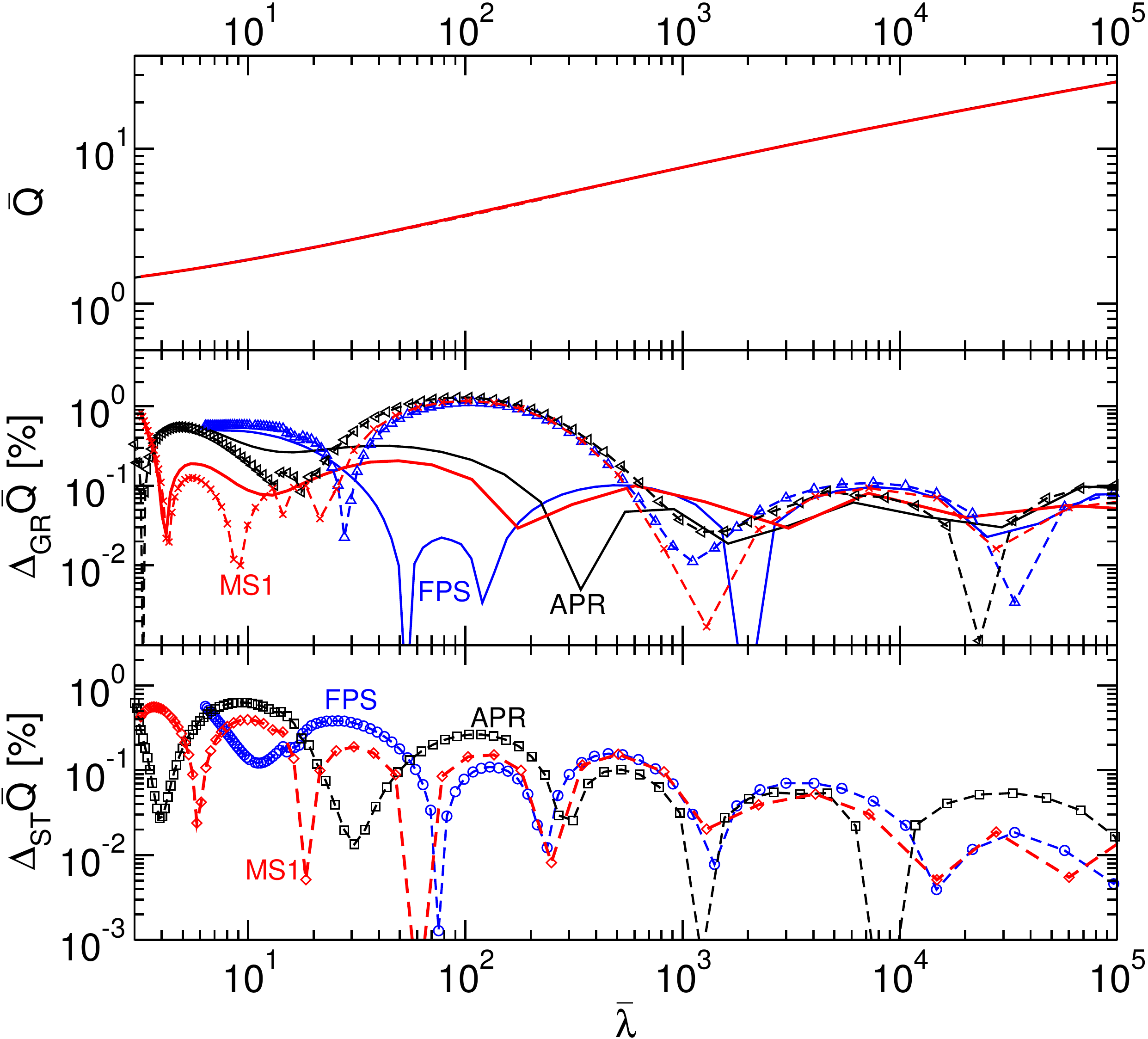}\\
\end{tabular}
\caption{EOS-independent relations $\bar{I}(\bar{\lambda})$ (left) and
  $\bar{Q}(\bar{\lambda})$ (right). Solid linestyles refer to GR;
  dashed linestyles, to scalarized stars in a scalar-tensor theory that
  is only marginally allowed by current binary pulsar tests.  In each
  panel, the top inset shows the relation itself; the middle and
  bottom insets show deviations from universality, as measured by the
  residual $\Delta X= 100[X/X_{\rm fit}-1]$. $\Delta_{GR}X$ means that
  the universal relation is obtained by fitting only pure GR
  solutions; $\Delta_{ST}X$ means that the fit is obtained only from
  scalarized solutions. The top panels show that both residuals are
  always smaller than $2\%$, and typically smaller than $1\%$, for
  scalar-tensor theories that are marginally ruled out by binary
  pulsar observations. [From~\cite{Pani:2014jra}.]
\label{fig:ILoveQST}}
\end{center}
\end{figure*}

Pani and Berti~\cite{Pani:2014jra} studied the I-Love-Q relations
using an extension of the Hartle-Thorne
formalism~\cite{Hartle:1967he,Hartle:1968ht} to scalar-tensor
theories. Working at second order in the spin, they focused on
theories that allow for spontaneous
scalarization~\cite{Damour:1993hw,Damour:1996ke,Barausse:2012da,Shibata:2013pra,Palenzuela:2013hsa}. As
shown in Figure~\ref{fig:ILoveQST}, they found that the universal
relations hold also in these theories, but they are essentially
indistinguishable from the GR relations for values of the coupling
parameters consistent with binary pulsar observations, even when
spontaneous scalarization occurs.
Doneva et al.~\cite{Doneva:2014faa} extended this work to rapidly
rotating stars, showing that deviations in the I-Q relation get larger
for fast rotation rates. If one considers theory parameters that are
consistent with observations, these deviations are still too small to
discriminate GR from scalar-tensor theories with spontaneous
scalarization via I-Love-Q-type tests.
In~\cite{Pappas:2014gca}, Pappas and Sotiriou extended the
formalism of Geroch and Hansen~\cite{Geroch:1970cd,Hansen:1974zz},
providing a way to define and compute multipole moments of stationary,
asymptotically flat spacetimes in scalar-tensor gravity. This approach
may allow us to look for deviations from the three-hair relations in
scalar-tensor theories of gravity, and, if such deviations occur, to
use them as a tool to discriminate these theories from GR.

\begin{figure}[t!]
\begin{center}
\begin{tabular}{cc}
\includegraphics[width=.7\textwidth, angle=0, clip=true]{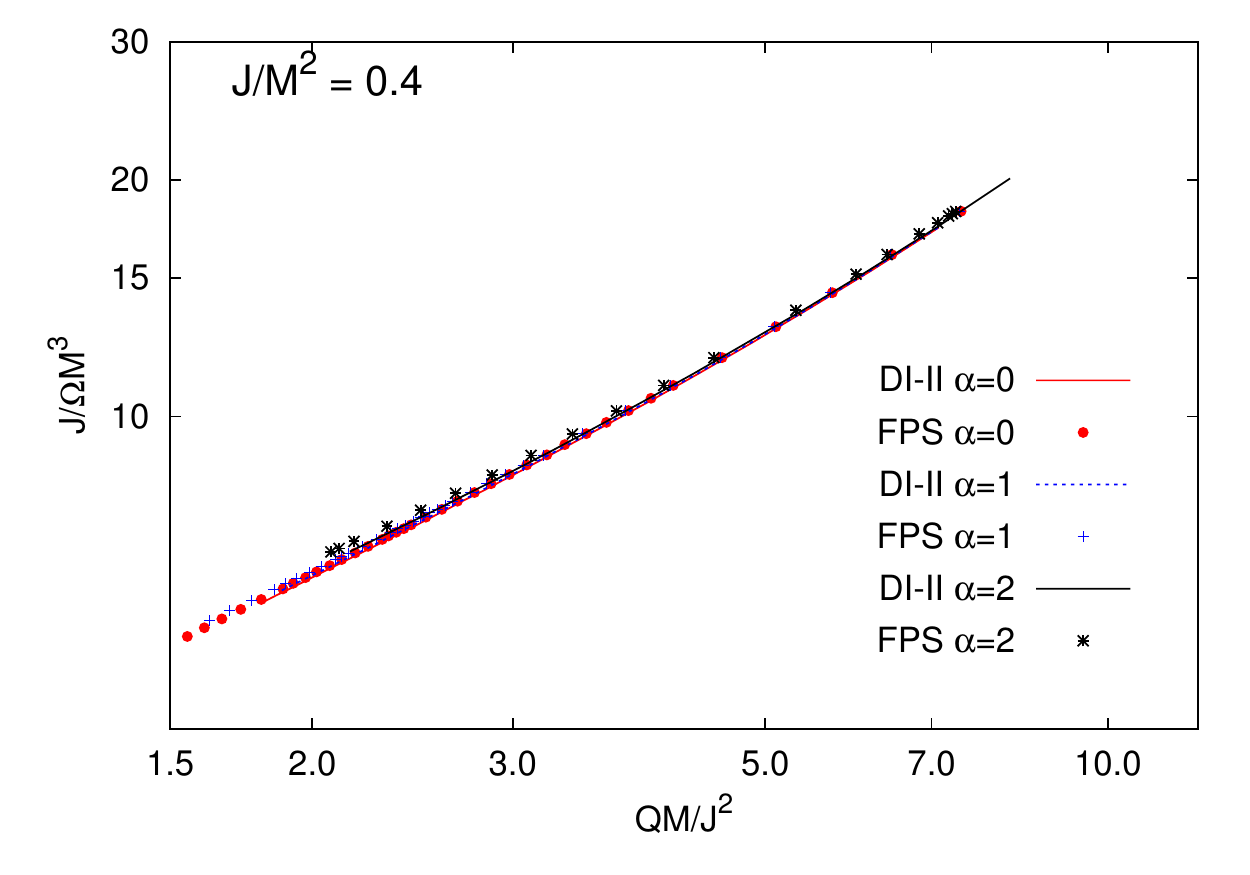}\\
\end{tabular}
\end{center}
\caption{The universal relation between the (scaled) moment of inertia
  and quadrupole moment for rapidly spinning NSs with angular momentum
  $J=0.4M^2$ in EdGB gravity using two different EOSs, namely 
  DI-II EOS~\cite{Diaz-Alonso:1985} and an approximation to the FPS
  EOS~\cite{Haensel:2004nu}. [From~\cite{Kleihaus:2014lba}.]}
\label{KKMfig3}
\end{figure}

Ref.~\cite{Kleihaus:2014lba} found a universal I-Q relation for
rapidly rotating NSs in EdGB gravity, that is shown in
Figure~\ref{KKMfig3} for a fixed dimensionless angular momentum $\chi
\equiv J/M^2=0.4$. The figure shows that the EOS dependence in EdGB
theory is (once again) weak: the relation remains similar to GR
even for a theory which is marginally ruled out ($\alpha=2$).

Staykov et al.~\cite{Staykov:2014mwa} constructed slowly rotating equilibrium models of NSs and strange stars (i.e.,
compact stars made up of deconfined up, down and strange quarks) in $f(R)$ 
gravity at first order in a slow-rotation
approximation. For large values of the coupling parameter of the theory, the NS moment of inertia can be up to $30\%$
larger than its GR counterpart. This deviation is much higher than the induced change in the maximum mass, and this
allows for a breaking of the EOS degeneracy.

\noindent
\paragraph{Other universality relations and tests of GR.}

Other universal relations besides the I-Love-Q relations can be used,
in principle, to test GR.  One approach is to consider the relation
between the tidal Love number and the NS compactness $C$ originally
found in~\cite{Damour:2009vw,Postnikov:2010yn,Hinderer:2009ca}, as
shown in the right panel of Figure~\ref{fig:I-Love-error}. This
relation is not as EOS-independent as the I-Love-Q relations, but it
can be treated as effectively universal if the EOS variation is
smaller than the observational errors on the Love number and on the
compactness.  Such a relation can be more useful than the I-Love
relation, as it may take more than 10 years to measure $I$ from the
double binary pulsar observations~\cite{Kramer:2009zza}, while the NS
compactness (or the mass and radius) have already been constrained
using type-I X-ray bursters with photospheric radius expansion
~\cite{Ozel:2008kb,Ozel:2010fw,Ozel:2011ht,Guver:2008gc,Guver:2010td,Guver:2013xa,2011A&A...527A.139S,Zamfir:2011ht,Steiner:2010fz}
and thermal spectra from transient low-mass X-ray binaries
~\cite{Becker:2002xx,Gendre:2003pw,Guillot:2010zp,Gendre:2002rx,Catuneanu:2013pz,2012MNRAS.423.1556S,2002ApJ...578..405R,
  2009MNRAS.392..665G,1996Natur.379..233W,Walter:2010ht,Pons:2001px,Rybicki:2005id,2007ApJ...671..727W,2003A&A...399.1109B,
  2007MNRAS.375..821H,Steiner:2010fz,Lattimer:2013hma}. Assume that
the tidal Love number can be measured to 75\% accuracy with future GW
observations (see Section~\ref{sec:sub:stellarmass}), and assume in
addition that the NS compactness is measured to 5\% accuracy with
future X-ray observations by, e.g., NICER~\cite{2012SPIE.8443E..13G}
or LOFT~\cite{2012AAS...21924906R,Feroci:2012qh}.
One can then draw an error box in the Love-compactness plane, similar
to that in the I-Love case (cf.~the right panel of
Figure~\ref{fig:I-Love-error}). The EOS-induced variation is smaller
than this hypothetical error box, while the QS relation lies outside
of the error box, suggesting that the relation is effectively
universal for NSs. Then a specific modified theory of gravity can be
ruled out (at least in principle) if the Love-compactness relation in
that theory is inconsistent with the error box.

The tidal Love number $\lambda$ discussed so far is, in technical
jargon, the $\ell=2$, electric-type Love number. Other Love numbers
exist, including the so-called magnetic-type and shape Love
numbers~\cite{Damour:2009vw,Binnington:2009bb,Landry:2014jka}. Yagi~\cite{Yagi:2013sva}
found additional universal relations among different types of Love
numbers. In principle these relations can be used to test GR, but in
practice it seems very difficult to measure any two Love numbers
independently from future GW observations, even if one considers
third-generation interferometers such as the Einstein
Telescope (ET)~\cite{Sathyaprakash:2009xt,Punturo:2010zz,Sathyaprakash:2012jk,Broeck:2014cwa}.
Chan et al.~\cite{Chan:2014kua} studied multipolar universal relations among 
the NS f-mode frequency $f_\ell$ and the electric-type tidal Love number $\lambda_{\ell'}$.
They found that the universality is stronger for $\ell = \ell'$ than $\ell \neq \ell'$.

Another proposal is to use a universal relation between the radius of
a 0.5$M_\odot$ NS and the neutron skin thickness of
${}^{208}$Pb~\cite{Carriere:2002bx} to constrain certain modified
theories of gravity. Sotani~\cite{Sotani:2014goa} showed that the GR
and EiBI relations are distinguishable if the radius of such low-mass
NSs is measured with an accuracy of a few percent.

\subsection{Neutron star sensitivities in modified gravity}
\label{sec:sensitivities}
In preparation for our discussion of the dynamics of compact binaries,
that will be the main topic of Chapter~\ref{sec:CB}, we conclude this
chapter with a brief introduction to the so-called ``sensitivities''
of extended self-gravitating objects in modified gravity.

In many extensions of GR the strong equivalence principle is violated
due to the presence of additional fields that mediate the
gravitational interaction. Self-gravitating objects (and in
particular compact objects) are not test particles, and violations of
the strong equivalence principle imply that the local value of the
gravitational constant depends on the additional dynamical
field(s). When self-gravitating bodies move in regions of spacetime
where the extra fields are not constant, their internal gravitational
energy, and therefore their total mass-energy, will change.

This effect can be described by endowing extended bodies in modified
gravity with a macroscopic property (the ``sensitivity'') measuring
how the body's internal energy depends on the additional field(s)~\cite{Eardley:1975}. The sensitivity can be expected to be larger (and
more relevant for, e.g., binary dynamics) for compact objects, such as
NSs, because their self-gravity is stronger.

\paragraph{Scalar-tensor theories.}

How can we represent the stress-energy tensor for an extended object
in scalar-tensor theories?  Eardley~\cite{Eardley:1975} showed that it
can be modeled as a normal point-particle stress-energy tensor, as
long as we promote the (constant) mass of each body to a function of
$\phi$: $M_{\scriptscriptstyle A} = M_{\scriptscriptstyle A}(\phi)$.

More precisely, in order to describe the orbital dynamics of a widely
separated binary system perturbatively in the size-to-separation
ratio, it is useful to integrate out length scales smaller than the
size of the bodies. We obtain an EFT in which the bodies are
represented by point particles, whose dynamics are governed by an
effective matter action
\begin{equation}\label{sctens_pp_action}
S_{\rm m} = \sum_A
\int_{\Gamma_{\scriptscriptstyle A}} \mathcal{L}_{\scriptscriptstyle A} ds_{\scriptscriptstyle A} \,.
\end{equation}
Here $A$ is an index that labels the bodies,
$\Gamma_{\scriptscriptstyle A}$ is the world-line of body $A$, and
$ds_{\scriptscriptstyle A}$ is the differential arclength
along it. The effective Lagrangian $\mathcal{L}_{\scriptscriptstyle
  A}$ encodes information about the internal structure of body $A$,
and admits a derivative expansion
\begin{equation}
\mathcal{L}_{\scriptscriptstyle A} = -M_{\scriptscriptstyle A}(\phi) + \dots \,,
\end{equation}
where the leading-order term describes the dependence of the
mass-energy on the local scalar value, and subleading terms are
discussed in the appendix of~\cite{Damour:1998jk}. For bodies with
negligible gravitational binding energy $M_{\scriptscriptstyle
  A}(\phi)$ is independent of $\phi$, and so
Eq.~(\ref{sctens_pp_action}) reduces to the action of a collection of
test particles, which move along geodesics of the Jordan-frame metric.

As will be discussed in Chapter~\ref{sec:CB}, the theoretical
predictions for the orbital dynamics and emission of radiation in
compact binaries depend on $M_{\scriptscriptstyle A}(\phi_0)$ and
successive derivatives $M_{\scriptscriptstyle A}^{(k)}(\phi_0)$, where
$\phi_0$ is the asymptotic value of $\phi$ far away from the binary
system.  In particular, the sensitivity of body $A$ is defined as

\begin{equation}
s_A \equiv \left(\frac{d\ln M_{\scriptscriptstyle A}(\phi)}{d\ln \phi}\right)_{\phi=\phi_0} \,. \label{sensitivity}
\end{equation}
Higher-order derivatives of $M_{\scriptscriptstyle A}(\phi)$ are used
to define higher-order sensitivities, e.g.~$s'_{\scriptscriptstyle A}$
and $s''_{\scriptscriptstyle A}$.
  
The first detailed calculation of the NS sensitivities in
scalar-tensor theories was carried out by
Zaglauer~\cite{Zaglauer:1992bp}. The sensitivities are related to the
gravitational binding energy, and therefore they depend on the
microphysics of the specific self-gravitating object we consider. For
ordinary stars and for white dwarfs the sensitivities are of order
$\approx 10^{-6}$ and $\approx 10^{-4}$, respectively, because the
gravitational binding energy of these objects is small. For compact
objects (such as NSs and QSs) the sensitivities can be of the order of
$0.1$, and nonlinear phenomena (like spontaneous scalarization) can
even produce arbitrarily large sensitivities.  A special case are BHs:
their mass in the Einstein frame is
constant~\cite{Damour:1992we,Jacobson:1999vr}, so BH sensitivities are
constant for a given scalar-tensor theory ($s_{\rm BH}=1/2$ in
Bergmann-Wagoner theories).

\paragraph{Quadratic gravity.}

Just like scalar-tensor theories, also quadratic gravity theories
generically violate the strong equivalence principle, and this
violation can be described in terms of sensitivities. Due to an
effective coupling between the matter fields and the scalar field, the
observable NS properties depend on the local value of the scalar field
near the object.

Yagi et al.~\cite{Yagi:2011xp,Yagi:2013mbt} computed the sensitivities
of compact objects in dCS theory. An important qualitative difference
with scalar-tensor theories consists in the fact that in quadratic
gravity BHs can carry a nontrivial scalar charge.  BHs in dCS gravity
have hair (and a nontrivial sensitivity) only if they are spinning; in
EdGB gravity, this happens even for static BHs. On the other hand, the
sensitivity of nonspinning NSs in EdGB gravity is vanishing, as argued
at the beginning of Section~\ref{sec:NSs/quadratic-gravity}
(cf.~Appendix A of~\cite{Yagi:2011xp} for details). This is another
example of how scalar-tensor theories and quadratic gravity are in
some sense ``orthogonal'' in the context of the structure and dynamics
of compact objects.

The sensitivity (more precisely, the rescaled scalar dipole
charge $\bar{\mu}$: cf.~\cite{Yagi:2013mbt} for definitions
and more details) of spinning NSs in dCS theory is shown in the top
inset of the bottom-left panel of Figure~\ref{fig:NSsCS}. To the best of
our knowledge, the analogous calculation for spinning NSs in EdGB
gravity is not available in the literature at the time of writing.

\paragraph{Lorentz-violating theories.}
In Lorentz-violating theories the sensitivities (also known as \AE
ther or khronon ``charges'') characterize the amount of violation of
the strong equivalence principle due to the effective coupling between
matter and the \AE ther or khronon field in the strong-gravity
regime. Due to this coupling, the compact object's structure, binding
energy and gravitational mass will be functions of the motion relative
to the \AE ther or khronon. The action describing the motion of a
strongly gravitating body with gravitational mass $\tilde m$ is given,
in the point-particle approximation, by~\cite{Foster:2007gr}
\be 
\label{eq:actionpp}
S_{\rm pp} = - \int d\tau \; \tilde{m} (\gamma) = - \tilde{m}  \int d\tau \; 
\left\{1 + \sigma \left(1- \gamma \right) + {\cal{O}}\left[\left(1-\gamma \right)^{2}\right] \right\}\,,
\ee
where $\gamma \equiv u_{\m} U^{\m}$ represents the Lorentz factor of
the body relative to the \AE ther, and $d\tau$ is the proper time
along the particle's trajectory.  In the second equality, we performed
a slow-motion PN expansion with $\gamma \ll 1$ and $\tilde{m} \equiv
\tilde m (1)$. The sensitivity parameter $\sigma$ is defined by
\be
\sigma \equiv - \left.\frac{d \ln \tilde{m} (\gamma)}{d \ln \gamma}
\right|_{\gamma = 1}\,.\label{sensitiv_lv}
\ee
For weakly gravitating objects, $\sigma \approx 0$. More in general,
Refs.~\cite{Yagi:2013qpa,Yagi:2013ava} showed that the sensitivity of
a star, whatever its compactness, can be extracted from the asymptotic
behavior of the metric describing the star to \textit{first} order in
a perturbative expansion in the velocity relative to the \AE ther or
khronon. These \textit{slowly moving} stellar solutions have been
discussed in Section~\ref{sec:NS_LV}, and it is possible to show that
the sensitivities can be mapped to the parameter $A$ appearing in
those solutions [cf.~Eq.~\eqref{eq:k1-asympt}]. Indeed, one can show
that the sensitivity in Einstein-\AE ther theory is
\begin{align}
\label{eq:sensitivity-AE}
\sigma_{\sAE} &= \frac{2 c_{1} \left(2 A - 4 - \alpha_{1}^{\sAE}\right)}{c_{-} \left(8 + \alpha_{1}^{\sAE}\right)}\,,
\end{align}
while in khronometric theory one has
\be
\sigma^{\mathrm{kh}} = \frac{2 A - 4 - \alpha_{1}^{\mathrm{kh}}}{8 + \alpha_{1}^{\mathrm{kh}}}\,.
\ee
Here $\alpha^{\sAE}_{1}$ and $\alpha^{\mathrm{kh}}_{1}$ are the
weak-field PPN parameters given in terms of the theory's coupling
constants (cf.~Section~\ref{lv-theories}).  In the weak-field limit,
one can show that the sensitivity scales as~\cite{Foster:2007gr}
\be
\label{eq:weak-field-AE-s}
s_{\sAE}^{\mathrm{wf}} = \left( \alpha_1^{\sAE} - \frac{2}{3} \alpha_2^{\sAE} \right)
\frac{\Omega}{M} + \mathcal{O} \left( \frac{\Omega^2}{M^2}
\right)
\ee
in Einstein-\AE ther theory, where $\Omega$ denotes the gravitational
binding energy. One obtains a similar expression in khronometric
theory by replacing $\alpha_{1,2}^{\sAE}$ with
$\alpha_{1,2}^{\mathrm{kh}}$.

\begin{figure*}[h]
 \capstart
 \begin{center}
\begin{tabular}{r l}
\includegraphics[width=0.45\textwidth,clip=true]{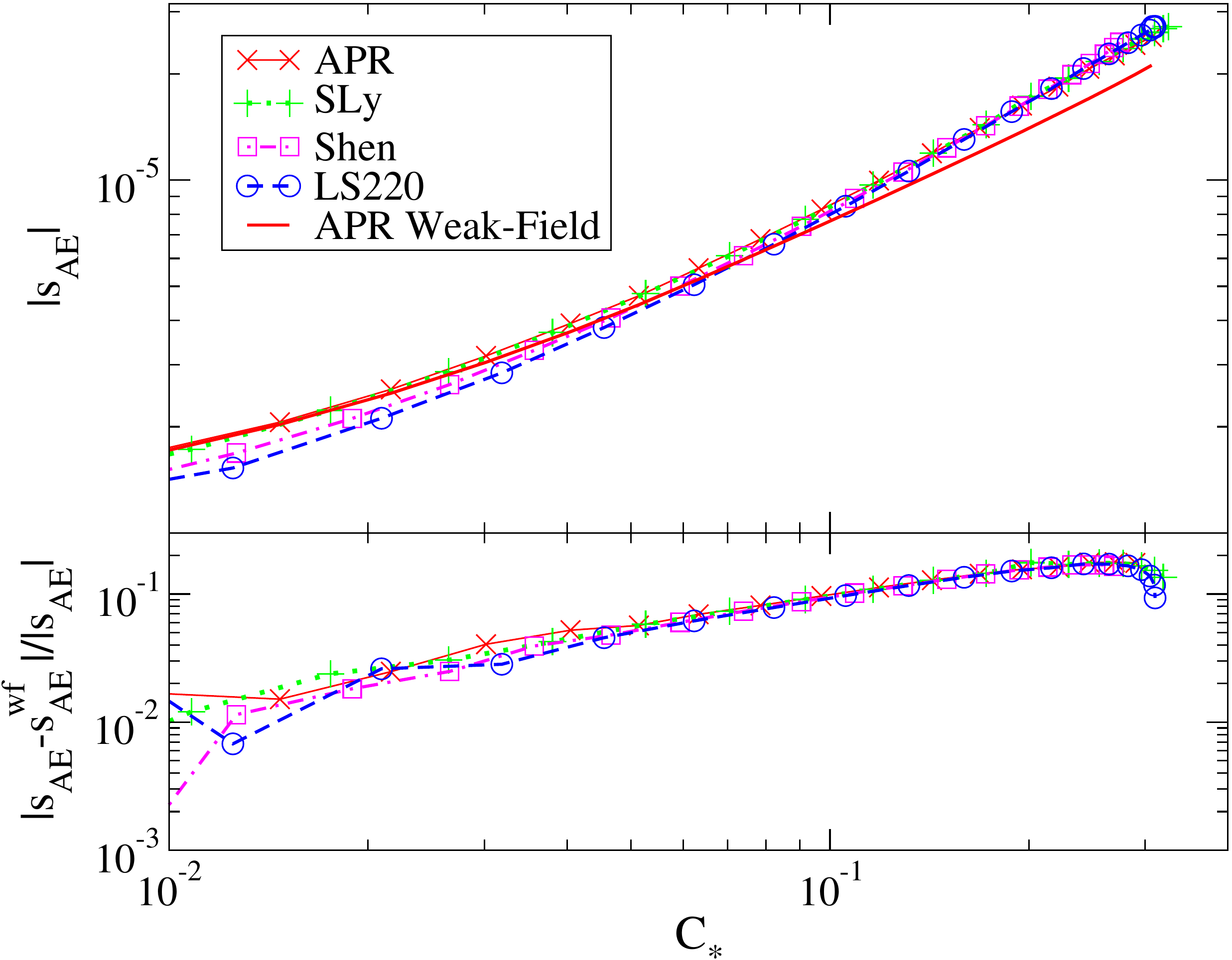}
\includegraphics[width=0.55\textwidth,clip=true]{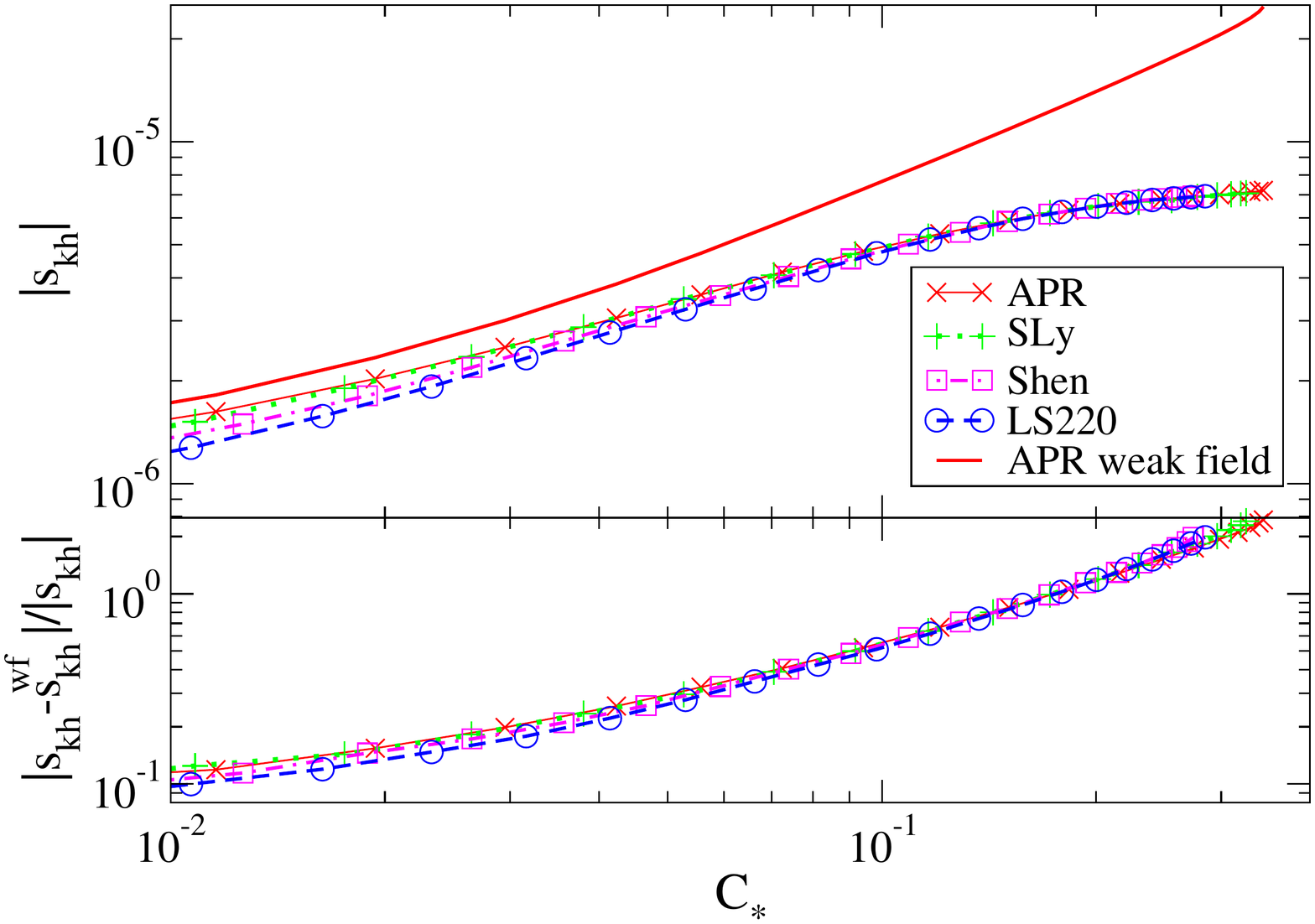}
\end{tabular}
\caption{\label{fig:sens-alpha1} (Left) The top panel shows the
  absolute magnitude of the NS sensitivity in Einstein-\AE ther~theory
  against the NS compactness for various EOSs, together with the
  weak-field expression in Eq.~\eqref{eq:weak-field-AE-s} with the APR
  EOS. The bottom panel presents the fractional difference between the
  sensitivity and the weak-field expression in
  Eq.~\eqref{eq:weak-field-AE-s}. The PPN parameters $\alpha_1^{\sAE}$
  and $\alpha_2^{\sAE}$ saturate the Solar System constraint, and
  $c_+=c_-=10^{-4}$. Observe that the weak-field result becomes
  inaccurate for realistic NS compactnesses, and that the relation is
  EOS-insensitive. (Right) Same as the left panels, but for
  khronometric theory with $\beta =
  10^{-4}$. [From~\cite{Yagi:2013ava}.]}
\end{center}
\end{figure*}

The top-left panel of Figure~\ref{fig:sens-alpha1} presents the
sensitivity as a function of the NS compactness for various EOSs~\cite{Yagi:2013qpa,Yagi:2013ava} . We
also show the weak-field sensitivity for the APR EOS, as defined in
Eq.~\eqref{eq:weak-field-AE-s}.  The PPN parameters
$\alpha_{1,2}^{\sAE}$ are chosen to saturate the Solar System bounds,
and the plot assumes $c_+=c_-=10^{-4}$. The bottom-left panel shows
the fractional difference between the sensitivity defined in
Eq.~\eqref{eq:sensitivity-AE} and the weak-field sensitivity for each
EOS: the two tend to the same limit for small compactness, and their
fractional difference is of $\sim 20\%$ at most in the
large-compactness limit. Observe also that the relation between the NS
sensitivity and compactness is insensitive to the EOS.
Similarly, the right panels of
Figure~\ref{fig:sens-alpha1} present the NS sensitivity (and the
fractional difference from the weak-field limit), with the PPN parameters $\alpha_{1,2}^{\mathrm{kh}}$  again
chosen to saturate the Solar System constraints, and $\beta =
10^{-4}$. As in Einstein-\AE ther theory, the weak-field and
strong-field sensitivities have a common limit as the compactness
decreases, but in the large-compactness regime their fractional
difference can exceed 100\%. The relation between the NS sensitivity
and compactness in khronometric theory is also insensitive to the EOS.

\subsection{Open problems}\label{op:NSs}

For the reader's convenience, here we list some important open
problems in the context of strong-field tests of gravity using NSs:
\begin{itemize}
  \item Except for a few special cases, the properties of NSs in the
    most general scalar-tensor theory with second-order equations of
    motion (Horndeski gravity) have not been explored, even in the
    static case.
 \item As discussed in Section~\ref{NS_fofR}, the very existence of
   compact stars in $f(R)$ gravity is still a matter of debate. The
   generic consensus seems to be that while it is hard to construct NS
   equilibrium configurations in $f(R)$ gravity from a numerical point
   of view, there is no fundamental obstacle to their existence (but
   see~\cite{Ganguly:2013taa} for a different point of view). Either
   way, NS configurations with realistic values of the physical
   parameters have never been constructed in viable $f(R)$ models.
 \item There are no calculations of fast-rotating NSs in dCS gravity.
 \item NS sensitivities have been computed only in scalar-tensor
   theories, quadratic gravity and Lorentz-violating theories, but not
   in other theories (such as massive gravity theories).
 \item Despite the vast literature on the Vainshtein effect, there is
   essentially no phenomenological study of NSs in massive gravity and
   Galileon theories, even for static models.
 \item Gravitational collapse in scalar-tensor theories has been
   considered under very idealized assumptions for the
   microphysics. Relativistic gravitational collapse in quadratic
   gravity has not been studied yet. The analysis of relativistic
   collapse in EiBI gravity and Palatini $f({\cal R})$ gravity is
   crucial to clarify the issue of BH formation and singularity
   avoidance of these theories.
 \item The appearance of curvature singularities near the surface of
   macroscopical
   objects~\cite{Barausse:2007pn,Barausse:2007ys,Barausse:2008nm,Pani:2012qd,Pani:2013qfa}
   in theories with auxiliary fields might be avoided using the
   arguments put forward in~\cite{Kim:2013nna}, but it is important to
   understand whether such effects are generic, and to devise tests to
   discriminate these theories from GR.
 \item Studies of universal relations for NSs and QSs are not
   complete, even within GR. For example, there are no studies for
   differentially rotating stars. The experimental interest of
   differential rotation is probably limited, but differential
   rotation may provide a way to understand universality breaking,
   because it is expected to produce variations in the eccentricity of
   isodensity contours~\cite{Stein:2013ofa,Yagi:2014qua}. Another
   topic that deserves further investigation are magnetic
   fields. Ref.~\cite{2014MNRAS.438L..71H} found that the universality
   is lost for NSs with large magnetic fields, but it would be
   interesting to see if it can be somehow restored for fixed
   \emph{dimensionless} magnetic field strength (e.g.~by normalizing
   the magnetic field strength with the NS mass or radius, as is the
   case for rapidly rotating
   NSs~\cite{Pappas:2013naa,Chakrabarti:2013tca,Yagi:2014bxa}).
\item Tests of GR with universal relations also deserve more study.
  Theories that have not been considered in the context of universal
  relations include quadratic gravity, Einstein-\AE ther and
  Ho\v{r}ava gravity. Static and slowly rotating NS solutions at first
  order in the slow-rotation parameter~\cite{Pani:2011xm}, as well as
  fast rotating solutions~\cite{Kleihaus:2014lba}, are known for EdGB
  gravity, whereas the properties of spinning NSs in dCS theory are
  known only up at second order in
  rotation~\cite{Yagi:2013bca}. Static NS solutions have been
  constructed for Einstein-\AE ther gravity
  in~\cite{Eling:2006df,Eling:2007xh,Greenwald:2009kp}. Finally,
  slowly moving NS solutions were constructed for Einstein-\AE ther
  and Ho\v{r}ava gravity in~\cite{Yagi:2013qpa,Yagi:2013ava}. These
  works should be extended to at least quadratic order in rotation,
  because the quadrupole moment is a second-order quantity. It would
  also be interesting to consider massive scalar-tensor theories and
  $f(R)$ gravity.
\item Universal relations among multipole moments are useful to
  measure the NS mass and radius with future X-ray
  observations~\cite{Baubock:2013gna,Psaltis:2013fha}. Using slowly
  rotating NS solutions and ray-tracing algorithms, it will be
  possible to compute X-ray pulse profiles from a rotating hot spot on
  the NS surface in modified theories of
  gravity~\cite{Psaltis:2010ww,Baubock:2011ke}.  The number of
  parameters describing the profile can be reduced using the modified
  universal relations; then one could fit for the model parameters
  (such as the stellar mass and radius) together with the coupling
  constants in the modified theories. This analysis would reveal
  whether future X-ray observations with
  NICER~\cite{2012SPIE.8443E..13G} and
  LOFT~\cite{2012AAS...21924906R,Feroci:2012qh} could constrain
  deviations from GR (cf.~\cite{Lo:2013ava}).
\end{itemize}

\clearpage
\section{Compact binaries}
\label{sec:CB}

Binary systems containing BHs and/or NSs have played a crucial role in
testing our understanding of strong-field gravitational physics (as we
shall see in Chapter~\ref{sec:BP}), and they will play an even more
decisive role in the future, when GW observations will become a
reality (Chapter~\ref{sec:GW}).  The two-body problem in GR has been a
major focus of numerical and analytical work over the past few decades
because of its relevance for GW detection, and there are several
excellent reviews on this topic. We refer the reader
to~\cite{Blanchet:2013haa,Damour:2013hea,Buonanno:2014aza,PW:2014} for
overviews of analytical calculations within PN theory and the
effective-one-body formalism, and
to~\cite{Pretorius:2007nq,Centrella:2010mx,Sperhake:2011xk,Pfeiffer:2012pc,Hannam:2013pra,Lehner:2014asa}
for summaries of recent numerical work on compact binary mergers. In
this chapter we focus on our current understanding of binary dynamics
in some popular extensions of GR.

\subsection{Scalar-tensor theories}
\label{sec:binaries/scal-tens-theor}
Two complementary approaches are used to model compact binaries in GR:
analytical calculations (usually PN expansions) and numerical
relativity simulations. The first approach is valid at large
separations, and for orbital velocities that are small compared to the
speed of light; the second is necessary when the components of the
binary system get close and merge. Both approaches have been extended
to scalar-tensor theories and will be discussed in this section. We
will focus mostly on the best studied case where the gravitational
interaction is mediated by the metric and a single scalar field
(see~\cite{Damour:1992we} for a pioneering study of binary dynamics in
tensor-multiscalar theories).

\subsubsection{Analytical calculations}\label{sec:pn_st}

Comprehensive studies of compact binaries in scalar-tensor theory were
carried out by Eardley~\cite{Eardley:1975}, Will and
Zaglauer~\cite{Will:1977wq,Will:1989sk} and Damour and
Esposito-Far\`{e}se~\cite{Damour:1992we,Damour:1996ke,Damour:1998jk},
among others. At present, the most accurate description of the orbital
motion and GW emission in scalar-tensor theory is due to work by
Mirshekari and Will~\cite{Mirshekari:2013vb} and
Lang~\cite{Lang:2013fna,Lang:2014osa}, who derived the equations of
motion and the radiation flux considering a single
scalar field with vanishing potential (and therefore vanishing
mass). The relevant field equations for this theory are given in
Eq.~\eqref{ST:Jordan}, with $U(\phi) = 0$.
Following Eardley~\cite{Eardley:1975}, Mirshekari, Will and Lang
describe the orbital motion of the binary in terms of the NS
sensitivities (reviewed in Section~\ref{sec:sensitivities}).

The calculation considers only the inspiral phase of the binary's
evolution. It is therefore appropriate to use the PN approximation, an
expansion in powers of $v/c \sim (Gm/rc^2)^{1/2}$.  The scalar-tensor
equations are solved by adapting the ``direct integration of the
Einstein equations'' (DIRE) method developed by Will, Wiseman, and
Pati~\cite{Will:1996zj,Pati:2000vt,Pati:2002ux}, which has proven
successful in GR and which has been extended to scalar-tensor theory (see~\cite{PW:2014} for a pedagogical
introduction). To begin with, it is convenient to consider a rescaled
version of the scalar field $\phi$: $\varphi \equiv \phi/\phi_0$, where
$\phi_0$ is the value of $\phi$ at infinity (assumed to be
constant).\footnote{The rescaled (Jordan-frame) field, denoted by
  $\varphi$ in this section, should not be confused with the
  Einstein-frame field, denoted by the same symbol elsewhere in this
  review.} By introducing a new tensorial quantity
\begin{equation}
\tilde{h}^{\mu \nu} \equiv \eta^{\mu \nu} -  \sqrt{-\tilde{g}}\tilde{g}^{\mu \nu} \,
\end{equation}
(where $\eta^{\mu \nu}$ is the inverse Minkowski metric,
$\tilde{g}_{\mu \nu} \equiv \varphi g_{\mu \nu}$, and $\tilde{g}$ is
its determinant) and choosing the gauge condition
$\tilde{h}^{\mu \nu}_{\hphantom{\mu \nu},\nu} = 0$, the
field equations reduce to two {\em flat-spacetime} wave
equations,
\begin{align}
\Box_\eta \tilde{h}^{\mu \nu} &= -16\pi \tau^{\mu \nu} \, , \\
\Box_\eta \varphi &= -8\pi \tau_s \, ,
\end{align}
where the sources $\tau^{\mu \nu}$ and $\tau_s$ on the right-hand side
contain terms depending not only on the matter stress-energy tensor $T^{\mu
  \nu}$, but also on the fields $\tilde{h}^{\mu \nu}$ and $\varphi$.
The formal solution of these ``relaxed'' Einstein equations can be
written down using the usual flat-spacetime retarded Green's function,
\begin{subequations}
\label{eq:retardedGF}
\begin{align}
\tilde{h}^{\mu \nu}(t,\mathbf{x}) &= 4\int \frac{\tau^{\mu \nu}(t',\mathbf{x}')\delta(t'-t+|\mathbf{x}-\mathbf{x}'|)}{|\mathbf{x}-\mathbf{x}'|}d^4x' \, ,\label{eq:hintegral}\\
\varphi(t,\mathbf{x}) &= 2\int \frac{\tau_s(t',\mathbf{x}'  )\delta(t'-t+|\mathbf{x}-\mathbf{x}'|)}{|\mathbf{x}-\mathbf{x}'|}d^4x' \, .
\label{eq:phiintegral}
\end{align}
\end{subequations}

The main qualitative difference with electromagnetism is that the
sources in these integrals do not have compact support. In the DIRE
method (and extensions thereof), spacetime is split into two regions, the ``near
zone'' and the ``radiation zone.'' In the near zone close to the source (at
distances smaller than the typical gravitational wavelength $\lambda$,
$|\mathbf{x}'|<\mathcal{R}$ where $\mathcal{R}\sim \lambda$), the
integral is calculated using a slow-motion approximation via the
usual PN expansion in powers of $v/c$.  In the radiation zone far from
the source (at distances $|\mathbf{x}'|>\mathcal{R}$), a change of
variables is used to evaluate the integral. The integration procedure
is different depending on whether the field point $\mathbf{x}$ itself
lies in the near zone or the radiation zone, so there are four different
classes of integrals.
In general, the integrals will produce terms which depend on the arbitrary quantity
$\mathcal{R}$, but it is safe to ignore these terms because any
$\mathcal{R}$-dependent contributions from near-zone integrals must be
completely canceled by contributions from radiation-zone integrals.

The first step to understanding compact binary systems is to find the
equations of motion for the bodies.  Mirshekari and Will~\cite{Mirshekari:2013vb} carried out this calculation up to 2.5PN
order, or $\mathcal{O}((v/c)^5)$.  Their procedure requires evaluating the
integrals \eqref{eq:retardedGF} at field points in the near zone,
where the bodies are located.  The procedure is
iterative: The lowest-order source, comprising only the compact object
stress-energy, is used to find the lowest-order fields.  These are
then substituted in to find the next-highest-order source, which can
then be used to find the next-highest fields, and so on.  Once the
fields, and thus the metric, have been calculated to the necessary
order, the equations of motion are found from the geodesic equations
(with a slight modification due to the $\phi$-dependence of mass in
the Eardley approach).  Schematically, the relative acceleration $\mathbf{a} \equiv \mathbf{a}_1-\mathbf{a}_2$ takes the form
\begin{align}
a^i =& -\frac{G\alpha m}{r^2}\hat{n}^i+\frac{G\alpha m}{r^2}(A_\text{PN}\hat{n}^i+B_\text{PN}\dot{r}v^i)+\frac{8}{5}\eta\frac{(G\alpha m)^2}{r^3}(A_\text{1.5PN}\dot{r}\hat{n}^i-B_\text{1.5PN}v^i) \nonumber\\
&{}+\frac{G\alpha m}{r^2}(A_\text{2PN}\hat{n}^i+B_\text{2PN}\dot{r}v^i) \, ,
\end{align}
where $m \equiv m_1+m_2$, $\eta \equiv m_1m_2/m^2$, $r$ is the orbital separation,
$\mathbf{\hat{n}}$ is a unit vector pointing from body 2 to body 1,
and $\mathbf{v} \equiv \mathbf{v}_1-\mathbf{v}_2$ is the relative
velocity.  The (typically time-dependent) coefficients $A_\text{PN}$,
$B_\text{PN}$, $A_\text{1.5PN}$, $B_\text{1.5PN}$, $A_\text{2PN}$, and
$B_\text{2PN}$ are given in~\cite{Mirshekari:2013vb}.  We use the
symbol $G$ to represent the combination
$(4+2\omega_0)/[\phi_0(3+2\omega_0)]$ [with $\omega_0 \equiv
\omega(\phi_0)$] because it appears in the metric component $g_{00}$
in the same manner as the gravitational constant $G$ in GR.  However,
the coupling in the Newtonian piece of the equations of motion is not
simply $G$ but $G\alpha$, where
\begin{equation}
\alpha \equiv \frac{3+2\omega_0}{4+2\omega_0}+\frac{(1-2s_1)(1-2s_2)}{4+2\omega_0} \,
\end{equation}
and $s_i$ ($i=1\,,2$) are the sensitivities of the two objects,
defined in Eq.~\eqref{sensitivity}. Another important deviation from
GR is the presence of a 1.5PN radiation-reaction contribution to the
equations of motion.  In GR, radiation reaction begins at 2.5PN order,
with the lowest-order quadrupole radiation contribution.  In
scalar-tensor theory, radiation reaction begins at 1.5PN order, due to
the presence of scalar dipole radiation.

Many other deviations from GR occur within the $A$ and $B$
coefficients: see~\cite{Mirshekari:2013vb} for details.  Although the
number of deviations is large, they can all be characterized using a
fairly small number of parameters, all combinations of $\phi_0$, the
Taylor coefficients of $\omega(\phi)$, and the sensitivities $s_A$,
$s_A'$, and $s_A''$.  The deviations are considerably simplified if
one object in the system is taken to be a BH (with the other being a
NS).  Then the motion of the system is indistinguishable from the
motion in GR up to 1PN order.  All deviations beyond 1PN order depend
only on a single parameter, which itself depends on $\omega_0$ and
the sensitivity of the NS.  Unfortunately, this parameter does not depend on any
more details of the scalar-tensor theory; if measured, it alone could
not be used to distinguish between Brans-Dicke theory and a more
general scalar-tensor theory.

The equations of motion simplify even more radically for a binary BH
system: They are {\em identical} to the equations of motion in GR,
except for an unobservable mass rescaling.
This result is a generalization to binary systems of
``no-scalar-hair'' theorems that apply to single
BHs~\cite{Hawking:1972qk}.  For generic mass ratio, Mirshekari
and Will proved this ``generalized no-hair theorem'' up to 2.5PN
order, but they conjectured that it should hold at all PN orders.
Indeed, Yunes et al.~have shown that the equations of motion are the
same as in GR at any PN order if one considers an extreme mass-ratio
system and works to lowest order in the mass
ratio~\cite{Yunes:2011aa}, and the conjecture is also supported by
numerical relativity studies~\cite{Healy:2011ef,Berti:2013gfa} (see
Section~\ref{sec:nr_st}).  This ``generalized no-hair theorem'' for
binary BHs depends on some crucial assumptions: vanishing scalar
potential, asymptotically constant value of the scalar field, and
vanishing matter content.  If any one of these assumptions breaks down, the BH binary's behavior will differ from GR.

The next step is the calculation of gravitational radiation.  The
tensor part of the radiation, encoded in $\tilde{h}^{ij}$, was
computed up to 2PN order by Lang~\cite{Lang:2013fna}.  The procedure
requires evaluating Eq.~\eqref{eq:hintegral} for field points in the
``far-away zone,'' a subset of the radiation zone which is very far
($R \equiv |\mathbf{x}| \gg \mathcal{R}$) from the source.  When integrating over source points in the near zone,
the main step is the calculation of certain moments of the source
$\tau^{ij}$, known as ``Epstein-Wagoner moments.'' The first of these,
the quadrupole moment, generates GW contributions at 0PN, 1PN, 1.5PN,
and 2PN orders.  (The 1.5PN order contribution does not occur in GR
and is a direct result of scalar dipole radiation in this theory.)
The next moment is the octupole moment, which generates GWs at 0.5PN,
1.5PN, and 2PN orders.  In all, Epstein-Wagoner moments with up to 6
indices are required.  The 6-index moment contributes only 2PN GWs.

The final expression for the tensor waves is considerably more
complicated than its GR equivalent; however, all deviations depend on
the same small number of parameters that characterize the equations of
motion.  Most deviations appear as modifications to GR terms, except
for entirely new terms which depend on the existence of a scalar
dipole moment.  The tensor
waves also show the same behavior as the equations of motion in
special cases.  For BH-NS systems, the waveform is indistinguishable
from GR up to 1PN order; deviations at higher order depend only on the
single parameter described earlier.  For BH-BH systems, the waveform is completely indistinguishable from GR.

Scalar radiation has recently been computed by Lang~\cite{Lang:2014osa}
using a very similar procedure. The near-zone contribution requires
the calculation of ``scalar multipole moments'' similar to the
Epstein-Wagoner moments, but involving $\tau_s$ instead of
$\tau^{ij}$.  In this case, the lowest-order moment is not the
quadrupole, but the monopole.  Using the standard definition of PN
orders, in which ``0PN'' waves are generated by the tensor quadrupole,
the scalar monopole moment generates a scalar field at $-1$PN order.
This field, however, turns out to be time-independent and not
wavelike.  The dipole moment generates the lowest-order scalar waves,
which are of $-0.5$PN order:
\begin{equation}
\varphi = \frac{4G\mu\alpha^{1/2}}{R}\zeta\mathcal{S}_-(\mathbf{\hat{N}}\cdot \mathbf{v})
 \, ,
\end{equation}
where $\mu \equiv m_1 m_2/m$ is the reduced mass, $\mathbf{\hat{N}}
\equiv \mathbf{x}/R$ is the direction from the source to the detector,
$\zeta \equiv 1/(4+2\omega_0)$, and
\begin{equation}
\mathcal{S}_- \equiv \alpha^{-1/2}(s_2-s_1) \, .
\end{equation}

Calculating the radiation up to 2PN order requires knowledge of the
monopole moment to 3PN order (relative to itself) and knowledge of the
dipole moment to 2.5PN order.  Just constructing the 3PN expansion of
the source $\tau_s$ is a challenging process.  Evaluating the
resulting integrals is even more difficult.  For these reasons,
Lang~\cite{Lang:2014osa} computes the scalar waveform only to 1.5PN
order, with the 2PN result saved for future work.  The 1.5PN waveform
turns out to be described by the same set of parameters that describes
the 2.5PN equations of motion and the 2PN tensor waveform.  Other
similarities include the vanishing of the scalar waveform for binary
BH systems (so that it is indistinguishable from GR) and tremendous
simplifications in the mixed BH-NS case. 

The tensor and scalar waveforms can be used to compute the total
energy carried off to infinity using the expressions
\begin{align}
\frac{dE_T}{dt} &= \frac{R^2}{32\pi}\phi_0\oint \dot{\tilde{h}}_\text{TT}^{ij}\dot{\tilde{h}}_\text{TT}^{ij} d^2\Omega \, , \label{eq:tensorfluxeqn}\\
\frac{dE_S}{dt} &= \frac{R^2}{32\pi}\phi_0(4\omega_0+6)\oint \dot{\varphi}^2 d^2\Omega \, ,
\label{eq:scalarfluxeqn}
\end{align}
for the tensor and scalar fluxes, respectively.  Here TT refers to the
transverse-traceless projection of the tensor.  The existence of a
$-0.5$PN piece of the scalar waveform means that the scalar waveform
must generally be known to $(N+1/2)$-th PN order to find the flux at
$N$th PN order.  Lang~\cite{Lang:2014osa} computes this flux to 1PN
order.  The result is
\begin{equation}
\frac{dE}{dt} = \dot{E}_{-1}+\dot{E}_0+\dot{E}_{0.5}+\dot{E}_1 \, ,
\label{eq:totalflux}
\end{equation}
where
\begin{subequations}
\begin{align}
\dot{E}_{-1} &= \frac{4}{3}\frac{\mu\eta}{r}\left(\frac{G\alpha m}{r}\right)^3\zeta\mathcal{S}_-^2 \, ,\label{eq:minus1PNflux}\\
\begin{split}
\dot{E}_0 &= \frac{8}{15}\frac{\mu\eta}{r}\left(\frac{G\alpha m}{r}\right)^3\left\{\frac{G\alpha m}{r}\left[-2\frac{\delta m}{m}\zeta \mathcal{S}_+\mathcal{S}_- \right.\right.\\
&\left.\qquad \quad +\left(-23+\eta-10\bar{\gamma}-10\bar{\beta}_++10\frac{\delta m}{m}\bar{\beta}_-\right)\zeta\mathcal{S}_-^2\right] \\
&\qquad +v^2\left[12+6\bar{\gamma}+2\zeta\mathcal{S}_+^2+2\frac{\delta m}{m}\zeta\mathcal{S}_+\mathcal{S}_-+(6-\eta+5\bar{\gamma})\zeta\mathcal{S}_-^2\right. \\
&\left.\qquad \quad -\frac{10}{\bar{\gamma}}\frac{\delta m}{m}\zeta\mathcal{S}_-(\mathcal{S}_+\bar{\beta}_++\mathcal{S}_-\bar{\beta}_-)+\frac{10}{\bar{\gamma}}\zeta\mathcal{S}_-(\mathcal{S}_-\bar{\beta}_++\mathcal{S}_+\bar{\beta}_-)\right]\\
&\qquad +\dot{r}^2\left[-11-\frac{11}{2}\bar{\gamma}+\frac{23}{2}\zeta\mathcal{S}_+^2-8\frac{\delta m}{m}\zeta\mathcal{S}_+\mathcal{S}_-+\left(-\frac{37}{2}+9\eta-10\bar{\gamma}\right)\zeta\mathcal{S}_-^2 \right.\\
&\qquad \quad -\frac{80}{\bar{\gamma}}\zeta\mathcal{S}_+(\mathcal{S}_+\bar{\beta}_++\mathcal{S}_-\bar{\beta}_-)+\frac{30}{\bar{\gamma}}\frac{\delta m}{m}\zeta\mathcal{S}_-(\mathcal{S}_+\bar{\beta}_++\mathcal{S}_-\bar{\beta}_-) \\
&\left.\left.\qquad \quad-\frac{10}{\bar{\gamma}}\zeta\mathcal{S}_-(\mathcal{S}_-\bar{\beta}_++\mathcal{S}_+\bar{\beta}_-)+\frac{120}{\bar{\gamma}^2}\zeta(\mathcal{S}_+\bar{\beta}_++\mathcal{S}_-\bar{\beta}_-)^2\right]\right\} \, ,
\label{eq:0PNflux}
\end{split}
\\
\begin{split}
\dot{E}_{0.5} &= -\frac{16}{9}\frac{\mu\eta}{r}\left(\frac{G\alpha m}{r}\right)^3(\zeta\mathcal{S}_-)^2\left(\mathcal{S}_+^2+2\frac{\delta m}{m}\mathcal{S}_+\mathcal{S}_-+\mathcal{S}_-^2\right)\frac{G\alpha m}{r}\dot{r} \\
&\quad-\frac{16}{3}\frac{\mu \eta}{r}\left(1+\frac{1}{2}\bar{\gamma}\right)\zeta \mathcal{S}_-^2\left\{2\left(\frac{G\alpha m}{r}\right)^3\frac{G\alpha m}{r}\dot{r} \right. \\
&\qquad +\frac{(G\alpha m)^3}{r} \hat{n}^k\int_0^\infty ds\ \left[\frac{G\alpha m}{r^4}\left(\left(3v^2-15\dot{r}^2-2\frac{G\alpha m}{r}\right)\hat{n}^k+6\dot{r}v^k\right)\right]_{\tau-s} \\
&\left.\qquad \quad \times \ln \frac{s}{2R+s}\right\} \, ,
\label{eq:RZflux}
\end{split}
\end{align}
\label{eq:fluxdetails}
\end{subequations}
and $\dot{E}_1$ is given in~\cite{Lang:2014osa}.  Here we define $\delta m \equiv m_1-m_2$, 
\begin{subequations}
\begin{align}
\mathcal{S}_+ &\equiv \alpha^{-1/2}(1-s_1-s_2) \, ,\\
\bar{\gamma} &\equiv -2\alpha^{-1}\zeta(1-2s_1)(1-2s_2) \, , \\
\bar{\beta}_\pm &= \frac{1}{2}(\bar{\beta}_1 \pm \bar{\beta}_2) \, , \\
\bar{\beta}_1 &\equiv \alpha^{-2}\zeta(1-2s_2)^2(\lambda_1(1-2s_1)+2\zeta s_1') \, , \\
\bar{\beta}_2 &\equiv \alpha^{-2}\zeta(1-2s_1)^2(\lambda_1(1-2s_2)+2\zeta s_2') \, ,\\
\intertext{and}
\lambda_1 &\equiv \frac{(d\omega/d\varphi)_0\zeta}{3+2\omega_0} \, . 
\end{align}
\end{subequations}
We also note that the subscript $\tau-s$ in \eqref{eq:RZflux} means that the quantity should evaluated at time $\tau-s$, where $\tau \equiv t-R$ is the retarded time.  Equations \eqref{eq:totalflux} and \eqref{eq:fluxdetails} can be used to determine the phase evolution of a
binary, the last step in producing a usable waveform for data analysis
studies.

While future work in this area will certainly involve extending the
current calculation to higher PN order, it may also be interesting to
investigate theories with multiple scalars or a potential.
A derivation of the quadrupole-order flux in tensor-multiscalar
theories, that agrees with Lang's results in the single-scalar limit,
can be found in~\cite{Damour:1992we}.
The current state-of-the-art calculation for compact binaries in the
massive Brans-Dicke theory was performed by Alsing et
al.~\cite{Alsing:2011er} (see
also~\cite{Krause:1994ar,Perivolaropoulos:2009ak}).  In the notation
used by Lang, and correcting a mistake in~\cite{Alsing:2011er}, they
found that the lowest-order flux is given by
\begin{equation}
\dot{E} = \frac{4}{3}\frac{\mu\eta}{r}\left(\frac{G\alpha m}{r}\right)^3\zeta\mathcal{S}_-^2\left[\frac{\omega^2-m_s^2}{\omega^2}\right]^{3/2}\Theta(\omega-m_s) \, ,
\label{eq:Alsingflux}
\end{equation}
where $\omega$ is the orbital frequency, $m_s$ is the mass of the
scalar field, and $\Theta$ is the Heaviside function.
In massive Brans-Dicke theory, scalar dipole radiation is emitted only
when $\omega > m_s$.  Alsing et al.~continued their calculation to 1PN
order; however, those terms are incomplete and we do not list them
here.

\subsubsection{Numerical relativity simulations}\label{sec:nr_st}

Numerical relativity (the use of numerical simulations to solve
Einstein's equations in full generality in the nonlinear regime) is
the most powerful tool at our disposal to understand strong
gravity. Numerical relativity had a 40-year long gestation
\cite{Sperhake:2014wpa}, and the
main motivation behind its development was the description of
high-energy astrophysical phenomena in the framework of GR. In recent
years the theory has been extended beyond GR and it found unexpected
applications in many other fields, ranging from high-energy physics to
solid-state physics~\cite{Cardoso:2014uka}.

Even within GR, obtaining numerically stable and accurate time
evolutions in the absence of high degrees of symmetry is a daunting
task that requires an understanding of many complex issues, such as
the well-posedness of the evolution system, the construction of
initial data and gauge conditions~\cite{Alcubierre:2008,Baumgarte2010}.
These same questions arise also in all proposed extensions of GR, and
at present they remain unanswered for most of the theories discussed
in this review.

Scalar-tensor theories represent a notable exception, because they can
be formulated in close analogy to GR. As discussed in
Section~\ref{subsec:ST}, the action of scalar-tensor theories in the
Einstein frame is the same as the Einstein-Hilbert action, except for
a minimal coupling with the scalar field in the gravitational sector;
a nonminimal coupling with the scalar field only appears in the matter
sector.

The field equations in the Einstein frame -- Eqs.~\eqref{ST:Einstein},
that we reproduce here for the reader's convenience  --
are
\begin{subequations}
  \label{ST:Einsteinbis}
  \begin{align}
  G^\star_{\mu \nu} &=2\left(\partial_\mu\varphi\partial_\nu\varphi-
  \frac{1}{2}g^\star_{\mu\nu}\partial_\sigma\varphi\partial^\sigma\varphi\right)-
  \frac{1}{2}g^\star_{\mu\nu}V(\varphi)+8\pi T^\star_{\mu\nu}\,,\label{eq:tensoreqnEbis} \\
\Box_{g^\star} \varphi &=-4\pi\alpha(\varphi)T^\star+\frac{1}{4}\frac{dV}{d\varphi}\label{eq:scalareqnEbis}\,.
  \end{align}
  \end{subequations}
In the Jordan frame the scalar field is minimally coupled to matter,
free particles follow geodesics of the spacetime metric and the
stress-energy tensor $T_{\mu\nu}$ of a given matter source (e.g., a
perfect fluid) has formally the same expression as in GR. In the
Einstein frame the stress-energy tensor is
\begin{equation}
  T^{\star~\nu}_{\mu}=A^4(\varphi)T^{~~\nu}_{\mu}\,,
  \label{transftmn}
\end{equation}
where $A(\varphi)$ is the conformal factor (see
Section~\ref{subsec:ST}). Therefore, as mentioned above, matter fields
are coupled with $\varphi$ in the Einstein frame.  Energy-momentum
conservation in the Jordan frame, $\nabla_{\mu}T^{\mu \alpha}=0$,
translates to the Einstein-frame condition
\begin{equation}
  \nabla_{g^\star}^{\mu} T^\star_{\mu\alpha} = \alpha(\varphi)T^\star\partial_\alpha \varphi\,.
  \label{eq:evolTE}
\end{equation}

The time evolution of a physical system can then be obtained by
solving Eqs.~(\ref{eq:tensoreqnEbis}), (\ref{eq:scalareqnEbis}) and
(\ref{eq:evolTE}).  Except for the addition of a minimally coupled
scalar field and -- when matter is present -- for the nonminimal
coupling of $\varphi$ with the stress-energy tensor in
Eq.~(\ref{transftmn}), this system of equations is identical to the
field equations of GR. This is the reason why scalar-tensor theories
of gravity can be attacked using relatively minor generalizations of
the numerical codes developed for GR. The evolution of the scalar
field $\varphi$ is dictated by Eq.~(\ref{eq:scalareqnEbis}), a wave
equation that manifestly preserves any hyperbolicity properties that
are satisfied when Eqs.~(\ref{eq:tensoreqnEbis}) and (\ref{eq:evolTE})
are formulated as an initial-value problem.

Salgado et al.~\cite{Salgado:2005hx,Salgado:2008xh} showed that a
strongly hyperbolic formulation can be obtained also in the physical
(Jordan) frame. However, the Einstein frame is exceptionally
convenient for applications to vacuum spacetimes. The reason is that
in vacuum ($T^{\star\,\alpha}{}_{\beta}=0$) the evolution equations
\eqref{ST:Einsteinbis} are independent of the coupling function
$A(\varphi)$. Therefore a single numerical evolution represents a
whole class of theories characterized by different functional forms of
$A(\varphi)$ for a given potential $V(\varphi)$.  Different choices of $A(\varphi)$
result in different physical predictions (e.g.~in terms of
gravitational waveforms), but all of these predictions can be
calculated by {\em post-processing} data from one and the same
numerical simulation. This would not be possible in the Jordan frame,
where the coupling function explicitly appears in the system of
equations that are numerically evolved in time. At least for vacuum
spacetimes, the Einstein frame allows for a considerable reduction in
the computational cost of exploring different scalar-tensor theories.

Early numerical studies of gravitational systems in scalar-tensor
theory focused on gravitational collapse in spherical symmetry, a 1+1
dimensional problem involving only time and one radial
coordinate. These studies explored dust collapse in Brans-Dicke theory~\cite{Shibata:1994qd,Scheel:1994yn,Scheel:1994yr}, the collapse and
stability of NSs~\cite{Novak:1997hw,Novak:1998rk}, and stellar
core-collapse~\cite{Novak:1999jg} in more general scalar-tensor
theories, with particular focus on the spontaneous scalarization
phenomenon~\cite{Damour:1993hw}. The recent breakthroughs in numerical
relativity have opened up the realm of compact binary simulations in
scalar-tensor theories of gravity. We now summarize the main findings
for BH-BH and NS-NS binaries.

\begin{figure}
  \capstart
  \centering
  \includegraphics[width=0.7\textwidth]{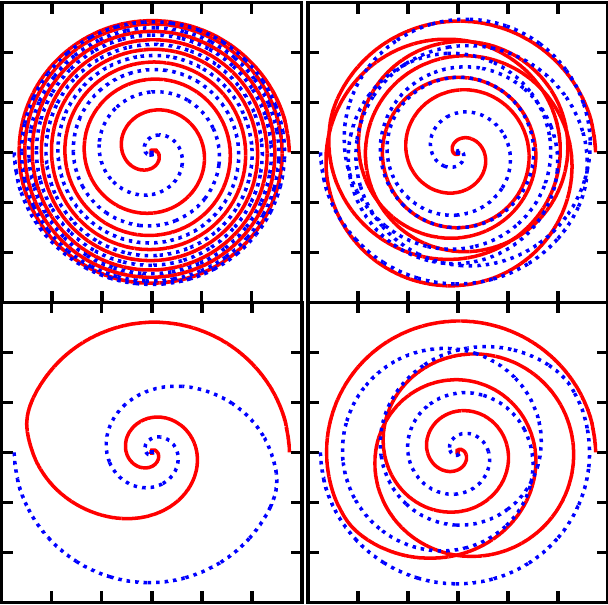}
  \caption{BH trajectories in the Einstein frame, assuming that the
    BHs are in a scalar field bubble. The
    upper-left panel corresponds to the GR limit; upper-right and
    lower-left panels correspond to different initial amplitudes of the scalar
    field. In the lower-right panel the evolution occurs in the
    presence of a nonzero quartic potential. [From~\cite{Healy:2011ef}.] }
  \label{fig:trajectories}
\end{figure}

\paragraph{Black hole binaries.}
\label{sec:NRblackholes}

For BH-BH binaries, scalar-tensor theories represent conceptually simple
modifications of GR.  A downside of this simplicity is that
introducing nontrivial BH binary dynamics in these theories (where by
``nontrivial'' we mean dynamics differing from pure GR) requires
somewhat contrived scenarios.

One obvious solution of the field equations \eqref{ST:Einsteinbis} in
the vacuum case ($T^{\alpha}{}_{\beta}=T^{\star\,\alpha}{}_{\beta}=0$)
is the GR solution for the metric, plus a constant scalar field. This
was realized a long time ago, and led to various no-hair theorems
stating that {\it stationary} BH solutions in Brans-Dicke theory are
the same as in GR (see
e.g.~\cite{Hawking:1972qk,1971ApJ...166L..35T,Chase:1970},
and~\cite{Chrusciel:2012jk} for a review). These results have recently
been extended to Bergmann-Wagoner theories~\cite{Sotiriou:2011dz} (see
Section~\ref{sec:ST_BHs}). Moreover, as discussed in
Section~\ref{sec:pn_st},
the dynamics of BH binaries in scalar-tensor theories was shown to be
indistinguishable from GR at all PN orders in the extreme mass-ratio
limit, and up to 2.5PN order in the equations of motion in PN theory.
This ``generalized no-hair theorem'' relies on the following
assumptions: (1) the spacetime contains no matter, (2) the potential
$V(\varphi)$ vanishes, (3) the scalar-tensor action is truncated at
second order in the derivative expansion, and (4) the metric is
asymptotically flat and the scalar field is asymptotically constant.
Deviations from GR in the radiation from BH binaries can occur if we
violate any of these four assumptions.

The most obvious way to obtain nontrivial dynamics is to violate
hypothesis (1), i.e.~to consider configurations involving matter, such
as NS-NS binaries. Leaving this possibility aside for the moment,
another way out of the no-hair theorems was suggested by Horbatsch \&
Burgess~\cite{Horbatsch:2011ye}: if the scalar field is not
asymptotically stationary, the BHs in a binary could retain scalar
hair~\cite{Jacobson:1999vr} and emit dipole radiation, as long as
their masses are not exactly equal. The introduction of higher-order
derivatives in the action would also violate the hypotheses of the
generalized no-hair theorem, but it would lead to substantially more
complicated equations, whose well-posedness remains unclear at present
(cf.~Section~1 of~\cite{Berti:2013gfa}).

\begin{figure}[t]
\begin{center}
\begin{tabular}{ll}
  \includegraphics[width=0.4\textwidth,clip=true]{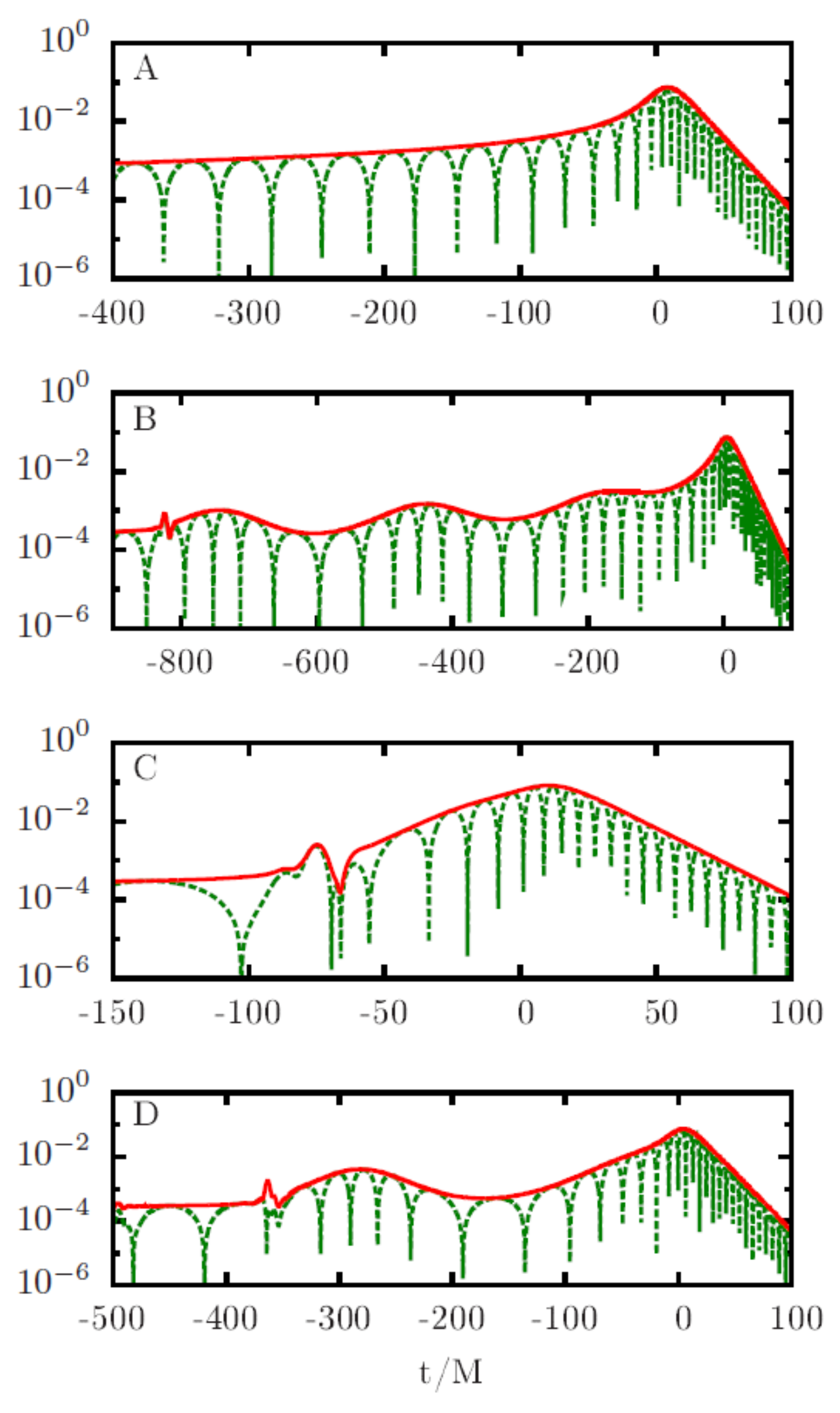}&
  \includegraphics[width=0.5\textwidth,clip=true]{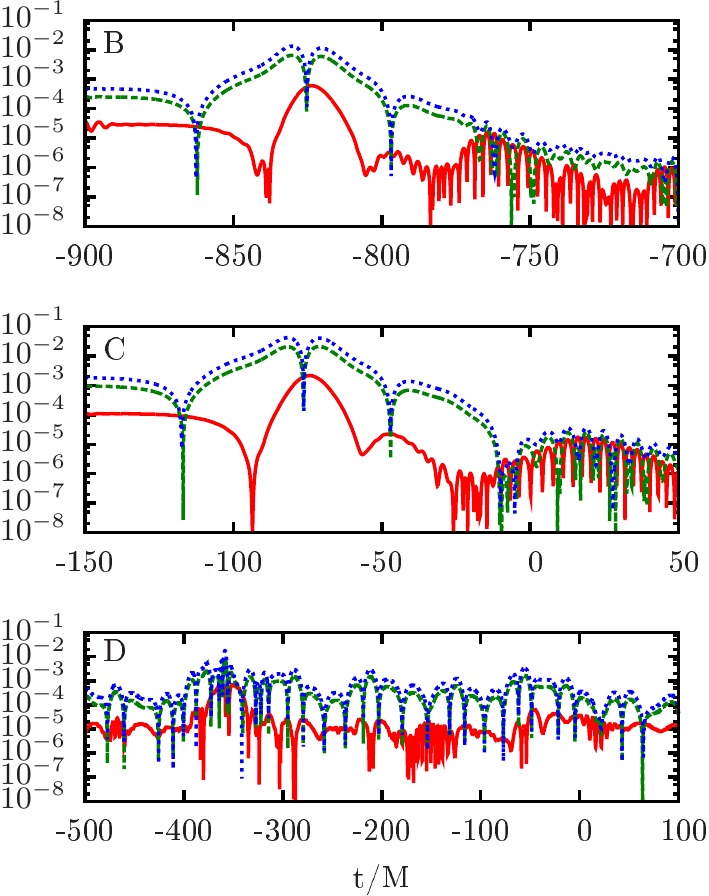}\\
\end{tabular}
  \caption{Left: The $l=m=2$ multipole of the complex Weyl scalar
    $\Psi_4$ (the dashed line corresponds to the real part of
    $\Psi_4$, and the solid line to its modulus). Right: The $l=m=0$
    multipole of the so-called breathing mode $r\,M\,\widetilde \Phi_{22}$,
    where $r$ is the extraction radius and $M$ is the total mass of
    the binary; solid, dashed and dotted lines correspond to
    different values of the parameter $\beta_0$ (see Eq.~(\ref{DEalphabeta})).
    Merger occurs at $t/M=0$. [From~\cite{Healy:2011ef}.]}
  \label{fig:strain}
\end{center}
\end{figure}

Healy et al.~\cite{Healy:2011ef} investigated whether generalized
no-hair theorems carry over to the nonlinear regime, i.e.,~whether the
dynamics of BH binaries during the late inspiral and merger is the
same as in GR. Their results show that the dynamics can differ if the
scalar field evolves: the scalar field triggers
energy loss that leads to a difference in the GW polarizations. In their
evolutions, nontrivial dynamics is triggered by placing the BHs inside
a scalar field ``bubble,'' which in some cases includes a nonvanishing
scalar field potential. As the bubble collapses, the BHs accrete the
scalar field and grow. The increase in mass of the BHs has a
dramatic effect on the binary dynamics.

Figure~\ref{fig:trajectories} shows the BH trajectories (in the
Einstein frame) for the four cases considered in their study (cases A,
B, C and D from top left to bottom right). Case A represents the
binary evolution in GR.  Cases B, C and D differ in the initial amplitude of
the scalar field in the bubble, and case D furthermore contains a
nonvanishing potential term: see Table~1 in~\cite{Healy:2011ef} for
details. The left panel of Figure~\ref{fig:strain} shows the $l=m=2$
multipole of the Weyl scalar $\Psi_4$ (roughly speaking, the second
time derivative of the GW signal) for each of the 4 initial
configurations. There are obvious differences between the various time
evolutions of $\Psi_4$, in particular when compared against GR (case
A). The $l=m=0$ multipole of the breathing mode
$\widetilde
\Phi_{22}$ in the Jordan frame is shown for
cases B, C and D in the right panel of Figure~\ref{fig:strain}. Notice
that the inclusion of a potential term (as in case D) introduces longer lived
dynamics in the scalar field mode; see also
Section~\ref{subsec-bh-exotic} for long-term evolutions of the
post-merger phase.

In summary, the study of Healy et al.~\cite{Healy:2011ef} supports the
view that an evolving scalar field is required to bypass the
generalized no-hair theorems for BH binaries. Inhomogeneities in the
initial scalar field configuration could provide such a mechanism.
This particular study considered a BH in a scalar field bubble, but
the conclusions can be carried over to more generic scenarios.  For
the effects to be observable, the merging BHs must accrete enough
scalar field to change their masses and modify the binary evolution.

A different mechanism to circumvent the no-hair theorems was
considered by Berti et al.~\cite{Berti:2013gfa}, who relaxed
assumption (4) in the list above by introducing non-asymptotically
flat or constant boundary conditions. Conceptually, the main
motivation for relaxing this assumption comes from cosmological
considerations.  Inhomogenous scalar fields have been considered in
cosmological models as an alternative to dark matter~\cite{Sahni:1999qe,Hu:2000ke}, and also as models of supermassive
boson stars~\cite{Macedo:2013qea}.
For scalar-field profiles that vary on a lengthscale much larger than
the BH binary separation, one effectively has a configuration with an
approximately constant scalar-field gradient at large separation from
the binary.
\begin{figure}
\capstart{}
  \includegraphics[height=120pt,clip=true]{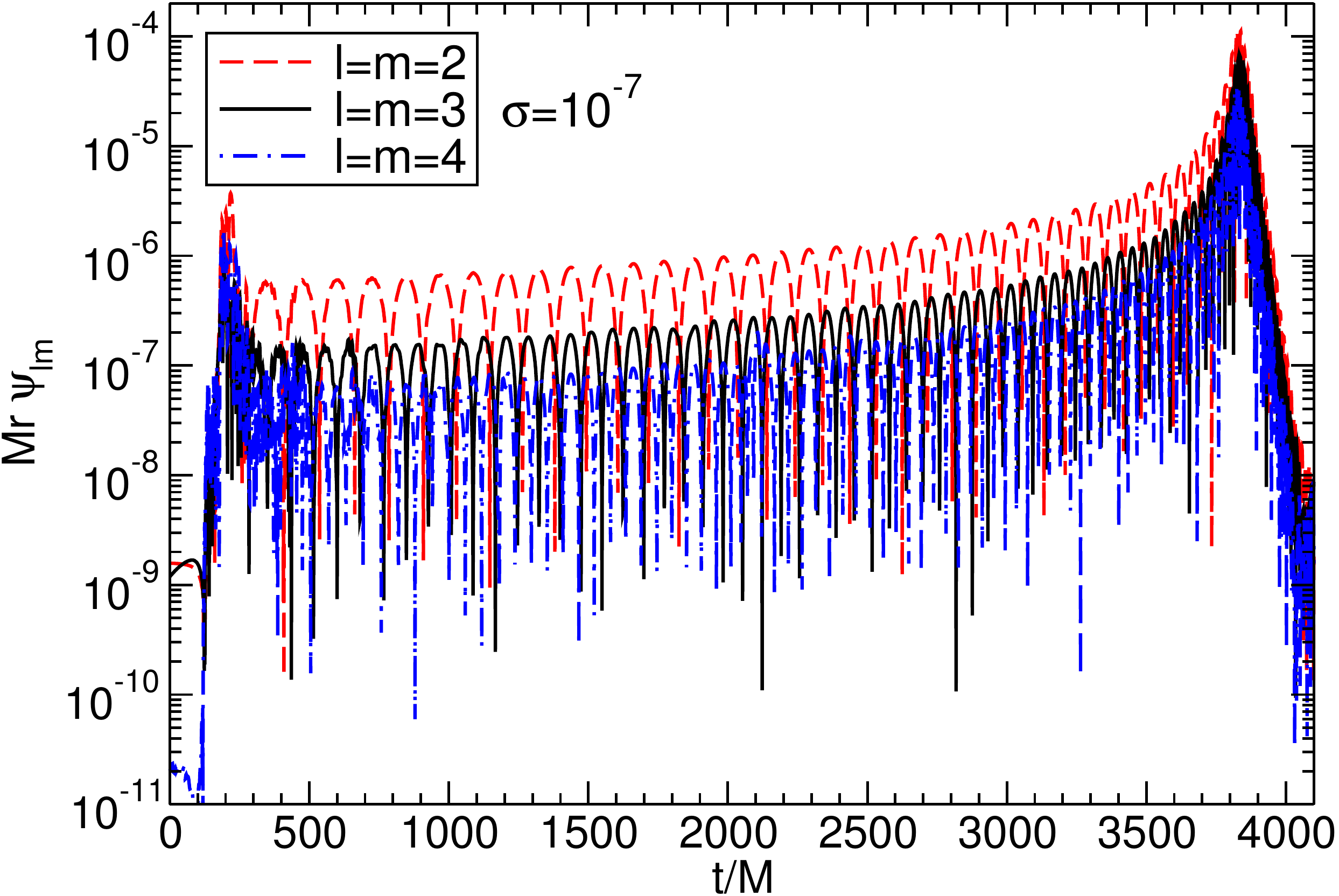}
  \includegraphics[height=120pt,clip=true]{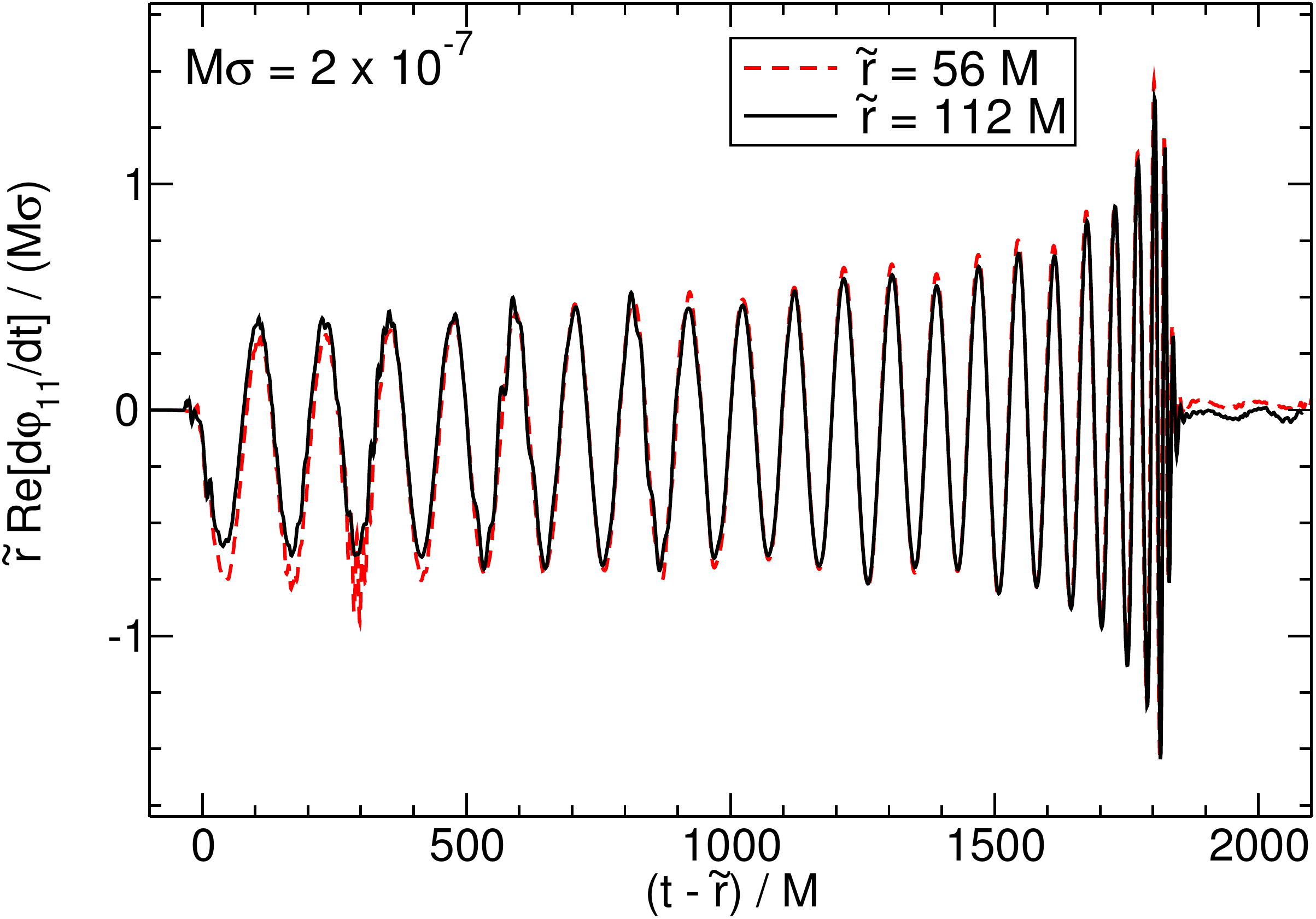}
  \caption{Numerical results for a 10-orbit inspiral of a nonspinning
    BH binary of mass ratio $3:1$ inside a scalar-field gradient of
    magnitude $M\sigma = 2 \times 10^{-7}$.  Left: Real part of the
    spin-weighted spheroidal harmonic components of the Newman-Penrose
    scalar $\Psi_4$ extracted at coordinate radius $\tilde{r}=56~M$
    for harmonic indices $l=m$ (the imaginary part is identical up to
    a phase shift). Right: Time-derivative of the scalar field at the
    largest and smallest extraction radii, rescaled by radius and
    shifted in time. [From~\cite{Berti:2013gfa}.]}
  \label{fig:bbhs2m07}
\end{figure}

Ref.~\cite{Berti:2013gfa} considered the quasi-circular inspiral of a
nonspinning BH binary (with mass ratio $3:1$) in a scalar-field
gradient of magnitude $M\sigma = 2\times 10^{-7}$ perpendicular to the
orbital angular momentum vector.  The three lowest multipoles of the
Newman-Penrose scalar $\Psi_4$ extracted from the Einstein metric are
shown in the left panel of Figure~\ref{fig:bbhs2m07}. These multipoles
are effectively indistinguishable from their GR counterparts
(cf.~Figure~5 of~\cite{Berti:2013gfa}). A nonvanishing ``background''
scalar field does, however, lead to the (mostly dipolar) emission of
scalar radiation, which is not present in GR.  The time derivative of
the real part of the dipole contribution is shown in the right panel
of Figure~\ref{fig:bbhs2m07}, and it displays the expected $1/r$
fall-off behavior. The oscillation frequency of this dipole mode is
{\em twice} the orbital frequency. At first glance, it may appear
surprising to see an $m=1$ multipole oscillating at twice the orbital
frequency (rather than at the orbital frequency). A simple
calculation, however, reveals that this feature is a consequence of
the interaction of the orbital motion with an $m=1$ background field:
cf.~the discussion around~Eqs.~(36)-(38) of~\cite{Berti:2013gfa}. In
summary, these simulations demonstrate that non-asymptotically flat
boundary conditions (here imposed in the form of a constant scalar-field 
gradient) provide a mechanism to generate scalar radiation in BH
inspirals in scalar-tensor theories of gravity, thus circumventing the
no-hair theorems. Unfortunately, there is little hope to observe
scalar radiation of this nature in the near future for cosmologically
realistic values of the scalar-field gradients.

\paragraph{Neutron star binaries.}
\label{sec:NRneutronstars}
\begin{figure}
\centering
\includegraphics[width=5.cm,angle=0]{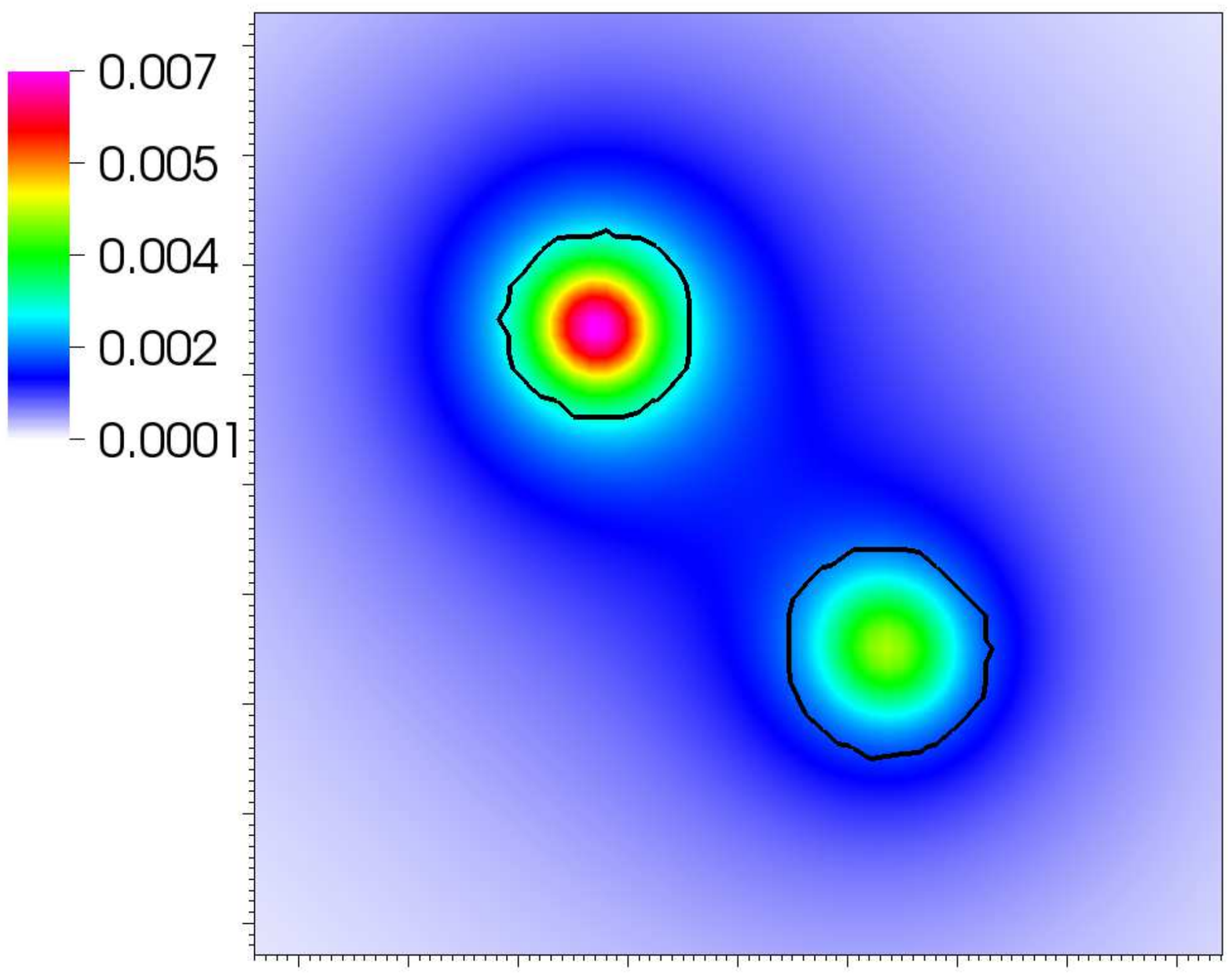}
\includegraphics[width=5.cm,angle=0]{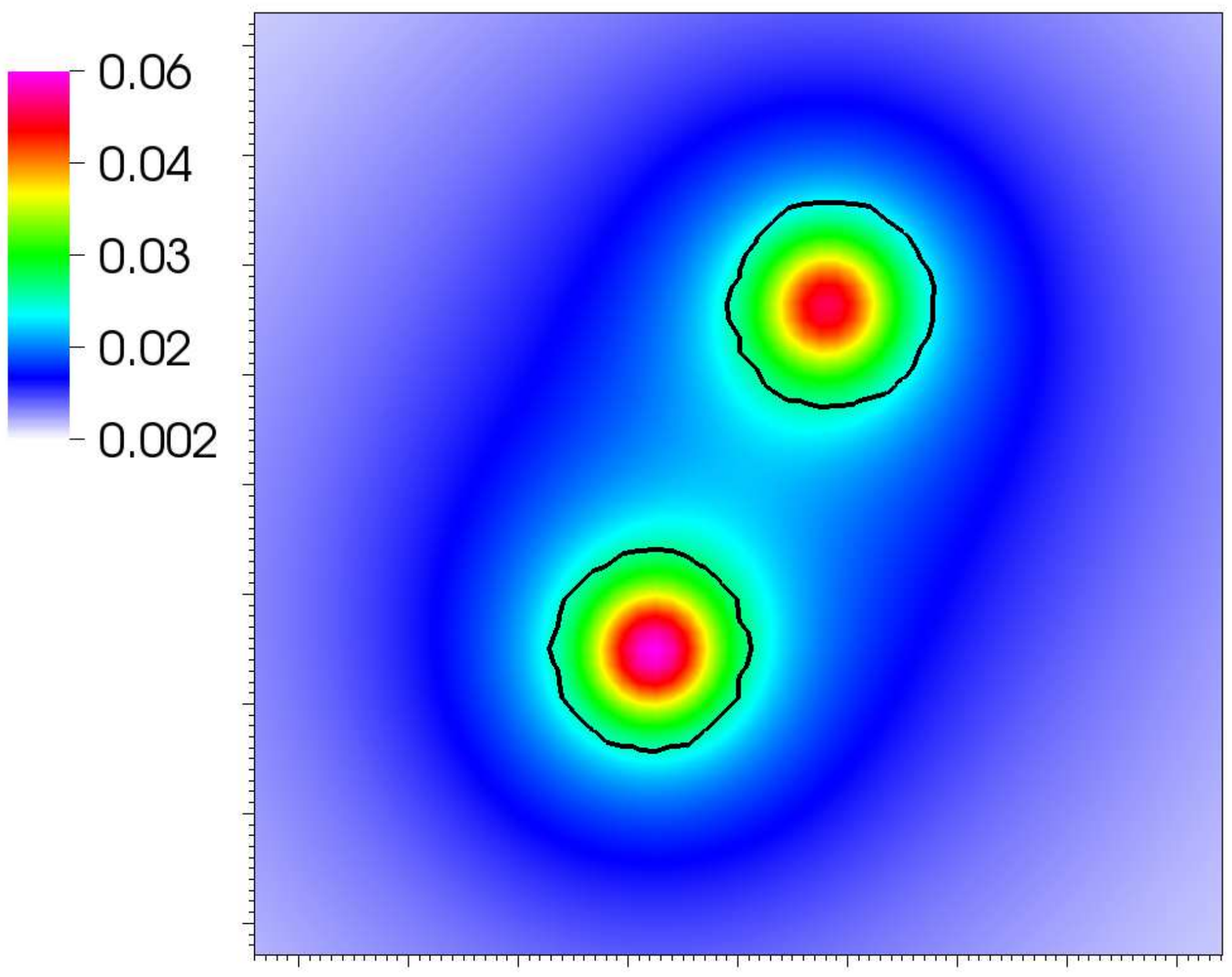}
\vskip -1.6cm
\includegraphics[width=5.cm,angle=0]{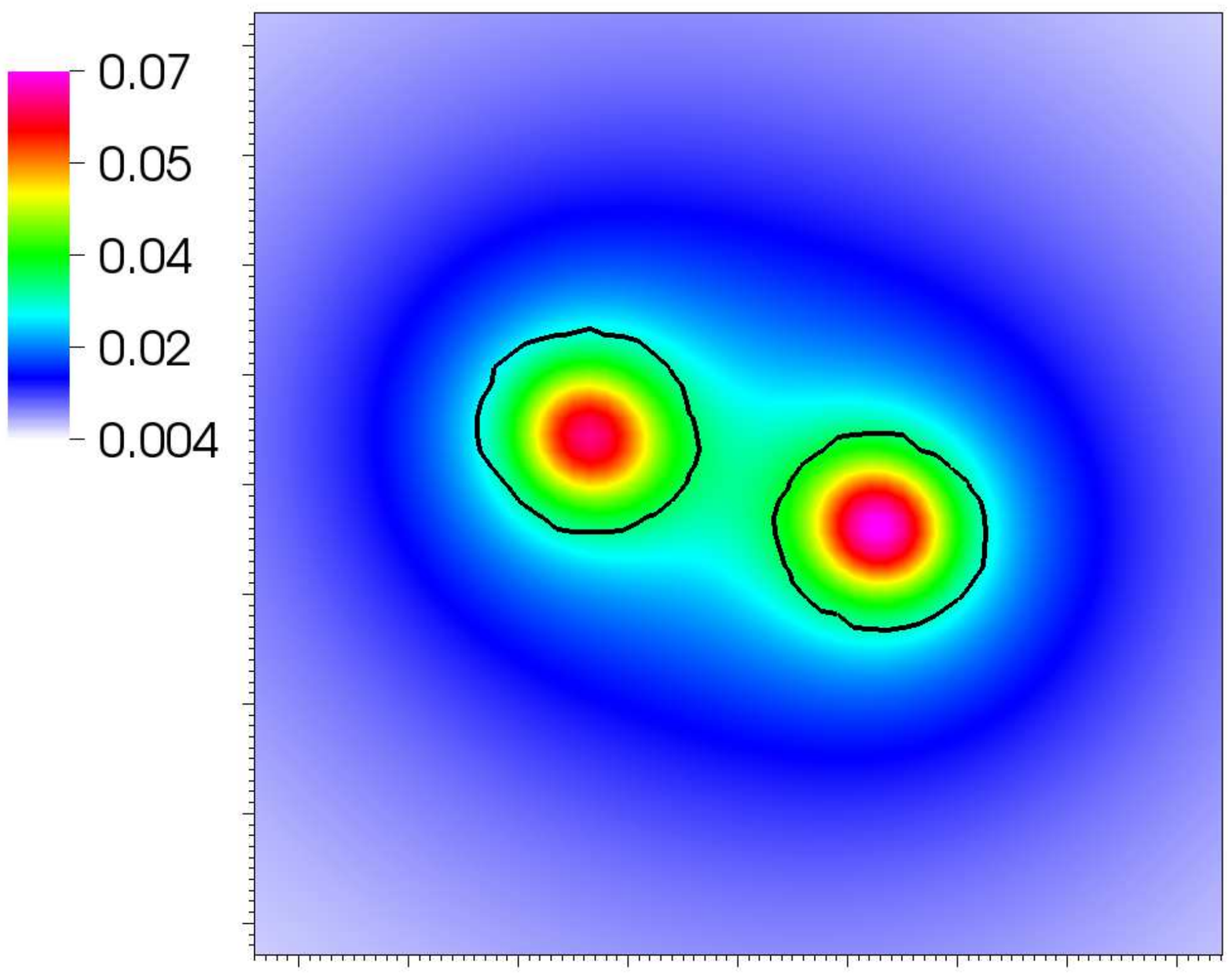}
\includegraphics[width=5.cm,angle=0]{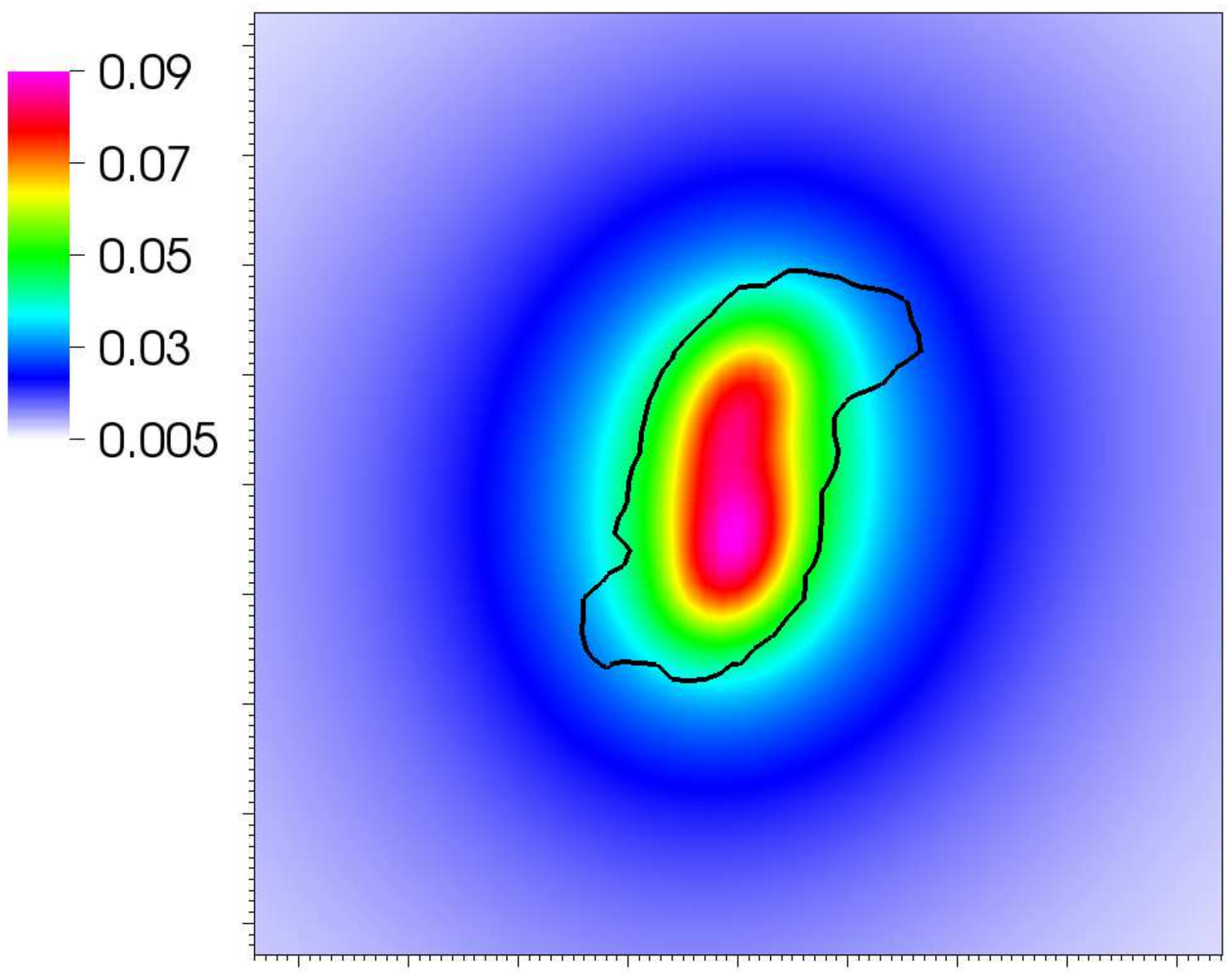}
\vskip -1.cm
\caption{Snapshots of the scalar field value (color code) and the
  stellar surfaces (solid black line) at $t=[1.8,3.1,4.0,5.3]$ms for
  a binary system of NSs with gravitational masses $(1.64,\,1.74)
  M_\odot$ in isolation, and a scalar-tensor theory with $\beta_0 =
  -4.5$. [From~\cite{Barausse:2012da}.]
\label{snapshots}}
\end{figure}

The dynamics of scalar-tensor theories of gravity is different from GR
whenever the spacetime contains matter sources.  This is evident in
the Einstein frame, where the stress-energy tensor explicitly depends
on the scalar field [see Eq.~(\ref{transftmn})], but of course it is
also true in the Jordan frame, by virtue of the physical equivalence
of the two frames (cf.~Section~\ref{subsec:ST}).
Violations of the strong equivalence principle mean that
self-gravitating objects follow trajectories that depend on their
internal composition/structure: this is the well known ``Nordtvedt
effect''~\cite{Nordtvedt:1968qr,Roll:1964rd,Eardley:1975}.

For generic (Bergmann-Wagoner) scalar-tensor theories, the
dimensionless coupling $\alpha(\varphi)$ between the scalar field and matter
depends on the local value of the scalar field, and it can be Taylor
expanded as (see Eq.~(\ref{DEalphabeta}))
\begin{equation}
  {\alpha}( \varphi )\equiv \f{d\ln A(\varphi)}{d\varphi}
        = \alpha_0
  +{\beta_0} (\varphi-\varphi_0)
        + {\cal O} ({\varphi})^2,
        \label{coupling}
\end{equation}
where $\alpha_0=1/\sqrt{3+2 \omega_{BD}}$ and $\beta_0$ are
dimensionless constants, and $\varphi_0$ is the asymptotic value of
the scalar field.
As discussed in Section~\ref{sec:NS_ST}, the constant $\alpha_0$ is
severely constrained by Solar System experiments ($\omega_{BD} >
40\,000$, or $\alpha_0<3.5\times
10^{-3}$)~\cite{Will:2014xja}. Observations of binary pulsars imply
${\beta_0}\gtrsim-4.52$, because for sufficiently negative values of
$\beta_0$ ($\beta_0\lesssim-4.35$ for a static NS) spontaneous
scalarization would set in (see Section~\ref{sec:NS_ST}), and the
motion of NSs in binary systems would be affected in ways that are
severely constrained by binary pulsar data (see
Section~\ref{subsec:pulsars}). Note however that these constraints are
somewhat degenerate with the EOS~\cite{Shibata:2013pra}, i.e., a
different EOS changes the value of $\beta_0$ below which dynamical
scalarization appears.

\begin{figure}
\capstart
  \centering
  \includegraphics[height=3.4cm,width=5.77cm,angle=0]{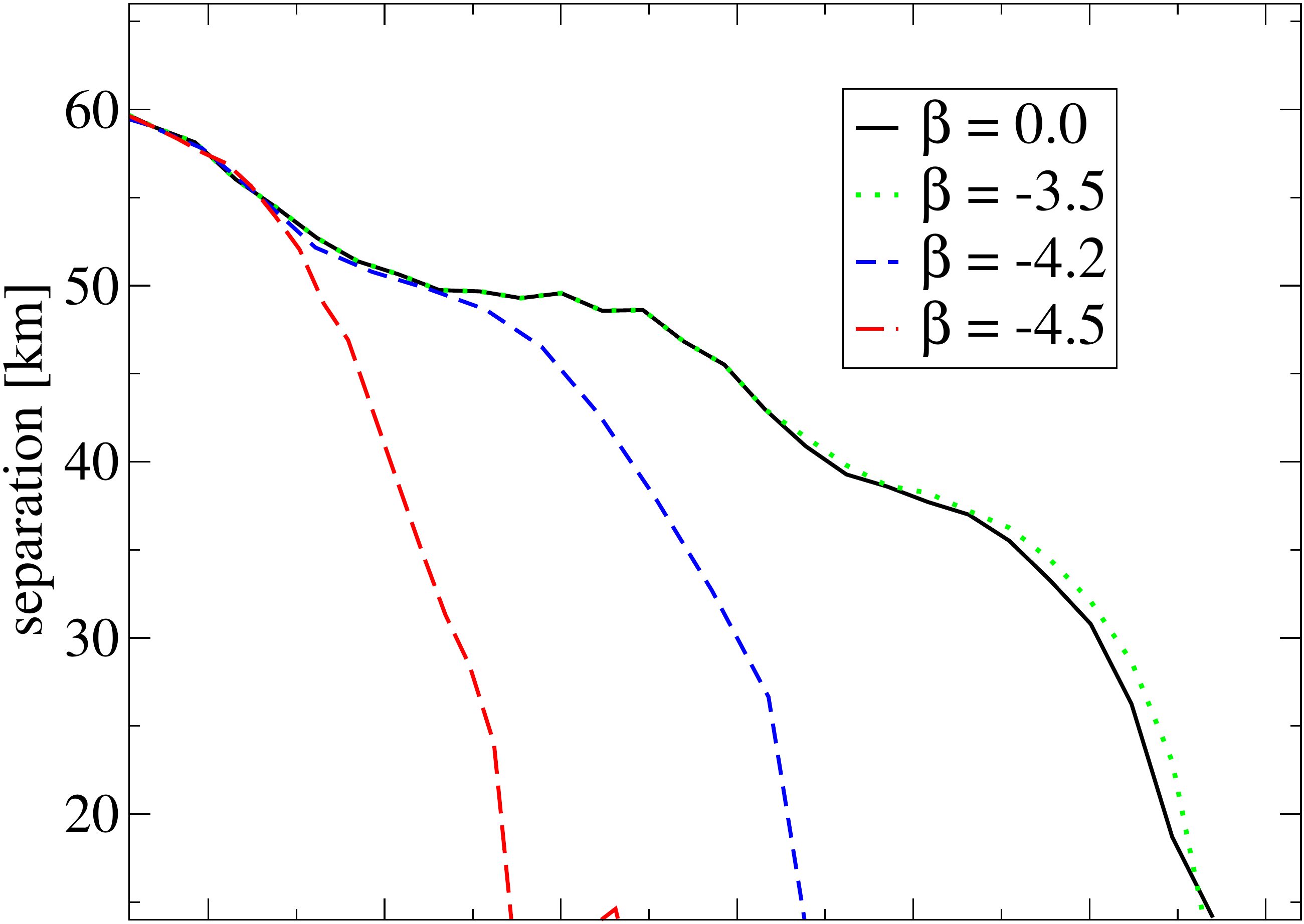}\\
  \vskip 0.028cm
  \hskip -0.25 cm\includegraphics[height=3.4cm,width=6.cm,angle=0]{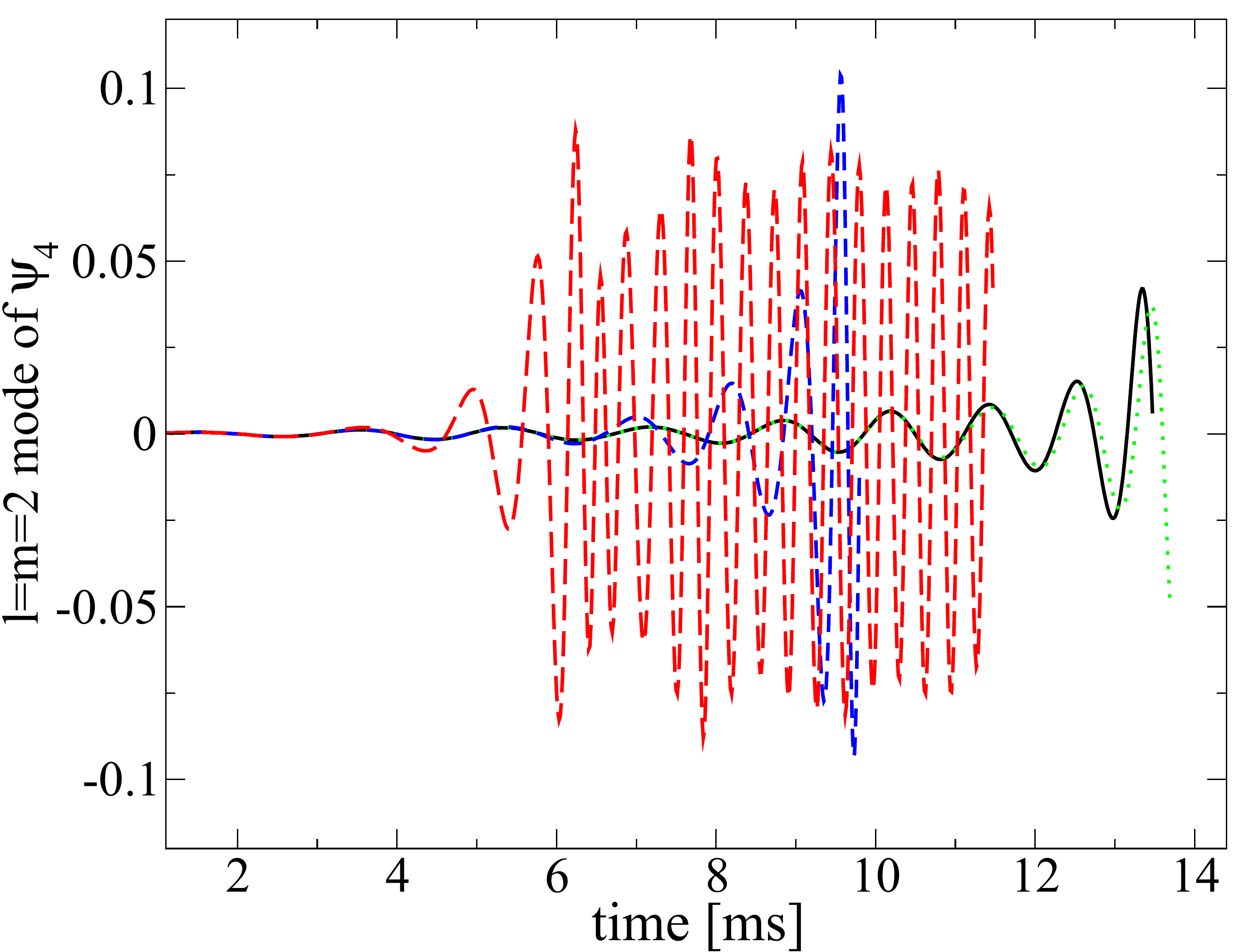}
  \caption{The separation and the dominant mode of the $\Psi_4$
    curvature scalar (which encodes the detector's response to spin-2
    GWs) for a binary system of NSs with gravitational masses $\{
    1.64, 1.74\} M_\odot$ in isolation, and for different values of
    $\beta_0$. [From~\cite{Barausse:2012da}.]
  \label{fig:separation_gw}}
\end{figure}

\begin{figure}
\centering
\includegraphics[width=0.6\textwidth,angle=0]{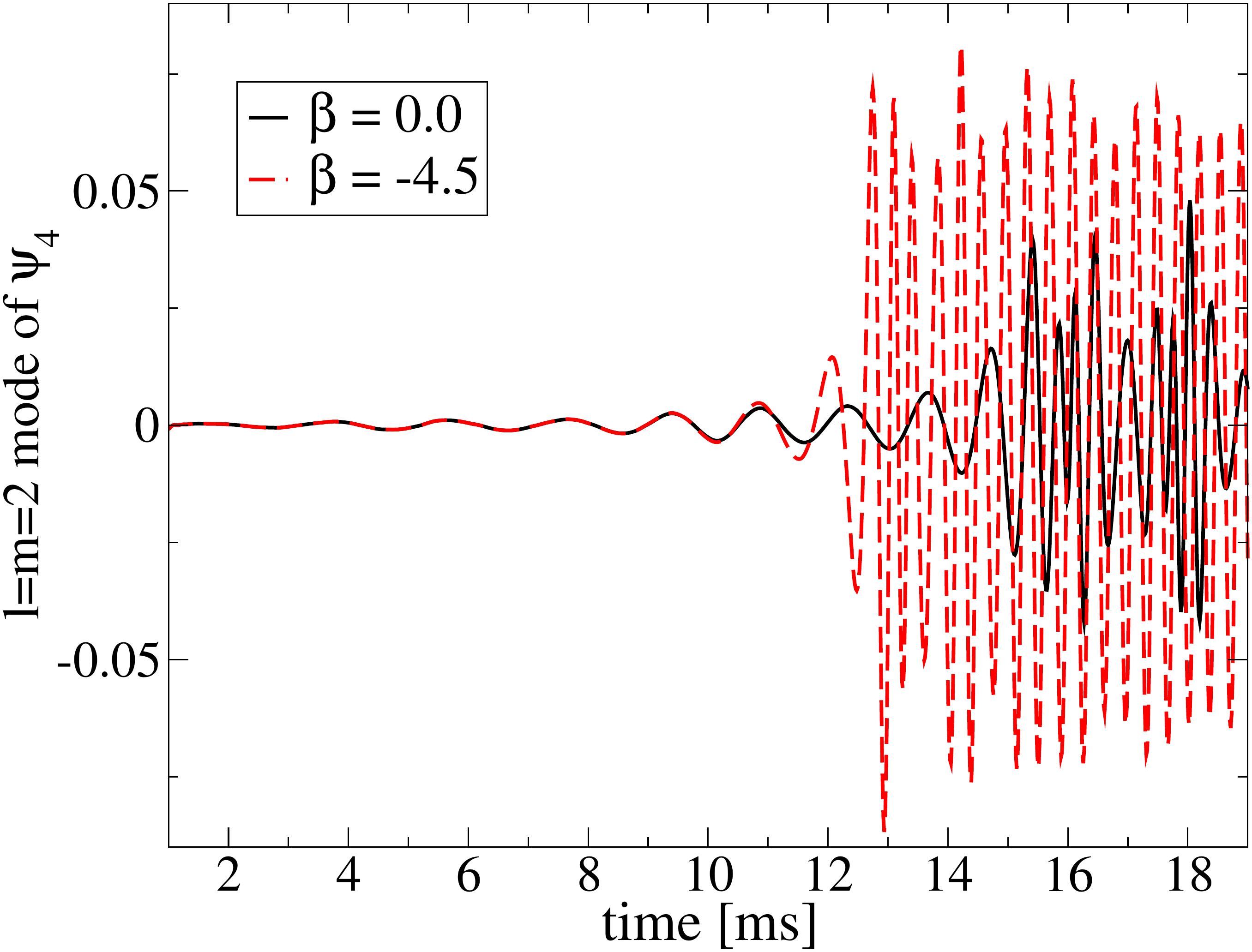}
\caption{The dominant mode of the $\Psi_4$ scalar, for $\beta_0=-4.5$
  and an equal-mass binary with gravitational masses of $1.51 M_\odot$
  in isolation. Note that neither star undergoes spontaneous
  scalarization in isolation, but the GW signal is still different
  from GR in the late inspiral/plunge because of the dynamical
  scalarization of the system.
\label{fig:gw2} [Adapted from~\cite{Barausse:2012da}.]}
\end{figure}

Recently, Refs.~\cite{Barausse:2012da} (using fully relativistic
numerical simulations, performed in the Einstein frame)
and~\cite{Palenzuela:2013hsa} (using semi-analytical arguments)
discovered a phenomenon similar to spontaneous scalarization in the
late stages of the evolution of NS-NS binaries: ``dynamical
scalarization.'' Their results were independently confirmed by Shibata
et al.~\cite{Shibata:2013pra}.  Even in cases in which the individual
NSs would \textit{not} spontaneously scalarize in isolation, the
scalar field inside each star grows when the binary separation
decreases to about $50-60~{\rm km}$
(cf. e.g. Fig.~\ref{snapshots}). This growth has a strong effect on
the binary dynamics, and produces an earlier plunge than in GR
(cf.~the upper panel of Figure~\ref{fig:separation_gw}). The plunge is
followed by the formation of a rotating bar-like matter configuration,
which sheds angular momentum in GWs before collapsing to a BH.  The
resulting gravitational waveforms are significantly different from GR
at frequencies $\sim 500-600~{\rm Hz}$, as shown in the lower panel of
Figure~\ref{fig:separation_gw}, as well as in Fig.~\ref{fig:gw2} for a
different system. Deviations at even lower frequencies are also
possible for certain binary systems and theory
parameters~\cite{Palenzuela:2013hsa,Sampson:2014qqa}.  Therefore, the
effects of dynamical scalarization are in principle detectable (at
least in some cases) with Advanced LIGO, Advanced VIRGO and KAGRA, for
values of the coupling parameters $\omega_0$ and $\beta$ that are
still allowed by all existing Solar System and binary pulsar tests, as
recently shown in~\cite{Sampson:2014qqa}.  Deviations away from GR may
also be observable in the electromagnetic signal (driven by
magnetosphere interactions prior to merger) from binaries of
magnetized NSs~\cite{Ponce:2014hha}. While these deviations are
subtle, they might provide a way in which measurements of
electromagnetic counterparts to GW sources can increase the confidence
with which GR will be confirmed (or ruled out) by GW observations.

\subsection{$f(R)$ theories}

Compact binaries in $f(R)$ gravity have been studied by considering
perturbative corrections to the Einstein-Hilbert action
(i.e.~$f(R)=R+aR^2$ with $aR\ll1$) and linear perturbations of
Minkowski spacetime. Berry and Gair~\cite{Berry:2011pb} used this
approach to compute the stress-energy pseudotensor and the parameters
of the PPN expansion. Within the same framework, but exploiting the
equivalence between $f(R)$ gravity and scalar-tensor theories, Naf and
Jetzer~\cite{Naf:2010zy} studied corrections to the periastron
precession for compact binaries, and in a follow-up
work~\cite{Naf:2011za} they also computed corrections to the GW flux
formula up to $\mathcal{O}((aR)^2)$. The flux formula derived
in~\cite{Naf:2011za} predicts that the binary would produce monopole
and dipole radiation in addition to the ordinary quadrupolar
radiation;
these contributions are expected to dominate the non-GR part of the
flux, because they enter at the lowest orders in a PN expansion.
De Laurentis and Capozziello~\cite{DeLaurentis:2011tp} derived a flux
formula at $\mathcal{O}(aR)$ in a similar perturbative expansion (see
also~\cite{DeLaurentis:2013zv}).
At this order, gravitational radiation does not contain monopole or
dipole contributions, but the quadrupole contribution has a correction
linear in $f^{\prime\prime} = 2a$.
To the best of our knowledge, no calculation of the sensitivities in
the context of $f(R)$ gravity has been performed yet. The results
should be qualitatively similar to those in scalar-tensor theory, due
to the equivalence between the two formulations.  

At the moment of writing there are no numerical investigations of
compact binaries in $f(R)$ gravity, but this is not due to pathologies
in the theory. In fact, $f(R)$ gravity is equivalent to a special
scalar-tensor theory, and as such it inherits the well-posedness
properties of scalar-tensor theories (see
Section~\ref{sec:nr_st}). Preliminary work on the feasibility of
numerical relativity simulations in $f(R)$ gravity can be found
in~\cite{Paschalidis:2011ww}.

\subsection{Quadratic gravity}
\label{sec:binary-quadratic-gravity}
\newcommand{\PPE}{{\mbox{\tiny PPE}}} The compact binary problem in
quadratic gravity theories has been studied
in~\cite{Yagi:2011xp,Yagi:2013mbt}. In~\cite{Yagi:2011xp}, both
parity-even and parity-odd theories were studied, for quasicircular
orbits consisting of objects with yet-undetermined scalar monopole
moments (parity-even theories) or scalar dipole moments (parity-odd
theories). At that time, moments were only known for BHs, along with
the result (see Section~\ref{sec:NSs/quadratic-gravity}) that NSs have
no $1/r$ scalar hair in EdGB or dCS gravity. In~\cite{Yagi:2013mbt},
the authors focused on dCS. They first constructed numerical solutions
for NSs to second order in rotation (these solutions include the
leading dipole piece of the scalar field solution; see
Section~\ref{sec:NSs/quadratic-gravity} and specifically
Figure~\ref{fig:NSsCS}). With these dipole moments in hand, they were
then able to study eccentric binaries consisting of either BHs or NSs.

There are four dominant physical corrections that arise in the compact
binary problem in quadratic gravity theories.
\begin{enumerate}
\item The scalar field solutions sourced by both compact objects
  interact with each other, just as electric charges or magnetic
  dipoles interact through the electromagnetic field (scalar pole-pole
  interaction). This modifies the binding energy of the binary and
  hence the Kepler relation (orbital frequency as a function of
  separation).
\item All metric multipole moments are shifted. The mass monopole
  shift and mass-current dipole shift are unobservable, i.e.~these
  shifts are absorbed back into the definition of physical mass and
  spin angular momentum. However, higher moments' shifts can not be
  absorbed and so they affect the motion. These also correct the
  binding energy, the Kepler relation, and cause additional
  precession.
\item The scalar field is dynamical, and sourced by a configuration of
  scalar monopoles or dipoles (plus higher moments) which are
  orbiting. Thus, there may be a time-varying scalar dipole or
  quadrupole, which sources scalar radiation. This radiation carries
  away energy and thus the system inspirals at a different rate.
\item Finally, the GWs emitted by the system are also corrected. The
  change in the gravitational waveform leads to another change in the
  energy flux, and thus an additional correction to the inspiral rate.
\end{enumerate}

Not all of the effects listed above can be directly physically observed. The
three primary observables relevant to the compact binary problem are:
\begin{enumerate}
\item[1.] The correction to the precession of pericenter, $\delta\langle\dot{\omega}\rangle$.
\item[2.] The correction to the orbital decay $\delta\dot{P}_{b}$, or equivalently, the
  correction to the energy flux $\delta\dot{E}$.
\item[3.] The modification to the gravitational waveform, which can be
  parametrized via the parametrized post-Einsteinian (PPE) parameters 
$(\alpha_{\PPE},a_{\PPE})$ for
  the amplitude and $(\beta_{\PPE},b_{\PPE})$ for the phase of the
  waveform (see Section~\ref{sec:paramtests} for more on the PPE 
parameterization).
\end{enumerate}
The first two of these corrections are observables for pulsar timing,
and the third is the observable for GW detection.  As the authors
of~\cite{Yagi:2011xp} only considered quasicircular orbits, they did
not compute $\delta\langle\dot{\omega}\rangle$.

Almost the entire compact binary problem can be recovered by modeling
compact objects as point particles with scalar hair. This amounts to
replacing the scalar field's source term (in its equation of motion)
with an \emph{effective} source term $\tau_{\mathrm{eff}}$, which recovers the correct
far-field solution~\cite{Yagi:2011xp}. There are some effects which
are not captured, but they were shown to be subdominant. The
effective equation of motion for the scalar field $\phi$ is then
\begin{equation}
  \square\phi = \tau_{\mathrm{eff}} =
  -4\pi q_{1}\delta^{(3)}(\mathbf{x}-\mathbf{x}_{1}) + (1 \leftrightarrow 2)
\end{equation}
when the bodies have scalar monopole charges $q_{A}$ $(A=1\,,2)$, as
in the case of EdGB or other parity-even theories; or
\begin{equation}
  \square\phi = +4\pi
  \mu^{i}_{1}\pd_{i}\delta^{(3)}(\mathbf{x}-\mathbf{x}_{1}) + (1 \leftrightarrow 2)
\end{equation}
when the bodies have scalar dipole hair $\mu_{A}^{i}$ (the
generalization to general $\ell$-pole is covered
in~\cite{Stein:2013wza}). The quantities $q_{A}$ or $\mu_{A}^{i}$ are
found from the strong-field matching procedure, and are proportional
to the small parameters $\alpha_i$ defined in
Eq.~\eqref{eq:quadratic_expansion}.
Thus, all of the above effects depend on powers of the $\alpha_i$'s,
which are the physical parameters that can be constrained from
observations.

The scalar pole-pole interaction, correction (i), can be computed by
``integrating out'' the scalar field at the level of the action. This
gives, for the case of dCS (and correcting a sign error
in~\cite{Yagi:2013mbt}), an interaction potential
\begin{equation}
\label{eq:U-int-dCS}
U_\mathrm{int}
= 4\pi \frac{1}{r_{12}^{3}} \left[ 3(\mu_{1}\cdot n_{12})
  (\mu_{2}\cdot n_{12}) - (\mu_{1}\cdot\mu_{2})  \right] \,,
\end{equation}
where $r_{12}$ is the distance between the two bodies,
and $n^i_{12}=(x_1^i-x_2^i)/r_{12}$.

The general pole-pole interaction (again with a sign error) is given
in~\cite{Stein:2013wza}. The dependence on $r_{12}$ is always as
$r_{12}^{-1-s-t}$, where $s$ and $t$ are the $\ell$'s of the dominant
scalar moments of the two bodies, e.g.~$s=0$ for monopole, $s=1$ for
dipole, etcetera. Notice that for the case of a BH-BH binary in EdGB,
where $s=t=0$, the shift in binding energy is $1/r$, just as in the
Kepler binding energy, and so this effect is completely absorbed by
redefining the gravitational constant $G$.

The corrected metric multipole moments [correction (ii)] come from the
strong-field matching calculation.

Computing corrections (iii) and (iv) now requires the far-field
radiative parts of the scalar and the metric, as sourced by the
dynamics of the binary. The far-field solution comes from the
post-Minkowskian expansion of the retarded Green function for
$\square$, e.g.
\begin{align}
\label{eq:vartheta-FZ}
\phi^{\rm FZ} =&\, \frac{1}{r} \sum_{m} \frac{1}{m!}
 \frac{\partial^m}{\partial t^m} \int_{\mathcal{M}} 
  \frac{-\tau'_{\mathrm{eff}}}{4\pi} ({n}_{j} \; {x}'^{j})^m d^3x' \,,
\end{align}
where $\mathcal{M}$ denotes a $t-r=\text{constant}$ hypersurface.

There is a similar expression for the GW correction. The
orbit-averaged energy flux then comes from inserting $\phi^{FZ}$ into
the stress-energy tensor $T_{\mu\nu}$ and evaluating
\begin{align}
\label{eq:Edotdefinition}
\dot{E}^{(\varphi)} = \lim_{r\to\infty} \int_{S^2_r} 
\left<T^{(\varphi)}_{ti} n^i \right>_{\omega}  r^2 d\Omega
\end{align}
on a 2-sphere at $r\to\infty$ (here $\langle\cdot\rangle_{\omega}$
denotes the orbit-average operation). Again, there is a similar
expression for the corrected energy flux from GWs.

As seen in Eq.~\eqref{eq:vartheta-FZ}, the radiative moments of
$\phi$ come directly from time derivatives of the multipole
moments of $\tau$, and we must study which moment will dominate. The
leading-in-$\alpha_i$ part of these corrections can be computed from the
leading PN motion of the binary, i.e.~simply Keplerian motion.

For a system containing a BH in EdGB, scalar dipole radiation will
dominate, and is a pre-Newtonian effect (as in scalar-tensor theory,
it contains fewer powers of $v^{2}$ than the GR quadrupolar energy
flux). For circular orbits, this is~\cite{Yagi:2011xp}
\begin{equation}
\label{eq:Edot-vartheta-EdGB}
\delta \dot{E}^{(\phi)} = -\frac{4\pi}{3}{1\over m^4}(m_2 q_1 -m_1 q_2)^2 v^8
\end{equation}
For either BH or NS systems in dCS, the variation of the scalar dipole
moment occurs on the spin-precession timescale, while the variation of
the scalar quadrupole moment occurs on the orbital timescale, so the
latter should dominate. For a binary with semimajor axis $a$ and
eccentricity $e$, this quadrupole contribution is~\cite{Yagi:2013mbt}
\begin{equation}
\label{eq:Edot-vartheta-dCS}
\delta\dot{E}^{(\phi)} = -\frac{5}{768}\frac{\alpha_{4}^{2}}{\pi m^{4}}\left(\frac{m}{a}
\right)^{7} \frac{ 2 \Delta_{1}^{2}f_{1}(e) + 2\Delta_{2}^{2}
  f_{2}(e)+\Delta_{3}^{2}f_{3(e)}}{(1-e^{2})^{11/2}}\,,
\end{equation}
where $f_{i}(e)=1+\ldots$ are $\mathcal{O}(1)$ polynomials of degree
6, $\Delta_{i}$ is an $\mathcal{O}(1)$ vector which depends on the
bodies' scalar dipole moments (and hence their spin vectors), and
where we have accounted for a difference in the convention for
$\alpha_i$ between~\cite{Yagi:2013mbt} and
Eq.~\eqref{eq:action_quadratic}.

Eqs.~\eqref{eq:Edot-vartheta-EdGB} and~\eqref{eq:Edot-vartheta-dCS}
are correction (iii) listed above. In EdGB, since correction (iii) was
pre-Newtonian, the GW effect (iv) is subdominant, so it was not computed
in~\cite{Yagi:2011xp}. However, in dCS, correction (iv) is of the same
PN order, and it is given by~\cite{Yagi:2013mbt}
\begin{align}
\label{eq:Edot-h-dCS}
\delta\dot{E}^{(\mathfrak{h})} =
-\frac{15}{16} \frac{\alpha_{4}^{2}}{\pi m^4}
\eta\left( \frac{m}{a} \right)^{7}
\chi_{1}\chi_{2}\bar{\mu}_{1}\bar{\mu}_{2}C_{1}^{3}C_{2}^{3} 
\frac{g_{1}(e)\hat{S}_{1}^{x}\hat{S}_{2}^{x} + g_{2}(e)\hat{S}_{1}^{y}\hat{S}_{2}^{y}
  + 2 g_{3}(e)\hat{S}_{1}^{z}\hat{S}_{2}^{z} }{(1-e^{2})^{11/2}}\,,
\end{align}
where $\eta=\mu/m$ is the symmetric mass ratio, $g_{i}(e)=1+\ldots$
are $\mathcal{O}(1)$ polynomials of degree 6, $\chi_{A}$ is the
dimensionless spin angular momentum of body $A$, $\bar{\mu}_{A}$ is the
dimensionless magnitude of the scalar dipole moment, $C_{A}$ is the
dimensionless compactness of body $A$, $\hat{S}^{i}_{A}$ is the
normalized spin vector of body $A$, and $z$ is the direction orthogonal to the orbital plane.

We now turn to the observable signatures of the above effects. 
A detailed pulsar timing model does not exist, but it is still
possible to compute the averaged additional precession of pericenter,
$\delta\langle\dot{\omega}\rangle$. The standard way to do this
averaged calculation is with the Gauss perturbation
method~\cite{will1993theory,Yagi:2013mbt}. First, the perturbing
acceleration is decomposed into components in an orthonormal frame
which co-rotates with the binary, with two axes aligned with $n_{12}$
and the unit angular momentum vector $\hat{L}$. The components are then inserted into standard formulae
which are averaged over the orbital phase of the
binary. Because~\cite{Yagi:2011xp} focused on quasi-circular orbits,
this result was only computed in~\cite{Yagi:2013mbt}. The additional
precession arising from the effect (i) (the scalar pole-pole
interaction) is
\begin{align}
\delta\langle\dot{\omega}\rangle_\phi &= \frac{75}{256}
\frac{1}{\mu}\frac{\alpha_{4}^{2}}{\pi m^4}
\frac{\chi_{1}\chi_{2}}{(1-e^{2})^{2}} C_1^3 C_2^3 \bar{\mu}_1
\bar{\mu}_2 \left( \frac{m}{a}  \right)^{7/2}
\left\{
\frac{1}{2}\left(\hat{S}_{1,x}\hat{S}_{2,x}
 +\hat{S}_{1,y}\hat{S}_{2,y}\right)
-\hat{S}_{1,z}\hat{S}_{2,z} \right.\nonumber\\
&\qquad\left.{}-\cot\iota \left[
\hat{S}_{1,z}\left(
\hat{S}_{2,x}\sin\omega + \hat{S}_{2,y}\cos\omega
\right) 
\right]
\right\}+(1\leftrightarrow 2)\,,
\end{align}
where $\iota$ is the orbital inclination.
The same calculation can be repeated for the perturbing acceleration
arising from effect (ii), the shift in the metric quadrupole moment of
each body, $Q_1$, $Q_2$. This latter effect is estimated to be more important than
$\delta\langle\dot{\omega}\rangle_{\phi}$, and it is given by
\begin{multline}
\delta\langle \dot\omega\rangle_h =
\frac{3}{a^{7/2}\sqrt{m}(1-e^{2})^2}
Q_{1} \Big[ -1 + \frac{3}{2}\left(
    \hat{S}_{1,x}^{2}+\hat{S}_{1,y}^{2} \right)\\
  -\hat{S}_{1,z}\cot\iota \left( \hat{S}_{1,x}\sin\omega +
    \hat{S}_{1,y}\cos\omega \right)
 \Big] +(1\leftrightarrow 2)\,.
\end{multline}

The rate of a binary pulsar's pericenter precession, $\dot{\omega}$,
is measured with much more precision than the rate of orbital decay,
$\dot{P}_{b}$. Thus, observable (i) is much better for placing
constraints than observable (ii). Regardless, it is algebraically
straightforward to combine the Kepler binding energy, the shift in the
binding energy [e.g.~Eq.~\eqref{eq:U-int-dCS}], and the shift in the
energy flux
[e.g.~Eq.~\eqref{eq:Edot-vartheta-EdGB},~\eqref{eq:Edot-vartheta-dCS},
  or~\eqref{eq:Edot-h-dCS}] to find the leading-order in $\alpha_i$
correction to $\delta\dot{P}_{b}$. This has been computed for a
circular BH-BH binary in dCS in~\cite{Yagi:2012vf} (this is not
sufficient for pulsar timing constraints, but the calculation is very
similar), giving $\dot{f} = \dot{f}_{GR} ( 1+ \delta C u^4 )$, where
$u\equiv(\pi m f)^{1/3}$, $f$ is the GW frequency (twice the orbital
frequency) and
\begin{align}
\label{eq:dCS-delta-C}
\delta C ={}& \frac{313345}{1107456} \frac{\alpha_{4}^{2}}{\pi m^{4}} \frac{m^2}{m_1^2} \chi_1^2 \left[ 1 - \frac{186607}{62669} \left( \hat{\bm{S}}_1 \cdot \hat{\bm{L}} \right)^2 \right] \nn \\
& {}+ \frac{99625}{316416} \frac{\alpha_{4}^{2}}{\pi m^{4}} \frac{\chi_1 \chi_2}{\eta} \left[ \left( \hat{\bm{S}}_1 \cdot \hat{\bm{S}}_2 \right) - \frac{8327}{3985} \left( \hat{\bm{S}}_1 \cdot \hat{\bm{L}} \right) \left( \hat{\bm{S}}_2 \cdot \hat{\bm{L}} \right) \right] + (1 \leftrightarrow 2)\,.
\end{align}

Finally, we come to observable (iii), which is the shift in the
gravitational waveform, relevant to ground-based detectors such as
Advanced LIGO. By the time a binary gets into the frequency band
relevant to these detectors, it is assumed that most eccentricity will
have been damped out, and so for quadratic theories these
calculations have only been done for quasi-circular orbits (see
Section~\ref{sec:paramtests}). Since the phase of a gravitational
waveform is measured with much more precision than the amplitude, most
of the attention has been paid to the phase. This is parameterized in
the PPE formalism in terms of $\beta_{\PPE}$ and $b_{\PPE}$ via
\begin{equation}
\label{eq:PPE-phase}
\Psi_{\PPE} = \Psi_{\rm GR} + \beta_{\PPE} (\pi \mathcal{M} f)^{b_{\PPE}}\,,
\end{equation}
where the chirp mass is $\mathcal{M}\equiv m\eta^{3/5}$.

For even-parity theories, the dominant physical effect comes from
(iii), the dipolar scalar radiation. This was computed for BH-BH binaries
in~\cite{Yagi:2011xp} as
\begin{equation}
  \beta_\PPE = - \frac{5}{7168} \frac{\alpha_{3}^{2}}{\pi m^{4}} \frac{\delta 
m^2}{m^2} \eta^{-18/5}\,,\qquad
  b_{\PPE} = -7/3\,.
\end{equation}
Meanwhile, effects (i)-(iv) all contribute for the dCS
calculation. In~\cite{Yagi:2012vf}, this was computed for circular
BH-BH binaries as
\begin{equation}
\beta_{\PPE} = - \frac{15}{64} \; \delta C \; \eta^{-4/5}, \qquad
b_{\PPE} = -\frac{1}{3}\,,
\end{equation}
where $\delta C$ was given in Eq.~\eqref{eq:dCS-delta-C}.

\subsection{Lorentz-violating theories}
\label{lv-bins}

As discussed in Section~\ref{sec:sensitivities}, in Lorentz-violating
theories strongly gravitating objects are characterized by
``sensitivities'' (or \AE ther or khronon ``charges'').  As a result,
the motion of these objects does not follow geodesics of the
background geometry, but rather depends on the numerical values of the
sensitivities, and thus ultimately on the object's nature. This means
that the universality of free fall and the strong equivalence
principle are violated in these theories. More specifically, the
sensitivities enter the equations of motion already at Newtonian
level, where the acceleration of the body $A$ in a binary system is
given by
\begin{equation}
\label{Newtonian-a}
\dot{v}_A^i= -\frac{G_N \tilde{m}_B \hat n_{AB}^i}{(1+\sigma_A) r_{AB}^2}\,,
\end{equation}
with $r_{AB}=|\boldsymbol{x}_A-\boldsymbol{x}_B|$ and $\hat
n_{AB}^i=(x^i_A-x^i_B)/r_{AB}$, and $\sigma_A$ the sensitivity parameter of 
body $A$ (see Eq.~(\ref{sensitiv_lv})).
This expression can be re-written
as~\cite{Foster:2007gr}
\begin{equation}
\label{eq:New_active}
\dot{v}_A^i= -\frac{{\cal G} {m}_B \hat n_{AB}^i}{r_{AB}^2}\,,
\end{equation}
where one defines the {\textit{active}} gravitational masses
\be
\label{active-mass}
m_B\equiv\tilde{m}_B (1+\sigma_B)
\ee
and the ``effective'' gravitational constant
\be
\label{calG}
{\cal G}\equiv\frac{G_N}{(1+\sigma_A) (1+\sigma_B)}\,.
\ee
Similarly, the sensitivities appear in the equations of motion at
higher PN orders in the conservative dynamics~\cite{Foster:2007gr}.

The sensitivities enter also in the dissipative sector of the motion
of binary systems, causing both the emission of dipolar fluxes, as
well as modifications of the quadrupolar emission of GWs that takes
place already in GR. More specifically, the most relevant quantity for
binary systems (and in particular for binary pulsars) is the rate of
change of the orbital period.  For systems whose orbital dynamics is
determined by Eq.~\eqref{eq:New_active}, denoting the orbital period
by $P_{b}$ and the binary's binding energy by $E_{b}$, this quantity
can be expressed as
\be
\label{Pdot-eq}
\frac{\dot{P_{b}}}{P_{b}} = - \frac{3}{2} \frac{\dot{E}_{b}}{E_{b}}\,,
\ee
which can be further manipulated by writing the binding energy's
rate of change in terms of the total flux of energy ${\cal{F}}$
carried away by GWs, i.e.~$\dot{E}_{b} = - {\cal{F}}\,$.
This flux can be calculated explicitly from the sensitivities and the
binary's orbital parameters, yielding~\cite{Yagi:2013qpa,Yagi:2013ava}
\be
\frac{\dot{P}_{b}}{P_{b}} = - \frac{192 \pi}{5} \left(\frac{2 \pi G m}{P_{b}}\right)^{5/3} 
\frac{\mu}{m\,P_{b}}  \left<\mathcal{A}\right>\,,
\label{flux-AE}
\ee
where as usual $\mu = m_{1} m_{2}/m$ is the reduced mass, $m = m_{1} +
m_{2}$ is the total mass, and we have defined
\begin{align}
\label{A-AE}
\left<\mathcal{A}\right> &=
\frac{5}{32} \left(s_1 - s_2\right)^2 \mathcal{A}_{4}  \left( \frac{P_b}{2 \pi G m} \right)^{2/3} 
\nn \\
&+  \left[\left(1 - s_{1}\right) \left(1 - s_{2}\right)\right]^{2/3}  
\left( \mathcal{A}_1 + \mathcal{S} \mathcal{A}_2 + \mathcal{S}^{2} \mathcal{A}_3  \right) 
\nn \\
&+ {\cal{O}}(1/c^{2})\,.
\end{align}
Also, $s_{A}=\sigma_A/(1+\sigma_A)$ are rescaled sensitivities,
${\cal{S}} = m_{1} s_{2}/m + m_{2} s_{1}/m$ and
$(\mathcal{A}_{1},\mathcal{A}_{2},\mathcal{A}_{3},\mathcal{A}_{4})$
are functions of the coupling constants [$(c_{+},c_{-})$ in
Einstein-\AE ther theory and $(\beta,\lambda)$ in khronometric
gravity].
 
 Two comments are in order at this stage. First, in the GR limit one
 obtains $\mathcal{A}=1$, thus recovering the usual quadrupole
 formula. Second, for widely separated systems (such as all observed
 binary pulsars) the decay rate of the orbital period is dominated by
 the terms appearing at the lowest PN order, i.e.~with the least powers
 of $G m/P_{b}$.  Therefore, the last term in Eq.~\eqref{A-AE} (the
 dipolar emission term) dominates the orbital decay rate unless $s_{1}
 - s_{2} \approx 0$. This provides a way to constrain Lorentz
 violations in gravity with white dwarf-pulsar systems
 (cf.~Section~\ref{sec:pulsar_LV}). Nevertheless, in the case $s_1\approx
 s_2$ (relevant for instance for the relativistic double pulsar,
 cf.~Section~\ref{sec:pulsar_LV}), even though dipolar emission is
 suppressed, the sensitivities still produce changes to the
 quadrupole formula of GR (i.e., $\mathcal{A}\neq1$).

Hansen et al.~\cite{Hansen:2014ewa} computed the characteristics of
gravitational radiation from NS binary inspirals using as a starting
point the energy flux described above.  The evolution of the orbital
frequency $F$ is related to the energy flux via
\begin{equation}
\label{eq:Fdot-LV}
\dot F(u) = \dot F_\mathrm{GR}(u) [1 + \delta_{\dot{F}}(u) ]\,,
\end{equation}
where $\dot F_\mathrm{GR}$ is the GR prediction, $u \equiv (2 \pi
\mathcal{G M} F)^{1/3}$ and
\begin{equation}
\delta_{\dot{F}}(u) = \frac{7}{4} \eta^{2/5} \dot E_\mathrm{-1PN} u^{-2} + \dot E_\mathrm{0PN}
\end{equation}
is the Lorentz-violating correction to the evolution of $F$. $\dot
E_\mathrm{-1PN}$ and $\dot E_\mathrm{0PN}$ represent the
Lorentz-violating correction to the energy flux due to the dipolar and
quadrupolar radiation, respectively. They are given by
\begin{equation}
\dot E_\mathrm{-1PN} = -\frac{5}{56} \mathcal{G} (s_1 - s_2)^2 \mathcal{A}_4\,, 
\quad  
\dot E_\mathrm{0PN} = \mathcal{G} (\mathcal{A}_1 + \mathcal{S} \mathcal{A}_2 + \mathcal{S}^2 \mathcal{A}_3)-1\,.
\end{equation}
From Eq.~\eqref{eq:Fdot-LV} one can calculate the gravitational
waveform in Fourier space using the stationary-phase approximation. In
particular, the phase is given by
\be
\Psi = \Psi_\mathrm{GR} - \frac{3}{128} u^{-5} \left[ \dot E_\mathrm{-1PN} \eta^{2/5} u^{-2} + \dot E_\mathrm{0PN}
+ \mathcal{O}(c^{-2}) \right]\,. 
\ee

\subsection{Massive gravity}
Radiation from binary pulsars in the inspiral phase was studied for
the Cubic Galileon in~\cite{deRham:2012fw} and for the general case
in~\cite{deRham:2012fg}. The calculation was done by approximating the
time dependence of the source as small: $T=T_0 + \delta T$, where $T_0
= -M\delta^3(\vec{x})$ and $\delta T(\vec{x},t)$ carries the time
dependence of the inspiraling pulsars. Then upon splitting $\pi=\pi_0
+ \phi$, the background profile $\pi_0$ sourced by $T_0$ is
responsible for Vainshtein screening. The radiation in the Galileon
$\phi$ sourced by $\delta T$ in the background of $\pi_0$ was then
computed using the effective action techniques proposed
by~\cite{Goldberger:2004jt}.

For the Cubic Galileon, the dominant channel is the quadrupole
($\ell=2$)\footnote{Because the Galileon is a scalar mode, there is also
  monopole radiation. However the monopole is suppressed relative to
  the quadrupole order effect because the monopole enters as a
  relativistic correction: $P^{\ell =
    0}_{\rm cubic}/P^{\ell=2}_{\rm cubic}\sim v \sim 10^{-3}$,
  see~\cite{deRham:2012fw}.}. The power radiated is given by
\begin{equation}
\frac{P^{\ell =2}_{\rm cubic}}{P^{\ell =2}_{\rm GR}} \sim v^{-1} (\Om r_V)^{-3/2},
\end{equation}
where $v$ is the velocity of the pulsar, $\Omega$ its orbital angular
velocity and $r_V$ is the Vainshtein radius (see
Section~\ref{subsec:quadratic}). Using parameters from the
Hulse-Taylor pulsar as a fiducial example~\cite{Weisberg:2004hi}
yields $P_{\rm cubic}/P_{\rm GR} \sim10^{-7}$, well below the
observational precision $\sigma\sim10^{-3}$. It is interesting to
note, however, that the time dependence in the system makes the
Vainshtein screening less effective compared to the static case; the
force on the two pulsars is suppressed by $(\bar{r}/r_V)^{3/2}$, where
$\bar{r}$ is the separation.

For the Quartic Galileon the situation is more subtle. For a given
multipole, there is more Vainshtein suppression,
$P^\ell_{\rm quartic}/P^\ell_{\rm GR} \sim v^{-2} (\Om r_V)^{-2}$. However,
in this case the approximation of a spherically symmetric background
is not good, because higher order multipoles are not suppressed
effectively. More work is needed to understand this case, e.g.~by
taking the time dependence into account in the background.

More recently, Narikawa et al.~\cite{Narikawa:2014fua} reported on the
prospects for GW detection from coalescing compact binaries in some
models of bimetric massive gravity~\cite{Hassan:2011zd}. They find
that, in a certain region of the parameter space, the gravitational
waveform can display large-amplitude modulations induced by the
interference between two modes. The peak amplitude can be up to an
order of magnitude larger than its GR value at a given frequency, and
such frequency depends on the parameters of the theory. By using
Bayesian methods (cf.~Section~\ref{sec:paramestimation}), Narikawa et
al.~evaluate the detectability of these deviations in the waveforms by
an advanced laser interferometer, finding that there is a region of
the parameter space of the bimetric gravity theory where the
deviations can be significant. The detectable region depends on the
specific model, but typically corresponds to a graviton mass
$\mu\gtrsim 10^{-22} {\rm eV}$~\cite{Narikawa:2014fua}. Remarkably,
this value overlaps with the bounds on the graviton mass derived
through the superradiant instability of supermassive Kerr BHs in
massive gravity~\cite{PDG,Brito:2013wya} (see
Section~\ref{sec:superradiance}). It is notable that comparable bounds
could follow from GW observations of \emph{stellar}-mass objects.

\subsection{Open problems}\label{op:binaries}
\begin{itemize}
 \item It has been shown that the dynamics of a BH binary system (with
   a nonextreme mass ratio) in Bergmann-Wagoner theory coincides with
   that of GR up to $2.5$ PN order. Does this result hold at all PN
   orders? If it does not, at which PN order does it break down?

 \item Can we extend numerical relativity to modified theories of
   gravity other than scalar-tensor theory? What is the signature of
   nonlinear effects in the late inspiral and merger?

\item How do spontaneous scalarization and dynamical scalarization
  generalize to Horndeski theories or tensor-multiscalar theories? Are
  there similar nonlinear effects that could produce sensible
  deviations from GR in quadratic gravity theories, Lorentz violating
  theories or massive gravity?
\end{itemize}

\clearpage
\section{Binary pulsar and cosmological tests of general relativity}
\label{sec:BP}

\subsection{Tests of gravity from radio pulsars}
\label{subsec:pulsars}

\paragraph{Overview.} 

Before the 1970s, precision tests for gravity theories were
constrained to the weak-field, slow-motion environment of the Solar
System. In terms of relativistic equations of motion, the Solar System
gave access to the first order corrections to Newtonian dynamics,
notably the well-measured anomalous precession of Mercury's orbit and
the deflection of light by the Sun~\cite{Will:2014xja}.  Testing
anything beyond the first PN contributions, like the emission of GWs,
was for a long time out of reach.

The discovery of the first radio pulsar in a binary system,
PSR~B1913+16, by Russell Hulse and Joseph Taylor in the summer of
1974~\cite{Hulse:1974eb} initiated a completely new field for testing
relativistic gravity. For the first time, the back reaction of GW
emission on the binary motion could be studied, which gave the first
evidence for the existence of GWs as predicted by Einstein's
theory. Furthermore, the Hulse-Taylor pulsar provided the first test
bed for the gravitational interaction of strongly self-gravitating
bodies.

To date, there are a number of known radio pulsars that can be utilized
for precision tests of gravity. Depending on their orbital properties and
their companion, these pulsars provide tests for various different
aspects of gravity, for instance (see~\cite{Wex:2014nva}, and references 
therein):
\begin{itemize}
\item GR's quadrupole formula for GW emission. The best test (with
  accuracy better than 0.1\%) comes from the Double Pulsar.
\item Emission of dipolar radiation. The best bounds come from
  relativistic pulsar-white dwarf systems, most notably PSR~J1738+0333
  and PSR~J0348+0432, with constraints of order $10^{-5}$.
\item Limits on the violation of the universality of free fall for
  strongly self-gravitating bodies [i.e., tests of the strong
    equivalence principle (SEP)]. The best tests (better than 1\%) are
  provided by wide pulsar-white dwarf systems.
\item Limits on the violation of local Lorentz invariance of gravity
  from binary as well as isolated pulsars. In this context, the bounds
  are better than $10^{-8}$ for some effects.
\end{itemize}

Besides tests of specific theories (GR, scalar-tensor gravity, vector-tensor
gravity, TeVeS, etcetera) these and other pulsars allow for generic constraints 
on deviations of gravity from GR in the quasi-stationary strong-field regime, and
in the generation of GWs, in particular systems where the
(effective gravitating) masses of the system can be determined in a 
theory-independent way.

In the future, the development of new instruments and larger radio
telescopes, like the Square Kilometre Array (SKA)\footnote{%
  http://www.skatelescope.org/}, will greatly enhance our capabilities
to test gravity with radio pulsars. On the one hand, the greatly
improved timing precision will allow for better and new tests with
existing systems.  On the other hand, new instruments and survey
techniques promise the discovery of new ``gravity labs,'' like a
pulsar-BH system. Furthermore, a SKA based pulsar timing array that
utilizes several hundred millisecond pulsars to form a ``multi-armed''
GW detector will, for the first time, provide tests in the nano-Hz GW
band.

\paragraph{Main results.} Radio pulsars are rotating NSs that emit
beams of radio waves along their magnetic poles, and due to their rotation act as
``cosmic light-houses.'' Some of the ``recycled''
pulsars have rotational stabilities that are
comparable to the stability of the best atomic clocks on Earth~\cite{Hobbs:2012apa}.
Presently there are more than 2000 radio pulsars known, where about 10\% of  
them are members of binary systems~\cite{Manchester:2004bp}. 
The observation of the rotational phase of a pulsar 
with a radiotelescope is basically a ``clock-comparison'' experiment in a 
spacetime that contains the binary system with the ``pulsar clock'' on the one 
hand, and the radio telescope with its atomic clock as part of the Solar System 
on the other hand. The world-line of the pulsar and the world-line of the 
telescope are connected by the radio signals of the pulsar, propagating 
through curved spacetime. By this one directly probes the solutions of field 
equations of different gravity theories.
The technique used is the so-called {\it pulsar timing}, which basically
consists of measuring the exact arrival time of pulses at the radio telescope on
Earth and fitting an appropriate timing model to these arrival times to obtain
a phase-coherent solution, that accounts for every observed rotation of the
pulsar. For some pulsars, that phase-coherent solution stretches over several
decades, allowing for an extremely precise measurement of the key parameters. For
instance, some pulsar periods are known to atto-seconds precision, and some
orbital periods of binary pulsars have uncertainties below a micro-second (see
Table~1 of~\cite{Kramer:2012zz} and references therein). 

For binary pulsar experiments that test the quasi-stationary
strong-field regime and the GW damping, a phenomenological
parametrization, the so-called parametrized post-Keplerian (PPK)
formalism, has been introduced by Damour~\cite{1988grra.conf..315D}
and extended by Damour and Taylor~\cite{Damour:1991rd}. The PPK
formalism generically parametrizes the observable relativistic effects
that can be extracted independently from binary pulsar timing and
pulse-structure data (``post-Keplerian (PK)
parameters''). Consequently, the PPK formalism allows us to obtain
theory-independent information from binary pulsar observations by
fitting for a set of Keplerian and PK parameters. The most important
PK parameters are~\cite{1986AIHS...44..263D}:
\begin{itemize}
\item $\dot\omega$: Relativistic precession of periastron.
\item $\gamma$: Amplitude of the time dilation of the ``pulsar clock''
      (compared to an averaged time). It is a combination of the second-order
      Doppler shift and the redshift caused by the gravitational field of the
      companion.
\item $r$ and $s$: {\it Range} and {\it shape} of the Shapiro delay, caused by 
      the gravitational field of the companion.
\item $\dot{P}_{\rm b}$: Change in the orbital period due to GW damping.
\end{itemize}
Depending on the nature of the companion and the size of the orbit, different 
aspects of relativistic gravity can be tested with binary pulsars. In the following, we briefly highlight the most important systems and some of the tests
that have been performed with them.  

\vspace{2mm}

\noindent
{\bf PSR~B1913+16} (Hulse-Taylor pulsar) was the first binary pulsar to be 
discovered~\cite{Hulse:1974eb}. 
It is in a 7.8 hour, high-eccentricity ($e = 0.62$) orbit
with another NS. This system allows the measurement of three PK
parameters ($\dot\omega$, $\gamma$, $\dot{P}_{\rm b}$), and led to the 
first proof of the existence of GWs as predicted by GR~\cite{Taylor:1979zz}. Presently, it gives a 0.2\% verification of GR's 
quadrupole formula,
a precision which currently, however, cannot be further improved due to 
uncertainties in the distance to PSR~B1913+16~\cite{Weisberg:2010zz}.

\vspace{2mm}

\begin{figure}
\capstart
\begin{center}
\includegraphics[scale=0.5]{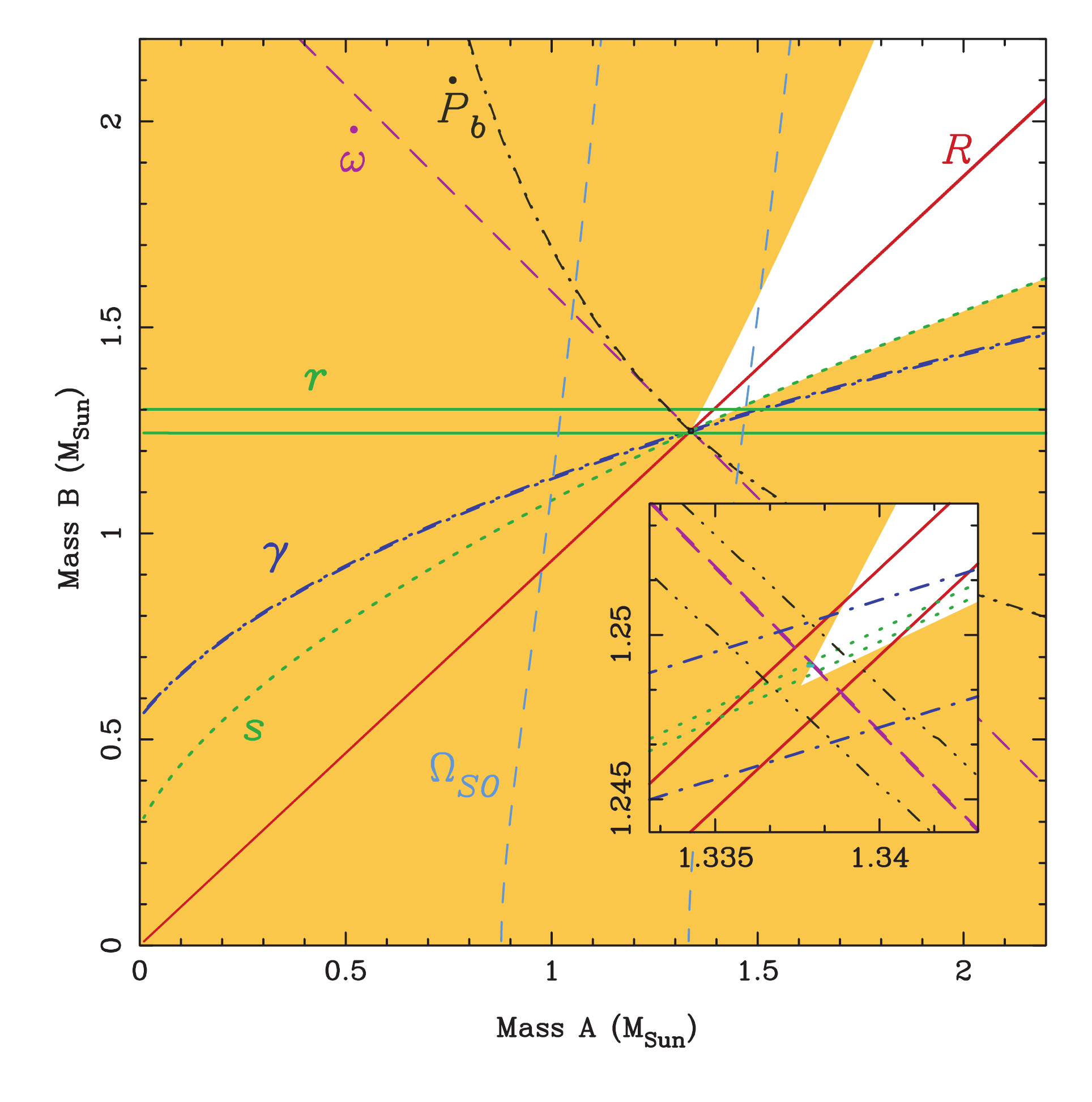}
\end{center}
\caption{GR mass-mass diagram for the Double pulsar with six PK parameters
($\dot\omega$, $\gamma$, $\dot{P}_{\rm b}$, $r$, $s$, $\Omega_{\rm B}$)
and the mass ratio ($R$). All constraints agree on a small common region (see 
inset), meaning that GR has passed this test of several relativistic effects
(quasi-stationary strong-field as well as radiative). [Figure courtesy of 
Michael~Kramer.]
\label{fig:mm0737}}
\end{figure}

\noindent
{\bf PSR~J0737$-$3039A/B} (Double Pulsar) was the first, and so far
only, binary pulsar system found to consist of two active radio
pulsars~\cite{Burgay:2003jj,Lyne:2004cj}. Pulsar A is a fast-rotating
(23\,ms) pulsar in a mildly eccentric ($e = 0.088$), 2.5 hour orbit.
Until 2008 its companion (pulsar B) was also visible as an active
radio pulsar with a rotational period of about 2.8 seconds. The timing
of both pulsars allowed an immediate determination of the mass ratio
from the projected semi-major axes of the two pulsar orbits. In the
Double Pulsar all PK parameters listed above have been measured, some
of them with exquisite precision. Most importantly, the change in the
orbital period $\dot{P}_{\rm b}$ due to GW damping has by now been
tested to agree with the quadrupole formula of GR to better than
0.1\%, giving the best test for the existence of GWs as predicted by
GR~\cite{Kramer:2012zz}. As a result of geodetic precession, pulsar B
has in the meantime turned away from our line-of-sight and is no
longer visible~\cite{Perera:2010sp}. Due to the high inclination of
the orbital plane (close to 90 degrees), pulsar A is getting eclipsed
by the plasma-filled magnetosphere of pulsar B every 2$\frac{1}{2}$
hours, for about 30 seconds around superior conjunction. Changes in
the eclipse pattern could be used to determine the rate of geodetic
precession, $\Omega_{\rm B}$, with a precision of about
13\%~\cite{Breton:2008xy}. The obtained value is in good agreement
with GR. All these tests are summarized in form of a mass-mass diagram
in Figure~\ref{fig:mm0737}.

\vspace{2mm}

\noindent
{\bf PSR~J1738+0333} is a pulsar in a nearly circular ($e \sim 3 \times 
10^{-7}$), 8.5 hour orbit with an optically bright
white dwarf. High-resolution spectroscopy allowed the determination of the
white dwarf mass (based on white dwarf models) and the motion of the white 
dwarf around the common center of mass (Doppler shifts). In combination with the
timing observations, one immediately gets the masses of the system, valid for
a large class of gravity theories: $m_{\rm p} = 1.47^{+0.07}_{-0.06}\,M_\odot$ 
and $m_{\rm c} = 0.181^{+0.007}_{-0.006}\,M_\odot$~\cite{Antoniadis:2012vy}. 
The corresponding change in the orbital period due to 
the emission of GWs as predicted by GR agrees
well with the observed value ($\sim 13$\% precision)~\cite{Freire:2012mg}. Although the precision 
is significantly worse than in the GW test with the Double Pulsar, the high
asymmetry in the compactness/binding energy of the bodies in the system 
makes this test particularly sensitive to gravitational dipolar radiation.
As a consequence, PSR~J1738+0333 provides the best test for scalar-tensor gravity 
for most values of $\beta_0$, in particular for the range $\beta_0 \gtrsim 3$ 
(see Figure~\ref{fig:stg}). 

The small eccentricity and comparably (unmeasured) high rate of
periastron advance ($\sim 1.6$\,deg/yr) allows for a test of
preferred-frame effects in the orbital motion, and in particular sets
the best limit on the strong-field generalization of the PPN parameter
$\alpha_1$: $\hat\alpha_1 = -0.4_{-3.1}^{+3.7} \times 10^{-5}$ at 95\%
confidence level\footnote{The hat denotes the fact that $\alpha_1$ can
  be modified by strong-field contributions: see e.g.~Eqs.~(194) and
  (196) in~\cite{Yagi:2013ava}.}~\cite{Shao:2012eg}.
An important piece of information for this test is the motion of the
system with respect to the frame of reference defined by the cosmic
microwave background, which is known for the PSR~J1738+0333 system,
since one also has the systemic radial velocity derived from the
spectroscopy of the white dwarf.

\vspace{2mm}

\begin{figure}
\capstart
\begin{center}
  \includegraphics[scale=0.35]{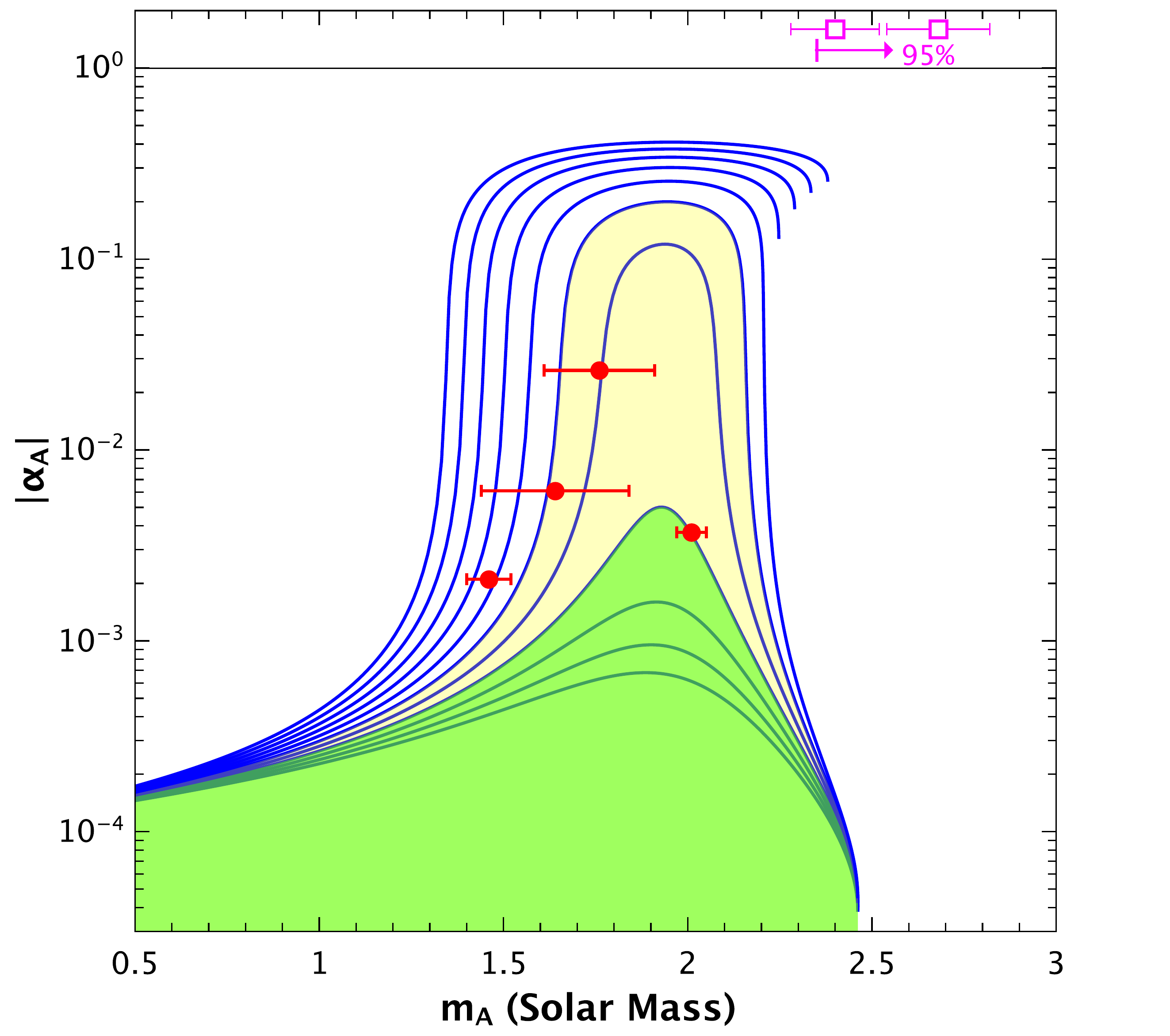}
\end{center}
  \caption{Upper limits on the effective scalar coupling $\alpha_A$
    (red dots) for four different pulsars (increasing in mass):
    PSR~J1738+0333, PSR~J1012+5307, PSR~J0437$-$4715, and
    PSR~J0348+0432. The blue lines are plotted for $\alpha_0=10^{-4}$, while 
$\beta_0$ runs from $-5.0$ to $-4.0$
    in steps of 0.1, with the largest $|\alpha_A|$ reached for
    $\beta_0 = -5.0$. Without PSR~J0348+0432 the green and yellow
    areas would be allowed, and massive NSs could be highly
    scalarized. This is now excluded with the limit from
    PSR~J0348+0432. The pink measurements at the top of the figure are
    (less robust) mass constraints for three high-mass pulsars
    candidates. All three of them do not allow for a gravity test, but
    indicate the need for a rather stiff equation of state.
\label{fig:stg0348}}
\end{figure}

\begin{figure}
\capstart
\begin{center}
  \includegraphics[scale=0.35]{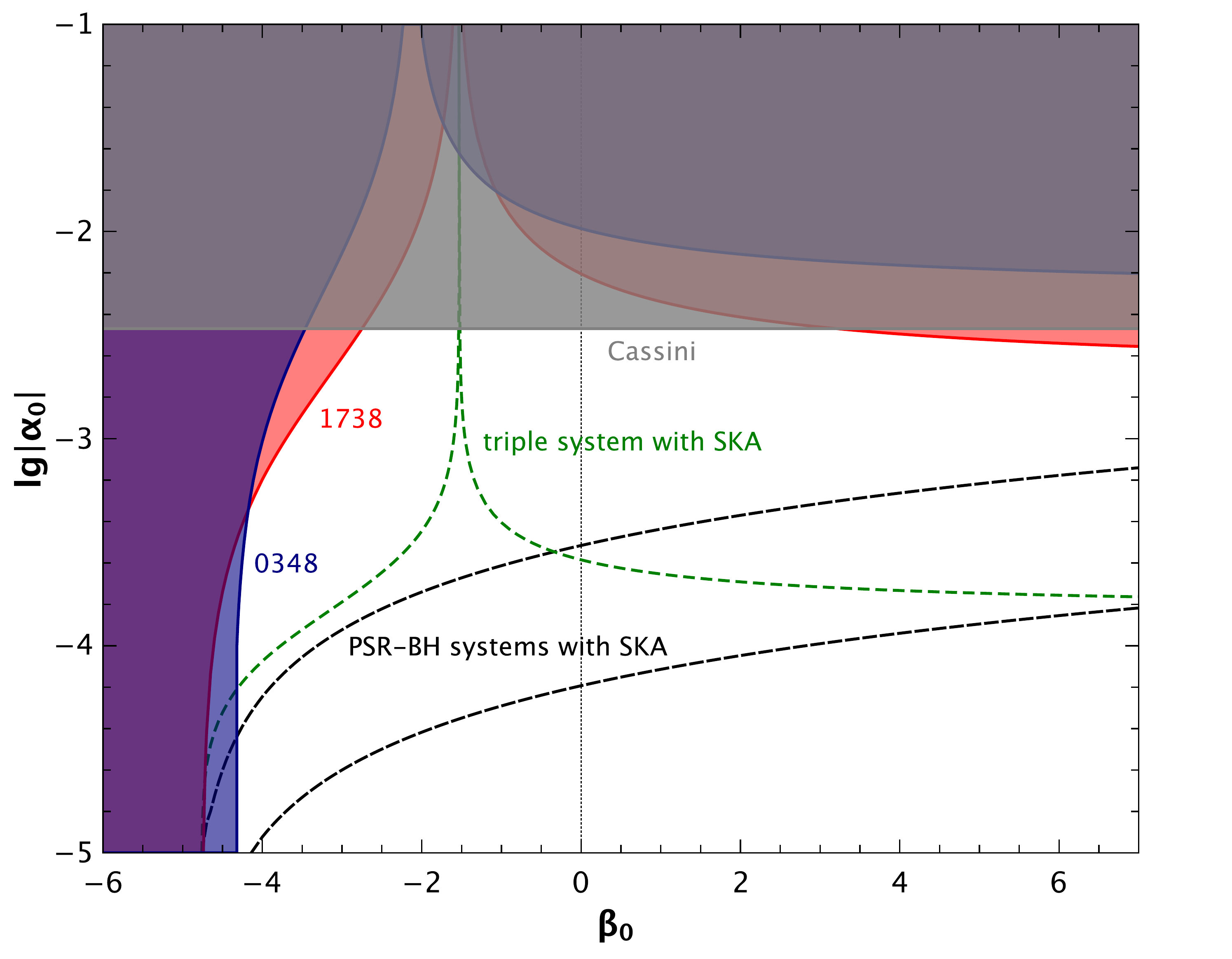}
\end{center}
  \caption{Constraints in the $\alpha_0$-$\beta_0$ plane of scalar-tensor 
theories by present experiments (Cassini, 
PSR~J1738+0333, PSR~J0348+0432) and selected future tests (dashed curves). The
black dashed lines show upper limits from two different (hypothetical) 
pulsar-BH systems, with $P_{\rm b} = 5$\,d, $e = 0.8$ (upper) and 
$P_{\rm b} = 0.5$\,d, $e = 0.1$ (lower), observed with the SKA. The dashed green 
line shows the expected upper limits from timing the triple-system pulsar 
(PSR~J0337+1715) with the SKA. Calculations are based on a stiff 
equation of state, and therefore conservative. For negative $\beta_0$,
PSR~J0348+0432 is the most constraining system, due to its high mass. The 
vertical line at $\beta_0 = 0$ corresponds to Jordan-Fierz-Brans-Dicke
gravity. This figure is an updated version of previous studies by
Damour and Esposito-Far{\`e}se (see \cite{Esposito-Farese:2011cha} and
references therein). \label{fig:stg}}
\end{figure}

\noindent
{\bf PSR~J0348+0432} is the first, and so far only (unambiguously identified) 
massive pulsar in a relativistic orbit. The pulsar has a mass of $2.01 \pm 
0.04\,M_\odot$ and is in a 2.5 hour orbit with a low-mass white dwarf~\cite{Antoniadis:2013pzd}. 
The white-dwarf companion is optically bright, and therefore allows for 
high-resolution spectroscopy, which, like for PSR~J1738+0333, was used to 
determine the masses of the white dwarf and the pulsar. In parallel, 
high-precision timing led to the measurement of an orbital decay of 
$\dot{P}_{\rm b} = -8.6 \pm 1.4 \, \mu{\rm s/yr}$. The measured value is in good
agreement with GR, and excludes any significant contribution from dipolar 
radiation: $|\alpha_{\rm PSR} - \alpha_0| < 0.005$ with 95\% 
confidence \footnote{The quantity $\alpha_{\rm PSR}$ denotes the effective 
(e.g.~scalar) coupling of the pulsar, and $\alpha_0$ the effective coupling of 
weakly self-gravitating bodies (e.g.~white dwarfs).}~\cite{Antoniadis:2013pzd}. 
Although that limit is weaker than the one of PSR~J1738+0333 
above, it is independently important, since PSR~J0348+0432 has a 
significantly larger gravitational binding energy than other binary pulsars
used to test gravity theories. Consequently, PSR~J0348+0432 constrains deviations
from GR that would only occur in the strong internal fields of more massive 
NSs, like certain types of spontaneous scalarization (see 
Figure~\ref{fig:stg0348}).  

\vspace{2mm}

Even binary-pulsar systems that are not very relativistic ($P_{\rm b}
\gtrsim 1$\,day) can be used to test gravity theories.  For instance,
a violation of the strong equivalence principle (SEP) is best tested
with pulsars in a wide orbit with a white dwarf companion. If the
strongly self-gravitating pulsar felt a different acceleration in the
gravitational field of the Milky Way than the white dwarf, a
polarizing force would change the orbital eccentricity in a
characteristic way. Damour and Sch\"afer suggested a statistical test,
based on Galactic small-eccentricity binary pulsars to constrain a
violation of SEP by strongly self-gravitating
bodies~\cite{Damour:1991rq}. The best constraint is based on an
ensemble of small-eccentricity binary pulsars, and gives $|\Delta_{\rm
  SEP}| < 0.0056$ with 95\%
confidence~\cite{Stairs:2005hu}.\footnote{Note that the somewhat better
  limit given by~\cite{Gonzalez:2011kt} is heavily based on a binary
  pulsar that does not fulfill the criteria for the Damour-Sch\"afer
  test, and is therefore clearly less robust.} The newly
discovered
millisecond pulsar PSR~J0337+1715 in a stellar triple
system~\cite{Ransom:2014xla}, where an inner pulsar-white dwarf binary
($P_{\rm b} = 1.63$\,d) is orbited by an outer white dwarf in just 327
days, promises a much better test of the
SEP~\cite{Freire:2012nb,Ransom:2014xla}. In particular, with the help
of future radio telescopes like the SKA, this system could provide one
of the most constraining tests for scalar-tensor gravity (see
Figure~\ref{fig:stg}). However pure strong-field deviations from GR,
like spontaneous scalarization in scalar-tensor gravity, cannot be
tested with the triple system pulsar, due to the weak-field nature of
the other two bodies.

\paragraph{Binary pulsar constraints on Lorentz violations in gravity.}
\label{sec:pulsar_LV}
As reviewed in Section~\ref{lv-theories}, Solar System tests impose tight
constraints on \textit{some} combinations of the coupling parameters
of both Einstein-\AE ther theory and khronometric gravity. More
specifically, once the constraints on the preferred-frame parameters
$\alpha_1$ and $\alpha_2$ are imposed (i.e.~when one requires
$|\alpha_1|\lesssim 10^{-4}$ and $|\alpha_2|\lesssim
10^{-7}$~\cite{Will:2014xja}), both theories remain characterized (to
leading order in $\alpha_1$ and $\alpha_2$) by only two dimensionless
coupling constants, which we denote by $c_+$ and $c_{-}$ in
Einstein-\AE ther theory and $\lambda$ and $\beta$ in khronometric
gravity (see Section~\ref{lv-theories} for details).

The parameter space for these coupling constants is further
constrained by theoretical requirements (absence of ghosts and
gradient instabilities), as well as by the requirement that
gravitational excitations propagate with speeds larger than the speed
of light, in order to avoid the production of gravitational Cherenkov
radiation by photons and relativistic particles~\cite{Elliott:2005va}.
Also, in the case of khronometric gravity, further constraints arise
from the requirements that the observed primordial element abundances
match the predictions of Big Bang
nucleosynthesis~\cite{Carroll:2004ai,Yagi:2013qpa,Yagi:2013ava}.  Note
that Big Bang nucleosynthesis gives much weaker constraints on
Einstein-\AE ther theory~\cite{Carroll:2004ai,Jacobson:2008aj}.

\begin{figure}[htb]
\capstart
\begin{center}
\begin{tabular}{lr}
\includegraphics[width=6.2cm,clip=true]{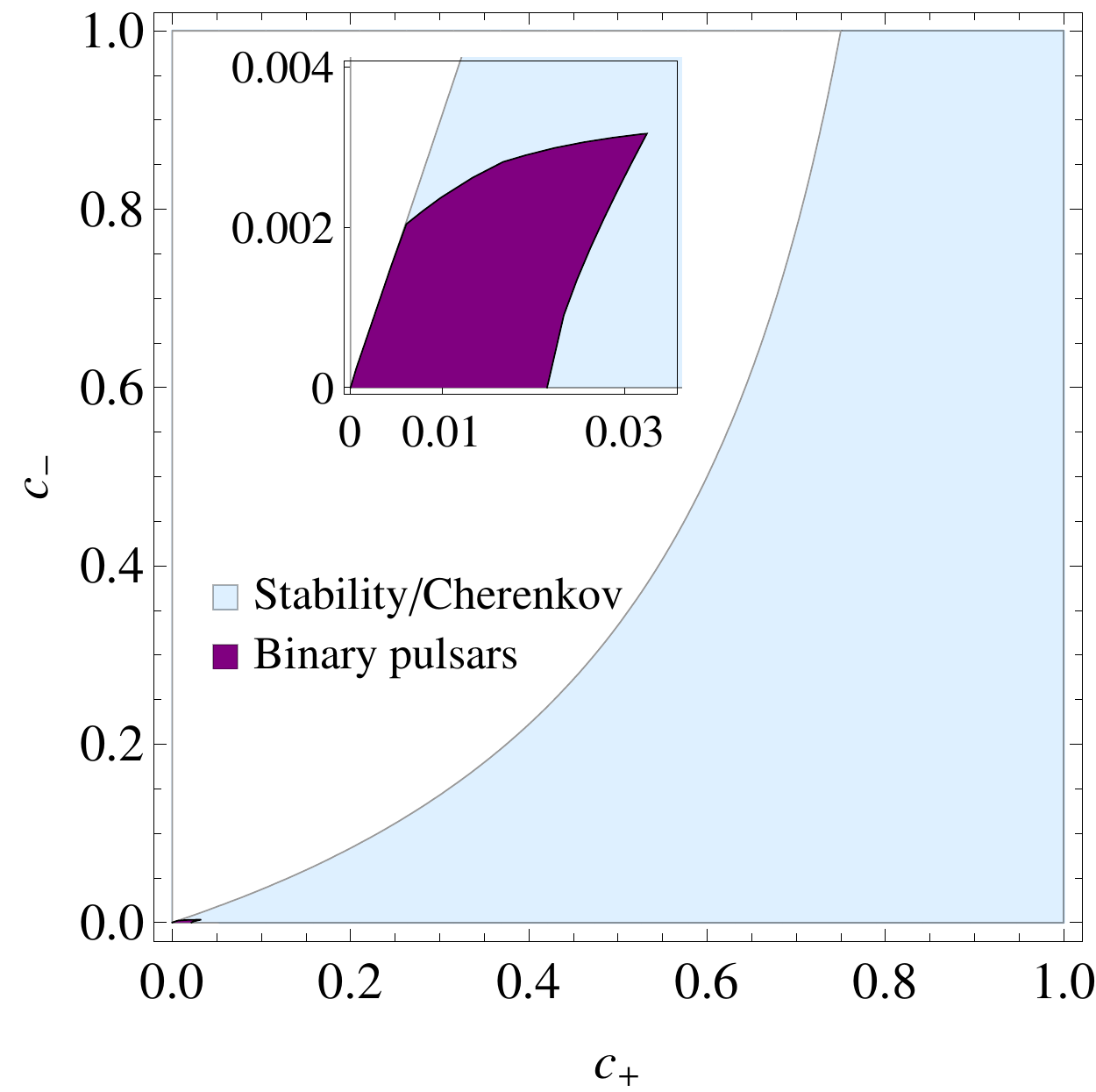}
& \includegraphics[width=6.2cm,clip=true]{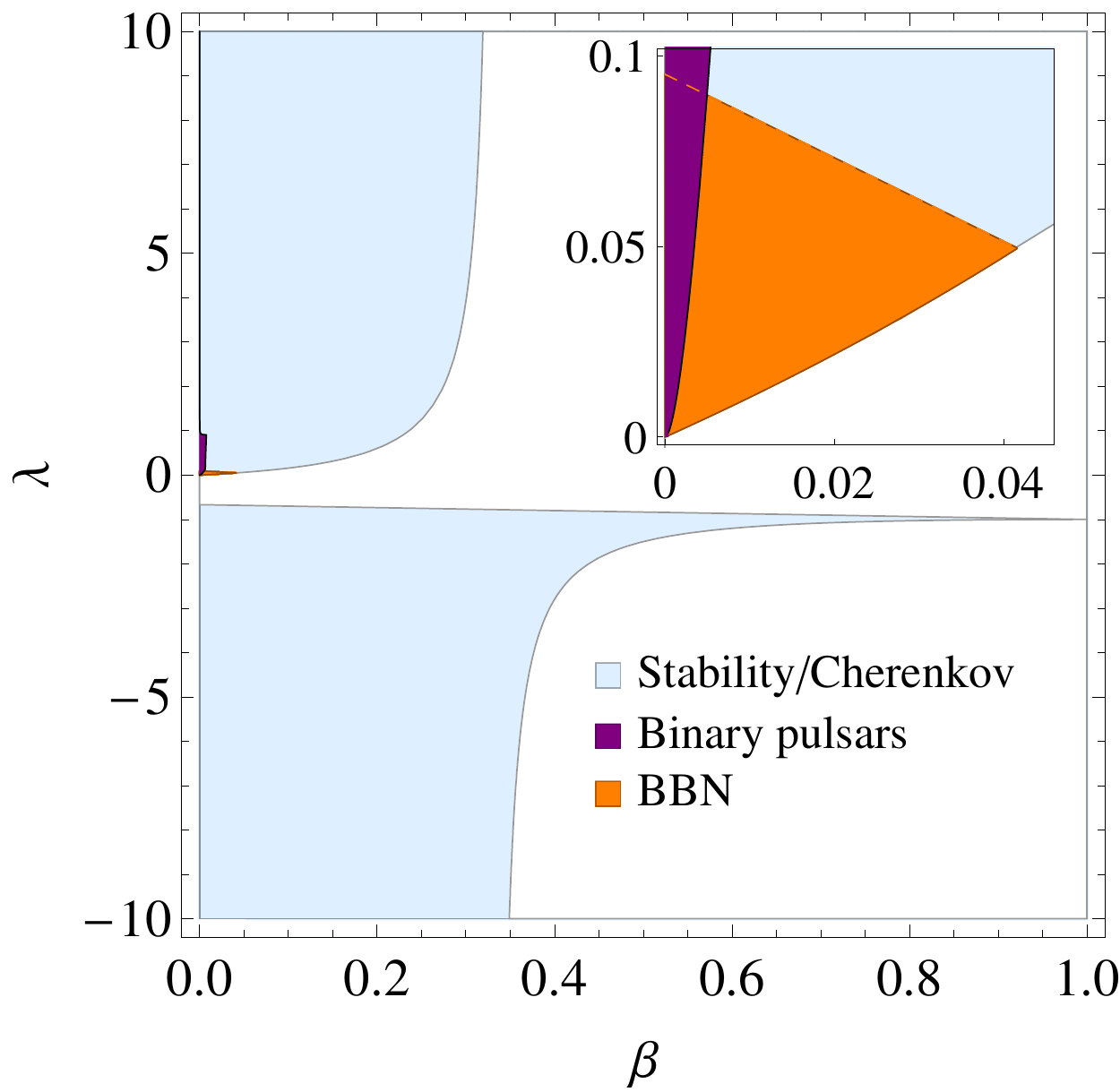}
\end{tabular}
\caption{Experimental constraints on Einstein-\AE ther (left) and
  khronometric theory (right). [Adapted
    from~\cite{Yagi:2013qpa,Yagi:2013ava}.]
\label{fig:LVconstraints}}
\end{center}
\end{figure}

All the above constraints are summarized in
Figure~\ref{fig:LVconstraints}. Also, as discussed in
Section~\ref{lv-bins}, competitive constraints can be placed on the
coupling parameters by requiring that the rate of change of
binary-pulsar systems matches the observations (as well as by
requiring that isolated pulsars show no anomalous precession, which
would be induced by Lorentz
violations)~\cite{Yagi:2013qpa,Yagi:2013ava}.  In particular, the
purple region in Figure~\ref{fig:LVconstraints} is obtained by imposing
agreement with the observations of PSR J1141--6545~\cite{bhat}, PSR
J0348+0432~\cite{Antoniadis:2013pzd}, PSR
J0737--3039~\cite{kramer-double-pulsar} and PSR
J1738+0333~\cite{Antoniadis:2012vy,Freire:2012mg}.
Finally, we should stress that additional constraints on Lorentz violations in
gravity come from cosmological observations such as the CMB and the large-scale structure,
\textit{if} a direct coupling between the Lorentz-violating field and the Dark Sector is introduced~\cite{Audren:2013dwa,Audren:2014hza}.

\subsubsection{Open problems}\label{op:pulsars}

In the near future, radio astronomy will benefit from the operation of
new radio telescopes with significantly larger collecting area. By the
end of 2016 the Five-hundred-meter Aperture Spherical radio Telescope
(FAST)~\cite{Nan:2011um} should see ``first light,'' and in the early
2020s the SKA should reach its design sensitivity, greatly enhancing
timing precision of known pulsars (up to a factor of 100 for many of
them) and increasing the number of known pulsars by an order of
magnitude~\cite{Lazio:2013mea}. In terms of gravity tests with
pulsars, this means a leap in the precision of current tests, various
qualitatively new tests with already known systems, and the discovery
of many new ``pulsar laboratories.''

Concerning the known systems, for the first time we will be able to
test relativistic effects in binary pulsars beyond the leading
order. For instance, in the double pulsar we will be able to test the
mass-octupole and current-quadrupole corrections to the quadrupole
formula~\cite{Blanchet:1989cu,Kramer:2009zza}. In the double pulsar,
we should also be able to extract the Lense-Thirring drag from the
total $\dot\omega$, and by this measure the moment-of-inertia of
pulsar A: see Eq.~(5.23) of \cite{Damour:1988mr}.  Since the moment of
inertia of a NS depends on its compactness and therefore on the
equation of state of NS matter, this measurement will test competing
NS matter models, and give insight into the properties of matter at
very high densities
($\sim 10^{15} \, {\rm
  g\,cm^{-3}}$)~\cite{Lattimer:2004nj,Kramer:2009zza}.

Qualitatively, new tests could come from the discovery of a
pulsar-BH system, either a pulsar in orbit with a stellar mass
BH or in orbit around the supermassive BH in the
center of our Galaxy. If intermediate-mass BHs do exist in
some of the dense cores of globular clusters, this might be a third
option to find such a test system. The ranging capability that comes
with the timing of a pulsar would provide a unique probe for the
BH spacetime, and allow for tests of the frame dragging and
the no-hair theorem~\cite{Wex:1998wt,Liu:2011ae,Wex:2012au,Liu:2014uka}. But even for
theories that predict the same BHs as GR, a pulsar-BH
system could be a unique ``laboratory'' (see Figure~\ref{fig:stg}).

A completely different type of test could come from pulsar timing arrays, which
are presently used in the effort to detect nano-Hz GWs~\cite{IPTA):2013lea}.
With the timing capabilities of the SKA, one can hope to probe the polarization
and propagation properties of these long-wavelength GWs~\cite{Lee:2013sxl}, or even study the evolution of a single supermassive 
BH binary~\cite{Corbin:2010kt,Lee:2011et}.

\subsection{Testing general relativity with cosmology}
\label{subsec:cosmology}

The past three decades have witnessed remarkable progress in relativistic cosmology. Cosmological data constraining the expansion rate of the Universe, allied with accurate maps of large-scale structure and the cosmic microwave background, have been used to place ever-tightening constraints on cosmological parameters~\cite{Ade:2013zuv}. In parallel with observational progress,
publicly available tools, that can be used to predict cosmological observables for general relativistic cosmologies and their extensions,
have been developed.
The field is alive and well with a plethora of new, extremely powerful experiments and observatories in the planning stage that will come to fruition at the end of this decade.

Crucial to the successes of modern cosmology has been the realization that the morphology and evolution of large-scale structures -- that is, the cosmic web of galaxies, clusters, voids and filaments -- should contain a wealth of information about various fundamental properties of the Universe~\cite{Peacock:1999ye}. What was initially a niche area developed by Peebles, Zel'dovich, Silk and others has become a powerhouse of exploration and one of the leading areas of research in modern astrophysics. The theory and observation of anisotropies in the cosmic microwave background are the jewels in the crown of this field of research, and have been used to place precise constraints on properties such as the geometry of the Universe, its material constituents and the initial conditions that seeded structure. A bizarre picture of the Universe has emerged in which, if GR is assumed to be the underlying theory of gravity, $96\%$ of the material content of the Universe is in a dark sector -- dark matter and dark energy. A plethora of models has been proposed that try to account for this dark sector~\cite{Copeland:2006wr}.

Given that the evidence for the dark sector is purely gravitational, there have been, in the past few years, attempts at finding theories in which it can arise from modifications of gravity~\cite{Clifton:2011jh}. These modifications affect the expansion rate of the Universe, but should also affect gravitational clustering in a way that might be distinguishable from GR. This idea has led to the proposal of a number of different theories (as discussed at length in Section~\ref{sec:alt-th}) but has also focused observational programs to  target deviations from GR. It is fair to say that the cosmological tests of modified gravity are now the primary driver for future surveys of large scale structure.

\subsubsection{Theory: the linear regime.}
Faced with the proliferation of ever-more exotic gravity theories, the conventional process of testing models on an individual basis is not the optimal way of testing GR on cosmological scales. Not only is the theory population too large to tackle (and growing still), but progress is slowed by the increasing mathematical complexity of the most popular ideas. It was for these reasons that, several years ago, a number of groups turned to the strategy of devising model-independent tests of gravity. Effectively, one attempts to build a template for what viable beyond-GR theories can look like, in terms of a minimal set of unknown parameters and functions. One then builds up a ``translation dictionary,'' that is, a correspondence of how the parameters involved in the general formalism relate to the parameters of specific theories. With this dictionary in hand, testing the unified framework provides an efficient way to test many theories simultaneously. However, the framework is more general than this, because it also covers regions of parameter space for which no corresponding theory has been established yet. Constraints on this part of the parameter space can be used to guide the direction of \textit{future} theoretical work.

While the PPN formalism~\cite{Will:2011nz} is an example of one such model-independent framework, it is limited to small-scale tests of gravity, say, in the Solar System or using compact object binaries. There is a need for an analogous formalism that can be applied to cosmological observables. This is a formidable task, so most work to date has focused on formalisms that capture the regime of linear cosmological perturbation theory only.

Let us consider, for a moment, the properties we would like a cosmological equivalent to PPN to have:
\begin{enumerate}
\item It should encapsulate a large portion of the existing theory space.
\item Existing theories should map onto it \textit{exactly}, rather than only as an approximation.
\item The parameters of the formalism should have physical meaning, rather than simply denoting mathematical terms.
\item The parameters (or combinations thereof) should be constrainable by near-future data.
\item The formalism should achieve all of the above with the minimum number of new parameters and/or free functions.
\end{enumerate}
There is no obvious, unique way to meet all of the above criteria. Instead, three species of parametrizations have been put forward in response to the challenge. One can think of these three formulations as corresponding to three manifestations of a gravity theory: its fundamental action, the field equations derived from that action, and the combinations of those field equations which directly influence observable quantities. We will now describe each of these formalisms briefly in turn.

\paragraph{Action-based approaches.}
To derive linearized gravitational field equations, one needs an action that is quadratic in perturbations. One way to parametrize gravity is to construct the most general quadratic action with a given field content that is consistent with some desirable symmetries. Placing a restriction on the derivative order of the field equations terminates the potentially infinite series of terms that could be constructed. A coefficient of appropriate mass dimensions is assigned to each term; these represent the ``dials'' of the formalism that can be tuned to match a specific theory.

This concept can, to a certain extent, be thought of as an EFT for modified gravity (cf.~Section~\ref{sec:EFT}), although the analogy with particle physics EFTs should not be taken too far. 
To date, nearly all work has focused on actions constructed from the metric and 
a single scalar degree of freedom~\cite{Battye:2012eu,Battye:2013aaa}, though a 
bi-scalar case recently appeared in~\cite{Gergely:2014rna}.

Such parametrized actions can quickly grow to contain large numbers of terms. However, the authors of~\cite{Gubitosi:2012hu,Bloomfield:2012ff,Gleyzes:2013ooa, Gleyzes:2014rba,Gao:2014soa} have made use of a clever device to simplify the procedure. They first consider the situation as viewed from the \textit{unitary gauge}, in which the time coordinate is chosen such as to eliminate perturbations of the scalar field. As a result of having used up one gauge freedom in this way, the metric is left with three spin-0 perturbations instead of the usual two (after gauge fixing); one might say that the metric has ``eaten'' the scalar field.

Given the preferred space-time slicing, it is then natural to reformulate metric perturbations in the ADM formalism. An example of a resulting action is~\cite{Gleyzes:2013ooa}:
\begin{align}
  S_{\mathrm{EFT}}={}&\int d^4x\sqrt{-g}\,\Bigg[\frac{M_P^2}{2}f(\eta)R-\Lambda(\eta)-c(\eta)g^{00}+\frac{M_2^4}{2}\left(\delta g^{00}\right)^2 \nonumber \\
&{}-\frac{m_1^3}{2}\delta K\delta g^{00}
-\frac{\bar{M}_2^2}{2}\delta K^2-\frac{\bar{M}_3^2}{2}\delta K^\mu_\nu\delta K_\mu^\nu+\frac{\mu_1^2}{2} {^{(3)}R}\delta g^{00} \nonumber \\
&{}+\frac{\bar{m}_5}{2} {^{(3)}R}\,\delta K+\frac{\lambda_1}{2} {^{(3)}R}^2+\frac{\lambda_2}{2} {^{(3)}R^\mu_\nu}\, {^{(3)}R}_\mu^\nu\Bigg]\,,
\end{align}
where $K_{\mu\nu}$ is the extrinsic curvature of the spatial hypersurfaces defined by the foliation, and $^{(3)}R$ is their three-dimensional intrinsic curvature. The metric component $g^{00}$ is related to the usual ADM lapse function by $g^{00}=-1/N^2$. Performing a Stuckelberg transformation~\cite{Cheung:2007st} on the time coordinate will exit the unitary gauge and cause the scalar field to reappear in the action. The action can then be varied in the usual manner.

The advantage of this EFT-inspired approach is that it directly parametrizes the 
action, and therefore, constraints on the EFT coefficients would have direct 
implications for which kinds of theories are allowed by the data. The 
disadvantages are that i) it is necessary to fix the field content allowed by 
the parametrization, and ii) the combinations of EFT parameters that filter 
through to observable quantities (say, the modified growth rate or weak lensing 
kernel) are cumbersome combinations of the action-level parameters. With regards 
to i), we emphasize that nearly all work to date has focused on a single scalar 
field. Clearly this covers only part of the broad space of gravity theories.

\paragraph{Field equation approaches.}
An alternative method is to directly parametrize the linearized gravitational field equations, the dynamical tools of a theory. After some consideration, one realizes that there are only three kinds of new terms that can appear in the field equations: perturbations of the metric, perturbations of the matter stress-energy tensor, and perturbations of any new degrees of freedom that a theory might add to GR.

 An example in this category is the Parametrized Post-Friedmann formalism (PPF)~\cite{Baker:2012zs}. For simplicity, we consider here the parametrization of the spin-0 field equations, restricting them to be of second order in time derivatives. We will not show the full framework, but as a representative example the extended 00--component of the field equations in the conformal Newtonian gauge is:
\begin{align}
-a^2\delta G_0^0={}&8\pi G_N a^2 \rho_M\delta_M+A_0(k,a)k^2\Phi+F_0(k,a)k^2\Gamma \nonumber\\
&{}+\alpha_0(k,a)k^2\chi+\alpha_1(k,a)k^2\dot\chi\,,
\end{align}
where $\Gamma=(\dot\Phi+H\Psi)/k$, $\delta G^\mu_\nu$ is the usual linearized Einstein tensor and $\chi$ is a template variable representing a new spin-0 degree of freedom; only one new degree of freedom is shown above, but in principle more could be added. The new degree of freedom does not have to be a simple scalar field: for example, when the parametrization is used to describe Einstein-\AE ther theory, $\chi$ represents the spatial spin-0 perturbation of a vector.

$A_0(k,a),\,F_0(k,a),\,\alpha_0(k,a)$ and $\alpha_1(k,a)$ are the free functions of time and scale that act as the ``dials'' of the parametrization in this case. In fact, the scale-dependence of $A_0\ldots\alpha_1$ is fixed by Lorentz symmetry and the derivative order of the parametrization -- they can only contain powers like $k^{2n}$. 

The other components of the spin-0 field equations follow an analogous pattern, see~\cite{Baker:2012zs} for details. Naively, it appears that 22 free functions are needed to map out all possible extensions to the field equations. In reality, this is too much freedom -- not all of these 22 PPF coefficients are independent. By connecting the PPF approach to the action-based parametrizations described above, one determines that only $\sim 5-9$ independent functions are needed to describe a simple scalar field theory (the exact number depends on assumptions made about Lorentz symmetry and whether one fixes the background expansion rate or not). The advantage of field equation approaches such as PPF is that they can encapsulate numerous gravity theories without requiring any assumptions about field content. The disadvantage is that, at face value, the parametrization contains redundant freedoms that could cause problematic degeneracies for a constraint analysis.

\paragraph{Quasistatic approaches.}
The final approach to parametrizing deviations from GR is the simplest, and therefore the easiest to constrain, but also the least comprehensive from a formal standpoint. It makes use of the \textit{quasistatic approximation}. To implement this, we focus on a restricted range of distance scales that are considered to be significantly smaller than the cosmological horizon, but sufficiently large that~\cite{Silvestri:2013ne}:
\begin{enumerate}
\item[(1)] Perturbations are in the linear regime.
\item[(2)] $H/k\ll 1$, so any term in the field equations containing this 
prefactor can be dropped.
\item[(3)] The time derivatives of perturbations are negligible compared to 
their spatial derivatives on these scales.
\end{enumerate}
In GR, it can be shown that (3) follows as a consequence of (2). In modified 
theories this is no longer necessarily the case; instead it must be 
\textit{assumed} that (3) applies to any new fields introduced by a theory, as 
well as to the metric itself. The majority of theory-specific simulations to 
date support this set of 
assumptions~\cite{Oyaizu:2008sr,Noller:2013wca,Brax:2013mua,Brax:2012nk}, but 
see~\cite{Llinares:2013qbh, Comelli:2014fg} for counter-examples.

By making the above approximations (and taking appropriate combinations of the field equations), one finds that in the quasistatic regime many theories of gravity can be reduced to the simple form:
\begin{align}
\nabla^2\Psi&=4\pi G_N a^2 \mu(k,a) \rho_M\Delta_M\,,\\
\gamma(k,a)&=\frac{\Phi}{\Psi}\,,
\end{align}
where $\Delta_M$ is a gauge-invariant matter density perturbation and, as in the previous subsection, $\mu(k,a)$ and $\gamma(k,a)$ are functions of time with fixed scale dependence of the form $k^{2n}$ (in the majority of cases). In particular, one can show that in the case of Horndeski gravity, which is the most general theory of a single scalar field with second-order field equations, $\mu$ and $\gamma$ have the form~\cite{DeFelice:2011hq, Amendola:2012ky}:
\begin{align}
\mu(k,a)&=h_1(a)\left(\frac{1+h_5(a)k^2}{1+h_3(a) k^2}\right)\,,\label{HDmu}\\
\gamma(k,a)&=h_2(a)\left(\frac{1+h_4(a)k^2}{1+h_5(a)k^2}\label{HDgamma}\right)\,,
\end{align}
where $h_i(a)$ are pure functions of time.

This parametrized Poisson equation and ``slip relation'' (the ratio of $\Phi$ and $\Psi$) are all that is needed to start calculating observable quantities, such as galaxy weak lensing and the growth rate of large-scale structure~\cite{Baker:2014tc,Leonard:2015hha}. Example forecasts for constraints on $\mu$ and $\gamma$ with the Large Synoptic Survey Telescope~\cite{LSST} can be found in~\cite{Hojjati:2013xqa}. Recent measurements using the 6dF peculiar velocity survey did not find any evidence for scale dependence in the growth rate of large-scale structure~\cite{Johnson:2014kaa}, which places a lower limit on any new mass scale involved in theories described by Eqs.~(\ref{HDmu}) and (\ref{HDgamma})~\cite{Baker:2014tc}.

The advantage of the quasistatic parametrization is clearly its simplicity and direct connection to observations. Its chief disadvantage is that it is an approximation with a limited regime of applicability, and does not exactly match the form of the underlying space of gravity theories. This obscures attempts to work out what constraints on $\mu$ and $\gamma$ really mean for a particular theory of interest.\newline

\subsubsection{Theory: the nonlinear regime.}\label{subsec-cosm-nonlinear}
Calculating the effects of modified gravity becomes significantly harder as we move into the nonlinear regime. Local (laboratory and Solar System) experiments place strong constraints on any deviation from GR: the results of such experiments require any new gravitational fifth force to be either very weakly coupled to matter or very short-ranged in the environments where the experiments have been performed. To avoid these strong experimental constraints, and at the same time give rise to interesting observable cosmological signatures, a screening mechanism is required. By screening we mean a way of hiding the modifications of gravity in our local (high-density) environments where high-precision gravity experiments have been performed, while at the same time allowing for potentially large deviations in regions of spacetime where the average density is much lower. We will briefly explain how screening works and which theories have some form of screening.

Most of the known screening mechanisms for scalar-tensor theories are encompassed by the general (Einstein-frame) Lagrangian:
\be
\mathcal{L} = \frac{R}{2}M_{\rm Pl}^2 + \mathcal{L}(\phi,\partial\phi,\partial\partial\phi) + \mathcal{L}_m(A^2(\phi)g_{\mu\nu},\psi_m)\,.
\ee
The matter fields are coupled to a metric $\tilde{g}_{\mu\nu} = g_{\mu\nu}A^2(\phi)$ that is conformally related to the space-time metric $g_{\mu\nu}$. In the nonrelativistic limit, such a theory gives rise to a fifth force given by:
\be
\vec{F}_\phi = \frac{\beta(\phi)}{M_{\rm Pl}}\vec{\nabla}\phi,~~~~\beta(\phi) \equiv \frac{d\log A}{d\phi}M_{\rm Pl}\,.
\ee
If the field equation for $\phi$ is linear, then the superposition principle is in play and screening cannot be achieved, so a fundamental requirement for a screening mechanism to work is nonlinear field equations. There are three ways of achieving this: with a self-interacting potential, in the coupling to matter or in the kinetic terms. To see the different ways screening can emerge we expand the Lagrangian about a field value $\phi_0$:
\be
\mathcal{L} \simeq \frac{R}{2}M_{\rm Pl}^2 + Z^{\mu\nu}(\phi_0)\delta\phi_{,\mu}\delta\phi_{,\nu} +m^2(\phi_0)\delta\phi + \frac{\beta(\phi_0)\rho_m}{M_{\rm Pl}} + ...
\ee
In a low-density environment, the field sits at some value $\phi_0 = \phi_A$ and the scalar field produces a fifth force on a test mass with strength $\alpha \propto \beta^2(\phi_A)$ relative to the gravitational force. Consider now a high-density region of space where $\phi_0 = \phi_B \not= \phi_A$. One way to reduce the effect of the fifth force is by having the field $\phi$ acquire a large local mass $m(\phi_B) \gg m(\phi_A)$, which implies a very short interaction range -- this is the chameleon mechanism~\cite{Khoury:2003rn,Mota:2006fz}. If the matter coupling is small, $\beta(\phi_B) \ll \beta(\phi_A)$, the fifth force will be suppressed -- the symmetron mechanism~\cite{Hinterbichler:2010es}. Lastly, the condition $|Z^{\mu\nu}(\phi_B)| \gg |Z^{\mu\nu}(\phi_A)|$ leads, after canonical normalization, to a weakened matter source and therefore also a weakened fifth force -- the Vainshtein mechanism~\cite{Vainshtein:1972sx,Deffayet:2009wt}.

In the nonlinear regime of structure formation the screening effect will be in operation and must be taken into account to obtain reliable theory predictions. Linear perturbation theory is unable to account for screening, as it is a purely nonlinear effect, and to understand structure formation in screened theories one is therefore led to N-body simulations. In such simulations one solves for the full evolution of the scalar field in the simulation box just as one does for the metric potential, though one can often apply the quasi-static approximation (see previous section) to simplify the field equation. The N-body equations of motion are given by: i) the particle displacement equation:
\be
\ddot{\bf x} + \left(2H+\frac{\beta(\phi)\dot{\phi}}{M_{\rm Pl}}\right)\dot{\bf x} = -\frac{1}{a^2}\vec{\nabla}\Phi - \frac{1}{a^2}\frac{\beta(\phi)}{M_{\rm Pl}}\vec{\nabla}\phi\,,
\ee
where the last term represents the scalar fifth force, and ii) the Poisson equation for the metric potential:
\be
\nabla^2\Phi = 4\pi G_N a^2\delta\rho_{\rm eff}\,,
\ee
where $\delta\rho_{\rm eff}$ is the perturbed total effective energy density, which contains contributions from matter and modifications to the Einstein tensor due to modified gravity. Lastly we have the field equation for $\phi$, which is model-dependent. Due to the nonlinearities in this equation, the method of choice for solving it is (Newton-Gauss-Seidel) relaxation. Accurately solving the field equation for the scalar field is by far the most challenging and time-consuming part of such simulations. To date several different models with different kinds of screening mechanism have been simulated, including
\begin{itemize}
\item Chameleon screening: Chameleon models~\cite{Brax:2013mua} and $f(R)$ gravity~\cite{Oyaizu:2008sr,Llinares:2013jza,Li:2011vk,Puchwein:2013lza}.
\item Symmetron screening: Symmetron models~\cite{Davis:2011pj,Brax:2012nk}.
\item Vainshtein screening: Dvali-Gabadadze-Porrati (DGP)~\cite{Schmidt:2009sv,Schmidt:2009sg,Li:2013nua} and Galileon models~\cite{Barreira:2013eea}.
\end{itemize}
The results from such simulations have so far mostly been used to map out potential signatures rather than computing explicit constraints. An exception is Ref.~\cite{Schmidt:2009am}, which constrains $f(R)$ gravity using cluster abundances, however the constraints found are not yet compatible with those found from taking local experiments into account.

\begin{figure}
\capstart
\includegraphics[scale=0.5,bb=0 130 0 700]{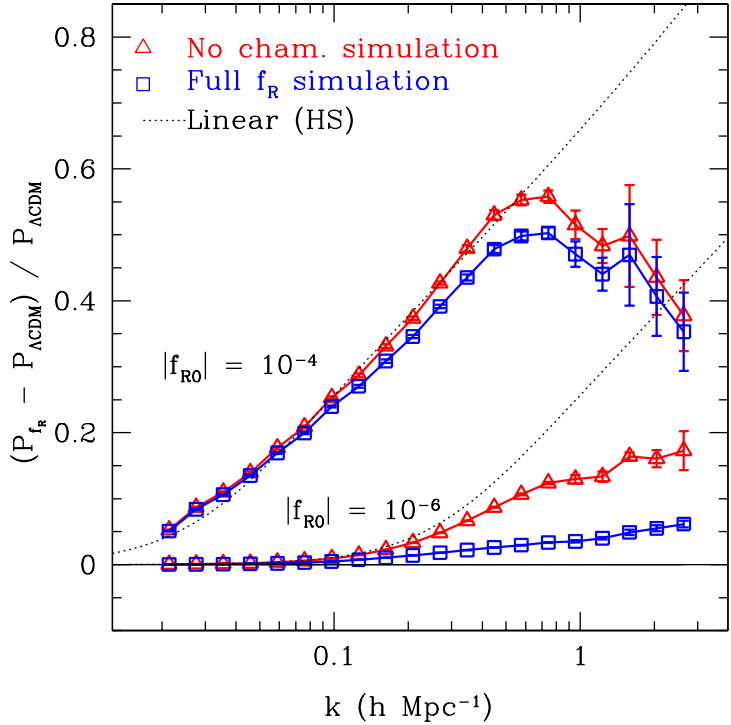}
\caption{Relative power spectrum enhancement over $\Lambda$CDM at $a = 1$ for 
the full $f(R)$ simulation compared with the no-chameleon (simulations with a 
linearized field equation) simulations and linear perturbation theory. At high 
$k$, linear theory overestimates the relative enhancement. Without the chameleon 
screening mechanism, power is sharply enhanced on scales smaller than the 
Compton scale in the background. For $|f_{R0}| = 10^{-6}$ the screening is very 
effective and the fifth force in the simulation is strongly suppressed. For 
$|f_{R0}| = 10^{-4}$, this suppression is nearly absent except for a residual 
effect from the chameleon at high redshift. [From~\cite{Schmidt:2008tn}.]}
\label{fig:fofrpower}
\end{figure}

The first key observable where the effects of modified gravity are seen is the matter power spectrum. When measured relative to $\Lambda$CDM, one typically finds an enhancement: a bump around $k\sim 1~h/$Mpc, as seen in Figure~\ref{fig:fofrpower}. N-body simulations have shown the importance of taking screening into account when making predictions: linear theory (or simulations with a linearized field equation for the scalar field) produces way too much clustering, and the power spectrum on nonlinear scales can be off by several tens of percent from the true result, as seen in Figure~\ref{fig:fofrpower}. Different screening mechanisms, different models and also different model parameters can give rise to very different results, making the construction of a model-independent parametrization, as presented above for the linear regime, hard to achieve. However, a parametrization valid for a certain sub-class of models (of the chameleon type) has been proposed~\cite{Brax:2012gr}.

\begin{figure}
  \capstart
    \includegraphics[scale=0.5,bb=0 130 0 720]{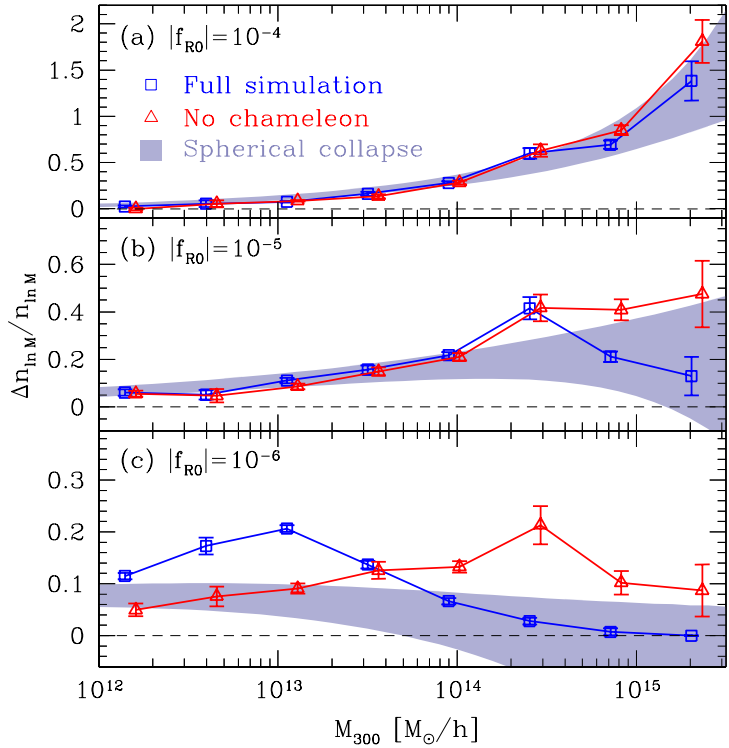}
\caption{Relative mass-function enhancement over $\Lambda$CDM at $a = 1$ for the 
full $f(R)$ simulation compared with the no-chameleon (simulations with a 
linearized field equation) simulations and the spherical collapse model. For  
the lowest value ($|f_{R0}|=10^{-4}$) the screening mechanism is not much in 
play, and the mass function is enhanced in the high-mass end. As we go to 
smaller values of $|f_{R0}|$ the screening mechanism becomes more and more 
effective, and the mass function is only enhanced for mid-sized halo masses. We 
also see the importance of solving the full field equation: the no-chameleon 
simulation significantly overestimates the mass function in the high-mass end 
for $|f_{R0}|=10^{-6}$. [From~\cite{Schmidt:2008tn}.]}
\label{fig:dndmfofr}
\end{figure}

Another key observable is the halo mass function, which is also enhanced relative to $\Lambda$CDM, see Figure~\ref{fig:dndmfofr}. When the screening mechanism is working effectively, the enhancement is typically found for low to mid-size halos ($M \sim 10^{12}-10^{14}~M_{\rm sun}/h$). A key property of all known screening mechanisms is that they are more effective for more massive halos. This implies that the mass function converges to that of the underlying cosmological model in the high-mass end ($M \gtrsim 10^{15}~M_{\rm sun}/h$). If the screening mechanism is not very effective then this does not have to happen, and the mass function is enhanced even for the largest halo masses.

Results from N-body simulations have also revealed several smoking-gun signatures of modified gravity. One particularly interesting signature is the difference between dynamical and lensing masses~\cite{Schmidt:2010jr}. The mass contained within a dark-matter halo can be found by either gravitational lensing measurements or by some measurement that relies on the dynamics of test masses. Gravitational lensing is determined by the sum of the two metric potentials $\Phi+\Psi$, and the inferred lensing mass in many theories with screening gives the same result as in GR. Dynamical masses, on the other hand, are affected by the fifth force, and will consequently be different from the GR versions. Combining lensing and dynamical mass measurements can therefore probe modified gravity. Additionally we have the effect that the amount of screening will depend on the density of the environment a certain massive object lies in. This will give rise to an environment dependence on dynamical mass estimates, and serves as a smoking gun signal for theories with screening. In principle, this effect could also distinguish between different types of screening~\cite{Zhao:2011cu}. 

Quite often, the strongest effects of modified gravity (measured relative to $\Lambda$CDM) are found in the velocity field. Interesting signatures that have been found here include the low-order moments of the pairwise velocity distribution~\cite{Hellwing:2014nma} and the full phase space around galaxy clusters~\cite{Lam:2012by}.

There are issues related to the nonlinear regime that need to be better understood before we can fully exploit its potential as a probe of modified gravity. For example, when it comes to the matter power spectrum, the modified gravity signal found in simulations is degenerate with several other effects, such as massive neutrinos~\cite{Shim:2014uta} and baryonic feedback processes, that are not fully understood at the moment. One way around this issue is to look for observables that are not significantly affected by baryonic physics, or combining observables that allow us to break the degeneracies. 

\subsubsection{Observations, current and future.}
There has been remarkable progress in constraining cosmological perturbations, and it is useful to summarize the data sets either in hand or expected over the next decade. The observables of choice are:
\begin{itemize}
\item Anisotropies in the CMB; the main statistic is the angular power spectrum of fluctuations, $C_\ell$. The current and future experiments are: WMAP, PLANCK, ACT, SPT, ACTPol, SPTPol, Spider, Polarbear, BICEP2, Keck Array.
\item Surveys cataloguing the angular positions and redshifts of individual galaxies leading to the power spectrum of fluctuations, $P(k)$, or the two-point correlation function, $\xi(r)$.  The current and future experiments are: BOSS, DES, Weave, HETDEX, eBOSS, MS-DESI, LSST, Euclid, SKA, Chime, Baobab, MEERKAT, ASKAP.
\item Weak lensing; images of distant galaxies will be distorted and correlated by intervening gravitational potential wells, leading to statistics such as the convergence power spectrum, $C^{\kappa}_\ell$. The current and future experiments are: DES, RCS, KIDS, HSC, LSST, Euclid, SKA.
\item Peculiar velocities; by measuring redshifts {\it and} radial distances of galaxies and clusters it is possible to reconstruct a radially projected map of large-scale motions. Progress in this field will primarily come for the latter method with Planck.
\end{itemize}

Cosmological data has been used to place constraints on standard scalar-tensor theories~\cite{Avilez:2013dxa} and related theories (such as $f(R)$~\cite{Lombriser:2010mp}, Galileons and more general Horndeski theories~\cite{Barreira:2012kk}), Einstein-Aether theories~\cite{Zuntz:2008zz,Koivisto:2008xf}, braneworld models (specifically DGP) and specific massive gravity models (see e.g.~\cite{Akrami:2013ffa,Konnig:2014xva} and references therein). There have been preliminary attempts at placing constraints on more generalized parametrizations. If one restricts oneself to the quasi-static functions, $\mu$ and $\gamma$, constraints are found to be very dependent on assumptions about time- and scale-dependence~\cite{Bean:2010zq,Hojjati:2013xqa}. So, for example, if these functions are assumed to be constant, constraints are found at the percent level, while freeing up the time evolution (but assuming scale independence) gives constraints of order $50-100\%$.

\subsubsection{Open problems}\label{op:cosm}

Cosmological observations will constrain GR on length scales which are fifteen orders of magnitude greater than current constraints. Current and future data will give us a unique opportunity to do so with remarkable precision. We currently have an excellent understanding of what happens in the linear regime and how it maps onto an incredibly broad family of models. A few self-consistent formulations of how to parametrize GR currently exist which mirror the PPN approach, the workhorse for testing gravity on Solar System scales. Observations will not be able to constrain all the parameters in these approaches, but it should be possible to constrain some of them reasonably well in the quasi-static regime.

There have been attempts at developing such a general framework on nonlinear scales. The inclusion of screening mechanisms has been an essential aspect of this approach. Unlike with the linear regime, a number of theoretical issues remain: how to efficiently and accurately model the nonlinear effects, how to incorporate the uncertainties that arise from baryonic physics -- most notably from supernovae and active galactic nuclei feedback -- and the more general problem of bias (or how galaxies trace the density field). These are hard problems that need to be understood and solved if we are to use the (abundant) information on smaller length scales.

There have been a number of suggestions for how to test gravity on galactic and cluster scales which may be promising. These involve the transverse Doppler gravitational redshifts in galaxy clusters~\cite{Zhao:2013nka}, the motion of BHs embedded in galaxies~\cite{Hui:2012jb}, constraints from distance indicators~\cite{Jain:2012tn}, galactic brightness~\cite{Davis:2011qf}, cluster abundances~\cite{Ferraro:2010gh} and cold tidal streams in galaxies~\cite{Penarrubia:2012ab}. These would add to, and complement, the constraints arising from large-scale structure.

In conclusion of this chapter we mention the results of the BICEP2 collaboration~\cite{Ade:2014xna}. Measurements of the CMB B-mode polarization at 150 GHz by the BICEP2 experiment were initially found consistent with a $\Lambda$CDM cosmological model plus a spectrum of tensor modes, described by a tensor-to-scalar ratio $r\approx0.2$. If confirmed, this result would have provided an independent confirmation of the existence of GWs, and placed restrictions on viable inflationary potentials.

However, the interpretation of the measurement in terms of primordial GWs was 
quickly challenged by several authors (see e.g.~\cite{Flauger:2014qra}). 
Subsequent polarization data from the ESA Planck satellite~\cite{Adam:2015wua} 
(in seven frequency bands between 30 GHz and 353 GHz) revealed the signal 
contribution from galactic dust in the BICEP2 fields to be much more 
significant than accounted for by the foreground subtraction performed 
in~\cite{Ade:2014xna}. 
A joint analysis of data from BICEP2, the Keck Array and the ESA Planck satellite reduced the initial detection of $r$ to an upper limit of $r<0.12$ at $95\%$ confidence (using the standard primordial spectrum pivot scale of $0.05$~Mpc$^{-1}$) \cite{Ade:2015tva}. This is consistent with the upper limit obtained from CMB data alone, $r<0.11$ at 95$\%$ confidence (pivot scale $0.002$~Mpc$^{-1}$ \cite{Ade:2015lrj}; the constraint relaxes to $r<0.15$ if the tensor and scalar spectral indices are allowed to be scale-dependent).

As such, primordial GWs remain undetected. The BICEP2 and Keck Array continue to take data, now observing at both 150 GHz and 100 GHz. Upgrades of the Atacama Cosmology Telescope (``Advanced ACTPol'' \cite{Calabrese:2014gwa}) and the South Pole Telescope (``SPT-3G'' \cite{Benson:2014qhw}) have been proposed to improve the constraint on $r$.

\clearpage
\section{Gravitational wave tests}
\label{sec:GW}

This chapter will focus on GW-based tests of GR.  There are several
comprehensive reviews on the topic, including two excellent and recent
{\em Living Reviews in Relativity} articles: one by Yunes and
Siemens~\cite{Yunes:2013dva} (focusing on Earth-based detectors and
Pulsar Timing Arrays) and the other by Gair et al.~\cite{Gair:2012nm}
(focusing on future space-based detectors). We find it unnecessary it
to reproduce that material in this review.

The consensus is that the ideal astrophysical systems to test
strong-field GR are compact objects in merging binaries. Close to
merger, the component bodies can reach speeds $v$ very close to the
speed of light ($v/c \sim 0.5$).  By contrast, the component bodies of
radio binary pulsars like PSR J0737$-$3039 are moving at a small fraction of
the speed of light ($v/c \lesssim 10^{-3}$). The leading-order
dissipative dynamics in compact binaries, governed by the backreaction
due to GW emission, is given by the quadrupole formula -- a correction
of order $(v/c)^5$ beyond Newtonian gravity. While precise timing of
radio pulses has made it possible to measure this leading-order
effect, PN corrections to the quadrupole formula are completely
negligible in binary pulsars, but they will be very important in the
frequency band where GW detectors are most sensitive. Therefore, GW
observations of coalescing BHs and NSs should probe the strong-field
dissipative dynamics to unprecedented levels of precision and
facilitate new tests of strong-field gravity.
A thorough introduction to the large and rapidly evolving literature
on compact binary waveform modeling within GR (and to the associated
systematic errors of theoretical nature) is beyond the scope of this
review. We refer the reader to one of the several recent reviews on
this
topic~\cite{Pretorius:2007nq,Centrella:2010mx,Sperhake:2011xk,Pfeiffer:2012pc,Hannam:2013pra,Lehner:2014asa}.

In this chapter we survey the broad range of physics and scales probed
by GW tests (Section~\ref{sec:TestConcepts}); the key ideas behind GW
measurements (Section~\ref{sec:paramestimation}) and GW-based tests of
GR (Section~\ref{sec:paramtests}); the details of one concrete
procedure to compare GR with proposed alternatives, i.e., the TIGER
pipeline (Section~\ref{sec:tiger}); and the challenges and
opportunities posed by theoretical and astrophysical systematics
(Section~\ref{sec:systematics}) in the context of Earth-based and
space-based detectors.

\subsection{Science opportunities}\label{sec:TestConcepts}

There are several ways in which GW observations could provide smoking
guns of modified gravitational dynamics in the strong-field
regime. Here we list some of them:

\noindent \emph{Tests of modified dynamics of binary mergers.}  The
most immediate opportunities for GW tests of GR are expected to be
provided by Earth-based detectors observing the inspiral and merger of
compact binaries.
The main target of Earth-based GW detectors like Advanced LIGO and
Advanced Virgo is the inspiral and merger of NSs and moderate-mass BHs
with masses in the range $[1,\,10^3] M_\odot$. As described in
Section~\ref{sec:CB}, these GW signals are very sensitive to the phase
evolution of the binary. Therefore they provide an excellent probe of
the compact binary's parameters and properties, and (if present) of
modifications to the underlying gravitational theory. The conceptual
foundations of these tests, as well as some details of their
implementation in a GW data analysis pipeline, are described in detail
in Sections~\ref{sec:paramestimation} and \ref{sec:paramtests} below.

Though potentially powerful, tests of GR based on GW observations from
compact binaries are limited by theoretical and astrophysical
systematics.  Even within GR, only a small submanifold of
astrophysically plausible scenarios has been exhaustively explored
through inspiral, merger, and
ringdown~\cite{Pretorius:2007nq,Centrella:2010mx,Sperhake:2011xk,Pfeiffer:2012pc,Hannam:2013pra,Lehner:2014asa},
particularly when unknown microphysics is involved.  For most of the
astrophysical parameter space, the natural framework for strong-field
tests is the application of robust but approximate models for inspiral
(e.g., PN waveforms) and merger (e.g., phenomenological models or QNM
decompositions).  As described in Section~\ref{sec:CB}, the
construction of these models is particularly challenging for tests of
gravity, as in principle we need comparably accurate calculations of
the waveforms from binary BHs and NSs.  To quantify the acceptable
amount of theoretical error permitted to observe deviations from GR, one
must possess sufficiently accurate solutions in both GR and modified
gravity.  That said, as described in Sections~\ref{sec:NSs}
and~\ref{sec:CB}, general principles (like the the no-hair relations
for BHs and the I-Love-Q relations for NSs) provide a promising
framework for enabling robust strong-field tests of gravity,
particularly for binary NSs, where a knowledge of the merger portion
of the signal is not critical.  Again, particularly for binary NSs,
the dominant effect of many modified-gravity theories seems to be
parametrized by a handful of parameters, as explored most
comprehensively for the simple case of quasicircular nonspinning
inspiral.  We will see below (Section~\ref{sec:paramestimation}) that
using these parametrized signal models into a conventional Bayesian
model selection and parameter estimation framework enables
straighforward tests to identify deviations from GR.

\noindent \emph{Tests of the spacetime geometry around massive compact
  objects.}  The orbits of compact objects plunging into massive BHs
are very complex and capture the nonlinear dynamics of gravity. The
full multipolar structure of the BH spacetime is imprinted in the
emitted signal. From the inspiral signals of an extreme mass-ratio
binary one can map out the geometry of the massive objects that reside
in galactic nuclei and test if it agrees with the Kerr
geometry~\cite{Ryan:1997hg,Barack:2006pq,AmaroSeoane:2007aw,Gair:2008bx,Gair:2012nm}.
The key point is that in the case of BHs the source multipole moments
all depend only on two parameters, mass and spin. By measuring more
than two multipoles one can check if the spacetime geometry is
described by the Kerr metric or if these objects have extra
``hair''~\cite{Ryan:1995wh,Ryan:1997hg}. The multipole moments and the
dissipative dynamics of the system also allow us to test whether the
central object possesses an event
horizon~\cite{Kesden:2004qx,Pani:2010em,Macedo:2013qea}.

An alternative approach to test the Kerr hypothesis consists of
looking at BH binary mergers and checking whether the QNM frequencies
of the merger remnant are consistent with the predictions of
GR~\cite{Berti:2005ys,Berti:2007zu,Berti:2009kk}. Again, all of these
complex frequencies depend only on the mass and spin of the final BH;
therefore measuring (say) two oscillation frequencies and one damping
time yields one test of GR, and further measurements provide
additional consistency checks.  The presence of an additional
modified-theory parameter characterizing QNMs may also be inferred (at
least in principle) by observing ringdown signals with eLISA or a
third-generation Earth-based detector, such as
ET~\cite{Gossan:2011ha}.

\noindent \emph{Tests of the dark energy equation of state.}
Compact binary sources are natural  standard 
candles~\cite{Schutz:1986gp} or, more appropriately, {\em standard sirens.} 
As discussed towards the end of Section~\ref{subsec:cosmology},
for inspiraling binaries one can measure the source's luminosity
distance from GW observations alone, although weak gravitational
lensing would bias distance measurements of individual
sources~\cite{Holz:2005df}. Methods have been proposed to correct for
the bias in the case of eLISA~\cite{Shapiro:2009sr}, which,
unfortunately, won't solve the problem completely, but a large
population of events, as in the case of ET, can average out lensing
biases~\cite{Sathyaprakash:2009xt}.  If the host galaxy of a merger
event is identified and its redshift measured, then we can use a
population of binary coalescence events to infer cosmological
parameters.  Indeed, GW observations might also measure the redshift
of host galaxies.  Tidal effects in NS binaries depend on the density
of the NS and not just on its compactness. It turns out that the tidal
effect can be used to determine the source's redshift provided the NS
EOS is known~\cite{Messenger:2011gi,Messenger:2013fya}. Statistical
methods that do not rely on host redshifts have also been proposed to
measure cosmological
parameters~\cite{DelPozzo:2012zz,Taylor:2011fs,Taylor:2012db}.
Alternatively, one can use the relation between the luminosity
distance and redshift drift to probe
cosmology~\cite{Seto:2001qf,Takahashi:2004yr,Nishizawa:2011eq,Nishizawa:2012vk}.
In principle, both quantities can be measured from GW observations
alone.  One can use such a relation to distinguish inhomogeneous
universe models from
$\Lambda$CDM~\cite{Loeb:1998bu,Quartin:2009xr,Yagi:2011bt,Yagi:2012vx}.

Advanced LIGO and Virgo will be able to determine the Hubble parameter
to an accuracy of 3\%~\cite{Dalal:2006qt}.  ET should measure the dark
energy EOS parameter to within a few
percent~\cite{Sathyaprakash:2009xt}, and its variation with redshift
to within about 20\%~\cite{Zhao:2010sz}.  By the time ET operates it
is likely that most cosmological parameters will have been measured
with good accuracy. Nevertheless, it will be very interesting, and
important, to have a completely independent way of verifying those
numbers.

\noindent \emph{Tests of alternative polarizations.} In contrast to
GR, where GWs are characterized by two polarization modes (the
``plus'' and ``cross'' polarizations), in alternative theories of
gravity GWs could have up to four additional modes. For example, a
scalar-tensor theory of gravity such as Brans-Dicke theory
\cite{Fierz:1956zz,Brans:1961sx} has a transverse scalar mode in
addition to the plus and cross polarizations. Such a mode causes
``breathing'' deformation of the test masses transverse to the
direction of propagation of the waves \cite{Will:2014xja}, as opposed
to the differential or quadrupolar deformation caused by the plus and
cross polarizations.  Interferometric detectors are built to
explicitly take advantage of the differential displacements; even so,
the presence of extra modes in a transient signal could be detected by
a network consisting of a suitably large number of detectors. For
example, a network of three noncollocated detectors can infer the
presence of an additional scalar mode~\cite{1985PhRvD..31.2480K,
  1989PhRvD..40.3884G,2006PhRvD..74h2005C,2008ApJ...685.1304L,Lee:2013sxl,
  2012CQGra..29g5011A, 2012PhRvD..86b2004C,Isi:2015cva}
(see~\cite{Yunes:2013dva} for a review).  A single detector, however,
should suffice for a continuous signal, since it can sample the
different polarizations as the detector changes its orientation
relative to the source as the Earth revolves around the Sun.

Though difficult to distinguish from instrumental sources of excess
power, the absence of these extra polarizations would weakly constrain
alternative theories of gravity.  In particular, Hayama and Nishizawa
have proposed a theory-independent method to reconstruct an arbitrary
number of polarization modes using the time-series data of an advanced
detector network~\cite{Hayama:2012au}. They consider GWs from a
supernova simulation and use a network of four detectors to recover
all the polarization modes. Future GW observations could also help
place further constraints on scalar-tensor theory if they fail to
detect scalar modes. The challenge, however, is to produce accurate
waveforms in alternative theories of gravity for different GW sources
such as supernovae, compact binary coalescences, etcetera.

\subsection{Parameter estimation and model selection}\label{sec:paramestimation} 

A brief introduction to the basic principles of GW data analysis is
useful to understand how GR can be tested within a Bayesian model
selection framework.

Given a set of compact binary merger observations from GW detectors,
the distribution of parameters $\vec{\theta}$ consistent with the data
$d$ can be inferred by comparing the predicted GW strain in each
instrument with the data.
Let $h(\vec{\theta})$ be the waveform family associated with the model
$H$.
If the detector noise $n$ is Gaussian and stationary,
noting that $n = d - h$, then
\begin{equation}
p(d | H, \vec{\theta}) = \mathcal{N}\,e^{-\frac{1}{2} \langle d - h | d - h 
\rangle},
\label{unmargevidence}
\end{equation} 
where $\mathcal{N}$ is a data-realization-independent normalization
constant, and the inner product between the real-valued time series
$A(t)$ and $B(t)$ is defined as
\begin{equation}
\langle A | B \rangle \equiv 4 \Re \int_0^\infty df\,\frac{\tilde{A}^\ast(f) \tilde{B}(f)}{S_n(f)}.
\end{equation}
Here a tilde indicates the Fourier transform, and $S_n(f)$ is the
one-sided noise power spectral density (for more details, see
e.g.~\cite{Maggiore2007b,Creighton:2011zz,Veitch:2014wba}).

Bayesian inference is used to turn the templates $h(\vec{\theta})$ and
detector data $d$ into posterior probability distributions
$p(\vec{\theta}|d,H)$ for the physical parameters which describe the
observation via Bayes' theorem:
\begin{equation}\label{eq:bayes}
p(\vec{\theta}|d,H) = \frac{p(d|\vec{\theta},H)p(\vec{\theta}|H)}{p(d|H)}\,,
\end{equation}
where $p(d|\vec{\theta},H)$ is the \emph{likelihood} or
``goodness-of-fit'' statistic, $p(\vec{\theta}|H)$ is the \emph{prior}
distribution for the model parameters, and $p(d|H)$ is the
\emph{marginalized likelihood} or \emph{evidence} for the model or
``hypothesis'' $H$.  We use the notation that $p(a|b)$ is the
conditional probability density of $a$, assuming that $b$ is
true. Notice that every term in Eq.~\eqref{eq:bayes} is conditional on
our model $H$ being correct. The model $H$ contains all of our
assumptions about the signal, including the physics of gravitational
radiation production and propagation, the astrophysics of how these
systems form (thereby guiding our choice of parameterization and prior
distributions), the response of our detectors to incident GWs, and the
instrumental noise with which the signal must compete.
More succinctly: in the matched filtering paradigm, conclusions drawn
from the data are conditional on our model for the data -- including
both the GWs and the detectors -- being correct.

Bayes' theorem provides the conceptual foundation for parameter
estimation and model selection in general, and for GW tests of GR in
particular.  General hypothesis tests follow by applying Bayes'
theorem \eqref{eq:bayes} to a collection of hypotheses $\{H_1,\ldots,
H_N\}$:
\begin{equation}
p(H_k | d) = \frac{p(d | H_k)\,p(H_k )}{p(d)}.
\label{eq:bayes:hypothesis}
\end{equation}
where $p(H_k)$ are prior probabilities for the models $H_k$.  The
\emph{evidence} $p(d|H)$ can be computed by integrating the likelihood
$p(d|\vec{\theta}_k,H_k)$ of the parameters $\vec{\theta}_k$ times the
prior probabilities $p(\theta_k|H_k)$ for these parameters within the
model $H_k$:
\begin{equation}
\label{margevidence}
p(d|H_k)\equiv \int
d\vec{\theta}_k p(d|\vec{\theta}_k,H_k)p(\theta_k|H_k).
\end{equation}
The probability of the data $p(d)$ depends on all the hypotheses under
consideration: $p(d) = \sum_k p(d|H_k) p(H_k)$.  That said, in
practice this overall probability never appears, since models are
compared using \emph{odds ratios} $O_{ij}$ between probabilities for
two different hypotheses:
\begin{equation}
O_{ij}  = \frac{p(H_i|d)}{p(H_j|d)}
\label{eq:oddsRatio}
\end{equation}

In the context of testing GR, we consider multiple models: (a) a model
$H_{\rm GR}$, where GR is correct; and (b) one or more ``modified GR''
models $H_{\rm modGR}$ including additional parameters in the signal
model, and usually reducing to GR in a suitable limit.  For the
purposes of model selection, these parameters can -- but need not --
be connected to an underlying physical theory (e.g., a Lagrangian).
In practice, the most-often used ``extended'' model parameters
describe ad-hoc changes to the GW orbital dynamics (e.g., phase).

Ideally, one of our models $H$ perfectly matches reality.  In this
case the posterior has some finite width, which, to first order,
scales as the inverse of the signal-to-noise ratio (SNR).  The width
of the posterior is the statistical error of the model: for example,
the 90\% credible interval of $p(\vec{\theta}|d,H)$ will contain the
``true'' model parameters 90\% of the time.  As the SNR increases
(either by finding closer sources or improving the sensitivity of the
detector) credible intervals shrink and the distributions (typically)
become more Gaussian, allowing for more precise statements to be made
about the observations.

In practice, our models $H$ will not fully describe the source physics
and the detector.  The best-fit model and parameters will be
systematically biased, as optimization invites parameters to flex away
from the true values in an attempt to overcome our model's
shortcomings in matching the data.  Systematic errors are less
strongly dependent on the signal strength.  Hence, relative to
statistical errors, the impact of systematic errors increases strongly
with signal amplitude and it becomes important as instruments become
more sensitive (see e.g.~\cite{Berti:2006ew,Cutler:2007mi}).  In other
words, the potentially most informative signals require the greatest
care.

\subsection{Direct versus parametrized tests of gravity}\label{sec:paramtests}
The approaches developed to test GR with compact binaries come in two
flavors, that we will call ``direct'' and ``parametrized'' tests.
While both approaches adopt a parametrized model to characterize
deviations from GR, they are distinguished by the role of the null
hypothesis and the significance of non-GR parameters.

\paragraph{Direct tests with inspiral waves.}
A direct test assumes that GR is the correct theory of gravity,
introduces a systematic (but not necessarily physical) modification to
the dynamics and to the resulting GW signal, and evaluates support for
the null hypothesis.
An example of the direct approach is as follows. At leading order in
the PN approximation, the observed strain amplitude from a
nonspinning binary moving on a quasi-circular orbit is given
by~\cite{Blanchet:2013haa}
\begin{equation}
h(t) = \frac{4 G {\cal C} \eta M}{c^4R}\, [GM\omega (t)]^{2/3}\, \cos2\phi(t),\quad
\end{equation}
where as usual $M$ and $\eta$ denote the total mass and symmetric mass
ratio; ${\cal C}<1$ is a number that depends on the position of the
source on the sky, the position of the orbital plane with respect to
the line of sight, the polarization of the waves and the distance to
the binary; $\phi(t)$ and $\omega(t)$ are the orbital phase and
frequency, respectively, obtained by solving the balance equation
$\dot\omega = - {\cal L}/(dE/d\omega),$ where $E(v)$ and ${\cal L}(v)$
are the gravitational binding energy (per unit mass) and GW luminosity
of the system.  It is often convenient to work with the Fourier
transform of the waveform, which, in the stationary-phase
approximation, is given by~\cite{Sathyaprakash:1991az,
  Dhurandhar:1992mw,Finn:1992xs}
\begin{align}
\label{phaseSPA1}
H(f) & =  \sqrt{\frac{5\eta}{24}}\, \frac{{\cal C}(GM)^{5/6}}{\pi^{2/3}c^{3/2}R}\, 
f^{-7/6}\, e^{i {\psi_{\rm SPA}}(f)}\,,\\
\psi_{\rm SPA}(f) & =  2\pi f t_c-\phi_c-\frac{\pi}{4}
+{\frac{3}{128\,\eta\, v_f^5}}\;\sum_{k=0}^{7}\psi_{(k/2)\mathrm{PN}}\,v_f^k\,,
\label{phaseSPA2}
\end{align}
where we denote $v_f= (\pi G M f)^{1/3}$, and the subscript ``SPA''
obviously stands for ``stationary-phase approximation.'' The
coefficients $\psi_{(k/2)\mathrm{PN}}$'s $(k=0,\ldots,7)$ in the
Fourier phase are computed in a PN series through the equation
\begin{equation}
\psi(f) = 2 \pi f\,t_c - \phi_c - \frac{\pi}{4} + 2 \int_{v}^{v_c} (v_c^3-v^3)\,\frac{E'(v)}{{\cal L}(v)}\,dv\,,
\label{eq:fourier phasing}
\end{equation}
by expanding the binary center-of-mass energy $E$ and the GW
luminosity ${\cal L}$ through the appropriate PN order; $t_c$ and
$\phi_c$ are the time and phase at coalescence and $v_c=(\pi G M
f_c)^{1/3}$, with $f_c$ the cutoff frequency.

Inspiral phasing is currently known to 3.5 PN order (i.e., order
$(v/c)^7$) beyond the leading-order quadrupole formula. If spins are
negligible, which would be a good approximation for binary NSs, all
the PN coefficients $\psi_k$ in Eq.\,(\ref{eq:fourier phasing}) (9 of
them including logarithmic terms) depend only on the component masses.
More in general, the PN coefficients in the GW phasing of a coalescing
binary only depend on the component masses and spins, and therefore
only a limited number of them are independent, so that a very generic
test could be to check for consistency between the measured
coefficients.
A first proposal in this direction was put forward by Arun et
al.~\cite{2006CQGra..23L..37A,2006PhRvD..74b4006A,Mishra:2010tp}.
As described in Section~\ref{sec:tiger} below, one approach treats
three of the coefficients (say $\psi_0,$ $\psi_2$ and $\psi_3$) as
independent, and asks if the measurement of the third is consistent
with the first two~\cite{Blanchet:1995xs,Mishra:2010tp}. The
conclusion is that, for Advanced LIGO/Virgo, a single loud event with
SNR$> 20$ would detect a departure of $\psi_3$ from its GR value by
about 2.5\%~\cite{Li:2011cg}.

A drawback of such an approach is that it is difficult to combine
information from multiple sources in this way so as to arrive at a
stronger test: deviations in the PN coefficients can be different from
one source to the next if they themselves depend on masses, spins, and
whatever additional charges may be present in an alternative theory of
gravity.
This problem can be circumvented by searching for GR violations using
Bayesian \textit{model selection}, rather than parameter
estimation. For instance, one could introduce parametrized
deformations in the waveforms predicted by GR, and compare the
resulting waveform model with the GR prediction. Del Pozzo et
al.~\cite{2011PhRvD..83h2002D} adopted this approach (using a single
extra parameter) in the context of binary inspiral, and Gossan et
al.~\cite{Gossan:2011ha} used a similar strategy in the context of
ringdown, considering multiple additional parameters.
With model selection it becomes possible to combine information from
multiple sources and build up evidence for or against GR, even if
deviations manifest themselves in a different way for each source. On
the other hand, when the ``non-GR'' model is insufficiently
parsimonious -- i.e., it has too many additional free parameters -- it
may be penalized if the true theory involves only a small number of
parameters.

\paragraph{Direct tests with quasinormal modes.}
\label{sec:test_qnm}

The BH remnant resulting from the merger of two compact objects
(either BHs or NSs) is initially highly deformed, but it soon settles
down to a quiescent state by emitting ringdown radiation, which
consists of a superposition of QNMs~\cite{Vishveshwara:1970cc} (for a
review see, e.g.,~\cite{Berti:2009kk}). A Kerr BH is characterized
only by its mass and angular momentum, and so are the complex
frequencies of its QNM oscillations~\cite{Ruffini:1971,Leaver:1985ax},
although the relative amplitudes of the modes depend on the specific
details of the excitation.

Detection of the characteristic ringdown GW signal of a BH would,
therefore, allow a direct test of the no-hair
theorem~\cite{Dreyer:2003bv}, and hence GR, through the comparison of
frequencies and decay times of these modes with the predictions of GR
for a BH with certain mass and spin.  In practice, the detection and
discrimination of multiple modes is essential, as it is first
necessary to infer the mass and spin of the BH before checking for
consistency between the modes.  If any of the modes have some
parameter dependence, other than mass and spin, then the mass and spin
obtained from these modes will not be consistent with that obtained
from the others, and thus the source of emission must be different
from a Kerr BH.
Departures from GR in the QNM spectrum can be encoded in extra
parameters to be identified by Bayesian model selection or parameter
estimation.

The idea of treating BHs as ``gravitational atoms,'' and their QNM
spectra as the GW analog of atomic lines, dates back to a seminal
paper by Detweiler~\cite{Detweiler:1978ge}.  Dreyer et al.~introduced
a formalism for testing GR with QNMs~\cite{Dreyer:2003bv},
and made a concrete suggestion to test the no-hair theorem through the
measurement of more than one mode. Berti et
al.~\cite{Berti:2005ys,Berti:2007zu} investigated the accuracy of
measurement of individual mode parameters using a Fisher matrix
analysis and estimated the resolvability of individual modes in the
complete signal as a function of signal-to-noise ratio (SNR). They
concluded that the presence of a second mode can be inferred as long
as the SNR is larger than a critical value, under the assumption that
the presence of a ringdown signal has been confirmed and the
parameters of the dominant modes are reliably measured. The critical
SNR depends on the mass ratio of the progenitor binary, but an SNR of
20 should suffice if the mass ratio of the progenitor binary is
$q=m_1/m_2 \gtrsim 2$. Kamaretsos et al.~showed that using BH ringdown
signals following a nonspinning binary merger, it might be possible to
recover the mass ratio of the progenitor binary from the relative
amplitudes of the QNMs~\cite{Kamaretsos:2011um}.

Using a limited set of sources, Gossan et al.~\cite{Gossan:2011ha}
conducted a proof-of-concept Bayesian model selection calculation for
modified gravity using the ringdown signal.  Specifically, they
applied Bayesian model selection to obtain a more robust and
quantitative measure of the consistency of the data with GR, as
opposed to a generalized theory where the mode parameters depended on
an extra parameter other than the BH mass and spin (i.e., ``hairy''
BHs).  Using this technique, for the sources in their catalog, they
can measure deviations at the 10\% level in the fundamental $l=m=2$
frequency parameter $\hat{\omega}_{22}$ out to $\simeq 6$\,Gpc for a
$500\,M_{\odot}$ source with ET.  With a space-based detector like
LISA it is possible to measure deviations at the 10\% level at
$6$\,Gpc with a $10^{6}\,M_{\odot}$ source, and at the 0.6\% level at
$z \sim 5$ with a $10^8\,M_{\odot}$ source.  This proof-of-concept
calculation adopted aggressive simplifying assumptions (a known
source location and source orientation) and explored only a handful of
candidate sources, very specific source location, orientation and mass
ratio of the progenitor binary.

More recently, eliminating these highly simplifying assumptions and
using improved waveform models, the TIGER pipeline (discussed in
Section~\ref{sec:tiger} below) showed that modifications to gravity
could be identified using as few as 10 astrophysically plausible
sources seen by ET~\cite{2014PhRvD..90f4009M}.  Specifically, these
authors used a QNM model that is matched to numerical simulations of
coalescing BH binaries where BH spins are aligned with the orbital
angular momentum. Their model consists of a superposition of the four
dominant modes $(\ell,m)=(2,2), (2,1), (3,3)$ and $(4,4)$ used in
Kamaretsos et al.~\cite{Kamaretsos:2012bs}, where the mode amplitudes
are given by
\begin{equation}
A_{22}(\eta) = 0.864 \eta, 
\end{equation}
\begin{equation}
A_{21}(\eta) = 0.43\,\left[ \sqrt{1-4\eta} - \chi_{\rm eff} \right] A_{22}(\eta),
\end{equation}
\begin{equation}
A_{33}(\eta) = 0.44(1-4\eta)^{0.45} A_{22}(\eta), 
\end{equation}
\begin{equation}
A_{44}(\eta) = \left[ 5.4 (\eta - 0.22)^2 + 0.04 \right] A_{22}(\eta),
\end{equation}
and $\chi_{\rm eff}$ is a single effective spin parameter, that is a
specific combination of the two progenitor BH dimensionless spins
$(\chi_1,\chi_2)$ weighted by their masses $(m_1,m_2)$:
\begin{equation}
\chi_{\rm eff} = \frac{1}{2}\left( \sqrt{1-4\eta}\,\chi_1 + \chi_- \right), \quad \quad
\chi_- = \frac{m_1\chi_1 - m_2\chi_2}{M_{\rm in}}. \label{chiMinus} 
\end{equation}
Here $M_{\rm in}$ is the initial total mass of the binary, assumed
equal to the mass of the final BH.  The frequencies $\omega_{lm}$ and
damping times $\tau_{lm}$ are related through the \emph{quality
  factors} $Q_{lm} = \omega_{lm} \tau_{lm}/2$:
\begin{equation}
M\omega = f_1 + f_2 (1 - j)^{f_3},\quad\quad
Q = q_1 + q_2 (1 - j)^{q_3},
\end{equation}
and the coefficients $f_1$, $f_2$, $f_3$, $q_1$, $q_2$, $q_3$ were
fitted to QNM data in~\cite{Berti:2006cc}. The spin $j$ of the remnant
BH can also be written in terms of the binary component masses $(m_1,
m_2)$ and spins $(\vec{\chi}_1, \vec{\chi}_2)$ (see
e.g.~\cite{Barausse:2009uz}).

Denote the free parameters corresponding to the
hypothesis that GR is correct by $\vec{\theta}_{\rm GR}$.
For the non-GR hypothesis, one can use one or more deviations of the
QNM frequencies and damping times. Meidam et
al.~\cite{2014PhRvD..90f4009M} considered the following different
hypotheses and corresponding additional parameters:
$H_1 \leftrightarrow \{\vec{\theta}_{\rm GR}, \delta\hat{\omega}_{22}\}, 
H_2 \leftrightarrow  \{\vec{\theta}_{\rm GR}, \delta\hat{\omega}_{33}\},
H_3 \leftrightarrow  \{\vec{\theta}_{\rm GR}, \delta\hat{\tau}_{22}\}, 
H_{12} \leftrightarrow \{\vec{\theta}_{\rm GR}, \delta\hat{\omega}_{22},\delta\hat{\omega}_{33}\}, 
H_{13} \leftrightarrow \{\vec{\theta}_{\rm GR}, \delta\hat{\omega}_{22}, \delta\hat{\tau}_{22}\},  
H_{23} \leftrightarrow \{\vec{\theta}_{\rm GR}, \delta\hat{\omega}_{33}, \delta\hat{\tau}_{22}\}, 
H_{123} \leftrightarrow  \{\vec{\theta}_{\rm GR}, \delta\hat{\omega}_{22}, \delta\hat{\omega}_{33}, 
\delta\hat{\tau}_{22}\}. $
Here $\delta\hat{\omega}_{22}, \delta\hat{\omega}_{33}$ and
$\delta\hat{\tau}_{22}$ are deviations of the two dominant mode
frequencies and of the dominant mode's damping time from their GR
values. These are, in essence, extra ``hair'' parameters of the
remnant BH. Using such a model and combining the different hypotheses,
Meidam et al.~concluded that deviations from GR at the level of 10\%
in $\delta\hat{\omega}_{22}$ and $\delta\hat{\omega}_{33}$ could be
inferred with $\sim 20$ intermediate-mass binary BH merger events seen
up to a distance of 60 Gpc. To confirm deviation of the same order in
$\delta\hat{\tau}_{22}$ would require $\sim 50$ detections. If one
assumes that GR is the correct theory, then $\sim 20$ detections would
be good enough to constrain $\delta\hat{\omega}_{22}$ and
$\delta\hat{\omega}_{33}$ to within 0.5\% and $\delta\hat{\tau}_{22}$
to within 5\% of their GR values (cf.~Figure~8
of~\cite{2014PhRvD..90f4009M}).

Just like inspiral tests, ringdown-based tests can also be limited by
theoretical and astrophysical systematics.  The presence of precessing
spins is well known to significantly alter the relative QNM amplitude.
Indeed, several authors have argued that ringdown radiation can be
used to draw inferences about the pre-merger
spins~\cite{Kamaretsos:2011um,O'Shaughnessy:2012ay,2013PhRvD..88b4040P}.
Additionally, many authors adopt simplifying assumptions about the
angular distribution of QNMs, even though the asymptotic behavior at
late time is well known to involve a superposition of several
spin-weighted spherical harmonics~\cite{London:2014cma}.  These
simplifications are expected to impact inferences about the mass
ratio, inclination, spins, amplitude of precession, and presence or
absence of GR modifications, and need to be included in future studies.

\paragraph{Parametrized tests.}

Yunes and Pretorius proposed a model-independent way of testing
alternative theories of gravity~\cite{Yunes:2009ke}.
The basic idea is to use as a template for the inspiraling, weak-field
regime the phasing formula $H(f) (1+\alpha f^{a/3}) e^{i\beta
  f^{b/3}},$ $H(f)$ being the GR template of
Eq.~\eqref{phaseSPA1}. The coefficients $\alpha$, $\beta$, $a$ and $b$
are called ``parametrized post-Einsteinian'' (PPE)
parameters~\cite{Yunes:2009ke}, in analogy with the PPN parameters
that measure deviations from GR in the weak-field, slow-motion
regime~\cite{Will:2014xja}. As in the PPN framework, the PPE parameters
depend on the specific theory of gravity, and their values are known
in some theories. The PPE framework (or extensions thereof) can
capture the predictions of most extensions of GR discussed in this
review, and it has been generalized to include more generic
corrections to the waveform, such as the presence of higher harmonics,
modified GW propagation effects and ringdown radiation.  We refer the
interested reader to Section~5.3.4 of the recent review by Yunes and
Siemens~\cite{Yunes:2013dva}.

An advantage of parametrized, PPE-like searches is that they would
avoid what Yunes and collaborators call ``fundamental bias,'' by
allowing the data to select the correct theory of gravity through a
systematic study of statistically significant anomalies.
A more subtle concern is that if GR happened to be wrong in the
strong-field regime and we observe low-SNR signals (as expected for
the first GW detections in Advanced LIGO/Virgo), GR-based GW detection
templates could still extract the signal with the wrong parameters,
without being able to identify that there is a non-GR anomaly in the
data. Vallisneri and Yunes~\cite{Vallisneri:2013rc} found that this
insidious ``stealth bias'' is indeed possible in a certain region of
parameter space.

A practical limitation of the PPE framework is that, if deviations
from GR are present in the weak-field regime, they might be already
well constrained in a PPN sense by pulsar timing observations. By
contrast, if GR deviations affect only the strong-field regime (i.e.,
the very last stages of inspiral, merger and ringdown), their signal
might not be a small deviation away from the GR template.  In this
scenario, only direct measurements of strong-field sources (e.g., GWs
from merging binaries) would enable us to detect and constrain the
parameters of the theory.
Alternative theories of gravity of this kind might be difficult to
conceive, but scalar-tensor theories with dynamical scalarization are
an interesting example~(see Sections~\ref{sec:NS_ST}
and~\ref{sec:binaries/scal-tens-theor}).

\subsubsection{Implementation of direct tests: the TIGER pipeline}
\label{sec:tiger}

The TIGER (Test Infrastructure for GEneral Relativity) data analysis pipeline
provides a  ``direct test'' of GR via the GW coalescence waveform~\cite{Li:2011cg,Li:2011vx,VanDenBroeck2014,Agathos:2013oma,2014PhRvD..89h2001A,2014PhRvD..90f4009M}.
Broadly speaking, the framework  assesses the evidence for  ``generic'' deviations from GR in the signal.  
Starting from waveform deformations characterized by an (in principle)
arbitrarily large number of additional parameters $\delta\xi_1,
\delta\xi_2, \ldots, \delta\xi_N$ (assumed to be zero if GR is
correct), the algorithm asks the question: ``do one or more of the
$\delta\xi_i$ $(i = 1, 2, \ldots, N)$ differ from zero?'' 
Although the method is not tied to any particular part of the
coalescence process (indeed, as anticipated in
section~\ref{sec:paramtests}, it was applied to ringdown
tests~\cite{2014PhRvD..90f4009M}), in this section we will focus on
the case of inspirals, for which a robust analysis pipeline is
available \cite{2014PhRvD..89h2001A}. Parametrized deformations can
be introduced in the PN coefficients $\psi_i$ of the GW
phase~\cite{2006CQGra..23L..37A,Li:2011cg}, e.g.~by setting
\begin{equation}
\psi_i = \psi_i^{\rm GR}(m_1, m_2, \vec{S}_1, \vec{S}_2)\,\left[1 + \delta\xi_i \right],
\label{phasecoeffshifts}
\end{equation}
where $\psi_i^{\rm GR}(m_1, m_2, \vec{S}_1, \vec{S}_2)$ describes the
functional dependence of the PN coefficients on the component masses
($m_1$, $m_2$) and spins ($\vec{S}_1$, $\vec{S}_2$) that is predicted
by GR.  
Given GW observations, TIGER compares two hypotheses:
\begin{itemize} 
\item The GR (null) hypothesis, denoted by $H_{\rm GR}$, is the hypothesis
  that GR is the correct theory of gravity.
\item The hypothesis $H_{\rm modGR}$ assumes that there is some
  deviation from GR. On first consideration, one might be inclined to
  make this the negation of the GR hypothesis.  However, $H_{\rm
    modGR}$ would then be associated with a family of waveforms that
  can not be parametrized by a finite number of parameters: literally
  any waveform outside the family predicted by GR would be allowed. In
  practice, only a finite number of deformation parameters can be
  considered. Call these $\delta\xi_1, \delta\xi_2, \ldots,
  \delta\xi_N$ as above, with the understanding that the GR waveform
  corresponds to $\delta\xi_i = 0$ for $i = 1, \ldots, N$. We then
  define $H_{\rm modGR}$ as follows:
\begin{quote}
  $H_{\mathrm{modGR}}$ is the hypothesis that one or more of the
  deformation parameters $\delta\xi_i$ in the waveform is different
  from zero, without specifying which.
\end{quote}
\end{itemize}
The odds ratio of interest is then:
\begin{equation}
O^{\rm modGR}_{\rm GR} \equiv \frac{p\left( H_{\rm modGR} | d \right)}{p\left( 
H_{\rm GR} | d \right)}.
\label{Otiger}
\end{equation}
In principle the odds ratio
has a straightforward interpretation: if
$O^{\rm modGR}_{\rm GR} < 1$, GR is favored by the data; if
$O^{\rm modGR}_{\rm GR} > 1$, the data tell us that a different theory of 
gravity is favored.

This Bayesian method has a number of
attractive features: any waveform model
or any particular part of the coalescence process can be employed;
information from multiple sources can trivially be combined; sources with small or marginal SNRs
can be included without penalty; and the Bayesian Occam's razor naturally accounts for (and where appropriate, penalizes) increasing model
dimension. Crucially, as illustrated below, the approach can identify the presence of deviations from GR that are not included in the
chosen parametrized waveform family.   

\paragraph{Frequentist approach and nontrivial systematics.}
Like all parametrized tests, the TIGER pipeline can be  limited by observational, astrophysical, and theoretical
systematics.  Because 
exact and generic binary merger solutions are generally unavailable, to make progress, the TIGER pipeline has adopted a phenomenological and
frequentist approach, usually adopting simplified waveform models (e.g., neglecting spin precession).  Real
detector noise is nonGaussian and could mimic a violation of GR in a sufficiently large sample of events.  For this
reason, the pipeline does not adhere to a strictly Bayesian interpretation of these odds ratios; instead, the odds ratios
are used as frequentist statistics.  

\begin{figure}[htp]
	\centering
	\includegraphics[width=0.75\textwidth]{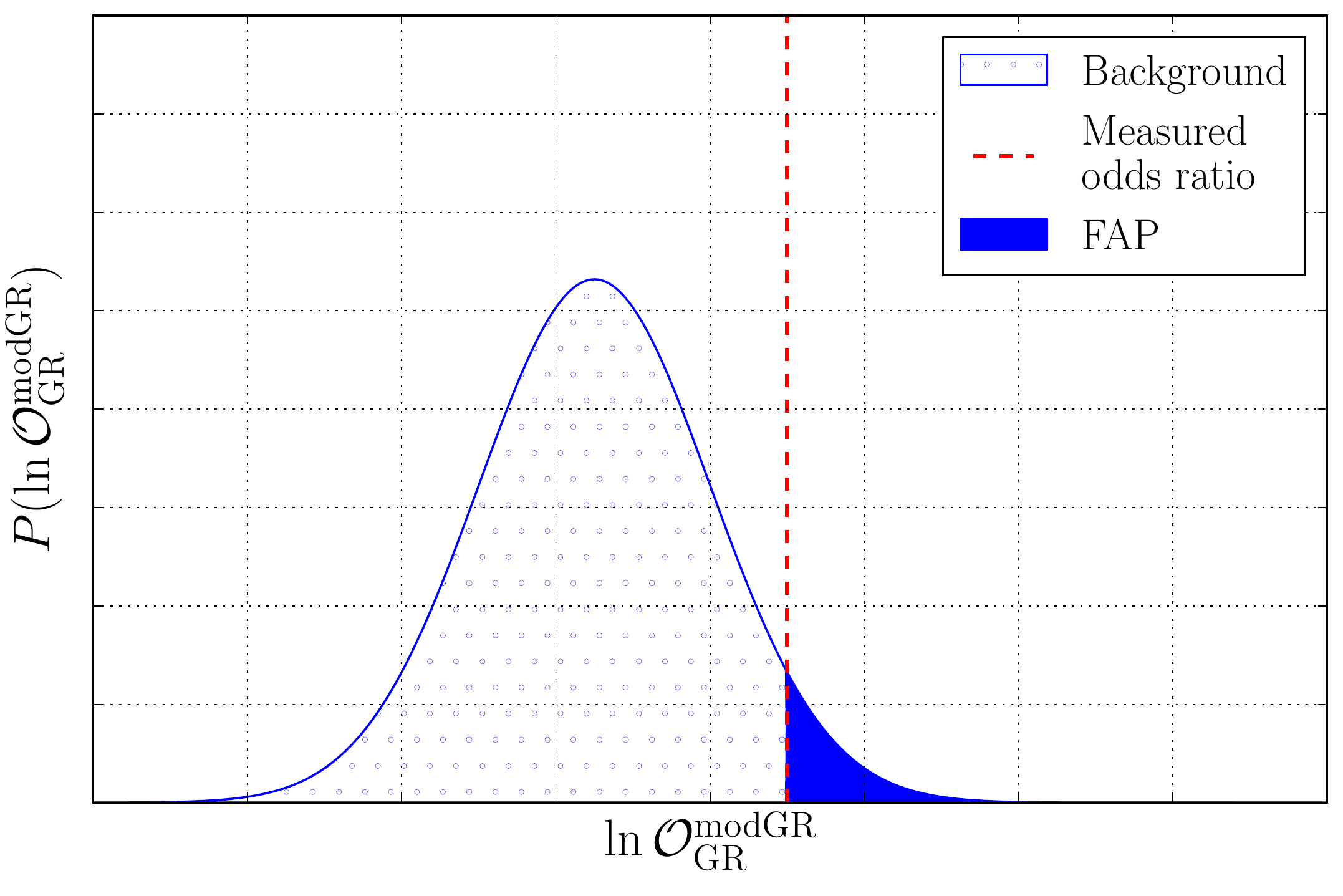}
	\caption{Example of a odds ratio background (blue curve),
          measured odds ratio (red dashed line), and the false alarm
          probability associated to the measured odds ratio (blue
          shaded area).}
	\label{fig:tigFAPExample}
\end{figure}

To assign a significance to the measured odds ratio as a statistic,
the pipeline evaluates the background distribution of the odds ratio.  
Specifically, it evaluates the 
 odds ratios for a large number of (catalogs of) simulated signals that are put 
in the noise, each of them being in accordance with GR.\footnote{For a 
comparison
  of analytic and numerical methods, see~\cite{DelPozzo:2014cla}.}
Figure~\ref{fig:tigFAPExample} schematically illustrates how a (normalized)
background distribution can be used to assign a false alarm
probability to the odds ratio computed from the actual, measured signals: it
is simply the area under the background distribution for the range of odds
ratios that are larger than the measured one.
We note that in practice it is convenient to work with the logarithm of
the odds ratio, $\ln O^{\rm modGR}_{\rm GR}$.

Finally, to assess the effectiveness of the pipeline as a test for the presence of modified gravity, the background
distribution is compared with a foreground distribution: the distribution of 
the detection statistic for a
distribution of sources, given a particular modification of gravity.  
The degree of overlap
between foreground and background then tells us how easy or
difficult it will be to confidently discover the given violation of
GR.  This can be formalized by introducing the notion of
efficiency, defined as the fraction of the foreground that is
above a pre-determined fraction (e.g.~95\%) of the
background. For details, see~\cite{Li:2011cg,Li:2011vx,2014PhRvD..89h2001A}.

\paragraph{Decomposing and evaluating the odds ratio.}
By construction,  no waveform model is associated with $H_{\rm
  modGR}$.   Instead, the hypothesis can be broken up into auxiliary
hypotheses, each of which \textit{does} come with a concrete waveform
model:
\begin{quote}
\item $H_{i_1 i_2 \ldots i_k}$ is the hypothesis that the parameters
  $\{\delta\xi_{i_1}, \delta\xi_{i_2}, \ldots, \delta\xi_{i_k}\}$
  differ from zero, but all the other $\delta\xi_j$, $j \notin \{i_1,
  i_2, \ldots, i_k\}$, are zero.
\end{quote}
In terms of these hypotheses, $H_{\rm modGR}$ is the logical ``or'' of all these auxiliary
hypotheses:
\begin{equation}
H_{\rm modGR} = \bigvee_{i_1 < i_2 < \ldots i_k; k \leq N} H_{i_1 i_2 \ldots i_k}. 
\end{equation}
Note that the hypotheses $H_{i_1 i_2 \ldots i_k}$ and $H_{j_1 j_2 \ldots j_l}$
for $\left\{i_1 i_2 \ldots i_k\right\}\neq\left\{j_1 j_2 \ldots j_l\right\}$
are \textit{mutually exclusive}, or logically disjoint. This implies that the
probability of the union of the auxiliary hypotheses equals the sum of the
probabilities associated to individual auxiliary hypotheses. The odds ratio in
Eq.~(\ref{Otiger}) then becomes
\begin{align}
O^{\rm modGR}_{\rm GR} = \sum_{k=1}^N \sum_{i _1 < i_2 < \ldots i_k} \frac{p(H_{i_1 i_2 \ldots i_k}
| d)}{p(H_{\rm GR} | d)}.
\label{Osumof}
\end{align}
If there are multiple detected sources -- which will
together be referred to as a \textit{catalog} -- in stretches of data
$d_1, d_2, \ldots, d_\mathcal{N}$, then the definition of the odds
ratio (\ref{Otiger}) can trivially be generalized, and a result like
Eq.~(\ref{Osumof}) will again hold:
\begin{align}
\mathcal{O}^{\rm modGR}_{\rm GR} &\equiv \frac{p\left( H_{\rm modGR} | d_1, d_2, 
\ldots, d_\mathcal{N} \right)}{p\left( H_{\rm GR} | d_1, d_2, \ldots, 
d_\mathcal{N} \right)} \nonumber\\
&= \sum_{k=1}^N \sum_{i _1 < i_2 < \ldots i_k} \frac{p(H_{i_1 i_2 \ldots i_k} | 
d_1, d_2, \ldots, d_\mathcal{N})}{p(H_{\rm GR} |d_1, d_2, \ldots, 
d_\mathcal{N})}.
\label{Osumofcatalog}
\end{align}
We refer the reader to Ref.~\cite{Li:2011cg} for further discussions on the
calculation of the odds ratio and the computational details.

\paragraph{Application to NS-NS binaries.}

We here demonstrate that TIGER is an effective test for a broad
spectrum of modifications to gravity, using binary NS sources.
Specifically, we show results from simulations where the sources had
component masses in the NS-NS range, $m_1, m_2 \in [1, 2]\,M_\odot$,
and were distributed uniformly in co-moving volume, with arbitrary
orientations.  Having used astrophysically realistic populations of
sources and assuming Advanced LIGO and Advanced Virgo operating at
design sensitivity, these examples demonstrate the ability to identify
modifications to gravity in an astrophysically plausible scenario,
where most sources have low SNR and are close to the detection
threshold.

\begin{figure}[thp!]
	\centering
	\includegraphics[width=0.7\textwidth]{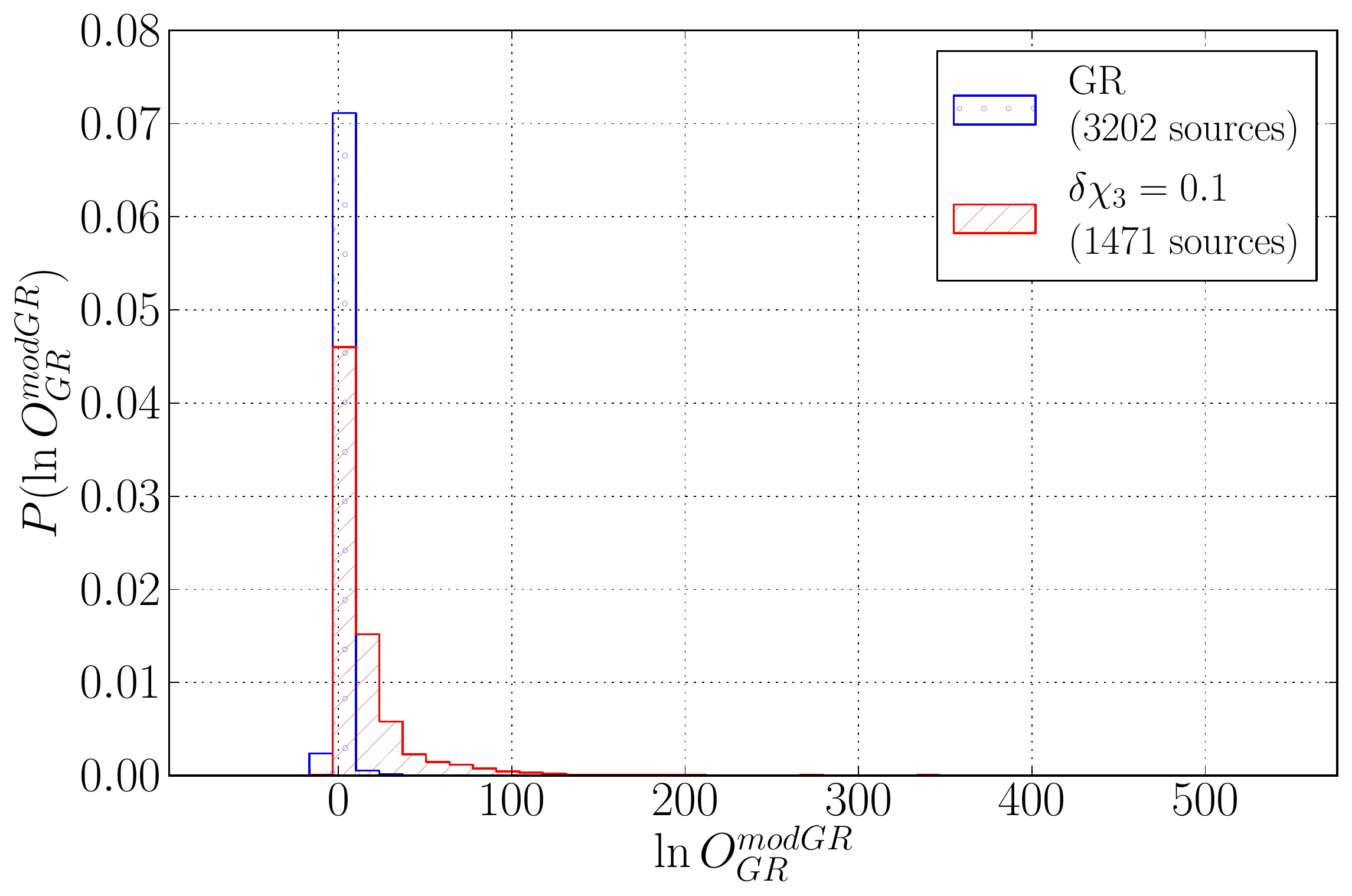}\\
	\includegraphics[width=0.7\textwidth]{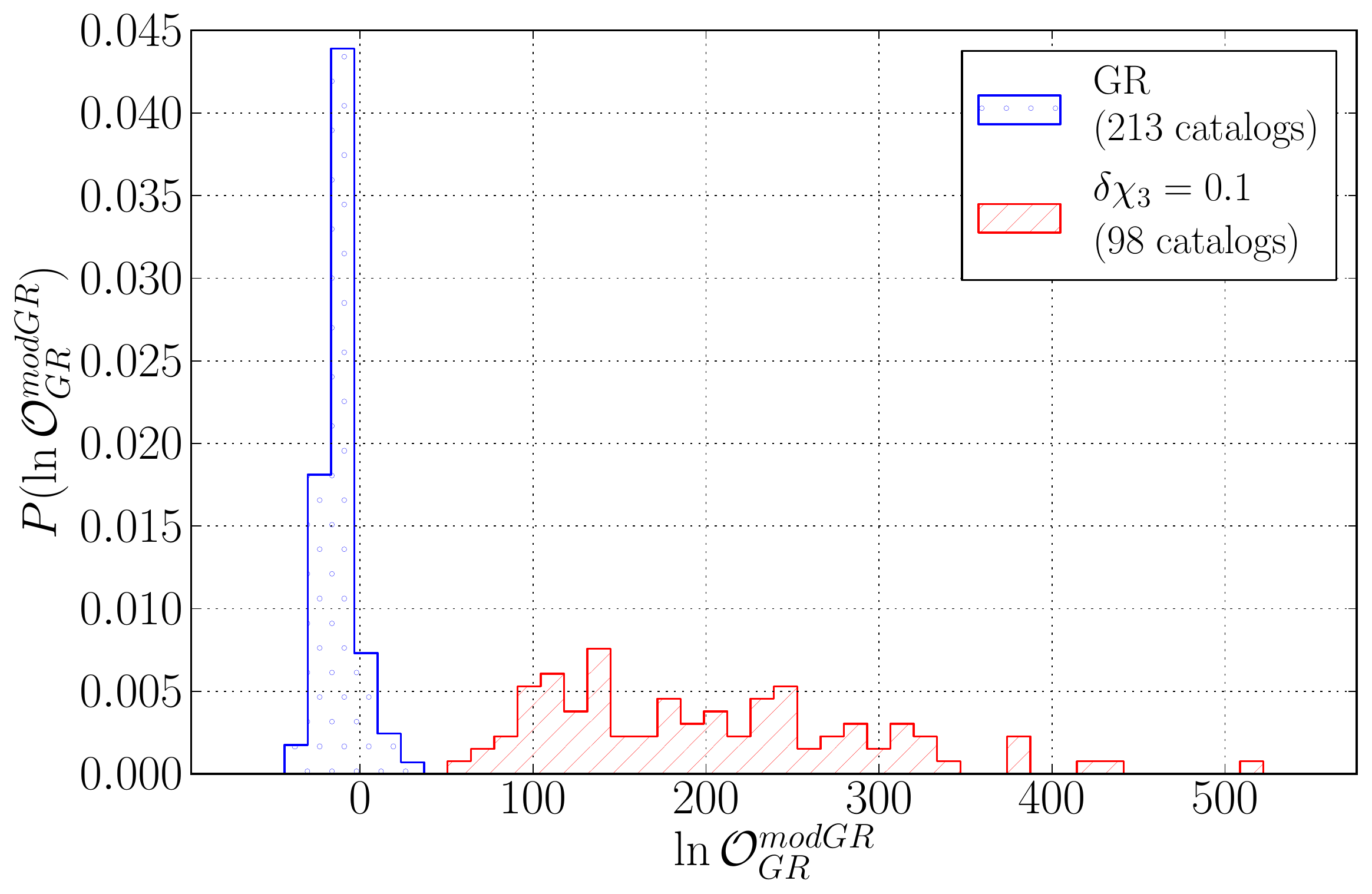}
	\caption{Background (blue) and foreground (red) for deviation
          of $10\%$ in the 1.5PN phase coefficient $\psi_3$. The top
          plot shows the odds ratio distributions for single sources,
          whereas the bottom plot shows the results for catalogs of 15
          sources each.}
	\label{fig:dchi3_10pc}
\end{figure}

For these sources, TIGER is sensitive to deviations in the PN phase
coefficients even at high PN order, where no other astrophysical
bounds (e.g., from the binary pulsar~\cite{2010PhRvD..82h2002Y})
exist.  For example, consider a deviation in the phase coefficient
$\psi_3$ at 1.5PN, which is the lowest order at which the dynamical
nonlinearity of GR manifests itself through the so-called tail
terms~\cite{1994CQGra..11.2807B,Blanchet:1995xs}. For the calculation
of the odds ratio, the auxiliary hypotheses corresponded to shifts in
different PN phase coefficients, as in Eq.~(\ref{phasecoeffshifts}).
Figure~\ref{fig:dchi3_10pc} shows the results for a constant 10\%
relative shift in $\psi_3$, which is far beyond the reach of any
current observation.
Even with a single detection (top panel), one could plausibly detect
the deviation. When we combine information from 15 sources (bottom
panel) the result is a complete separation between the background and
foreground, which means that the given deviation can be detected with
near-certainty. 

\begin{figure}[htp!]
	\centering
	\includegraphics[width=0.7\textwidth]{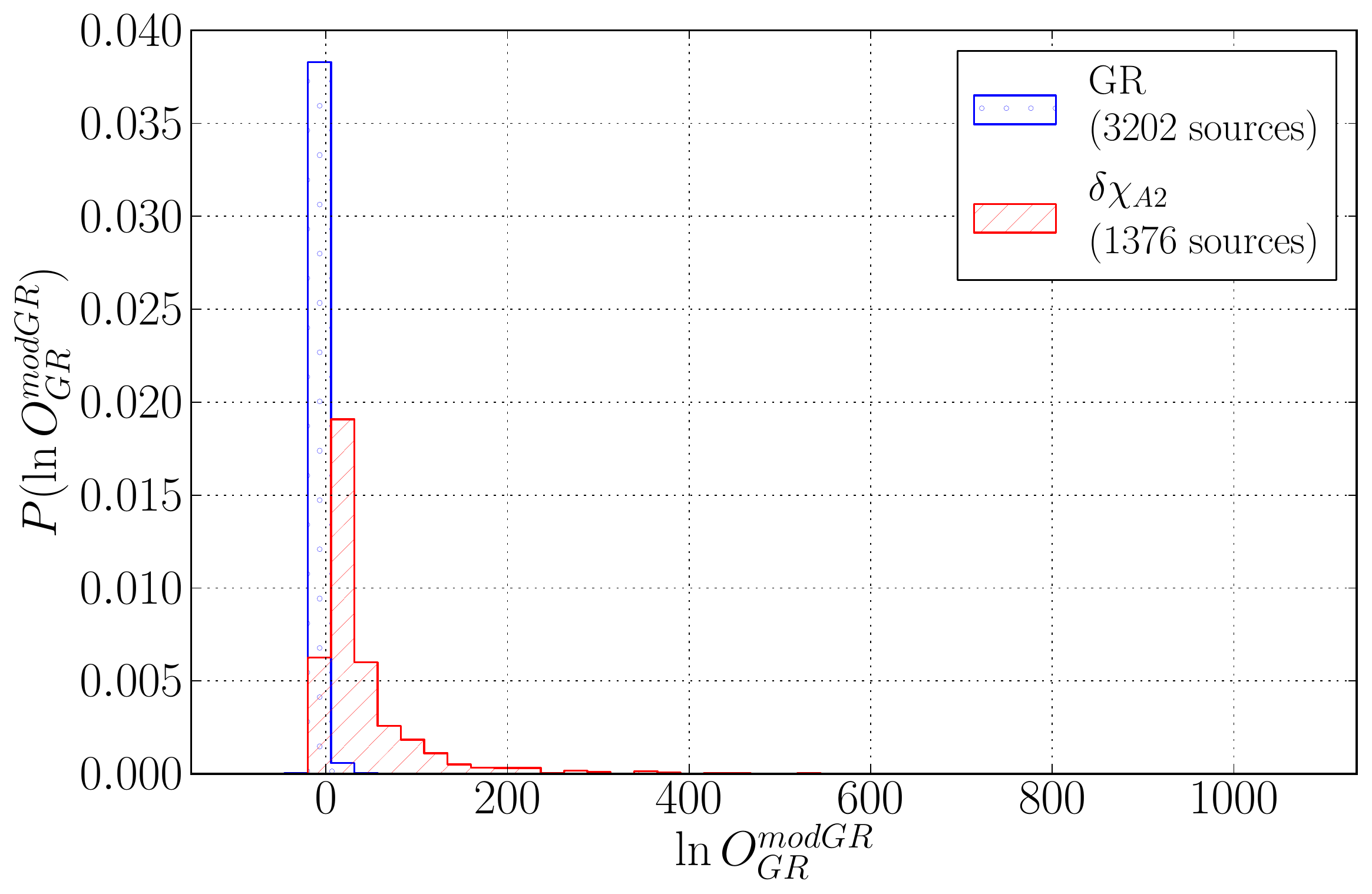}\\
	\includegraphics[width=0.7\textwidth]{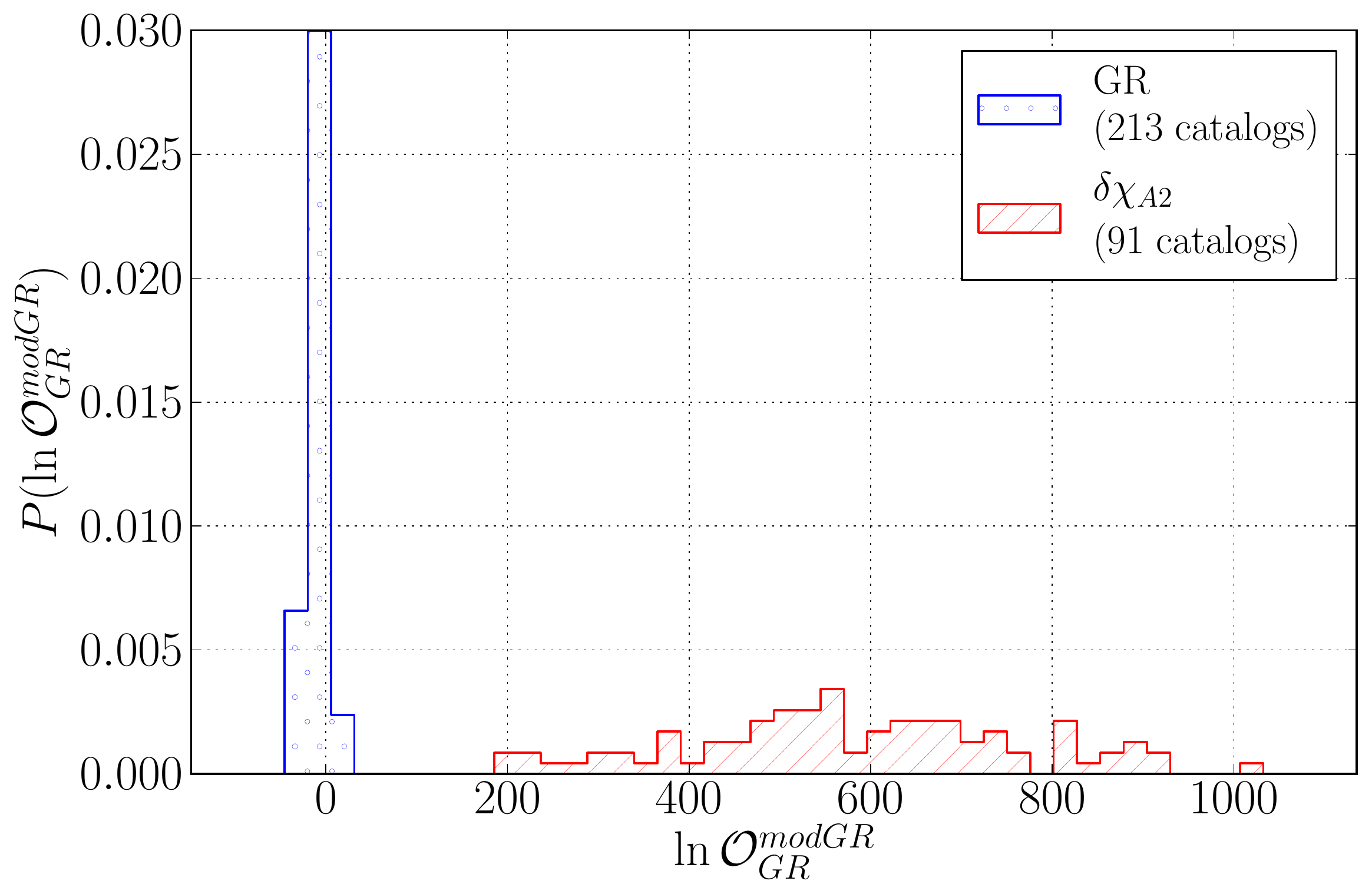}
	\caption{Background (blue) and the foreground (red) for a deviation as 
		in Eq.~(\ref{eq:dchiA2}). The top plot show the results for single sources,
		whereas the bottom plot shows the results for catalogs of 15 sources.
	}
	\label{fig:dchiA2}
\end{figure}

Despite having adopted a parametrized phase model for the PN phase
$\Psi(f)$, TIGER is also demonstrably sensitive to modifications of
gravity outside its model space.  In the present example, these would
be GR violations that do not take the form of simple shifts in the
phase coefficients, as in Eq.~(\ref{phasecoeffshifts}). As an extreme
example, we can consider a deviation of the form
\begin{equation}
	\Psi^{\rm GR}(M,\eta; f) \rightarrow \Psi^{\rm GR}(M,\eta;f) + \frac{3}{128 \eta} (\pi M f)^{-2+M/(3M_\odot)},
	\label{eq:dchiA2}
\end{equation}
where $\Psi^{\rm GR}(M,\eta; f)$ is the GR phase in the
frequency domain.
For a system with $m_1 = m_2 = 1.5\,M_\odot$, the change in phase at
$f = 150$ Hz is about the same as for a 10\% shift in $\psi_3$.
Figure~\ref{fig:dchiA2} shows the distribution of odds ratio for both
single sources and catalogs of 15 sources.
The trend is similar to that seen in Figure~\ref{fig:dchi3_10pc}: there
is a moderate separation between background and foreground for single
sources, and a complete separation in the case of 15 sources per
catalog.

Other specific kinds of deviations were considered in~\cite{Li:2011cg,Li:2011vx}.  So far, all results have indicated that
TIGER should be able to discern very generic deviations from
GR. Moreover, they show that the results are strengthened by the
combination of multiple sources, even if the majority of the sources
are near the detection threshold. An open problem is how to identify
the exact underlying nature of a GR violation, should one be found:
see the discussions in~\cite{Yunes:2009ke,2011PhRvD..83h2002D,2014PhRvD..89b2002V,2014PhRvD..89f4037S}.

\paragraph{Assessing theoretical, astrophysical, and instrumental systematics.}

As described below, in general both theoretical and astrophysical
systematics are a common concern for any parameter estimation
pipeline.  On the one hand, the PN approximation has limited ability
to accurately evaluate the waveform for any set of parameters.  As
described in Section~\ref{sec:sub:stellarmass}, these PN uncertainties
have little impact on TIGER's conclusions.  On the other hand, TIGER
has adopted a simplified (nonprecessing, point-particle) model for
binary NSs.  To address these astrophysical systematics, tests have
been conducted~\cite{2014PhRvD..89h2001A} to gauge the effects of
ignoring the misalignment of the spins of the component objects and
ignoring the tidal deformability of NSs.
The impact on the background distribution of all of these effects was
found to be negligible compared to the deviations from GR considered
in the tests (cf.~Figures~\ref{fig:dchi3_10pc}
and~\ref{fig:dchiA2}). This does not guarantee that the background
distribution in the advanced detector era will be robust against these
effects, but it does give us some confidence that sensitive tests of
GR can be performed.

Another spurious effect that could impact a test of GR is the
calibration uncertainty of the detector. The strain reported for data
analysis is conditioned on calibration measurements. However, these
calibration measurements are subject to errors, which propagate to
uncertainties in the strain. To probe the influence of this effect on
the background distribution, calibration measurement uncertainties can
be modeled using existing LIGO/Virgo
data~\cite{2012PhRvD..85f4034V,Vitale:2014owa}. Background distributions
are then calculated with and without the introduction of these
calibration errors.
\begin{figure}[htp!]
	\centering
	\includegraphics[width=0.7\textwidth]{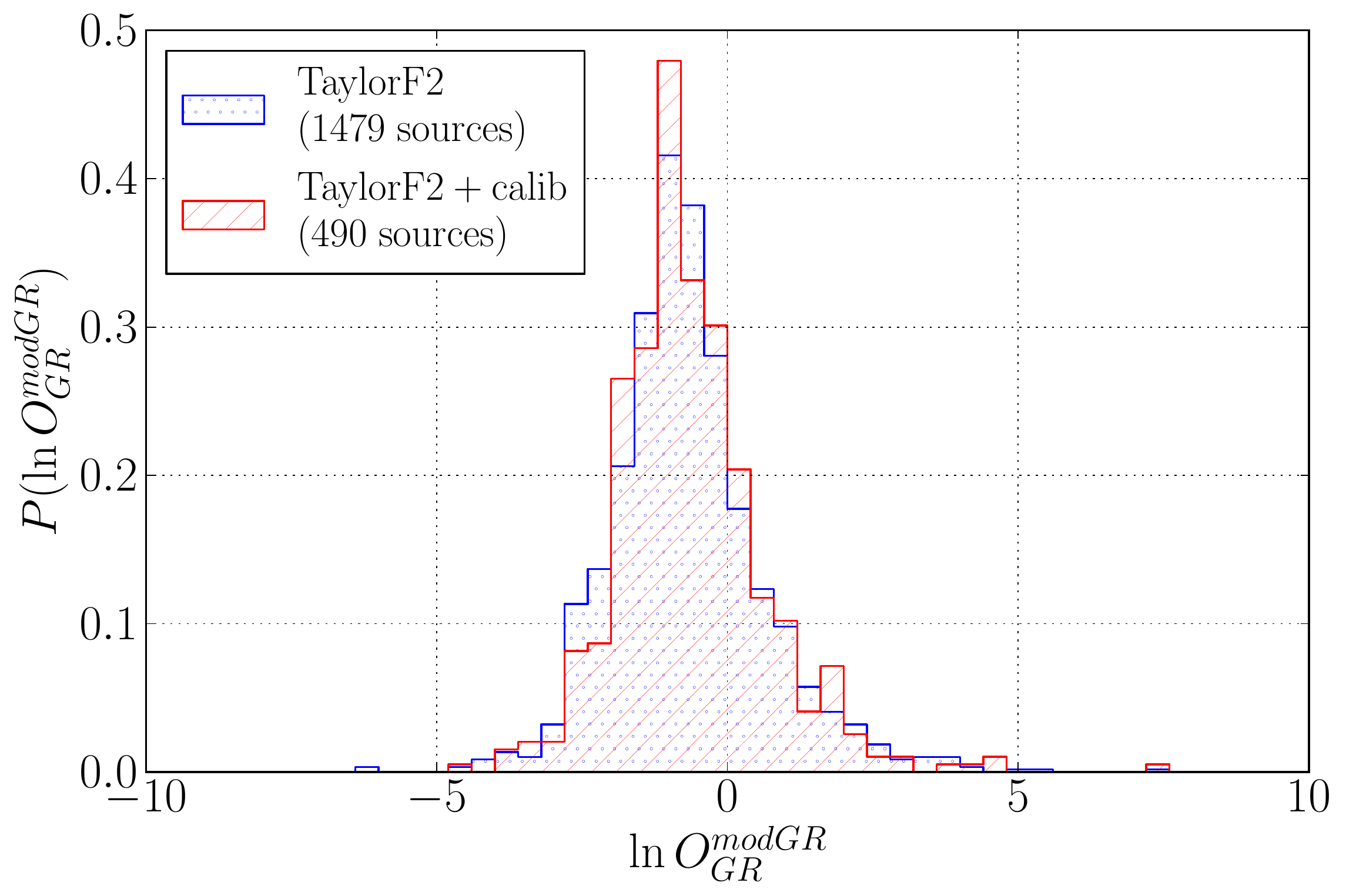}
	\caption{Single-source background distributions where the
          simulated detector outputs are generated with (red) and
          without (blue) detector calibration uncertainties.}
	\label{fig:calibrationError}
\end{figure}
The results for the single-source background distribution are shown in
Figure~\ref{fig:calibrationError}: the effects of detector calibration
uncertainty are limited.  Although it is impossible to predict the
behavior of the calibration system for Advanced LIGO/Virgo, there is
no indication that calibration uncertainty will have a significant
effect on the background distribution.

The robustness tests described above and the additional ones
in~\cite{2014PhRvD..89h2001A} used simulated stationary, Gaussian
noise. By contrast, real data will contain departures from
stationarity through short-duration ``glitches'' caused by various
instrumental and environmental noise sources. Preliminary studies of
the effects of glitches on the background
indicate that, with the implementation of appropriate instrumental
``vetoes'' (such as those used in the detection effort itself),
glitches will not pose any real problems.

\subsection{Waveform and astrophysical systematics}
\label{sec:systematics}

Estimates of the accuracy of GW detectors in measuring binary
parameters, including hypothetical effects from extensions of GR,
often assume that the sources are isolated point particles in
quasicircular orbits in vacuum, inspiraling and radiating exclusively
due to gravitational radiation reaction.  Both for stellar-mass
objects and supermassive BHs, these assumptions neglect astrophysical
realities, ignoring the possibility of nonzero eccentricity and spin
precession, the impact of the surrounding environment, and the
composition of the object in the case of NSs.

Additionally, these calculations assume that GR waveforms are
perfectly known. In practice, particularly for a wide range of
astrophysically plausible sources, the solution to the binary problem
in GR is known only approximately.

In this section we will first address the impact of astrophysical and
waveform systematics on stellar-mass BHs and NSs, and then the
corresponding challenges for supermassive BHs.

\subsubsection{Stellar mass objects}
\label{sec:sub:stellarmass}

For stellar-mass compact binaries, the assumptions listed above
neglect the delicate but nontrivial impact of composition and initial
conditions, modifying the expected signal at both high and low
frequencies.
Just as these effects mimic or mask our ability to measure parameters
(see e.g.~\cite{Favata:2013rwa}), they can also mimic or mask
modifications to GR, weakening our ability to test it with real
astrophysical systems.
Conversely, astrophysical processes like precessing spins and
eccentricity~\cite{2014ApJ...784...71S,O'Leary:2007qa,2011MNRAS.416..133D,Morscher:2012se,2014ApJ...781...45A,2014MNRAS.439.1079A}
introduce more structure in the gravitational waveforms, potentially
enabling stronger and new constraints on both astrophysics and
modified gravity.

As outlined above, existing proposals to test modified theories of
gravity with second-generation GW detectors such as Advanced LIGO and
Advanced Virgo rely on carefully monitoring the phase evolution of the
leading-order GW harmonic (see
e.g.~\cite{DelPozzo:2013ala,2014PhRvD..89f4037S}).  NS-NS binaries are
the preferred laboratory for testing GR because (a) they have been
observed in the electromagnetic spectrum, and therefore (unlike
stellar-mass BH binaries) their rates are constrained by
observations~\cite{Abadie:2010cf,Dominik:2014yma}; (b) most of the
SNR is accumulated during the inspiral, which is
both analytically tractable and less sensitive to systematic effects,
due to uncertainties in the modeling of the late inspiral and merger;
and (c) for most formation scenarios, astrophysics suggests that the
gravitational waveforms will be relatively simple (e.g., the
eccentricity of the orbit and the spin of NSs is small).
Individual events with large SNR could prove
definitive, but some techniques to test GR build statistical
confidence by ``stacking'' many individual events, searching for a
common signature~\cite{Damour:2012yf,DelPozzo:2013ala,Wade:2014vqa}.
Astrophysical effects not included in these models, if sufficiently
common, might mimic or mask the effects of modified gravity.
Conversely, the near-universal ``I-Love-Q'' relationships discussed in
Section~\ref{subsec:ILQ} suggest that, to leading order, EOS-dependent
effects can be encapsulated in a handful of
parameters~\cite{Yagi:2013bca,Yagi:2013awa,Yagi:2014bxa}, potentially
enabling direct tests of strong-field gravity even in the presence of
nontrivial and poorly constrained matter physics.

According to the above discussion, a test of strong-field gravity
would seem to require a detailed understanding of all possible
astrophysical influences, including the nuclear EOS and tidal
deformability of NSs.  In fact, as discussed in \S~\ref{subsec:ILQ},
strong relationships exist between matter-sourced ``tidal'' multipoles
of different orders, both in GR and in a broad class of
modified-gravity theories.  Specifically, the Q-Love relation may help
us to break the degeneracy between the NS spin and quadrupole moment
in GW observations.

To see this, let us consider GWs from a NS-NS binary.  The quadrupole
moment first enters at 2PN order in the waveform
phase~\cite{Poisson:1997ha,Mikoczi:2005dn}, together with the NS
spin-spin coupling
term~\cite{Kidder:1992fr,Mikoczi:2005dn,Arun:2008kb}, which leads to a
strong degeneracy between the NS quadrupole moment and spin. On the
other hand, the NS finite-size effect enters first at 5PN
order~\cite{Flanagan:2007ix} through the tidal Love number, which can
be measured with second-generation ground-based interferometers such
as Advanced
LIGO~\cite{Flanagan:2007ix,Read:2009yp,Hinderer:2009ca,Lackey:2011vz,Damour:2012yf,Lackey:2013axa,Read:2013zra,
Favata:2013rwa,Yagi:2013baa,Yagi:2013sva,Wade:2014vqa}. The Q-Love
relation allows us to express the quadrupole moment in terms of the
Love number, leading to degeneracy breaking.
This observation suggests that strong-field gravity could be coarsely
tested even \emph{without knowing the nuclear EOS}.

Since (with notable exceptions, such as scalar-tensor theories and
EiBI theory) the functional form of the I-Love-Q relations depends on
the underlying gravitational theory, if one can measure {\em any two}
of the I-Love-Q quantities independently, one can in principle perform
a model-independent consistency test of GR or test a specific
alternative theory. These tests could exploit combined observations of
binary pulsars in the electromagnetic spectrum and binary inspirals in
the GW spectrum. For example, future observations of the double binary
pulsar~\cite{Burgay:2003jj,Lyne:2004cj} may measure the NS moment of
inertia to an accuracy $\sim
10\%$~\cite{Lattimer:2004nj,Kramer:2009zza}, and the tidal Love number
may be measured to an accuracy $\sim 60\%$ with GW
observations~\cite{Yagi:2013awa}. These measurements would identifiy a
measurement point with an error box in the I-Love plane (see the left
panel of Figure~\ref{fig:I-Love-error}). If the I-Love relation is
modified from GR in a specific alternative theory, such a theory can
only be valid if the ``modified I-Love relation'' is consistent with
the error box.
A problem with this idea is that (in the example considered here) the
double binary pulsar and the hypothetical GW observations would target
different NSs, while the universal relations are valid for any single
star. However, since the parameter that is fixed for different models
satisfying the universal relation is the NS mass (or compactness), the
relations would still hold if the NSs in the two systems have the same
mass. This is not too unlikely, given that NSs have a rather peaked
mass distribution~\cite{Lattimer:2010zz,Steiner:2012xt}.
Moreover, even if the two NSs have different masses, it turns out that
the EOS-universality is preserved to good accuracy for NS models with
fixed mass ratio~\cite{Yagi:2013awa}.

\begin{figure}[tb]
\capstart
\begin{center}
\includegraphics[width=8cm]{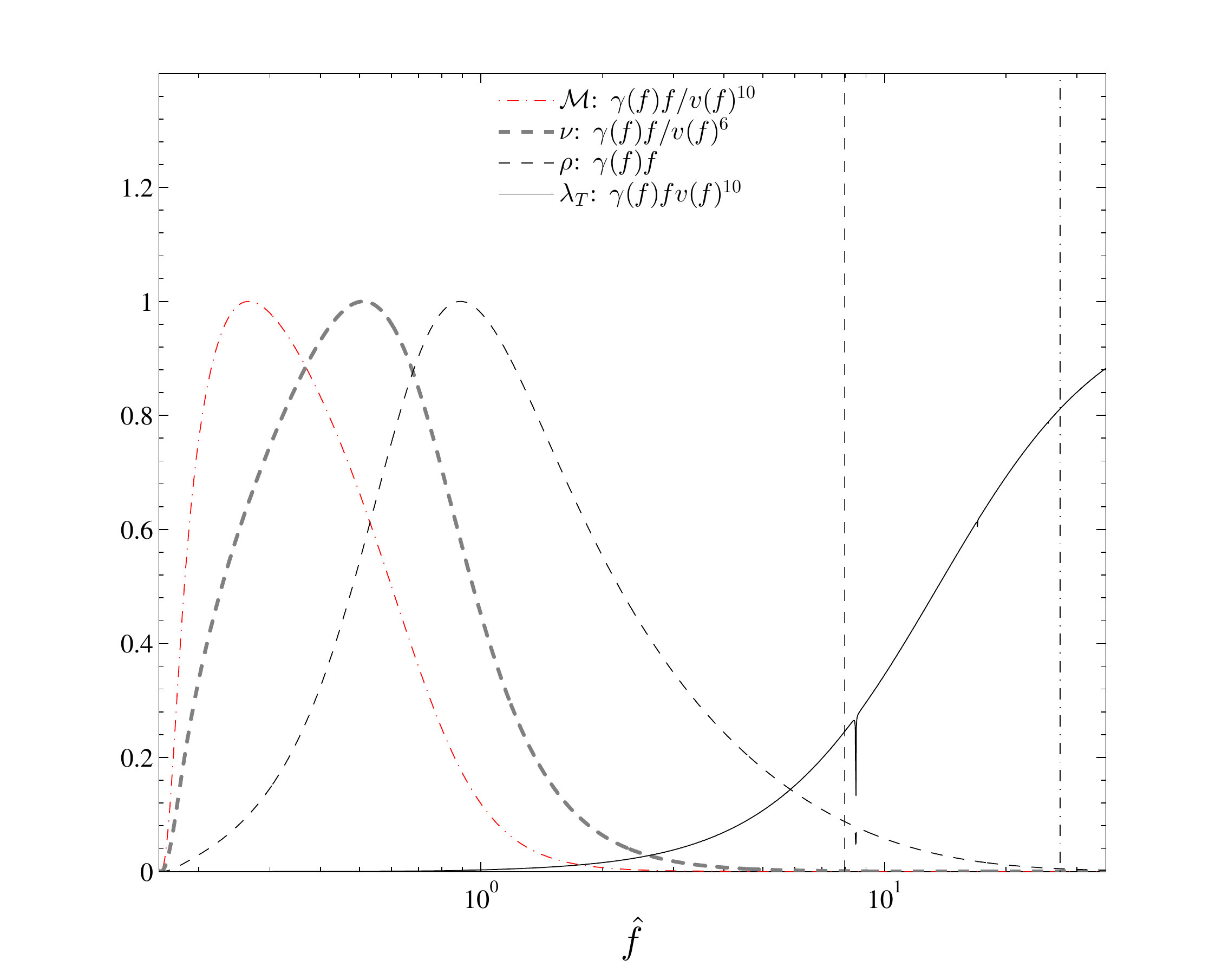}
\caption{Integrands, per frequency octave, of the integrals determining the SNR
$\rho$ and the measurement accuracy of the chirp mass ${{\cal M}}$,
the symmetric mass ratio (denoted by $\nu$ in the figure legend, and
by $\eta$ elsewhere in this review), and the tidal parameter
$\lambda_T\propto \kappa_2^T$. These integrands are plotted as a
function of the rescaled frequency $\hat{f}=f/(56.56 {\rm Hz})$ for a
typical $1.4M_\odot + 1.4M_\odot$ binary of two NSs with equal
compactness $M/R=0.1645$ ($M$ and $R$ being the mass and the
radius of each star). While most of the SNR is gathered around
frequencies $\hat{f}\sim 1$, the measurability of ${\cal M}$ and
$\eta$ is concentrated towards lower frequencies ($\hat{f}<1$), and
that of the tidal parameter $\lambda_T$ gets its largest contribution
from the late inspiral up to the merger. The rightmost vertical line
indicates the merger frequency, while the leftmost vertical line
corresponds to a GW frequency of 450~Hz. [From~\cite{Damour:2012yf}.]}
\label{fig:DamourNagarVillain}
\end{center}
\end{figure}

To exploit these opportunities, then, requires care in identifying
possible confusing astrophysical degrees of freedom, their expected
magnitude, and methods to robustly disentangle them from the
signatures of modified gravity.  A useful way to illustrate how
different astrophysical degrees of freedom come into play in different
frequency bands is shown in Figure~\ref{fig:DamourNagarVillain}.
Using a Fisher matrix approximation to parameter estimation accuracy,
this figure illustrates that different frequency bands of the GW
signal encode information about different astrophysical parameters,
namely the chirp mass, the symmetric mass ratio and the tidal (Love)
parameter.
Extending this argument to
eccentricity~\cite{Yunes:2009yz,1995PhRvD..52.2089K,2011PhRvD..84l4007G,Huerta:2014eca},
precession~\cite{2014PhRvD..89d4021L} and merger, one finds the
following hierarchy: information on small residual eccentricity and
chirp mass is encoded at low frequencies (being tied to the overall
number of cycles); information on mass ratio and spin is encoded at
intermediate frequencies; finally, information on tidal interactions
and strong-field effects comes mostly from the highest frequencies.
Similarly -- and roughly speaking -- different modifications to
gravity also predominantly occur at different frequencies, i.e., at
different PN orders (cf.~Section~\ref{sec:paramtests}).
Qualitatively, each frequency scale couples strongly to itself and
neighboring scales; for example, mass ratio and (aligned) spins are
strongly degenerate~\cite{Berti:2004bd}.
For this physical reason, we expect that confusing effects of
astrophysical phenomena are principally entangled with modifications
to gravity that dominate at the corresponding frequency interval or PN
order (along with all other strongly coupled degrees of freedom).
However, because each process has distinctive radiation content (e.g.,
higher harmonics; precession-induced modulations), astrophysical
calculations within GR suggest that these degeneracies can be broken.
This will require more precise modeling of gravitational waveforms in
various proposed extensions of GR, and further study to quantify
precisely what beyond-GR properties are accessible to experiment in
the presence of all astrophysical parameters that may contaminate the
signal.

\begin{figure}[tb]
	\centering
	\includegraphics[width=0.7\textwidth]{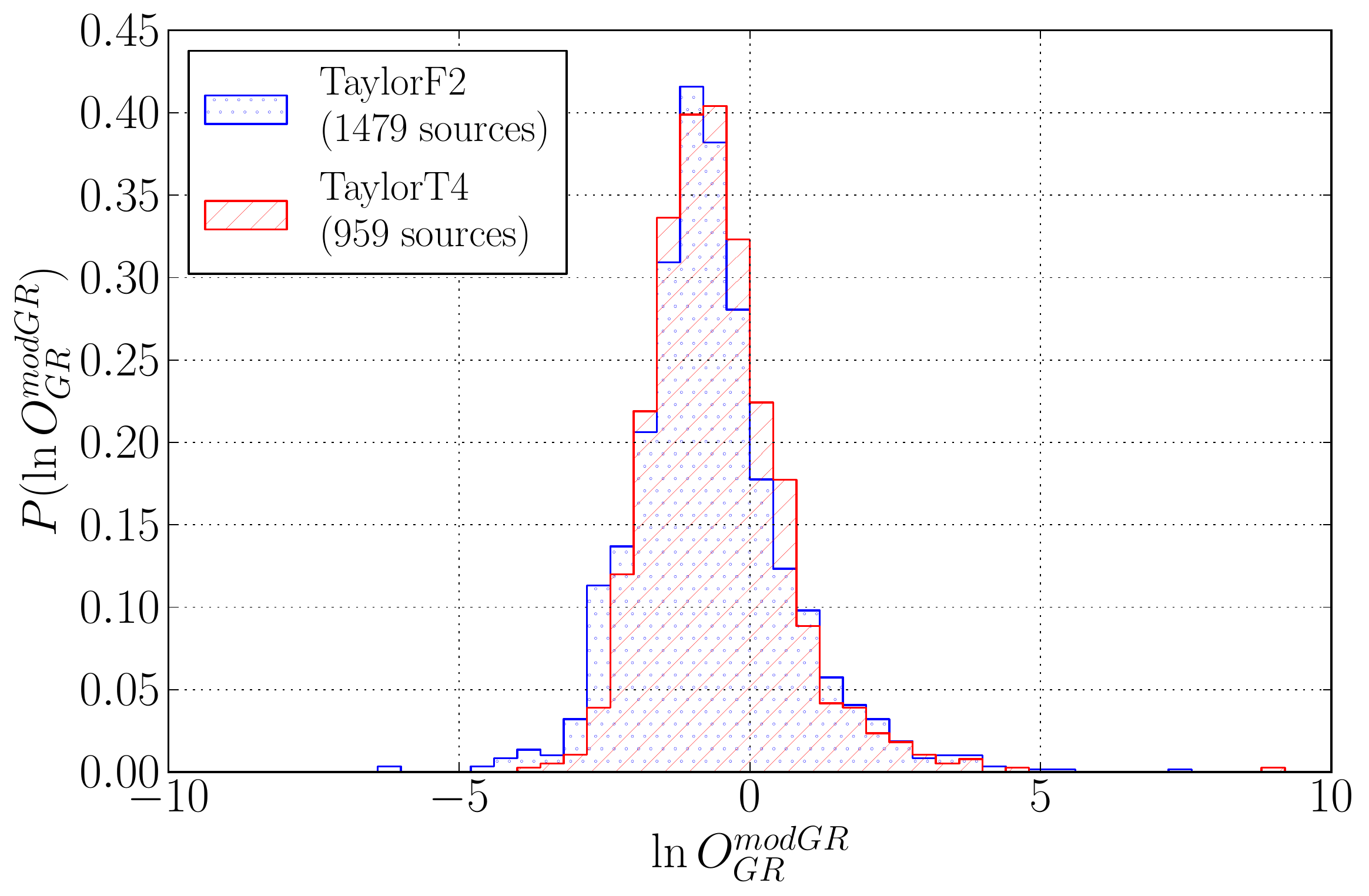}
	\caption{Single-source background distributions where the
          simulated sources are generated by the \texttt{TaylorF2}
          waveform approximant (blue) or by the \texttt{TaylorT4}
          approximant (red).}
	\label{fig:waveformMismatch}
\end{figure}

In the case of NS-NS coalescence, accurate waveforms have been
available for some time now. As shown in~\cite{Buonanno:2009zt}, in
the NS-NS mass regime the PN waveform approximants agree with each
other and with effective-one-body waveforms to a high degree of
accuracy. Nevertheless, it was checked explicitly that (at least for
NS-NS sources) the remaining small differences between approximants
will not cause one to declare a violation of GR when none is
present. Figure~\ref{fig:waveformMismatch} shows single-source
background distributions where the sources are simulated using
the \texttt{TaylorF2} waveform family (red) and the
\texttt{TaylorT4} waveform family (blue).
We see that the impact of the waveform mismatch on the background
distribution indeed appears to be minimal.

To summarize, binary NSs provide a relatively clean laboratory to
investigate modifications to strong-field GR.  Unfortunately, in these
systems, non-GR deviations in the GW signal could be degenerate with
tidal effects, eccentricity and spin.
While tidal effects are strong and EOS-dependent at late times, tidal
interactions seem to be relatively universal, potentially enabling
tests of GR without detailed knowledge of the properties of nuclear
matter.

\paragraph{Higher-mass objects: Future directions.}
Quasicircular NS-NS binaries are not the only target of Earth-based GW
interferometers. Binaries containing BHs have larger total mass than
NS-NS binaries, and their GW signal encodes more nonlinear
strong-field dynamics in the LIGO sensitivity band. Fortunately, even
though massive binaries may be intrinsically rare, GW detectors are
expected to be much more likely to observe
them~\cite{Abadie:2010cf,Dominik:2014yma}.

NS-NS coalescences are relatively simple from the observational point
of view: it is mostly the inspiral part that is visible in the
detectors' sensitive frequency band, and NSs in binaries are expected
to have small spins.  By contrast, for NS-BH and BH-BH coalescences
the full process of inspiral, merger, and ringdown will be visible.
Moreover, BH spins will likely be large and strongly precessing. Hence
NS-BH and BH-BH events are dynamically much richer than NS-NS, but
this also makes it more challenging to use them in tests of GR. One
problem has been the unavailability of accurate and faithful
(semi-)analytical waveform models, though recently there has been
great progress in that regard. For instance, Pan et al.~arrived at an
effective-one-body model with full precessing spins that appears to
have great faithfulness with waveforms obtained from large-scale
numerical simulations~\cite{2014PhRvD..89f1501P}, although it is still
rather costly to generate on a computer. Other options could include
(improvements of) the phenomenological time-domain
\texttt{PhenSpin} waveforms of Sturani et al.~\cite{Sturani:2010yv,Sturani:2010ju}, or the frequency-domain (and
hence computationally cheap) waveforms for precessing spins developed
both for inspiral \cite{2014PhRvD..89d4021L} and for
inspiral-merger-ringdown~\cite{Hannam:2013oca,Schmidt:2014iyl}.
The application of these waveforms to parameter estimation and model
selection of alternative theories of gravity is an exciting research
topic for the future.

\subsubsection{Supermassive black holes}
\label{sec:sub:supermassive}

In addition to the ``intrinsic'' corrections due to uncertainties on
the source composition and on the binary parameters, ``environmental''
effects such as accretion disks, electromagnetic fields, galactic
plasma and dark matter distributions around compact objects may also
play a role in GW detection and parameter estimation. These effects
are usually dismissed on the grounds that they are believed to be
negligible for the typical targets of GW detectors. Nonetheless,
unmodeled environmental effects on the ``vacuum'' gravitational
waveforms predicted by GR may degrade the SNR, the accuracy of
parameter estimation, and our ability to perform tests of
gravitational theories and of astrophysical models.

A careful examination of the environment's impact on GW observables is
mandatory to assess whether GW astrophysics can become a precision
discipline in the space-based detector era. Investigations in this
direction started only recently, and this section is devoted to a
brief summary of the main findings.
The upshot is that GW sources are among the ``cleanest'' astrophysical
systems: environmental effects are typically too small to affect the
detection of GW signals and the estimation of the source
parameters. The few and rather extreme cases in which environmental
effects can leave a detectable imprint should be seen as
opportunities, in the sense that (if such effects are adequately
modeled) GW observations may be used to study the behavior of matter
around compact objects, as routinely done in the electromagnetic
spectrum. These conclusions confirm the enormous potential of GW
astronomy and justify the excitement for future observations of GWs
from compact objects.

\paragraph{Environmental corrections for eLISA sources.}
\label{sec:environment}

\begin{table}[t]
\capstart
\centering
\begin{tabular}{ll|ll}
 \hline \noalign{\smallskip}\hline \noalign{\smallskip}
	\multicolumn{2}{c|}{Correction}             		  &$|\delta_{\rm per}|$	      & $|\delta_\varphi| [{\rm rads}]$ \\ \hline
\multirow{3}{*}{\rotatebox[origin=c]{90}{\parbox[t]{1cm}{\centering thin\\ disks}}}
&  planetary migration 		  &   ---		      & $10^4$ \\
 &  dyn. friction/accretion	  &   ---		      & $10^{2}$ \\
	    &  gravitational pull  	          & $10^{-8}$	      	      &	$10^{-3}$ \\ 
\hline
	    &  magnetic field        		  & $10^{-8}$	              & $10^{-4}$ \\
	    &  electric charge 		       		  & $10^{-7} $	              &	$10^{-2} $ \\
	    &  gas accretion   		  & $10^{-8}$	              &	$10^{-2}$ \\ 
	    & cosmological effects	&$10^{-31}$	              &	$10^{-26}$ \\ 
\hline
\multirow{2}{*}{\rotatebox[origin=c]{90}{\parbox[t]{.7cm}{\centering thick\\ disks}}}
&  dyn. friction/accretion		  & ---	              &	$10^{-9}$ \\  
	    &  gravitational pull		  & $10^{-16}$	      &	$10^{-11}$ \\
\hline
\multirow{3}{*}{\rotatebox[origin=c]{90}{\parbox[t]{1cm}{\centering dark\\ matter}}}
&  accretion               		  & ---		              &	$10^{-8}\rho_3^{\rm DM}$ \\
	    &  dynamical friction  		  & ---		              &	$10^{-14}\rho_3^{\rm DM}$\\
	    &  gravitational pull	          & $10^{-21}\rho_3^{\rm DM}$ &	$10^{-16}\rho_3^{\rm DM}$ \\
\noalign{\smallskip}\hline \noalign{\smallskip}\hline
\end{tabular}
\caption{Upper limit on (1) the relative correction to the periastron
  shift of an EMRI ($\delta_{\rm per}$), and (2) the absolute
  corrections to the GW phase ($\delta_\varphi$) due to a variety of
  environmental effects, over a typical eLISA mission duration of one
  year.  We consider two BHs with masses
  $(10M_{\odot},M=10^6M_{\odot}$) on a quasicircular inspiral ending
  at the innermost stable circular orbit (ISCO) $r=6 M$, whereas
  the periastron shift is computed at $r=10 M$.  Dissipative
  effects such as GW radiation reaction, dynamical friction and
  hydrodynamic drag from accretion produce negligible $\delta_{\rm
    per}$, and therefore they are not shown.  The scaling with all
  relevant parameters can be found
  in~\cite{Barausse:2014tra}. [Adapted from~\cite{Barausse:2014pra}.]
}
\label{tab:inspiral}
\end{table}
A detailed analysis of the impact of the environment on GW observables
for an eLISA-like space detector has been recently presented
in~\cite{Barausse:2014tra} and summarized
in~\cite{Barausse:2014pra}. The authors have modeled the effects of
electromagnetic fields, cosmological evolution, accretion disks and
dark matter for the inspiral, merger and ringdown of BH binaries. The
main results of this study are summarized in Tables~\ref{tab:inspiral}
and~\ref{tab:ringdown}, which show the environmental corrections to
the periastron shift and GW phase of a typical quasicircular EMRI
(Table~\ref{tab:inspiral}) and to the ringdown modes of a massive BH
(Table~\ref{tab:ringdown}).

As for the inspiral, the largest corrections come from the presence of
geometrically thin, radiatively efficient accretion disks, which have
been described using a Shakura-Sunyaev disk model with viscosity
parameter $\alpha=0.1$ and Eddington ratio $f_{\rm Edd}=1$. During the
binary's inspiral, a gaseous disk affects the orbital evolution in
three ways~\cite{Barausse:2007dy,Barausse:2007ph,Macedo:2013qea}: (i)
through its own gravitational field that modifies the trajectories of
the inspiral, (ii) through accretion of gas that changes the masses
and the spins of the compact objects, and (iii) through the
gravitational interaction of the compact objects with their own wake
in the gaseous medium, which produces dynamical friction and leads to
planetary-like migration~\cite{Yunes:2011ws,Kocsis:2011dr}.  As shown
in Table~\ref{tab:inspiral}, for thin disks such effects introduce
corrections of very different magnitude, with the gravitational pull
of the disk being negligible, whereas planetary migration, dynamical
friction and accretion can introduce large corrections, which can even
dominate over gravitational radiation reaction at separations larger
than $\sim 60$ gravitational radii~\cite{Barausse:2014pra}.
In fact, these corrections may be large enough to affect estimates of the source parameters, prevent accurate tests of GR, and possibly even affect the detectability of the signal.

However, radiatively efficient thin disks are mainly expected in AGNs at very high redshifts, while only EMRIs at $z\lesssim 1$ will be detectable by eLISA~\cite{AmaroSeoane:2012km}. In the local Universe most galactic nuclei
are quiescent rather than active, so that only a few percent of the EMRIs detected by eLISA are expected to be significantly affected by thin-disk environmental corrections~\cite{Barausse:2014pra}. On the other hand, BHs in quiescent nuclei are expected to be surrounded by thick, radiatively inefficient disks, which, as shown in Table~\ref{tab:inspiral}, introduce small relative corrections to the GW observables (of the order of $10^{-9}$ in the most conservative scenario).

Electromagnetic fields and cosmological effects are also negligible. Because of the uncertainties on the value of galactic magnetic fields $B$ and on the charge $q M$ of massive BHs, the upper limits shown in Table~\ref{tab:inspiral} were obtained using the rather extreme reference values of $q=Q/M=10^{-3}$ and $B=10^{8}\,{\rm Gauss}$, so that actual corrections are expected to be smaller in less extreme (and more realistic) situations. 

The effects of dark matter depend strongly on the assumptions made for the dark matter distribution near the galactic center and, more specifically, on the steepness of the dark matter profile near massive BHs. Dark matter ``spikes'' produced by the adiabatic growth of massive BHs~\cite{Gondolo:1999ef} can be efficiently destroyed by various mechanisms (including mergers, dark matter scattering off stars and off-center formation of the BH seeds)~\cite{Merritt:2002vj,Bertone:2005hw,Ullio:2001fb}, so that the actual dark matter profiles are believed to display a shallow slope, rather than a spike. In this case, a typical reference value for the dark matter density near BH binaries is $\rho_{\rm DM}\sim10^3M_{\odot}/{\rm pc}^{3}\sim4\times 10^{4}{\rm GeV}/{\rm cm^3}$.
Using the normalization $\rho_3^{\rm DM}=\rho_{\rm DM}/(10^3 M_\odot/{\rm pc}^3)$, Table~\ref{tab:inspiral} shows that any effect of dark matter (including accretion, dynamical friction and gravitational pull) is negligible. A possible exception are EMRIs in satellite galaxies that have never undergone mergers, so that dark matter spikes with densities as large as $\rho_{\rm DM}\sim 10^{12} M_\odot /{\rm pc}^3$ may survive in these systems~\cite{Bertone:2005xz}.

\begin{table}[t]
\capstart
\centering
\begin{tabular}{l|ll}
 \hline \noalign{\smallskip}\hline \noalign{\smallskip}

Correction           	   	& $|\delta_R|[\%]$ 	&$|\delta_I|[\%]$   	\\
\hline
spherical near-horizon distribution
	&$0.05$	             	&$0.03$             	\\
ring at ISCO	 	       	&$0.01$	             	&$0.01$	            \\
electric charge 		               	&$10^{-5} $   	&$10^{-6}$    	\\
magnetic field           	&$10^{-8}$  	&$10^{-7}$	\\
gas accretion              	&$10^{-11} $  	&$10^{-11}$     \\
dark matter halos	 	       	&$10^{-21}\rho_3^{\rm DM}$ 	&$10^{-21}\rho_3^{\rm DM}$   	\\
cosmological effects 		&$10^{-32}$ 	&$10^{-32}$   	\\
\noalign{\smallskip}\hline \noalign{\smallskip}\hline
\end{tabular}
\caption{Upper limits on the environmental corrections to the BH
  ringdown frequencies. $\delta_R$ and $\delta_I$ denote the
  deviations in the real and imaginary parts of the QNM
  frequencies due to environmental effects, relative to the case of an
  isolated BH with the same total mass.  The (rather extreme)
  reference values for the environmental effects are given in the
  text, and the scaling with the parameters for each effect can be
  found in~\cite{Barausse:2014tra}. [From~\cite{Barausse:2014pra}.]}
\label{tab:ringdown}
\end{table}
Similar considerations apply to the environmental corrections to the ringdown frequencies of a single massive BH. As discussed above, ringdown tests can be used to estimate the BH mass and spin within fractions of a percent when the object is assumed to be in isolation~\cite{Berti:2009kk}. Table~\ref{tab:ringdown} shows that environmental effects do not change this prospect, the relative corrections being at most of the order of $0.01\%$, and usually even smaller. These results were obtained for a nonspinning BH, and degeneracy with spin effects would make it even harder to detect imprints of the environment.

Overall, EMRI detection, ringdown tests and parameter estimation with eLISA should only be marginally affected by the environment. The detectability of these effects in the most optimistic scenarios would depend on the actual SNR and require a more sophisticated modeling (e.g.~including spin, eccentricity, tidal effects, etcetera) than the ones currently developed.

\paragraph{Ringdown modes versus quasinormal modes.}
The presence of matter is the prototypical example of an interesting phenomenon of ``mode camouflage'' which is not widely recognized in BH physics. 
Specifically, for matter configurations localized far from the BH or
very close to its horizon, the deviations from the isolated BH QNMs
(as defined by the poles of the relevant Green's function) are
\emph{arbitrarily large}, even for vanishingly small matter
densities. This surprising phenomenon was discovered using a
thin-shell toy model in~\cite{Leung:1999rh}, but it is actually very
generic~\cite{Barausse:2014pra,Barausse:2014tra}: it occurs for
several matter distributions, for BHs surrounded by light bosonic
fields~\cite{Dolan:2007mj,Pani:2012vp,Pani:2012bp,Witek:2012tr,Brito:2013wya},
for ultracompact horizonless objects~\cite{Cardoso:2014sna}, and it
may also have important implications for detecting GW signatures of
``firewalls'' near BH horizons~\cite{Almheiri:2013hfa} (see
also~\cite{Braunstein:2009my} for a similar earlier proposal).
This effect would have two important corollaries: (i) deviations from
the standard Kerr QNMs may be very large, and thus signatures of new
physics would be easily detectable; however (ii) any arbitrarily small
deviation from the isolated BH case could destroy the mode spectrum,
thus making it essentially impossible to use the Kerr modes as a basis
for tests of gravity.

This seems in constrast with the results of Table~\ref{tab:ringdown}
and with our previous discussion of ringdown tests. In fact, the
situation is much less dramatic. Although the QNM spectrum of various
dirty BHs and BH mimickers is totally different from the isolated BH
case, nevertheless the ringdown waveforms at early and intermediate
times are dominated by the QNMs of the pure, isolated BH geometry,
whereas the modes of the composite system get excited only at late
times and with very small
amplitudes~\cite{Barausse:2014pra,Barausse:2014tra}.  In other words,
for various deformed BH spacetimes \emph{the ringdown modes that
  dominate the waveform are not necessarily the same as the poles of
  the relevant Green's function}.  While poles of the Green's function
can be dramatically different, only the modes are of direct interest
for GW astronomy.  This ensures that current GW ringdown searches --
which {\it assume} the source is described by the pure Kerr
geometry~\cite{Abbott:2009km,Aasi:2014bqj} -- are most likely to
perform well under all circumstances.

\paragraph{Intrinsic limits to tests of gravity theories.}
The clean nature of compact binaries as GW sources, as highlighted by the small impact of environmental effects in most situations, is good news for testing strong-field gravity in the GW spectrum. In order to avoid mistaking environmental effects for deviations from GR, it is essential to understand the impact of the environment on strong-field tests of GR, the latter usually assuming that the sources are isolated and that GWs propagate in vacuum.

Indeed, any environmental effect will provide an \emph{intrinsic
  limit} to the precision of GR tests. Beyond-GR modifications that
introduce effects smaller than environmental perturbations will be
very hard to detect, unless the environment is precisely modeled.  For
instance, EMRIs in thin-disk environments are not good laboratories
for tests of GR, because in these systems astrophysical effects such
as planetary migration, dynamical friction and accretion can be even
more important than GW emission, as previously discussed.

On the other hand, the vast majority of GW sources for terrestrial and space detectors turns out to be extremely clean.
Using the estimates in Tables~\ref{tab:inspiral}
and~\ref{tab:ringdown}, Refs.~\cite{Barausse:2014pra,Barausse:2014tra}
computed the intrinsic lower bounds due to the environment on the
coupling parameters of a large class of modified theories of gravity.
It turns out that the environmental lower limits are much less
stringent than current observational constraints, and even less
stringent than projected bounds that will be placed with GW
detectors. In other words, environmental effects are too small to
affect the accuracy of GW tests of GR in the foreseeable future.

\clearpage

\section{Discussion and conclusions}

The theoretical necessity to unify GR with quantum mechanics and the
puzzling implications of cosmological measurements led to an explosion
of activity in the field of modified gravity. This major area of
research has been summarized by many outstanding reviews in the recent
past~\cite{Sotiriou:2008rp,DeFelice:2010aj,Nojiri:2010wj,Capozziello:2011et,Clifton:2011jh,Hinterbichler:2011tt,deRham:2014zqa,Joyce:2014kja}. Unlike
most of these reviews, our focus here was on the astrophysical and
phenomenological implications of modified gravity in the {\em strong-field 
regime}. Any attempt at completeness would be foolish, given
the amount of literature on strong gravity and compact objects.
Our hope is that our collective effort will be seen as a useful,
practical map for both novices and experts in strong-field gravity.

In these concluding remarks we wish to highlight some parts of this
review that should be particularly valuable as roadmaps for future
research. Table~\ref{tab:theories} gives a simple overview of key
references on modified gravity theories whose strong-field behavior
has been investigated in some depth. Tables~\ref{tab:BHsummary} and
\ref{tab:NSsummary} collect work on compact object solutions in these
theories and on their stability. Each question mark in those tables
(and there are many!) could lead to a good Ph.D. project.

At the end of each chapter we have summarized important questions that
require further investigation. We hope that some readers of this
review will pick up the gauntlet and shed light on our current
understanding of modified gravity (Section~\ref{op:theories}), the
structure and stability of BHs (Section~\ref{op:BHs}) and NSs
(Section~\ref{op:NSs}), the dynamics of compact binaries
(Section~\ref{op:binaries}), binary pulsar tests
(Section~\ref{op:pulsars}) and cosmology (Section~\ref{op:cosm}).

We have not even bothered writing down a list of open problems when it
comes to GW tests. The fact that Advanced LIGO and Advanced Virgo are
coming online in 2015, the year when we celebrate the centenary of
Einstein's milestone accomplishment, is particularly meaningful and
inspiring. We hope that when we start listening to GWs, the Universe
will amaze us and confuse us, as it has so many times in the past when
we turned our gaze in new directions; and that out of the confusion will
emerge new understanding.

\clearpage

\section*{Acknowledgments}

This review was conceived during a workshop funded by the
FP7-PEOPLE-2011-IRSES Grant No.~295189 ``NRHEP'' \cite{TestGR14}.
We thank Marco Cavagli\`a, Neil Cornish, Lu\'is Crispino and Nicol\'as
Yunes for attending the workshop and for useful discussions.
We are also grateful to T\'erence Delsate and Claudia de Rham for
comments, and to Alessandro Nagar, Thibault Damour, Loic Villain,
Michael Kramer and Fabian Schmidt for allowing us to use their
figures.
This work has been supported by the H2020-MSCA-RISE-2015 Grant
No.~690904 `StronGrHEP', the European Union's FP7 ERC Starting Grant
``The dynamics of black holes: testing the limits of Einstein's theory'' grant 
agreement no.~DyBHo--256667, H2020 ERC Consolidator Grant ``Matter and 
strong-field gravity: New frontiers in Einstein's theory'' grant agreement 
no.~MaGRaTh--646597,
the FP7-PEOPLE-2011-CIG Grant No.~293412 ``CBHEO,''
the FP7-PEOPLE-2011-CIG Grant PCIG11-GA-2012-321608 ``GALFORMBHS,''
the NSF Grants No.~PHY-1055103, PHY-1260995, PHY-1306069 and PHY-1300903,
the NASA Grant NNX13AH44G,
the ERC-2011-StG Grant No.~279363--HiDGR, the FP7/2007-2013 ERC Grant 
No.~306425 ``Challenging General Relativity,'' the DFG Research Training Group 
1620 ``Models of Gravity'' FP7-PEOPLE-2011-IRSES Grant No.~606096,
the STFC GR Consolidator Grant No.~ST/L000636/1,
the FCT-Portugal projects PTDC/FIS/116625/2010,
CERN/FP/116341/2010, CERN/FP/123593/2011, IF/00293/2013, 
IF/00797/2014/CP1214/CT0012, and CIDMA strategic funding UID/MAT/04106/2013,
the Marie Curie IEF contracts aStronGR-2011-298297 and AstroGRAphy-2013-623439,
the COST Action MP1304 ``NewCompStar,''
a UIUC Fortner Fellowship,
the S\~ao Paulo Research Foundation (FAPESP) under grants 2011/11973-4 and 2013/14754-7,
the NSF XSEDE Grant No.~PHY-090003,
the Cosmos system, part of DiRAC, funded by STFC and BIS under
Grant Nos.~ST/K00333X/1, ST/H008586/1, ST/J001341/1 and ST/J005673/1,
and
the CESGA-ICTS Grant No.~249.
Computations have been performed on
the ``Baltasar Sete-Sois'' cluster at IST,
the ``venus'' cluster at YITP,
the COSMOS supercomputer,
the Trestles cluster at SDSC,
the Kraken cluster at NICS,
and
Finis Terrae at CESGA.
T.~Baker is supported by All Souls College, Oxford.
D.~Doneva would like to thank the Alexander von Humboldt Foundation
for support.
P.~G.~Ferreira acknowledges support from STFC, BIPAC and Oxford Martin
School.
D.~Gerosa is supported by the UK STFC and the Isaac Newton Studentship
of the University of Cambridge.
J.~Kunz acknowledges support from DFG Research Training Group 1620 “Models of 
Gravity” and FP7-PEOPLE-2011-IRSES Grant No.~606096.
A.~Matas would like to thank Claudia de Rham and Andrew Tolley for many
useful conversations and support. A.~Matas is supported by an NSF-GRFP
fellowship.
B.~S.~Sathyaprakash acknowledges the support of the LIGO Visitor Program 
through the National Science Foundation award PHY-0757058 and STFC grant 
ST/J000345/1.
L.~C.~Stein acknowledges that support for this work was provided by
the NASA through Einstein Postdoctoral Fellowship Award Number
PF2-130101 issued by the Chandra X-ray Observatory Center, which is
operated by the Smithsonian Astrophysical Observatory for and on
behalf of the National Aeronautics Space Administration under contract
NAS8-03060.
This research was supported in part by Perimeter Institute for
Theoretical Physics. Research at Perimeter Institute is supported by
the Government of Canada through Industry Canada and by the Province
of Ontario through the Ministry of Economic Development \& Innovation.

\clearpage
\appendix

\clearpage

\section*{References}

\bibliographystyle{iopart-num}
\bibliography{biblio}

\providecommand{\newblock}{}
\begin{thebibliography}{100}
\expandafter\ifx\csname url\endcsname\relax
  \def\url#1{{\tt #1}}\fi
\expandafter\ifx\csname urlprefix\endcsname\relax\def\urlprefix{URL }\fi
\providecommand{\eprint}[2][]{\href{http://arxiv.org/abs/#2}{arXiv:#2}}

\bibitem{PW:2014}
Poisson E and Will C~M 2014 {\em Gravity: Newtonian, Post-Newtonian,
  Relativistic\/} (Cambridge University Press)

\bibitem{Will:2014xja}
Will C~M 2014 {\em Living Rev. Relativ.\/} {\bf 17} 4 [\eprint{1403.7377}]

\bibitem{Stelle:1976gc}
Stelle K 1977 {\em Phys. Rev.\/} {\bf D16} 953--969

\bibitem{Hawking:1969sw}
Hawking S and Penrose R 1970 {\em Proc.Roy.Soc.Lond.\/} {\bf A314} 529--548

\bibitem{Weinberg:1988cp}
Weinberg S 1989 {\em Rev. Mod. Phys.\/} {\bf 61} 1--23

\bibitem{Deser:1969wk}
Deser S 1970 {\em Gen. Relativ. Gravit.\/} {\bf 1} 9--18
  [\eprint{gr-qc/0411023}]

\bibitem{Wald:1986bj}
Wald R~M 1986 {\em Phys. Rev.\/} {\bf D33} 3613

\bibitem{Wex:2014nva}
Wex N 2014 {Testing Relativistic Gravity with Radio Pulsars} {\em Frontiers in
  Relativistic Celestial Mechanics\/} vol~2 ed Kopeikin S (De Gruyter) ISBN
  9783110345667 [\eprint{1402.5594}]

\bibitem{Baker:2014zba}
Baker T, Psaltis D and Skordis C 2015 {\em Astrophys. J.\/} {\bf 802} 63
  [\eprint{1412.3455}]

\bibitem{Psaltis:2008bb}
Psaltis D 2008 {\em Living Rev. Relativ.\/} {\bf 11} [\eprint{0806.1531}]
  \urlprefix\url{http://www.livingreviews.org/lrr-2008-9}

\bibitem{Sotiriou:2008rp}
Sotiriou T~P and Faraoni V 2010 {\em Rev. Mod. Phys.\/} {\bf 82} 451--497
  [\eprint{0805.1726}]

\bibitem{DeFelice:2010aj}
De~Felice A and Tsujikawa S 2010 {\em Living Rev. Relativ.\/} {\bf 13} 3
  [\eprint{1002.4928}]

\bibitem{Nojiri:2010wj}
Nojiri S and Odintsov S~D 2011 {\em Phys. Rept.\/} {\bf 505} 59--144
  [\eprint{1011.0544}]

\bibitem{Capozziello:2011et}
Capozziello S and De~Laurentis M 2011 {\em Phys. Rept.\/} {\bf 509} 167--321
  [\eprint{1108.6266}]

\bibitem{Clifton:2011jh}
Clifton T, Ferreira P~G, Padilla A and Skordis C 2012 {\em Phys. Rept.\/} {\bf
  513} 1--189 [\eprint{1106.2476}]

\bibitem{Hinterbichler:2011tt}
Hinterbichler K 2012 {\em Rev. Mod. Phys.\/} {\bf 84} 671--710
  [\eprint{1105.3735}]

\bibitem{deRham:2014zqa}
de~Rham C 2014 {\em Living Rev. Relativ.\/} {\bf 17} 7 [\eprint{1401.4173}]

\bibitem{Joyce:2014kja}
Joyce A, Jain B, Khoury J and Trodden M 2015 {\em Phys. Rept.\/} {\bf 568}
  1--98 [\eprint{1407.0059}]

\bibitem{Burgess:2003jk}
Burgess C 2004 {\em Living Rev. Relativ.\/} {\bf 7} 5--56
  [\eprint{gr-qc/0311082}]

\bibitem{Burgess:2007pt}
Burgess C 2007 {\em Ann. Rev. Nucl. Part. Sci.\/} {\bf 57} 329--362
  [\eprint{hep-th/0701053}]

\bibitem{Woodard:2006nt}
Woodard R~P 2007 {\em Lect. Notes Phys.\/} {\bf 720} 403--433
  [\eprint{astro-ph/0601672}]

\bibitem{Narayan:2005ie}
Narayan R 2005 {\em New J. Phys.\/} {\bf 7} 199 [\eprint{gr-qc/0506078}]

\bibitem{Narayan:2013gca}
Narayan R and McClintock J~E 2015 {Observational Evidence for Black Holes} {\em
  General Relativity and Gravitation: A Centennial Perspective\/} ed Ashtekar
  A, Berger B, Isenberg J and MacCallum M~A~H (Cambridge University Press) ISBN
  9781107037311 [\eprint{1312.6698}]

\bibitem{Abramowicz:2002vt}
Abramowicz M~A, Kluzniak W and Lasota J~P 2002 {\em Astron. Astrophys.\/} {\bf
  396} L31--L34 [\eprint{astro-ph/0207270}]

\bibitem{Johannsen:2013rqa}
Johannsen T 2013 {\em Phys. Rev.\/} {\bf D87} 124017 [\eprint{1304.7786}]

\bibitem{Damour:1993hw}
Damour T and Esposito-Far{\`e}se G 1993 {\em Phys. Rev. Lett.\/} {\bf 70}
  2220--2223

\bibitem{Yagi:2013bca}
Yagi K and Yunes N 2013 {\em Science\/} {\bf 341} 365--368 [\eprint{1302.4499}]

\bibitem{Pappas:2013naa}
Pappas G and Apostolatos T~A 2014 {\em Phys. Rev. Lett.\/} {\bf 112} 121101
  [\eprint{1311.5508}]

\bibitem{Yagi:2014bxa}
Yagi K, Kyutoku K, Pappas G, Yunes N and Apostolatos T~A 2014 {\em Phys.
  Rev.\/} {\bf D89} 124013 [\eprint{1403.6243}]

\bibitem{1992Natur.355..132T}
{Taylor} J~H, {Wolszczan} A, {Damour} T and {Weisberg} J~M 1992 {\em Nature\/}
  {\bf 355} 132--136

\bibitem{Yunes:2013dva}
Yunes N and Siemens X 2013 {\em Living Rev. Relativ.\/} {\bf 16} 9
  [\eprint{1304.3473}]

\bibitem{Gair:2012nm}
Gair J~R, Vallisneri M, Larson S~L and Baker J~G 2013 {\em Living Rev.
  Relativ.\/} {\bf 16} 7 [\eprint{1212.5575}]

\bibitem{MTW}
Misner C, Thorne K and Wheeler J 1973 {\em {Gravitation}\/} (San Francisco: W.
  H. Freeman) ISBN 9780716703440

\bibitem{Salgado:2008xh}
Salgado M, Rio D~M~d, Alcubierre M and Nunez D 2008 {\em Phys. Rev.\/} {\bf
  D77} 104010 [\eprint{0801.2372}]

\bibitem{Bertotti:2003rm}
Bertotti B, Iess L and Tortora P 2003 {\em Nature\/} {\bf 425} 374

\bibitem{Alsing:2011er}
Alsing J, Berti E, Will C~M and Zaglauer H 2012 {\em Phys. Rev.\/} {\bf D85}
  064041 [\eprint{1112.4903}]

\bibitem{Freire:2012mg}
Freire P~C, Wex N, Esposito-Farese G, Verbiest J~P, Bailes M {\em et~al.\/}
  2012 {\em Mon. Not. R. Astron. Soc.\/} {\bf 423} 3328 [\eprint{1205.1450}]

\bibitem{Choquet-Bruhat:2009xil}
Choquet-Bruhat Y 2009 {\em {General Relativity and the Einstein Equations}\/}
  (Oxford University Press)

\bibitem{Damour:1992we}
Damour T and Esposito-Far{\'e}se G 1992 {\em Class. Quantum Grav.\/} {\bf 9}
  2093--2176

\bibitem{LanahanTremblay:2007sg}
Lanahan-Tremblay N and Faraoni V 2007 {\em Class. Quantum Grav.\/} {\bf 24}
  5667--5680 [\eprint{0709.4414}]

\bibitem{Paschalidis:2011ww}
Paschalidis V, Halataei S~M, Shapiro S~L and Sawicki I 2011 {\em Class. Quantum
  Grav.\/} {\bf 28} 085006 [\eprint{1103.0984}]

\bibitem{Berry:2011pb}
Berry C~P and Gair J~R 2011 {\em Phys. Rev.\/} {\bf D83} 104022
  [\eprint{1104.0819}]

\bibitem{Yagi:2012gp}
Yagi K 2012 {\em Phys. Rev.\/} {\bf D86} 081504 [\eprint{1204.4524}]

\bibitem{Delsate:2014hba}
Delsate T, Hilditch D and Witek H 2015 {\em Phys. Rev.\/} {\bf D91} 024027
  [\eprint{1407.6727}]

\bibitem{AliHaimoud:2011fw}
Ali-Haimoud Y and Chen Y 2011 {\em Phys. Rev.\/} {\bf D84} 124033
  [\eprint{1110.5329}]

\bibitem{Foster:2005dk}
Foster B~Z and Jacobson T 2006 {\em Phys. Rev.\/} {\bf D73} 064015
  [\eprint{gr-qc/0509083}]

\bibitem{Jacobson:2008aj}
Jacobson T 2007 {\em PoS\/} {\bf QG-PH} 020 [\eprint{0801.1547}]

\bibitem{Yagi:2013qpa}
Yagi K, Blas D, Yunes N and Barausse E 2014 {\em Phys. Rev. Lett.\/} {\bf 112}
  161101 [\eprint{1307.6219}]

\bibitem{Yagi:2013ava}
Yagi K, Blas D, Barausse E and Yunes N 2014 {\em Phys. Rev.\/} {\bf D89} 084067
  [\eprint{1311.7144}]

\bibitem{Blas:2010hb}
Blas D, Pujolas O and Sibiryakov S 2011 {\em JHEP\/} {\bf 1104} 018
  [\eprint{1007.3503}]

\bibitem{Blas:2011zd}
Blas D and Sanctuary H 2011 {\em Phys. Rev.\/} {\bf D84} 064004
  [\eprint{1105.5149}]

\bibitem{Coelho:2013dya}
Coelho F~S, Herdeiro C, Hirano S and Sato Y 2014 {\em Phys. Rev.\/} {\bf D90}
  064040 [\eprint{1307.4598}]

\bibitem{deRham:2012fw}
de~Rham C, Tolley A~J and Wesley D~H 2013 {\em Phys. Rev.\/} {\bf D87} 044025
  [\eprint{1208.0580}]

\bibitem{Iorio:2012cm}
Iorio L and Saridakis E~N 2012 {\em Mon. Not. R. Astron. Soc.\/} {\bf 427} 1555
  [\eprint{1203.5781}]

\bibitem{Hawking:1972qk}
Hawking S 1972 {\em Commun. Math. Phys.\/} {\bf 25} 167--171

\bibitem{Heusler:1995qj}
Heusler M 1995 {\em Class. Quantum Grav.\/} {\bf 12} 2021--2036
  [\eprint{gr-qc/9503053}]

\bibitem{Jacobson:1999vr}
Jacobson T 1999 {\em Phys. Rev. Lett.\/} {\bf 83} 2699--2702
  [\eprint{astro-ph/9905303}]

\bibitem{HeuslerBook}
Heusler M 1996 {\em {Black Hole Uniqueness Theorems}\/} (Cambridge: Cambridge
  University Press)

\bibitem{Sotiriou:2011dz}
Sotiriou T~P and Faraoni V 2012 {\em Phys. Rev. Lett.\/} {\bf 108} 081103
  [\eprint{1109.6324}]

\bibitem{Graham:2014ina}
Graham A~A~H and Jha R 2014 {\em Phys. Rev.\/} {\bf D90} 041501
  [\eprint{1407.6573}]

\bibitem{Anabalon:2014lea}
Anabalon A, Bičák J and Saavedra J 2014 {\em Phys. Rev.\/} {\bf D90} 124055
  [\eprint{1405.7893}]

\bibitem{Damour:1976kh}
Damour T, Deruelle N and Ruffini R 1976 {\em Lett. Nuovo Cim.\/} {\bf 15}
  257--262

\bibitem{Detweiler:1980uk}
Detweiler S~L 1980 {\em Phys. Rev.\/} {\bf D22} 2323--2326

\bibitem{Zouros:1979iw}
Zouros T and Eardley D 1979 {\em Annals Phys.\/} {\bf 118} 139--155

\bibitem{Cardoso:2004nk}
Cardoso V, Dias O~J~C, Lemos J~P~S and Yoshida S 2004 {\em Phys. Rev.\/} {\bf
  D70} 044039 [\eprint{hep-th/0404096}]

\bibitem{Shlapentokh-Rothman:2013ysa}
Shlapentokh-Rothman Y 2014 {\em Commun. Math. Phys.\/} {\bf 329} 859--891
  [\eprint{1302.3448}]

\bibitem{Cardoso:2013krh}
Cardoso V 2013 {\em Gen. Relativ. Gravit.\/} {\bf 45} 2079--2097
  [\eprint{1307.0038}]

\bibitem{Herdeiro:2014goa}
Herdeiro C~A~R and Radu E 2014 {\em Phys. Rev. Lett.\/} {\bf 112} 221101
  [\eprint{1403.2757}]

\bibitem{Herdeiro:2015gia}
Herdeiro C and Radu E 2015 {\em Class. Quantum Grav.\/} {\bf 32} 144001
  [\eprint{1501.04319}]

\bibitem{Hersh:1985hz}
Hersh J and Ove R 1985 {\em Phys. Lett.\/} {\bf B156} 305

\bibitem{Nzioki:2014oaa}
Nzioki A~M, Goswami R and Dunsby P~K~S 2014  [\eprint{1408.0152}]

\bibitem{Mignemi:1992nt}
Mignemi S and Stewart N 1993 {\em Phys. Rev.\/} {\bf D47} 5259--5269
  [\eprint{hep-th/9212146}]

\bibitem{Kanti:1995vq}
Kanti P, Mavromatos N, Rizos J, Tamvakis K and Winstanley E 1996 {\em Phys.
  Rev.\/} {\bf D54} 5049--5058 [\eprint{hep-th/9511071}]

\bibitem{Yunes:2011we}
Yunes N and Stein L~C 2011 {\em Phys. Rev.\/} {\bf D83} 104002
  [\eprint{1101.2921}]

\bibitem{Pani:2009wy}
Pani P and Cardoso V 2009 {\em Phys. Rev.\/} {\bf D79} 084031
  [\eprint{0902.1569}]

\bibitem{Ayzenberg:2014aka}
Ayzenberg D and Yunes N 2014 {\em Phys. Rev.\/} {\bf D90} 044066
  [\eprint{1405.2133}]

\bibitem{Kleihaus:2011tg}
Kleihaus B, Kunz J and Radu E 2011 {\em Phys. Rev. Lett.\/} {\bf 106} 151104
  [\eprint{1101.2868}]

\bibitem{Torii:1998gm}
Torii T and Maeda K~i 1998 {\em Phys. Rev.\/} {\bf D58} 084004

\bibitem{Ayzenberg:2013wua}
Ayzenberg D, Yagi K and Yunes N 2014 {\em Phys. Rev.\/} {\bf D89} 044023
  [\eprint{1310.6392}]

\bibitem{Pani:2011gy}
Pani P, Macedo C~F, Crispino L~C and Cardoso V 2011 {\em Phys. Rev.\/} {\bf
  D84} 087501 [\eprint{1109.3996}]

\bibitem{Maselli:2014fca}
Maselli A, Gualtieri L, Pani P, Stella L and Ferrari V 2015 {\em Astrophys.
  J.\/} {\bf 801} 115 [\eprint{1412.3473}]

\bibitem{Kleihaus:2014lba}
Kleihaus B, Kunz J and Mojica S 2014 {\em Phys. Rev.\/} {\bf D90} 061501
  [\eprint{1407.6884}]

\bibitem{Yunes:2009hc}
Yunes N and Pretorius F 2009 {\em Phys. Rev.\/} {\bf D79} 084043
  [\eprint{0902.4669}]

\bibitem{Konno:2009kg}
Konno K, Matsuyama T and Tanda S 2009 {\em Prog. Theor. Phys.\/} {\bf 122}
  561--568 [\eprint{0902.4767}]

\bibitem{Yagi:2012ya}
Yagi K, Yunes N and Tanaka T 2012 {\em Phys. Rev.\/} {\bf D86} 044037
  [\eprint{1206.6130}]

\bibitem{Stein:2014xba}
Stein L~C 2014 {\em Phys. Rev.\/} {\bf D90} 044061 [\eprint{1407.2350}]

\bibitem{Cardoso:2009pk}
Cardoso V and Gualtieri L 2009 {\em Phys. Rev.\/} {\bf D80} 064008
  [\eprint{0907.5008}]

\bibitem{Molina:2010fb}
Molina C, Pani P, Cardoso V and Gualtieri L 2010 {\em Phys. Rev.\/} {\bf D81}
  124021 [\eprint{1004.4007}]

\bibitem{Garfinkle:2010zx}
Garfinkle D, Pretorius F and Yunes N 2010 {\em Phys. Rev.\/} {\bf D82} 041501
  [\eprint{1007.2429}]

\bibitem{Konno:2014qua}
Konno K and Takahashi R 2014 {\em Phys. Rev.\/} {\bf D90} 064011
  [\eprint{1406.0957}]

\bibitem{Vincent:2013uea}
Vincent F 2013 {\em Class. Quantum Grav.\/} {\bf 31} 025010
  [\eprint{1311.3251}]

\bibitem{Sotiriou:2013qea}
Sotiriou T~P and Zhou S~Y 2014 {\em Phys. Rev. Lett.\/} {\bf 112} 251102
  [\eprint{1312.3622}]

\bibitem{Sotiriou:2014pfa}
Sotiriou T~P and Zhou S~Y 2014 {\em Phys. Rev.\/} {\bf D90} 124063
  [\eprint{1408.1698}]

\bibitem{Babichev:2013cya}
Babichev E and Charmousis C 2014 {\em JHEP\/} {\bf 1408} 106
  [\eprint{1312.3204}]

\bibitem{Kobayashi:2012kh}
Kobayashi T, Motohashi H and Suyama T 2012 {\em Phys. Rev.\/} {\bf D85} 084025
  [\eprint{1202.4893}]

\bibitem{Kobayashi:2014wsa}
Kobayashi T, Motohashi H and Suyama T 2014 {\em Phys. Rev.\/} {\bf D89} 084042
  [\eprint{1402.6740}]

\bibitem{Eling:2006ec}
Eling C and Jacobson T 2006 {\em Class. Quantum Grav.\/} {\bf 23} 5643--5660
  [\eprint{gr-qc/0604088}]

\bibitem{Barausse:2011pu}
Barausse E, Jacobson T and Sotiriou T~P 2011 {\em Phys. Rev.\/} {\bf D83}
  124043 [\eprint{1104.2889}]

\bibitem{Barausse:2013nwa}
Barausse E and Sotiriou T~P 2013 {\em Class. Quantum Grav.\/} {\bf 30} 244010
  [\eprint{1307.3359}]

\bibitem{Wang:2012nv}
Wang A 2013 {\em Phys. Rev. Lett.\/} {\bf 110} 091101 [\eprint{1212.1876}]

\bibitem{Barausse:2012qh}
Barausse E and Sotiriou T~P 2013 {\em Phys. Rev.\/} {\bf D87} 087504
  [\eprint{1212.1334}]

\bibitem{Blas:2011ni}
Blas D and Sibiryakov S 2011 {\em Phys. Rev.\/} {\bf D84} 124043
  [\eprint{1110.2195}]

\bibitem{Herdeiro:2011im}
Herdeiro C, Hirano S and Sato Y 2011 {\em Phys. Rev.\/} {\bf D84} 124048
  [\eprint{1110.0832}]

\bibitem{Coelho:2013zq}
Coelho F~S, Herdeiro C and Wang M 2013 {\em Phys. Rev.\/} {\bf D87} 047502
  [\eprint{1301.1070}]

\bibitem{Brito:2013xaa}
Brito R, Cardoso V and Pani P 2013 {\em Phys. Rev.\/} {\bf D88} 064006
  [\eprint{1309.0818}]

\bibitem{Babichev:2014fka}
Babichev E and Fabbri A 2014 {\em JHEP\/} {\bf 1407} 016 [\eprint{1405.0581}]

\bibitem{Babichev:2014tfa}
Babichev E and Fabbri A 2014 {\em Phys. Rev.\/} {\bf D90} 084019
  [\eprint{1406.6096}]

\bibitem{Volkov:2014ooa}
Volkov M~S 2015 {\em Lect. Notes Phys.\/} {\bf 892} 161--180
  [\eprint{1405.1742}]

\bibitem{Babichev:2013una}
Babichev E and Fabbri A 2013 {\em Class. Quantum Grav.\/} {\bf 30} 152001
  [\eprint{1304.5992}]

\bibitem{Brito:2013wya}
Brito R, Cardoso V and Pani P 2013 {\em Phys. Rev.\/} {\bf D88} 023514
  [\eprint{1304.6725}]

\bibitem{Brito:2013yxa}
Brito R, Cardoso V and Pani P 2013 {\em Phys. Rev.\/} {\bf D87} 124024
  [\eprint{1306.0908}]

\bibitem{Babichev:2014oua}
Babichev E and Fabbri A 2014 {\em Phys. Rev.\/} {\bf D89} 081502
  [\eprint{1401.6871}]

\bibitem{Hui:2012qt}
Hui L and Nicolis A 2013 {\em Phys. Rev. Lett.\/} {\bf 110} 241104
  [\eprint{1202.1296}]

\bibitem{Salmona:1967zz}
Salmona A 1967 {\em Phys. Rev.\/} {\bf 154} 1218--1223

\bibitem{1974GReGr...5..663H}
{Hillebrandt} W and {Heintzmann} H 1974 {\em Gen. Relativ. Gravit.\/} {\bf 5}
  663--672

\bibitem{Damour:1996ke}
Damour T and Esposito-Farese G 1996 {\em Phys. Rev.\/} {\bf D54} 1474--1491
  [\eprint{gr-qc/9602056}]

\bibitem{Tsuchida:1998jw}
Tsuchida T, Kawamura G and Watanabe K 1998 {\em Prog. Theor. Phys.\/} {\bf 100}
  291--313 [\eprint{gr-qc/9802049}]

\bibitem{Salgado:1998sg}
Salgado M, Sudarsky D and Nucamendi U 1998 {\em Phys. Rev.\/} {\bf D58} 124003
  [\eprint{gr-qc/9806070}]

\bibitem{Sotani:2012eb}
Sotani H 2012 {\em Phys. Rev.\/} {\bf D86} 124036 [\eprint{1211.6986}]

\bibitem{Pani:2014jra}
Pani P and Berti E 2014 {\em Phys. Rev.\/} {\bf D90} 024025
  [\eprint{1405.4547}]

\bibitem{Doneva:2013qva}
Doneva D~D, Yazadjiev S~S, Stergioulas N and Kokkotas K~D 2013 {\em Phys.
  Rev.\/} {\bf D88} 084060 [\eprint{1309.0605}]

\bibitem{Doneva:2014uma}
Doneva D~D, Yazadjiev S~S, Stergioulas N, Kokkotas K~D and Athanasiadis T~M
  2014 {\em Phys. Rev.\/} {\bf D90} 044004 [\eprint{1405.6976}]

\bibitem{Doneva:2014faa}
Doneva D~D, Yazadjiev S~S, Staykov K~V and Kokkotas K~D 2014 {\em Phys. Rev.\/}
  {\bf D90} 104021 [\eprint{1408.1641}]

\bibitem{Scheel:1994yr}
Scheel M~A, Shapiro S~L and Teukolsky S~A 1995 {\em Phys. Rev.\/} {\bf D51}
  4208--4235

\bibitem{Scheel:1994yn}
Scheel M~A, Shapiro S~L and Teukolsky S~A 1995 {\em Phys. Rev.\/} {\bf D51}
  4236--4249 [\eprint{gr-qc/9411026}]

\bibitem{Shibata:1994qd}
Shibata M, Nakao K and Nakamura T 1994 {\em Phys. Rev.\/} {\bf D50} 7304--7317

\bibitem{Harada:1996wt}
Harada T, Chiba T, Nakao K~i and Nakamura T 1997 {\em Phys. Rev.\/} {\bf D55}
  2024--2037 [\eprint{gr-qc/9611031}]

\bibitem{Novak:1997hw}
Novak J 1998 {\em Phys. Rev.\/} {\bf D57} 4789--4801 [\eprint{gr-qc/9707041}]

\bibitem{Novak:1998rk}
Novak J 1998 {\em Phys. Rev.\/} {\bf D58} 064019 [\eprint{gr-qc/9806022}]

\bibitem{Kerimo:1998qu}
Kerimo J and Kalligas D 1998 {\em Phys. Rev.\/} {\bf D58} 104002

\bibitem{Novak:1999jg}
Novak J and Ibanez J~M 2000 {\em Astrophys. J.\/} {\bf 533} 392--405
  [\eprint{astro-ph/9911298}]

\bibitem{Zaglauer:1992bp}
Zaglauer H 1992 {\em Astrophys. J.\/} {\bf 393} 685--696

\bibitem{1969ApJ...155..999N}
{Nutku} Y 1969 {\em \apj\/} {\bf 155} 999

\bibitem{Harada:1997mr}
Harada T 1997 {\em Prog. Theor. Phys.\/} {\bf 98} 359--379
  [\eprint{gr-qc/9706014}]

\bibitem{Harada:1998}
Harada T 1998 {\em Phys. Rev.\/} {\bf D57} 4802 [\eprint{gr-qc/9801049}]

\bibitem{Sotani:2004rq}
Sotani H and Kokkotas K~D 2004 {\em Phys. Rev.\/} {\bf D70} 084026
  [\eprint{gr-qc/0409066}]

\bibitem{Sotani:2005qx}
Sotani H and Kokkotas K~D 2005 {\em Phys. Rev.\/} {\bf D71} 124038
  [\eprint{gr-qc/0506060}]

\bibitem{Lima:2010na}
Lima W~C, Matsas G~E and Vanzella D~A 2010 {\em Phys. Rev. Lett.\/} {\bf 105}
  151102 [\eprint{1009.1771}]

\bibitem{Pani:2010vc}
Pani P, Cardoso V, Berti E, Read J and Salgado M 2011 {\em Phys. Rev.\/} {\bf
  D83} 081501 [\eprint{1012.1343}]

\bibitem{Mendes:2013ija}
Mendes R~F, Matsas G~E and Vanzella D~A 2014 {\em Phys. Rev.\/} {\bf D89}
  047503 [\eprint{1310.2185}]

\bibitem{Landulfo:2014wra}
Landulfo A~G~S, Lima W~C~C, Matsas G~E~A and Vanzella D~A~T 2015 {\em Phys.
  Rev.\/} {\bf D91} 024011 [\eprint{1410.2274}]

\bibitem{Sotani:2014tua}
Sotani H 2014 {\em Phys. Rev.\/} {\bf D89} 064031 [\eprint{1402.5699}]

\bibitem{Silva:2014ora}
Silva H~O, Sotani H, Berti E and Horbatsch M 2014 {\em Phys. Rev.\/} {\bf D90}
  124044 [\eprint{1410.2511}]

\bibitem{DeDeo:2004kk}
DeDeo S and Psaltis D 2004  [\eprint{astro-ph/0405067}]

\bibitem{Kobayashi:2008tq}
Kobayashi T and Maeda K 2008 {\em Phys. Rev.\/} {\bf D78} 064019
  [\eprint{0807.2503}]

\bibitem{Cooney:2009rr}
Cooney A, DeDeo S and Psaltis D 2010 {\em Phys. Rev.\/} {\bf D82} 064033
  [\eprint{0910.5480}]

\bibitem{Upadhye:2009kt}
Upadhye A and Hu W 2009 {\em Phys. Rev.\/} {\bf D80} 064002
  [\eprint{0905.4055}]

\bibitem{Babichev:2009td}
Babichev E and Langlois D 2009 {\em Phys. Rev.\/} {\bf D80} 121501
  [\eprint{0904.1382}]

\bibitem{Babichev:2009fi}
Babichev E and Langlois D 2010 {\em Phys. Rev.\/} {\bf D81} 124051
  [\eprint{0911.1297}]

\bibitem{Miranda:2009rs}
Miranda V, Joras S~E, Waga I and Quartin M 2009 {\em Phys. Rev. Lett.\/} {\bf
  102} 221101 [\eprint{0905.1941}]

\bibitem{Jaime:2010kn}
Jaime L~G, Patino L and Salgado M 2011 {\em Phys. Rev.\/} {\bf D83} 024039
  [\eprint{1006.5747}]

\bibitem{Astashenok:2013vza}
Astashenok A~V, Capozziello S and Odintsov S~D 2013 {\em JCAP\/} {\bf 1312} 040
  [\eprint{1309.1978}]

\bibitem{Astashenok:2014pua}
Astashenok A~V, Capozziello S and Odintsov S~D 2014 {\em Phys. Rev.\/} {\bf
  D89} 103509 [\eprint{1401.4546}]

\bibitem{Astashenok:2014gda}
Astashenok A~V, Capozziello S and Odintsov S~D 2015 {\em Astrophys. Space
  Sci.\/} {\bf 355} 333--341 [\eprint{1405.6663}]

\bibitem{Yazadjiev:2014cza}
Yazadjiev S~S, Doneva D~D, Kokkotas K~D and Staykov K~V 2014 {\em JCAP\/} {\bf
  1406} 003 [\eprint{1402.4469}]

\bibitem{Arapoglu:2010rz}
Arapoglu A~S, Deliduman C and Eksi K~Y 2011 {\em JCAP\/} {\bf 1107} 020
  [\eprint{1003.3179}]

\bibitem{Alavirad:2013paa}
Alavirad H and Weller J~M 2013 {\em Phys. Rev.\/} {\bf D88} 124034
  [\eprint{1307.7977}]

\bibitem{Staykov:2014mwa}
Staykov K~V, Doneva D~D, Yazadjiev S~S and Kokkotas K~D 2014 {\em JCAP\/} {\bf
  1410} 006 [\eprint{1407.2180}]

\bibitem{Yazadjiev:2015zia}
Yazadjiev S~S, Doneva D~D and Kokkotas K~D 2015 {\em Phys. Rev.\/} {\bf D91}
  084018 [\eprint{1501.04591}]

\bibitem{Cembranos:2012fd}
Cembranos J, de~la Cruz-Dombriz A and Montes~Nunez B 2012 {\em JCAP\/} {\bf
  1204} 021 [\eprint{1201.1289}]

\bibitem{Borisov:2011fu}
Borisov A, Jain B and Zhang P 2012 {\em Phys. Rev.\/} {\bf D85} 063518
  [\eprint{1102.4839}]

\bibitem{Seifert:2007fr}
Seifert M~D 2007 {\em Phys. Rev.\/} {\bf D76} 064002 [\eprint{gr-qc/0703060}]

\bibitem{Kainulainen:2008pr}
Kainulainen K and Sunhede D 2008 {\em Phys. Rev.\/} {\bf D78} 063511
  [\eprint{0803.0867}]

\bibitem{Pani:2011xm}
Pani P, Berti E, Cardoso V and Read J 2011 {\em Phys. Rev.\/} {\bf D84} 104035
  [\eprint{1109.0928}]

\bibitem{Yunes:2009ch}
Yunes N, Psaltis D, Ozel F and Loeb A 2010 {\em Phys. Rev.\/} {\bf D81} 064020
  [\eprint{0912.2736}]

\bibitem{Yagi:2013mbt}
Yagi K, Stein L~C, Yunes N and Tanaka T 2013 {\em Phys. Rev.\/} {\bf D87}
  084058 [\eprint{1302.1918}]

\bibitem{Yagi:2013awa}
Yagi K and Yunes N 2013 {\em Phys. Rev.\/} {\bf D88} 023009
  [\eprint{1303.1528}]

\bibitem{Eling:2006df}
Eling C and Jacobson T 2006 {\em Class. Quantum Grav.\/} {\bf 23} 5625--5642
  [\eprint{gr-qc/0603058}]

\bibitem{Eling:2007xh}
Eling C, Jacobson T and Miller M~C 2007 {\em Phys. Rev.\/} {\bf D76} 042003
  [\eprint{0705.1565}]

\bibitem{Garfinkle:2007bk}
Garfinkle D, Eling C and Jacobson T 2007 {\em Phys. Rev.\/} {\bf D76} 024003
  [\eprint{gr-qc/0703093}]

\bibitem{Greenwald:2009kp}
Greenwald J, Papazoglou A and Wang A 2010 {\em Phys. Rev.\/} {\bf D81} 084046
  [\eprint{0912.0011}]

\bibitem{Babichev:2010jd}
Babichev E, Deffayet C and Ziour R 2010 {\em Phys. Rev.\/} {\bf D82} 104008
  [\eprint{1007.4506}]

\bibitem{Gruzinov:2011mm}
Gruzinov A and Mirbabayi M 2011 {\em Phys. Rev.\/} {\bf D84} 124019
  [\eprint{1106.2551}]

\bibitem{Chagoya:2014fza}
Chagoya J, Koyama K, Niz G and Tasinato G 2014 {\em JCAP\/} {\bf 1410} 055
  [\eprint{1407.7744}]

\bibitem{Bellini:2012qn}
Bellini E, Bartolo N and Matarrese S 2012 {\em JCAP\/} {\bf 1206} 019
  [\eprint{1202.2712}]

\bibitem{Barreira:2013xea}
Barreira A, Li B, Baugh C~M and Pascoli S 2013 {\em JCAP\/} {\bf 1311} 056
  [\eprint{1308.3699}]

\bibitem{Kainulainen:2006wz}
Kainulainen K, Reijonen V and Sunhede D 2007 {\em Phys. Rev.\/} {\bf D76}
  043503 [\eprint{gr-qc/0611132}]

\bibitem{Kainulainen:2007bt}
Kainulainen K, Piilonen J, Reijonen V and Sunhede D 2007 {\em Phys. Rev.\/}
  {\bf D76} 024020 [\eprint{0704.2729}]

\bibitem{Barausse:2007pn}
Barausse E, Sotiriou T~P and Miller J~C 2008 {\em Class. Quantum Grav.\/} {\bf
  25} 062001 [\eprint{gr-qc/0703132}]

\bibitem{Barausse:2007ys}
Barausse E, Sotiriou T and Miller J 2008 {\em Class. Quantum Grav.\/} {\bf 25}
  105008 [\eprint{0712.1141}]

\bibitem{Olmo:2008pv}
Olmo G~J 2008 {\em Phys. Rev.\/} {\bf D78} 104026 [\eprint{0810.3593}]

\bibitem{Pani:2011mg}
Pani P, Cardoso V and Delsate T 2011 {\em Phys. Rev. Lett.\/} {\bf 107} 031101
  [\eprint{1106.3569}]

\bibitem{Pani:2012qb}
Pani P, Delsate T and Cardoso V 2012 {\em Phys. Rev.\/} {\bf D85} 084020
  [\eprint{1201.2814}]

\bibitem{Pani:2012qd}
Pani P and Sotiriou T~P 2012 {\em Phys. Rev. Lett.\/} {\bf 109} 251102
  [\eprint{1209.2972}]

\bibitem{Sham:2013sya}
Sham Y, Leung P and Lin L 2013 {\em Phys. Rev.\/} {\bf D87} 061503
  [\eprint{1304.0550}]

\bibitem{Harko:2013wka}
Harko T, Lobo F~S~N, Mak M and Sushkov S~V 2013 {\em Phys. Rev.\/} {\bf D88}
  044032 [\eprint{1305.6770}]

\bibitem{Sotani:2014goa}
Sotani H 2014 {\em Phys. Rev.\/} {\bf D89} 104005 [\eprint{1404.5369}]

\bibitem{Casanellas:2011kf}
Casanellas J, Pani P, Lopes I and Cardoso V 2012 {\em Astrophys. J.\/} {\bf
  745} 15 [\eprint{1109.0249}]

\bibitem{Sham:2012qi}
Sham Y~H, Lin L~M and Leung P 2012 {\em Phys. Rev.\/} {\bf D86} 064015
  [\eprint{1208.1314}]

\bibitem{Sotani:2014xoa}
Sotani H 2014 {\em Phys. Rev.\/} {\bf D89} 124037 [\eprint{1406.3097}]

\bibitem{Lovelock:1971yv}
Lovelock D 1971 {\em J. Math. Phys.\/} {\bf 12} 498--501

\bibitem{Lovelock:1972vz}
Lovelock D 1972 {\em J. Math. Phys.\/} {\bf 13} 874--876

\bibitem{Cartan:1922}
Cartan E 1922 {\em Journal de math{\'e}matiques pures et appliqu{\'e}es\/} {\bf
  1} 141--204 \urlprefix\url{http://eudml.org/doc/235383}

\bibitem{Sotiriou:2007zu}
Sotiriou T~P, Liberati S and Faraoni V 2008 {\em Int. J. Mod. Phys.\/} {\bf
  D17} 399--423 [\eprint{0707.2748}]

\bibitem{DiCasola:2013yga}
Di~Casola E, Liberati S and Sonego S 2014 {\em Phys. Rev.\/} {\bf D89} 084053
  [\eprint{1401.0030}]

\bibitem{Sotiriou:2015lxa}
Sotiriou T~P 2015 {\em Lect. Notes Phys.\/} {\bf 892} 3--24
  [\eprint{1404.2955}]

\bibitem{Pani:2013qfa}
Pani P, Sotiriou T~P and Vernieri D 2013 {\em Phys. Rev.\/} {\bf D88} 121502
  [\eprint{1306.1835}]

\bibitem{Banados:2010ix}
Banados M and Ferreira P~G 2010 {\em Phys. Rev. Lett.\/} {\bf 105} 011101
  [\eprint{1006.1769}]

\bibitem{Olmo:2011uz}
Olmo G~J 2011 {\em Int. J. Mod. Phys.\/} {\bf D20} 413--462
  [\eprint{1101.3864}]

\bibitem{Horava:2009uw}
Horava P 2009 {\em Phys. Rev.\/} {\bf D79} 084008 [\eprint{0901.3775}]

\bibitem{Jacobson:2005bg}
Jacobson T, Liberati S and Mattingly D 2006 {\em Annals Phys.\/} {\bf 321}
  150--196 [\eprint{astro-ph/0505267}]

\bibitem{ChoquetBruhat:1988dw}
Choquet-Bruhat Y 1988 {\em J. Math. Phys.\/} {\bf 29} 1891--1895

\bibitem{Reall:2014pwa}
Reall H, Tanahashi N and Way B 2014 {\em Class. Quantum Grav.\/} {\bf 31}
  205005 [\eprint{1406.3379}]

\bibitem{Willison:2014era}
Willison S 2015 {\em Class. Quantum Grav.\/} {\bf 32} 022001
  [\eprint{1409.6656}]

\bibitem{Kanti:2004nr}
Kanti P 2004 {\em Int. J. Mod. Phys.\/} {\bf A19} 4899--4951
  [\eprint{hep-ph/0402168}]

\bibitem{Bertolami:2007gv}
Bertolami O, Boehmer C~G, Harko T and Lobo F~S 2007 {\em Phys. Rev.\/} {\bf
  D75} 104016 [\eprint{0704.1733}]

\bibitem{Schlamminger:2007ht}
Schlamminger S, Choi K~Y, Wagner T, Gundlach J and Adelberger E 2008 {\em Phys.
  Rev. Lett.\/} {\bf 100} 041101 [\eprint{0712.0607}]

\bibitem{Burrage:2014oza}
Burrage C, Copeland E~J and Hinds E 2015 {\em JCAP\/} {\bf 1503} 042
  [\eprint{1408.1409}]

\bibitem{Hamilton:2015zga}
Hamilton P, Jaffe M, Haslinger P, Simmons Q, Müller H and Khoury J 2015 {\em
  Science\/} {\bf 349} 849--851 [\eprint{1502.03888}]

\bibitem{Steffen:2010ze}
Steffen J~H {\em et~al.\/} (GammeV) 2010 {\em Phys. Rev. Lett.\/} {\bf 105}
  261803 [\eprint{1010.0988}]

\bibitem{Jain:2012tn}
Jain B, Vikram V and Sakstein J 2013 {\em Astrophys. J.\/} {\bf 779} 39
  [\eprint{1204.6044}]

\bibitem{Chiba:1997ms}
Chiba T, Harada T and Nakao K~i 1997 {\em Prog. Theor. Phys. Suppl.\/} {\bf
  128} 335--372

\bibitem{2003sttg.book.....F}
{Fujii} Y and {Maeda} K~I 2003 {\em {The Scalar-Tensor Theory of
  Gravitation}\/} (Cambridge University Press)

\bibitem{Faraoni:2004pi}
Faraoni V 2004 {\em {Cosmology in scalar tensor gravity}\/} ({Springer})

\bibitem{polchinski1998string}
Polchinski J 1998 {\em String theory\/} (Cambridge university press)

\bibitem{Duff:1994tn}
Duff M~J 1995 {Kaluza-Klein theory in perspective} {\em {The Oscar Klein
  Centenary: Proceedings of the Symposium 19-21 September 1994 Stockholm,
  Sweden }\/} ed Lindstrom U (World Scientific) ISBN 9810223323
  [\eprint{hep-th/9410046}]

\bibitem{Randall:1999ee}
Randall L and Sundrum R 1999 {\em Phys. Rev. Lett.\/} {\bf 83} 3370--3373
  [\eprint{hep-ph/9905221}]

\bibitem{Randall:1999vf}
Randall L and Sundrum R 1999 {\em Phys. Rev. Lett.\/} {\bf 83} 4690--4693
  [\eprint{hep-th/9906064}]

\bibitem{Bergmann:1968ve}
Bergmann P~G 1968 {\em Int. J. Theor. Phys.\/} {\bf 1} 25--36

\bibitem{Wagoner:1970vr}
Wagoner R~V 1970 {\em Phys. Rev.\/} {\bf D1} 3209--3216

\bibitem{Jordan:1959eg}
Jordan P 1959 {\em Z. Phys.\/} {\bf 157} 112--121

\bibitem{Fierz:1956zz}
Fierz M 1956 {\em Helv. Phys. Acta\/} {\bf 29} 128--134

\bibitem{Brans:1961sx}
Brans C and Dicke R 1961 {\em Phys. Rev.\/} {\bf 124} 925--935

\bibitem{Goenner:2012cq}
Goenner H 2012 {\em Gen. Relativ. Gravit.\/} {\bf 44} 2077--2097
  [\eprint{1204.3455}]

\bibitem{Brans:2008zz}
Brans C~H 2008 {\em AIP Conf. Proc.\/} {\bf 1083} 34--46

\bibitem{Brans:2005ra}
Brans C~H 2005  [\eprint{gr-qc/0506063}]

\bibitem{Flanagan:2004bz}
Flanagan E~E 2004 {\em Class. Quantum Grav.\/} {\bf 21} 3817
  [\eprint{gr-qc/0403063}]

\bibitem{Eardley:1975}
Eardley D~M 1975 {\em Astrophys. J.\/} {\bf 196} L59--L62

\bibitem{will1993theory}
Will C 1993 {\em Theory and Experiment in Gravitational Physics\/} (Cambridge
  University Press) ISBN 9780521439732

\bibitem{Damour:1995kt}
Damour T and Esposito-Farese G 1996 {\em Phys. Rev.\/} {\bf D53} 5541--5578
  [\eprint{gr-qc/9506063}]

\bibitem{Will:1989sk}
Will C~M and Zaglauer H~W 1989 {\em Astrophys. J.\/} {\bf 346} 366

\bibitem{Cardoso:2011xi}
Cardoso V, Chakrabarti S, Pani P, Berti E and Gualtieri L 2011 {\em Phys. Rev.
  Lett.\/} {\bf 107} 241101 [\eprint{1109.6021}]

\bibitem{Mirshekari:2013vb}
Mirshekari S and Will C~M 2013 {\em Phys. Rev.\/} {\bf D87} 084070
  [\eprint{1301.4680}]

\bibitem{Yunes:2011aa}
Yunes N, Pani P and Cardoso V 2012 {\em Phys. Rev.\/} {\bf D85} 102003
  [\eprint{1112.3351}]

\bibitem{Horndeski:1974wa}
Horndeski G~W 1974 {\em Int. J. Theor. Phys.\/} {\bf 10} 363--384

\bibitem{Deffayet:2011gz}
Deffayet C, Gao X, Steer D and Zahariade G 2011 {\em Phys. Rev.\/} {\bf D84}
  064039 [\eprint{1103.3260}]

\bibitem{Carroll:2000fy}
Carroll S~M 2001 {\em Living Rev. Relativ.\/} {\bf 4} 1
  [\eprint{astro-ph/0004075}]

\bibitem{Martin:2012bt}
Martin J 2012 {\em Comptes Rendus Physique\/} {\bf 13} 566--665
  [\eprint{1205.3365}]

\bibitem{Schmidt:2006jt}
Schmidt H~J 2006 {\em eConf\/} {\bf C0602061} 12 [\eprint{gr-qc/0602017}]

\bibitem{Teyssandier:1983zz}
Teyssandier P and Tourrenc P 1983 {\em J. Math. Phys.\/} {\bf 24} 2793

\bibitem{Barrow:1988xi}
Barrow J~D 1988 {\em Nucl. Phys.\/} {\bf B296} 697--709

\bibitem{Barrow:1988xh}
Barrow J~D and Cotsakis S 1988 {\em Phys. Lett.\/} {\bf B214} 515--518

\bibitem{Wands:1993uu}
Wands D 1994 {\em Class. Quantum Grav.\/} {\bf 11} 269--280
  [\eprint{gr-qc/9307034}]

\bibitem{O'Hanlon:1972ya}
O'Hanlon J 1972 {\em Phys. Rev. Lett.\/} {\bf 29} 137--138

\bibitem{Fiziev:1999zt}
Fiziev P 2000 {\em Mod. Phys. Lett.\/} {\bf A15} 1977 [\eprint{gr-qc/9911037}]

\bibitem{Sotiriou:2006hs}
Sotiriou T~P 2006 {\em Class. Quantum Grav.\/} {\bf 23} 5117--5128
  [\eprint{gr-qc/0604028}]

\bibitem{Biswas:2011ar}
Biswas T, Gerwick E, Koivisto T and Mazumdar A 2012 {\em Phys. Rev. Lett.\/}
  {\bf 108} 031101 [\eprint{1110.5249}]

\bibitem{Talaganis:2014ida}
Talaganis S, Biswas T and Mazumdar A 2015 {\em Class. Quant. Grav.\/} {\bf 32}
  215017 [\eprint{1412.3467}]

\bibitem{Moura:2006pz}
Moura F and Schiappa R 2007 {\em Class. Quantum Grav.\/} {\bf 24} 361--386
  [\eprint{hep-th/0605001}]

\bibitem{ashtekar1989cp}
Ashtekar A, Balachandran A and Jo S 1989 {\em Int. J. Mod. Phys.\/} {\bf A4}
  1493--1514

\bibitem{Chen:2012au}
Chen T~j, Fasiello M, Lim E~A and Tolley A~J 2013 {\em JCAP\/} {\bf 1302} 042
  [\eprint{1209.0583}]

\bibitem{Chen:2013aha}
Chen T~j and Lim E~A 2014 {\em JCAP\/} {\bf 1405} 010 [\eprint{1311.3189}]

\bibitem{Psaltis:2007cw}
Psaltis D, Perrodin D, Dienes K~R and Mocioiu I 2008 {\em Phys. Rev. Lett.\/}
  {\bf 100} 091101 [\eprint{0710.4564}]

\bibitem{Yunes:2007ss}
Yunes N and Sopuerta C~F 2008 {\em Phys. Rev.\/} {\bf D77} 064007
  [\eprint{0712.1028}]

\bibitem{Barausse:2008xv}
Barausse E and Sotiriou T~P 2008 {\em Phys. Rev. Lett.\/} {\bf 101} 099001
  [\eprint{0803.3433}]

\bibitem{Gross:1986mw}
Gross D~J and Sloan J~H 1987 {\em Nucl. Phys.\/} {\bf B291} 41--89

\bibitem{Kobayashi:2011nu}
Kobayashi T, Yamaguchi M and Yokoyama J 2011 {\em Prog. Theor. Phys.\/} {\bf
  126} 511--529 [\eprint{1105.5723}]

\bibitem{Alexander:2009tp}
Alexander S and Yunes N 2009 {\em Phys. Rept.\/} {\bf 480} 1--55
  [\eprint{0907.2562}]

\bibitem{Grumiller:2007rv}
Grumiller D and Yunes N 2008 {\em Phys. Rev.\/} {\bf D77} 044015
  [\eprint{0711.1868}]

\bibitem{Everitt:2011hp}
Everitt C, DeBra D, Parkinson B, Turneaure J, Conklin J {\em et~al.\/} 2011
  {\em Phys. Rev. Lett.\/} {\bf 106} 221101 [\eprint{1105.3456}]

\bibitem{Canizares:2012ji}
Canizares P, Gair J and Sopuerta C 2012 {\em J. Phys. Conf. Ser.\/} {\bf 363}
  012019 [\eprint{1206.0322}]

\bibitem{Yagi:2012vf}
Yagi K, Yunes N and Tanaka T 2012 {\em Phys. Rev. Lett.\/} {\bf 109} 251105
  [\eprint{1208.5102}]

\bibitem{Shafee:2005ef}
Shafee R, McClintock J~E, Narayan R, Davis S~W, Li L~X and Remillard R~A 2006
  {\em Astrophys. J.\/} {\bf 636} L113--L116 [\eprint{astro-ph/0508302}]

\bibitem{Kostelecky:2003fs}
Kostelecky V~A 2004 {\em Phys. Rev.\/} {\bf D69} 105009
  [\eprint{hep-th/0312310}]

\bibitem{Kostelecky:2008ts}
Kostelecky V~A and Russell N 2011 {\em Rev. Mod. Phys.\/} {\bf 83} 11--31
  [\eprint{0801.0287}]

\bibitem{Mattingly:2005re}
Mattingly D 2005 {\em Living Rev. Relativ.\/} {\bf 8} 5
  [\eprint{gr-qc/0502097}]

\bibitem{Liberati:2013xla}
Liberati S 2013 {\em Class. Quantum Grav.\/} {\bf 30} 133001
  [\eprint{1304.5795}]

\bibitem{Blas:2014aca}
Blas D and Lim E 2015 {\em Int. J. Mod. Phys.\/} {\bf D23} 1443009
  [\eprint{1412.4828}]

\bibitem{Colladay:1998fq}
Colladay D and Kostelecky V~A 1998 {\em Phys. Rev.\/} {\bf D58} 116002
  [\eprint{hep-ph/9809521}]

\bibitem{Bailey:2006fd}
Bailey Q~G and Kostelecky V~A 2006 {\em Phys. Rev.\/} {\bf D74} 045001
  [\eprint{gr-qc/0603030}]

\bibitem{Kostelecky:2010ze}
Kostelecky V~A and Tasson J~D 2011 {\em Phys. Rev.\/} {\bf D83} 016013
  [\eprint{1006.4106}]

\bibitem{Shao:2014oha}
Shao L 2014 {\em Phys. Rev. Lett.\/} {\bf 112} 111103 [\eprint{1402.6452}]

\bibitem{Shao:2014bfa}
Shao L 2014 {\em Phys. Rev.\/} {\bf D90} 122009 [\eprint{1412.2320}]

\bibitem{Jacobson:2000xp}
Jacobson T and Mattingly D 2001 {\em Phys. Rev.\/} {\bf D64} 024028
  [\eprint{gr-qc/0007031}]

\bibitem{Eling:2004dk}
Eling C, Jacobson T and Mattingly D 2006 {Einstein-Aether theory} {\em
  {Deserfest: A celebration of the life and works of Stanley Deser}\/} ed Liu
  J~T, Duff M~J, Stelle K~S and Woodard R~P (World Scientific) pp 163--179 ISBN
  9789812560827 [\eprint{gr-qc/0410001}]

\bibitem{Carroll:2004ai}
Carroll S~M and Lim E~A 2004 {\em Phys. Rev.\/} {\bf D70} 123525
  [\eprint{hep-th/0407149}]

\bibitem{Jacobson:2004ts}
Jacobson T and Mattingly D 2004 {\em Phys. Rev.\/} {\bf D70} 024003
  [\eprint{gr-qc/0402005}]

\bibitem{Elliott:2005va}
Elliott J~W, Moore G~D and Stoica H 2005 {\em JHEP\/} {\bf 0508} 066
  [\eprint{hep-ph/0505211}]

\bibitem{Blas:2009qj}
Blas D, Pujolas O and Sibiryakov S 2010 {\em Phys. Rev. Lett.\/} {\bf 104}
  181302 [\eprint{0909.3525}]

\bibitem{Jacobson:2010mx}
Jacobson T 2010 {\em Phys. Rev.\/} {\bf D81} 101502 [\eprint{1001.4823}]

\bibitem{Barausse:2012ny}
Barausse E and Sotiriou T~P 2012 {\em Phys. Rev. Lett.\/} {\bf 109} 181101
  erratum-ibid.\ {\bf 110}, 039902 (2013) [\eprint{1207.6370}]

\bibitem{Liberati:2012jf}
Liberati S, Maccione L and Sotiriou T~P 2012 {\em Phys. Rev. Lett.\/} {\bf 109}
  151602 [\eprint{1207.0670}]

\bibitem{Froggatt:1991ft}
Froggatt C and Nielsen H~B 1991 {\em {Origin of symmetries}\/} (World
  Scientific)

\bibitem{GrootNibbelink:2004za}
Groot~Nibbelink S and Pospelov M 2005 {\em Phys. Rev. Lett.\/} {\bf 94} 081601
  [\eprint{hep-ph/0404271}]

\bibitem{Chadha:1982qq}
Chadha S and Nielsen H~B 1983 {\em Nucl. Phys.\/} {\bf B217} 125

\bibitem{Bednik:2013nxa}
Bednik G, Pujolàs O and Sibiryakov S 2013 {\em JHEP\/} {\bf 1311} 064
  [\eprint{1305.0011}]

\bibitem{Pospelov:2010mp}
Pospelov M and Shang Y 2012 {\em Phys. Rev.\/} {\bf D85} 105001
  [\eprint{1010.5249}]

\bibitem{Papazoglou:2009fj}
Papazoglou A and Sotiriou T~P 2010 {\em Phys. Lett.\/} {\bf B685} 197--200
  [\eprint{0911.1299}]

\bibitem{Kimpton:2010xi}
Kimpton I and Padilla A 2010 {\em JHEP\/} {\bf 1007} 014 [\eprint{1003.5666}]

\bibitem{Blas:2009ck}
Blas D, Pujolas O and Sibiryakov S 2010 {\em Phys. Lett.\/} {\bf B688} 350--355
  [\eprint{0912.0550}]

\bibitem{Herdeiro:2011km}
Herdeiro C and Hirano S 2012 {\em JCAP\/} {\bf 1205} 031 [\eprint{1109.1468}]

\bibitem{Arnowitt:1960es}
Arnowitt R~L, Deser S and Misner C~W 1960 {\em Phys. Rev.\/} {\bf 117}
  1595--1602

\bibitem{Coelho:2012xi}
Coelho F~S, Herdeiro C, Hirano S and Sato Y 2012 {\em Phys. Rev.\/} {\bf D86}
  064009 [\eprint{1205.6850}]

\bibitem{Alcubierre:2008}
{Alcubierre} M 2008 {\em {Introduction to 3+1 Numerical Relativity}\/} (Oxford
  University Press)

\bibitem{Pauli:1939xp}
Pauli W and Fierz M 1939 {\em Helv. Phys. Acta\/} {\bf 12} 297--300

\bibitem{Boulware:1973my}
Boulware D and Deser S 1972 {\em Phys. Rev.\/} {\bf D6} 3368--3382

\bibitem{deRham:2010ik}
de~Rham C and Gabadadze G 2010 {\em Phys. Rev.\/} {\bf D82} 044020
  [\eprint{1007.0443}]

\bibitem{deRham:2010kj}
de~Rham C, Gabadadze G and Tolley A~J 2011 {\em Phys. Rev. Lett.\/} {\bf 106}
  231101 [\eprint{1011.1232}]

\bibitem{Nibbelink:2006sz}
Groot~Nibbelink S, Peloso M and Sexton M 2007 {\em Eur. Phys. J.\/} {\bf C51}
  741--752 [\eprint{hep-th/0610169}]

\bibitem{Hinterbichler:2012cn}
Hinterbichler K and Rosen R~A 2012 {\em JHEP\/} {\bf 1207} 047
  [\eprint{1203.5783}]

\bibitem{Hassan:2011tf}
Hassan S, Rosen R~A and Schmidt-May A 2012 {\em JHEP\/} {\bf 1202} 026
  [\eprint{1109.3230}]

\bibitem{Hassan:2011hr}
Hassan S and Rosen R~A 2012 {\em Phys. Rev. Lett.\/} {\bf 108} 041101
  [\eprint{1106.3344}]

\bibitem{Bernard:2014bfa}
Bernard L, Deffayet C and von Strauss M 2015 {\em Phys. Rev.\/} {\bf D91}
  104013 [\eprint{1410.8302}]

\bibitem{Hassan:2011zd}
Hassan S~F and Rosen R~A 2012 {\em JHEP\/} {\bf 02} 126 [\eprint{1109.3515}]

\bibitem{Nicolis:2008in}
Nicolis A, Rattazzi R and Trincherini E 2009 {\em Phys. Rev.\/} {\bf D79}
  064036 [\eprint{0811.2197}]

\bibitem{vanDam:1970vg}
van Dam H and Veltman M 1970 {\em Nucl. Phys.\/} {\bf B22} 397--411

\bibitem{Zakharov:1970cc}
Zakharov V 1970 {\em JETP Lett.\/} {\bf 12} 312

\bibitem{Vainshtein:1972sx}
Vainshtein A 1972 {\em Phys. Lett.\/} {\bf B39} 393--394

\bibitem{Babichev:2013usa}
Babichev E and Deffayet C 2013 {\em Class. Quantum Grav.\/} {\bf 30} 184001
  [\eprint{1304.7240}]

\bibitem{deRham:2012fg}
de~Rham C, Matas A and Tolley A~J 2013 {\em Phys. Rev.\/} {\bf D87} 064024
  [\eprint{1212.5212}]

\bibitem{Chow:2009fm}
Chow N and Khoury J 2009 {\em Phys. Rev.\/} {\bf D80} 024037
  [\eprint{0905.1325}]

\bibitem{deRham:2011by}
de~Rham C and Heisenberg L 2011 {\em Phys. Rev.\/} {\bf D84} 043503
  [\eprint{1106.3312}]

\bibitem{Li:2013tda}
Li B, Barreira A, Baugh C~M, Hellwing W~A, Koyama K {\em et~al.\/} 2013 {\em
  JCAP\/} {\bf 1311} 012 [\eprint{1308.3491}]

\bibitem{Li:2013nua}
Li B, Zhao G~b and Koyama K 2013 {\em JCAP\/} {\bf 1305} 023
  [\eprint{1303.0008}]

\bibitem{Hiramatsu:2012xj}
Hiramatsu T, Hu W, Koyama K and Schmidt F 2013 {\em Phys. Rev.\/} {\bf D87}
  063525 [\eprint{1209.3364}]

\bibitem{Falck:2014jwa}
Falck B, Koyama K, Zhao G~b and Li B 2014 {\em JCAP\/} {\bf 1407} 058
  [\eprint{1404.2206}]

\bibitem{Falck:2015rsa}
Falck B, Koyama K and Zhao G~b 2015 {\em JCAP\/} {\bf 1507} 049
  [\eprint{1503.06673}]

\bibitem{Nicolis:2004qq}
Nicolis A and Rattazzi R 2004 {\em JHEP\/} {\bf 0406} 059
  [\eprint{hep-th/0404159}]

\bibitem{Brito:2014ifa}
Brito R, Terrana A, Johnson M and Cardoso V 2014 {\em Phys. Rev.\/} {\bf D90}
  124035 [\eprint{1409.0886}]

\bibitem{Jaccard:2013gla}
Jaccard M, Maggiore M and Mitsou E 2013 {\em Phys. Rev.\/} {\bf D88} 044033
  [\eprint{1305.3034}]

\bibitem{Maggiore:2013mea}
Maggiore M 2014 {\em Phys. Rev.\/} {\bf D89} 043008 [\eprint{1307.3898}]

\bibitem{Dvali:2002vf}
Dvali G, Gruzinov A and Zaldarriaga M 2003 {\em Phys. Rev.\/} {\bf D68} 024012
  [\eprint{hep-ph/0212069}]

\bibitem{Lue:2002sw}
Lue A and Starkman G 2003 {\em Phys. Rev.\/} {\bf D67} 064002
  [\eprint{astro-ph/0212083}]

\bibitem{Tseytlin:1999dj}
Tseytlin A~A 1999 {Born-Infeld action, supersymmetry and string theory} {\em
  The Many Faces of the Superworld: Yuri Golfand Memorial Volume\/} ed Shifman
  M~A (World Scientific) ISBN 9810242069 [\eprint{hep-th/9908105}]

\bibitem{Deser:1998rj}
Deser S and Gibbons G 1998 {\em Class. Quantum Grav.\/} {\bf 15} L35--L39
  [\eprint{hep-th/9803049}]

\bibitem{Vollick:2003qp}
Vollick D~N 2004 {\em Phys. Rev.\/} {\bf D69} 064030 [\eprint{gr-qc/0309101}]

\bibitem{Olmo:2013gqa}
Olmo G~J, Rubiera-Garcia D and Sanchis-Alepuz H 2014 {\em Eur. Phys. J.\/} {\bf
  C74} 2804 [\eprint{1311.0815}]

\bibitem{Barragan:2009sq}
Barragan C, Olmo G~J and Sanchis-Alepuz H 2009 {\em Phys. Rev.\/} {\bf D80}
  024016 [\eprint{0907.0318}]

\bibitem{Delsate:2012ky}
Delsate T and Steinhoff J 2012 {\em Phys. Rev. Lett.\/} {\bf 109} 021101
  [\eprint{1201.4989}]

\bibitem{Barausse:2008nm}
Barausse E, Sotiriou T~P and Miller J~C 2008 {\em EAS Publ.Ser.\/} {\bf 30}
  189--192 [\eprint{0801.4852}]

\bibitem{Sotiriou:2008dh}
Sotiriou T~P 2008 {\em Phys. Lett.\/} {\bf B664} 225--228 [\eprint{0805.1160}]

\bibitem{Kim:2013nna}
Kim H~C 2014 {\em Phys. Rev.\/} {\bf D89} 064001 [\eprint{1312.0705}]

\bibitem{Flanagan:2003rb}
Flanagan E~E 2004 {\em Phys. Rev. Lett.\/} {\bf 92} 071101
  [\eprint{astro-ph/0308111}]

\bibitem{Vollick:2003ic}
Vollick D~N 2004 {\em Class. Quantum Grav.\/} {\bf 21} 3813--3816
  [\eprint{gr-qc/0312041}]

\bibitem{Li:2008fa}
Li B, Mota D~F and Shaw D~J 2008 {\em Phys. Rev.\/} {\bf D78} 064018
  [\eprint{0805.3428}]

\bibitem{Goldberger:2007hy}
Goldberger W~D 2007 {Les Houches lectures on effective field theories and
  gravitational radiation} {\em {Particle Physics and Cosmology: The Fabric of
  Spacetime. Les Houches, France, July 31-August 25, 2006}\/} ed Bernardeau F,
  Grojean C and Dalibard J (Elsevier) ISBN 9780444530073
  [\eprint{hep-ph/0701129}]

\bibitem{Donoghue:2012zc}
Donoghue J~F 2012 {\em AIP Conf. Proc.\/} {\bf 1483} 73--94
  [\eprint{1209.3511}]

\bibitem{Polchinski:2006gy}
Polchinski J 2007 {The Cosmological Constant and the String Landscape} {\em
  {The Quantum Structure of Space and Time}\/} ed Gross D, Henneaux M and
  Sevrin A (World Scientific) pp 216--236 ISBN 9789812569523
  [\eprint{hep-th/0603249}]

\bibitem{Burgess:2013ara}
Burgess C~P 2015 {The Cosmological Constant Problem: Why it's hard to get Dark
  Energy from Micro-physics} {\em {Post-Planck Cosmology. Les Houches, France,
  July 8-August 2, 2013}\/} ed Deffayet C, Peter P, Wandelt B, Zaldarriaga M
  and Cugliandolo L~F (Oxford University Press) ISBN 0198728859
  [\eprint{1309.4133}]

\bibitem{DeWitt:1967yk}
DeWitt B~S 1967 {\em Phys. Rev.\/} {\bf 160} 1113--1148

\bibitem{DeWitt:1967ub}
DeWitt B~S 1967 {\em Phys. Rev.\/} {\bf 162} 1195--1239

\bibitem{DeWitt:1967uc}
DeWitt B~S 1967 {\em Phys. Rev.\/} {\bf 162} 1239--1256

\bibitem{Dunbar:1994bn}
Dunbar D~C and Norridge P~S 1995 {\em Nucl. Phys.\/} {\bf B433} 181--208
  [\eprint{hep-th/9408014}]

\bibitem{Weinberg:1965nx}
Weinberg S 1965 {\em Phys. Rev.\/} {\bf 140} B516--B524

\bibitem{Donoghue:1999qh}
Donoghue J~F and Torma T 1999 {\em Phys. Rev.\/} {\bf D60} 024003
  [\eprint{hep-th/9901156}]

\bibitem{Weinberg:1978kz}
Weinberg S 1979 {\em Physica\/} {\bf A96} 327

\bibitem{Burgess:2009ri}
Burgess C~P 2012 {Effective Theories and Modifications of Gravity} {\em
  {Proceedings, Foundations of Space and Time: Reflections on Quantum
  Gravity}\/} ed Murugan J, Weltman A and Ellis G~F~R (Cambridge University
  Press) pp 50--68 ISBN 9780521114400 [\eprint{0912.4295}]

\bibitem{Burgess:2014lwa}
Burgess C and Williams M 2014 {\em JHEP\/} {\bf 1408} 074 [\eprint{1404.2236}]

\bibitem{Clifton:2012ry}
Clifton T, Dunsby P, Goswami R and Nzioki A~M 2013 {\em Phys. Rev.\/} {\bf D87}
  063517 [\eprint{1210.0730}]

\bibitem{Clifton:2015ira}
Clifton T and Dunsby P~K 2015 {\em Phys. Rev.\/} {\bf D91} 103528
  [\eprint{1501.04004}]

\bibitem{Mannheim:1988dj}
Mannheim P~D and Kazanas D 1989 {\em Astrophys. J.\/} {\bf 342} 635--638

\bibitem{Varieschi:2014ata}
Varieschi G~U 2014 {\em Gen. Relativ. Gravit.\/} {\bf 46} 1741
  [\eprint{1401.6503}]

\bibitem{Visser:2009fg}
Visser M 2009 {\em Phys. Rev.\/} {\bf D80} 025011 [\eprint{0902.0590}]

\bibitem{D'Amico:2011jj}
D'Amico G, de~Rham C, Dubovsky S, Gabadadze G, Pirtskhalava D {\em et~al.\/}
  2011 {\em Phys. Rev.\/} {\bf D84} 124046 [\eprint{1108.5231}]

\bibitem{Gratia:2012wt}
Gratia P, Hu W and Wyman M 2012 {\em Phys. Rev.\/} {\bf D86} 061504
  [\eprint{1205.4241}]

\bibitem{Fasiello:2013woa}
Fasiello M and Tolley A~J 2013 {\em JCAP\/} {\bf 1312} 002 [\eprint{1308.1647}]

\bibitem{DeFelice:2014nja}
De~Felice A, Gümrükçüoğlu A~E, Mukohyama S, Tanahashi N and Tanaka T 2014
  {\em JCAP\/} {\bf 1406} 037 [\eprint{1404.0008}]

\bibitem{D'Amico:2012zv}
D'Amico G, Gabadadze G, Hui L and Pirtskhalava D 2013 {\em Phys. Rev.\/} {\bf
  D87} 064037 [\eprint{1206.4253}]

\bibitem{DeFelice:2013tsa}
De~Felice A and Mukohyama S 2014 {\em Phys. Lett.\/} {\bf B728} 622--625
  [\eprint{1306.5502}]

\bibitem{Sotiriou:2006qn}
Sotiriou T~P and Liberati S 2007 {\em Annals Phys.\/} {\bf 322} 935--966
  [\eprint{gr-qc/0604006}]

\bibitem{Bekenstein:1996pn}
Bekenstein J~D 1997 {Black hole hair: 25 - years after} {\em {Proceedings of
  the Second International A.D. Sakharov Conference on Physics: Moscow, Russia
  20-24 May 1996}\/} ed Dremin I~M and Semikhatov A~M (World Scientific) ISBN
  9810228619 [\eprint{gr-qc/9605059}]

\bibitem{Carter:1997im}
Carter B 1999 {Has the black hole equilibrium problem been solved?} {\em
  {Proceedings, 8th Marcel Grossmann meeting, Jerusalem, Israel, June 22-27,
  1997.}\/} ed Piran T (World Scientific) pp 136--155 ISBN 9810237936
  [\eprint{gr-qc/9712038}]

\bibitem{Chrusciel:2012jk}
Chrusciel P~T, Costa J~L and Heusler M 2012 {\em Living Rev. Relativ.\/} {\bf
  15} 7 [\eprint{1205.6112}]

\bibitem{Robinson}
Robinson D 2009 {\em {Four decades of black holes uniqueness theorems}\/}
  (Cambridge University Press)

\bibitem{Gibbons:1975kk}
Gibbons G~W 1975 {\em Commun. Math. Phys.\/} {\bf 44} 245--264

\bibitem{1969ApJ...157..869G}
{Goldreich} P and {Julian} W~H 1969 {\em \apj\/} {\bf 157} 869

\bibitem{1975ApJ...196...51R}
{Ruderman} M~A and {Sutherland} P~G 1975 {\em \apj\/} {\bf 196} 51--72

\bibitem{Blandford:1977ds}
Blandford R~D and Znajek R~L 1977 {\em Mon. Not. R. Astron. Soc.\/} {\bf 179}
  433--456

\bibitem{Kerr:1963ud}
Kerr R~P 1963 {\em Phys. Rev. Lett.\/} {\bf 11} 237--238

\bibitem{Rhoades:1974fn}
Rhoades Clifford~E J and Ruffini R~J 1974 {\em Phys. Rev. Lett.\/} {\bf 32}
  324--327

\bibitem{Teukolsky:2014vca}
Teukolsky S~A 2015 {\em Class. Quantum Grav.\/} {\bf 32} 124006
  [\eprint{1410.2130}]

\bibitem{Shapiro:1983du}
Shapiro S and Teukolsky S 1983 {\em {Black holes, white dwarfs, and neutron
  stars: The physics of compact objects}\/} (Wiley)

\bibitem{MTB}
Chandrasekhar S 1983 {\em The Mathematical Theory of Black Holes\/} (New York:
  Oxford University Press)

\bibitem{Frolov:1998wf}
Frolov V~P and Novikov I~D 1998 {\em {Black hole physics: Basic concepts and
  new developments}\/} (Dordrecht: Kluwer Academic)

\bibitem{Teukolsky:1972my}
Teukolsky S~A 1972 {\em Phys. Rev. Lett.\/} {\bf 29} 1114--1118

\bibitem{Teukolsky:1973ha}
Teukolsky S~A 1973 {\em Astrophys. J.\/} {\bf 185} 635--647

\bibitem{Press:1973zz}
Press W~H and Teukolsky S~A 1973 {\em Astrophys. J.\/} {\bf 185} 649--674

\bibitem{Teukolsky:1974yv}
Teukolsky S~A and Press W~H 1974 {\em Astrophys. J.\/} {\bf 193} 443--461

\bibitem{Whiting:1988vc}
Whiting B~F 1989 {\em J. Math. Phys.\/} {\bf 30} 1301

\bibitem{pressringdown}
Press W~H 1971 {\em Astrophys. J.\/} {\bf 170} L105

\bibitem{Leaver:1985ax}
Leaver E~W 1985 {\em Proc. R. Soc. London, Ser. A\/} {\bf 402} 285--298

\bibitem{Kokkotas:1999bd}
Kokkotas K~D and Schmidt B~G 1999 {\em Living Rev. Relativ.\/} {\bf 2} 2
  [\eprint{gr-qc/9909058}]

\bibitem{Nollert:1999ji}
Nollert H~P 1999 {\em Class. Quantum Grav.\/} {\bf 16} R159--R216

\bibitem{Berti:2009kk}
Berti E, Cardoso V and Starinets A~O 2009 {\em Class. Quantum Grav.\/} {\bf 26}
  163001 [\eprint{0905.2975}]

\bibitem{Konoplya:2011qq}
Konoplya R and Zhidenko A 2011 {\em Rev. Mod. Phys.\/} {\bf 83} 793--836
  [\eprint{1102.4014}]

\bibitem{Kay:1987ax}
Kay B~S and Wald R~M 1987 {\em Class. Quantum Grav.\/} {\bf 4} 893--898

\bibitem{Dafermos:2008en}
Dafermos M and Rodnianski I 2013 {\em Clay Math.Proc.\/} {\bf 17} 97--205
  [\eprint{0811.0354}]

\bibitem{Dafermos:2009uq}
Dafermos M and Rodnianski I 2010 {A New physical-space approach to decay for
  the wave equation with applications to black hole spacetimes} {\em {XVIth
  International Congress on Mathematical Physics}\/} ed Exner P (World
  Scientific) pp 421--433 ISBN 9789814304627 [\eprint{0910.4957}]

\bibitem{Dafermos:2010hb}
Dafermos M and Rodnianski I 2010  [\eprint{1010.5132}]

\bibitem{Dafermos:2014cua}
Dafermos M, Rodnianski I and Shlapentokh-Rothman Y 2014  [\eprint{1402.7034}]

\bibitem{Dafermos:2014jwa}
Dafermos M, Rodnianski I and Shlapentokh-Rothman Y 2014  [\eprint{1412.8379}]

\bibitem{Aretakis:2012ei}
Aretakis S 2015 {\em Adv. Theor. Math. Phys.\/} {\bf 19} 507--530
  [\eprint{1206.6598}]

\bibitem{Lucietti:2012sf}
Lucietti J and Reall H~S 2012 {\em Phys. Rev.\/} {\bf D86} 104030
  [\eprint{1208.1437}]

\bibitem{Yang:2012pj}
Yang H, Zhang F, Zimmerman A, Nichols D~A, Berti E {\em et~al.\/} 2013 {\em
  Phys. Rev.\/} {\bf D87} 041502 [\eprint{1212.3271}]

\bibitem{Yang:2013uba}
Yang H, Zimmerman A, Zenginoğlu A, Zhang F, Berti E {\em et~al.\/} 2013 {\em
  Phys. Rev.\/} {\bf D88} 044047 [\eprint{1307.8086}]

\bibitem{Cook:2014cta}
Cook G~B and Zalutsky M 2014 {\em Phys. Rev.\/} {\bf D90} 124021
  [\eprint{1410.7698}]

\bibitem{1971ApJ...166L..35T}
{Thorne} K~S and {Dykla} J~J 1971 {\em Astrophys. J. Lett.\/} {\bf 166} L35

\bibitem{Chase:1970}
Chase J 1970 {\em Commun. Math. Phys.\/} {\bf 19} 276--288

\bibitem{Bekenstein:1995un}
Bekenstein J 1995 {\em Phys. Rev.\/} {\bf D51} 6608--6611

\bibitem{zeldovich1}
Zel'dovich Y~B 1971 {\em Pis'ma Zh. Eksp. Teor. Fiz.\/} {\bf 14} 270

\bibitem{zeldovich2}
Zel'dovich Y~B 1972 {\em Zh. Eksp. Teor. Fiz\/} {\bf 62} 2076

\bibitem{Press:1972zz}
Press W~H and Teukolsky S~A 1972 {\em Nature\/} {\bf 238} 211--212

\bibitem{Witek:2012tr}
Witek H, Cardoso V, Ishibashi A and Sperhake U 2013 {\em Phys. Rev.\/} {\bf
  D87} 043513 [\eprint{1212.0551}]

\bibitem{Okawa:2014nda}
Okawa H, Witek H and Cardoso V 2014 {\em Phys. Rev.\/} {\bf D89} 104032
  [\eprint{1401.1548}]

\bibitem{Brito:2014wla}
Brito R, Cardoso V and Pani P 2015 {\em Class. Quantum Grav.\/} {\bf 32} 134001
  [\eprint{1411.0686}]

\bibitem{Pena:1997cy}
Pena I and Sudarsky D 1997 {\em Class. Quantum Grav.\/} {\bf 14} 3131--3134

\bibitem{Hod:2012px}
Hod S 2012 {\em Phys. Rev.\/} {\bf D86} 104026 [\eprint{1211.3202}]

\bibitem{Hod:2013zza}
Hod S 2013 {\em Eur. Phys. J.\/} {\bf C73} 2378 [\eprint{1311.5298}]

\bibitem{Benone:2014ssa}
Benone C~L, Crispino L~C, Herdeiro C and Radu E 2014 {\em Phys. Rev.\/} {\bf
  D90} 104024 [\eprint{1409.1593}]

\bibitem{Herdeiro:2014pka}
Herdeiro C, Radu E and Runarsson H 2014 {\em Phys. Lett.\/} {\bf B739} 302--307
  [\eprint{1409.2877}]

\bibitem{Herdeiro:2014ima}
Herdeiro C~A~R and Radu E 2014 {\em Int. J. Mod. Phys.\/} {\bf D23} 1442014
  [\eprint{1405.3696}]

\bibitem{Yoshida:1997qf}
Yoshida S and Eriguchi Y 1997 {\em Phys. Rev.\/} {\bf D56} 762--771

\bibitem{Kleihaus:2005me}
Kleihaus B, Kunz J and List M 2005 {\em Phys. Rev.\/} {\bf D72} 064002
  [\eprint{gr-qc/0505143}]

\bibitem{Ryan:1996nk}
Ryan F~D 1997 {\em Phys. Rev.\/} {\bf D55} 6081--6091

\bibitem{Herdeiro:2014jaa}
Herdeiro C and Radu E 2014 {\em Phys. Rev.\/} {\bf D89} 124018
  [\eprint{1406.1225}]

\bibitem{ref:webpage}
{Webpages with numerical data and Mathematica notebooks: \\
  \url{http://www.phy.olemiss.edu/~berti/research/} \\
  \url{http://centra.tecnico.ulisboa.pt/network/grit/files/} }

\bibitem{Cardoso:2013fwa}
Cardoso V, Carucci I~P, Pani P and Sotiriou T~P 2013 {\em Phys. Rev. Lett.\/}
  {\bf 111} 111101 [\eprint{1308.6587}]

\bibitem{Cardoso:2013opa}
Cardoso V, Carucci I~P, Pani P and Sotiriou T~P 2013 {\em Phys. Rev.\/} {\bf
  D88} 044056 [\eprint{1305.6936}]

\bibitem{Davis:2014tea}
Davis A~C, Gregory R, Jha R and Muir J 2014 {\em JCAP\/} {\bf 1408} 033
  [\eprint{1402.4737}]

\bibitem{Brax:2012gr}
Brax P, Davis A~C, Li B and Winther H~A 2012 {\em Phys. Rev.\/} {\bf D86}
  044015 [\eprint{1203.4812}]

\bibitem{Maselli:2013mva}
Maselli A, Cardoso V, Ferrari V, Gualtieri L and Pani P 2013 {\em Phys. Rev.\/}
  {\bf D88} 023007 [\eprint{1304.2052}]

\bibitem{Cardoso:2008bp}
Cardoso V, Miranda A~S, Berti E, Witek H and Zanchin V~T 2009 {\em Phys.
  Rev.\/} {\bf D79} 064016 [\eprint{0812.1806}]

\bibitem{Geroch:1970cd}
Geroch R~P 1970 {\em J. Math. Phys.\/} {\bf 11} 2580--2588

\bibitem{Hansen:1974zz}
Hansen R 1974 {\em J. Math. Phys.\/} {\bf 15} 46--52

\bibitem{0264-9381-7-10-012}
Hoenselaers C and Perjes Z 1990 {\em Classical and Quantum Gravity\/} {\bf 7}
  1819

\bibitem{Sotiriou:2004ud}
Sotiriou T~P and Apostolatos T~A 2004 {\em Class. Quantum Grav.\/} {\bf 21}
  5727--5733 [\eprint{gr-qc/0407064}]

\bibitem{Torii:1996yi}
Torii T, Yajima H and Maeda K~i 1997 {\em Phys. Rev.\/} {\bf D55} 739--753
  [\eprint{gr-qc/9606034}]

\bibitem{Ryan:1995wh}
Ryan F~D 1995 {\em Phys. Rev.\/} {\bf D52} 5707--5718

\bibitem{Feroci:2012qh}
Feroci M, Herder J, Bozzo E, Barret D, Brandt S {\em et~al.\/} 2012 {\em
  Proc.SPIE\/} {\bf 8443} 84432D [\eprint{1209.1497}]

\bibitem{Pappas:2012nt}
Pappas G 2012 {\em Mon. Not. R. Astron. Soc.\/} {\bf 422} 2581--2589
  [\eprint{1201.6071}]

\bibitem{Kanti:1997br}
Kanti P, Mavromatos N~E, Rizos J, Tamvakis K and Winstanley E 1998 {\em Phys.
  Rev.\/} {\bf D57} 6255--6264 [\eprint{hep-th/9703192}]

\bibitem{Bambi:2011jq}
Bambi C and Barausse E 2011 {\em Astrophys. J.\/} {\bf 731} 121
  [\eprint{1012.2007}]

\bibitem{Seoane:2013qna}
Seoane P~A {\em et~al.\/} (eLISA Collaboration) 2013  [\eprint{1305.5720}]

\bibitem{Berglund:2012bu}
Berglund P, Bhattacharyya J and Mattingly D 2012 {\em Phys. Rev.\/} {\bf D85}
  124019 [\eprint{1202.4497}]

\bibitem{Cropp:2013sea}
Cropp B, Liberati S, Mohd A and Visser M 2014 {\em Phys. Rev.\/} {\bf D89}
  064061 [\eprint{1312.0405}]

\bibitem{DyBHo:web}
Gravity group, CENTRA/IST, Lisbon:
  \href{http://blackholes.ist.utl.pt/?page=Files}{http://blackholes.ist.utl.pt/?page=Files}

\bibitem{EddingtonBook}
{Eddington, A S} 1954 {\em {The Mathematical Theory of Relativity}\/}
  (Cambridge: Cambridge University Press)

\bibitem{Will:1972zz}
Will C~M and Nordtvedt Kenneth J 1972 {\em Astrophys. J.\/} {\bf 177} 757

\bibitem{Nordtvedt:1972zz}
Nordtvedt K~J and Will C~M 1972 {\em Astrophys. J.\/} {\bf 177} 775--792

\bibitem{Cardoso:2014rha}
Cardoso V, Pani P and Rico J 2014 {\em Phys. Rev.\/} {\bf D89} 064007
  [\eprint{1401.0528}]

\bibitem{Rico_thesis}
Rico J 2013 {\em {The Kerr black hole hypothesis: a review of methods and
  results}\/} Master's thesis Instituto Superior T{\'e}cnico
  \urlprefix\url{http://blackholes.ist.utl.pt/fp-content/attachs/thesis_joaorico.pdf}

\bibitem{Collins:2004ex}
Collins N~A and Hughes S~A 2004 {\em Phys. Rev.\/} {\bf D69} 124022
  [\eprint{gr-qc/0402063}]

\bibitem{Vigeland:2009pr}
Vigeland S~J and Hughes S~A 2010 {\em Phys. Rev.\/} {\bf D81} 024030
  [\eprint{0911.1756}]

\bibitem{Gair:2011ym}
Gair J and Yunes N 2011 {\em Phys. Rev.\/} {\bf D84} 064016
  [\eprint{1106.6313}]

\bibitem{Vigeland:2011ji}
Vigeland S, Yunes N and Stein L 2011 {\em Phys. Rev.\/} {\bf D83} 104027
  [\eprint{1102.3706}]

\bibitem{Glampedakis:2005cf}
Glampedakis K and Babak S 2006 {\em Class. Quantum Grav.\/} {\bf 23} 4167--4188
  [\eprint{gr-qc/0510057}]

\bibitem{1992CQGra...9.2477M}
{Manko} V~S and {Novikov} I~D 1992 {\em Class. Quantum Grav.\/} {\bf 9}
  2477--2487

\bibitem{Gair:2007kr}
Gair J~R, Li C and Mandel I 2008 {\em Phys. Rev.\/} {\bf D77} 024035
  [\eprint{0708.0628}]

\bibitem{Newman:1965tw}
Newman E and Janis A 1965 {\em J. Math. Phys.\/} {\bf 6} 915--917

\bibitem{Johannsen:2011dh}
Johannsen T and Psaltis D 2011 {\em Phys. Rev.\/} {\bf D83} 124015
  [\eprint{1105.3191}]

\bibitem{Hansen:2013owa}
Hansen D and Yunes N 2013 {\em Phys. Rev.\/} {\bf D88} 104020
  [\eprint{1308.6631}]

\bibitem{Bambi:2014sfa}
Bambi C 2014 {\em Phys. Rev.\/} {\bf D90} 047503 [\eprint{1408.0690}]

\bibitem{Bambi:2014mla}
Bambi C 2015 {\em Class. Quantum Grav.\/} {\bf 32} 065005 [\eprint{1409.0310}]

\bibitem{Rezzolla:2014mua}
Rezzolla L and Zhidenko A 2014 {\em Phys. Rev.\/} {\bf D90} 084009
  [\eprint{1407.3086}]

\bibitem{Broderick:2005xa}
Broderick A~E and Narayan R 2006 {\em Astrophys. J.\/} {\bf 638} L21--L24
  [\eprint{astro-ph/0512211}]

\bibitem{Liebling:2012fv}
Liebling S~L and Palenzuela C 2012 {\em Living Rev. Relativ.\/} {\bf 15} 6
  [\eprint{1202.5809}]

\bibitem{Macedo:2013qea}
Macedo C~F, Pani P, Cardoso V and Crispino L~C 2013 {\em Astrophys. J.\/} {\bf
  774} 48 [\eprint{1302.2646}]

\bibitem{Mazur:2001fv}
Mazur P~O and Mottola E 2001  [\eprint{gr-qc/0109035}]

\bibitem{Gimon:2007ur}
Gimon E~G and Horava P 2009 {\em Phys. Lett.\/} {\bf B672} 299--302
  [\eprint{0706.2873}]

\bibitem{Berti:2005ys}
Berti E, Cardoso V and Will C~M 2006 {\em Phys. Rev.\/} {\bf D73} 064030
  [\eprint{gr-qc/0512160}]

\bibitem{Berti:2006qt}
Berti E and Cardoso V 2006 {\em Int. J. Mod. Phys.\/} {\bf D15} 2209--2216
  [\eprint{gr-qc/0605101}]

\bibitem{Kesden:2004qx}
Kesden M, Gair J and Kamionkowski M 2005 {\em Phys. Rev.\/} {\bf D71} 044015
  [\eprint{astro-ph/0411478}]

\bibitem{Pani:2009ss}
Pani P, Berti E, Cardoso V, Chen Y and Norte R 2009 {\em Phys. Rev.\/} {\bf
  D80} 124047 [\eprint{0909.0287}]

\bibitem{Barausse:2014tra}
Barausse E, Cardoso V and Pani P 2014 {\em Phys. Rev.\/} {\bf D89} 104059
  [\eprint{1404.7149}]

\bibitem{1978CMaPh..63..243F}
{Friedman} J~L 1978 {\em Commun. Math. Phys.\/} {\bf 63} 243--255

\bibitem{SchutzComins1978}
{Schutz} B~F and {Comins} N 1978 {\em \mnras\/} {\bf 182} 69--76

\bibitem{Cardoso:2014sna}
Cardoso V, Crispino L~C~B, Macedo C~F~B, Okawa H and Pani P 2014 {\em Phys.
  Rev.\/} {\bf D90} 044069 [\eprint{1406.5510}]

\bibitem{Cardoso:2007az}
Cardoso V, Pani P, Cadoni M and Cavaglia M 2008 {\em Phys. Rev.\/} {\bf D77}
  124044 [\eprint{0709.0532}]

\bibitem{Chirenti:2008pf}
Chirenti C~B and Rezzolla L 2008 {\em Phys. Rev.\/} {\bf D78} 084011
  [\eprint{0808.4080}]

\bibitem{Cardoso:2008kj}
Cardoso V, Pani P, Cadoni M and Cavagli{\`a} M 2008 {\em Class. Quantum
  Grav.\/} {\bf 25} 195010 [\eprint{0808.1615}]

\bibitem{Pani:2010jz}
Pani P, Barausse E, Berti E and Cardoso V 2010 {\em Phys. Rev.\/} {\bf D82}
  044009 [\eprint{1006.1863}]

\bibitem{Keir:2014oka}
Keir J 2014  [\eprint{1404.7036}]

\bibitem{Johannsen:2015qca}
Johannsen T 2013 {\em Astrophys. J.\/} {\bf 777} 170 [\eprint{1501.02814}]

\bibitem{Lu:2014zja}
Lu R~S, Broderick A~E, Baron F, Monnier J~D, Fish V~L {\em et~al.\/} 2014 {\em
  Astrophys. J.\/} {\bf 788} 120 [\eprint{1404.7095}]

\bibitem{GRAVITY}
{Eisenhauer} F {\em et~al.\/} 2008 {GRAVITY: getting to the event horizon of
  Sgr A*} {\em Society of Photo-Optical Instrumentation Engineers (SPIE)
  Conference Series\/} vol 7013 [\eprint{0808.0063}]

\bibitem{Gundlach:2007gc}
Gundlach C and Martin-Garcia J~M 2007 {\em Living Rev. Relativ.\/} {\bf 10} 5
  [\eprint{0711.4620}]

\bibitem{Bizon:2011gg}
Bizon P and Rostworowski A 2011 {\em Phys. Rev. Lett.\/} {\bf 107} 031102
  [\eprint{1104.3702}]

\bibitem{Dias:2011ss}
Dias O~J, Horowitz G~T and Santos J~E 2012 {\em Class. Quantum Grav.\/} {\bf
  29} 194002 [\eprint{1109.1825}]

\bibitem{Buchel:2012uh}
Buchel A, Lehner L and Liebling S~L 2012 {\em Phys. Rev.\/} {\bf D86} 123011
  [\eprint{1210.0890}]

\bibitem{Bizon:2013xha}
Bizoń P and Jalmuzna J 2013 {\em Phys. Rev. Lett.\/} {\bf 111} 041102
  [\eprint{1306.0317}]

\bibitem{Buchel:2013uba}
Buchel A, Liebling S~L and Lehner L 2013 {\em Phys. Rev.\/} {\bf D87} 123006
  [\eprint{1304.4166}]

\bibitem{Dias:2012tq}
Dias O~J, Horowitz G~T, Marolf D and Santos J~E 2012 {\em Class. Quantum
  Grav.\/} {\bf 29} 235019 [\eprint{1208.5772}]

\bibitem{Maliborski:2013jca}
Maliborski M and Rostworowski A 2013 {\em Phys. Rev. Lett.\/} {\bf 111} 051102
  [\eprint{1303.3186}]

\bibitem{Seidel:1991zh}
Seidel E and Suen W~M 1991 {\em Phys. Rev. Lett.\/} {\bf 66} 1659--1662

\bibitem{Seidel:1993zk}
Seidel E and Suen W~M 1994 {\em Phys. Rev. Lett.\/} {\bf 72} 2516--2519
  [\eprint{gr-qc/9309015}]

\bibitem{Page:2003rd}
Page D~N 2004 {\em Phys. Rev.\/} {\bf D70} 023002 [\eprint{gr-qc/0310006}]

\bibitem{Cardoso:2005vk}
Cardoso V and Yoshida S 2005 {\em JHEP\/} {\bf 0507} 009
  [\eprint{hep-th/0502206}]

\bibitem{Dolan:2007mj}
Dolan S~R 2007 {\em Phys. Rev.\/} {\bf D76} 084001 [\eprint{0705.2880}]

\bibitem{Pani:2012vp}
Pani P, Cardoso V, Gualtieri L, Berti E and Ishibashi A 2012 {\em Phys. Rev.
  Lett.\/} {\bf 109} 131102 [\eprint{1209.0465}]

\bibitem{Pani:2012bp}
Pani P, Cardoso V, Gualtieri L, Berti E and Ishibashi A 2012 {\em Phys. Rev.\/}
  {\bf D86} 104017 [\eprint{1209.0773}]

\bibitem{Choptuik:1992jv}
Choptuik M~W 1993 {\em Phys. Rev. Lett.\/} {\bf 70} 9--12

\bibitem{Brady:1997fj}
Brady P~R, Chambers C~M and Goncalves S~M 1997 {\em Phys. Rev.\/} {\bf D56}
  6057--6061 [\eprint{gr-qc/9709014}]

\bibitem{Okawa:2013jba}
Okawa H, Cardoso V and Pani P 2014 {\em Phys. Rev.\/} {\bf D89} 041502
  [\eprint{1311.1235}]

\bibitem{Grandclement:2011wz}
Grandclement P, Fodor G and Forgacs P 2011 {\em Phys. Rev.\/} {\bf D84} 065037
  [\eprint{1107.2791}]

\bibitem{Okawa:2014nea}
Okawa H, Cardoso V and Pani P 2014 {\em Phys. Rev.\/} {\bf D90} 104032
  [\eprint{1409.0533}]

\bibitem{Brito:2015oca}
Brito R, Cardoso V and Pani P 2015 {\em {Superradiance}\/} (Springer) ISBN
  9783319189994 [\eprint{1501.06570}]

\bibitem{Rosa:2011my}
Rosa J~G and Dolan S~R 2012 {\em Phys. Rev.\/} {\bf D85} 044043
  [\eprint{1110.4494}]

\bibitem{Arvanitaki:2009fg}
Arvanitaki A, Dimopoulos S, Dubovsky S, Kaloper N and March-Russell J 2010 {\em
  Phys. Rev.\/} {\bf D81} 123530 [\eprint{0905.4720}]

\bibitem{Peccei:1977hh}
Peccei R~D and Quinn H~R 1977 {\em Phys. Rev. Lett.\/} {\bf 38} 1440--1443

\bibitem{McClintock:2009as}
McClintock J~E and Remillard R~A 2009  [\eprint{0902.3488}]

\bibitem{Reynolds:2013qqa}
Reynolds C~S 2014 {\em Space Sci.Rev.\/} {\bf 183} 277--294
  [\eprint{1302.3260}]

\bibitem{AmaroSeoane:2012km}
Amaro-Seoane P, Aoudia S, Babak S, Binetruy P, Berti E {\em et~al.\/} 2013 {\em
  GW Notes\/} {\bf 6} 4--110 [\eprint{1201.3621}]

\bibitem{PDG}
Olive K {\em et~al.\/} (Particle Data Group) 2014 {\em Chin. Phys.\/} {\bf C38}
  090001

\bibitem{Arvanitaki:2010sy}
Arvanitaki A and Dubovsky S 2011 {\em Phys. Rev.\/} {\bf D83} 044026
  [\eprint{1004.3558}]

\bibitem{Arvanitaki:2014wva}
Arvanitaki A, Baryakhtar M and Huang X 2015 {\em Phys. Rev.\/} {\bf D91} 084011
  [\eprint{1411.2263}]

\bibitem{Kodama:2011zc}
Kodama H and Yoshino H 2012 {\em Int. J. Mod. Phys. Conf. Ser.\/} {\bf 7}
  84--115 [\eprint{1108.1365}]

\bibitem{Brenneman:2011wz}
Brenneman L, Reynolds C, Nowak M, Reis R, Trippe M {\em et~al.\/} 2011 {\em
  Astrophys. J.\/} {\bf 736} 103 [\eprint{1104.1172}]

\bibitem{East:2013mfa}
East W~E, Ramazanoğlu F~M and Pretorius F 2014 {\em Phys. Rev.\/} {\bf D89}
  061503 [\eprint{1312.4529}]

\bibitem{Aasi:2013wya}
Aasi J {\em et~al.\/} (LIGO Scientific Collaboration, Virgo Collaboration) 2013
   [\eprint{1304.0670}]

\bibitem{Rosa:2012uz}
Rosa J~G 2013 {\em JHEP\/} {\bf 1302} 014 [\eprint{1209.4211}]

\bibitem{Rosa:2015hoa}
Rosa J~G 2015 {\em Phys. Lett.\/} {\bf B749} 226--230 [\eprint{1501.07605}]

\bibitem{Yoshino:2012kn}
Yoshino H and Kodama H 2012 {\em Prog. Theor. Phys.\/} {\bf 128} 153--190
  [\eprint{1203.5070}]

\bibitem{Lattimer:2004pg}
Lattimer J and Prakash M 2004 {\em Science\/} {\bf 304} 536--542
  [\eprint{astro-ph/0405262}]

\bibitem{Lattimer:2006xb}
Lattimer J~M and Prakash M 2007 {\em Phys. Rept.\/} {\bf 442} 109--165
  [\eprint{astro-ph/0612440}]

\bibitem{Ozel:2010fw}
Ozel F, Baym G and Guver T 2010 {\em Phys. Rev.\/} {\bf D82} 101301
  [\eprint{1002.3153}]

\bibitem{Steiner:2010fz}
Steiner A~W, Lattimer J~M and Brown E~F 2010 {\em Astrophys. J.\/} {\bf 722}
  33--54 [\eprint{1005.0811}]

\bibitem{Steiner:2012xt}
Steiner A~W, Lattimer J~M and Brown E~F 2013 {\em Astrophys. J.\/} {\bf 765} L5
  [\eprint{1205.6871}]

\bibitem{Hebeler:2013nza}
Hebeler K, Lattimer J, Pethick C and Schwenk A 2013 {\em Astrophys. J.\/} {\bf
  773} 11 [\eprint{1303.4662}]

\bibitem{Psaltis:2013fha}
Psaltis D, Özel F and Chakrabarty D 2014 {\em Astrophys. J.\/} {\bf 787} 136
  [\eprint{1311.1571}]

\bibitem{Stergioulas:2003yp}
Stergioulas N 2003 {\em Living Rev. Relativ.\/} {\bf 6} 3
  [\eprint{gr-qc/0302034}]

\bibitem{FriedmanStergioulas}
{Friedman} J~L and {Stergioulas} N 2013 {\em {Rotating Relativistic Stars}\/}
  (Cambridge: Cambridge University Press)

\bibitem{Hartle:1967he}
Hartle J~B 1967 {\em Astrophys. J.\/} {\bf 150} 1005--1029

\bibitem{Hartle:1968ht}
{Hartle} J~B and {Thorne} K~S 1968 {\em \apj\/} {\bf 153} 807

\bibitem{Berti:2004ny}
Berti E, White F, Maniopoulou A and Bruni M 2005 {\em Mon. Not. R. Astron.
  Soc.\/} {\bf 358} 923--938 [\eprint{gr-qc/0405146}]

\bibitem{Benhar:2005gi}
Benhar O, Ferrari V, Gualtieri L and Marassi S 2005 {\em Phys. Rev.\/} {\bf
  D72} 044028 [\eprint{gr-qc/0504068}]

\bibitem{1989nos..book.....U}
{Unno} W, {Osaki} Y, {Ando} H, {Saio} H and {Shibahashi} H 1989 {\em {Nonradial
  oscillations of stars}\/} (Tokyo: University of Tokyo Press) ISBN
  9784130661034

\bibitem{Ferrari:2007dd}
Ferrari V and Gualtieri L 2008 {\em Gen. Relativ. Gravit.\/} {\bf 40} 945--970
  [\eprint{0709.0657}]

\bibitem{Andersson:2009yt}
Andersson N, Ferrari V, Jones D, Kokkotas K, Krishnan B {\em et~al.\/} 2011
  {\em Gen. Relativ. Gravit.\/} {\bf 43} 409--436 [\eprint{0912.0384}]

\bibitem{Andersson:1997rn}
Andersson N and Kokkotas K~D 1998 {\em Mon. Not. R. Astron. Soc.\/} {\bf 299}
  1059--1068 [\eprint{gr-qc/9711088}]

\bibitem{Benhar:2004xg}
Benhar O, Ferrari V and Gualtieri L 2004 {\em Phys. Rev.\/} {\bf D70} 124015
  [\eprint{astro-ph/0407529}]

\bibitem{Chandrasekhar:1992pr}
Chandrasekhar S 1970 {\em Phys. Rev. Lett.\/} {\bf 24} 611--615

\bibitem{Friedman:1978hf}
Friedman J and Schutz B~F 1978 {\em Astrophys. J.\/} {\bf 222} 281

\bibitem{Andersson:2000mf}
Andersson N and Kokkotas K~D 2001 {\em Int. J. Mod. Phys.\/} {\bf D10} 381--442
  [\eprint{gr-qc/0010102}]

\bibitem{Andersson:2006nr}
Andersson N and Comer G 2007 {\em Living Rev. Relativ.\/} {\bf 10} 1
  [\eprint{gr-qc/0605010}]

\bibitem{Horbatsch:2010hj}
Horbatsch M and Burgess C 2011 {\em JCAP\/} {\bf 1108} 027 [\eprint{1006.4411}]

\bibitem{Stefanov:2008}
Stefanov I~Z, Yazadjiev S~S and Todorov M~D 2008 {\em Mod. Phys. Lett.\/} {\bf
  A23} 2915 [\eprint{0708.4141}]

\bibitem{Doneva:2010}
Doneva D~D, Yazadjiev S~S, Kokkotas K~D and Stefanov I~Z 2010 {\em Phys.
  Rev.\/} {\bf D82} 064030 [\eprint{1007.1767}]

\bibitem{Damour:1998jk}
Damour T and Esposito-Farese G 1998 {\em Phys. Rev.\/} {\bf D58} 042001
  [\eprint{gr-qc/9803031}]

\bibitem{Silva:2014fca}
Silva H~O, Macedo C~F~B, Berti E and Crispino L~C~B 2015 {\em Class. Quant.
  Grav.\/} {\bf 32} 145008 [\eprint{1411.6286}]

\bibitem{Barausse:2012da}
Barausse E, Palenzuela C, Ponce M and Lehner L 2013 {\em Phys. Rev.\/} {\bf
  D87} 081506 [\eprint{1212.5053}]

\bibitem{Palenzuela:2013hsa}
Palenzuela C, Barausse E, Ponce M and Lehner L 2014 {\em Phys. Rev.\/} {\bf
  D89} 044024 [\eprint{1310.4481}]

\bibitem{Shibata:2013pra}
Shibata M, Taniguchi K, Okawa H and Buonanno A 2014 {\em Phys. Rev.\/} {\bf
  D89} 084005 [\eprint{1310.0627}]

\bibitem{Taniguchi:2014fqa}
Taniguchi K, Shibata M and Buonanno A 2015 {\em Phys. Rev.\/} {\bf D91} 024033
  [\eprint{1410.0738}]

\bibitem{Frolov:2008uf}
Frolov A~V 2008 {\em Phys. Rev. Lett.\/} {\bf 101} 061103 [\eprint{0803.2500}]

\bibitem{Khoury:2003aq}
Khoury J and Weltman A 2004 {\em Phys. Rev. Lett.\/} {\bf 93} 171104
  [\eprint{astro-ph/0309300}]

\bibitem{Gubser:2004uf}
Gubser S~S and Khoury J 2004 {\em Phys. Rev.\/} {\bf D70} 104001
  [\eprint{hep-ph/0405231}]

\bibitem{Starobinsky:2007hu}
Starobinsky A~A 2007 {\em JETP Lett.\/} {\bf 86} 157--163 [\eprint{0706.2041}]

\bibitem{Hu:2007nk}
Hu W and Sawicki I 2007 {\em Phys. Rev.\/} {\bf D76} 064004
  [\eprint{0705.1158}]

\bibitem{Yagi:2011xp}
Yagi K, Stein L~C, Yunes N and Tanaka T 2012 {\em Phys. Rev.\/} {\bf D85}
  064022 [\eprint{1110.5950}]

\bibitem{Demorest:2010bx}
Demorest P, Pennucci T, Ransom S, Roberts M and Hessels J 2010 {\em Nature\/}
  {\bf 467} 1081--1083 [\eprint{1010.5788}]

\bibitem{Antoniadis:2013pzd}
Antoniadis J, Freire P~C, Wex N, Tauris T~M, Lynch R~S {\em et~al.\/} 2013 {\em
  Science\/} {\bf 340} 6131 [\eprint{1304.6875}]

\bibitem{Lattimer:2004nj}
Lattimer J~M and Schutz B~F 2005 {\em Astrophys. J.\/} {\bf 629} 979--984
  [\eprint{astro-ph/0411470}]

\bibitem{Diaz-Alonso:1985}
Diaz-Alonso J and Iba\~{n}ez Cabanell J 1985 {\em Astrophys. J.\/} {\bf 291}
  308--318

\bibitem{Haensel:2004nu}
Haensel P and Potekhin A~Y 2004 {\em Astron. Astrophys.\/} {\bf 428} 191--197
  [\eprint{astro-ph/0408324}]

\bibitem{Akmal:1998cf}
Akmal A, Pandharipande V and Ravenhall D 1998 {\em Phys. Rev.\/} {\bf C58}
  1804--1828 [\eprint{nucl-th/9804027}]

\bibitem{2001A&A...380..151D}
{Douchin} F and {Haensel} P 2001 {\em \aap\/} {\bf 380} 151--167
  [\eprint{astro-ph/0111092}]

\bibitem{Shen:1998gq}
Shen H, Toki H, Oyamatsu K and Sumiyoshi K 1998 {\em Nucl. Phys.\/} {\bf A637}
  435--450 [\eprint{nucl-th/9805035}]

\bibitem{Shen:1998by}
Shen H, Toki H, Oyamatsu K and Sumiyoshi K 1998 {\em Prog. Theor. Phys.\/} {\bf
  100} 1013 [\eprint{nucl-th/9806095}]

\bibitem{Lattimer:1991nc}
Lattimer J~M and Swesty F~D 1991 {\em Nucl. Phys.\/} {\bf A535} 331--376

\bibitem{Damour:2002gp}
Damour T, Kogan I~I and Papazoglou A 2003 {\em Phys. Rev.\/} {\bf D67} 064009
  [\eprint{hep-th/0212155}]

\bibitem{Babichev:2009jt}
Babichev E, Deffayet C and Ziour R 2009 {\em Phys. Rev. Lett.\/} {\bf 103}
  201102 [\eprint{0907.4103}]

\bibitem{Read:2008iy}
Read J~S, Lackey B~D, Owen B~J and Friedman J~L 2009 {\em Phys. Rev.\/} {\bf
  D79} 124032 [\eprint{0812.2163}]

\bibitem{Avelino:2012ge}
Avelino P 2012 {\em Phys. Rev.\/} {\bf D85} 104053 [\eprint{1201.2544}]

\bibitem{Harko:2013aya}
Harko T, Lobo F~S~N, Mak M~K and Sushkov S~V 2015 {\em Mod. Phys. Lett.\/} {\bf
  A30} 1550190 [\eprint{1307.1883}]

\bibitem{Stairs:2003yja}
Stairs I~H 2003 {\em Living Rev. Relativ.\/} {\bf 6} 5

\bibitem{Benhar:1998au}
Benhar O, Berti E and Ferrari V 1999 {\em Mon. Not. R. Astron. Soc.\/} {\bf
  310} 797--803 [\eprint{gr-qc/9901037}]

\bibitem{Tsui:2004qd}
Tsui L and Leung P 2005 {\em Mon. Not. R. Astron. Soc.\/} {\bf 357} 1029--1037
  [\eprint{gr-qc/0412024}]

\bibitem{Gaertig2010}
{Gaertig} E and {Kokkotas} K~D 2011 {\em \prd\/} {\bf 83} 064031
  [\eprint{1005.5228}]

\bibitem{Doneva2013}
{Doneva} D~D, {Gaertig} E, {Kokkotas} K~D and {Kr{\"u}ger} C 2013 {\em \prd\/}
  {\bf 88} 044052 [\eprint{1305.7197}]

\bibitem{Lau:2009bu}
Lau H, Leung P and Lin L 2010 {\em Astrophys. J.\/} {\bf 714} 1234--1238
  [\eprint{0911.0131}]

\bibitem{1994ApJ...424..846R}
{Ravenhall} D~G and {Pethick} C~J 1994 {\em \apj\/} {\bf 424} 846--851

\bibitem{Lattimer:2000nx}
Lattimer J and Prakash M 2001 {\em Astrophys. J.\/} {\bf 550} 426
  [\eprint{astro-ph/0002232}]

\bibitem{Bejger:2002ty}
Bejger M and Haensel P 2002 {\em Astron. Astrophys.\/} {\bf 396} 917
  [\eprint{astro-ph/0209151}]

\bibitem{2013MNRAS.433.1903U}
{Urbanec} M, {Miller} J~C and {Stuchl{\'{\i}}k} Z 2013 {\em \mnras\/} {\bf 433}
  1903--1909 [\eprint{1301.5925}]

\bibitem{Lattimer:1989zz}
Lattimer J~M and Yahil A 1989 {\em Astrophys. J.\/} {\bf 340} 426--434

\bibitem{Prakash:1996xs}
Prakash M, Bombaci I, Prakash M, Ellis P~J, Lattimer J~M {\em et~al.\/} 1997
  {\em Phys. Rept.\/} {\bf 280} 1--77 [\eprint{nucl-th/9603042}]

\bibitem{Haensel:2009wa}
Haensel P, Zdunik J, Bejger M and Lattimer J 2009 {\em Astron. Astrophys.\/}
  {\bf 502} 605--610 [\eprint{0901.1268}]

\bibitem{1975NuPhA.237..507P}
{Pandharipande} V~R and {Smith} R~A 1975 {\em Nuclear Physics A\/} {\bf 237}
  507--532

\bibitem{Prakash:1995uw}
Prakash M, Cooke J and Lattimer J 1995 {\em Phys. Rev.\/} {\bf D52} 661--665

\bibitem{Lattimer:2012xj}
Lattimer J~M and Lim Y 2013 {\em Astrophys. J.\/} {\bf 771} 51
  [\eprint{1203.4286}]

\bibitem{2014MNRAS.438L..71H}
{Haskell} B, {Ciolfi} R, {Pannarale} F and {Rezzolla} L 2014 {\em \mnras\/}
  {\bf 438} L71--L75 [\eprint{1309.3885}]

\bibitem{Doneva:2013rha}
Doneva D~D, Yazadjiev S~S, Stergioulas N and Kokkotas K~D 2013 {\em Astrophys.
  J.\/} {\bf 781} L6 [\eprint{1310.7436}]

\bibitem{Stergioulas:1994ea}
Stergioulas N and Friedman J 1995 {\em Astrophys. J.\/} {\bf 444} 306
  [\eprint{astro-ph/9411032}]

\bibitem{Chakrabarti:2013tca}
Chakrabarti S, Delsate T, Gürlebeck N and Steinhoff J 2014 {\em Phys. Rev.
  Lett.\/} {\bf 112} 201102 [\eprint{1311.6509}]

\bibitem{Martinon:2014uua}
Martinon G, Maselli A, Gualtieri L and Ferrari V 2014 {\em Phys. Rev.\/} {\bf
  D90} 064026 [\eprint{1406.7661}]

\bibitem{Stein:2013ofa}
Stein L~C, Yagi K and Yunes N 2014 {\em Astrophys. J.\/} {\bf 788} 15
  [\eprint{1312.4532}]

\bibitem{Lai:1993ve}
Lai D, Rasio F~A and Shapiro S~L 1993 {\em Astrophys. J. Suppl.\/} {\bf 88}
  205--252

\bibitem{Chatziioannou:2014tha}
Chatziioannou K, Yagi K and Yunes N 2014 {\em Phys. Rev.\/} {\bf D90} 064030
  [\eprint{1406.7135}]

\bibitem{1978SvA....22..711S}
{Seidov} Z~F and {Kuzakhmedov} R~K 1978 {\em Sov. Astron.\/} {\bf 22} 711

\bibitem{Yagi:2014qua}
Yagi K, Stein L~C, Pappas G, Yunes N and Apostolatos T~A 2014 {\em Phys.
  Rev.\/} {\bf D90} 063010 [\eprint{1406.7587}]

\bibitem{Psaltis:2013zja}
Psaltis D and Özel F 2014 {\em Astrophys. J.\/} {\bf 792} 87
  [\eprint{1305.6615}]

\bibitem{Baubock:2013gna}
Bauböck M, Berti E, Psaltis D and Özel F 2013 {\em Astrophys. J.\/} {\bf 777}
  68 [\eprint{1306.0569}]

\bibitem{2012SPIE.8443E..13G}
{Gendreau} K~C, {Arzoumanian} Z and {Okajima} T 2012 {The Neutron star Interior
  Composition ExploreR (NICER): an Explorer mission of opportunity for soft
  x-ray timing spectroscopy} {\em Society of Photo-Optical Instrumentation
  Engineers (SPIE) Conference Series\/} vol 8443

\bibitem{2012AAS...21924906R}
{Ray} P~S, {Feroci} M, {den Herder} J, {Bozzo} E, {Stella} L and {LOFT
  Collaboration} 2012 {The Large Observatory for X-ray Timing (LOFT): An ESA
  M-class Mission Concept} {\em American Astronomical Society Meeting Abstracts
  \#219\/} p 249.06

\bibitem{Lo:2013ava}
Lo K~H, Coleman~Miller M, Bhattacharyya S and Lamb F~K 2013 {\em Astrophys.
  J.\/} {\bf 776} 19 [\eprint{1304.2330}]

\bibitem{jackiw:2003:cmo}
Jackiw R and Pi S~Y 2003 {\em Phys. Rev.\/} {\bf D68} 104012
  [\eprint{gr-qc/0308071}]

\bibitem{Sham:2013cya}
Sham Y~H, Lin L~M and Leung P 2014 {\em Astrophys. J.\/} {\bf 781} 66
  [\eprint{1312.1011}]

\bibitem{Pappas:2014gca}
Pappas G and Sotiriou T~P 2015 {\em Phys. Rev.\/} {\bf D91} 044011
  [\eprint{1412.3494}]

\bibitem{Damour:2009vw}
Damour T and Nagar A 2009 {\em Phys. Rev.\/} {\bf D80} 084035
  [\eprint{0906.0096}]

\bibitem{Postnikov:2010yn}
Postnikov S, Prakash M and Lattimer J~M 2010 {\em Phys. Rev.\/} {\bf D82}
  024016 [\eprint{1004.5098}]

\bibitem{Hinderer:2009ca}
Hinderer T, Lackey B~D, Lang R~N and Read J~S 2010 {\em Phys. Rev.\/} {\bf D81}
  123016 [\eprint{0911.3535}]

\bibitem{Kramer:2009zza}
Kramer M and Wex N 2009 {\em Class. Quantum Grav.\/} {\bf 26} 073001

\bibitem{Ozel:2008kb}
Ozel F, Guver T and Psaltis D 2009 {\em Astrophys. J.\/} {\bf 693} 1775--1779
  [\eprint{0810.1521}]

\bibitem{Ozel:2011ht}
Ozel F, Gould A and Guver T 2012 {\em Astrophys. J.\/} {\bf 748} 5
  [\eprint{1104.5027}]

\bibitem{Guver:2008gc}
Guver T, Ozel F, Cabrera-Lavers A and Wroblewski P 2010 {\em Astrophys. J.\/}
  {\bf 712} 964--973 [\eprint{0811.3979}]

\bibitem{Guver:2010td}
Guver T, Wroblewski P, Camarota L and Ozel F 2010 {\em Astrophys. J.\/} {\bf
  719} 1807 [\eprint{1002.3825}]

\bibitem{Guver:2013xa}
Guver T and Ozel F 2013 {\em Astrophys. J.\/} {\bf 765} L1 [\eprint{1301.0831}]

\bibitem{2011A&A...527A.139S}
{Suleimanov} V, {Poutanen} J and {Werner} K 2011 {\em Astron. Astrophys.\/}
  {\bf 527} A139 [\eprint{1009.6147}]

\bibitem{Zamfir:2011ht}
Zamfir M, Cumming A and Galloway D~K 2012 {\em Astrophys. J.\/} {\bf 749} 69
  [\eprint{1111.0347}]

\bibitem{Becker:2002xx}
Becker W, Swartz D~A, Pavlov G~G, Elsner R~F, Grindlay J {\em et~al.\/} 2003
  {\em Astrophys. J.\/} {\bf 594} 798--811 [\eprint{astro-ph/0211468}]

\bibitem{Gendre:2003pw}
Gendre B, Barret D and Webb N~A 2003 {\em Astron. Astrophys.\/} {\bf 403}
  L11--L14 [\eprint{astro-ph/0303471}]

\bibitem{Guillot:2010zp}
Guillot S, Rutledge R~E and Brown E~F 2011 {\em Astrophys. J.\/} {\bf 732} 88
  [\eprint{1007.2415}]

\bibitem{Gendre:2002rx}
Gendre B, Barret D and Webb N~A 2003 {\em Astron. Astrophys.\/} {\bf 400}
  521--532 [\eprint{astro-ph/0212173}]

\bibitem{Catuneanu:2013pz}
Catuneanu A, Heinke C~O, Sivakoff G~R, Ho W~C and Servillat M 2013 {\em
  Astrophys. J.\/} {\bf 764} 145 [\eprint{1301.3768}]

\bibitem{2012MNRAS.423.1556S}
{Servillat} M, {Heinke} C~O, {Ho} W~C~G, {Grindlay} J~E, {Hong} J, {van den
  Berg} M and {Bogdanov} S 2012 {\em Mon. Not. R. Astron. Soc.\/} {\bf 423}
  1556--1561 [\eprint{1203.5807}]

\bibitem{2002ApJ...578..405R}
{Rutledge} R~E, {Bildsten} L, {Brown} E~F, {Pavlov} G~G and {Zavlin} V~E 2002
  {\em \apj\/} {\bf 578} 405--412 [\eprint{astro-ph/0105405}]

\bibitem{2009MNRAS.392..665G}
{Guillot} S, {Rutledge} R~E, {Bildsten} L, {Brown} E~F, {Pavlov} G~G and
  {Zavlin} V~E 2009 {\em Mon. Not. R. Astron. Soc.\/} {\bf 392} 665--681
  [\eprint{0808.1305}]

\bibitem{1996Natur.379..233W}
{Walter} F~M, {Wolk} S~J and {Neuh{\"a}user} R 1996 {\em \nat\/} {\bf 379}
  233--235

\bibitem{Walter:2010ht}
Walter F, Eisenbeiss T, Lattimer J, Kim B, Hambaryan V {\em et~al.\/} 2010 {\em
  Astrophys. J.\/} {\bf 724} 669--677 [\eprint{1008.1709}]

\bibitem{Pons:2001px}
Pons J~A, Walter F~M, Lattimer J~M, Prakash M, Neuhauser R {\em et~al.\/} 2002
  {\em Astrophys. J.\/} {\bf 564} 981--1006 [\eprint{astro-ph/0107404}]

\bibitem{Rybicki:2005id}
Heinke C~O, Rybicki G~B, Narayan R and Grindlay J~E 2006 {\em Astrophys. J.\/}
  {\bf 644} 1090--1103 [\eprint{astro-ph/0506563}]

\bibitem{2007ApJ...671..727W}
{Webb} N~A and {Barret} D 2007 {\em \apj\/} {\bf 671} 727--733
  [\eprint{0708.3816}]

\bibitem{2003A&A...399.1109B}
{Burwitz} V, {Haberl} F, {Neuh{\"a}user} R, {Predehl} P, {Tr{\"u}mper} J and
  {Zavlin} V~E 2003 {\em Astron. Astrophys.\/} {\bf 399} 1109--1114
  [\eprint{astro-ph/0211536}]

\bibitem{2007MNRAS.375..821H}
{Ho} W~C~G, {Kaplan} D~L, {Chang} P, {van Adelsberg} M and {Potekhin} A~Y 2007
  {\em Mon. Not. R. Astron. Soc.\/} {\bf 375} 821--830
  [\eprint{astro-ph/0612145}]

\bibitem{Lattimer:2013hma}
Lattimer J~M and Steiner A~W 2014 {\em Astrophys. J.\/} {\bf 784} 123
  [\eprint{1305.3242}]

\bibitem{Binnington:2009bb}
Binnington T and Poisson E 2009 {\em Phys. Rev.\/} {\bf D80} 084018
  [\eprint{0906.1366}]

\bibitem{Landry:2014jka}
Landry P and Poisson E 2014 {\em Phys. Rev.\/} {\bf D89} 124011
  [\eprint{1404.6798}]

\bibitem{Yagi:2013sva}
Yagi K 2014 {\em Phys. Rev.\/} {\bf D89} 043011 [\eprint{1311.0872}]

\bibitem{Sathyaprakash:2009xt}
Sathyaprakash B, Schutz B and Van Den~Broeck C 2010 {\em Class. Quantum
  Grav.\/} {\bf 27} 215006 [\eprint{0906.4151}]

\bibitem{Punturo:2010zz}
Punturo M, Abernathy M, Acernese F, Allen B, Andersson N {\em et~al.\/} 2010
  {\em Class. Quantum Grav.\/} {\bf 27} 194002

\bibitem{Sathyaprakash:2012jk}
Sathyaprakash B, Abernathy M, Acernese F, Ajith P, Allen B {\em et~al.\/} 2012
  {\em Class. Quantum Grav.\/} {\bf 29} 124013 [\eprint{1206.0331}]

\bibitem{Broeck:2014cwa}
Van Den~Broeck C 2014 {\em J. Phys. Conf. Ser.\/} {\bf 484} 012008

\bibitem{Chan:2014kua}
Chan T, Sham Y~H, Leung P and Lin L~M 2014 {\em Phys. Rev.\/} {\bf D90} 124023
  [\eprint{1408.3789}]

\bibitem{Carriere:2002bx}
Carriere J, Horowitz C and Piekarewicz J 2003 {\em Astrophys. J.\/} {\bf 593}
  463--471 [\eprint{nucl-th/0211015}]

\bibitem{Foster:2007gr}
Foster B~Z 2007 {\em Phys. Rev.\/} {\bf D76} 084033 [\eprint{0706.0704}]

\bibitem{Ganguly:2013taa}
Ganguly A, Gannouji R, Goswami R and Ray S 2014 {\em Phys. Rev.\/} {\bf D89}
  064019 [\eprint{1309.3279}]

\bibitem{Psaltis:2010ww}
Psaltis D and Johannsen T 2012 {\em Astrophys. J.\/} {\bf 745} 1
  [\eprint{1011.4078}]

\bibitem{Baubock:2011ke}
Baubock M, Psaltis D, Ozel F and Johannsen T 2012 {\em Astrophys. J.\/} {\bf
  753} 175 [\eprint{1110.4389}]

\bibitem{Blanchet:2013haa}
Blanchet L 2014 {\em Living Rev. Relativ.\/} {\bf 17} 2 [\eprint{1310.1528}]

\bibitem{Damour:2013hea}
Damour T 2014 {The general relativistic two body problem} {\em Frontiers in
  Relativistic Celestial Mechanics\/} vol~1 ed Kopeikin S (De Gruyter) ISBN
  9783110337495 [\eprint{1312.3505}]

\bibitem{Buonanno:2014aza}
Buonanno A and Sathyaprakash B 2015 {Sources of Gravitational Waves: Theory and
  Observations} {\em General Relativity and Gravitation: A Centennial
  Perspective\/} ed Ashtekar A, Berger B, Isenberg J and MacCallum M~A~H
  (Cambridge University Press) ISBN 9781107037311 [\eprint{1410.7832}]

\bibitem{Pretorius:2007nq}
Pretorius F 2007 {\em {Binary Black Hole Coalescence}\/} (New York: Springer)
  chap~9 [\eprint{0710.1338}]

\bibitem{Centrella:2010mx}
Centrella J, Baker J~G, Kelly B~J and van Meter J~R 2010 {\em Rev. Mod.
  Phys.\/} {\bf 82} 3069 [\eprint{1010.5260}]

\bibitem{Sperhake:2011xk}
Sperhake U, Berti E and Cardoso V 2013 {\em Comptes Rendus Physique\/} {\bf 14}
  306--317 [\eprint{1107.2819}]

\bibitem{Pfeiffer:2012pc}
Pfeiffer H~P 2012 {\em Class. Quantum Grav.\/} {\bf 29} 124004
  [\eprint{1203.5166}]

\bibitem{Hannam:2013pra}
Hannam M 2014 {\em Gen. Relativ. Gravit.\/} {\bf 46} 1767 [\eprint{1312.3641}]

\bibitem{Lehner:2014asa}
Lehner L and Pretorius F 2014 {\em Ann. Rev. Astron. Astrophys.\/} {\bf 52} 661
  [\eprint{1405.4840}]

\bibitem{Will:1977wq}
Will C 1977 {\em Astrophys. J.\/} {\bf 214} 826--839

\bibitem{Lang:2013fna}
Lang R~N 2014 {\em Phys. Rev.\/} {\bf D89} 084014 [\eprint{1310.3320}]

\bibitem{Lang:2014osa}
Lang R~N 2015 {\em Phys. Rev.\/} {\bf D91} 084027 [\eprint{1411.3073}]

\bibitem{Will:1996zj}
Will C~M and Wiseman A~G 1996 {\em Phys. Rev.\/} {\bf D54} 4813--4848
  [\eprint{gr-qc/9608012}]

\bibitem{Pati:2000vt}
Pati M~E and Will C~M 2000 {\em Phys. Rev.\/} {\bf D62} 124015
  [\eprint{gr-qc/0007087}]

\bibitem{Pati:2002ux}
Pati M~E and Will C~M 2002 {\em Phys. Rev.\/} {\bf D65} 104008
  [\eprint{gr-qc/0201001}]

\bibitem{Healy:2011ef}
Healy J, Bode T, Haas R, Pazos E, Laguna P {\em et~al.\/} 2012 {\em Class.
  Quantum Grav.\/} {\bf 29} 232002 [\eprint{1112.3928}]

\bibitem{Berti:2013gfa}
Berti E, Cardoso V, Gualtieri L, Horbatsch M and Sperhake U 2013 {\em Phys.
  Rev.\/} {\bf D87} 124020 [\eprint{1304.2836}]

\bibitem{Krause:1994ar}
Krause D, Kloor H~T and Fischbach E 1994 {\em Phys. Rev.\/} {\bf D49}
  6892--6906

\bibitem{Perivolaropoulos:2009ak}
Perivolaropoulos L 2010 {\em Phys. Rev.\/} {\bf D81} 047501
  [\eprint{0911.3401}]

\bibitem{Sperhake:2014wpa}
Sperhake U 2015 {\em Class. Quantum Grav.\/} {\bf 32} 124011
  [\eprint{1411.3997}]

\bibitem{Cardoso:2014uka}
Cardoso V, Gualtieri L, Herdeiro C and Sperhake U 2015 {\em Living Rev.
  Relativ.\/} {\bf 18} [\eprint{1409.0014}]
  \urlprefix\url{http://www.livingreviews.org/lrr-2015-1}

\bibitem{Baumgarte2010}
Baumgarte T~W and Shapiro S~L 2010 {\em {Numerical Relativity}\/} (Cambridge
  University Press)

\bibitem{Salgado:2005hx}
Salgado M 2006 {\em Class. Quantum Grav.\/} {\bf 23} 4719--4742
  [\eprint{gr-qc/0509001}]

\bibitem{Horbatsch:2011ye}
Horbatsch M and Burgess C 2012 {\em JCAP\/} {\bf 1205} 010 [\eprint{1111.4009}]

\bibitem{Sahni:1999qe}
Sahni V and Wang L~M 2000 {\em Phys. Rev.\/} {\bf D62} 103517
  [\eprint{astro-ph/9910097}]

\bibitem{Hu:2000ke}
Hu W, Barkana R and Gruzinov A 2000 {\em Phys. Rev. Lett.\/} {\bf 85}
  1158--1161 [\eprint{astro-ph/0003365}]

\bibitem{Nordtvedt:1968qr}
Nordtvedt K 1968 {\em Phys. Rev.\/} {\bf 169} 1014--1016

\bibitem{Roll:1964rd}
Roll P~G, Krotkov R and Dicke R~H 1964 {\em Annals Phys.\/} {\bf 26} 442--517

\bibitem{Sampson:2014qqa}
Sampson L, Yunes N, Cornish N, Ponce M, Barausse E {\em et~al.\/} 2014 {\em
  Phys. Rev.\/} {\bf D90} 124091 [\eprint{1407.7038}]

\bibitem{Ponce:2014hha}
Ponce M, Palenzuela C, Barausse E and Lehner L 2015 {\em Phys. Rev.\/} {\bf
  D91} 084038 [\eprint{1410.0638}]

\bibitem{Naf:2010zy}
Naf J and Jetzer P 2010 {\em Phys. Rev.\/} {\bf D81} 104003
  [\eprint{1004.2014}]

\bibitem{Naf:2011za}
Naf J and Jetzer P 2011 {\em Phys. Rev.\/} {\bf D84} 024027
  [\eprint{1104.2200}]

\bibitem{DeLaurentis:2011tp}
De~Laurentis M and Capozziello S 2011 {\em Astropart. Phys.\/} {\bf 35}
  257--265 [\eprint{1104.1942}]

\bibitem{DeLaurentis:2013zv}
De~Laurentis M and De~Martino I 2013 {\em Mon. Not. R. Astron. Soc.\/} {\bf
  431} 741--748 [\eprint{1302.0220}]

\bibitem{Stein:2013wza}
Stein L~C and Yagi K 2014 {\em Phys. Rev.\/} {\bf D89} 044026
  [\eprint{1310.6743}]

\bibitem{Hansen:2014ewa}
Hansen D, Yunes N and Yagi K 2015 {\em Phys. Rev.\/} {\bf D91} 082003
  [\eprint{1412.4132}]

\bibitem{Goldberger:2004jt}
Goldberger W~D and Rothstein I~Z 2006 {\em Phys. Rev.\/} {\bf D73} 104029
  [\eprint{hep-th/0409156}]

\bibitem{Weisberg:2004hi}
Weisberg J~M and Taylor J~H 2005 {\em ASP Conf.Ser.\/} {\bf 328} 25
  [\eprint{astro-ph/0407149}]

\bibitem{Narikawa:2014fua}
Narikawa T, Ueno K, Tagoshi H, Tanaka T, Kanda N {\em et~al.\/} 2015 {\em Phys.
  Rev.\/} {\bf D91} 062007 [\eprint{1412.8074}]

\bibitem{Hulse:1974eb}
Hulse R and Taylor J 1975 {\em Astrophys. J.\/} {\bf 195} L51--L53

\bibitem{Hobbs:2012apa}
Hobbs G, Coles W, Manchester R, Keith M, Shannon R {\em et~al.\/} 2012 {\em
  Mon. Not. R. Astron. Soc.\/} {\bf 427} 2780–2787 [\eprint{1208.3560}]

\bibitem{Manchester:2004bp}
Manchester R~N, Hobbs G~B, Teoh A and Hobbs M 2005 {\em Astron. J.\/} {\bf 129}
  1993 [\eprint{astro-ph/0412641}]

\bibitem{Kramer:2012zz}
Kramer M 2012 {\em IAU Proc.\/} {\bf 8} 19--26 [\eprint{1211.2457}]

\bibitem{1988grra.conf..315D}
{Damour} T 1988 {Strong-field tests of general relativity and the binary
  pulsar.} {\em Proceedings of the 2nd Canadian Conference on General
  Relativity and Relativistic Astrophysics\/} ed {Coley} A, {Dyer} C and
  {Tupper} T pp 315--334

\bibitem{Damour:1991rd}
Damour T and Taylor J~H 1992 {\em Phys. Rev.\/} {\bf D45} 1840--1868

\bibitem{1986AIHS...44..263D}
{Damour} T and {Deruelle} N 1986 {\em Ann. Inst. Henri Poincar{\'e} Phys.
  Th{\'e}or.,\/} {\bf 44} 263--292

\bibitem{Taylor:1979zz}
Taylor J, Fowler L and McCulloch P 1979 {\em Nature\/} {\bf 277} 437--440

\bibitem{Weisberg:2010zz}
Weisberg J, Nice D and Taylor J 2010 {\em Astrophys. J.\/} {\bf 722} 1030--1034
  [\eprint{1011.0718}]

\bibitem{Burgay:2003jj}
Burgay M, D'Amico N, Possenti A, Manchester R, Lyne A {\em et~al.\/} 2003 {\em
  Nature\/} {\bf 426} 531--533 [\eprint{astro-ph/0312071}]

\bibitem{Lyne:2004cj}
Lyne A, Burgay M, Kramer M, Possenti A, Manchester R {\em et~al.\/} 2004 {\em
  Science\/} {\bf 303} 1153--1157 [\eprint{astro-ph/0401086}]

\bibitem{Perera:2010sp}
Perera B {\em et~al.\/} 2010 {\em Astrophys. J.\/} {\bf 721} 1193--1205
  [\eprint{1008.1097}]

\bibitem{Breton:2008xy}
Breton R~P, Kaspi V~M, Kramer Michael, McLaughlin M~A, Lyutikov M {\em
  et~al.\/} 2008 {\em Science\/} {\bf 321} 104--107 [\eprint{0807.2644}]

\bibitem{Antoniadis:2012vy}
Antoniadis J, van Kerkwijk M, Koester D, Freire P, Wex N {\em et~al.\/} 2012
  {\em Mon. Not. R. Astron. Soc.\/} {\bf 423} 3316 [\eprint{1204.3948}]

\bibitem{Shao:2012eg}
Shao L and Wex N 2012 {\em Class. Quantum Grav.\/} {\bf 29} 215018
  [\eprint{1209.4503}]

\bibitem{Esposito-Farese:2011cha}
Esposito-Farèse G 2011 {\em Fundam.Theor.Phys.\/} {\bf 162} 461--489

\bibitem{Damour:1991rq}
Damour T and Schaefer G 1991 {\em Phys. Rev. Lett.\/} {\bf 66} 2549--2552

\bibitem{Stairs:2005hu}
Stairs I~H, Faulkner A, Lyne A, Kramer M, Lorimer D {\em et~al.\/} 2005 {\em
  Astrophys. J.\/} {\bf 632} 1060--1068 [\eprint{astro-ph/0506188}]

\bibitem{Gonzalez:2011kt}
Gonzalez M, Stairs I, Ferdman R, Freire P, Nice D {\em et~al.\/} 2011 {\em
  Astrophys. J.\/} {\bf 743} 102 [\eprint{1109.5638}]

\bibitem{Ransom:2014xla}
{Ransom} S~M {\em et~al.\/} 2014 {\em \nat\/} {\bf 505} 520--524
  [\eprint{1401.0535}]

\bibitem{Freire:2012nb}
Freire P~C, Kramer M and Wex N 2012 {\em Class. Quantum Grav.\/} {\bf 29}
  184007 [\eprint{1205.3751}]

\bibitem{bhat}
Bhat N~R, Bailes M and Verbiest J~P 2008 {\em Phys. Rev.\/} {\bf D77} 124017
  [\eprint{0804.0956}]

\bibitem{kramer-double-pulsar}
Kramer M, Stairs I~H, Manchester R, McLaughlin M, Lyne A {\em et~al.\/} 2006
  {\em Science\/} {\bf 314} 97--102 [\eprint{astro-ph/0609417}]

\bibitem{Audren:2013dwa}
Audren B, Blas D, Lesgourgues J and Sibiryakov S 2013 {\em JCAP\/} {\bf 1308}
  039 [\eprint{1305.0009}]

\bibitem{Audren:2014hza}
Audren B, Blas D, Ivanov M, Lesgourgues J and Sibiryakov S 2015 {\em JCAP\/}
  {\bf 1503} 016 [\eprint{1410.6514}]

\bibitem{Nan:2011um}
Nan R, Li D, Jin C, Wang Q, Zhu L {\em et~al.\/} 2011 {\em Int. J. Mod.
  Phys.\/} {\bf D20} 989--1024 [\eprint{1105.3794}]

\bibitem{Lazio:2013mea}
Lazio T 2013 {\em Class. Quantum Grav.\/} {\bf 30} 224011

\bibitem{Blanchet:1989cu}
Blanchet L and Schaefer G 1989 {\em Mon. Not. R. Astron. Soc.\/} {\bf 239}
  845--867

\bibitem{Damour:1988mr}
Damour T and Sch{\"a}fer G 1988 {\em Nuovo Cim.\/} {\bf B101} 127

\bibitem{Wex:1998wt}
Wex N and Kopeikin S 1999 {\em Astrophys. J.\/} {\bf 514} 388
  [\eprint{astro-ph/9811052}]

\bibitem{Liu:2011ae}
Liu K, Wex N, Kramer M, Cordes J and Lazio T 2012 {\em Astrophys. J.\/} {\bf
  747} 1 [\eprint{1112.2151}]

\bibitem{Wex:2012au}
Wex N, Liu K, Eatough R, Kramer M, Cordes J {\em et~al.\/} 2012 {\em IAU
  Proc.\/} {\bf 8} 171--176 [\eprint{1210.7518}]

\bibitem{Liu:2014uka}
Liu K, Eatough R, Wex N and Kramer M 2014 {\em Mon. Not. R. Astron. Soc.\/}
  {\bf 445} 3115 [\eprint{1409.3882}]

\bibitem{IPTA):2013lea}
Manchester R 2013 {\em Class. Quantum Grav.\/} {\bf 30} 224010

\bibitem{Lee:2013sxl}
Lee K 2013 {\em Class. Quantum Grav.\/} {\bf 30} 224016 [\eprint{1404.2090}]

\bibitem{Corbin:2010kt}
Corbin V and Cornish N~J 2010  [\eprint{1008.1782}]

\bibitem{Lee:2011et}
Lee K, Wex N, Kramer M, Stappers B, Bassa C {\em et~al.\/} 2011 {\em Mon. Not.
  R. Astron. Soc.\/} {\bf 414} 3251 [\eprint{1103.0115}]

\bibitem{Ade:2013zuv}
Ade P {\em et~al.\/} (Planck Collaboration) 2014 {\em Astron. Astrophys.\/}
  {\bf 571} A16 [\eprint{1303.5076}]

\bibitem{Peacock:1999ye}
Peacock J 1999 {\em Cosmological Physics\/} Cambridge Astrophysics (Cambridge
  University Press) ISBN 9780521422703

\bibitem{Copeland:2006wr}
Copeland E~J, Sami M and Tsujikawa S 2006 {\em Int. J. Mod. Phys.\/} {\bf D15}
  1753--1936 [\eprint{hep-th/0603057}]

\bibitem{Will:2011nz}
Will C~M 2011 {\em Proc.Nat.Acad.Sci.\/} {\bf 108} 5938 [\eprint{1102.5192}]

\bibitem{Battye:2012eu}
Battye R~A and Pearson J~A 2012 {\em JCAP\/} {\bf 1207} 019
  [\eprint{1203.0398}]

\bibitem{Battye:2013aaa}
Battye R~A and Pearson J~A 2013 {\em Phys. Rev.\/} {\bf D88} 061301
  [\eprint{1306.1175}]

\bibitem{Gergely:2014rna}
Gergely L~A and Tsujikawa S 2014 {\em Phys. Rev.\/} {\bf D89} 064059
  [\eprint{1402.0553}]

\bibitem{Gubitosi:2012hu}
Gubitosi G, Piazza F and Vernizzi F 2013 {\em JCAP\/} {\bf 1302} 032
  [\eprint{1210.0201}]

\bibitem{Bloomfield:2012ff}
Bloomfield J~K, Flanagan E~E, Park M and Watson S 2013 {\em JCAP\/} {\bf 1308}
  010 [\eprint{1211.7054}]

\bibitem{Gleyzes:2013ooa}
Gleyzes J, Langlois D, Piazza F and Vernizzi F 2013 {\em JCAP\/} {\bf 1308} 025
  [\eprint{1304.4840}]

\bibitem{Gleyzes:2014rba}
Gleyzes J, Langlois D and Vernizzi F 2015 {\em Int. J. Mod. Phys.\/} {\bf D23}
  1443010 [\eprint{1411.3712}]

\bibitem{Gao:2014soa}
Gao X 2014 {\em Phys. Rev.\/} {\bf D90} 081501 [\eprint{1406.0822}]

\bibitem{Cheung:2007st}
Cheung C, Fitzpatrick A~L, Kaplan J, Senatore L and Creminelli P 2008 {\em
  JHEP\/} {\bf 0803} 014 [\eprint{0709.0293}]

\bibitem{Baker:2012zs}
Baker T, Ferreira P~G and Skordis C 2013 {\em Phys. Rev.\/} {\bf D87} 024015
  [\eprint{1209.2117}]

\bibitem{Silvestri:2013ne}
Silvestri A, Pogosian L and Buniy R~V 2013 {\em Phys. Rev.\/} {\bf D87} 104015
  [\eprint{1302.1193}]

\bibitem{Oyaizu:2008sr}
Oyaizu H 2008 {\em Phys. Rev.\/} {\bf D78} 123523 [\eprint{0807.2449}]

\bibitem{Noller:2013wca}
Noller J, von Braun-Bates F and Ferreira P~G 2014 {\em Phys. Rev.\/} {\bf D89}
  023521 [\eprint{1310.3266}]

\bibitem{Brax:2013mua}
Brax P, Davis A~C, Li B, Winther H~A and Zhao G~b 2013 {\em JCAP\/} {\bf 1304}
  029 [\eprint{1303.0007}]

\bibitem{Brax:2012nk}
Brax P, Davis A~C, Li B, Winther H~A and Zhao G~b 2012 {\em JCAP\/} {\bf 1210}
  002 [\eprint{1206.3568}]

\bibitem{Llinares:2013qbh}
Llinares C and Mota D 2013 {\em Phys. Rev. Lett.\/} {\bf 110} 161101
  [\eprint{1302.1774}]

\bibitem{Comelli:2014fg}
{Comelli} D, {Crisostomi} M and {Pilo} L 2014 {\em \prd\/} {\bf 90} 084003
  [\eprint{1403.5679}]

\bibitem{DeFelice:2011hq}
De~Felice A, Kobayashi T and Tsujikawa S 2011 {\em Phys. Lett.\/} {\bf B706}
  123--133 [\eprint{1108.4242}]

\bibitem{Amendola:2012ky}
Amendola L, Kunz M, Motta M, Saltas I~D and Sawicki I 2013 {\em Phys. Rev.\/}
  {\bf D87} 023501 [\eprint{1210.0439}]

\bibitem{Baker:2014tc}
Baker T, Ferreira P~G, Leonard C~D and Motta M 2014 {\em Phys. Rev.\/} {\bf
  D90} 124030 [\eprint{1409.8284}]

\bibitem{Leonard:2015hha}
Leonard C~D, Baker T and Ferreira P~G 2015 {\em Phys. Rev.\/} {\bf D91} 083504
  [\eprint{1501.03509}]

\bibitem{LSST}
http://www.lsst.org/lsst/

\bibitem{Hojjati:2013xqa}
Hojjati A, Pogosian L, Silvestri A and Zhao G~b 2014 {\em Phys. Rev.\/} {\bf
  D89} 083505 [\eprint{1312.5309}]

\bibitem{Johnson:2014kaa}
Johnson A, Blake C, Koda J, Ma Y~Z, Colless M {\em et~al.\/} 2014 {\em Mon.
  Not. R. Astron. Soc.\/} {\bf 444} 3926 [\eprint{1404.3799}]

\bibitem{Khoury:2003rn}
Khoury J and Weltman A 2004 {\em Phys. Rev.\/} {\bf D69} 044026
  [\eprint{astro-ph/0309411}]

\bibitem{Mota:2006fz}
Mota D~F and Shaw D~J 2007 {\em Phys. Rev.\/} {\bf D75} 063501
  [\eprint{hep-ph/0608078}]

\bibitem{Hinterbichler:2010es}
Hinterbichler K and Khoury J 2010 {\em Phys. Rev. Lett.\/} {\bf 104} 231301
  [\eprint{1001.4525}]

\bibitem{Deffayet:2009wt}
Deffayet C, Esposito-Farese G and Vikman A 2009 {\em Phys. Rev.\/} {\bf D79}
  084003 [\eprint{0901.1314}]

\bibitem{Llinares:2013jza}
Llinares C, Mota D~F and Winther H~A 2014 {\em Astron. Astrophys.\/} {\bf 562}
  A78 [\eprint{1307.6748}]

\bibitem{Li:2011vk}
Li B, Zhao G~b, Teyssier R and Koyama K 2012 {\em JCAP\/} {\bf 1201} 051
  [\eprint{1110.1379}]

\bibitem{Puchwein:2013lza}
Puchwein E, Baldi M and Springel V 2013 {\em Mon. Not. R. Astron. Soc.\/} {\bf
  436} 348 [\eprint{1305.2418}]

\bibitem{Davis:2011pj}
Davis A~C, Li B, Mota D~F and Winther H~A 2012 {\em Astrophys. J.\/} {\bf 748}
  61 [\eprint{1108.3081}]

\bibitem{Schmidt:2009sv}
Schmidt F 2009 {\em Phys. Rev.\/} {\bf D80} 123003 [\eprint{0910.0235}]

\bibitem{Schmidt:2009sg}
Schmidt F 2009 {\em Phys. Rev.\/} {\bf D80} 043001 [\eprint{0905.0858}]

\bibitem{Barreira:2013eea}
Barreira A, Li B, Hellwing W~A, Baugh C~M and Pascoli S 2013 {\em JCAP\/} {\bf
  1310} 027 [\eprint{1306.3219}]

\bibitem{Schmidt:2009am}
Schmidt F, Vikhlinin A and Hu W 2009 {\em Phys. Rev.\/} {\bf D80} 083505
  [\eprint{0908.2457}]

\bibitem{Schmidt:2008tn}
Schmidt F, Lima M~V, Oyaizu H and Hu W 2009 {\em Phys. Rev.\/} {\bf D79} 083518
  [\eprint{0812.0545}]

\bibitem{Schmidt:2010jr}
Schmidt F 2010 {\em Phys. Rev.\/} {\bf D81} 103002 [\eprint{1003.0409}]

\bibitem{Zhao:2011cu}
Zhao G~b, Li B and Koyama K 2011 {\em Phys. Rev. Lett.\/} {\bf 107} 071303
  [\eprint{1105.0922}]

\bibitem{Hellwing:2014nma}
Hellwing W~A, Barreira A, Frenk C~S, Li B and Cole S 2014 {\em Phys. Rev.
  Lett.\/} {\bf 112} 221102 [\eprint{1401.0706}]

\bibitem{Lam:2012by}
Lam T~Y, Nishimichi T, Schmidt F and Takada M 2012 {\em Phys. Rev. Lett.\/}
  {\bf 109} 051301 [\eprint{1202.4501}]

\bibitem{Shim:2014uta}
Shim J, Lee J and Baldi M 2014  [\eprint{1404.3639}]

\bibitem{Avilez:2013dxa}
Avilez A and Skordis C 2014 {\em Phys. Rev. Lett.\/} {\bf 113} 011101
  [\eprint{1303.4330}]

\bibitem{Lombriser:2010mp}
Lombriser L, Slosar A, Seljak U and Hu W 2012 {\em Phys. Rev.\/} {\bf D85}
  124038 [\eprint{1003.3009}]

\bibitem{Barreira:2012kk}
Barreira A, Li B, Baugh C~M and Pascoli S 2012 {\em Phys. Rev.\/} {\bf D86}
  124016 [\eprint{1208.0600}]

\bibitem{Zuntz:2008zz}
Zuntz J~A, Ferreira P and Zlosnik T 2008 {\em Phys. Rev. Lett.\/} {\bf 101}
  261102 [\eprint{0808.1824}]

\bibitem{Koivisto:2008xf}
Koivisto T and Mota D~F 2008 {\em JCAP\/} {\bf 0808} 021 [\eprint{0805.4229}]

\bibitem{Akrami:2013ffa}
Akrami Y, Koivisto T~S, Mota D~F and Sandstad M 2013 {\em JCAP\/} {\bf 1310}
  046 [\eprint{1306.0004}]

\bibitem{Konnig:2014xva}
Koennig F, Akrami Y, Amendola L, Motta M and Solomon A~R 2014 {\em Phys.
  Rev.\/} {\bf D90} 124014 [\eprint{1407.4331}]

\bibitem{Bean:2010zq}
Bean R and Tangmatitham M 2010 {\em Phys. Rev.\/} {\bf D81} 083534
  [\eprint{1002.4197}]

\bibitem{Zhao:2013nka}
Zhao H, Peacock J~A and Li B 2013 {\em Phys. Rev.\/} {\bf D88} 043013

\bibitem{Hui:2012jb}
Hui L and Nicolis A 2012 {\em Phys. Rev. Lett.\/} {\bf 109} 051304
  [\eprint{1201.1508}]

\bibitem{Davis:2011qf}
Davis A~C, Lim E~A, Sakstein J and Shaw D 2012 {\em Phys. Rev.\/} {\bf D85}
  123006 [\eprint{1102.5278}]

\bibitem{Ferraro:2010gh}
Ferraro S, Schmidt F and Hu W 2011 {\em Phys. Rev.\/} {\bf D83} 063503
  [\eprint{1011.0992}]

\bibitem{Penarrubia:2012ab}
Penarrubia J, Koposov S~E and Walker M~G 2012 {\em Astrophys. J.\/} {\bf 760} 2
  [\eprint{1209.2126}]

\bibitem{Ade:2014xna}
Ade P {\em et~al.\/} (BICEP2 Collaboration) 2014 {\em Phys. Rev. Lett.\/} {\bf
  112} 241101 [\eprint{1403.3985}]

\bibitem{Flauger:2014qra}
Flauger R, Hill J~C and Spergel D~N 2014 {\em JCAP\/} {\bf 1408} 039
  [\eprint{1405.7351}]

\bibitem{Adam:2015wua}
Adam R {\em et~al.\/} (Planck) 2015  [\eprint{1502.01588}]

\bibitem{Ade:2015tva}
Ade P {\em et~al.\/} (BICEP2, Planck) 2015 {\em Phys. Rev. Lett.\/} {\bf 114}
  101301 [\eprint{1502.00612}]

\bibitem{Ade:2015lrj}
Ade P {\em et~al.\/} (Planck) 2015  [\eprint{1502.02114}]

\bibitem{Calabrese:2014gwa}
Calabrese E, Hložek R, Battaglia N, Bond J~R, de~Bernardis F {\em et~al.\/}
  2014 {\em JCAP\/} {\bf 1408} 010 [\eprint{1406.4794}]

\bibitem{Benson:2014qhw}
Benson B {\em et~al.\/} (SPT-3G) 2014 {\em Proc.SPIE Int.Soc.Opt.Eng.\/} {\bf
  9153} 91531P [\eprint{1407.2973}]

\bibitem{Ryan:1997hg}
Ryan F~D 1997 {\em Phys. Rev.\/} {\bf D56} 1845--1855

\bibitem{Barack:2006pq}
Barack L and Cutler C 2007 {\em Phys. Rev.\/} {\bf D75} 042003
  [\eprint{gr-qc/0612029}]

\bibitem{AmaroSeoane:2007aw}
Amaro-Seoane P, Gair J~R, Freitag M, Miller M~C, Mandel I, Cutler C~J and Babak
  S 2007 {\em Class. Quantum Grav.\/} {\bf 24} R113--R169
  [\eprint{astro-ph/0703495}]

\bibitem{Gair:2008bx}
Gair J~R 2009 {\em Class. Quantum Grav.\/} {\bf 26} 094034 [\eprint{0811.0188}]

\bibitem{Pani:2010em}
Pani P, Berti E, Cardoso V, Chen Y and Norte R 2010 {\em Phys. Rev.\/} {\bf
  D81} 084011 [\eprint{1001.3031}]

\bibitem{Berti:2007zu}
Berti E, Cardoso J, Cardoso V and Cavagli{\`a} M 2007 {\em Phys. Rev.\/} {\bf
  D76} 104044 [\eprint{0707.1202}]

\bibitem{Gossan:2011ha}
Gossan S, Veitch J and Sathyaprakash B 2012 {\em Phys. Rev.\/} {\bf D85} 124056
  [\eprint{1111.5819}]

\bibitem{Schutz:1986gp}
Schutz B~F 1986 {\em Nature\/} {\bf 323} 310--311

\bibitem{Holz:2005df}
Holz D~E and Hughes S~A 2005 {\em Astrophys. J.\/} {\bf 629} 15--22
  [\eprint{astro-ph/0504616}]

\bibitem{Shapiro:2009sr}
Shapiro C, Bacon D, Hendry M and Hoyle B 2010 {\em Mon. Not. R. Astron. Soc.\/}
  {\bf 404} 858--866 [\eprint{0907.3635}]

\bibitem{Messenger:2011gi}
Messenger C and Read J 2012 {\em Phys. Rev. Lett.\/} {\bf 108} 091101
  [\eprint{1107.5725}]

\bibitem{Messenger:2013fya}
Messenger C, Takami K, Gossan S, Rezzolla L and Sathyaprakash B 2014 {\em Phys.
  Rev.\/} {\bf X4} 041004 [\eprint{1312.1862}]

\bibitem{DelPozzo:2012zz}
Del~Pozzo W 2012 {\em Phys. Rev.\/} {\bf D86} 043011

\bibitem{Taylor:2011fs}
Taylor S~R, Gair J~R and Mandel I 2012 {\em Phys. Rev.\/} {\bf D85} 023535
  [\eprint{1108.5161}]

\bibitem{Taylor:2012db}
Taylor S~R and Gair J~R 2012 {\em Phys. Rev.\/} {\bf D86} 023502
  [\eprint{1204.6739}]

\bibitem{Seto:2001qf}
Seto N, Kawamura S and Nakamura T 2001 {\em Phys. Rev. Lett.\/} {\bf 87} 221103
  [\eprint{astro-ph/0108011}]

\bibitem{Takahashi:2004yr}
Takahashi R and Nakamura T 2005 {\em Prog. Theor. Phys.\/} {\bf 113} 63--71
  [\eprint{astro-ph/0408547}]

\bibitem{Nishizawa:2011eq}
Nishizawa A, Yagi K, Taruya A and Tanaka T 2012 {\em Phys. Rev.\/} {\bf D85}
  044047 [\eprint{1110.2865}]

\bibitem{Nishizawa:2012vk}
Nishizawa A, Yagi K, Taruya A and Tanaka T 2012 {\em J. Phys. Conf. Ser.\/}
  {\bf 363} 012052 [\eprint{1204.2877}]

\bibitem{Loeb:1998bu}
Loeb A 1998 {\em Astrophys. J.\/} {\bf 499} L111--L114
  [\eprint{astro-ph/9802122}]

\bibitem{Quartin:2009xr}
Quartin M and Amendola L 2010 {\em Phys. Rev.\/} {\bf D81} 043522
  [\eprint{0909.4954}]

\bibitem{Yagi:2011bt}
Yagi K, Nishizawa A and Yoo C~M 2012 {\em JCAP\/} {\bf 1204} 031
  [\eprint{1112.6040}]

\bibitem{Yagi:2012vx}
Yagi K, Nishizawa A and Yoo C~M 2012 {\em J. Phys. Conf. Ser.\/} {\bf 363}
  012056 [\eprint{1204.1670}]

\bibitem{Dalal:2006qt}
Dalal N, Holz D~E, Hughes S~A and Jain B 2006 {\em Phys. Rev.\/} {\bf D74}
  063006 [\eprint{astro-ph/0601275}]

\bibitem{Zhao:2010sz}
Zhao W, Van Den~Broeck C, Baskaran D and Li T~G~F 2011 {\em Phys. Rev.\/} {\bf
  D83} 023005 [\eprint{1009.0206}]

\bibitem{1985PhRvD..31.2480K}
{Krisher} T~P and {Will} C~M 1985 {\em \prd\/} {\bf 31} 2480--2487

\bibitem{1989PhRvD..40.3884G}
{G{\"u}rsel} Y and {Tinto} M 1989 {\em \prd\/} {\bf 40} 3884--3938

\bibitem{2006PhRvD..74h2005C}
{Chatterji} S, {Lazzarini} A, {Stein} L, {Sutton} P~J, {Searle} A and {Tinto} M
  2006 {\em \prd\/} {\bf 74} 082005 [\eprint{gr-qc/0605002}]

\bibitem{2008ApJ...685.1304L}
{Lee} K~J, {Jenet} F~A and {Price} R~H 2008 {\em \apj\/} {\bf 685} 1304--1319

\bibitem{2012CQGra..29g5011A}
{Arun} K~G 2012 {\em Class. Quantum Grav.\/} {\bf 29} 075011
  [\eprint{1202.5911}]

\bibitem{2012PhRvD..86b2004C}
{Chatziioannou} K, {Yunes} N and {Cornish} N 2012 {\em \prd\/} {\bf 86} 022004
  [\eprint{1204.2585}]

\bibitem{Isi:2015cva}
Isi M, Weinstein A~J, Mead C and Pitkin M 2015 {\em Phys. Rev.\/} {\bf D91}
  082002 [\eprint{1502.00333}]

\bibitem{Hayama:2012au}
Hayama K and Nishizawa A 2013 {\em Phys. Rev.\/} {\bf D87} 062003
  [\eprint{1208.4596}]

\bibitem{Maggiore2007b}
Maggiore M 2007 {\em Gravitational Waves\/} (Oxford University Press) ISBN
  9780198570745

\bibitem{Creighton:2011zz}
Creighton J~D and Anderson W~G 2011 {\em {Gravitational-wave physics and
  astronomy: An introduction to theory, experiment and data analysis}\/}
  (Weinheim, Germany: Wiley-VCH)

\bibitem{Veitch:2014wba}
Veitch J, Raymond V, Farr B, Farr W, Graff P {\em et~al.\/} 2015 {\em Phys.
  Rev.\/} {\bf D91} 042003 [\eprint{1409.7215}]

\bibitem{Berti:2006ew}
Berti E 2006 {\em Class. Quantum Grav.\/} {\bf 23} S785--S798
  [\eprint{astro-ph/0602470}]

\bibitem{Cutler:2007mi}
Cutler C and Vallisneri M 2007 {\em Phys. Rev.\/} {\bf D76} 104018
  [\eprint{0707.2982}]

\bibitem{Sathyaprakash:1991az}
Sathyaprakash B and Dhurandhar S 1991 {\em Phys. Rev.\/} {\bf D44} 3819--3834

\bibitem{Dhurandhar:1992mw}
Dhurandhar S and Sathyaprakash B 1994 {\em Phys. Rev.\/} {\bf D49} 1707--1722

\bibitem{Finn:1992xs}
Finn L~S and Chernoff D~F 1993 {\em Phys. Rev.\/} {\bf D47} 2198--2219
  [\eprint{gr-qc/9301003}]

\bibitem{2006CQGra..23L..37A}
{Arun} K~G, {Iyer} B~R, {Qusailah} M~S~S and {Sathyaprakash} B~S 2006 {\em
  Class. Quantum Grav.\/} {\bf 23} L37--L43 [\eprint{gr-qc/0604018}]

\bibitem{2006PhRvD..74b4006A}
{Arun} K~G, {Iyer} B~R, {Qusailah} M~S~S and {Sathyaprakash} B~S 2006 {\em
  \prd\/} {\bf 74} 024006 [\eprint{gr-qc/0604067}]

\bibitem{Mishra:2010tp}
Mishra C~K, Arun K, Iyer B~R and Sathyaprakash B 2010 {\em Phys. Rev.\/} {\bf
  D82} 064010

\bibitem{Blanchet:1995xs}
Blanchet L and Sathyaprakash B~S 1995 {\em Phys. Rev. Lett.\/} {\bf 74}
  1067--1070

\bibitem{Li:2011cg}
Li T~G~F, Del~Pozzo W, Vitale S, Van Den~Broeck C, Agathos M {\em et~al.\/}
  2012 {\em Phys. Rev.\/} {\bf D85} 082003 [\eprint{1110.0530}]

\bibitem{2011PhRvD..83h2002D}
{Del Pozzo} W, {Veitch} J and {Vecchio} A 2011 {\em \prd\/} {\bf 83} 082002
  [\eprint{1101.1391}]

\bibitem{Vishveshwara:1970cc}
Vishveshwara C 1970 {\em Phys. Rev.\/} {\bf D1} 2870--2879

\bibitem{Ruffini:1971}
Ruffini R~J and Wheeler J~A 1971 {\em Physics Today\/} {\bf 24} 30

\bibitem{Dreyer:2003bv}
Dreyer O, Kelly B~J, Krishnan B, Finn L~S, Garrison D and Lopez-Aleman R 2004
  {\em Class. Quantum Grav.\/} {\bf 21} 787--804 [\eprint{gr-qc/0309007}]

\bibitem{Detweiler:1978ge}
Detweiler S~L 1978 {\em Astrophys. J.\/} {\bf 225} 687--693

\bibitem{Kamaretsos:2011um}
Kamaretsos I, Hannam M, Husa S and Sathyaprakash B~S 2012 {\em Phys. Rev.\/}
  {\bf D85} 024018 [\eprint{1107.0854}]

\bibitem{2014PhRvD..90f4009M}
{Meidam} J, {Agathos} M, {Van Den Broeck} C, {Veitch} J and {Sathyaprakash} B~S
  2014 {\em \prd\/} {\bf 90} 064009 [\eprint{1406.3201}]

\bibitem{Kamaretsos:2012bs}
Kamaretsos I, Hannam M and Sathyaprakash B 2012 {\em Phys. Rev. Lett.\/} {\bf
  109} 141102 [\eprint{1207.0399}]

\bibitem{Berti:2006cc}
Berti E, Cardoso V and Will C~M 2006 {\em AIP Conf. Proc.\/} {\bf 873} 82--88

\bibitem{Barausse:2009uz}
Barausse E and Rezzolla L 2009 {\em Astrophys. J. Lett.\/} {\bf 704} L40--L44
  [\eprint{0904.2577}]

\bibitem{O'Shaughnessy:2012ay}
O'Shaughnessy R, London L, Healy J and Shoemaker D 2013 {\em Phys. Rev.\/} {\bf
  D87} 044038 [\eprint{1209.3712}]

\bibitem{2013PhRvD..88b4040P}
{Pekowsky} L, {O'Shaughnessy} R, {Healy} J and {Shoemaker} D 2013 {\em \prd\/}
  {\bf 88} 024040 [\eprint{1304.3176}]

\bibitem{London:2014cma}
London L, Shoemaker D and Healy J 2014 {\em Phys. Rev.\/} {\bf D90} 124032
  [\eprint{1404.3197}]

\bibitem{Yunes:2009ke}
Yunes N and Pretorius F 2009 {\em Phys. Rev.\/} {\bf D80} 122003
  [\eprint{0909.3328}]

\bibitem{Vallisneri:2013rc}
Vallisneri M and Yunes N 2013 {\em Phys. Rev.\/} {\bf D87} 102002
  [\eprint{1301.2627}]

\bibitem{Li:2011vx}
Li T~G~F, Del~Pozzo W, Vitale S, Van Den~Broeck C, Agathos M {\em et~al.\/}
  2012 {\em J. Phys. Conf. Ser.\/} {\bf 363} 012028 [\eprint{1111.5274}]

\bibitem{VanDenBroeck2014}
{Van Den Broeck} C 2014 {\em Probing dynamical spacetimes with gravitational
  waves\/} Springer Handbook of Spacetime (Springer Verlag)

\bibitem{Agathos:2013oma}
Agathos M, Del~Pozzo W, Li T~G~F, Van Den~Broeck C, Veitch J and Vitale S 2015
  {Testing general relativity using gravitational waves from binary neutron
  stars: Effect of spins} {\em {Proceedings, 13th Marcel Grossmann Meeting on
  Recent Developments in Theoretical and Experimental General Relativity,
  Astrophysics, and Relativistic Field Theories (MG13)}\/} pp 1710--1712
  [\eprint{1305.2963}]

\bibitem{2014PhRvD..89h2001A}
{Agathos} M, {Del Pozzo} W, {Li} T~G~F, {Van Den Broeck} C, {Veitch} J and
  {Vitale} S 2014 {\em \prd\/} {\bf 89} 082001 [\eprint{1311.0420}]

\bibitem{DelPozzo:2014cla}
Del~Pozzo W, Grover K, Mandel I and Vecchio A 2014 {\em Class. Quantum Grav.\/}
  {\bf 31} 205006 [\eprint{1408.2356}]

\bibitem{2010PhRvD..82h2002Y}
{Yunes} N and {Hughes} S~A 2010 {\em \prd\/} {\bf 82} 082002
  [\eprint{1007.1995}]

\bibitem{1994CQGra..11.2807B}
{Blanchet} L and {Sathyaprakash} B~S 1994 {\em Class. Quantum Grav.\/} {\bf 11}
  2807--2831

\bibitem{2014PhRvD..89b2002V}
{Vitale} S and {Del Pozzo} W 2014 {\em \prd\/} {\bf 89} 022002
  [\eprint{1311.2057}]

\bibitem{2014PhRvD..89f4037S}
{Sampson} L, {Cornish} N and {Yunes} N 2014 {\em \prd\/} {\bf 89} 064037
  [\eprint{1311.4898}]

\bibitem{2012PhRvD..85f4034V}
{Vitale} S, {Del Pozzo} W, {Li} T~G~F, {Van Den Broeck} C, {Mandel} I, {Aylott}
  B and {Veitch} J 2012 {\em \prd\/} {\bf 85} 064034 [\eprint{1111.3044}]

\bibitem{Vitale:2014owa}
Vitale S, Del~Pozzo W, Li T, Van Den~Broeck C, Aylott B {\em et~al.\/} 2014
  {\em J. Phys. Conf. Ser.\/} {\bf 484} 012026

\bibitem{Favata:2013rwa}
Favata M 2014 {\em Phys. Rev. Lett.\/} {\bf 112} 101101 [\eprint{1310.8288}]

\bibitem{2014ApJ...784...71S}
{Samsing} J, {MacLeod} M and {Ramirez-Ruiz} E 2014 {\em \apj\/} {\bf 784} 71
  [\eprint{1308.2964}]

\bibitem{O'Leary:2007qa}
O'Leary R, O'Shaughnessy R and Rasio F 2007 {\em Phys. Rev.\/} {\bf D76} 061504
  [\eprint{astro-ph/0701887}]

\bibitem{2011MNRAS.416..133D}
{Downing} J~M~B, {Benacquista} M~J, {Giersz} M and {Spurzem} R 2011 {\em
  \mnras\/} {\bf 416} 133--147 [\eprint{1008.5060}]

\bibitem{Morscher:2012se}
Morscher M, Umbreit S, Farr W~M and Rasio F~A 2013 {\em Astrophys. J.\/} {\bf
  763} L15 [\eprint{1211.3372}]

\bibitem{2014ApJ...781...45A}
{Antonini} F, {Murray} N and {Mikkola} S 2014 {\em \apj\/} {\bf 781} 45
  [\eprint{1308.3674}]

\bibitem{2014MNRAS.439.1079A}
{Antognini} J~M, {Shappee} B~J, {Thompson} T~A and {Amaro-Seoane} P 2014 {\em
  \mnras\/} {\bf 439} 1079--1091 [\eprint{1308.5682}]

\bibitem{DelPozzo:2013ala}
Del~Pozzo W, Li T~G~F, Agathos M, Van Den~Broeck C and Vitale S 2013 {\em Phys.
  Rev. Lett.\/} {\bf 111} 071101 [\eprint{1307.8338}]

\bibitem{Abadie:2010cf}
Abadie J {\em et~al.\/} (LIGO Scientific) 2010 {\em Class. Quantum Grav.\/}
  {\bf 27} 173001 [\eprint{1003.2480}]

\bibitem{Dominik:2014yma}
Dominik M, Berti E, O'Shaughnessy R, Mandel I, Belczynski K, Fryer C, Holz D,
  Bulik T and Pannarale F 2015 {\em Astrophys. J.\/} {\bf 806} 263
  [\eprint{1405.7016}]

\bibitem{Damour:2012yf}
Damour T, Nagar A and Villain L 2012 {\em Phys. Rev.\/} {\bf D85} 123007
  [\eprint{1203.4352}]

\bibitem{Wade:2014vqa}
Wade L, Creighton J~D~E, Ochsner E, Lackey B~D, Farr B~F {\em et~al.\/} 2014
  {\em Phys. Rev.\/} {\bf D89} 103012 [\eprint{1402.5156}]

\bibitem{Poisson:1997ha}
Poisson E 1998 {\em Phys. Rev.\/} {\bf D57} 5287--5290 [\eprint{gr-qc/9709032}]

\bibitem{Mikoczi:2005dn}
Mikoczi B, Vasuth M and Gergely L~A 2005 {\em Phys. Rev.\/} {\bf D71} 124043
  [\eprint{astro-ph/0504538}]

\bibitem{Kidder:1992fr}
Kidder L~E, Will C~M and Wiseman A~G 1993 {\em Phys. Rev.\/} {\bf D47}
  4183--4187 [\eprint{gr-qc/9211025}]

\bibitem{Arun:2008kb}
Arun K~G, Buonanno A, Faye G and Ochsner E 2009 {\em Phys. Rev.\/} {\bf D79}
  104023 [\eprint{0810.5336}]

\bibitem{Flanagan:2007ix}
Flanagan {\'E}~{\'E} and Hinderer T 2008 {\em Phys. Rev.\/} {\bf D77} 021502
  [\eprint{0709.1915}]

\bibitem{Read:2009yp}
Read J~S, Markakis C, Shibata M, Uryu K, Creighton J~D {\em et~al.\/} 2009 {\em
  Phys. Rev.\/} {\bf D79} 124033 [\eprint{0901.3258}]

\bibitem{Lackey:2011vz}
Lackey B~D, Kyutoku K, Shibata M, Brady P~R and Friedman J~L 2012 {\em Phys.
  Rev.\/} {\bf D85} 044061 [\eprint{1109.3402}]

\bibitem{Lackey:2013axa}
Lackey B~D, Kyutoku K, Shibata M, Brady P~R and Friedman J~L 2014 {\em Phys.
  Rev.\/} {\bf D89} 043009 [\eprint{1303.6298}]

\bibitem{Read:2013zra}
Read J~S, Baiotti L, Creighton J~D~E, Friedman J~L, Giacomazzo B {\em et~al.\/}
  2013 {\em Phys. Rev.\/} {\bf D88} 044042 [\eprint{1306.4065}]

\bibitem{Yagi:2013baa}
Yagi K and Yunes N 2014 {\em Phys. Rev.\/} {\bf D89} 021303
  [\eprint{1310.8358}]

\bibitem{Lattimer:2010zz}
Lattimer J~M 2010 {\em Prog. Theor. Phys. Suppl.\/} {\bf 186} 1--8

\bibitem{Yunes:2009yz}
Yunes N, Arun K, Berti E and Will C~M 2009 {\em Phys. Rev.\/} {\bf D80} 084001
  [\eprint{0906.0313}]

\bibitem{1995PhRvD..52.2089K}
{Kr{\'o}lak} A, {Kokkotas} K~D and {Sch{\"a}fer} G 1995 {\em \prd\/} {\bf 52}
  2089--2111 [\eprint{gr-qc/9503013}]

\bibitem{2011PhRvD..84l4007G}
{Gopakumar} A and {Sch{\"a}fer} G 2011 {\em \prd\/} {\bf 84} 124007

\bibitem{Huerta:2014eca}
Huerta E, Kumar P, McWilliams S~T, O'Shaughnessy R and Yunes N 2014 {\em Phys.
  Rev.\/} {\bf D90} 084016 [\eprint{1408.3406}]

\bibitem{2014PhRvD..89d4021L}
{Lundgren} A and {O'Shaughnessy} R 2014 {\em \prd\/} {\bf 89} 044021
  [\eprint{1304.3332}]

\bibitem{Berti:2004bd}
Berti E, Buonanno A and Will C~M 2005 {\em Phys. Rev.\/} {\bf D71} 084025
  [\eprint{gr-qc/0411129}]

\bibitem{Buonanno:2009zt}
Buonanno A, Iyer B, Ochsner E, Pan Y and Sathyaprakash B~S 2009 {\em Phys.
  Rev.\/} {\bf D80} 084043 [\eprint{0907.0700}]

\bibitem{2014PhRvD..89f1501P}
{Pan} Y, {Buonanno} A, {Taracchini} A, {Boyle} M, {Kidder} L~E, {Mrou{\'e}}
  A~H, {Pfeiffer} H~P, {Scheel} M~A, {Szil{\'a}gyi} B and {Zenginoglu} A 2014
  {\em \prd\/} {\bf 89} 061501 [\eprint{1311.2565}]

\bibitem{Sturani:2010yv}
Sturani R, Fischetti S, Cadonati L, Guidi G~M, Healy J, Shoemaker D~M and
  Vicer{\'e} A 2010 {\em J. Phys.: Conf. Ser.\/} {\bf 243} 012007
  [\eprint{1005.0551}]

\bibitem{Sturani:2010ju}
Sturani R, Fischetti S, Cadonati L, Guidi G, Healy J {\em et~al.\/} 2010
  [\eprint{1012.5172}]

\bibitem{Hannam:2013oca}
Hannam M, Schmidt P, Bohé A, Haegel L, Husa S {\em et~al.\/} 2014 {\em Phys.
  Rev. Lett.\/} {\bf 113} 151101 [\eprint{1308.3271}]

\bibitem{Schmidt:2014iyl}
Schmidt P, Ohme F and Hannam M 2015 {\em Phys. Rev.\/} {\bf D91} 024043
  [\eprint{1408.1810}]

\bibitem{Barausse:2014pra}
Barausse E, Cardoso V and Pani P 2015 {\em J. Phys. Conf. Ser.\/} {\bf 610}
  012044 [\eprint{1404.7140}]

\bibitem{Barausse:2007dy}
Barausse E and Rezzolla L 2008 {\em \prd\/} {\bf 77} 104027
  [\eprint{0711.4558}]

\bibitem{Barausse:2007ph}
Barausse E 2007 {\em Mon. Not. R. Astron. Soc.\/} {\bf 382} 826--834
  [\eprint{0709.0211}]

\bibitem{Yunes:2011ws}
Yunes N, Kocsis B, Loeb A and Haiman Z 2011 {\em Phys. Rev. Lett.\/} {\bf 107}
  171103 [\eprint{1103.4609}]

\bibitem{Kocsis:2011dr}
Kocsis B, Yunes N and Loeb A 2011 {\em \prd\/} {\bf 84} 024032
  [\eprint{1104.2322}]

\bibitem{Gondolo:1999ef}
Gondolo P and Silk J 1999 {\em Phys. Rev. Lett.\/} {\bf 83} 1719--1722
  [\eprint{astro-ph/9906391}]

\bibitem{Merritt:2002vj}
Merritt D, Milosavljevic M, Verde L and Jimenez R 2002 {\em Phys. Rev. Lett.\/}
  {\bf 88} 191301 [\eprint{astro-ph/0201376}]

\bibitem{Bertone:2005hw}
Bertone G and Merritt D 2005 {\em Phys. Rev.\/} {\bf D72} 103502
  [\eprint{astro-ph/0501555}]

\bibitem{Ullio:2001fb}
Ullio P, Zhao H and Kamionkowski M 2001 {\em Phys. Rev.\/} {\bf D64} 043504
  [\eprint{astro-ph/0101481}]

\bibitem{Bertone:2005xz}
Bertone G, Zentner A~R and Silk J 2005 {\em Phys. Rev.\/} {\bf D72} 103517
  [\eprint{astro-ph/0509565}]

\bibitem{Leung:1999rh}
Leung P~T, Liu Y~T, Suen W~M, Tam C~Y and Young K 1997 {\em Phys. Rev. Lett.\/}
  {\bf 78} 2894--2897 [\eprint{gr-qc/9903031}]

\bibitem{Almheiri:2013hfa}
Almheiri A, Marolf D, Polchinski J, Stanford D and Sully J 2013 {\em JHEP\/}
  {\bf 1309} 018 [\eprint{1304.6483}]

\bibitem{Braunstein:2009my}
Braunstein S~L, Pirandola S and Zyczkowski K 2013 {\em Phys. Rev. Lett.\/} {\bf
  110} 101301 [\eprint{0907.1190}]

\bibitem{Abbott:2009km}
Abbott B~P {\em et~al.\/} (LIGO Scientific) 2009 {\em Phys. Rev.\/} {\bf D80}
  062001 [\eprint{0905.1654}]

\bibitem{Aasi:2014bqj}
Aasi J {\em et~al.\/} (LIGO Scientific Collaboration, VIRGO Collaboration) 2014
  {\em Phys. Rev.\/} {\bf D89} 102006 [\eprint{1403.5306}]

\bibitem{TestGR14}
{Webpage of the workshop ``Testing General Relativity with Astrophysical
  Observations'' (Oxford, MS, January 6--10 2014): \\
  \url{http://www.phy.olemiss.edu/TestGR2014/} }

\end{thebibliography}

\end{document}